\documentclass[rmp,aps,floatfix,twocolumn,unsortedaddress]{revtex4}
\usepackage{epsfig}
\begin{document}


\title{Spintronics: Fundamentals and applications}

\author{Igor \v{Z}uti\'{c}\footnote{Electronic address: igorz@physics.umd.edu.
Present address: Center for Computational Materials Science,
Naval Research Laboratory, Washington, D.C. 20375, USA}}
\affiliation{Condensed Matter Theory Center,
Department of Physics, University of Maryland at College
Park, College Park, Maryland 20742-4111, USA}

\author{Jaroslav Fabian\footnote{Electronic address: 
jaroslav.fabian@uni.graz.at}}
\affiliation{Institute for Theoretical Physics, Karl-Franzens University,
Universit\"atsplatz 5, 8010 Graz, Austria }

\author{S. Das Sarma}
\affiliation{Condensed Matter Theory Center,
Department of Physics, University of Maryland at College
Park, College Park, Maryland 20742-4111, USA}

\begin{abstract}
{Spintronics, or spin electronics, involves the study of active control
and manipulation of spin degrees of freedom in solid-state systems.
This article reviews
the current status of this subject, including both recent advances
and well-established results.
The primary focus is on the basic physical principles underlying the
generation of carrier spin polarization, spin dynamics, and spin-polarized
transport in semiconductors and metals.
Spin transport differs from charge transport in that spin is
a nonconserved quantity in solids due to spin-orbit and hyperfine coupling.
The authors discuss in detail spin decoherence mechanisms in metals and 
semiconductors. Various theories of spin injection
and spin-polarized transport are applied to hybrid structures relevant to
spin-based devices and fundamental studies of materials properties.
Experimental work is reviewed with the emphasis on projected applications, 
in which
external electric and magnetic fields and illumination by light
will be used to control spin and charge dynamics to create new functionalities 
not feasible or ineffective with conventional electronics.}
\end{abstract}
\maketitle

\setcounter{tocdepth}{4}
\tableofcontents

\section{\label{sec:I} Introduction}

\subsection{\label{sec:IA} Overview}

Spintronics is a multidisciplinary field
whose central theme is the
active manipulation of spin degrees of freedom in 
solid-state systems.\footnote{While there are proposals for spintronic
devices based on deoxyribonucleic acid (DNA) molecules
\cite{Zwolak2002:APL}, the whole device, which includes 
electrodes, voltage/current source, etc., is still a solid-state system.}
In this article the term spin stands for either the spin of a single
electron $\bf s$, which can be detected by its magnetic moment 
$-g \mu_B {\bf s}$
($\mu_B$ is the Bohr magneton and $g$ is the electron $g$ factor, in a
solid generally different from the free electron value of 
$g_0=2.0023$),
or the average spin of an ensemble of electrons, manifested by magnetization.
The control of spin is then a control of either the population and
the phase of the spin of an ensemble of particles, or a coherent
spin manipulation of a single or a few-spin system.
The goal of spintronics is to understand the interaction
between the particle spin and its solid-state environments
and to make useful devices using the acquired knowledge. 
Fundamental studies of spintronics include investigations of spin 
transport in electronic materials, as well as 
understanding spin dynamics and spin relaxation. 
Typical questions that are posed are 
(a) what is an effective way to polarize
a spin system?
(b) how long is the system able to remember its spin
orientation? (c) how 
can spin be detected? 

Generation of spin polarization usually means creating a nonequilibrium
spin population. This can be achieved in several ways. While
traditionally spin has been oriented using optical techniques in which 
circularly polarized photons transfer their angular momenta
to electrons, for device applications electrical spin injection
is more desirable. In electrical spin injection a 
magnetic electrode is connected
to the sample. When the current drives spin-polarized electrons
from the electrode to the sample, nonequilibrium spin accumulates
there. The rate of spin accumulation depends
on spin relaxation, 
the
process of bringing the accumulated spin population 
back to equilibrium. There are several relevant mechanisms
of spin relaxation, most involving spin-orbit coupling to  provide
the spin-dependent potential, in combination with momentum
scattering providing a randomizing force. Typical time scales
for spin relaxation in electronic systems are 
measured in 
nanoseconds, 
while the range is from pico to microseconds.
Spin detection, also 
part of a generic spintronic scheme, typically
relies on sensing the changes in the signals caused by the
presence of nonequilibrium spin in the system. The common
goal in many spintronic devices is to maximize the 
spin detection sensitivity  to the point it detects 
not the spin itself, but changes in the spin states.  

\begin{figure}
\centerline{\psfig{file=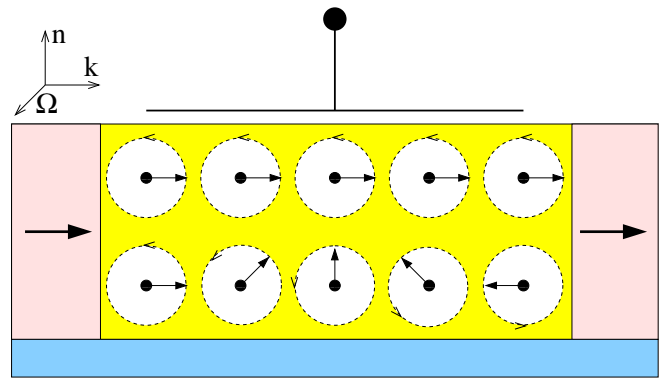,width=1\linewidth,angle=0}}
\caption{Scheme of the Datta-Das spin field-effect transistor (SFET). 
The source (spin injector) and the drain (spin detector) are ferromagnetic 
metals or semiconductors,
with parallel magnetic moments. 
The injected spin-polarized electrons with wave vector ${\bf k}$ 
move ballistically
along a quasi-one-dimensional channel formed by, for example, 
an InGaAs/InAlAs heterojunction in a plane normal to ${\bf n}$. Electron spins
precess about the precession vector ${\bf \Omega}$, which arises from  
spin-orbit coupling and which is defined by the structure and 
the materials properties of the channel. The magnitude
of ${\bf \Omega}$ is tunable by the gate voltage $V_G$ 
at the top of the channel.  The current is large if the electron 
spin at the drain points in the initial direction (top row), for
example, if the precession period is much larger than the 
time of flight, and small if the direction is reversed (bottom).
}
\label{fig:DD}
\end{figure}

Let us illustrate the generic spintronic scheme on a prototypical
device, the Datta-Das spin field effect
transistor (SFET) \cite{Datta1990:APL}, depicted in Fig.~\ref{fig:DD}. 
The scheme shows the structure of the usual FET, 
with a drain, a source, a narrow channel, and a gate for controlling the 
current.
The gate either allows 
the current to flow (ON) or does not (OFF). The
spin transistor is similar in that the result is also a
control of the charge current through the narrow channel. 
The difference, however, is in the physical realization
of the current control. In the Datta-Das SFET the source
and the drain are ferromagnets
acting as the injector and detector of the electron spin. 
The drain injects electrons with spins parallel to the
transport direction. The electrons are transported ballistically
through the channel. When they arrive at the drain, their spin is 
detected. In a simplified picture, the electron can enter 
he drain (ON) if its spin 
points in the same direction as the spin of the drain. Otherwise
it is scattered away (OFF). The role of the gate is to generate an
effective magnetic field 
(in the direction of $\bf\Omega$ in Fig.~\ref{fig:DD}), 
arising from the 
spin-orbit coupling in the substrate material,
from the confinement geometry of the transport channel, and the
electrostatic potential of the gate.
This 
effective magnetic field causes
the electron spins to precess. By modifying the voltage, one can cause
the precession to lead to either parallel or antiparallel 
(or anything between) electron spin at the drain, effectively
controlling the current. 

Even though the name {\it spintronics} is rather novel,\footnote{The term
was coined by S. A. Wolf in 1996, as a name for a 
DARPA initiative for novel magnetic materials and devices.} 
contemporary research in spintronics relies closely on a long
tradition of results obtained in diverse areas of physics (for
example, magnetism, semiconductor physics, superconductivity, optics,
and mesoscopic physics) and establishes new connections between its
different subfields \cite{Rashba2002:JS,Zutic2002:JS}.
We review here both well-established results and the physical
principles relevant to the present and future applications. 
Our strategy is to give a comprehensive view of what
has been accomplished, focusing in detail on a few
selected topics that we believe are representative for
the broader subject within which they appear. For example, 
while discussing the generation of spin polarization, we survey
many experimental and theoretical studies of both optical orientation
and electrical spin injection and present a detailed and
self-contained formalism of electrical spin injection. Similarly,
when we discuss spin relaxation, we give a catalog 
of important work, while studying spin relaxation in 
the cases of Al and GaAs as representative of the whole field.
Finally, in the section on spin devices we give detailed
physical principles of several selected devices, such as, 
for example, the above-mentioned Datta-Das SFET.

There have been many other reviews written on spintronics, 
most focusing on a particular aspect of the field. We divide
them here, for an easier orientation, into two groups,
those that cover the emerging applications\footnote{Reviews on
emerging application include those of 
\cite{Wolf2000:IEEE,
DasSarma2000:SM,DasSarma2000:IEEE,DasSarma2000:DRC,%
DasSarma2001:SSC,DasSarma2001:AS,Zutic2002:JS,Rashba2002:JS,Wolf2001:S,%
Oestreich2002:SST,Zutic2002:PROC}.} 
and those covering already well-established schemes 
and materials\footnote{Established schemes and materials are
reviewed by 
\cite{Prinz1995:PT,Prinz1998:S,Gregg1997:JMMM,Bass1999:JMMM,%
Ansermet1998:JPCM,Gijs1997:AP,Tedrow1994:PR,Daughton1999:JPDAP,Stiles2003:P}.}
The latter group, often described as {\it magnetoelectronics} 
typically covers paramagnetic and ferromagnetic metals and insulators,
which utilize magnetoresistive effects, 
realized, for example, as
magnetic read heads in computer hard drives, nonvolatile magnetic
random access memory (MRAM), and circuit isolators
\cite{Wang2002:JAP}. These more established aspects of spintronics
have been also addressed in several books\footnote{See, for example, the
books of 
\cite{Hartmann:2000,Parkin:2002,Shinjo:2002,Maekawa:2002,Hirota:2002,%
Levy:2002,Chtchelkanova:2003,Ziese:2001}}
and will be discussed in another review,\footnote{In preparation
by S. S. P. Parkin for Review Modern Physics.} complementary to
ours.

Spintronics also benefits from a large class of emerging materials, such as
ferromagnetic semiconductors \cite{Ohno1998:S,Pearton2003:JAP},
organic semiconductors \cite{Dediu2002:SSC}, organic ferromagnets 
\cite{Pejakovic2002:PRL,Epstein2003:MRS},
high temperature superconductors \cite{Goldman1999:JMMM},
and carbon nanotubes \cite{Tsukagoshi1999:N,Zhao2002:APL},
which can bring novel functionalities to the traditional devices.
There is a continuing need for fundamental studies before the
potential of spintronic applications is fully realized.

After an overview, Sec.~\ref{sec:I} covers some basic historical and 
background material, part of which has already  been extensively covered 
in the context of magnetoelectronics
and will not be discussed further in this review. Techniques 
for generating spin polarization, focusing on optical spin orientation and
electrical spin injection, are described in Sec.~\ref{sec:II}. 
The underlying mechanisms responsible for the
loss of spin orientation and coherence, which impose fundamental limits on 
the length  and time scales in spintronic devices, are addressed in 
Sec.~\ref{sec:III}.
Spintronic applications and devices, with the emphasis on 
those based on semiconductors, are discussed in Sec.~\ref{sec:IV}.
The review concludes with a look at future prospects
in Sec.~\ref{sec:V}
and with the table (Tab. II) listing the most common abbreviations 
used in the text.

\subsection{\label{sec:IB} History and background}

\subsubsection{\label{sec:IB1} 
Spin-polarized transport and magnetoresistive effects}

In a pioneering work, \textcite{Mott1936:PRCa,Mott1936:PRCb} provided
a basis for our understanding of spin-polarized transport.  Mott
sought an explanation for an unusual behavior of resistance in
ferromagnetic metals. He realized that at sufficiently low
temperatures, where magnon scattering becomes vanishingly small,
electrons of majority and minority spin, with magnetic moment parallel
and antiparallel to the magnetization of a ferromagnet, respectively,
do not mix in the scattering processes. The conductivity can then be
expressed as the sum of two independent and unequal parts for two
different spin projections--the current in ferromagnets is
spin polarized.  This is also known as the two-current model and has
been extended by \textcite{Campbell1967:PM,Fert1968:PRL}. It 
continues, in its modifications, to provide an explanation for various
magnetoresistive phenomena \cite{Valet1993:PRB}.

Tunneling measurements played a key role in early experimental work on
spin-polarized transport. Studying N/F/N junctions, where N was a
nonmagnetic\footnote{Unless explicitly specified, we shall use the
terms ``nonmagnetic'' and ``paramagnetic'' interchangeably, i.e.,
assume that they both refer to a material with no long-range
ferromagnetic order and with Zeeman-split carrier spin subbands in
an applied magnetic field.} metal and F was an Eu-based ferromagnetic
semiconductor \cite{Kasuya1968:RMP,Nagaev:1983}, revealed that I-V
curves could be modified by an applied magnetic field
\cite{Esaki1967:PRL} and show potential for developing a solid-state
spin-filter. When unpolarized current is passed across a 
ferromagnetic semiconductor, the current becomes spin-polarized
\cite{Moodera1988:PRL,Hao1990:PRB}.

\begin{figure}
\centerline{\psfig{file=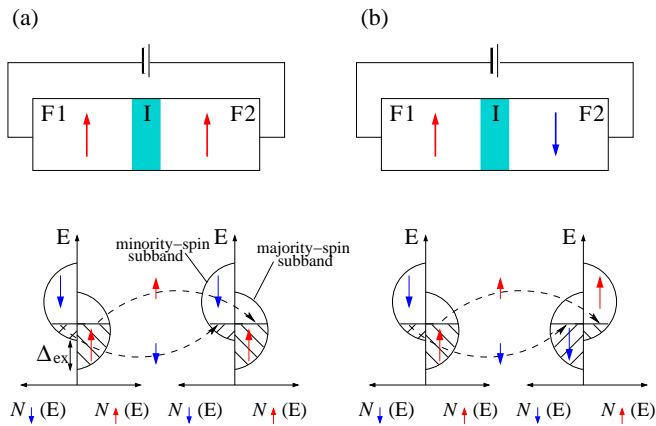,width=\linewidth,angle=0}}
\caption{Schematic illustration of electron tunneling in
ferromagnet/insulator/ferromagnet (F/I/F) tunnel junctions: 
(a) Parallel and (b) antiparallel orientation of magnetizations 
with the corresponding spin-resolved density of the d states
in ferromagnetic metals that have exchange spin splitting $\Delta_{ex}$.
Arrows in the two ferromagnetic regions are determined by the
majority-spin subband. Dashed lines depict spin-conserved tunneling.
}
\label{intro:1}
\end{figure}

A series of experiments
\cite{Tedrow1971:PRLa,Tedrow1973:PRB,Tedrow1994:PR} in
ferromagnet/insulator/superconductor (F/I/S) junctions has unambiguously
proved that the tunneling current remains spin-polarized even outside
of the ferromagnetic region.\footnote{It has been shown
that electrons photoemitted from ferromagnetic gadolinium remain 
spin polarized \cite{Busch1969:PRL}.}
The Zeeman-split quasiparticle density of
states in a superconductor \cite{Tedrow1970:PRL,Fulde1973:AP}
was used as a detector of spin polarization of conduction electrons in
various magnetic materials.  
\textcite{Julliere1975:PL} measured  
tunneling conductance of F/I/F junctions, where I was 
an amorphous Ge. 
By adopting the
\textcite{Tedrow1971:PRLa,Tedrow1973:PRB} analysis 
of the tunneling
conductance from F/I/S to the F/I/F junctions,  
\textcite{Julliere1975:PL} formulated a model for a change of
conductance between the parallel ($\uparrow \uparrow$) and
antiparallel ($\uparrow \downarrow$) magnetization in the two
ferromagnetic regions F1 and F2, as depicted in Fig.~\ref{intro:1}.
The corresponding tunneling magnetoresistance\footnote{Starting with
\textcite{Julliere1975:PL} an equivalent expression $(G_{\uparrow
\uparrow}-G_{\uparrow \downarrow})/G_{\uparrow \uparrow}$ has also also
used by different authors and often referred to as junction
magnetoresistance (JMR) \cite{Moodera1999:JMMM}.}  
(TMR) in an 
F/I/F 
magnetic tunnel   
junction (MTJ) is defined as 
\begin{equation}
TMR= \frac{\Delta R}{R_{\uparrow \uparrow}}
=\frac{R_{\uparrow \downarrow}-R_{\uparrow \uparrow}}{R_{\uparrow \uparrow}}  
=\frac{G_{\uparrow \uparrow}-G_{\uparrow \downarrow}}{G_{\uparrow \downarrow}}, 
\label{eq:tmr}
\end{equation}
where conductance G and resistance R=1/G are labeled by the relative
orientations of the magnetizations in F1 and F2 (it is possible
to change the relative orientations, between $\uparrow \uparrow$ and
$\uparrow \downarrow$, even at small applied magnetic fields $\sim$ 10
G). 
TMR is a particular manifestation of a magnetoresistance (MR) that 
yields a change of electrical resistance in the presence of an  
external magnetic field.\footnote{The concept of TMR was proposed
independently by R. C. Barker in 1975 [see \textcite{Meservey1983:JMMM}] 
and by \textcite{Slonczewski1976:IBM}, 
who envisioned its use for 
magnetic bubble memory \cite{Parkin:2002}.}
Historically, the anisotropic
MR in bulk ferromagnets such as Fe and Ni was discovered
first, dating back the to experiments of Lord Kelvin
\cite{Thomson1857:PRSL}.  Due to spin-orbit interaction, electrical
resistivity changes with the relative direction of the charge current
(for example, parallel or perpendicular) with respect to the direction
of magnetization.

Within Julli{\`{e}}re's model, which assumes constant tunneling matrix elements
and that electrons tunnel without spin flip, Eq.~(\ref{eq:tmr}) yields
\begin{equation}
TMR=\frac{2P_1 P_2}{1-P_1 P_2},
\label{eq:julliere}
\end{equation}
where the polarization $P_i=({\cal N}_{M i} - {\cal N}_{m i})/
({\cal N}_{M i} + {\cal N}_{m i})$
is expressed in terms of
the spin-resolved density of states ${\cal N}_{M i}$ and 
${\cal N}_{m i}$, for majority and minority spin in F$_i$, respectively. 
Conductance in Eq.~(\ref{eq:tmr}) can then 
be expressed as
\cite{Maekawa1982:IEEE} $G_{\uparrow \uparrow} \sim {\cal N}_{M 1}
{\cal N}_{M 2}+ {\cal N}_{m 1} {\cal N}_{m 2}$ and $G_{\uparrow
\downarrow} \sim {\cal N}_{M 1} {\cal N}_{m 2}+ {\cal N}_{m 1}
{\cal N}_{M 2}$ to give Eq.~(\ref{eq:julliere}).\footnote{In
~\ref{sec:IV} we address some limitations of the Julli{\`{e}}re's
model and its potential ambiguities to identify precisely which spin
polarization is actually measured.}  While the early results of
\textcite{Julliere1975:PL} were not confirmed, TMR at 4.2 K was
observed using NiO as a tunnel barrier by \textcite{Maekawa1982:IEEE}.

\begin{figure}
\centerline{\psfig{file=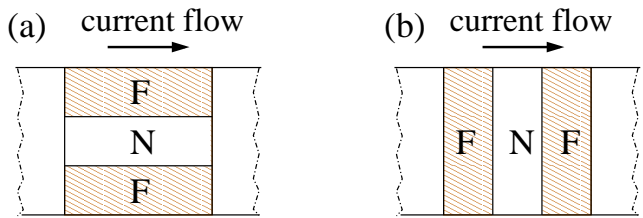,width=0.33\linewidth,angle=-90}}
\caption{Schematic illustration of (a) the current in plane (CIP)
(b) the current perpendicular to the plane (CPP) giant magnetoresistance
geometry.} 
\label{gmr:1}
\end{figure}

The prediction of Julli{\`{e}}re's model illustrates the spin-valve effect:
the resistance of a device can be changed by manipulating the relative
orientation of the magnetizations {\bf M$_1$} and {\bf M$_2$}, in
F1 and F2, respectively.  
Such  orientation 
can be preserved even in the absence of a power
supply and the spin-valve effect,\footnote{The term was coined by
\textcite{Dieny1991:PRB} in the context of GMR,
by invoking an analogy with the physics of the TMR.}  later
discovered in multilayer structures displaying the giant
magnetoresistance (GMR) effect,\footnote{The term ``giant'' reflected the
magnitude of the effect 
(more than $\sim 10$ \%), as compared to the better known anisotropic
magnetoresistance ($\sim 1$ \%).} \cite{Baibich1988:PRL,Binasch1989:PRB} can be
used for nonvolatile memory applications
\cite{Hartmann:2000,Parkin:2002,Hirota:2002}. GMR structures are often
classified according whether the current flows parallel (CIP) or
perpendicular (CPP) to the interfaces between the different layers,
as depicted in Fig.~\ref{gmr:1}.  Most of the GMR applications use the
CIP geometry, while the CPP version, first realized by
\cite{Pratt1991:PRL}, is easier to analyze theoretically
\cite{Gijs1997:AP,Levy:2002} and relates to the physics of the TMR
effect \cite{Mathon1997:PRB}. The size of magnetoresistance in the GMR
structures can be expressed analogously to  Eq.~(\ref{eq:tmr}), where
parallel and antiparallel orientations of the magnetizations in the
two ferromagnetic regions are often denoted by ``P'' and ``AP,''
respectively (instead of $\uparrow \uparrow$ and $\uparrow
\downarrow$). 
Realization of a large room temperature GMR 
\cite{Parkin1991:PRL,Parkin1991:APL} enabled a quick transition from basic 
physics to commercial applications in magnetic recording \cite{Parkin2003:PIEEE}.

One of the keys to the success of the
MR-based-applications is their ability to 
control\footnote{For example, with small magnetic field \cite{Parkin:2002} 
or at high
switching speeds \cite{Schumacher2003:PRLa,Schumacher2003:PRLb}.} the
relative orientation of {\bf M$_1$} and {\bf M$_2$}. An interesting
realization of such control was proposed independently by 
\textcite{Berger1996:PRB} and \textcite{Slonczewski1996:JMMM}. 
While in GMR or TMR structures the relative orientation of magnetizations
will affect the flow of spin-polarized current, they predicted
a reverse effect. The flow of spin-polarized current can transfer angular
momentum from carriers to ferromagnet and alter the orientation 
of the corresponding magnetization, even in the absence of an applied
magnetic field.
This phenomenon, known as spin-transfer
torque, has since been extensively studied both theoretically and 
experimentally
\cite{Bazily1998:PRB,Tsoi1998:PRLa,Myers1999:S,Sun2000:PRB,
Stiles2002:PRB,Wanital2000:PRB}
and current-induced magnetization reversal was demonstrated at room 
temperature \cite{Katine2000:PRL}. 
It was also shown that the magnetic field generated by passing the
current through a CPP GMR device could produce room temperature
magnetization reversal \cite{Bussman1999:APL}.
In the context of ferromagnetic semiconductors additional control of
magnetization was demonstrated optically (by shining light)
\cite{Koshihara1997:PRL,Oiwa2002:PRL,Boukari2002:PRL} and electrically
(by applying gate voltage)
\cite{Ohno2000:N,Park2002:S,Boukari2002:PRL} to perform switching
between the ferromagnetic and paramagnetic states.

Julli{\`{e}}re's model also justifies the  
continued quest for highly spin-polarized materials -- they
would provide large magnetoresistive effects, desirable for device 
applications. In an extreme case, spins would be completely polarized even in
the absence of magnetic field. Numerical support for the existence of such
materials--the so called half-metallic ferromagnets\footnote{Near the Fermi
level they behave as metals only for one spin, the density of states 
vanishes completely for the other spin.} was provided 
by \textcite{deGroot1983:PRL},
and these materials were reviewed by \textcite{Pickett2001:PT}.
In addition to ferromagnets, such as CrO$_2$ \cite{Soulen1998:S,Parker2002:PRL}
and manganite perovskites \cite{Park1998:N}, there is evidence for high
spin polarization in III-V ferromagnetic semiconductors like 
(Ga,Mn)As \cite{Braden2003:P,Panguluri2003:P}.
The challenge remains
to preserve such spin polarization above room temperature and in junctions
with other materials, since the surface (interface) and bulk magnetic
properties can be significantly different 
\cite{Falicov1990:JMR,Fisher1967:PRL,Mills1971:PRB}. 

While many existing spintronic applications
\cite{Hartmann:2000,Hirota:2002} are based on the GMR effects, the
discovery of large room-temperature TMR
\cite{Moodera1995:PRL,Miyazaki1995:JMMM} has renewed interest in the
study of magnetic tunnel junctions 
which are now the basis for the
several MRAM prototypes\footnote{Realization of the early MRAM
proposals used the effect of anisotropic magnetoresistance
\cite{Pohm1987:IEEETM,Pohm1988:IEEETM}.} 
\cite{Parkin1999:JAP,Tehrani2000:IEEE}. 
Future generations of magnetic read heads are expected to  
use MTJ's instead of CIP GMR.
To improve the switching
performance of related devices it is important to reduce the junction
resistance, which determines the RC time constant of the MTJ cell.
Consequently, semiconductors, which would provide a lower tunneling
barrier than the usually employed oxides, are being investigated both as the
non-ferromagnetic region in MTJ's and as the basis for an
all-semiconductor junction that would demonstrate large TMR at low
temperatures \cite{Tanaka2002:SST,Tanaka2001:PRL}. Another
desirable property of semiconductors has been demonstrated by
the extraordinary large room-temperature MR in hybrid
structures with metals reaching 750 000\% at a magnetic field of 4 T 
\cite{Solin2000:S} which could lead to improved magnetic
read heads \cite{Solin2002:APL,Moussa2003:JAP}. MR effects of
similar magnitude have also been found in hybrid metal/semiconductor
granular films \cite{Akinaga2002:SST}. Another 
approach to obtaining large room-temperature magnetoresistance 
($>100$\% at $B\sim 100$ G)
is to fabricate ferromagnetic regions separated by 
a nanosize contact. For simplicity, such a structure could be thought of 
as the limiting case of the CPP GMR scheme in Fig.~\ref{gmr:1}(b).
This behavior, also known as ballistic magnetoresistance, has already 
been studied in a large number of materials and geometries 
\cite{Garcia1999:PRL,Tatara1999:PRL,Imamura2000:PRL,Chung2002:PRL,%
Bruno1999:PRL,Versluijs2002:PRL}.

\subsubsection{\label{sec:IB2} 
Spin injection and optical orientation}

Many  materials in their ferromagnetic state can have a substantial degree of 
{\it equilibrium} carrier spin polarization. However, as illustrated
in Fig.~\ref{fig:DD}, this alone is 
usually  not sufficient for spintronic applications, which typically 
require current 
flow and/or manipulation of the {\it nonequilibrium} spin 
(polarization).\footnote{Important exceptions are tunneling devices operating 
at low bias
and near {\it equilibrium} spin. Equilibrium polarization 
and the current flow can be potentially realized, for example,
in spin-triplet superconductors and thin-film ferromagnets 
\cite{Konig2001:PRL},
accompanied by dissipationless spin currents. 
Using an analogy with the quantum Hall effect, it has been suggested that the 
spin-orbit interaction could lead to dissipationless spin currents in
hole-doped semiconductors \cite{Murakami2003:P}. 
\textcite{Rashba2003:PRB} has pointed out that similar dissipationless spin
currents in thermodynamic equilibrium, due to spin-orbit interaction,
are not transport currents which could be employed for transporting spins
and spin injection.
It is also 
instructive to compare several earlier proposals that use spin-orbit coupling 
to generate spin currents, discussed in Sec.~\ref{sec:IIA}.
} 
The importance of generating {\it nonequilibrium} spin is not limited to 
device applications; it can also be used as a sensitive spectroscopic
tool to study a wide variety of fundamental properties ranging from spin-orbit 
and hyperfine interactions \cite{Meier:1984} to the pairing symmetry of high 
temperature superconductors 
\cite{Tsuei2000:RMP,Vasko1997:PRL,Wei1999:JAP,Ngai2003:P} 
and the
creation of  spin-polarized beams to measure parity violation in high energy 
physics \cite{Pierce:1984}.

Nonequilibrium spin is the result of some source
of pumping arising from transport, optical, or resonance methods. 
Once the pumping is turned off the spin will return to its equilibrium value.
While for most applications it is desirable to 
have long spin relaxation times, it has been demonstrated that short spin 
relaxation times are useful in the implementation of fast switching 
\cite{Nishikawa1995:APL}. 

\begin{figure}
\centerline{\psfig{file=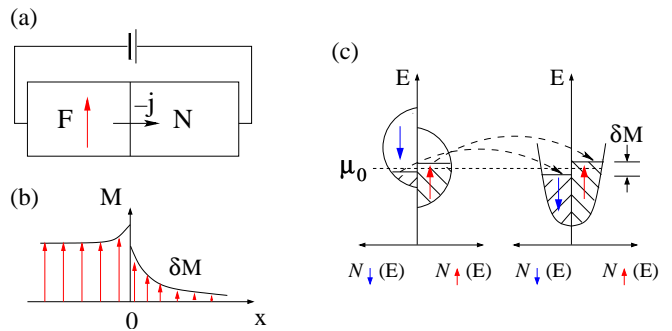,width=\linewidth,angle=0}}
\caption{Pedagogical illustration of the concept of electrical
spin injection from a ferromagnet (F) into a normal metal (N).
Electrons flow from F to N:
(a) schematic device geometry; (b) magnetization M as a function of
position. Nonequilibrium magnetization $\delta M$ (spin accumulation)
is injected into a normal metal;
(c) contribution of 
different spin-resolved densities of states to 
charge and spin transport across the F/N interface.
Unequal filled levels in the density of states depict spin-resolved
electrochemical potentials different from the equilibrium value
$\mu_0$.}
\label{intro:2}
\end{figure}

Electrical spin injection, an example of a transport method for generating 
nonequilibrium spin, has already been realized
experimentally by
\textcite{Clark1963:PRL}, 
who drove a direct current
through a sample of InSb in the presence of
constant applied magnetic filed.
The principle was based on the Feher effect,\footnote{The importance and 
possible applications
of the Feher effect \cite{Feher1959:PRL} to polarize electrons was discussed
by \cite{DasSarma2000:IEEE,Suhl2002:P}.}
in which the hyperfine coupling between the electron and nuclear spins, 
together with
different temperatures representing electron velocity  and  
electron spin populations, is responsible for the dynamical
nuclear polarization \cite{Slichter:1989}.\footnote{Such an effect can be 
thought of as a generalization of the 
Overhauser effect \cite{Overhauser1953b:PR} in which the use of 
a resonant microwave excitation causes 
the spin relaxation of the nonequilibrium  
electron 
population through hyperfine coupling to lead to the spin polarization of 
nuclei.
\textcite{Feher1959:PRL} suggested several other methods, instead 
of microwave excitation, that could 
produce a  
nonequilibrium electron population and yield a dynamical polarization of 
nuclei [see also
\textcite{Weger1963:PR}].} 
Motivated by the work of \textcite{Clark1963:PRL},
\textcite{Tedrow1971:PRLa,Tedrow1973:PRB}, and the principle of
optical orientation \cite{Meier:1984},
\textcite{Aronov1976:JETPL,Aronov1976:SPJETP} 
and \textcite{Aronov1976:SPS} established several key concepts
in electrical spin injection from ferromagnets into metals,
semiconductors\footnote{In an earlier
work spin injection of minority carriers was proposed in a 
ferromagnet/insulator/$p$-type semiconductor structure. Measuring polarization
of electroluminescence was suggested as a technique for detecting injection
of polarized carriers in a semiconductor \cite{Scifres1973:SSC}.}
and superconductors. 
When a charge current flowed across 
the F/N junction (Fig.~\ref{intro:2}) 
\textcite{Aronov1976:JETPL} predicted 
that spin-polarized carriers in a ferromagnet would
contribute to the net current of magnetization entering the nonmagnetic
region and would lead to nonequilibrium magnetization $\delta M$, 
depicted in Fig.~\ref{intro:2}(b), 
with the spatial extent
given by the spin diffusion length
\cite{Aronov1976:SPS,Aronov1976:JETPL}.\footnote{Supporting the findings of 
\textcite{Clark1963:PRL}, Aronov calculated
that the electrical spin injection would polarize nuclei and lead to a 
measurable effect in the electron spin resonance (ESR). Several decades
later related experiments on spin injection are also examining other
implications of dynamical nuclear polarization 
\cite{Johnson2000:APL,Strand2003:PRL}.} 
Such $\delta M$, which is also
equivalent to a {\it nonequilibrium} spin accumulation, was first measured 
in metals by \textcite{Johnson1985:PRL,Johnson1988:PRBb}.
In the steady state $\delta M$ is realized as the balance between spins 
added by 
the magnetization current and spins removed by spin 
relaxation.\footnote{The spin diffusion length 
is an important quantity
for CPP GMR. The thickness of the N region in Fig.~\ref{gmr:1}
should not exceed the spin diffusion length, otherwise the 
information on the orientation of the magnetization in F1 will
not be transferred to the F2 region.}

Generation of nonequilibrium spin polarization and spin accumulation
is also possible by optical methods known as optical orientation or
optical pumping. In optical orientation, the angular momentum of absorbed 
circularly polarized 
light is transferred to the medium. Electron orbital momenta are directly 
oriented by light and through spin-orbit interaction electron spins 
become polarized. In ~\ref{sec:IIB} we focus on the optical orientation in 
semiconductors, a well-established technique \cite{Meier:1984}.
In a pioneering work \textcite{Lampel1968:PRL} demonstrated 
that spins in silicon can be optically oriented (polarized). 
This technique is derived from optical pumping proposed by 
\textcite{Kastler1950:JDP} in which optical irradiation changes the relative
populations within the  Zeeman and the hyperfine levels of the ground states 
of atoms.
While there are similarities with previous studies of free atoms 
\cite{Cohen-Tannoudji:1966,Happer1972:RMP},
optical orientation in semiconductors has important differences related
to the strong coupling between the electron and nuclear
spin and macroscopic number of particles 
\cite{Paget1977:PRB,Hermann1985:APF,Meier:1984}.  
Polarized nuclei can exert large magnetic 
fields ($\sim 5$ T) 
 on electrons. 
In bulk III-V semiconductors, such as GaAs, optical orientation
can lead to 50\% polarization of electron density which could be further 
enhanced 
in quantum structures of reduced dimensionality or by applying a strain.
A simple reversal in the polarization of the illuminating light (from
positive to negative helicity) also reverses the sign of the electron density
polarization. Combining these properties of optical orientation with the 
semiconductors tailored to have a negative electron affinity allows 
photoemission of spin-polarized electrons to be used as a powerful detection 
technique in high-energy physics and for investigating surface 
magnetism \cite{Pierce:1984}.

\section{\label{sec:II} Generation of spin polarization}

\subsection{\label{sec:IIA} Introduction}

Transport, optical, and resonance methods (as well as their combination) 
 have all been used to create nonequilibrium spin. 
After introducing the concept of spin polarization in solid-state systems
we give a pedagogical picture of electrical spin injection and
detection of polarized carriers. While electrical spin injection
and optical orientation will be discussed in more detail later in
this section, we also survey here several other techniques for polarizing
carriers. 

Spin polarization not only of electrons, but also of holes, nuclei, 
and excitations can be defined as 
\begin{equation}
P_X=X_s/X,
\label{eq:polar}
\end{equation}                                                                     
the ratio of the difference 
$X_s=X_\lambda-X_{-\lambda}$ and the sum $X=X_\lambda+X_{-\lambda}$, 
of the spin-resolved   
$\lambda$ components for a particular quantity $X$. 
To avoid ambiguity as to what precisely is meant by spin polarization 
both the choice of the spin-resolved components and the relevant
physical quantity $X$ need to be specified.
Conventionally, $\lambda$ is taken to be $\uparrow$ or $+$
(numerical value +1) for spin up, $\downarrow$ or $-$ (numerical
value -1) for spin down, with respect to the chosen axis of 
quantization.\footnote{For example, 
along the spin angular momentum, applied magnetic field, magnetization, or 
direction of light propagation.}
In ferromagnetic metals it is customary to refer to $\uparrow$ ($\downarrow$) 
as carriers with magnetic moment parallel (antiparallel) to the magnetization 
or, 
equivalently, as carriers with majority (minority) spin \cite{Tedrow1973:PRB}. 
In semiconductors the terms majority and minority usually refer to relative 
populations of the carriers while $\uparrow$ or $+$ and $\downarrow$ or $-$
correspond to the quantum numbers $m_j$ with respect to the $z$-axis taken 
along the direction of the light propagation or along the applied magnetic 
field \cite{Meier:1984,Jonker2003:P}.
It is important to emphasize that both the magnitude and the sign of
the spin polarization in Eq.~(\ref{eq:polar}) depends of the choice of 
$X$, relevant to the detection technique employed, say optical vs. transport 
and bulk vs. surface measurements \cite{Mazin1999:PRL,Jonker2003:P}.
Even 
in the same homogeneous material the measured $P_X$ can vary for
different $X$, and it is crucial to identify which physical 
quantity---charge current, carrier density, conductivity, 
or the density of states---is being measured experimentally.

\begin{figure}
\centerline{\psfig{file=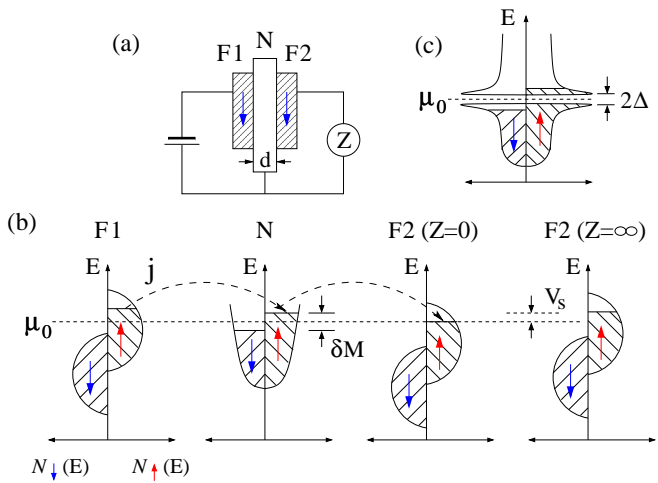,width=\linewidth,angle=0}} 
\caption{Spin injection, spin accumulation, and spin detection:
(a) two idealized completely polarized ferromagnets F1 and F2
(the spin down density of states ${\cal N}_\downarrow$ is zero
at the energy of electrochemical potential $E=\mu_0$) 
with parallel
magnetizations are separated by the nonmagnetic region N;
(b) density-of-states diagrams for spin injection from F1 into N,
accompanied by the spin accumulation--generation of the nonequilibrium
magnetization $\delta M$. At F2
in the limit of low impedance ($Z$=0) 
spin is detected
by measuring the spin-polarized
current across the N/F2 interface. In the limit of high impedance 
($Z=\infty$) spin is detected
by measuring the voltage $V_s\sim \delta M$ developed across the N/F2 
interface;
(c) spin accumulation in a device in which  
a superconductor (with the superconducting gap $\Delta$) is
occupying the region between F1 and F2.}
\label{inj:1}
\end{figure}                                                                                       
The spin polarization of electrical current or carrier density, 
generated in a nonmagnetic region,
is typically used to describe the efficiency of electrical spin injection.
\textcite{Silsbee1980:BMR}  suggested that the nonequilibrium 
density polarization in the N region, or equivalently the
nonequilibrium magnetization, acts as the source of spin electromotive force 
(EMF) and produces a measurable ``spin-coupled'' 
voltage $V_s \propto \delta M$. Using this concept, also referred to
as  {\it spin-charge coupling},
\textcite{Silsbee1980:BMR} proposed a detection technique  
consisting of two ferromagnets F1 and F2 (see Fig.~\ref{inj:1})
separated by a nonmagnetic region.\footnote{A similar geometry was also 
proposed independently by \textcite{deGroot1983:PA}, where F1 and F2 were two
half-metallic ferromagnets with the goal of implementing  spin-based
devices to amplify and/or switch current.} 
F1 serves as the spin injector (spin aligner)
and F2 as the spin detector. This could be called the polarizer-analyzer 
method, 
the optical counterpart of the transmission of light through 
two optical linear polarizers.
From  Fig.~\ref{inj:1} it follows that
the reversal of the magnetization
direction in one of the ferromagnets would lead either to 
$V_s \rightarrow -V_s$,
in an open circuit (in the limit of large impedance Z), or to the reversal of
charge current $j \rightarrow -j$, in a short circuit (at small Z), 
a consequence
of Silsbee-Johnson spin-charge coupling 
\cite{Silsbee1980:BMR,Johnson1987:PRB,Johnson1988:PRBa}.
Correspondingly, as discussed in the following sections,
the spin injection could be detected through the spin accumulation signal 
either as a voltage or the resistance change when the magnetizations 
in F1 and F2 are changed from parallel to antiparallel alignment. 

Since the experiments demonstrating
the spin accumulation of conduction electrons
in metals \cite{Johnson1985:PRL}, spin injection has been realized
in a wide range of materials. While in Sec.~\ref{sec:IIC} we focus on related
theoretical work motivated by potential applications, experiments on
spin injection have also stimulated proposals for examining the fundamental
properties of electronic systems.\footnote{For example, studies          
probing the  spin-charge separation
in the non-Fermi liquids have been proposed by
\cite{Kivelson1990:PRB,Zhao1995:PRB,Si1997:PRL,Si1998:PRL,Balents2000:PRL,Balents2001:PRB}. 
Spin and charge are carried by
separate excitations and can lead to spatially separated spin and 
charge currents \cite{Kivelson1990:PRB}.}

The generation of nonequilibrium spin polarization has a long
tradition in magnetic resonance methods
\cite{Abragam:1961,Slichter:1989}. However, transport methods
to generate carrier spin polarization are not limited to electrical
spin injection. For example, they also include scattering of unpolarized
electrons in the presence of spin-orbit coupling 
\cite{Mott:1965,Kessler:1976} and in materials that lack the
inversion symmetry \cite{Levitov1985:SPJETP}, adiabatic 
\cite{Mucciolo2002:PRL,Sharma2003:PRB,Watson2003:PRL} and nonadiabatic  
quantum spin pumping \cite{Zheng2002:P} [for an instructive description
of parametric pumping see \cite{Brouwer1998:PRB}],
and the proximity effects \cite{Ciuti2002:PRL}.  

It would be interesting to know what the limits are on
the magnitude of various spin polarizations. Could we have a completely
polarized current [$P_j\rightarrow \infty$, see Eq.~(\ref{eq:polar})],
with only a spin current ($j_\uparrow-j_\downarrow$) and no charge current
($j_\uparrow+j_\downarrow=0$)?
While it is tempting to
recall the Stern-Gerlach experiment and try to set up magnetic
drift through  inhomogeneous magnets \cite{Kessler:1976}, 
this would most likely work
only as a transient effect \cite{Fabian2002b:PRB}. It was proposed
already by \textcite{Dyakonov1971:PL,Dyakonov1971:JETPLa} that a
transverse spin current (and transverse spin polarization in a
closed sample) would form as a result of spin-orbit coupling-induced
skew scattering in the presence of a longitudinal electric field.
This interesting effect, also called the {\it spin Hall effect} 
\cite{Hirsch1999:PRL,Zhang2000:PRL},
has yet to be demonstrated. An alternative scheme for producing pure
spin currents was proposed by \textcite{Bhat2000:PRL},
motivated by the experimental demonstration of
phase coherent control of charge currents 
\cite{Atanasov1996:PRL,Hache1997:PRL}
and carrier population \cite{Fraser1999:PRL}. 
A quantum-mechanical
interference between one- and two-photon absorptions of orthogonal
linear polarizations creates an opposite ballistic flow of spin up and
spin down electrons in a semiconductor. Only a spin
current can flow, without a charge current, as
demonstrated by \textcite{Stevens2003:PRL} and \textcite{Hubner2003:PRL}, 
who were able to achieve coherent 
control of the spin current direction and magnitude by the
polarization and the relative phase of two exciting laser light fields.

Charge current also can be driven by circularly polarized
light \cite{Ivchenko:1997}. 
Using the principles of optical orientation (see Sec.~\ref{sec:IB2}
and 
further discussion in Sec.~\ref{sec:IIB}) in semiconductors of reduced 
dimensionality or lower symmetry, 
both  the direction and  the magnitude 
of a generated charge current 
can be controlled by
circular polarization of the light. 
This is called the circular photo-voltaic effect \cite{Ganichev2003:JPCM},
which can be viewed as a transfer of the angular momentum of photons to 
directed motion of electrons. This could also be called a spin
corkscrew effect, since a nice mechanical analog is a corkscrew
whose rotation generates linear directed motion. A related effect,
in which the photocurrent is driven, is called the spin-galvanic effect
\cite{Ganichev2003:JPCM}. The current here is causes by the difference
in spin-flip scattering rates for electrons with different spin states
in some systems with broken inversion symmetry.
A comprehensive survey of the related effects 
from the circular photo-galvanic effect \cite{Asnin1979:SSC}
to recent demonstrations in semiconductor quantum wells 
\cite{Ganichev2001:PRL,Ganichev2002:PRL,Ganichev2002:N,Ganichev2003:PRB}]
is given by \textcite{Ganichev2003:JPCM}.

There is a wide range of recent theoretical proposals for devices
that would give rise to a spin electromotive force 
\cite{Long2002:APL,Zutic2001:APL,Zutic2001:PRB,Governale2003:PRB,%
Brataas2002:PRB,Malshukov2003:PRB,Ting2003:APL}, 
often
referred to as spin(-polarized) pumps, cells, or batteries. However,
even when it is feasible to generate pure spin current, this does 
not directly imply that 
it would be dissipationless. 
In the context of superconductors, it has been 
shown that Joule heating can arise from pure spin current 
flowing through a Josephson junction \cite{Takahashi2001:JAP}.

\subsection{\label{sec:IIB} Optical spin orientation}

\begin{figure}
\centerline{\psfig{file=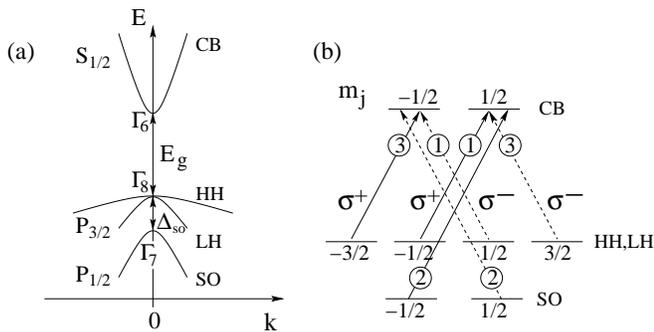,width=\linewidth,angle=0}}
\caption{(a) Interband transitions in GaAs: 
(a) schematic band structure of GaAs near the center 
of the Brillouin zone ($\Gamma$ point), where E$_g$ 
is the band gap and $\Delta_{so}$ is
the spin-orbit splitting; 
CB, conduction band; HH, valence heavy hole; LH, light hole; SO,
spin-orbit-split-off 
subbands; $\Gamma_{6,7,8}$ are the corresponding symmetries at the $k=0$ 
point, more precisely, the irreducible 
representations of the tetrahedron group $T_d$ \cite{Ivchenko:1997};
(b) selection rules for interband transitions between 
$m_j$ sublevels 
for circularly polarized light $\sigma^+$ 
and $\sigma^-$ (positive and negative helicity). The circled numbers
denote the relative transition intensities that apply for both
excitations (depicted by the arrows) and radiative recombinations.} 
\label{oo:1}
\end{figure}

In a semiconductor the photoexcited spin-polarized electrons and holes 
exist for the time $\tau$ before they recombine. If a fraction of the 
carriers' initial 
orientation survives longer than the recombination time, that is, 
if $\tau < \tau_s$, \footnote{In Si this condition is not
fulfilled. Instead of measuring the luminescence polarization, 
\textcite{Lampel1968:PRL} has used NMR to detect optical spin orientation.}
where $\tau_s$ is the spin relaxation time (see Sec.~\ref{sec:III}), 
the luminescence  (recombination radiation) will be partially polarized.  
By measuring the circular polarization of the luminescence it is 
possible to study the spin dynamics of the nonequilibrium carriers 
in semiconductors \cite{Oestreich2002:SST} and to extract such 
useful quantities as the spin orientation, the recombination time, or  
the spin relaxation time of the carriers
\cite{Parsons1969:PRL,Garbuzov1971:ZhETF,Ekimov1970:ZhETF,Meier:1984}.

We illustrate the basic principles of optical orientation
by the example of GaAs which is representative of a large
class of III-V and II-VI zincblende semiconductors. The
band structure is depicted in Fig.~\ref{oo:1}(a).
The band gap is $E_g=1.52$ eV at $T=0$  $K$, while the spin 
split-off band is separated from the light and 
heavy hole bands by $\Delta_{so}=0.34$ eV. 
We denote the Bloch states according to the total angular
momentum $J$ and its projection onto the positive
$z$ axis $m_j$: $|J,m_j \rangle$. Expressing the wave
functions with the symmetry of $s$, $p_x$, $p_y$, and $p_z$
orbitals as $|S\rangle $, $|X\rangle $, $|Y\rangle$, and $|Z\rangle$, 
respectively, the band wave functions can be written as listed 
in Table~\ref{tab:1} 
\cite{Pierce1976:PRB} [with minor typos removed, 
see also \cite{Kittel:1963}]. 

To obtain the excitation (or recombination) probabilities  
consider photons arriving in the $z$ direction. 
Let $\sigma^{\pm}$ represent the helicity of the exciting light. 
When we represent the dipole operator corresponding to the 
$\sigma^{\pm}$ optical transitions as \footnote{For an outgoing light 
in the $-z$ direction the helicities are reversed.}
$\propto (X\pm iY) \propto Y^{\pm 1}_1$, where $Y^m_l$  is the spherical
harmonic, it follows from Table~\ref{tab:1} that  
\begin{equation}
\label{eq:three}
\frac{|\langle 1/2,-1/2 | Y^1_1 |3/2, -3/2 \rangle|^2}
{|\langle 1/2,1/2 | Y^1_1 |3/2, -1/2 \rangle|^2}=3
\end{equation}
for the relative intensity of the $\sigma^+$ transition
between the heavy ($|m_j=3/2|$) and the light ($|m_j=1/2|$) hole subbands 
and the conduction band.
Other transitions are analogous. The relative transition rates
are indicated  in Fig.~\ref{oo:1}(b). The same selection rules apply
to the optical orientation of shallow impurities 
\cite{Parsons1969:PRL,Ekimov1970:ZhETF}.

\begin{table}
\begin{tabular}{lll}
\hline
\hline
& &  \\
Symmetry   &  $|J,m_j \rangle$ & Wave function  \\
& &  \\
\hline 
$\Gamma_6$ & $|1/2,1/2 \rangle$ &  $|S \! \uparrow  \rangle$  \\
 & $|1/2,-1/2 \rangle$ &  $|S \! \downarrow  \rangle$  \\
\hline
$\Gamma_7$ & $|1/2,1/2 \rangle$ &  
$|-(1/3)^{1/2}[\, (X+iY) \! \downarrow  - Z \! \uparrow] \, \rangle$  \\
 & $|1/2,-1/2 \rangle$ &  
$|(1/3)^{1/2}[\, (X-iY) \! \uparrow  + Z \! \downarrow]  \, \rangle$  \\
\hline
$\Gamma_8$ & $|3/2,3/2 \rangle$ &  
$|(1/2)^{1/2}(X+iY) \! \uparrow \rangle$  \\
 & $|3/2,1/2 \rangle$ &  
$|(1/6)^{1/2}[\, (X+iY) \! \downarrow  + 2 Z \! \uparrow] \, \rangle$  \\
 & $|3/2,-1/2 \rangle$ &  
$|-(1/6)^{1/2}[\, (X-iY) \! \uparrow  - 2 Z \! \downarrow] \, \rangle$  \\
 & $|3/2,-3/2 \rangle$ &  
$|(1/2)^{1/2}(X-iY) \! \downarrow \rangle$  \\
\hline
\hline
\end{tabular}
\caption{Angular and spin part of the wave function at $\Gamma$.}
\label{tab:1}
\end{table}
  
The spin polarization of the excited electrons\footnote{Although holes
are initially polarized too, they lose spin orientation very
fast, on the time scale of the momentum relaxation time 
(see Sec.~\ref{sec:IIID1})
However, it was suggested that manipulating hole spin by short electric
field pulses, between momentum scattering events, could be useful for
ultrafast spintronics applications \cite{Dargys2002:PRB}.
}.
depends on the photon energy $\hbar\omega$. For $\hbar\omega$ between $E_g$ 
and $E_g+\Delta_{so}$, only the light and heavy hole subbands contribute. 
Denoting by $n_{+}$ and $n_{-}$ the density of electrons polarized 
parallel ($m_j=1/2$) and
antiparallel ($m_j=-1/2$) to the direction of light propagation, we
define the 
spin polarization as (see Sec.~\ref{sec:IIA})
\begin{equation} \label{eq:pn}
P_n=(n_+-n_-)/(n_++n_-).
\end{equation}
For our example of the zincblende structure, 
\begin{equation}
P_n=(1-3)/(3+1)=-1/2
\end{equation}
is the spin polarization at the moment of photoexcitation. The spin is 
oriented against the direction of light propagation, since there are more 
transitions from the heavy hole than from the light hole subbands. 
The circular polarization of the luminescence is defined as
\begin{equation}
\label{eq:circ}
P_{\rm circ}=(I^+ -I^-)/(I^+ + I^-),
\end{equation}
where $I^\pm$ is the radiation intensity for the helicity $\sigma^{\pm}$.
The polarization of the $\sigma^+$ photoluminescence is then
\begin{equation}
P_{\rm circ}=\frac{(n_+ + 3n_-)-(3n_+ + n_-)}{(n_+ + 3n_-)+(3n_+ + n_-)}
             =-\frac{P_n}{2}=\frac{1}{4}.
\label{eq:1/4}       
\end{equation}

If the excitation involves transitions from the spin split-off band, 
that is, if  $\hbar \omega \gg E_g + \Delta_{so}$, the electrons
will not be spin polarized ($P_n=P_{\rm circ}=0$), underlining the 
vital role of spin-orbit coupling for spin orientation.  On the
other hand, Fig.~\ref{oo:1} 
suggests that a removal of the heavy/light
hole degeneracy can substantially increase $P_n$ \cite{Dyakonov:1984}, 
up to the limit of complete spin polarization. 
An increase in $P_n$ and $P_{\rm circ}$
in GaAs strained due to a lattice mismatch with a substrate, or due
to confinement in quantum well heterostructures,
has indeed been demonstrated \cite{Oskotski1997:PLDS,Vasilev1993:SM},
detecting $P_n$ greater than 0.9. 

While photoexcitation with  circularly polarized light creates 
spin-polarized electrons, the nonequilibrium spin decays due to both 
carrier recombination and spin relaxation. The steady-state degree
of spin polarization depends on the balance between the spin 
excitation and decay. Sometimes a distinction is made 
\cite{Pierce1976:PRB,Meier:1984}
between the terms optical {\it spin orientation} and optical {\it spin 
pumping}. The former term is used in relation to the minority carriers
(such as electrons in p-doped samples) and represents the 
orientation of the excited carriers. The latter term is reserved for the
majority carriers (electrons in n-doped samples), representing
spin polarization of the ``ground'' state. Both spin orientation
and spin pumping were demonstrated in the early investigations
on p-GaSb \cite{Parsons1969:PRL} and  p- and n-Ga$_{0.7}$Al$_{0.3}$As 
\cite{Ekimov1970:ZhETF,Ekimov1971:JETPL,Zakharchenya1971:JETPL}. Unless
specified otherwise, we shall use the term optical orientation to 
describe both
spin orientation and spin pumping.

To derive the steady-state expressions for the spin polarization due to optical
orientation, consider the simple model of carrier recombination and spin 
relaxation
(see Sec.~\ref{sec:IVA4}) in a homogeneously doped semiconductor.
The balance between direct electron-hole recombination and optical 
pair creation can be written as 
\begin{equation} 
\label{eq:np}
r(np-n_0p_0)=G,
\end{equation}
where $r$ measures the recombination rate, the electron and hole densities are 
$n$ and
$p$, with index zero denoting the equilibrium values, 
and $G$ is the electron-hole
photoexcitation rate. Similarly, the balance between spin relaxation and 
spin generation is expressed by
\begin{equation} 
\label{eq:sp}
rsp+s/\tau_s=P_{n}(t=0) G,
\end{equation}
where $s=n_+-n_-$ 
is the electron spin density and $P_{n}(t=0)$ is the spin polarization
at the moment of photoexcitation, given by Eq.~(\ref{eq:pn}). 
Holes are assumed to lose their
spin orientation very fast, so they are treated as unpolarized. The first
term in Eq.~(\ref{eq:sp}) describes the disappearance of the spin density 
due to carrier recombination, while the second term describes the
intrinsic spin relaxation. From Eqs.~(\ref{eq:np}) and (\ref{eq:sp}) we 
obtain the
steady-state electron polarization as \cite{Zutic2001:PRB}
\begin{equation} \label{eq:p}
P_n=P_n(t=0)\frac{1-n_0p_0/np}{1+1/\tau_srp}.
\end{equation}

In a p-doped sample $p\approx p_0$, $n\gg n_0$, and Eq.~(\ref{eq:p}) gives 
\begin{equation}
\label{eq:orient}
P_n=P_n(t=0)/(1+\tau/\tau_s),
\end{equation}
where $\tau=1/rp_0$ is the electron lifetime.\footnote{After the illumination
is switched off, the electron spin density, or equivalently the nonequilibrium
magnetization, will decrease exponentially with the inverse time constant
$1/T_s=1/\tau+1/\tau_s$ \cite{Parsons1969:PRL}.}
The steady-state polarization is independent of the illumination intensity,
being reduced from the initial spin polarization $P_n(t=0)$.
\footnote{The effect of a finite length for the light absorption on $P_n$ 
is discussed 
by \textcite{Pierce:1984}. The absorption length $\alpha^{-1}$ is
typically a micron for GaAs.   It varies with frequency
roughly as $\alpha(\hbar \omega) \propto (\hbar \omega - E_g)^{1/2}$ 
\cite{Pankove:1971}.}
The polarization of the photoluminescence is 
$P_{\rm circ}=P_n(t=0) P_n$ \cite{Parsons1969:PRL}.
Early measurements of  $P_n=0.42 \pm 0.08$ in  GaSb \cite{Parsons1969:PRL} 
and 
$P_n=0.46 \pm 0.06$  in  Ga$_{0.7}$Al$_{0.3}$As \cite{Ekimov1970:ZhETF}
showed an effective spin orientation close to the maximum 
value of $P_n(t=0)=1/2$ for a bulk unstrained  zincblende structure,
indicating that $\tau/\tau_s \ll 1$. 

For spin pumping in an n-doped sample, where $n\approx n_0$ and $p\gg p_0$, 
Eqs.~(\ref{eq:np}) and (\ref{eq:p}) give \cite{Dyakonov1971:JETPL}
\begin{equation}
\label{eq:pump}
P_n=P_n(t=0)/(1+n_0/G\tau_s).
\end{equation}
In contrast to the previous case, the carrier (now hole) lifetime 
$\tau=1/rn_0$
has  no effect on $P_n$. However, $P_n$ depends on the
photoexcitation intensity $G$, as expected for a pumping process. 
The effective
carrier lifetime is $\tau_J=n_0/G$, 
where $J$ represents the intensity of the illuminating light. 
If it is comparable to or shorter than $\tau_s$, 
spin pumping is very effective. Spin pumping works because the photoexcited 
spin-polarized electrons do not need to recombine with holes. There
are plenty of unpolarized electrons in the conduction band available
for recombination. The spin is thus pumped in to the electron system.

When magnetic field {\bf B} is applied 
perpendicular to the axis of spin orientation (transverse magnetic field),
it will induce spin precession with the 
Larmor frequency $\Omega_L=\mu_B g B/\hbar$, where $\mu_B$ is the Bohr
magneton and $g$ is the electron $g$ factor.\footnote{In our convention
the $g$ factor of free electrons is positive, 
$g_0=2.0023$ \cite{Kittel:1996}.}
The spin precession, together with the random character of carrier generation
or diffusion, leads to the spin dephasing (see Sec.~\ref{sec:IIIA1}).
Consider spins excited by circularly polarized light (or by any means
of spin injection) at a steady rate. In a steady rate a balance between
nonequilibrium spin generated and spin relaxation is maintained, resulting
in a net magnetization. If a transverse magnetic field is applied,
the decrease of the steady-state magnetization can have two sources:
(a) spins which were excited at random time and (b) random diffusion
of spins towards a detection region. Consequently, spins precess along
the applied field acquiring random phases relative to those which were
excited or have arrived at different times.
As a result, the projection
of the electron spin along the exciting beam will decrease with the 
increase of transverse magnetic field, leading to depolarization of the  
luminescence. 
This is also known as the Hanle effect \cite{Hanle1924:ZP}, in analogy
to the depolarization of the resonance fluorescence of gases. The Hanle
effect was first measured in semiconductors by \textcite{Parsons1969:PRL}. 
The steady-state spin polarization of the precessing electron spin 
can be calculated by solving the Bloch-Torrey equations 
\cite{Bloch1946:PR,Torrey1956:PR}, 
Eqs.~(\ref{eq:relax:bloch1})--(\ref{eq:relax:bloch3}) describing the 
spin dynamics
of diffusing carriers.

In p-doped semiconductors the Hanle curve 
shows a Lorentzian decrease of the polarization \cite{Parsons1969:PRL},
$P_n(B)=P_n(B=0)/(1+\Omega_L T_s)^2$,
where $P_n(B=0)$ is the polarization at $B=0$ from Eq.~(\ref{eq:orient})
and $T_s^{-1}$ is the effective spin lifetime given by 
$1/T_s=1/\tau+1/\tau_s$;
see footnote 26.
Measurements of the Hanle curve in GaAlAs were used 
by  \textcite{Garbuzov1971:ZhETF}    
to separately determine both $\tau$ and $\tau_s$ at various temperatures.
The theory of the Hanle effect in n-doped semiconductors was developed by
\textcite{D'yakonov1976:FTP} who showed the  
non-Lorentzian decay of the luminescence for the
regimes  both of low ($\tau_J / \tau_s \gg  1$) 
and high ($\tau_J / \tau_s \ll 1$) intensity of the exciting light. 
At high fields $P_n(B)\propto 1/B^{1/2}$, consistent
with the experiments of \textcite{Vekua1976:FTP} in Ga$_{0.8}$Al$_{0.2}$As, 
showing a Hanle curve different from the usual $P_n(B)\propto 1/B^2$ Lorentzian 
behavior \cite{Dyakonov:1984}. Recent findings on the Hanle effect in
nonuniformly doped GaAs and reanalysis of some earlier studies are
given by \textcite{Dzihoev2003:FTT}.

\subsection{\label{sec:IIC} Theories of spin injection}

Reviews on spin injection have covered materials ranging from semiconductors 
to high temperature superconductors and have 
addressed the implications for device 
operation as well as for fundamental studies in solid 
state systems.\footnote{See, for example, 
\cite{Johnson2002:SST,Schmidt2002:SST,Jedema2002:JS,Wei2002:JS,%
Goldman1999:JMMM,Goldman2001:JS,Tang:2002,Johnson2001:JS,Osofsky2000:JS,%
Maekawa2001:MSE}}.  
In addition to degenerate conductors, examined in these works, we also give
results for nondegenerate semiconductors in which the violation of local 
charge neutrality, electric fields, and carrier band bending require
solving the Poisson equation. The notation introduced here emphasizes the
importance of different (and inequivalent) spin polarizations arising in 
spin injection.

\subsubsection{\label{sec:IIC1} F/N junction}

A theory of spin injection across a ferromagnet/normal metal (F/N) interface
was first offered by \textcite{Aronov1976:JETPL}. Early work also
included spin injection into a semiconductor (Sm) 
\cite{Aronov1976:SPS,Masterov1979:SPS}
and a superconductor (S) \cite{Aronov1976:SPJETP}. Spin injection in 
F/N junctions
was subsequently studied in detail by 
\textcite{Johnson1987:PRB,Johnson1988:PRBa},\footnote{Johnson and Silsbee base 
their approach on irreversible thermodynamics and consider also  the effects of 
a temperature gradient on spin-polarized transport, omitted in this section.} 
\textcite{vanSon1987:PRL},
\textcite{Valet1993:PRB}, \textcite{Hershfield1997:PRB}, and others.
Here we follow the approach of \textcite{Rashba2000:PRB,Rashba2002:EPJ} and 
consider
a steady-state\footnote{Even some dc spin injection experiments are actually
performed at low (audio-frequency) bias. Generalization to ac spin injection, 
with a harmonic time dependence, was studied by \textcite{Rashba2002:APL}.}
flow of electrons along the $x$ direction in a three-dimensional (3D) 
geometry consisting of a metallic ferromagnet (region $x<0$) and a 
paramagnetic metal or
a degenerate semiconductor (region $x>0$).

\begin{figure}
\centerline{\psfig{file=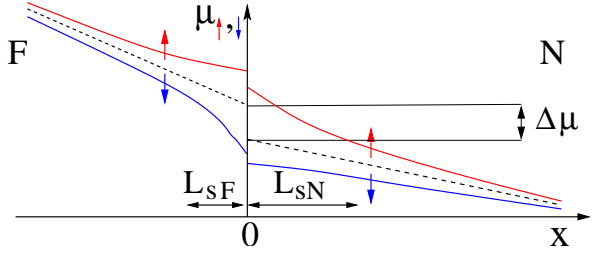,width=0.9\linewidth,angle=0}}
\caption{Spatial variation of the electrochemical potentials near a
spin-selective resistive interface at an F/N junction. At the interface $x=0$
both the spin-resolved electrochemical potentials ($\mu_\lambda$, 
$\lambda=\uparrow,\downarrow$, denoted with solid lines) and the average 
electrochemical potential ($\mu_{F}$, $\mu_{N}$, dashed lines) are 
discontinuous. The spin diffusion length $L_{sF}$ and $L_{sN}$
characterizes the decay of  $\mu_s=\mu_\uparrow - \mu_\downarrow$ 
(or equivalently the decay of spin accumulation and  
the nonequilibrium magnetization) away from the interface and into the bulk 
F and N regions, respectively.} 
\label{inj:2}
\end{figure}
The two regions, F and N, form a contact at $x=0$,
as depicted in Fig.~\ref{inj:2}.
The relative magnitudes of three characteristic resistances  
per unit area\footnote{For this simple
geometry various resistances have a common factor of the cross-sectional area, 
which
can be factored out. This is no longer possible for a more complicated geometry 
\cite{Takahashi2003:PRB}.} determine the degree of current polarization 
injected 
into a nonmagnetic material. These are the contact resistance $r_c$ 
and the two characteristic 
resistances $r_N$ $r_F$, each given by the ratio of the spin diffusion 
length and the 
effective bulk conductivity in the corresponding region.
Two limiting cases  correspond to the transparent limit, 
where $r_c\rightarrow 0$, and the low-transmission limit, where 
$r_c \gg r_N,r_F$.

Spin-resolved quantities are labeled by 
$\lambda=1$ or $\uparrow$ for spin up, $\lambda=-1$ or $\downarrow$ 
for spin down along the chosen quantization axis. 
For a free electron, spin angular momentum
and magnetic moment are in opposite directions, and what
precisely is denoted by ``spin up'' varies in the 
literature \cite{Jonker2003:P}.
Conventionally, in metallic systems 
\cite{Tedrow1973:PRB,Gijs1997:AP},
spin up refers to carriers with majority spin. This means that the
spin (angular momentum) of such carriers is
antiparallel to the magnetization.
Spin-resolved charge current (density)
in a diffusive regime can be expressed as 
\begin{equation}
\label{eq:jq}
j_\lambda=\sigma_\lambda \nabla \mu_\lambda,
\end{equation}
where $\sigma_\lambda$ is conductivity and the electrochemical potential is
\begin{equation}
\label{eq:elchem}
\mu_\lambda=(q D_\lambda/\sigma_\lambda) \delta n_\lambda-\phi,
\end{equation}
with $q$ proton charge, $D_\lambda$ diffusion coefficient, 
$\delta n_\lambda=n_\lambda-n_{\lambda 0}$ the change of electron density 
from the equilibrium value for spin $\lambda$, 
and $\phi$ the electric potential.~\footnote{More generally, 
for a noncollinear magnetization, 
$j_\lambda$ becomes a second-rank tensor 
\cite{Johnson1988:PRBa,Margulis1994:PB,Stiles2002:PRB}.}

In the steady state the continuity equation is
\begin{equation}
\label{eq:jcont}
\nabla j_\lambda=
\lambda q \left[\frac{\delta n_\lambda}{\tau_{\lambda-\lambda}}
               -\frac{\delta n_{-\lambda}}{\tau_{-\lambda \lambda}} \right],
\end{equation}
and $\tau_{\lambda\lambda'}$ is the average time for flipping a $\lambda$-spin
to $\lambda'$-spin.
For a degenerate conductor\footnote{In the nondegenerate case of
Boltzmann statistics, the Einstein relation implies 
that the ratio of the diffusion coefficient and the mobility is $k_BT/q$.} 
the Einstein relation 
is
\begin{equation}
\sigma_\lambda=q^2 {\cal N}_\lambda D_\lambda, 
\label{eq:einstein}
\end{equation}
where $\sigma=\sigma_\uparrow+\sigma_\downarrow$ and 
${\cal N}={\cal N}_\uparrow+{\cal N}_\downarrow$
is the density of states. Using a detailed balance 
${\cal N}_\uparrow/ \tau_{\uparrow \downarrow}=
{\cal N}_\downarrow/ \tau_{\downarrow \uparrow}$  
\cite{Hershfield1997:PRB,Kravchenko2002:JETP} 
together with Eqs.~(\ref{eq:elchem}) and (\ref{eq:einstein}),
the continuity equation can be expressed as
\begin{equation}
\label{eq:jq'}
\nabla j_\lambda=\lambda q^2 \frac{{\cal N}_\uparrow {\cal N}_\downarrow}
{{\cal N}_\uparrow+{\cal N}_\downarrow}
\frac{\mu_\lambda-\mu_{-\lambda}}{\tau_s},
\end{equation}
where $\tau_s=\tau_{\uparrow \downarrow} \tau_{\downarrow \uparrow}/ 
(\tau_{\uparrow \downarrow}+\tau_{\downarrow \uparrow})$ 
is the spin relaxation time. Equation (\ref{eq:jq'})  
implies the conservation of charge 
current $j=j_\uparrow+j_\downarrow=const.$, while the spin counterpart,
the difference of the spin-polarized currents 
$j_s=j_\uparrow-j_\downarrow$ is position dependent. 
Other ``spin quantities,'' $X_s$,
unless explicitly defined, are analogously expressed 
with the corresponding (spin) polarization given by $P_X=X_s/X$.
For example, the current polarization\footnote{This is 
also referred to as a
spin injection coefficient \cite{Rashba2000:PRB,Rashba2002:EPJ}.} 
$P_j=j_s/j$, generally different
from the density polarization
$P_n=(n_\uparrow-n_\downarrow)/n$, is related to the conductivity polarization 
$P_\sigma$ as 
\begin{equation}
\label{eq:Pj}
P_j=2(\sigma_\uparrow \sigma_\downarrow/\sigma) \nabla \mu_s/j+P_\sigma
\end{equation}
where $\mu_s=\mu_\uparrow-\mu_\downarrow$. In terms of the
average electrochemical potential $\mu=(\mu_\uparrow+\mu_\downarrow)/2$, 
$P_\sigma$ further satisfies
\begin{equation}
\label{eq:mu}
\nabla \mu=-P_\sigma \nabla \mu_s/2+j/\sigma.
\end{equation}
From Eqs.~(\ref{eq:elchem}) and (\ref{eq:jq'}) it follows that $\mu_s$
satisfies the diffusion equation
\cite{vanSon1987:PRL,Valet1993:PRB,Hershfield1997:PRB,Schmidt2000:PRB} 
\begin{equation}
\label{eq:mus}
\nabla ^2 \mu_s=\mu_s/L_s^2,
\end{equation}
where the spin diffusion length is $L_s=(\overline{D}\tau_s)^{1/2}$ with
the spin averaged diffusion coefficient 
$\overline{D}=(\sigma_\downarrow D_\uparrow 
+\sigma_\uparrow D_\downarrow)/\sigma
={\cal N}({\cal N}_\downarrow/D_\uparrow+{\cal N}_\uparrow/D_\downarrow)^{-1}$. 
Using Eq.~(\ref{eq:elchem}) and the local charge quasineutrality 
$\delta n_\uparrow+\delta n_\downarrow=0$
shows that $\mu_s$ is proportional to the nonequilibrium 
spin density $\delta s=\delta n_\uparrow - \delta n_\downarrow$ 
($s=s_0+\delta s=n_\uparrow-n_\downarrow$)
\begin{equation}
\label{eq:spin}
\mu_s=\frac{1}{2 q} \frac{{\cal N}_\uparrow+{\cal N}_\downarrow} 
{{\cal N}_\uparrow {\cal N}_\downarrow}
\delta s.
\end{equation}                                                                 
Correspondingly,  $\mu_s$ is often referred to as the (nonequilibrium) 
{\it spin 
accumulation}\footnote{Spin accumulation is also relevant to a number of
physical phenomena outside the scope of this article, for example, to the
tunneling rates in the quantum Hall regime 
\cite{MacDonald1999:PRL,Chan1999:PRL}.} 
and is used to explain the GMR effect in CPP structures 
\cite{Johnson1991:PRL,Valet1993:PRB,Hartmann:2000,Hirota:2002,Gijs1997:AP}. 

The preceding equations are simplified for the N region 
by noting that 
$\sigma_\lambda=\sigma/2$, $\sigma_s=0$, and 
$D_\lambda=\overline{D}$.
Quantities pertaining to a particular region are denoted by the index 
F or N.

Equation ~(\ref{eq:mus}) has also been used to study the diffusive 
spin-polarized transport and spin accumulation in  ferromagnet/superconductor 
structures \cite{Jedema1999:PRB}. 
Some care is needed to establish the appropriate
boundary conditions at the F/N interface.
In the absence of spin-flip scattering\footnote{The effects
of non-conserving interfacial scattering on spin injection were considered 
in \cite{Valet1993:PRB,Fert1996:PRB,Rashba2002:EPJ}.}
at the F/N interface (which can arise, for example, 
due to spin-orbit coupling or magnetic impurities) 
the spin current is continuous 
and thus $P_{jF}(0^-)=P_{jN}(0^+)\equiv P_j$ (omitting
$x=0^\pm$ for brevity, and superscripts $\pm$ in other quantities). 
These boundary conditions were used by 
\textcite{Aronov1976:SPS,Aronov1976:JETPL} 
without relating $P_j$ to the effect of the F/N contact or 
material parameters in the F region.

Unless the F/N contact is highly transparent, $\mu_\lambda$ is discontinuous
across the interface 
\cite{Johnson1988:PRL,Valet1993:PRB,Hershfield1997:PRB,Rashba2000:PRB} 
and the boundary condition is
\begin{equation}
\label{eq:jbc}
\ j_\lambda(0)=\Sigma_\lambda [\mu_{\lambda N}(0)-\mu_{\lambda F}(0)],
\end{equation}
where 
\begin{equation}
\label{eq:sigma}
\Sigma=\Sigma_\uparrow+\Sigma_\downarrow
\end{equation}
is the contact conductivity. For a free-electron model 
$\Sigma_\uparrow \neq \Sigma_\downarrow$ can be simply 
inferred from the effect of the exchange energy, which would 
yield spin-dependent Fermi wave vectors and transmission
coefficients. A microscopic determination of the 
corresponding contact resistance [see Eq.~(\ref{eq:rc})]
is complicated by the influence of disorder,
surface roughness, and different scattering mechanisms and
is usually obtained from  model calculations 
\cite{Schep1997:PRB,Stiles2000:PRB}. Continued work on
the first-principles calculation of F/N interfaces
\cite{Stiles1996:JAP,Erwin2002:PRB} is needed for 
a more detailed understanding of spin injection.
From Eqs.~(\ref{eq:jbc}) and (\ref{eq:sigma})  
it follows that
\begin{eqnarray}
\label{eq:musbc}
\mu_{sN}(0)-\mu_{sF}(0)=2r_c(P_j-P_\Sigma)j, \\
\label{eq:mubc}
\mu_N(0)-\mu_F(0)=r_c ( 1-P_\Sigma P_j ) j,
\end{eqnarray}
where the effective contact resistance is 
\begin{equation}
\label{eq:rc}
r_c=\Sigma/4\Sigma_\uparrow \Sigma_\downarrow.
\end{equation}
The decay of $\mu_s$, away from the interface, is characterized by the
corresponding spin diffusion length
\begin{equation}
\label{eq:decay}
\mu_{sF}=\mu_{sF}(0) e^{x/L_{sF}}, \quad
\mu_{sN}=\mu_{sN}(0) e^{-x/L_{sN}}.  
\end{equation}
A nonzero value for $\mu_{sN}(0)$ implies the existence of nonequilibrium
magnetization $\delta M$ in the N region (for noninteracting electrons
$q \mu_s=\mu_B \delta M/\chi$, where $\chi$ is 
the magnetic susceptibility).
Such  a $\delta M$, as a result of electrical spin injection, was proposed
by \textcite{Aronov1976:SPS} and first measured in metals by 
\textcite{Johnson1985:PRL}.

By applying Eq.~(\ref{eq:Pj}), separately, to the F and N regions, 
one can obtain the amplitude of spin accumulation in terms 
of the current and density of states spin polarization 
and the effective resistances $r_F$ and $r_N$, 
\begin{equation}
\label{eq:mus0}
\mu_{sF}(0)=2r_F \left [P_j-P_{\sigma F} \right ]j, \quad
\mu_{sN}(0)=-2r_N P_j j, 
\end{equation}
where
\begin{equation}
\label{eq:rs}
r_N=L_{sN}/\sigma_N, \quad
r_F=L_{sF} \sigma_F/(4\sigma_{\uparrow F}\sigma_{\downarrow F}).
\end{equation}
From Eqs.~(\ref{eq:mus0}) and 
(\ref{eq:musbc}) the current polarization can be obtained as 
\begin{equation}
\label{eq:gamma}
P_j=\left [r_c P_\Sigma+ r_F P_{\sigma F} \right] /r_{FN},
\end{equation}
where $r_{FN}=r_F+r_c+r_N$ is the effective equilibrium
resistance of the F/N junction. 
It is important to emphasize that a measured highly polarized current,
representing an efficient spin injection, does not itself imply
a large spin accumulation or a large density polarization, typically
measured by optical techniques. In contrast to the derivation of $P_j$
from Eq.~(\ref{eq:gamma}), determining $P_n$
requires using 
Poisson's
equation or a condition of the local charge 
quasineutrality.\footnote{Carrier density
will also be influenced by the effect of screening, which changes with
the dimensionality of the spin injection geometry \cite{Korenblum2002:PRL}.} 

It is useful to note\footnote{\textcite{Rashba2002:PC}.}
that Eq.~(\ref{eq:gamma}), written as Eq.~(18) in \cite{Rashba2000:PRB}
can be mapped to Eq.~(A11) from \cite{Johnson1987:PRB}, where it
was first derived.\footnote{The substitutions are $P_j \rightarrow \eta^*$, 
$P_\sigma \rightarrow p$,
$P_\Sigma \rightarrow \eta$, $r_c \rightarrow [G(\xi-\eta^2)]^{-1}$,
$r_N \rightarrow \delta_n/\sigma_n\zeta_n$,
$r_F \rightarrow \delta_f/\sigma_f(\zeta_f-p_f^2)$, 
$L_{sN,F} \rightarrow  \delta_{n,F}$,
and $n,f$ label N and F region, respectively. 
$\eta$, $\zeta_n$, and $\zeta_f$ are of the order of unity.
To ensure that resistances and the spin diffusion lengths in 
\textcite{Johnson1987:PRB} are positive, one must additionally have 
$(\xi-\eta^2)>0$ and
$(\zeta_i-p_i^2)>0$, $i=n,f$ (for normal and ferromagnetic regions, 
respectively). In particular, assuming $\xi=\zeta_n=\zeta_f=1$ 
a detailed correspondence between  Eq.~(\ref{eq:gamma}) and  Eq.~(A11) in 
\cite{Johnson1987:PRB} is recovered. For example,  
$r_c \rightarrow [G(\xi-\eta^2)]^{-1}$
yields Eq.~(\ref{eq:rc}), where $\Sigma \rightarrow G$.}
An equivalent form for $P_j$ in Eq.~(\ref{eq:gamma}) was obtained by 
\textcite{Hershfield1997:PRB} and for $r_c=0$ results from 
\textcite{vanSon1987:PRL} are recovered.

In contrast to normal metals \cite{Johnson1985:PRL,Johnson1988:PRBb} 
and superconductors, for which  injection has been reported in 
both  conventional 
\cite{Johnson1994:APL}, and high temperature superconductors 
\cite{Hass1994:PC,Vasko1997:PRL,Dong1997:APL,Yeh1999:PRB},
creating a substantial current polarization by direct electrical spin 
injection 
from a metallic ferromagnet
into a semiconductor proved to be more difficult
\cite{Monzon1999:JMMM,Hammar1999:PRL,Filip2000:PRB,Zhu2001:PRL}.

By examining  
Eq.~(\ref{eq:gamma}) we can both infer some possible limitations and
deduce several experimental strategies for effective spin injection
i.e. to increase $P_j$ into semiconductors.
For a perfect Ohmic contact $r_c=0$, the typical resistance mismatch 
$r_F \ll r_N$
(where F is a metallic ferromagnet)
implies inefficient spin injection with $P_j \approx r_F/r_N \ll 1$, 
referred to
as the {\it conductivity mismatch} problem by \textcite{Schmidt2000:PRB}. 
Even in  the absence of the resistive contacts, effective spin injection into 
a semiconductor can be achieved if the resistance mismatch is reduced 
by using for spin injectors either a magnetic semiconductor or a highly 
spin-polarized ferromagnet.\footnote{From Eq.~(\ref{eq:rs}) a half-metallic 
ferromagnet implies a large $r_F$.}  

While there was early experimental evidence \cite{Alvarado1992:PRL}
that employing resistive (tunneling) contacts could lead to 
an efficient spin injection\footnote{The influence of the resistive contacts on 
spin injection can also be inferred by 
explicitly considering resistive contacts
\cite{Johnson1987:PRB,Hershfield1997:PRB}.}
a systematic understanding was provided by \textcite{Rashba2000:PRB}
and supported with the subsequent experimental and theoretical studies 
\cite{Smith2001:PRB,Fert2001:PRB,Rashba2002:EPJ,Takahashi2003:PRB,%
Johnson2003:PRB,Johnson:2003}.
As can be seen from Eq.~(\ref{eq:gamma}) the spin-selective resistive 
contact $r_c \gg r_F,r_N$ (such as a tunnel or Schottky 
contact) would contribute to effective spin injection with 
$P_j\approx P_\Sigma$ 
being dominated by 
the effect
$r_c$ and not the ratio $r_F/r_N$.\footnote{A similar result was stated
previously by \textcite{Johnson1988:PRBa}.}
This limit is also instructive to illustrate the principle of 
spin filtering \cite{Esaki1967:PRL,Moodera1988:PRL,Hao1990:PRB,Filip2002:APL}.
In a spin-discriminating transport process
the resulting degree of spin polarization is changed. Consequently
the effect of spin filtering, similar to spin injection, leads  
to the generation of (nonequilibrium) spin 
polarization. \footnote{While most of the schemes resemble a CPP geometry 
[Fig.~\ref{gmr:1}(b)], there are also proposals for generating highly polarized 
currents in a CIP-like geometry [Fig.~\ref{gmr:1}(a)] 
\cite{Gurzhi2001:FNT,Gurzhi2003:P}.}
For example, 
at low temperature EuS and EuSe, discussed in 
Sec.~\ref{sec:IVC}, can act as spin-selective barriers.
In the extreme case,
initially spin-unpolarized carriers (say, injected from a nonmagnetic material)
via spin-filtering could attain a complete polarization.
For a strong spin-filtering contact $P_\Sigma > P_{\sigma F}$, the sign of the
spin accumulation (nonequilibrium magnetization) is reversed in the F and
N regions, near the interface [recall Eq.~(\ref{eq:musbc})], in contrast
to the behavior sketched in Fig.~\ref{inj:2}, where $\mu_{s F,N} > 0$.

The spin injection process alters the potential drop across the F/N interface
because differences of spin-dependent electrochemical potentials on
either side of the interface generate an effective resistance $\delta R$.
By integrating Eq.~(\ref{eq:mu}) for N and F regions, separately,
it follows  that $Rj=\mu_N(0)-\mu_F(0)+P_{\sigma F}\mu_{sF}(0)/2$,
where $R$ is the junction resistance.
Using Eqs.~(\ref{eq:mubc}), (\ref{eq:rs}), and (\ref{eq:gamma}) 
allows us to express 
$R=R_0+ \delta R$, where $R_0=1/\Sigma$ ($R_0=r_c$ if 
$\Sigma_\uparrow=\Sigma_\downarrow$) 
is the equilibrium resistance, in the absence of spin injection, and 
\begin{eqnarray} 
\delta R= [r_N ( r_F P^2_{\sigma F} +r_c P^2_{\Sigma} ) 
        +r_F r_c ( P_{\sigma F} - P_{\Sigma})^2]  / r_{FN}, 
\label{eq:delR} 
\end{eqnarray}
where $\delta R>0$ is the nonequilibrium resistance.
Petukhov has shown \cite{Jonker2003:MRS} that Eqs.~(\ref{eq:gamma})
and (\ref{eq:delR}) could be obtained by considering an equivalent
circuit scheme with two resistors 
$\tilde{R}_\uparrow$,
$\tilde{R}_\downarrow$ connected in parallel, where 
$\tilde{R}_\lambda=L_{sF}/\sigma_{\lambda F}+1/\Sigma_\lambda+2L_{sN}/\sigma_N$
and $\tilde{R}_\uparrow + \tilde{R}_\downarrow=4r_{FN}$.
For such a resistor scheme, by noting that 
$j_\uparrow \tilde{R}_\uparrow= j_\downarrow \tilde{R}_\downarrow$,
Eq.~(\ref{eq:gamma}) is obtained as 
$P_j=-P_{\tilde{R}}\equiv-(\tilde{R}_\uparrow-\tilde{R}_\downarrow)
/(\tilde{R}_\uparrow+\tilde{R}_\downarrow)$. $\delta R$ in
Eq.~({\ref{eq:delR}) is then obtained as the difference between the
total resistance of the nonequilibrium spin-accumulation region of the
length $L_{sF}+L_{sN}$ [given by the equivalent resistance 
$\tilde{R}_\uparrow \tilde{R}_\downarrow/ 
(\tilde{R}_\uparrow + \tilde{R}_\downarrow)$] and the equilibrium 
resistance for the same region, $L_{sF}/\sigma_F + L_{sN}/\sigma_N$.

The concept of the excess resistance $\delta R$ can also be explained as 
a consequence of the Silsbee-Johnson spin-charge coupling 
\textcite{Silsbee1980:BMR,Johnson1985:PRL,Johnson1987:PRB}
and illustrated by considering the simplified schemes in  
Figs.~\ref{inj:1} and \ref{inj:2}.
Accumulated spin 
near the F/N interface, 
together with a finite spin relaxation  and a finite spin diffusion,
impedes the 
flow of spins and acts as a ``spin bottleneck'' 
\cite{Johnson1991:PRL}. A rise of $\mu_{sN}$ must be accompanied by the rise of 
$\mu_{sF}$ [their precise alignment at the interface is given in 
Eq.~(\ref{eq:musbc})] 
or there will be a backflow of the nonequilibrium spin back into the F region. 
Because both spin and charge are carried by electrons 
in spin-charge coupling 
the backflow of spin 
driven by diffusion creates an additional resistance for the charge flow 
across the 
F/N interface. Based on an analogy with the charge transport across a 
clean N/superconductor (S) interface (see Sec.~\ref{sec:IVA3}) 
\textcite{vanSon1987:PRL} explained $\delta R$ by invoking
the consequences of current conversion from spin-polarized, at 
far to the left of the F/N interface, to completely unpolarized, at far right 
in the N region.

The increase in the total resistance with spin injection can be most
dramatic if the N region is taken to be a superconductor (S);
see Fig.~\ref{inj:1}(c).  
Spin injection depletes the superconducting condensate and can
result in the switching to a normal state of much higher resistance 
\cite{Vasko1997:PRL,Dong1997:APL,Yeh1999:PRB,Wei1999:JAP,Takahashi1999:PRL}.
A critical review of possible spurious effects 
in reported  experiments 
\textcite{Gim2001:JAP} 
has also stimulated the development of a novel detection technique
which uses scanning tunneling spectroscopy combined with 
pulsed quasiparticle  spin injection to minimize 
Joule heating \cite{Ngai2003:P} (see Sec.~\ref{sec:IVA1}).
In the S region the quasiparticle energy is 
$E_k=(\xi_k^2+\Delta^2)^{1/2}$, where $\xi_k$ is the single particle
excitation energy 
corresponding to the wave vector ${\bf k}$
and $\Delta$ is the superconducting gap
[see Fig.~\ref{inj:1}(c)]. Such a dispersion relation results in a
smaller diffusion coefficient and a longer spin flip time than
in the N region, 
while their product, the spin diffusion length, remains the same
\cite{Yamashita2002:PRB}. Consequently,
Eq.~(\ref{eq:mus}) also applies to  the diffusive 
spin-polarized transport and spin accumulation in  ferromagnet/superconductor 
structures \cite{Jedema1999:PRB,Yamashita2002:PRB}. 
Opening of the superconducting gap implies that a superconductor is a
low carrier system for spin, which is carried by quasiparticles 
\cite{Takahashi2003:PRB}.

In the preceding analysis, appropriate for bulk, homogeneous,
three-dimensional N and F regions and degenerate (semi)conductors,
Poisson's equation was not invoked and the local charge neutrality
$\delta n_\uparrow+\delta n_\downarrow$ was used only to derive
Eq.~(\ref{eq:spin}).\footnote{For spin injection in
nondegenerate semiconductors (with the carriers obeying the 
Boltzmann statistics)
there can be large effects due to 
built-in fields and deviation
from local charge neutrality, as discussed in Sec.~\ref{sec:IIC3}.}
Focusing on bulk samples in which both the size of the F and N regions and the
corresponding spin diffusion lengths are much larger than the Debye 
screening length, one can find that 
the quasineutrality condition, combined with the Eqs.~(\ref{eq:elchem})
and (\ref{eq:einstein}), yields
\begin{equation}
\label{eq:phi}
\phi=-\mu-P_{\cal N} \mu_s/2,
\end{equation}
where the density of states spin polarization of $P_{\cal N}$ vanishes 
in the N region.
At the contact $x=0$ there is a potential drop, even when $r_c=0$,
which can be evaluated from
Eqs.~(\ref{eq:mubc}) and (\ref{eq:phi}) as
\begin{equation}
\label{eq:contact}
\phi_N(0)-\phi_F(0)=-r_c [1-P_\Sigma P_j] j + P_{{\cal N}F}(0)\mu_{sF}(0)/2.
\end{equation}
The creation of nonequilibrium spin in the N region results in the 
spin EMF in the F/N structure 
which 
can be used to detect electrical spin injection,
as depicted in Fig.~\ref{inj:1}. Within a simplified semi-infinite geometry
for the F and N regions, we consider an effect of spin 
pumping in the N region, realized either by electrical spin injection from
another F region [as shown in Fig.~\ref{inj:1}(b)] or by optical pumping 
(see Sec.~\ref{sec:IIB}).  The resulting potential drop can calculated by
modifying $\mu_{sN}$ in Eq.~(\ref{eq:decay}), 
\begin{equation}
\mu_{sN}=\mu_{sN}(\infty)
+[\mu_{sN}(0)-\mu_{sN}(\infty)]e^{-x/L_{sN}},  
\end{equation}
where $\mu_{sN}(\infty)$ 
represents the effect of homogeneous spin pumping
in the N region. To calculate the open circuit voltage ($j=0$)  the continuity
of spin current at $x=0$ should be combined with the fact that $P_j j=j_s$. 
From Eq.~(\ref{eq:Pj}) it follows that
\begin{equation}
\label{eq:js0}
j_s(0)=2\frac{\sigma_\uparrow \sigma_\downarrow}{\sigma_F} 
\frac{\mu_{sF}(0)}{L_{sF}}=
-\frac{1}{2}\sigma_N \frac{\mu_{sN}(0)-\mu_{sN}(\infty)}{L_{sN}}, 
\end{equation}                                                            
while the discontinuity of $\mu_s$ in Eq.~(\ref{eq:musbc}) 
yields\footnote{A missprint
in $\mu_{sF}(0)$ from \textcite{Rashba2002:EPJ} has been corrected.}
\begin{eqnarray} \nonumber
\mu_{sF}(0)&=&(r_F/r_{FN})\mu_{sN}(\infty), \quad
j_s(0)=\mu_{sN}(\infty)/2 r_{FN}, \\ 
\label{eq:musNF}
\mu_{sN}(0)&=&[(r_c+r_F)/r_{FN}]\mu_{sN}(\infty).
\end{eqnarray} \nonumber
By substituting this solution into Eq.~(\ref{eq:contact}),
we can evaluate
the contact 
potential drop
can be evaluated as
\begin{equation}
\label{eq:contact2}
\phi_N(0)-\phi_F(0)= \left [ r_F P_{{\cal N}F}+r_c P_\Sigma \right ]
\mu_{sN}(\infty)/2 r_{FN}. 
\end{equation}
The total potential drop (recall $j=0$) at the F/N junction\footnote{A similar
potential drop was also calculated across
a ferromagnetic domain wall \cite{Dzero2003:PRB}.} is \cite{Rashba2002:EPJ}
\begin{equation}
\label{eq:drop}
\Delta \phi_{FN}=\phi_N(\infty)-\phi_F(-\infty)=P_j\mu_{sN}(\infty)/2. 
\end{equation}
where $P_j$ is given in Eq.~(\ref{eq:gamma}).
In the context of the spin-detection scheme from Fig.~\ref{inj:1} and high
impedance measurements at the N/F2 junction, the spin-coupled voltage $V_s$ 
\cite{Silsbee1980:BMR,Johnson1985:PRL} was also found to be proportional
to current polarization and the spin accumulation 
($\mu_s \propto \delta s \propto \delta M$) \cite{Johnson1988:JAP}.

\subsubsection{\label{sec:IIC2} F/N/F junction}
\label{F/N/F}

The above 
analysis of the F/N bilayer can be readily extended to the geometry
in which the infinite F regions are separated by an N region 
of thickness $d$. 
The quantities pertaining to the two ferromagnets
are defined as in the case of an F/N junction and labeled
by the superscripts L and R (left and right regions, respectively).
It follows from Eq.~(\ref{eq:Pj}), 
by assuming the continuity of the spin current at L,R, 
that the difference of the spin-resolved
electrochemical potential, responsible for the spin accumulation, is
\begin{eqnarray} 
\label{eq:musL}
\mu_{sF}^L&=&2 r_F^L (P_j^L-P_{\sigma F}^L) j e^{x/L_{sF}^L}, \quad x<0, \\ 
\nonumber
\mu_{sN}&=&2 r_N \left [ P_j^R \cosh(x/L_{sN}) 
-P_j^L\cosh[(d-x)/L_{sN}] \right  ]  \\   
&\times& j/\sinh(d/L_{sN}), \quad 0<x<d, \\ 
\label{eq:musN}
\mu_{sF}^R&=&-2 r_F^R (P_j^R-P_{\sigma F}^R) j e^{(d-x/L_{sF}^R)}, \quad x>d, 
\label{eq:musR}
\end{eqnarray}
where the current spin polarization $P_j^{L,R}$ at the two contacts
in the F/N/F geometry can be expressed \cite{Rashba2002:EPJ} in terms of 
the $P_j$ calculated for F/N junction with the infinite F and N regions
in Eq.~(\ref{eq:Pj}) and the appropriate effective resistances. 
By $P_{j\infty}^{L,R}$ we denote the $P_j$ calculated in Eq.~(\ref{eq:gamma})
for at left
and right contact (with the appropriate parameters for the F/N/F junction)
as if it were surrounded by the infinite F and N regions.
Analogously to the F/N junction, the consequence of the spin injection
is the increase of the resistance $R=R_0+\delta R$, as compared to the 
equilibrium value $R_0=(\Sigma^L)^{-1}+(\Sigma^R)^{-1}$. 
The nonequilibrium resistance $\delta R$ is also always
positive for spin-conserving contacts \cite{Rashba2002:EPJ,Rashba2000:PRB},
in agreement with experiments on all-semiconductor trilayer structures
\cite{Schmidt2001:PRL}; see Sec.~\ref{sec:IID3}.

Many applications 
based on magnetic multilayers rely on the spin-valve effect
in which
the resistance changes due to the relative orientations of the magnetization
in the two F regions. The geometry considered here is relevant for CPP GMR 
\cite{Parkin:2002,Bass1999:JMMM,Gijs1997:AP} and all-metallic
spin injection of \textcite{Johnson1985:PRL}.
In particular, the resistance change between 
antiparallel and parallel magnetization orientations in the two 
ferromagnets can
be expressed using current polarization of an infinite F/N junction 
$P_{j\infty}^{L,R}$
\cite{Rashba2002:EPJ}:
\begin{equation}
\label{eq:DelR}
\Delta R= R_{\uparrow \downarrow} - R_{\uparrow \uparrow} =
4 P_{j\infty}^L P_{j\infty}^R \frac{r_{FN}^L r_{FN}^R r_N}{{\cal D} 
\sinh(d/L_{sN})}, 
\end{equation}
where $r_{F}^{L,R}$, $r_c^{L,R}$, and $r_N$ are defined as in 
the case of an F/N junction
and
\begin{eqnarray}
\label{eq:det}
{\cal D}&=&(r_F^L+r_c^L)(r_c^R+r_F^R)  \\ \nonumber
&+& r_N^2 +r_N(r_F^L+r_c^L+r_c^R+r_F^R) \coth(d/L_{s N}).  
\end{eqnarray}
Up to a factor of 2, Eq.~(\ref{eq:DelR}) has also been obtained by 
\textcite{Hershfield1997:PRB} using Onsager relations. 
In the limit of a thin N region, 
$d/L_{sN} \rightarrow 0$, 
$\Delta R$ remains finite. In the opposite limit, for $d \gg L_{sN}$, 
\begin{equation}
\Delta R \sim  P_{j\infty}^L P_{j\infty}^R \exp(-d/L_{sN}).
\label{rd}
\end{equation}
For a symmetric F/N/F junction, 
where $r_{c,F}^{L}=r_{c,F}^{R}$, it follows that
\begin{equation}
\label{eq:DelRS}
\Delta R=\frac{4 r_N (r_c P_\Sigma + r_F P_{\sigma F})}
{ {\cal D} \sinh(d/L_{sN})}. 
\end{equation}

Considering the spin injection from F into a ballistic 
N region in the presence of diffusive interfacial scattering,
where the phase coherence is lost and the Boltzmann equation can be 
applied,
it is 
instructive to reconsider the effect of contact 
resistance \cite{Kravchenko2003:PRB}. We introduce the Sharvin resistance 
$R_{\rm Sharvin}$ 
\cite{Sharvin1965:ZETF}, arising in ballistic transport between the 
two infinite
regions connected by a contact (an orifice or a narrow and short constriction)
of  radius much smaller than the mean free path, $a \ll l$. 
In a 3D geometry the resistance is
\begin{equation}
\label{eq:sharvin}
R_{\rm Sharvin}=\frac{4 \rho l}{3 \pi a^2}=\left [ \frac{e^2}{h} 
\frac{k^2 A}{2 \pi} \right ]^{-1}, 
\end{equation}
where 
$h/e^2\approx 25.81$ k$\Omega$ is the quantum of resistance per spin, 
$A$ is the contact
area, and $k$ is the Fermi wave vector.
The opposite limit,  of  diffusive transport through the contact with
$a \gg l$, corresponds to the the Maxwell or Drude resistance
$R_{\rm Maxwel}=\rho/2 a$. The studies of intermediate cases provide an
interpolation scheme between the $R_{\rm Maxwell}$ and $R_{\rm Sharvin}$ 
for  various ratios of $a/l$
\cite{Wexler1966:PPSL,Jansen1980:JPCSSP,deJong1994:PRB,Nikolic1999:PRB}.
Following \textcite{Kravchenko2003:PRB} the effective contact resistance 
$r_c=r_{c\uparrow}+r_{c\downarrow}$ 
(recall that it is defined per unit area)
is obtained as
\begin{equation}
\label{eq:contact22}
r_{c\lambda}=(4 R_{\rm Sharvin}/A) (1-t^L_\lambda-t^R_\lambda)/t^L_\lambda,
\end{equation}
where $t^{L,R}_\alpha$ represent the transmission coefficients 
for the electrons 
reaching the contact from the left and from the right and satisfy
$t^L+t^R \le 1$.
For $r_c$ which would exceed the resistance of the N and F bulk regions 
the spin injection efficiency can attain 
$P_j\sim (r_{c \uparrow}-r_{c \downarrow})/r_c$
\cite{Kravchenko2003:PRB},
showing, similarly to the diffusive regime, the importance of the 
resistive contacts
to efficient spin injection. Connection with the results in the 
diffusive regime can be obtained \cite{Kravchenko2003:PRB}
by identifying 
$r_{c\lambda}=1/4 \Sigma_\lambda$, where the contact conductivity 
$\Sigma_\lambda$ was introduced in  Eq.~\ref{eq:sigma}. 

While most of the experimental results on spin injection are feasible in the 
diffusive regime, there are many theoretical studies treating the
ballistic case and phase-coherent transport both in F/N and
F/N/F junctions
\cite{Mireles2001:PRB,Matsuyama2002:PRB,Hu2001:PRB,Hu2001:PRL}.
Simple models in which the N region is a degenerate semiconductor often adopt an
approach developed first for charge transport in  junctions involving 
superconductors, discussed in Sec.~\ref{sec:IVA3}. Considering spin-orbit
coupling and the potential scattering at the F/N interface modeled by the 
$\delta$-function, \textcite{Hu2001:PRL} have examined ballistic
spin injection in the F/N junction. They show that even a 
spin-independent barrier 
can be used to enhance the spin injection and lead to an increase in the
conductance polarization. First-principles calculations were also used for 
ballistic spin injection from a ferromagnetic metal into a 
semiconductor \cite{Wunnicke2002:PRB,Mavropoulos2002:PRB,Zwierzycki2003:PRB}.
In the limit of coherent (specular) scattering\footnote{The 
wave-vector 
component 
along the interface is conserved during scattering.}
and high interfacial quality it was shown that 
different band structure in the F and the N regions would contribute to a 
significant 
contact resistance and  an efficient spin injection \cite{Zwierzycki2003:PRB}.

\subsubsection{\label{sec:IIC3} Spin injection through the space-charge region}

Interfaces making up a  semiconductor often
develop a space-charge region---a region of local macroscopic charges.
Typical examples are the Schottky contact and the depletion layer in 
{\it p-n} junctions. 
While phenomenological models, such as the one introduced 
in Sec.~\ref{sec:IIC1}, capture a remarkable wealth of spin injection physics, 
they carry little information about spin-dependent processes
right at the interfaces. Microscopic studies of spin-polarized transport and
spin-resolved tunneling through space-charge regions are still 
limited in scope. The difficulty lies in the need
to consider self-consistently simultaneous charge accumulation and
electric-field generation (through Poisson's equation), both affecting 
transport. 
Non-self-consistent analyses of a Schottky barrier spin injection 
were performed in \cite{Prins1995:JPCM,Albrecht2002:PRB,Albrecht2003:PRB}, 
while
\textcite{Osipov2003:P} proposed an efficient spin 
injection method using a $\delta$-doped Schottky contact.

Let us now consider spin injection through the depletion
layer in magnetic {\it p-n} junctions 
\cite{Zutic2002:PRL,Fabian2002:PRB,Zutic2003:APL}.
The physics is based on drift and diffusion\footnote{Tunneling or field 
emission becomes important, for example, 
in thin Schottky barriers
or in {\it p-n} junctions and heterostructures at large reverse biases 
\cite{Kohda2001:JJAP,Johnston-Halperin2002:PRB,vanDorpe2003:P}.} 
limited
by carrier recombination and spin relaxation, as described in more detail
in Sec.~\ref{sec:IVA4}. The transport equations are solved
self-consistently  with Poisson's equation, 
taking full account of electric field due to accumulated charges.
Additional examples of magnetic {\it p-n} junctions are discussed
in Sec.~\ref{sec:IVD}. 

The system is depicted in Fig.~\ref{deplete}. The {\it p-n} junction 
has a magnetic 
$n$-region\footnote{Equilibrium magnetization can be 
a consequence of doping with magnetic impurities, yielding large carrier 
$g$ factors, and applying magnetic
field, or of using a ferromagnetic semiconductor 
\cite{Ohno1998:S,Pearton2003:JAP}.}
with a net equilibrium electron spin $P_{n0}^R$, where $R$ stands
for the right (here $n$) region. Holes are assumed to be unpolarized. 
An important issue to be resolved is whether
there will be spin accumulation in the 
$p$-region 
if a forward bias is applied to the junction.
In other words, will spin be injected across the depletion layer? Naively 
the answer is yes, since spin is carried by electrons, but the result shown 
in Fig.~\ref{deplete} suggests a more complicated answer. 
At small biases there is
no spin injection. This is the normal limit of diode operation,
in which
where the
injected carrier density through the depletion region is still smaller 
than the equilibrium carrier density. Only with
bias increasing to the high-injection limit (typically above 1 V)  
is spin injected.
 
\begin{figure}
\centerline{\psfig{file=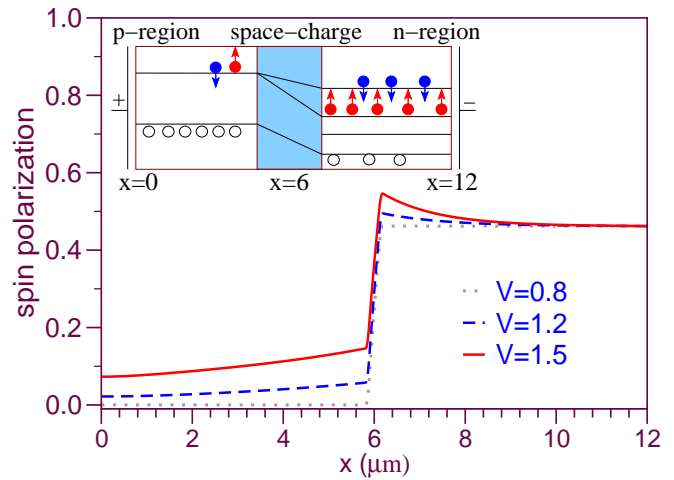,width=1\linewidth,angle=0}}
\caption{Spin injection through the space-charge region of a magnetic 
{\it p-n}
junction. The geometry 
is depicted in the inset, which shows  
a junction
with a spin-split conduction band in the 
$n$-region
with spin-polarized electrons (solid circles) and 
unpolarized holes (empty circles).
Under applied forward bias $V$ the charge current flows to the right. 
The curves, labeled by $V$, show the electron density polarization profiles 
$P_n(x)$ for the depicted geometry and GaAs materials parameters. 
The equilibrium density polarization in the $n$-region 
is about 0.5.
At low bias (0.8 V) 
there is no
spin injection. Spin injection, manifested by the
increase of $P_n$ in the 
$p$-region, 
appears only at large biases
(1.2 and 1.5 V), 
where it is driven by electric drift \cite{Zutic2002:PRL}.
Spin polarization of the current is discussed by  
\textcite{Zutic2001:APL,Fabian2002:PRB}.
Adapted from \onlinecite{Zutic2002:PRL}. 
}
\label{deplete}
\end{figure}

The explanation for the absence of spin injection at small biases 
and for nondegenerate doping levels (Boltzmann statistics are applicable)
is as follows.
On the $n$ side, there are more spin up than spin down electrons,
$n_\uparrow > n_\downarrow$.
If $2q\zeta$ is the spin splitting of the
conduction band, $n_\uparrow(\zeta)/n_\uparrow(\zeta=0)=\exp(q\zeta/k_BT)$.
Under a forward bias, electrons 
flow to the 
$p$-region. 
The flow is limited by thermal activation over
the barrier (given by the built-in electrostatic potential  
minus bias),
which is, for the spin up electrons, greater by $q\zeta$. For Boltzmann
statistics, the rate of transmission of spin up electrons
over the barrier is 
$\sim \exp(-q\zeta/k_BT)$.
Since current is proportional to both the carrier density and the transmission
rate, the two exponential factors cancel out. Similarly for spin down.
As a result, the spin-resolved current is unaffected by $2q\zeta$ 
and there is no spin current flowing through the depletion
layer. There is no spin accumulation.
Spin injection appears only at large biases, where it is driven
by electric drift leading to nonequilibrium 
spin population already in the $n$-region \cite{Fabian2002:PRB,Zutic2002:PRL}. 
In addition to spin injection, spin extraction has also been
predicted in magnetic {\it p-n} junctions with a magnetic
$p$-region 
\cite{Zutic2002:PRL}.  Under a large bias, spin is extracted 
(depleted) from the nonmagnetic $n$-region. 

Electric field in the bulk regions next to the space charge is important only 
at large biases. It affects not only spin density, but spin diffusion
as well. That spin injection efficiency can increase in the presence
of large electric fields due to an increase in the spin diffusion length 
(spin drag) was first shown by \textcite{Aronov1976:SPS},
and was later revisited by other authors.\footnote{See, for example,
\cite{Zutic2001:PRB,Fabian2002:PRB,Margulis1994:PB,Martin2003:PRB,%
Yu2002:PRBa,Flensberg2001:PRB,Vignale2003:SST,Bratkovsky2003:P}.} 
To be important, 
the electric field
needs to be very large,\footnote{The critical magnitude is
obtained by dividing a typical energy, such as the thermal or Fermi 
energy, by $q$ and by the spin diffusion length. At room temperature the thermal 
energy is 25 meV, while the spin diffusion length can be several
microns.} more than 100 V/cm 
at room temperature. While such large fields are usually present inside the
space-charge regions, they exist in the adjacent bulk regions only at
the high injection limit and  
affect transport and spin injection.
In addition to electric drift, magnetic drift,
in magnetically inhomogeneous semiconductors, can also enhance spin injection
\cite{Fabian2002:PRB}. 

The following formula was obtained for spin injection at small
biases \cite{Fabian2002:PRB}:
\begin{equation}
P_{n}^L=\frac{P_{n0}^L[1-(P_{n0}^R)^2]+
\delta P_{n}^R(1-P_{n0}^LP_{n0}^R)}
{1-(P_{n0}^R)^2+\delta P_{n}^R(P_{n0}^L-P_{n0}^R)},
\label{eq:aL}
\end{equation}
where $L$ (left) and $R$ (right) label the edges of the
space-charge (depletion) region of a {\it p-n} junction. Correspondingly, 
$\delta P_n^R$ represents  the nonequilibrium
electron  polarization, evaluated at $R$, arising from a 
spin source. The case
discussed in Fig.~\ref{deplete} is for $P_{n0}^L=\delta P_n^R=0$. Then
$P_n^L=0$, in accord with the result of no spin injection. For a homogeneous
equilibrium magnetization ($P_{n0}^L=P_{n0}^R$), 
$\delta P_{n}^L=\delta P_{n}^R$;
the nonequilibrium spin polarization is the same across the depletion layer.
Equation (\ref{eq:aL}) demonstrates that only {\it nonequilibrium} spin, 
already
present in the bulk region, can be transferred through the depletion
layer at small biases \cite{Zutic2001:PRB,Fabian2002:PRB}.  
Spin injection of nonequilibrium spin is also very effective if 
it proceeds from the 
$p$-region 
\cite{Zutic2001:PRB},
which is the case for a spin-polarized solar cell \cite{Zutic2001:APL}. 
The resulting spin 
accumulation in the $n$-region 
extends the spin diffusion range,
leading to spin amplification---increase of the spin population away from 
the spin source. 
These results were also confirmed in the junctions
with two differently doped $n$-regions 
\cite{Pershin2003:PRL,Pershin2003:Pb}.
Note, however, that  
the term ``spin polarization density'' used in
\textcite{Pershin2003:PRL,Pershin2003:Pb} 
is actually the spin density $s=n_\uparrow-n_\downarrow$,
not the spin polarization $P_n$. 

Theoretical understanding of spin injection has focused largely on 
spin density while neglecting spin phase, which is 
important for some 
proposed spintronic applications. The problem of
spin evolution in various transport modes (diffusion, tunneling,
thermionic emission) remains to be investigated. Particularly relevant
is the question of whether spin phase is conserved during spin injection.
\textcite{Malajovich2001:N} showed, by studying spin evolution 
in transport through a 
n-GaAs/n-ZnSe
heterostructure, 
that the phase can indeed be preserved.

\subsection{\label{sec:IID} Experiments on spin injection}

\subsubsection{\label{sec:IID1} Johnson-Silsbee spin injection}

The first spin polarization of electrons by electrical spin injection 
\cite{Johnson1985:PRL} was demonstrated in a ``bulk wire'' of 
aluminum on which 
an array of thin film permalloy (Py) pads (with 70 \% nickel and 30 \% iron)  
was deposited spaced in multiples of 50 $\mu$m, center to center 
\cite{Johnson1988:PRBb}
to serve as spin injectors and detectors. In one detection scheme
a single ferromagnetic pad was used as a spin injector while the distance
to the spin detector was altered by selecting different Py pads to
detect $V_s$ and through the spatial decay of this spin-coupled voltage 
infer $L_{sN}$.\footnote{The spin relaxation time in a ferromagnet is 
often assumed to be
very short.  Correspondingly, in the analysis of the experimental data, 
both the spin 
diffusion length and $\delta M$ are taken to vanish in the F region  
\cite{Silsbee1980:BMR,Johnson1985:PRL,Johnson1988:PRBa,Johnson1988:PRBb}.}
This procedure is illustrated in 
Fig.~\ref{injexp:1}, where 
the separation between the spin injector and detector $L_x$ is variable. 

\begin{figure}
\centerline{\psfig{file=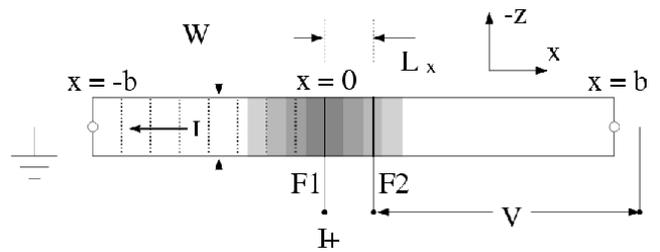,width=1.\linewidth,angle=0}}
\caption{Schematic top view of nonlocal, quasi-one-dimensional 
geometry used by \textcite{Johnson1985:PRL}: F1 and F2, 
the two metallic ferromagnets having magnetizations in the $x-z$ plane;  
dotted lines, equipotentials characterizing electrical current flow;
grey shading, diffusing population of nonequilibrium spin-polarized
electrons injected at $x=0$, with darker shades corresponding to higher density 
of polarized electrons. From \onlinecite{Johnson2002:SST}.} 
\label{injexp:1}
\end{figure}
\textcite{Johnson1985:PRL} 
point out that in the depicted geometry  
there
is no flow of the charge current for $x >0$ and that 
in the absence of 
nonequilibrium spins a voltage measurement between $x=L_x$ and $x=b$ gives zero.
Injected spin-polarized electrons will diffuse symmetrically (at low current
density the effect of electric fields can be neglected), and the 
measurement of voltage will give a spin-coupled signal $V_s$ related to the 
relative orientation of magnetizations in F1 and F2.\footnote{This method for
detecting the effects of spin injection is also referred to as a
potentiometric method.}
The results, corresponding to the polarizer-analyzer detection and the
geometry of Fig.~\ref{injexp:1}, 
are given in Fig.~\ref{injexp:2}. An in-plane field 
(${\bf B}$ $\|$ ${\bf \hat{z}}$), 
of a magnitude several times larger
than a typical field for magnetization reversal, $B_0 \approx 100$ G,
is applied to define the direction of magnetization in the injector 
and detector.  
As the field sweep is performed, from negative to positive values, 
at $B_{01}$ there is 
a reversal of magnetization in one of the ferromagnetic films accompanied 
by a sign change in the spin-coupled signal. As $B_z$ is further
increased, at approximately $B_{02}$, there is 
another reversal of magnetization,
resulting in parallel orientation of F1 and F2 and a $V_s$ 
of magnitude similar to that for the previous parallel orientation 
when $B_z < B_{01}$.

\begin{figure}
\centerline{\psfig{file=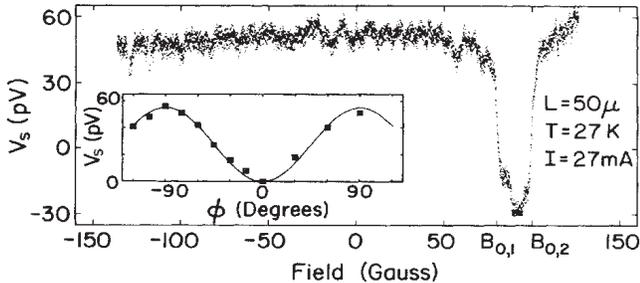,width=1.\linewidth,angle=0}}
\caption{Spin injection data from bulk Al wire sample.
Negative  magnetic field is applied parallel to the magnetization (-$z$ axis) 
in the two ferromagnetic regions. As the field is increased, at $B_{0,1}$ 
magnetization in one of the ferromagnetic regions is reversed, and at $B_{0,2}$
the magnetization in the other region is also reversed 
(both are along +$z$ axis).
Inset: amplitude of the observed Hanle signal as a function of orientation 
angle $\phi$ of magnetic field. From \onlinecite{Johnson1985:PRL}.} 
\label{injexp:2}
\end{figure}

A more effective detection of the spin injection 
is realized through measurements of the Hanle effect, 
also discussed Secs.~\ref{sec:IIB} and \ref{sec:IIIA2}, and described by 
the Bloch-Torrey equations \cite{Bloch1946:PR,Torrey1956:PR}
[see Eqs.(~\ref{eq:relax:bloch1})--(\ref{eq:relax:bloch3})].
The inset of Fig.~\ref{injexp:2} summarizes results from a
series of Hanle experiments on a single sample. For the
Hanle effect ${\bf B}$ must have a component perpendicular to the orientation 
axes
of the injected spins. Only projection of ${\bf B}$ perpendicular to 
the spin axis
applies a torque and dephases spins. The magnitude of ${\bf B}$, applied at
an angle $\phi$ to the $z$-axis in the $y-z$ plane, is small enough that the
magnetizations in ferromagnetic thin films remain in the $x-z$ plane (see 
Fig.~\ref{injexp:1}). If, at ${\bf B}=0$, injected nonequilibrium magnetization
is $\delta M(0) \hat{z}$ 
then at finite field $\delta {\bf M}$ precesses about
${\bf B}$ with the cone of angle $2\phi$. After averaging over several cycles,
only $\delta M(0) \cos \phi$, 
the component $\|$ ${\bf B}$, will survive.
The voltage detector\footnote{Recall from the discussion leading to 
Eq.~(\ref{eq:drop}) that the spin-coupled signal $\propto \delta M$.}
senses the remaining part of the magnetization projected on the axis
of the detector 
$\delta M(0) \cos \phi \times \cos\phi$ \cite{Johnson1988:PRBa}.
The predicted angular dependence for the  amplitude of the Hanle signal 
(proportional to the  depolarization of $\delta M$ in a finite field) 
$[\delta M(0) -\delta M(0) \cos^2\phi]$ 
is plotted in the inset together
with the measured data.\footnote{The range of the angle $\phi$, 
in the inset, is corrected from the one originally given in Fig.~3 of 
\textcite{Johnson1985:PRL}.} Results confirm the first application of
the Hanle effect to dc spin injection. 

The Hanle effect was also studied theoretically by solving
the Bloch-Torrey equations for an arbitrary orientation, characterized by the
angle $\alpha$, between the magnetization
in F1 and F2 \cite{Johnson1988:PRBa}. From the Hanle curve [$V_s(B_\bot)$]
measured at $T=4.3$ ($36.6$) K,  
the parameters $L_s=450$ ($180$) $\mu$m and 
$P_\Sigma=0.06$ ($0.08$) were extracted.\footnote{The fitting parameters are 
$\tau_s$, $P_\Sigma$, and $\alpha$
\cite{Johnson1988:PRBb}, and since the diffusion coefficient is obtained from 
Einstein's relation $L_s$ is known.}
This spin injection technique using a few pV resolution of
a  superconducting quantum interference device (SQUID) 
and with an 
estimated $P_\Sigma\approx 0.07$ provided an accuracy able to detect 
$P_n \approx 5\times 10^{-12}$, causing 
speculating on that
a single-spin sensitivity might be possible in  smaller samples 
\cite{Johnson1985:PRL,Johnson1988:PRBb}.
While in a good conductor, such as Al, the observed 
resistance  change 
$\Delta R$ was small ($\sim n\Omega$), 
the relative change at low temperatures and for 
$L_x \ll L_s$ was $\Delta R/R \approx 5 \%$, where $\Delta R$  
is defined as in Eq.~(\ref{eq:tmr}), determined by the relative orientation of 
the magnetization in F1 and F2, and $R$ is the
Ohmic resistance \cite{Johnson2002:SST}.
Analysis from Sec.~\ref{sec:IIC2} shows that the measurement 
of $\Delta R$ could 
be used to determine the product of injected current polarizations 
in the two F/N junctions.

The studies of spin injection were extended to the thin-film geometry, 
also known as the ``bipolar spin switch'' or ``Johnson spin transistor''
\cite{Johnson1993:S,Johnson1993:PRL} 
similar to the one depicted in Fig.~\ref{inj:1}(a). 
The measured spin-coupled signals
\footnote{$d \sim 100$ nm 
was much smaller then the separation between F1 and F2 in bulk Al wires 
\cite{Johnson1985:PRL}, and the amplitude of the Hanle effect
was about $10^4$ larger \cite{Johnson2002:SST}.}
in Au films were larger than the values obtained in bulk Al 
wires \cite{Johnson1985:PRL,Johnson1988:PRBb}.
A similar trend, 
$V_s\sim 1/d$, potentially important for applications,
was already anticipated by \textcite{Silsbee1980:BMR}.
The saturation of this increase can be 
inferred from Eqs.~(\ref{eq:DelR}) and (\ref{eq:det}) for 
$d \ll L_{sN}$ 
and has been discussed by \textcite{Hershfield1997:PRB} and
\textcite{Fert1996:PRB}. 

When polarizer-analyzer detection was used, one of the fitting parameters
from the measured data $P_\Sigma$ sometimes exceeded 1---which 
corresponds to complete
interfacial polarization. The origin of this discrepancy remains to 
be fully resolved
\cite{Hershfield1997:PRB,Fert1996:PRB,Johnson1993:PRL,Johnson2002:SST,Geux2000:APPA}.
Results obtained from the Hanle effect, on similar samples, gave the 
expected  $P_\Sigma <1$ values \cite{Johnson2002:SST}.\footnote{Theoretical 
estimates for $V_s$  from which $P_\Sigma > 1$ was inferred
are modified when one considers the Coulomb interaction and 
proximity effects---near the N/F interface the spin splitting 
of the carrier bands in the N region will be finite even at equilibrium.
Model calculations \cite{Chui1995:PRL,Chui1995:PRB}, 
which treat the F/N/F junction as a whole, show that the magnetic 
susceptibility 
$\chi$ in N can be much smaller than the free electron value and can increase 
the predicted 
$V_s \propto 1/\chi$. These corrections to the free-electron picture 
of an F/N/F junction
are smaller for larger $d$, as in the bulk-wire geometry of 
\textcite{Johnson1985:PRL},
where theoretical estimates of $V_s$ did not lead to $P_\Sigma > 1$.}  

A modification of the bipolar spin switch structure was used to demonstrate the
spin injection into a niobium film \cite{Johnson1994:APL},
realizing the theoretical assertion of \textcite{Aronov1976:SPJETP}
that nonequilibrium spin could be injected into a superconductor.
Two insulating Al$_2$O$_3$ films were inserted between F1 and F2 
(both made of Py) 
and a Nb film [see Fig.~\ref{inj:1}(b)]. The measurements were performed 
near the superconducting transition temperature $T_c$
with the data qualitatively similar, above and below $T_c$, to the spin-coupled
voltage, as obtained in the magnetic-field sweep from Fig.~\ref{injexp:2}.
The results were interpreted as support for enhanced  depletion of the 
superconducting condensate (and correspondingly the reduction of 
the critical current $I_c$) by spin-polarized  quasiparticles, as compared 
to the usual spin-unpolarized quasiparticle injection. Related measurements 
were recently performed in a CPP geometry \cite{Gu2002:PRB}, and the
penetration depth of the 
quasiparticle in the Nb films was measured to be $\sim 16$ nm, 
as compared to $2$ nm in 
\cite {Johnson1994:APL}. The corresponding temperature dependence of CPP GMR is
well explained by the theory of \textcite{Yamashita2003:PRB} and the 
modification
of Andreev reflection (see Sec.~\ref{sec:IVA3}) by spin polarization.

The spin injection technique of Johnson and Silsbee was also applied to 
semiconductors.
Initial experiments on using a metallic ferromagnet to inject spin into a 
two-dimensional
electron gas (2DEG) showed only a very low ($\sim 1 \%$) efficiency
\cite{Hammar1999:PRL}  
for which various
explanations were offered \cite{Monzon2000:PRL,vanWees2000:PRL,Hammar2000:PRL}. 
However, stimulated by the proposal of \textcite{Rashba2000:PRB}
to employ spin-selective diffusive contacts (Sec.~\ref{sec:IIC1}), 
the subsequent 
measurements have showed substantially more efficient spin injection 
into a 2DEG
after an insulating layer was inserted \cite{Hammar2001:APL,Hammar2002:PRL}. 
The geometry
employed is depicted in Fig.~\ref{injexp:1}. 
In interpreting 
the results, the spin-orbit coupling and the energy-independent density
of states 
at the Fermi level were taken into account \cite{Silsbee2001:PRB}. 
This topic is reviewed by \textcite{Tang:2002}. 

\begin{figure}
\centerline{\psfig{file=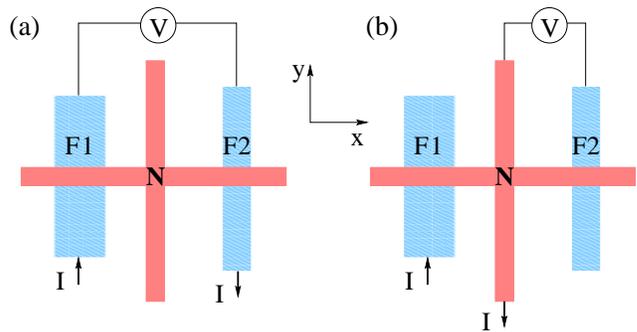,width=0.5\linewidth,angle=-90}}
\caption{Schematic representation of (a) local and (b) nonlocal geometry
used to measure the effects of spin injection and spin accumulation.}
\label{exp:3}
\end{figure} 

\subsubsection{\label{sec:IID2} Spin injection into metals}

An important part of the operation of  CPP GMR structures is the presence 
of nonequilibrium spin polarization in nonmagnetic metallic regions.
Studies of spin-injection parameters
in such systems have been reviewed by \textcite{Bass1999:JMMM} and
\textcite{Gijs1997:AP}.   
However, until recently, except for the work of Johnson and Silsbee,
there were few  other experimental studies directly concerned with 
spin injection into metals. A series of 
experiments, \cite{Jedema2001:N,Jedema2002:JS,Jedema2002:Na,Jedema2002:APL} 
at both low (4.2 K) and room temperature,
were performed using  
the van der Pauw geometry depicted in Fig.~\ref{exp:3}.
In various structures \cite{Jedema2002:T}
the two ferromagnetic regions (made of Py, Co, or Ni) 
were chosen to be of different sizes to provide different  
coercive fields, allowing an independent reversal of magnetization 
in F1 and F2.
The cross-shaped nonmagnetic region was made of Al or Cu \cite{Jedema2002:T}. 
Nonlocal measurements, similar to the approach shown in Figs.~\ref{inj:1} and
\ref{injexp:1} [discussed in \cite{Johnson1988:PRBb,Johnson1993:PRL}], were
shown to simplify the extraction of spurious effects (for example,
anisotropic magnetoresistance  and the Hall signal)
from effects intrinsic to spin injection, as compared
to the local or conventional  spin-valve geometry. 

In the first type of experiment the cross-shaped region was deposited directly
over the F region 
(Fig.~\ref{exp:3}), 
and the spin-coupled resistance $\Delta R$, defined
analogously to Eq.~(\ref{eq:tmr}),
was measured as a function of an in-plane magnetic filed. A theoretical 
analysis
\cite{Jedema2001:N, Jedema2002:JS} was performed assuming no interfacial
resistance ($r_c=0)$ and the continuity of the electrochemical 
potentials at the F/N interface (see Sec.~\ref{sec:IIC1}).
For a spin injection from Py into Cu, the maximum current polarization 
was obtained to be $P_j \approx 0.02$ at 4.2 K. The results for $\Delta R$
\cite{Jedema2001:N} scaled to the size of the samples
used by \textcite{Johnson1993:S,Johnson1993:PRL} were interpreted to be 
3-4 orders of magnitude smaller. As discussed in Secs.~\ref{sec:IIC1} and 
\ref{sec:IIC2}, the presence of interfacial
spin-selective resistance can substantially change the spin injection 
efficiency 
and influence the resistance mismatch between the F and N regions
[see Eq.~(\ref{eq:Pj})]. 
Estimates of how these considerations would  affect the results of 
\textcite{Jedema2001:N} 
were given by \textcite{Jedema2002:Nb}
as well as by others 
\cite{Takahashi2003:PRB,Johnson2003:PRB}, 
who analyzed 
the importance 
of multidimensional geometry. 
In addition to
comparing characteristic
values of contact resistance
obtained on different samples,\footnote{For example, the measured
resistance of clean F/N contacts in CPP GMR \cite{Bussmann1998:IEEETM} was 
used to infer that there is also a large contact resistance in 
all-metal spin injection experiments \cite{Johnson2002:N}.}
for a conclusive
understanding it will be crucial to have 
{\it in situ} measurements.

In analyzing data for the van der Pauw cross, a two-dimensional geometry 
has an important effect---while the electric current is following  
the paths depicted in Fig.~\ref{exp:3}, the spin current, through 
the diffusion of nonequilibrium spin,
would have similar flow in all four arms \cite{Johnson2002:SST}. This
is different from the usual (quasi)one-dimensional analysis in which 
spin and charge currents flow along the same paths. 
For a full
understanding of the van der Pauw cross geometry,  two-dimensional 
modeling might be necessary \cite{Johnson2003:PRB,Takahashi2003:PRB}.

In the second type of experiment, tunneling contacts were fabricated 
by inserting Al$_2$O$_3$ as an insulator into the regions where F1 and F2 
overlapped with the cross. By applying a transverse field B$_z$ 
(see Fig.~\ref{exp:3}) the precession of the injected nonequilibrium spin 
was controlled and the amplitude of the Hanle effect was measured
\cite{Jedema2002:Na,Jedema2002:APL},  as  outlined in Sec.~{\ref{sec:IID1}.
From Co/Al$_2$O$_3$/Al/Al$_2$O$_3$/Co structures
$L_s \approx 0.5$ $\mu$m was extracted at room temperature.
The analysis of the Hanle signal was performed  by averaging contributions
of different lifetimes \cite{Dyakonov:1984b}. This proved to be
equivalent to the 
\textcite{Johnson1988:PRBb}
solution to the Bloch-Torrey equations.

\subsubsection{\label{sec:IID3}  All-semiconductor spin injection}
\label{magneticSM}

\begin{figure}
\centerline{\psfig{file=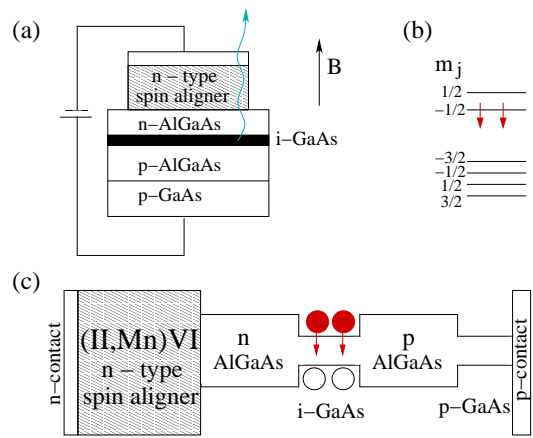,width=0.8\linewidth,angle=0}}
\caption{Schematic device geometry and band diagram of a spin LED:
(a) Recombination of
 spin-polarized electrons injected from
the (II,Mn)VI spin aligner and  unpolarized holes injected from the 
p-doped GaAs, in the intrinsic GaAs quantum well, 
producing circularly polarized  light; (b) conduction and valence bands of 
a spin aligner in an external magnetic field; 
(c) sketch of the corresponding
band edges and band offsets in the device geometry.
In the quantum well, spin down electrons and unpolarized holes
are depicted by solid and empty circles, respectively. 
Adapted from \onlinecite{Fiederling1999:N}.}
\label{injexp:3}
\end{figure} 

If a magnetic semiconductor could be used as a robust spin injector 
(spin aligner) 
into a  nonmagnetic semiconductor it would facilitate the integration of 
spintronics 
and semiconductor-based electronics. Comparable resistivities of magnetic 
and nonmagnetic semiconductors could provide efficient spin injection 
[see Eq.~(\ref{eq:gamma}), with $r_F\approx r_N$] even without using  
resistive contacts. Ultimately, for a wide range of applications and 
for compatibility with
complementary metal-oxide semiconductors (CMOS) 
\cite{Wong1999:PIEEE}, it would be 
desirable to be able to inject spin into silicon at room temperature.

Early studies \cite{Viglin1991:FTT,Osipov1990:PZETF,Viglin1997:PLDS,Osipov1998:PL}, 
which have since largely been ignored, used a Cr- and Eu-based 
chalcogenide ferromagnetic semiconductor (FSm) \cite{Nagaev:1983} as the spin 
injector.\footnote{These materials, while more difficult to fabricate 
than the subsequent 
class of III-V ferromagnetic semiconductors,  have desirable properties of 
providing injection of spin-polarized electrons 
(with spin lifetimes typically
much longer than for holes) and large spin splitting 
[$\sim 0.5$ eV at 4.2 K for  n-doped HgCr$_2$Se$_4$ \cite{Nagaev:1983}]
with nearly complete spin polarization 
and a Curie temperature $T_C$ of up to 130 K 
(HgCr$_2$Se$_4$) \cite{Osipov1998:PL}.}
The experiments were motivated
by the theoretical work of \textcite{Aronov1976:SPS,Aronov1976:SPJETP} 
predicting that the ESR signal, proportional to the steady-state
magnetization, would be changed by spin injection. The measurements of 
\textcite{Viglin1991:FTT,Osipov1990:PZETF} prompted a 
related prediction \cite{Margulis1994:PB} that spin injection could be 
detected through changes in electric dipole spin resonance (EDSR).
EDSR is the spin-flip resonance absorption for conduction electrons
at Zeeman frequency, which is excited by the electric-field 
vector of
an incident electromagnetic wave. The theory of EDSR, developed 
by \textcite{Rashba1961:SPSS} is extensively reviewed by
\textcite{Rashba:1991}.    

Ferromagnetic semiconductor 
spin injectors 
formed {\it p-n} and {\it n-n} heterostructures with a 
nonmagnetic semiconductor
InSb. The choice of InSb is  very suitable due to its large 
negative ($\sim -50$) $g$ factor \cite{McCombe1971:PRB}, 
for detecting the effects of spin injection through ESR. The observed absorption
and emission of microwave power \cite{Osipov1998:PL} was tuned by an 
applied magnetic field (from 35 GHz at $\approx$ 400 G up to 1.4 THz at 20 kG)
and only seen when electrons flowed  from FSm into an Sm region. 
The injection to the lower Zeeman level increased the ESR absorption,
while injection to the higher Zeeman level, leading to  population 
inversion,
generated  microwave emission.

The most recent experiments using semiconductor spin injectors can be grouped 
into two different classes, In one approach (II,Mn)VI paramagnetic 
semiconductors 
were employed as the spin aligners. 
These included
CdMnTe \cite{Oestreich1999:APL}, BeMnZnSe \cite{Fiederling1999:N}, and
ZnMnSe \cite{Jonker2000:PRB}.  
In the second approach 
ferromagnetic semiconductors like 
(Ga,Mn)As \cite{Ohno1999:N,Chun2002:PRB,Mattana2002:PRL} were used.
Both approaches were also employed to inject spins into 
CdSe/ZnSe \cite{Seufert2004:PRB} and InAs
\cite{Chye2002:PRB} quantum dots, respectively.

In (II,Mn)VI materials, at low Mn-concentration and at low temperatures,
there is a  giant Zeeman splitting $\Delta E = g^* \mu_B H$ 
\cite{Furdyna1988:JAP,Gaj:1988}
of the conduction  band,
in which $g^*$ is the effective electron  $g$ factor.
Such splitting arises due to 
{\it sp-d} exchange between the spins of
conduction electrons and the S=5/2 spins of the localized Mn$^{2+}$ ions.
The $g^*$ factor for $H \ne 0$
can exceed\footnote{At low temperatures ($\sim1$ K)
Cd$_{0.95}$Mn$_{0.05}$Se has $|g^*|>500$
\cite{Dietl:1994}, while in n-doped (In,Mn)As $|g^*|>100$
at 30 K \cite{Zudov2002:PRB}. Such large $g$ factors, in the
presence of a highly inhomogeneous magnetic filed could lead to the
charge carrier localization \cite{Berciu2003:PRL}.} 100
and is given by \cite{Furdyna1988:JAP,Brandt1984:AP}
\begin{equation}
g^*=g+\alpha M/(g_{Mn} \mu_B^2 H),
\label{eq:geff}
\end{equation}
where $g$ is the $H=0$ II-VI ``band'' value $g$, generally different from the
free-electron value, magnetization
 M$\propto$ $\langle S_z \rangle\propto B_s[(g_{Mn} \mu_B S H)/(k_B T)]$, B$_s$ 
is the Brillouin function 
\cite{Ashcroft:1976}, and $\alpha$ 
is the exchange integral for $s$-like $\Gamma_6$ electrons
(see Table I in Sec.~\ref{sec:IIB}), given by 
\cite{Furdyna1988:JAP} 
\begin{equation}
\alpha\equiv\langle S|J_{sp-d}|S \rangle/V_0,
\label{eq:alpha}
\end{equation}
where $J_{sp-d}$ is the electron-ion exchange coupling,
and V$_0$ is the volume of an elementary cell.
From Eqs.~(\ref{eq:geff}) and (\ref{eq:alpha}) it follows that 
$g^*=g^*(H)$
can even change its sign. 
Similar analysis applies also to $g$ factors of holes, with the 
Zeeman splitting of a valence band being typically several times larger
than that of a conduction band \cite{Brandt1984:AP}.

(II,Mn)VI materials can be incorporated in high quality heterostructures 
with different optically active III-V nonmagnetic semiconductors which,
by providing 
circularly polarized luminescence, can also serve
as spin detectors. In this case 
carriers
are excited by electrical means and we speak of 
electroluminescence  rather then photoluminescence.
The selection rules for the recombination light are
the same as discussed in Sec.~\ref{sec:IIB}.

Figure \ref{injexp:3} depicts a scheme for
realization of all-semiconductor electrical spin injection and  
optical detection \cite{Fiederling1999:N,Jonker2000:PRB}. 
Displayed is a spin light-emitting diode (LED) \cite{Jonker1999:PA}
in a Faraday geometry where both the applied B-field and the 
direction of propagation of the
emitted light 
lie along the growth direction.
Similar to a an ordinary
LED \cite{Sze:1981}, electrons and holes recombine
(in a quantum well or a {\it p-n} junction) and produce electroluminescence. 
However,
in a spin LED,  as a consequence of radiative recombination of spin-polarized 
carriers, the emitted light is circularly polarized. 
In experiments of  
\textcite{Fiederling1999:N,Jonker2000:PRB}, at B$\approx 1$ T, 
$T\approx 4$ K, 
and forward bias, electrons entering from the $n$-contact were almost 
completely polarized 
in the spin down state as they left the spin aligner and are injected 
across the (II,Mn)VI/AlGaAs interface. The electrons further traveled
(by drift and diffusion) to an intrinsic GaAs quantum well (QW) 
where they recombined with the unpolarized holes, which were injected from the 
p-doped GaAs.\footnote{The spatial separation and  spin relaxation
between the spin injection and the point of spin detection (in QW)
make a fully quantitative analysis of the injected polarization more difficult. 
It would be valuable to perform realistic calculations of a spin-polarized
transport and spin injection which would treat the whole spin LED as a single
entity.} 

The efficiency of  electrical spin injection across 
the (II,Mn)VI/AlGaAs interface was studied \cite{Fiederling1999:N}
using $P_{\rm circ}$ (defined in Sec.~\ref{sec:IIB}) of electroluminescence, 
as a function of 
B and  the thickness of the magnetic spin aligner 
(0 nm, 3 nm, and 300 nm, respectively).
$P_{\rm circ}$ increased with the
thickness of the magnetic layer, suggesting the finite spin relaxation time
needed for initially unpolarized electrons to relax into the lower 
(spin down) Zeeman level. 
The results of \textcite{Jonker2000:PRB} were similar to those of 
\textcite{Fiederling1999:N} 
for the thickest magnetic region. The behavior of $P_{\rm circ}$(B), up to the 
saturation value (B$\approx 3$ T), 
could be well explained by the magnetization
described with the Brillouin function \cite{Furdyna1988:JAP,Gaj:1988}, 
expected
for the (II,Mn)VI semiconductors. In Fig.~\ref{injexp:3} the injected spin down 
electrons are majority electrons with their magnetic
moments parallel to the applied magnetic field.
The principles of optical orientation discussed 
in Sec.~\ref{sec:IIB} and the 
selection 
rules for GaAs sketched in Fig.~\ref{oo:1} are used to infer $P_n$ in a QW.

For a QW of approximately the same width (150 nm) the conversion of 
$P_{\rm circ}$ to $P_n$  used by
\textcite{Fiederling1999:N}  differed by a factor of 2 
from that used by \textcite{Jonker2000:PRB}.
\textcite{Fiederling1999:N}
assumed that confinement effects were negligible, leading to the selection
rules for a bulk GaAs 
(recall $P_{\rm circ}=-P_n/2$, from Sec.~\ref{sec:IIB}).
The maximum $P_{\rm circ}\approx 43 \%$ was interpreted 
as implying nearly $90$\% 
polarized injected electrons. \textcite{Jonker2000:PRB} 
inferred $|P_n| \approx 50$\%, 
from  $P_{\rm circ}=-P_n$ \cite{Weisbuch:1991}, as a consequence of QW 
confinement and 
lifting of the degeneracy 
between light and heavy hole states 
in the valence band ($\approx 5-6$ meV), see Fig.~\ref{oo:1}.
Both results clearly demonstrated a 
robust low-temperature
spin injection using the spin LED's. 
Subsequent studies 
\cite{Park2000:APL,Jonker2001:APL,Stroud2002:PRL}
have supported the lifting of degeneracy 
between the light and heavy hole bands. The corresponding data are 
shown in Fig.~\ref{injexp:4}.
Similar values
of $P_{\rm circ}$ were also measured in a resonant tunneling diode based
on ZnMnSe \cite{Waag2001:JS,Gruber2001:APL}. Spin injection using the
spin LED's,
described above, is not limited to structures grown by molecular-beam epitaxy
(MBE). It is also feasible using air-exposed interfaces \cite{Park2000:APL}
similar to the actual fabrication conditions employed in conventional
electronics. 

\begin{figure} 
\centerline{\psfig{file=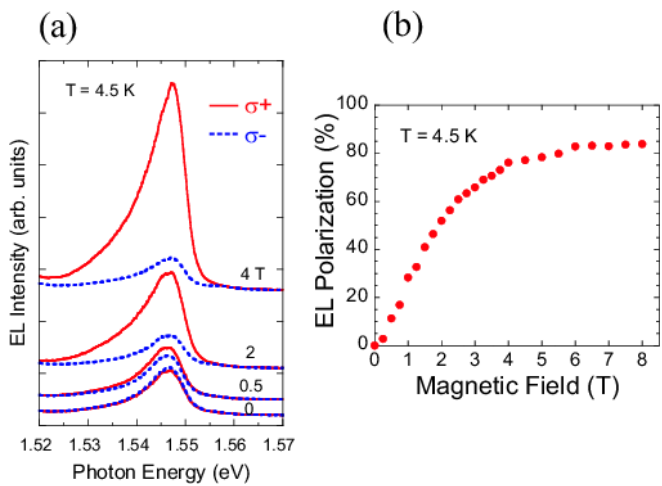,width=1.0\linewidth}}
\caption{(a)  Electroluminescence in a spin light-emitting diode (LED):
(a) Electroluminescence (EL) spectra from a surface-emitting spin LED 
with a Zn$_{0.94}$Mn$_{0.06}$Se contact for selected values of applied 
magnetic field, 
analyzed for  $\sigma^{\pm}$ (positive and negative helicity);
the magnetic field is applied along the 
surface normal (Faraday geometry) and the spectra are dominated by the heavy 
hole exciton;
(b)  magnetic field dependence of the EL circular polarization. 
Adapted from \onlinecite{Jonker2001:APL}.}         
\label{injexp:4}
\end{figure} 

The robustness of measured $P_{\rm circ}$ was studied by intentionally
changing the density of linear defects, from stacking faults at
the ZnMnSe/AlGaAs  
interface \cite{Stroud2002:PRL}. An approximate linear 
decrease of $P_{\rm circ}$ with the density of stacking faults
was shown to be
consistent with the influence of spin-orbit interaction as
modeled by Elliot-Yafet scattering (see Sec.~\ref{sec:IIIB1}) at the interface.
The nonspherically symmetric defect potential (entering the spin-orbit 
interaction) causes  a highly anisotropic loss of 
spin polarization.  At small angles to the axis of growth
[see Fig.~(\ref{injexp:3})], the probability of the spin flip
of an injected electron is very high,  leading to a small spin polarization.
These findings illustrate the importance of
interface quality and the effect of defects on the spin
injection efficiency, an issue not limited to semiconductor heterostructures.
Related information is currently being sought by spatial imaging of the spin 
polarization 
in spin LED's \cite{Thruber2002:APL,Thruber2003:JMR} 
using magnetic resonance
force microscopy \cite{Sidles1995:RMP}.

(III,Mn)V ferromagnetic semiconductors are also used to inject spin  
in a spin-LED structures as depicted in 
Fig.~\ref{injexp:3}. Spin injection can be achieved with no external
field, and the reports of high $T_C$ in some compounds suggest  
that all-semiconductor spin LED's could operate at room temperature.
The drawback, however, 
is that the most 
(III,Mn)V's have spin-polarized holes 
(rather than electrons) as the main carriers 
which, due to spin-orbit coupling, lose their polarization
very quickly after being injected into a nonmagnetic semiconductor. 
Consequently, results of the spin injection show only a small degree 
of hole polarization. 

In the experiment of \textcite{Ohno1999:N}, 
an intrinsic GaAs spacer of thickness $d$ was 
introduced
between the spin aligner (Ga,Mn)As 
and the (In,Ga)As quantum well. 
The electroluminescence  
in a QW was measured perpendicular to
the growth direction [the easy magnetization axis of (Ga,Mn)As and the
applied magnetic field were both perpendicular to the growth
direction]. The corresponding relation between the $P_{\rm circ}$ 
and
hole density  polarization 
$P_p$ is not
straightforward; 
the analysis was  performed only on the 
electroluminescence 
[for possible difficulties see \textcite{Fiederling2003:APL}].
A small measured signal ($P_{\rm circ}\sim 1 \%$ at $5$ K), 
consistent with the 
expectation for holes as the injected spin-polarized carriers,
was also obtained in an additional experiment \cite{Young2002:APL}.
$P_{\rm circ}$ was approximately independent of the GaAs thickness 
($d=20-420$ nm), a 
behavior which remains to be understood considering that the 
hole spins 
should relax fast \cite{Hilton2002:PRL} 
as they are
transfered across the nonmagnetic semiconductor.\footnote{A possible 
exception is QW, in which the effects of quantum confinement and 
quenching spin-orbit coupling lead to longer $\tau_s$.}
In contrast, for a repeated experiment \cite{Young2002:APL}
using a Faraday
geometry (as in Fig.~\ref{injexp:3}), with both measured electroluminescence  
and $B$ along the growth direction, the same change of thickness
$P_{\rm circ}$ was reduced from $7 \%$ to $0.5\%$. A highly
efficient spin injection of $P_n \approx 80\%$ in GaAs has been realized
using (Ga,Mn)As
as a spin injector in a Zener diode structure \cite{vanDorpe2003:P}. 
The detection
employed the technique of an oblique Hanle effect, discussed in 
the next section.

All-electrical spin injection studies of trilayer structures
(II,Mn)VI/II-VI/(II,Mn)VI have displayed up to 25\% MR at $B\approx5$ T and
$T=4$ K \cite{Schmidt2001:PRL}. A strong suppression of this MR signal
at applied bias of $\sim10$ mV was attributed to the nonlinear regime
of spin injection, in  which the effects of band bending 
and charge accumulation
at the (II,Mn)VI/II-VI interface were important \cite{Schmidt2002:P}.
It would be instructive to analyze these measurements by adopting 
the approach discussed in the context of magnetic {\it p-n} junctions
(Secs.~\ref{sec:IIC3} and \ref{sec:IVD}), which self-consistently
incorporates the effects of band bending and deviation from 
local charge neutrality.

\subsubsection{\label{sec:IID4} Metallic ferromagnet/semiconductor junctions}

A large family of  metallic ferromagnets, some of them highly spin polarized,
offer the 
possibility of spin injection at room temperature, even in the 
absence of
applied magnetic field. Spin injection into  (110) GaAs at room temperature 
has been already 
demonstrated 
using vacuum tunneling from a polycrystalline Ni STM tip and 
optical detection via circularly polarized luminescence 
\cite{Alvarado1992:PRL,Alvarado1995:PRL}. 
It was shown that the minority spin electrons (spin $\downarrow$ in 
the context of
metals; see Sec.~\ref{sec:IIA}) 
in 
Ni produced the dominant contribution to the tunneling current, and the    
resulting polarization was inferred to be $P_n=(-31 \pm 5.6)\%$ 
\cite{Alvarado1992:PRL}. 
Even thought the spin injection in  future spintronic 
devices will likely be implemented by some means other than 
vacuum tunneling, this result 
supports the importance of the tunneling contact for efficient 
spin injection, as discussed in Sec.~\ref{sec:IIC1}. 
Similar studies of spatially 
resolved spin injection,
sensitive to the topography of the GaAs surface, have employed 
a single-crystal 
Ni (100) 
tip \cite{LaBella2001:S}.  At 100 K nearly fully 
spin-polarized injection of electrons
was reported.
However,
further analyses 
of the measurements of 
$P_{\rm circ}$ have substantially reduced these estimates 
to $\approx 25$\% \cite{Egelhoff2002:S,LaBella2002:S}.

Direct spin injection from a ferromagnet into 
a 2DEG,\footnote{For a 
comprehensive review of 2DEG,  
see \textcite{Ando1982:RMP}.} 
motivated by the proposal of \textcite{Datta1990:APL} proposal,
initially  
showed only small effects \cite{Lee1999:JAP,Hammar1999:PRL,Gardelis1999:PRB}, 
with $\Delta R/R \sim 1\%$,
or effects within the noise \cite{Filip2000:PRB}. Such inefficiency could be 
attributed to the resistance mismatch in the F and N regions, discussed 
in Secs.~\ref{sec:IIC1} and \ref{sec:IIC2}. 
The possibility of spurious effects arising from the Hall and 
anisotropic magnetoresistance signals in
similar structures was suggested earlier  \cite{Monzon1997:APL}
as well as after the initial 
experiments \cite{Monzon2000:PRL,vanWees2000:PRL,Tang:2002}.
Control measurements have been performed to address these issues
\cite{Hammar2000:PRL,Hammar2000:PRB}. This debate about the 
presence/absence of spin injection effects via Ohmic contacts stimulated
further studies, 
but the experimental focus has since shifted to  other 
approaches.

Spin injection via Schottky contacts at room temperature was demonstrated
in a Fe/GaAs junction by \textcite{Zhu2001:PRL}, 
who reported 
detection of 
$P_{\rm circ} \approx 2\%$ using spin LED structures and optical detection as 
described in \ref{sec:IID3}.
These studies were extended \cite{Ramsteiner2002:PRB}
by using MBE to grow 
MnAs,  a ferromagnetic metal,  
on top of the GaAs to provide high-quality interfaces \cite{Tanaka2002:SST}. 
There was no preferential behavior for spin injection using different 
azimuthal orientations of the epitaxial MnAs layer, which could have been
expected from the symmetry between 
conduction-band wave functions in MnAs and GaAs.
The tunneling properties of a Schottky barrier
were discussed by 
\textcite{Meservey1982:JAP,Gibson1985:JAP,Prins1995:JPCM,Kreuzer2002:APL}.
The measured I-V curves display a complicated behavior 
\cite{Hirohata2001:PRB,Isakovic2001:PRB} which can be significantly affected
by the interface (midgap) states at the Schottky barrier \cite{Jonker1997:PRL}.
A theoretical explanation of this behavior is still lacking.  
 
\begin{figure} 
\centerline{\psfig{file=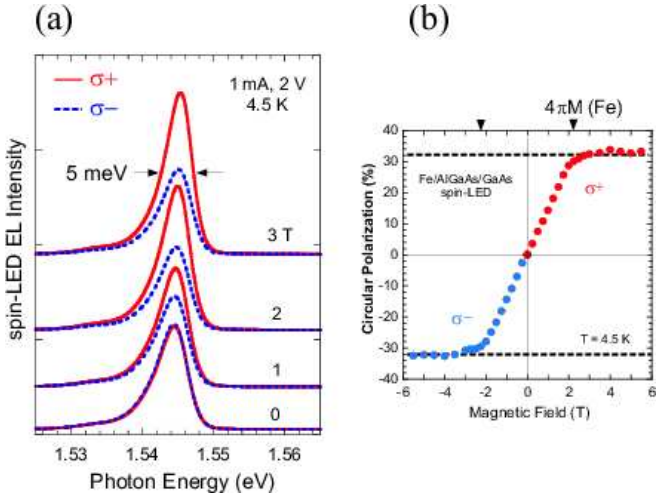,width=1.0\linewidth}}
\caption{Electroluminescence (EL) and polarization due to spin injection
from Fe Schottky contact: (a) EL spectra 
from a surface-emitting 
spin LED with an Fe/AlGaAs
Schottky tunnel contact  for selected values of applied magnetic field, 
analyzed for $\sigma^{\pm}$ 
circular polarization. The large difference in intensity
between these components indicates successful spin injection from the Fe 
into the GaAs
QW, and reveals an electron spin polarization in the QW of 32\%.  
The magnetic field is
applied along the surface normal (Faraday geometry).  
The spectra are dominated by the heavy
hole (HH) exciton.  Typical operating parameters are 1 mA and 2 V.
(b)  Magnetic-field dependence of the EL circular polarization of the 
HH exciton.  The
polarization tracks the hard axis magnetization of the Fe contact, 
and saturates at an
applied magnetic-field value 4$\pi$M = 2.2 T, at which the Fe 
magnetization is entirely along
the surface normal. From \onlinecite{Hanbicki2003:P}.}  
\label{injexp:5}
\end{figure} 

As discussed in Sec.~\ref{sec:IIC1}, tunnel contacts formed between a metallic 
ferromagnet and a semiconductor can provide effective spin injections.
Optical detection in spin LED structures, as discussed 
in Sec.~\ref{sec:IID3},
was used to show carrier polarization of $P_n \approx 30 \%$ using 
Fe as a spin injector \cite{Hanbicki2003:P}.
Experimental results are given in Fig.~\ref{injexp:5}.
Some care has to be taken in defining the efficiency of the spin injection,
normalized to the polarization of a ferromagnet, as used in related previous
experiments~\cite{Hanbicki2002:APLa,Jansen2002:APL,Hanbicki2002:APLb}. 
Furthermore,
there are often different conventions for defining the sign of $P_n$ 
\cite{Hanbicki2003:P},  used in the context of semiconductors and ferromagnetic 
metals, as discussed in Sec.~\ref{sec:IIA}. 
\textcite{Jiang2004:P} have demonstrated that MgO can be a suitable choice of an
insulator for highly efficient spin injection into GaAs. Spin LED with 
GaAs/AlGaAs quantum well was used to detect $P_n \approx 50 \%$
at 100 K injected from CoFe/MgO (100) tunnel injector. While quantum
well emission efficiency limits detection at higher temperatures ($>100$ K),
the same tunnel injector should also be suitable for efficient spin
injection even at room temperature.

An oblique Hanle effect \cite{D'yakonov1975:SPJETP} (see also 
Secs.~\ref{sec:IIB}
and \ref{sec:IID1})
was used \cite{Motsnyi2002:APL,Motsnyi2003:PRB,VanDorpe2002:P}
to detect spin injection, giving up to $P_n\approx 16 \%$, 
at room temperature.
The geometry used is similar to that sketched in Fig.~\ref{injexp:3}(a),
with an insulating layer (AlO$_x$) separating the ferromagnetic spin injector
and the (Al,Ga)As/GaAs spin LED.
The magnetization easy axis lies in the plane of 
the ferromagnet. 
An oblique
magnetic field is applied to give a net out-of-plane component of injected
spin which could contribute to the emission of circularly polarized light.
This approach allows one to apply a magnetic field several times smaller 
than would be 
needed to pull the magnetization out of plane
[for Fe it is $\approx 2$ T \cite{Hanbicki2002:APLa}] and to measure polarized
luminescence in a Faraday geometry. Furthermore, using standard measurements
of the Hanle curve, one can extract separately the spin lifetime
and carrier recombination time.

Hot-electron spin injection above the Schottky barrier is another method for 
using a high polarization of metallic ferromagnets  
to create a nonequilibrium 
spin in a semiconductor even at room temperatures. 
Typically such injection is performed in three-terminal, transistor-like
devices, as discussed in Sec.~\ref{sec:IVE3}.

Direct electrical spin injection has also been demonstrated in organic 
semiconductors \cite{Dediu2002:SSC}. MR measurements were performed in 
an F/N/F junction, where F is 
La$_{0.7}$Sr$_{0.3}$MnO$_3$ 
(LSMO), a colossal magnetoresistive manganite, and N is 
is sexithienyl (T$_6$), a  
$\pi$-conjugated rigid-rod oligomer organic semiconductor. The decrease of MR
with increasing thickness of the N region was used to infer 
$L_{sN} \sim 100$ nm at room temperature. The resulting spin 
diffusion length is a combination of low mobility, 
$\sim$ 10$^{-4}$cm$^2$V$^{-1}$s$^{-1}$
(about $\sim$ 10$^7$ times smaller than for the bulk GaAs),
and long spin relaxation times, $\sim$ $\mu$s,\footnote{This is
also a typical value
for other organic semiconductors \cite{Krinichnyi2000:SM}, 
a consequence of weak spin-orbit coupling
\cite{Davis2003:JAP}.} as compared to the usual III-V inorganic semiconductors.
Motivated by these findings \textcite{Xiong2004:N} have replaced one of
the LSMO electrodes by Co. Different coercive fields in the two ferromagnetic
electrodes allowed them to measure a spin-valve effect with
$\Delta R / R \sim 40 \%$ at 11 K.
Related theoretical studies of the ferromagnetic metal/conjugated polymer
interfaces were reported by \textcite{Xie2003:PRB}. 

\section{\label{sec:III} Spin relaxation and spin dephasing}

\subsection{\label{sec:IIIA} Introduction}

Spin relaxation and spin dephasing are processes that lead to spin 
equilibration
and are thus of great importance for spintronics. The fact that 
nonequilibrium
electronic spin in metals and semiconductors lives relatively long (typically a
nanosecond), allowing for spin-encoded information to travel macroscopic 
distances, is 
what makes spintronics a viable option for technology. After introducing the 
concepts of spin relaxation and spin 
dephasing times $T_1$ and $T_2$, 
respectively, 
which
are commonly called $\tau_s$ throughout this paper, we
discuss four major physical mechanisms responsible for spin equilibration in 
{\it nonmagnetic} electronic
systems: Elliott-Yafet, D'yakonov-Perel', Bir-Aronov-Pikus, and 
hyperfine-interaction 
processes. We then survey recent works on electronic spin relaxation in 
nonmagnetic metals and semiconductors, using the important examples of
Al 
and GaAs for illustration.

\subsubsection{\label{sec:IIIA1} $T_1$ and $T_2$}

Spin relaxation and spin dephasing of a spin ensemble are traditionally defined 
within the
framework of the Bloch-Torrey equations~\cite{Bloch1946:PR,Torrey1956:PR} 
for magnetization dynamics. For mobile electrons, spin 
relaxation time $T_1$ (often called longitudinal or spin-lattice time) and
spin dephasing time $T_2$ (also transverse or decoherence time) are
defined via the equations for the spin precession, decay, and diffusion of 
electronic 
magnetization $\bf M$ in an applied magnetic field ${\bf B}(t)=B_0\hat{\bf 
z}+{\bf B}_1(t)$, 
with a static longitudinal component $B_0$ (conventionally in the $\hat{z}$ 
direction) and, frequently, a transverse oscillating part ${\bf B}_1$ 
perpendicular to $\hat{\bf z}$
 ~\cite{Torrey1956:PR,Kaplan1959:PR}:
\begin{eqnarray} 
\label{eq:relax:bloch1}
\frac{\partial M_{x}}{\partial t}&=&\gamma ({\bf M} \times {\bf B})_x-
\frac{M_x}{T_2}+D\nabla^2 M_x, \\
\label{eq:relax:bloch2} 
\frac{\partial M_{y}}{\partial t}&=&\gamma ({\bf M} \times {\bf B})_y-
\frac{M_y}{T_2}+D\nabla^2 M_y, \\
\label{eq:relax:bloch3}
\frac{\partial M_{z}}{\partial t}&=&\gamma 
({\bf M} \times {\bf B})_z-\frac{M_z-M^0_z}{T_1}+D\nabla^2 M_z.
\end{eqnarray}
Here $\gamma=\mu_B g/\hbar$ is the electron gyromagnetic ratio 
($\mu_B$ is the Bohr 
magneton 
and $g$ 
is the electronic $g$ factor), $D$ is the diffusion coefficient (for simplicity
we assume an isotropic or a cubic solid with scalar $D$), and 
$M_z^0=\chi B_0$ is the thermal equilibrium magnetization with $\chi$ 
denoting the system's static susceptibility. 
The Bloch equations are phenomenological, describing quantitatively 
very well the dynamics of mobile electron spins (more properly, magnetization)
in experiments such as conduction electron spin resonance and optical 
orientation. 
Although relaxation and decoherence processes in a 
many-spin system are generally
too complex to be fully described by only two parameters, 
$T_1$ and
$T_2$ are nevertheless an extremely robust and convenient measure 
for quantifying 
such processes in many cases of interest. 
To obtain microscopic expressions for spin relaxation and dephasing times,
one starts with a microscopic description of the spin system (typically
using the density-matrix approach), derives the magnetization dynamics, 
and compares 
it 
with the Bloch equations to extract $T_1$ and $T_2$.

Time $T_1$ is the time it takes for the longitudinal magnetization to reach 
equilibrium. 
Equivalently, it is the time of thermal equilibration of the spin population 
with 
the lattice.  
In $T_1$ processes an energy has to 
be taken from the spin system, usually by phonons, to the lattice. 
Time $T_2$ is classically the time it takes for an 
ensemble of transverse electron spins, initially precessing in phase  
about the 
longitudinal
field, to lose their phase due to spatial and temporal fluctuations of the 
precessing frequencies. 
For an ensemble of mobile electrons the measured $T_1$ and $T_2$ come about by
averaging spin over the thermal distribution of electron momenta. Electrons
in different momentum states have not only different spin-flip characteristics,
but also slightly different $g$ factors and thus different precession 
frequency. This is analogous to precession frequencies fluctuations
of localized spins due to inhomogeneities in the static field $B_0$. 
However, since momentum scattering (analogous to intersite hopping or exchange 
interaction 
of localized spins) typically proceeds much faster than spin-flip scattering,
the $g$-factor-induced broadening is inhibited by motional 
narrowing\footnote{Motional (dynamical) 
narrowing is an inhibition of  phase change
by random fluctuations \cite{Slichter:1989}. 
Consider a spin rotating with frequency
$\omega_0$. The spin phase changes by 
$\Delta \phi = \omega_0 t$ 
over time $t$. If the spin is subject to a random force that 
makes spin precession equally likely clockwise and anticlockwise,
the average spin phase does not change, but the root-mean-square phase change
increases with time as 
$(\langle \Delta^2 \phi \rangle)^{1/2} \approx 
(\omega_0 \tau_c) (t/\tau_c)^{1/2}$,
where $\tau_c$ is the correlation time of the random force, or the average
time of spin precession in one direction. 
This is valid for rapid fluctuations, $\omega_0 \tau_c \ll 1$.
The phase relaxation time  
$t_\phi$ is defined as the time over which the phase 
fluctuations reach unity: $1/t_\phi=\omega_0^2 \tau_c$.}
and
need not be generally considered as contributing to $T_2$ 
[see, however,~\cite{Dupree1967:PSS}]. Indeed, motional narrowing of the 
$g$-{factor
fluctuations, $\delta g$, 
gives a contribution to $1/T_2$ of the order of 
$\Delta \omega^2\tau_p$, where the $B_0$-dependent precession frequency
spread is 
$\Delta \omega = (\delta g/g) \gamma B_0$ 
and $\tau_p$ is the momentum scattering time. For $B_0$ fields of the order 
1 T, 
scattering times of 1 ps, and $\delta g$ as large as 0.01, 
the ``inhomogeneous broadening'' is a microsecond, which is much 
more than 
the observed values for $T_2$. Spatial inhomogeneities of $B_0$, 
like those coming 
from hyperfine fields, are inhibited by motional narrowing, too, due to the
itinerant nature 
of electrons. For localized electrons (e.g., for donor states in 
semiconductors),
spatial inhomogeneities play an important role 
and are often observed to affect 
$T_2$. 
To describe such reversible phase losses, which can potentially be 
eliminated by spin-echo experiments, sometimes the symbol $T_2^*$ 
~\cite{Hu2001:lanl}
is used to describe spin dephasing of ensemble spins, while the symbol $T_2$ is
reserved for irreversible loss of the ensemble spin phase. 
In general, $T_2^* \le T_2$, 
although for conduction electrons to a very good approximation $T_2^*=T_2$. 

In isotropic and cubic solids $T_1=T_2$ if $\gamma B_0 \ll 1/\tau_c$, 
where $\tau_c$ is the so-called correlation or interaction time: $1/\tau_c$ is
the rate of change of the effective dephasing magnetic field 
(see footnote 70).  
Phase losses occur during time intervals of $\tau_c$.
As shown below, 
in electronic systems $\tau_c$ is given either by $\tau_p$ or by the
time of the interaction of electrons with phonons and holes. Those
times are typically smaller than a picosecond, so $T_1=T_2$ is fulfilled for 
magnetic
fields up to 
several tesla. The equality between the relaxation and 
dephasing
times was noticed first in the context of 
NMR~\cite{Bloch1946:PR,Wangsness1953:PR}
and later extended to electronic spin systems 
~\cite{Pines1955:PR,Andreev1958:JETP}.
A qualitative reason for $T_1=T_2$ is that if the phase acquires a random 
contribution 
during a short time interval $\tau_c$, the energy uncertainty of the spin 
levels 
determining the longitudinal spin is greater than the Zeeman splitting $\hbar 
\gamma B_0$ 
of the levels. The splitting then does not play a role in dephasing, and 
the dephasing field will act equally on the longitudinal and
transverse spin. In classical terms, spin that is 
oriented along the direction of the 
magnetic
field can precess a full period about the perpendicular fluctuating field, 
feeling
the same dephasing fields as transverse components. As the external field 
increases,
the precession angle of the longitudinal component is reduced, inhibiting 
dephasing.

The equality of the two times is very convenient for comparing
experiment and theory, since measurements usually yield $T_2$, 
while theoretically it is often more convenient to calculate $T_1$. In many
cases a single symbol $\tau_s$ is used for spin relaxation and dephasing
(and called indiscriminately either of these terms), 
if it does not matter what 
experimental situation is involved, or if one is working at small magnetic 
fields.\footnote{Sometimes one finds as spin relaxation
time $2\tau_s$. 
While this is correct for just one spin state, 
conventionally by spin relaxation we mean magnetization relaxation, 
in which each spin flip adds
to the equilibration of both spin up and spin down states, doubling the
rate for magnetization relaxation.} Throughout this paper we adopt this
notation to avoid unnecessary confusion.

If the system is anisotropic, the equality $T_1=T_2$ no longer holds, even in
the case of full motional narrowing of the spin-spin
interactions and $g$-factor 
broadening. Using simple qualitative analysis
\textcite{Yafet:1963} showed that, while there is no general relation
between the two times, the inequality $T_2 \le 2 T_1$ holds, and that
$T_2$ changes with the direction by at most a factor of 2. In the motionally
narrowed case this difference between $T_1$ and $T_2$ can be considered
as arising from the tensorial nature of the spin relaxation rate. Specific
examples of this will be discussed in studying spin relaxation in 
two-dimensional semiconductor heterostructures (Sec.~\ref{sec:IIIB2}).

Finally, we discuss the connection between $\tau_s$ and the single-spin 
decoherence 
time\footnote{To distinguish ensemble and individual spin dephasing, we use
the term decoherence in relation to single, or a few, spins.},
$\tau_{sc}$, which is the single-spin 
correlation time. Time $\tau_{sc}$ becomes important for applications 
in spin-based 
quantum computing \cite{Loss1998:PRA,Hu2000:PRA}, 
where spin coherence needs to last for at least $10^4-10^6$ gate
operations for the computation to be fault tolerant \cite{Preskill1998:PRSL}. 
The relative magnitudes of $\tau_s$ and $\tau_{sc}$ depend on many factors. 
Often
$\tau_{sc} < \tau_s$, as is the case for a direct exchange interaction causing 
single-spin decoherence, while contributing to ensemble dephasing only as a
dynamical averaging factor (the exchange interaction preserves total spin).  
An
analogy with momentum scattering may be helpful. Electron-electron collisions 
lead to
individual momentum equilibration while conserving the total momentum 
and hence 
do not contribute 
to charge current
(unless umklapp processes are taken into account).
A momentum scattering time obtained 
from conductivity (analogous to $\tau_s$) 
would thus
be very different from a single-state momentum time (analogous to 
$\tau_{sc}$).
It is not clear at present how much $\tau_s$ and $\tau_{sc}$ differ for 
different 
materials under different conditions, although it is often, with little 
justification, 
assumed that they are similar. As \textcite{Dzhioev2002:PRB} 
recently suggested,
$\tau_{sc}$ can be smaller than $\tau_s$ by several orders of magnitude 
in n-GaAs at low doping densities where electrons are localized in 
donor states, see also Sec.~\ref{sec:IIID3a}. 
We note that it is $\tau_s$ that is relevant for spintronics (spin transport) 
applications, while $\tau_{sc}$ is relevant for solid-state quantum computing. 
Much remains to be learned
about $\tau_{sc}$.

\subsubsection{\label{sec:IIIA2} Experimental probes}

Experiments detecting spin relaxation and decoherence of conduction electrons 
can be grouped into two
broad categories: (a) those measuring spectral characteristics of magnetization 
depolarization
and (b) those measuring time or space correlations of 
magnetization.

Experiments of type (a) are exemplified by conduction-electron spin 
resonance (CESR) 
and optical orientation 
in combination with the Hanle effect. CESR was the first technique used to 
detect the spin
of conduction electrons in metals ~\cite{Griswold1952:PR,Feher1955:PR}
and donor states in semiconductors like Si \cite{Feher1959b:PR}. 
In GaAs, spin resonance techniques are aided by other measurements, e.g., 
optical detection \cite{Weisbuch1977:PRB}, photoconductivity 
\cite{Stein1983:PRL},
or magnetoresistance \cite{Dobers1988:PRB}.
The technique measures signatures of resonant absorption of microwaves by a 
Zeeman-split
spin system of conduction electrons. Typically changes in surface 
impedance and 
in 
the transmission coefficient of the sample are observed. By comparing the 
absorption
resonance curve with theory~\cite{Dyson1955:PR,Kaplan1959:PR} 
one can obtain
both  the carrier $g$ factor and $T_2$.

Optical spin orientation~\cite{Meier:1984} is a method
of exciting spin-polarized carriers (electrons and holes) in direct-gap 
semiconductors like GaAs by absorption of circularly polarized light 
(see Sec.~\ref{sec:IIB}). 
The injected spin polarization can be detected by observing circularly 
polarized luminescence resulting from recombination of the spin-polarized 
electrons and 
holes.  
Since in a steady state the excited spin polarization depends not only on 
$\tau_s$
but also on the carrier recombination time, the Hanle effect,
polarization of luminescence by a transverse magnetic field 
(see Sec.~\ref{sec:IID1}),
is employed to deduce 
$\tau_s$ unambiguously.

The Hanle effect has also been a great tool for investigating $T_2$ in metals,
in connection with electrical spin injection. The advantage of optical 
orientation
over CESR is that, if carrier lifetime is known, zero-field 
(or zero-$g$-factor 
material) 
data can be measured. In addition, smaller $\tau_s$ values can be detected. 

Type (b) experiments measure magnetization in a time or space domain. 
The most important examples are the Johnson-Silsbee spin injection  experiment,
the time-resolved (pump-probe) photoluminescence (in which the probe
is used on the same principle as optical orientation),
and time-resolved
Faraday and (magneto-optic) Kerr rotation. The last two methods can
follow coherent (in the ensemble sense) dynamics of electron spin 
precession. 

Spin injection experiments \cite{Johnson1985:PRL} detect the amount of 
nonequilibrium magnetization by observing charge response 
(see Sec.~\ref{sec:IID1}). 
In the Johnson-Silsbee scheme, electrons
are first injected by electrical spin injection from a ferromagnetic
electrode into a metal or semiconductor. As the spin diffuses throughout the 
sample, another ferromagnetic electrode detects the amount of spin as a 
position-dependent quantity. The detection is by means of 
spin-charge coupling,
whereby an EMF appears across the paramagnet/ferromagnet
interface in proportion to the nonequilibrium magnetization 
\cite{Silsbee1980:BMR}. 
By fitting the
spatial dependence of magnetization to the exponential decay formula, 
one can extract the
spin diffusion length $L_s$ and thus the spin relaxation 
time $T_1=L_s^2/D$. 
The Hanle effect, too, can be used in combination with
Johnson-Silsbee spin injection, yielding directly $T_1=T_2$, which
agrees with $T_1$ determined from the measurement of $L_s$.
A precursor to the Hanle effect in spin injection was 
transmission-electron
spin resonance (TESR), in which
nonequilibrium electron spin excited in the skin 
layer of a metallic sample is transported to the other side, emitting 
microwave
radiation. In very clean samples and at low temperatures, electrons
ballistically propagating 
through
the sample experience Larmor precession
resulting in Larmor waves, seen as an oscillation of the transmitted
radiation amplitude with changing static magnetic field~\cite{Janossy1980:PRB}.

Time-resolved photoluminescence
detects, after creation of spin-polarized carriers, circular polarization of 
the recombination light. This technique was used, for example, 
to detect a 500 ps spin coherence time $T_2$ of free excitons in a 
GaAs quantum well \cite{Heberle1994:PRL}.
The Faraday and (magneto-optic) Kerr effects are the rotation of the 
polarization
plane of a linearly polarized light, transmitted through (Faraday) 
or reflected by
(Kerr)  a magnetized sample. The Kerr effect is more useful for 
thicker and nontransparent samples or for thin films fabricated on thick 
substrates.  The angle of rotation is proportional
to the amount of magnetization in the direction of the incident light. 
Pump (a circularly polarized laser pulse) and probe experiments employing 
magneto-optic effects can now follow, with 100 fs resolution, the evolution
of magnetization as it dephases in a transverse magnetic field. 
Using lasers for spin excitation has the
great advantage of not only detecting, 
but also 
manipulating spin dephasing, as shown using Faraday rotation on bulk 
GaAs and GaAs/ZnSe 
heterostructures \cite{Kikkawa2001:PE,Awschalom1999:PT}. 
The Kerr effect was used, for example, to investigate the spin dynamics of bulk 
CdTe \cite{Kimel2000:PRB}, 
and Faraday rotation was used to study spin coherence in nanocrystals of 
CdSe \cite{Gupta2002:PRB} and coherent control of spin dynamics of
excitons in GaAs quantum wells \cite{Heberle1996:IEEE}.

\subsection{\label{sec:IIIB} Mechanisms of spin relaxation}

Four  mechanisms for spin relaxation of conduction electrons have been found
relevant for metals and semiconductors: the Elliott-Yafet (EY),
D'yakonov-Perel' (DP), Bir-Aronov-Pikus (BIP), and hyperfine-interaction 
(HFI) 
mechanisms.\footnote{We do not 
consider magnetic scattering, that is, 
scattering due to an exchange interaction between conduction electrons
and magnetic impurities.} In the EY
mechanism electron spins relax because the electron wave functions normally
associated with a given spin have an admixture of the
opposite spin states, due to spin-orbit coupling induced by ions. The DP
mechanism explains spin dephasing in solids without a center of symmetry.
Spin dephasing occurs because electrons feel an effective
magnetic field, resulting from the lack of inversion symmetry and from the
spin-orbit interaction, which 
changes in random directions every time the electron
scatters to a different momentum states. The BIP mechanism is 
important for p-doped semiconductors, in which the electron-hole exchange 
interaction
gives rise to fluctuating local magnetic fields flipping electron spins. 
Finally,
in semiconductor heterostructures (quantum wells and quantum dots) based on
semiconductors with a nuclear magnetic moment, 
it is the hyperfine interaction 
of the electron spins and nuclear moments which dominates spin dephasing of 
localized or confined electron spins. An informal review of spin 
relaxation of conduction 
electrons can be found 
in \textcite{Fabian1999:JVST}.

\subsubsection{\label{sec:IIIB1} Elliott-Yafet mechanism}

\textcite{Elliott1954:PR} was the first to realize that conduction-electron 
spins can relax via
ordinary momentum scattering (such as by phonons or impurities) 
if the lattice ions induce spin-orbit
coupling in the electron wave function. In the presence 
of 
the spin-orbit interaction 
\begin{equation}\label{eq:relax:vso}
V_{so}=\frac{\hbar}{4 m^2 c^2} \nabla V_{sc} \times \hat{\bf p} \cdot 
{\bf \hat{\bf \sigma}},
\end{equation}
where $m$ is the free-electron mass, $V_{sc}$ is the scalar (spin-independent) 
periodic lattice 
potential, $\hat{\bf p}\equiv -i\hbar \nabla$
is the linear momentum operator, and ${\bf \hat{\bf \sigma}}$ 
are the Pauli matrices, 
single-electron (Bloch) wave functions in a solid are no longer the 
eigenstates of 
$\hat{\sigma}_z$, but rather a mixture of the Pauli spin up 
$|\uparrow \rangle$ 
and spin down $|\downarrow \rangle$
states. If the solid possesses a center of symmetry, the case of 
elemental metals which Elliott considered, the Bloch states 
of ``spin up'' and ``spin down'' electrons with the lattice momentum 
$\bf k$ and band index $n$  
can be written as \cite{Elliott1954:PR}
\begin{eqnarray}
\Psi_{{\bf k}n\uparrow}({\bf r})&=&\left [a_{{\bf k}n} ({\bf r}) 
|\uparrow\rangle
+b_{{\bf k}n} ({\bf r}) |\downarrow\rangle  \right ]e^{i{\bf k}\cdot {\bf r}}, 
\\
\Psi_{{\bf k}n\downarrow}({\bf r})&=&\left [a^*_{-{\bf k}n} ({\bf r}) 
|\downarrow\rangle  
-b^*_{-{\bf k}n} ({\bf r}) |\uparrow\rangle  \right ] e^{i{\bf k}\cdot {\bf r}}, 
\end{eqnarray}
where we write the explicit dependence of the complex 
lattice-periodic 
coefficients $a$ and $b$ 
on the radius vector $\bf r$.
The two Bloch states are degenerate: they are connected by spatial 
inversion and time reversal~\cite{Elliott1954:PR}. That it makes sense to call
$\Psi_{{\bf k}n\uparrow}$ and  $\Psi_{{\bf k}n\downarrow}$, respectively,
spin up 
and spin down 
states follows from the fact that in most cases the typical values of $|a|$ are 
close to unity,
while $|b| \ll 1$. 

Indeed, consider a band structure in the absence of spin-orbit coupling.
Turning $V_{so}$ on couples electron states of opposite 
Pauli spins which are 
of the same ${\bf k}$ (because $V_{so}$ has the period of the lattice), 
but different $n$. Because the spin-orbit interaction is normally much 
smaller than the
distance between the bands, perturbation theory gives 
\begin{equation} \label{eq:relax:lambda}
|b|\approx \lambda_{so}/\Delta E \ll 1,
\end{equation}
where $\Delta E$ is the energy distance between the band state in question and 
the
state (of the same momentum) 
in the nearest band, and $\lambda_{so}$ is the amplitude of the matrix element 
of $V_{so}$ 
between the two states. 
The spin-orbit coupling alone does not lead to spin relaxation. However, in 
combination 
with
 momentum scattering, the spin up [Eq.~(56)] and spin down  [Eq.~(57)] 
states
can couple and lead to spin relaxation.

Momentum scattering is typically caused by impurities (at low $T$) and phonons 
(at high
$T$). There is another important spin-flip scattering mechanism that involves 
phonons. 
A periodic, lattice ion-induced spin-orbit interaction is modified by phonons 
and can 
directly couple the (Pauli) spin up and spin down states. Such processes 
were first 
considered 
for a jellium model by \textcite{Overhauser1953:PR} [see also 
\cite{Grimaldi1997:PRB}], 
and for band structure systems by \textcite{Yafet:1963}. They must be combined
with the Elliott processes discussed above to form a consistent picture of 
phonon-induced spin relaxation, especially at low 
temperatures~\cite{Yafet:1963}, 
where the
two processes have similar magnitudes. At larger $T$,
phonon-modified $V_{so}$ 
is not 
important for polyvalent metals, whose spin relaxation is dominated by spin hot 
spots---states
with anomalously large $|b|$---as shown by the explicit calculation 
of \textcite{Fabian1999:PRL}. Spin hot spots are discussed in more detail in
Sec.~\ref{sec:IIIC}.  

We now give a recipe, useful for elemental metals and semiconductors, 
for calculating phonon-induced $1/\tau_s$ 
from the known band and phonon structure. 
The corresponding theory was systematically developed by \textcite{Yafet:1963}.
The spin relaxation rate due to phonon scattering according 
to the EY mechanism
can be expressed through the spin-flip Eliashberg function 
$\alpha_S^2F(\Omega)$
as ~\cite{Fabian1999:PRL}
\begin{equation} \label{eq:relax:ey}
1/\tau_s=8\pi T\int_{0}^{\infty} d\Omega \alpha_s^2F(\Omega)\partial 
N(\Omega)/\partial T, 
\end{equation}
where $N(\Omega)=[\exp(\hbar \Omega/k_B T)-1]^{-1}$ is the phonon distribution 
function.
The spin-flip Eliashberg function gives the contribution of the phonons with 
frequency
$\Omega$ to the spin-flip electron-phonon interaction, 
\begin{equation} \label{eq:relax:ef}
\alpha_s^2F(\Omega)=\frac{g_S}{2M\Omega}\sum_{\nu}\langle\langle 
g_{{\bf k}n\uparrow, {\bf k}'n'\downarrow}^{\nu}\delta (\omega_{{\bf q}\nu}-
\Omega)
\rangle_{{\bf k}n}\rangle _{{\bf k}'n'}.
\end{equation}
Here $g_S$ is the number of electron states per spin and atom at the Fermi 
level, 
$M$ is the ion mass, $\omega_{{\bf q}\nu}$ is the frequency of phonons
with momentum ${\bf q}={\bf k}-{\bf k}'$ and branch index $\nu$,
and the spin-flip matrix element is
\begin{equation} 
g_{{\bf k}n\uparrow, {\bf k}'n'\downarrow}^{\nu}\equiv |{\bf u}_{{\bf 
q}\nu}\cdot
\left (\Psi_{{\bf k}n\uparrow}, \nabla V \Psi_{{\bf k}'n'\downarrow} 
\right )  |^2,
\end{equation}
where ${\bf u}_{{\bf q}\nu}$ is the polarization of the phonon with momentum 
$\bf q$ and in branch $\nu$. 
The brackets $\langle ... \rangle_{{\bf k}n}$ in Eq.~(\ref{eq:relax:ef}) 
denote Fermi-surface averaging. The Bloch wave functions for this 
calculation are chosen 
to satisfy $(\Psi_{{\bf k}n\uparrow}, \hat {\sigma}_z
\Psi_{{\bf k}n\uparrow})=-(\Psi_{ {\bf k}n\downarrow}, \hat {\sigma}_z 
\Psi_{{\bf k}n\downarrow})$.
The periodic lattice-ion interaction $V$ contains both scalar and 
spin-orbit parts: $V=V_{sc}+V_{so}$.

There are two important relations giving  an order-of-magnitude estimate
of $\tau_s$, as well as 
its temperature dependence: the Elliott and the
Yafet relations.
The Elliott relation relates $\tau_s$ to the shift
$\Delta g$ of the electronic $g$ factor from the free-electron value of
$g_0=2.0023$~\cite{Elliott1954:PR}:
\begin{eqnarray} \label{eq:relax:elliott}
1/\tau_s \approx (\Delta g)^2/\tau_p,
\end{eqnarray}
where $\tau_p$ is the momentum relaxation time.
This relation follows from the fact that for a momentum scattering
interaction $V_i$ the spin-flip scattering probability in the
Born approximation is proportional to
$|(\Psi_{{\bf k}n \uparrow},
V_i \Psi_{{\bf k}'n' \downarrow})|^2 \approx |b|^2\times|(\Psi_{{\bf k}n
\uparrow},
V_i \Psi_{{\bf k}'n' \uparrow})|^2$. Realizing that the spin-conserving
scattering
probability gives the momentum relaxation rate, after a 
Fermi-surface averaging we
get the  estimate
\begin{equation} \label{eq:relax:aux}
1/\tau_s \approx \langle b^2 \rangle  /\tau_p.
\end{equation}

On the other hand, $\Delta g$ is determined by the expectation value of 
$\hat{l}_z$, 
the z-component of the orbital momentum 
in a Bloch state. Without the spin-orbit 
interaction this expectation value is 
zero. 
Considering the spin-orbit interaction to be a small parameter,
we find by 
perturbation theory $\Delta g \approx |b|$, 
which combines with Eq.~(\ref{eq:relax:aux}) to give the Elliott relation 
Eq.~(\ref{eq:relax:elliott}). An empirical test of the Elliott relation 
for simple
metals when spin relaxation is due to thermal phonons 
~\cite{Beuneu1978:PRB} gives the revised estimate 
\begin{equation} \label{eq:revisedE}
1/\tau_s \approx 10\times (\Delta g)^2 /\tau_p. 
\end{equation}

The Elliott relation is only a very rough estimate of $\tau_s$. The
experimentally relevant ratio  $\tau_p/\tau_s$ depends on the scattering
mechanism. The ratio is different for scattering off impurities,
boundaries, or phonons, although one would expect it to be
within an order of magnitude. For example, scattering by heavy impurities 
induces an
additional spin relaxation channel where spin flip is due to the 
spin-orbit interaction induced by the impurities. 
Equation (\ref{eq:relax:aux}) then
does not hold.  
Scattering by phonons is too complex to be simply equated 
with the ratio
$\tau_p/\tau_s$ 
for impurity or boundary scattering. However, the ratio is comparable
for scattering by light impurities and by boundaries. The ratio
$\tau_p/\tau_s$ for impurity and phonon scattering in Al and Cu is
compared by \textcite{Jedema2003:PRB}.

The Yafet relation is only qualitative, connecting the temperature 
dependence of 
$1/T_1$ with 
that of the resistivity $\rho$:
\begin{equation} \label{eq:relax:yafet}
1/T_1(T) \sim  \langle b^2 \rangle \rho (T),
\end{equation}
as follows directly from Eq.~(\ref{eq:relax:aux}) after realizing that 
$1/\tau_p(T)\sim \rho (T) $. By careful symmetry considerations 
\textcite{Yafet:1963} 
proved that 
$1/T_1 \sim T^5$ at low temperatures, similarly to the Bloch-Gr\"{u}neisen 
law for resistivity,
justifying Eq.~(\ref{eq:relax:yafet}) over a large temperature range.   
Yafet's $T^5$ law stems from the nontrivial fact that for spin-flip 
electron-phonon scattering 
$g_{{\bf k}n\uparrow,{\bf k}'n\downarrow}^\nu \sim
({\bf k}-{\bf k}')^4$ as ${\bf k} \rightarrow {\bf k}'$ ~\cite{Yafet:1963}, 
while only a quadratic dependence holds for the spin-conserving scattering. 
This corresponds to the long-wavelength
behavior $\alpha_S^2 F(\Omega)\sim \Omega^4$ of the spin-flip Eliashberg 
function. The Yafet relation was tested experimentally by 
\textcite{Monod1979:PRB}. This work led to a deeper understanding of spin 
relaxation processes
in polyvalent metals \cite{Silsbee1983:PRB,Fabian1998:PRL}.

Realistic calculations of the EY $\tau_s$ in semiconductors  
can be performed analytically using approximations of the 
band and phonon structures, as
most important states are usually  around high symmetry points. Here we give 
a formula for the spin relaxation of conduction electrons with energy 
$E_{\bf k}$
in the frequently studied case of III-V semiconductors 
~\cite{Chazalviel1975:PRB,Pikus:1984}:
\begin{equation} \label{eq:relax:chazalviel}
\frac{1}{\tau_s(E_{\bf k})}= A\left 
(\frac{\Delta_{so}}{E_g+\Delta_{so}}\right )^2\left 
(\frac{E_{\bf k}}{E_g}\right )^2\frac{1}
{\tau_p(E_{\bf k})}, 
\end{equation}
where $\tau_p(E_{\bf k})$ 
is the momentum scattering time at energy $E_{\bf k}$, 
$E_g$ is the energy gap,
and $\Delta_{so}$ is the spin-orbit splitting of the valence band 
(see Fig.~\ref{oo:1}). 
The Numerical factor
$A$, which is of order 1, depends on the dominant scattering 
mechanism (charge or neutral impurity, phonon, electron-hole).
Analytic formulas
for the EY mechanism due to electron-electron scattering 
are given by
\textcite{Boguslawski1980:SSC}.

Equation~(\ref{eq:relax:chazalviel}) shows that the EY mechanism is important 
for small-gap semiconductors with large spin-orbit splitting (the prototypical 
example is InSb). For degenerate electron densities
the spin relaxation time is given by Eq.~(\ref{eq:relax:chazalviel}) 
with $E_{\bf k}=E_F$,
while for nondegenerate densities the thermal averaging leads to a substitution 
of thermal energy $k_B T$ for $E_{\bf k}$ and thermal-averaged momentum 
relaxation
time for $\tau_p$. To estimate $\tau_s$ from 
Eq.~(\ref{eq:relax:chazalviel}) one needs to know $\tau_p$. 
It often suffices to know the doping or temperature dependence of $\tau_p$ 
to decide on the relevance of the EY mechanism \cite{Pikus:1984}.  

The temperature dependence of $1/\tau_s$ for metals and degenerate  
semiconductors follows the dependence of $1/\tau_p$. 
In metals this means a constant at low $T$ 
and a linear increase at large $T$. 
For nondegenerate semiconductors $1/\tau_s(T)\sim T^2/\tau_p(T)$. In the 
important case
of scattering by charged impurities ($\tau_p\sim T^{3/2}$) $1/\tau_s\sim 
T^{1/2}$. 
The magnetic field dependence of the EY spin relaxation  has not been
systematically investigated. At low temperatures, where cyclotron 
effects become important, 
one needs to average over cyclotron trajectories on the Fermi 
surface need to obtain $1/\tau_s$. 
We expect that such averaging leads, in general, to an 
increase in $1/T_1$, especially in systems where spin hot 
spots are important (see Sec.~\ref{sec:IIIC}).

\subsubsection{\label{sec:IIIB2} D'yakonov-Perel' mechanism}

An efficient mechanism of spin relaxation due to spin-orbit coupling
in systems lacking inversion symmetry 
was found by \textcite{Dyakonov1972:SPSS}. Without
inversion symmetry the momentum states of the spin up and spin down electrons 
are not degenerate: $E_{{\bf k}\uparrow}\ne E_{{\bf k}\downarrow}$.  
Kramer's theorem still dictates that $E_{{\bf k}\uparrow}= E_{-{\bf 
k}\downarrow}$. Most prominent examples of materials without inversion
symmetry 
come from groups
III-V (such as GaAs) and II-VI (ZnSe) semiconductors, where 
inversion symmetry is broken by the presence of two distinct atoms in the
Bravais lattice. Elemental semiconductors like Si possess inversion symmetry
in the bulk, so the DP mechanism does not apply to them. In heterostructures
the symmetry is broken by the presence of asymmetric confining potentials.  

Spin splittings 
induced by  
inversion asymmetry 
can be described by introducing an
intrinsic ${\bf k}$-dependent magnetic field ${\bf B}_{i}({\bf k})$
around which electron spins 
precess with Larmor frequency ${\bf \Omega}({\bf k})=(e/m) {\bf B}_i({\bf k})$.
The intrinsic field derives from the spin-orbit coupling in the band-structure.
The corresponding Hamiltonian term describing the precession of
electrons in the conduction band is
\begin{equation} \label{eq:HDP}
H({\bf{k}})=\frac{1}{2}\hbar {\bf \hat{\sigma}} \cdot{\bf \Omega}(\bf{k}),
\end{equation}
where ${\bf \hat{\sigma}}$ are the Pauli matrices. Momentum-dependent 
spin precession 
described
by $H$, together with momentum scattering 
characterized by momentum 
relaxation time $\tau_p$,\footnote{In the qualitative reasonings below we use 
$\tau_p$
instead of the effective correlation time $\tilde{\tau}$ for ${\bf \Omega}$ 
during momentum 
scattering; $\tilde\tau$ is defined later, in Eq.~(\ref{eq:relax:taul}).}
leads to spin dephasing. While the microscopic expression for 
$\bf \Omega({\bf k})$
needs to be obtained from the band structure, treating the effects of 
inversion asymmetry by introducing intrinsic precession helps to give a 
qualitative understanding
of spin dephasing. It is important to 
note, however, that the analogy with real
Larmor precession is not complete. An applied magnetic field induces 
a macroscopic spin polarization and magnetization, while $H$ of 
Eq.~(\ref{eq:HDP})
produces an equal number of spin up and spin down states.  

Two limiting 
cases can be considered: (i) $\tau_p\Omega_{av}\agt 1$ and (ii) 
$\tau_p\Omega_{av}\alt 1$, 
where $\Omega_{av}$ is an average magnitude of the intrinsic 
Larmor frequency $\Omega({\bf k})$ over the actual momentum
distribution. Case (i) corresponds to the situation in which
individual electron 
spins precess a full cycle before
being scattered to another momentum state. The total spin in this regime 
initially dephases 
reversibly due to the anisotropy in ${\bf \Omega}({\bf k})$. The spin dephasing 
rate,\footnote{The reversible decay need not be exponential.}
which depends on the
distribution of values of ${\bf \Omega}({\bf k})$, is in general
proportional to the width $\Delta \Omega$ of the distribution: 
$1/\tau_s \approx \Delta \Omega$. The spin is irreversibly lost after time 
$\tau_p$, when randomizing scattering takes place.

Case (ii) is what is usually meant by the D'yakonov-Perel' 
mechanism. This regime can be viewed from the point of view of individual 
electrons as a 
spin precession about fluctuating magnetic fields, whose magnitude and 
direction change randomly with the average time step of $\tau_p$. 
The electron spin rotates about 
the intrinsic field at an angle $\delta \phi = \Omega_{av}\tau_p$, 
before experiencing 
another field
and starting to rotate with a different speed and in a different direction. 
As a result, the spin phase follows a random walk: after time $t$, which 
amounts to
$t/\tau_p$ steps of the random walk, the phase
progresses by $\phi(t) \approx \delta\phi\sqrt{t/\tau_p}$. Defining $\tau_s$ as 
the time
at which $\phi(t)=1$, the usual motional narrowing result is obtained:
$1/\tau_s=\Omega_{av}^2 \tau_p$ (see footnote 70).
The faster the momentum relaxation, 
the slower 
the
spin dephasing. The difference between cases (i) and (ii) is that in 
case (ii)
the electron spins form an ensemble that directly samples the 
distribution of 
$\Omega({\bf k})$, while
in case (ii) it is the distribution of the {\it sums} 
of the intrinsic 
Larmor frequencies (the 
total 
phase of a spin after many steps consists of a sum of randomly selected 
frequencies
multiplied by $\tau_p$), which, according to the central limit theorem, has a 
significantly reduced variance. Both limits (i) and (ii) 
and the transition between
them have been experimentally demonstrated in n-GaAs/AlGaAs quantum wells by 
observing temporal spin oscillations over a large range of temperatures 
(and thus $\tau_p$)
 \cite{Brand2002:PRL}.

A more rigorous expression for $\tau_s$ in regime (ii) 
has been obtained by solving the kinetic rate equation for a spin-dependent 
density matrix 
\cite{Dyakonov1971:SPJETP,Dyakonov1972:SPSS}.
If the band structure is isotropic and scattering is both elastic and 
isotropic, 
evolution 
of the z-component of spin ${\bf s}$
is  
\cite{Pikus:1984}
\begin{equation}\label{eq:relax:sz}
\dot{s}_z =
-\tilde{\tau_l}\left [s_z \overline{\left ( \Omega^2 -\Omega_z^2 \right )}
-s_x \overline{\Omega_x\Omega_z}-s_y \overline{\Omega_y\Omega_z}\right ],
\end{equation}
where the bar denotes averaging over directions of $\bf k$. Analogous 
expressions
for $\dot{s}_x$ and $\dot{s}_y$ can be written by index permutation. The 
effective
momentum scattering time is introduced as
\begin{equation}\label{eq:relax:taul}
1/\tilde{\tau_l}=\int_{-1}^1 W(\theta)\left [1-P_l\left (\cos\theta\right 
)\right ] d\cos\theta,
\end{equation}
where $W(\theta)$ is the rate of momentum scattering through angle $\theta$ at 
energy
$E_{\bf k}$, and $P_l$ is the Legendre polynomial whose order $l$ 
is the power 
of $\bf k$ in
$\Omega({\bf k})$. [It is assumed that $\Omega({\bf k})\sim k^l$ in 
Eq.~(\ref{eq:relax:sz}).] 
 In two dimensions $P_l(\cos\theta)$ is replaced by $\cos(l\theta)$
in Eq.~(\ref{eq:relax:taul}) for the $l$th polar harmonic of 
$\bf \Omega({\bf k})$
~\cite{Pikus1995:PRB}.
Since it is useful to express the results in terms of the known momentum 
relaxation
times\footnote{In fact,  normal (not umklapp) electron-electron collisions 
should also be included in the effective spin randomization time 
$\tilde{\tau}$, though they 
do not
contribute to the momentum relaxation time which appears in the measured 
mobility~\cite{Glazov2002:JETPL,Glazov2003:P}.}
 $\tau_p=\tilde{\tau}_1$, one defines 
\footnote{ \textcite{Pikus:1984} 
initially define $\gamma_l$ as here, but 
later evaluate it, inconsistently, as the inverse $\gamma_l \rightarrow 
\gamma_l^{-1}$.} 
$\gamma_l=\tau_p/\tilde{\tau_l}$ to measure the effectiveness of momentum 
scattering
in randomizing Larmor frequencies; $\tilde{\tau}_l$ accounts for the relative 
angle between 
${\bf \Omega}$ before and after scattering. Generally $\gamma_l > 1$ for $l>1$, 
that is, momentum scattering is more effective in randomizing spins than 
in randomizing momentum.

Comparing with the Bloch-Torrey equations 
(\ref{eq:relax:bloch1})--(\ref{eq:relax:bloch3}), for $\bf{B}=0$ and no spin 
diffusion, we see that spin decay is described by the tensor 
$1/\tau_{s,ij}$ 
(here $i$ and $j$ are the Cartesian coordinates) whose diagonal 
$1/\tau_{s,ii}$
and off-diagonal $1/\tau_{s,ij}$, for $i\ne j$, terms
are
\begin{equation} \label{eq:relax:tauii}
1/\tau_{s,ii}=\gamma_l^{-1}\tau_p(\overline{\Omega^2}-
\overline{\Omega_i^2}),\,\,\,
1/\tau_{s,ij}=-\gamma_l^{-1}\tau_p\overline{\Omega_i\Omega_j}.
\end{equation}
In general, spin dephasing depends on the spin direction and on the 
dephasing rates of 
the perpendicular
spin components. Equations (\ref{eq:relax:tauii}) are valid for small magnetic 
fields, 
satisfying $\Omega_0 \tau_p \ll 1$, where $\Omega_0$ is the Larmor frequency of 
the
external field. 

The most important difference between the EY and the DP mechanism is their 
opposite 
dependence on $\tau_p$. While increased scattering intensity makes the EY 
mechanism 
more effective, it decreases the effectiveness of the
the DP processes. In a sense the two 
mechanisms are 
similar to collision broadening and motional narrowing in 
NMR \cite{Slichter:1989}. Indeed, in the EY 
process
the precession frequency is conserved between collisions and the loss of
phase occurs only in 
the short time {\it during} collision. The more 
collisions there are,
the greater is the loss of phase memory, in analogy with collision 
broadening of 
spectral
lines. On the other hand, in DP spin dephasing, spin phases are randomized 
{\it between} collisions, since electrons precess with different frequencies 
depending
on their momenta. Spin-independent collisions with impurities or phonons do not 
lead
to phase randomization during the collision itself, but help to establish the 
random-walk-like evolution of the phase, leading to motional narrowing.
While these two mechanisms coexist in systems lacking inversion symmetry,
their relative strength depends on many factors. Perhaps the most robust trend
is that the DP mechanism becomes more important with increasing band
gap and increasing temperature. 

In the rest of the section we apply Eq.~(\ref{eq:relax:tauii})
to the study of spin dephasing in bulk and two-dimensional III-V semiconductor 
heterostructures.

\paragraph{\label{sec:IIIB2a} Bulk III-V semiconductors.}

In bulk III-V semiconductors the intrinsic Larmor frequency vector of
Eq.~(\ref{eq:HDP}) 
due
to the
lack of inversion symmetry is \cite{Dyakonov1971:SPJETP}
\begin{equation} \label{eq:relax:dress}
{\bf \Omega}({\bf{k}})=\alpha\hbar^2(2m_c^3E_g)^{-1/2}\bf{\kappa},
\end{equation}
where 
\begin{eqnarray}
{\bf\kappa} = \left [k_x(k_y^2-k_z^2),k_y(k_z^2-k_x^2),
k_z(k_x^2-k_y^2)\right ].
\end{eqnarray}
Here $k_i$ are the lattice wave-vector components along the crystal principal 
axes.
The material-specific parameters are the band gap $E_g$ and the 
conduction electron mass
$m_c$; $\alpha$ is a dimensionless parameter specifying
the strength of the spin-orbit interaction.
The spin splitting described by Eq.~(\ref{eq:relax:dress})
is proportional to the cube of the lattice momentum, as was first found by 
\textcite{Dresselhaus1955:PR}. For GaAs $\alpha\approx 
0.07$~\cite{Marushchak1984:SPSS}. Spin splitting of conduction and
heavy and light holes in GaAs quantum wells, due to bulk inversion
asymmetry was calculated by \textcite{Rashba1988:PLA}.

Substituting Eq.~(\ref{eq:relax:dress}) for ${\bf\Omega}$ in 
Eq.~(\ref{eq:relax:tauii}), 
and using $\overline{\kappa_i\kappa_j}=(4/105)k^6\delta_{ij}$, we obtain the 
expected
result that the off-diagonal elements of $1/\tau_{s,ij}$ vanish for cubic 
systems
and the diagonal elements are all equal to~\cite{Pikus:1984} 
\begin{equation}\label{eq:relax:DP}
1/\tau_s(E_{\bf k})=
\frac{32}{105}\gamma_3^{-1}\tau_p(E_{\bf k}) \alpha^2\frac{E_{\bf 
k}^3}{\hbar^2 E_g}.
\end{equation}
The above expression describes DP spin dephasing of degenerate ($E_{\bf k}=E_F$) or 
hot\footnote{This is strictly true only if the spin relaxation of 
the hot electrons 
is faster
than energy relaxation by optical phonon emission, which is rarely the case. 
One has to consider either the spin relaxation at different energy levels 
during
the cascade process of optical phonon emission or, if the optical 
phonon emission is
particularly fast, spin relaxation only during the final stages of energy 
relaxation
by acoustic phonon emission (see \textcite{Pikus:1984}.} electrons
in bulk III-V semiconductors. For impurity scattering $\gamma_3\approx 
6$, for
acoustic phonons $\gamma_3\approx 1$, while for optical polar phonons 
$\gamma_3\approx 41/6$.
The temperature dependence of $1/\tau_s$ comes from the temperature dependence 
of 
$\tau_p$. For the important case of charged impurity scattering ($\tau_p\sim 
T^{3/2}$),
$1/\tau_s\sim T^{3/2}$. 

Compared to the EY expression, Eq.~(\ref{eq:relax:chazalviel}),
the DP spin dephasing increases much faster with 
increasing
electron energy and is expected to be dominant at large donor doping levels and
at high temperatures. The EY mechanism can be dominant in small-band-gap and 
large spin-orbit-splitting materials. 
The two mechanisms can also be easily distinguished by 
their
opposite dependence on momentum relaxation. Contrary to the EY mechanism, 
greater impurity density will decrease the importance of the DP processes. 
The  most frequently used ways
to distinguish between various methods of spin relaxation are comparing the
electron density (through the variation of the Fermi energy) 
and the temperature dependences of $1/\tau_s$ with the
theoretical estimates. Since the prefactors may vary with different scattering
mechanisms, it is best to deduce $\tau_p(E_{\bf k})$  and $\tau_p(T)$ 
from mobility measurements and use Eqs.~(\ref{eq:relax:chazalviel}) and 
(\ref{eq:relax:DP})
or the equations given below for $1/\tau_s(T)$.

Another interesting distinction between the two mechanisms is revealed 
by the 
dependence
of their spin diffusion length $L_s=\sqrt{D\tau_s}$ on momentum scattering. 
Since $D\sim \tau_p$, for EY $L_s\sim \tau_p$, while for DP $L_s$ does 
not depend on 
the momentum
scattering time and for a degenerate electron system should be  a constant 
independent of
$T$, of the order of $v_F/\Omega_{av}$.  
We do not know of an experimental verification of this distinction.

If the electrons obey nondegenerate statistics, which is the usual case of 
p-doped 
materials, thermal averaging over the Boltzmann distribution 
gives~\cite{Pikus:1984}. 
\begin{equation}\label{eq:relax:DPt}
1/\tau_s=Q \tau_m \alpha^2 \frac{(k_B T)^3}{\hbar^2 E_g},
\end{equation} 
where $\tau_m=\langle \tau_p(E_{\bf k}) E_{\bf k}\rangle/\langle E_{\bf k} 
\rangle$.
The coefficient $Q$, which is of order 1, is 
\begin{equation}
Q=\frac{16}{35}\gamma_3^{-1} \left ( \nu+\frac{7}{2} \right )
\left (\nu+\frac{5}{2} \right ),
\end{equation}
where the power law $\tau_p\sim E^{\nu}_{\bf k}$ is assumed for momentum 
relaxation time.
For scattering by ionized impurities $Q\approx 1.5$, while scattering by polar 
optical
or piezoelectric phonons gives $Q\approx 0.8$, and scattering by acoustic 
phonons 
(deformation potential) $Q \approx 2.7$~\cite{Pikus:1984}. The temperature 
behavior of
DP spin dephasing in nondegenerate samples is 
$1/\tau_s\sim T^3\tau_m(T)$. 
For 
scattering
by charged impurities $1/\tau_s\sim T^{9/2}$.

Application of longitudinal (to the initial spin direction) 
magnetic field suppresses the DP mechanism~\cite{Pikus:1984} for two 
reasons: (i) The $B$-field 
suppresses
precession along the transverse intrinsic 
fluctuating fields when $\Omega_L\tau_p > 1$, where $\Omega_L$ is 
the Larmor precession due to $B$. (ii)  $\Omega_{\bf k}$ 
is orbitally averaged,
which 
has a similar effect to averaging by random scattering,  when 
$\Omega_c\tau_p>1$, where $\Omega_c$ is the cyclotron frequency. 
Since for conduction electrons  $m_c \ll m_e$, it follows that
$\Omega_c \gg \Omega_L$, the orbital motion induced by $B$ is the cause for 
suppression
of spin relaxation in semiconductors.    
%

\paragraph{\label{sec:IIIB2b} Two-dimensional III-V semiconductor systems.}

In two-dimensional III-V semiconductor systems (quantum wells and 
heterostructures) there are 
two distinct Hamiltonian terms that contribute
to DP spin dephasing: the bulk inversion asymmetry term $H_{\rm BIA}$  
and the structure inversion asymmetry term, $H_{\rm SIA}$, which appears 
only in 
asymmetric systems. Both $H_{\rm BIA}$ and
$H_{\rm SIA}$ lead to spin splitting of the conduction band linear in 
${\bf k}$. The two 
terms, however, predict a different dependence of $\tau_s$ on the 
quantum-well 
orientation relative to the principal axes. Figure \ref{fig:omega} shows 
the vector field patterns of the intrinsic magnetic fields for both
bulk and spin inversion asymmetry.

The bulk inversion asymmetry term comes from the bulk Dresselhaus 
spin splitting, Eq.~(\ref{eq:relax:dress}). 
Treating wave vectors ${\bf k}$ in Eq.~(\ref{eq:relax:dress}) as operators
$\hat{\bf k}=-i\nabla$, and evaluating $\bf\Omega$ as the expectation value in 
the 
confined states, leads to momentum quantization along the confinement  
unit vector
${\bf n}$ of the quantum well (QW). In the following, ${\bf k}$ denotes the 
wave vector for a Bloch state propagating in the plane, and 
${k_n^2}\equiv\langle (\hat{\bf k}\cdot {\bf n})^2\rangle$ 
denotes the expectation value 
of the square 
of the component of the wave-number 
operator 
normal to the plane 
in the lowest-subband state. For a rectangular QW of width $a$, 
$k_n^2=(\pi/a)^2$.
For a triangular well of confining potential $eEz$, 
$k_n^2 \approx 0.7794 (2m_e E/\hbar^2)^{2/3}$ [see, for example,
\textcite{deSousa2003:PRBc}].
Quantum averaging of $\kappa$ can be done using the formula
\begin{equation}
\langle \hat{k}_i \hat{k}_j \hat{k}_l \rangle 
= k_n^2 \left (k_i n_j n_l + k_j n_l n_i + k_l n_i n_j  \right ) 
+ k_i k_j k_l. 
\end{equation}
This readily gives~\cite{Dyakonov1986:SPS}
\begin{equation}\label{eq:relax:kappa2d}
\kappa_x= k_n^2  \left [2 n_x\left (n_y k_y -n_z k_z \right ) + 
k_x \left ( n_y^2 -n_z^2 \right ) \right ],
\end{equation}
and similarly for $\kappa_y$ and $\kappa_z$ by index permutation. 
Terms cubic in $k$ were omitted from the above equation, assuming that for 
narrow QW's 
$k^2\ll k_n^2$. The explicit knowledge of $\kappa$ is useful in qualitative 
analysis
of spin dephasing for particular orientations of QW's.

The spin dephasing tensor $1/\tau_{s,ij}$, defined in 
Eq.~(\ref{eq:relax:tauii}), 
is readily evaluated using 
Eqs.~(\ref{eq:relax:dress}) and 
(\ref{eq:relax:kappa2d})\footnote{Averaging over the directions
of $\bf k$ in a plane perpendicular to $\bf n$ can be performed by using 
$\overline{k_ik_j}=
(k^2/2)(\delta_{ij}-n_in_j)$.}
\begin{eqnarray}\label{eq:relax:nu}
1/\tau_{s,ij}=\left (\delta_{ij}{\rm Tr} \hat{\nu} 
- \nu_{ij}  \right )/\tau_{s}^0(E_{\bf k}),  
\end{eqnarray}
where 
\begin{equation}\label{eq:relax:DK}
\frac{1}{\tau_{s}^0(E_{\bf k})}=\frac{\alpha^2 \hbar^2 
\left (k_n^2\right) ^2}{2 m_c^2 E_g} 
E_{\bf k}\tau_p(E_{\bf k}).
\end{equation}
The tensor $\hat{\nu}$ depends on the orientation of $\bf n$ with
respect to the principal crystal 
axes~\cite{Dyakonov1986:SPS}\footnote{A trivial
typo in $\nu_{xx}$ is corrected.} 
\begin{eqnarray}
\nu_{xx}&=&4n_x^2(n_y^2+n_z^2)-(n_y^2-n_z^2)^2(9n_x^2-1), \\
\nu_{xy}&=&n_xn_y\left[9(n_x^2-n_z^2)(n_y^2-n_z^2)-2(1-n_z^2)\right],
\end{eqnarray}
and analogously for other components. 

\begin{figure}
\centerline{\psfig{file=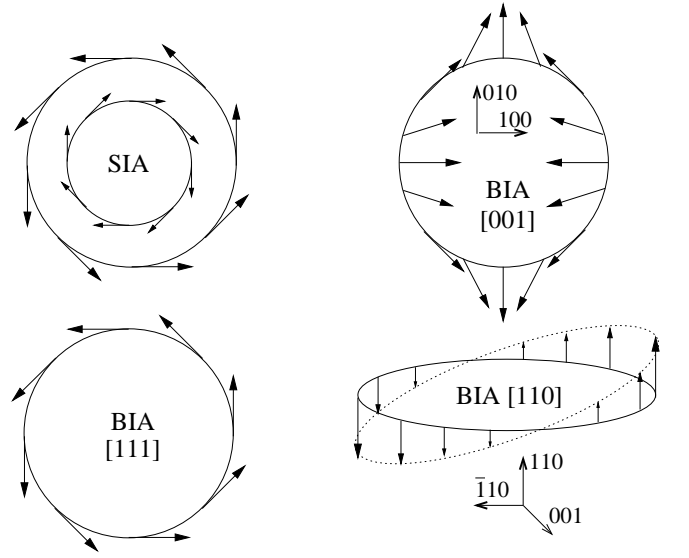,width=1\linewidth,angle=0}}
\caption{Vector fields ${\bf\Omega}({\bf k})\sim \kappa({\bf k})$ 
on the Fermi surface (here a circle) 
for the structure 
inversion asymmetry
(SIA)
and bulk 
inversion asymmetry (BIA). 
Since 
${\bf\Omega}({\bf k})$
is also the spin quantization axis, the vector pattern is also the 
pattern of the spin on 
the Fermi surface. As 
the opposite spins have different energies, the Fermi circle becomes 
two concentric circles with opposite spins. This is shown here only for 
the SIA case,
but the analogy extends to all examples. The field for BIA [110] lies 
perpendicular to
the plane, with the magnitude varying along the Fermi surface. All other 
cases have 
constant fields lying in the plane.  
}
\label{fig:omega}
\end{figure}

We follow \textcite{Dyakonov1986:SPS} in discussing the three important 
cases of 
[001], [111], and [110] quantum wells. For [001], 
\begin{equation}
{\bf\Omega}({\bf k})\sim\kappa=k_n^2 (-k_x,k_y,0).
\end{equation}
While the magnitude of ${\bf\Omega}({\bf k})$ is  
constant over the Fermi surface,
the directions follow a ``breathing'' pattern as shown in Fig.~\ref{fig:omega}. 
The spin relaxation times follow from 
Eq.~(\ref{eq:relax:DK}): 
$1/\tau_{s,xx}=1/\tau_{s,yy}=1/2\tau_{zz}=1/\tau_{s}^0$.
Defining $1/\tau_{s\parallel}$ and $1/\tau_{s,\perp}$ as spin dephasing 
rates of 
spins parallel and perpendicular to the plane, one obtains
\begin{equation}
1/\tau_{s,\parallel}=1/2\tau_{s,\perp}=1/\tau_{s}^0.
\end{equation}
As expected for the case of the in-plane field, the lifetime of a spin 
parallel to 
the plane is twice that of the a spin perpendicular to the plane.

For [111] QW's,
\begin{equation}
{\bf\Omega}({\bf k})\sim\kappa = 2/\sqrt{3} k_n^2 ({\bf k}\times {\bf n}).
\end{equation}
The intrinsic magnetic field lies in the plane, having a constant magnitude 
(refer to Fig.~\ref{fig:omega}). 
Spin relaxation rates are now 
$1/\tau_{s,ii}=16/9\tau_{s}^0$ and $1/\tau_{s,i\ne j}=4/9\tau_{s}^0$. 
By diagonalizing $1/\tau_{ij}$  we obtain
\begin{equation}
1/\tau_{s\parallel}=1/2\tau_{s,\perp}=4/3\tau_{s}^0.
\end{equation}
As for the [001] case, a perpendicular spin dephases twice as fast as a 
parallel one, since 
${\bf\Omega}({\bf k})$ lies in the plane.

The most interesting case is the [110] orientation for which 
$1/\tau_{s,xx}=1/\tau_{s,yy}=1/2\tau_{s,zz}=-1/\tau_{s,xy}=1/8\tau_{s}^0$. 
Other
off-diagonal components vanish. Diagonalizing the tensor  gives
\begin{equation}
1/\tau_{s,\parallel}=1/4\tau_{s}^0,\,\,\,1/\tau_{s,\perp}=0.
\end{equation}
The perpendicular spin does not dephase. This is due to the fact that $\kappa$, 
unlike the previous cases, is always normal to the plane 
(see Fig.~\ref{fig:omega}), 
and thus cannot
affect the precession of the perpendicular spin. Indeed, 
\begin{equation}
{\bf\Omega}({\bf k})\sim\kappa = k_n^2 \left (k_x/2) (-1,-1, 0 \right ),
\end{equation}
where it is used that ${\bf k}\cdot {\bf n}=0$. Spin dephasing in [110] QW's can 
still be 
due to the cubic terms in $k$ left out of Eq.~(\ref{eq:relax:kappa2d}) or to 
other spin relaxation mechanisms. Note that the magnitude of 
${\bf\Omega}({\bf k})$
changes along the Fermi surface. 
Electrons moving along [001] experience
little spin dephasing.

The structure inversion asymmetry term arises from the Bychkov-Rashba spin 
splitting ~\cite{Bychkov1984:JETPL,Bychkov1984:JPC,Rashba1960:SPSS} 
occurring in asymmetric QW's 
or in deformed bulk systems. The corresponding Hamiltonian is that
of Eq.~(\ref{eq:HDP}), with the precession vector
\begin{equation} \label{eq:relax:BR}
{\bf\Omega}({\bf k})=2\alpha_{BR} ({\bf k}\times {\bf n}). 
\end{equation}
Here $\alpha_{BR}$ is a parameter depending on spin-orbit coupling and the 
asymmetry
of the confining electrostatic potentials arising from the growth 
process of the
heterostructure. The splitting can also arise in nominally symmetric 
heterostructures
with fluctuations in doping density \cite{Sherman2003:APL}.  
The Bychkov-Rashba field always lies
in the plane, having a constant magnitude. As for the bulk inversion 
asymmetry case, the
structure inversion asymmetry leads to a splitting of the Fermi surface, 
according
to the direction of the spin pattern---parallel or antiparallel to 
$\bf{\Omega}({\bf k})$,
as shown in Fig.~\ref{fig:omega}. Perhaps the most appealing fact about 
structure
inversion asymmetry is that $\alpha_{BR}$ can be tuned electrostatically, 
potentially
providing an effective spin precession control without the need for magnetic 
fields 
\cite{Rashba2003:PRL,Levitov2003:PRB}.
This has led to one of the pioneering spintronic proposals 
by \textcite{Datta1990:APL} (see Sec.~\ref{sec:IVE1}). 
Note that for the [111] orientation the bulk
and structure inversion asymmetry terms have the same form.

Using the same procedure as for bulk inversion asymmetry, 
we describe the 
spin relaxation rate by Eq.~(\ref{eq:relax:nu}) as 
\begin{equation}
1/\tau_{s}^0=4\alpha_{BR}^2\frac{m_c}{\hbar^2} E_{\bf k} \tau_p
\end{equation}
and
\begin{equation}\label{eq:relax:nusia}
\nu_{ij}=1-n_in_j.
\end{equation}
Since the intrinsic precession vector $\sim {\bf \Omega}({\bf k})$ for the 
structure inversion
asymmetry always lies in the plane, 
a perpendicular spin should dephase 
twice as fast as a spin in the plane. Indeed,
by diagonalizing $1/\tau_{s,ij}$ one finds that 
\begin{equation}
1/\tau_{s,\parallel}=1/2\tau_{s,\perp}=1/\tau_s^0
\end{equation}
holds for all QW orientations ${\bf n}$. 
This interesting fact qualitatively
distinguishes structure from bulk inversion asymmetry and can be used in 
assessing the 
relative importance of
the Dresselhaus and Rashba spin splittings in III-V heterostructure systems. 
If bulk and structure inversion asymmetry are of similar importance, 
the interference terms from the cross 
product $\overline{\Omega_{\rm BIA} \Omega_{\rm SIA}}$ can lead to spin 
dephasing 
anisotropies within the plane, as was shown for [001] 
QW's  
in~ \cite{Averkiev1999:PRB,Kainz2003:PRB}. This plane anisotropy can be 
easily seen by adding
the corresponding vector fields in Fig.~\ref{fig:omega}. Another interesting
feature of bulk and structure inversion asymmetry fields is that injection
of electrons along a quasi-one-dimensional channel can lead to large relaxation
times for spins oriented along $\bf{\Omega}({\bf k})$, where ${\bf k}$ is
the wave vector for the states in the channel \cite{Hammar2002:PRL}.

Model spin dephasing calculations based on structure inversion asymmetry 
were carried out 
by \textcite{Pareek2002:PRB}. 
Calculations of $\tau_s$ based on the DP mechanism, 
with structural asymmetry due to doping fluctuations in the heterostructure 
interface were performed by \textcite{Sherman2003:APL}.

Research on spin inversion asymmetry is largely motivated by Datta-Das
spin field-effect transistor proposal (see Secs.~\ref{sec:IA} and
\ref{sec:IVE1}) in which $\alpha_{BR}$ is tailored by a gate.
This tailoring, however, has been controversial and the microscopic
origin of the Bychkov-Rashba Hamiltonian, and thus the interpretation
of experimental results on splitting in semiconductor heterostructures, 
has been debated. The Bychkov-Rashba Hamiltonian is often interpreted 
as arising from the electric field of the confining potential, assisted by
external bias, which acts on a moving electron in a transverse direction.
The relativistic transformation then gives rise to a magnetic field
(spin-orbit coupling) acting on the electron spin. The parameter 
$\alpha_{BR}$ is then assumed to be directly proportional to the 
confining electrical field. This is in general wrong, since the average
electric force acting on a confined particle of uniform effective mass
is zero.

The asymmetry that gives rise to structure inversion asymmetry is the
asymmetry in the band structure (including spin-orbit coupling)
parameters of a heterostructure, such as the effective mass, or the
asymmetry in the penetration of the electron wave function into the
barriers \cite{Silva1997:PRB}. The difficulty in understanding the 
influence of the external gates is caused by the lack of the
understanding of the influence of the gate field on the asymmetry
of the well. For a clear qualitative explanation of the involved
physics see \textcite{Pfeffer1999:PRB} and \textcite{Zawadzki2001:PRB}.
Band-structure {\bf $k \cdot p$} calculations of $\alpha_{BR}$ for
quantum wells in GaAs/AlGaAs heterostructures can be found in 
\textcite{Pfeffer1995:PRB}, \textcite{Pfeffer1997:PRB},
\textcite{Wissinger1998:PRB}, and \textcite{Kainz2003:PRB};
a calculation for InGaAs/InP quantum wells is reported in 
\textcite{Engels1997:PRB} and \textcite{Schapers1998:JAP},
in InSb/InAlSb asymmetric quantum well it can be found in 
\textcite{Pfeffer2003:PRB}, and in p-InAs metal-oxide-semiconductor
field-effect transistor channel in \textcite{Lamari2003:PRB}.
Adding to the controversy, \textcite{Majewski:2003} have recently
calculated the structure inversion asymmetry by local density functional
methods and concluded that the induced spin splitting arises from
microscopic electric fields in asymmetric atomic arrangements at the 
interfaces, so that a large Bychkov-Rashba term can be present in 
otherwise symmetric quantum wells with no common atom.

Interpretation of experimental data on structure inversion asymmetry is
difficult, especially at determining the zero magnetic field spin splitting
(usually seen in Shubnikov- de Haas oscillations), which is masked by
Zeeman splitting at finite fields. In addition, the splittings are small,
typically less than 1 meV. The Bychkov-Rashba parameter was measured in
GaSb/InAs/GaSb quantum wells 
($\alpha_{BR} \approx 0.9 \times 10^{-11}$ eV m for 75 $\AA$  
thick well)
by \textcite{Luo1990:PRB}; in InAlAs/InGaAs/InAlAs quantum wells (20 nm),
where also the gate voltage is obtained: $\alpha_{BR}$ ranged from
$10^{-11}$ eV m at the depleting voltage of -1 V, to $5 \times 10^{-12}$
eV m at +1.5 V. Weak antilocalization studies of InAlAs/InGaAs/InAlAs
quantum wells have recently been used to study electron density dependence
of $\alpha_{BR}$ by \textcite{Koga2002:PRL}. Gate dependence of $\alpha_{BR}$
was also measured in modulation-doped InP/InGaAs/InP quantum wells
\cite{Engels1997:PRB,Schapers1998:JAP}. The observed values are several
$10^{-12}$ eV m. On the other hand, there are experimental reports that
either fail to observe the expected spin splitting due to 
Bychkov-Rashba field, or interpret the splitting differently [see, for example,
\textcite{Brosig1999:PRB} and \textcite{Rowe2001:PRB}].
Furthermore, measurements of \textcite{Heida1998:PRB} show a constant
$\alpha_{BR} \approx 0.6 \times 10^{-11}$ eV m, independent of gate
voltage, in asymmetric AlSb/InAS/AlSb quantum wells, demonstrating that
control of $\alpha_{BR}$ may be difficult. In order to unify the
different views on what exactly the Bychkov-Rashba spin splitting means
and how the spin splitting can be tuned with gate voltage, more
experimental efforts need to be devoted to this interesting topic.

\subsubsection{\label{sec:IIIB3} Bir-Aronov-Pikus Mechanism} 

Spin relaxation of conduction electrons in p-doped semiconductors can also 
proceed through scattering, accompanied by spin exchange, by holes, 
as was first 
shown by \textcite{Bir1976:SPJETP}. 

The exchange interaction between electrons and holes is governed by the 
Hamiltonian
\begin{equation}
H= A{\bf S}\cdot {\bf J} \delta({\bf r}),
\end{equation}
where $A$ is proportional to the exchange integral between the conduction and 
valence states,
${\bf J}$ is the angular momentum operator for holes, ${\bf S}$ is the electron 
spin operator,
and ${\bf r}$ is the relative position of electrons and holes.

The spin-flip scattering probability depends on the state of the holes 
(degenerate or
nondegenerate, bound on acceptors or free, fast or slow). We present below the 
most 
frequently used formulas when assessing the relevance of the BAP mechanism. 
The formulas are 
valid for 
the usual cases of heavy holes $m_v \gg m_c$. For electron spin relaxation 
due to exchange with 
nondegenerate
holes, 
\begin{equation}\label{eq:relax:BAP1}
\frac{1}{\tau_s}= \frac{2}{\tau_0}N_a a_B^3 \frac{v_{\bf k}}{v_B} \left 
[\frac{p}{N_a}
|\psi(0)|^4 + \frac{5}{3}\frac{N_a-p}{N_a} \right],
\end{equation}
where $a_B$ is the exciton Bohr radius $a_B=\hbar^2\epsilon/e^2m_c$,
$p$ is the density of free holes, $\tau_0$ is an exchange splitting parameter:
$\hbar/\tau_0=(3\pi/64)\Delta_{\rm ex}^2/E_B$ (with $E_B$ denoting the Bohr 
exciton
energy, $E_B=\hbar^2/2m_ca_B^2$ and $\Delta_{\rm ex}$ the exchange splitting 
of the
excitonic ground state), and $v_B=\hbar/m_ca_B$; $|\psi(0)|^2$ is Sommerfeld's 
factor, which
enhances the free hole contribution. 
For an unscreened Coulomb potential
\begin{equation}
|\psi(0)|^2 = \frac{2\pi}{\kappa} \left 
[1-\exp\left(-\frac{2\pi}{\kappa}\right) 
\right ]^{-1},
\end{equation}
where $\kappa=E_{\bf k}/E_B$. For a completely screened 
potential $|\psi(0)|^2=1$.

If holes are degenerate and the electrons' velocity $v_{\bf k}$ 
is greater than the
Fermi velocity of  the holes', then  
\begin{equation}\label{eq:relax:BAP2}
\frac{1}{\tau_s}=\frac{3}{\tau_0}p a_B^3 \frac{v_{\bf k}}{v_B}\frac{k_B 
T}{E_{Fh}},
\end{equation}
where $E_{Fh}$ is the hole Fermi energy. For degenerate holes $|\psi(0)|^2$ is 
of order 1.
If electrons are thermalized, $v_{\bf k}$ needs to be replaced by the thermal 
velocity $v_e=(3k_B T/m_c)^{1/2}$.

The temperature dependence of $\tau_s$ is dominated by the temperature 
dependence of $|\psi(0)|^2$
as well as by $p$. The dependence on the 
acceptor density
is essentially $1/\tau_s \sim N_a$ for nondegenerate/bound holes from 
Eq.~(\ref{eq:relax:BAP1}) and $1/\tau_s \sim N_a^{1/3}$ for degenerate holes 
from 
Eq.~(\ref{eq:relax:BAP2}). In between, $1/\tau_s$ is only weakly dependent 
on $N_a$.
For GaAs $a_B\approx 114$ \AA, $E_B \approx 4.9$ meV, 
$v_B\approx 1.7 \times 10^7$ cm$\cdot$s$^{-1}$, 
$\tau_0\approx 1\times 10^{-8}$ s, and 
$\Delta_{\rm ex}\approx 4.7 \times 10^{-5}$ eV~\cite{Aronov1983:SPJETP}.

The BAP mechanism coexists with the the EY and DP mechanisms 
in p-doped materials lacking
inversion symmetry. The three mechanisms can be distinguished by their
unique density and temperature dependences. A general trend is that the
BAP dominates in heavily doped samples at small temperatures. At large
temperatures even for large acceptor densities, the DP mechanism can become
more important, due to its increased importance at large electron energies. 
Specific examples of the domain of importance for the three
mechanisms are discussed in Sec.~\ref{sec:IIID1}.  
Model calculations of BAP processes for electrons in p-doped bulk and quantum 
wells were
performed by \textcite{Maialle1996:PRB,Maialle1997:PRB}. 

Another potentially relevant mechanism for spin relaxation of donor-bound 
electrons in p-doped semiconductors is the
exchange interaction with holes bound to acceptors~\cite{Dyakonov1974:SPJETP}. 
The exchange interaction provides an effective magnetic field for electron 
spins 
to precess, leading to inhomogeneous dephasing. Both electron hopping and hole 
spin-flip motionally narrow the precession.

\subsubsection{\label{sec:IIIB4} Hyperfine-interaction mechanism}

The hyperfine interaction, which is the magnetic interaction between 
the magnetic moments of electrons and nuclei, provides an important 
mechanism~\cite{Dyakonov1974:SPJETP} for ensemble spin dephasing
and single-spin decoherence of localized electrons, such as those confined 
in quantum dots (QD) or bound on donors. The interaction is too weak to cause 
effective 
spin relaxation
of free electrons in metals or in bulk semiconductors~\cite{Overhauser1953:PR},
as it is strongly dynamically narrowed by the itinerant nature of 
electrons (see Sec.~\ref{sec:IIIA1}).
In addition to spin dephasing, the hyperfine interaction
is relevant for spintronics as a means to couple, in a controlled way, 
electron 
and nuclear spins \cite{Dyakonov:1984}.

Localized electrons are typically spread over many lattice sites 
($10^4$--$10^6$),
experiencing the combined magnetic moments of many nuclei. In GaAs all the 
lattice nuclei carry the
magnetic moment of 3/2 spin, while in Si the most abundant isotope, $^{28}$Si,
carries no spin and the hyperfine interaction is due to $^{29}$Si 
(natural abundance 4.67\%) 
or the frequent donor $^{31}$P, both of nuclear spin $1/2$. As a result, an 
electron bound
on a shallow donor in Si experiences only around 100 magnetic nuclei, and the 
effects of the hyperfine interaction are considerably smaller than in GaAs.

The effective Hamiltonian for the hyperfine interaction is the 
Fermi contact potential 
energy~\cite{Slichter:1989} 
\begin{equation}
H=\frac{8\pi}{3} \frac{\mu_0}{4\pi}g_0 \mu_B \sum_i \hbar \gamma_{n,i} 
{\bf S}\cdot{\bf I}_i
\delta ({\bf r}- {\bf R}_i),
\end{equation}
where $\mu_0$ is the vacuum permeability, $g_0=2.0023$ is the free-electron 
$g$ factor, $\mu_B$
is the Bohr magneton, $i$ is the label for nuclei at positions 
${\bf R}_i$, ${\bf S}$,  and ${\bf I}_i$ are, respectively,  electron and 
nuclear spin operators expressed in the units of $\hbar$, and $\gamma_{n,i}$ 
is the nuclear 
gyromagnetic ratio.
We stress that it is the electron $g$ factor $g_0$ and not the effective
$g$ that appears in the hyperfine interaction, Eq.~(96), 
as shown by \textcite{Yafet1961:JPCS} [see also \textcite{Paget1977:PRB}].
It follows that the spin of an electron in an orbital state $\psi({\bf r})$ 
experiences 
magnetic field 
\begin{equation}
{\bf B}_n=\frac{2\mu_0}{3} \frac{g_0}{g} \sum_i \hbar \gamma_{n,i} {\bf I}_i 
|\psi({\bf R}_i)|^2,
\end{equation}
where $g$ is the effective $g$ factor of the electron. The electron Zeeman 
splitting 
due to the average $B_n$  
corresponds to a field of  $\sim 1$ 
T or thermal energy of 1 K, for a complete nuclear 
polarization~\cite{Paget1977:PRB}.

There are three important regimes in which the hyperfine interaction leads to 
spin dephasing of 
localized
electrons:

(i) In the limit of small orbital and spin correlation between 
separated 
electron states and nuclear spin states, spatial variations in ${\bf B}_n$ 
lead to 
inhomogeneous dephasing of 
the spin ensemble, with the rate proportional to the r.m.s. of $B_n$, 
given by the 
corresponding
thermal or nonequilibrium distribution of the nuclear spins. 
Such inhomogeneous dephasing is seen by electron spin resonance (ESR) 
experiments 
on donor states both in Si \cite{Feher1959b:PR} and in GaAs \cite{Seck1997:PRB}. 
This
effect can be removed by spin echo experiments (in Si donor states performed, 
for example, by
 \textcite{Gordon1958:PR}). The spread in the Larmor precession period due to 
the variance in 
$B_n$ in GaAs is estimated to be around 1 ns 
\cite{Merkulov2002:PRB,Dzhioev2002:PRL}).

(ii) Temporal fluctuations in $B_n$, 
which can occur due to nuclear dipole-dipole 
interactions, lead to irreversible spin dephasing and decoherence of electron 
spins.
Such processes are sometimes referred to as spectral diffusion, since the 
electron Zeeman levels split by $B_n$ undergo random shifts \cite{Sousa2002:P}.
The typical time scale for the fluctuations in GaAs is 
given by the nuclear Larmor precession period in the field of neighboring 
nuclei and
is of order 100 $\mu$s~\cite{Merkulov2002:PRB}. Nuclear moments 
also precess (and orient) in the magnetic fields of polarized electrons, an 
effect important in 
optical orientation \cite{Meier:1984}, where the feedback from this precession 
can 
be directly observed through the modulated precession of electron spins. The 
time
scale for the Larmor precession of nuclear spins in hyperfine fields is 1 $\mu$s in 
GaAs 
\cite{Merkulov2002:PRB}, so this effect does not lead to motional narrowing of 
$B_n$;
electron spins precess many times before the nuclear spin flips.

(iii) In the presence of strong orbital correlations (electron hopping or 
recombination with
acceptor hole states) or spin (direct exchange interaction) between
neighboring electron states, spin precession due to $B_n$ is motionally 
narrowed.
While the direct spin exchange interaction does not cause ensemble spin 
relaxation 
(the total 
spin is preserved in spin flip-flops), it leads to individual spin decoherence, 
which 
can be much faster than what is inferred from $T_2$. This 
effect 
is much more pronounced in GaAs than in Si, since the donor states spread to 
greater
distances, and thus even in the low-doping limits ($\approx 10^{14}$ cm$^{-3}$ 
donors)
the exchange interaction can be rather large, masking the effects of temporal 
fluctuations
of $B_n$ (see Sec.~\ref{sec:IIID3}). 
Many useful parameters for evaluating effective magnetic fields and precession 
frequencies
due the HFI mechanism in GaAs are given by \textcite{Paget1977:PRB}.

Ensemble spin dephasing due to the HFI mechanism in an external magnetic 
field has been studied 
by \textcite{Dyakonov1974:SPJETP}, who found suppression of $1/\tau_s$ 
if the external field 
is greater than $B_n$ for regime (i), or a smaller Larmor precession 
period than the correlation time 
for random changes 
in $B_n$, in regime (iii).
due to the external field. 

Calculations of $\tau_s$ using the HFI mechanism  were performed for shallow 
donor states in Si 
at low 
temperatures and magnetic fields~\cite{Saikin2002:NL}, for electron spins in 
QD's~\cite{Merkulov2002:PRB,Semenov2002:P}, and even for the case of 
conduction 
electrons in semiconductors (revisited by 
\textcite{Pershin2003:P}). Spin relaxation processes due 
to phonon-assisted HFI  were investigated in GaAs QD's, but were found to be 
ineffective \cite{Erlingsson2001:PRB}. Unfortunately, there are still 
too few experimental 
data to make conclusions about the merits of specific models of the 
HFI mechanism. 

Spin decoherence times for single-electron spins were recently computed 
for  
case 
(ii) by \textcite{Khaetskii2002:PRL,Khaetskii2002:P},
who studied spin coherence time $\tau_{sc}$ of a single electron spin 
in the regime in which the electron Larmor period due to $B_n$ is much
shorter than the correlation time of the nuclear magnetic field fluctuations.
Realistic estimates 
of HFI spin dephasing in GaAs QD's were given 
by \textcite{Sousa2003:PRB,Sousa2002:P}, who offer
reasons why mechanism (ii)
should dominate spin decoherence in GaAs QD's of radius smaller than 100 nm.
For instance,  in a 50 nm wide QD the estimated $\tau_{sc}$
is $\approx 50$ $\mu$s, 
large enough 
for quantum computing applications (see Sec.~\ref{sec:IVF}).
This claim is supported by the recent measurement of the spin dephasing
time of about 60 ms of an isolated spin in a phosphorus donor in
isotopically pure $^{28}$Si, by spin echo measurements 
\cite{Tyryshkin2003:PRB}.

\subsection{\label{sec:IIIC} Spin relaxation in metals}

The spin relaxation time of conduction electrons 
in metals has been measured by both CESR and spin injection techniques. 
Typical values of $\tau_s$ were found to be 0.1 to 1 ns, but the range
of observed values is large, from pico- to microseconds. To our knowledge the 
longest $\tau_s$ reported for a metal--a microsecond--was found in high-purity
sodium below 10 K~\cite{Kolbe1971:PRB}. 

The majority of simple metals are believed to follow the EY mechanism of spin 
relaxation,
with the possible exception of Li~\cite{Damay1976:PRB}. This is
supported by several facts: 

(i) The EY processes give the right order of 
magnitude
for $\tau_s$ ~\cite{Elliott1954:PR,Yafet:1963},
while other known possible spin relaxation mechanisms lead to 
much greater $\tau_s$ than what is observed~\cite{Overhauser1953:PR}.

(ii) The temperature dependence of $\tau_s$ is consistent with
the EY mechanism: $1/\tau_s$ is constant at low temperatures, indicating 
impurity
spin-flip scattering, while at high temperatures $1/\tau_s$ grows linearly with 
increasing
$T$, consistent with phonon-induced spin relaxation. 

(iii) The Elliott relationship, 
Eq.~(\ref{eq:relax:elliott}), has been tested for many important metals and 
found 
to be valid over many orders of magnitude of $\Delta g$~\cite{Beuneu1978:PRB} 
(this
reference contains a useful collection of data for $\Delta g$). For the 
majority of metals tested 
(alkali and noble), a best fit gives the quantitative formula, 
Eq.~(\ref{eq:revisedE}) 
\cite{Beuneu1978:PRB}. 
(iv) The Yafet relation, Eq.~(\ref{eq:relax:yafet}), is satisfied for 
most metals 
with the known 
temperature dependence of $\tau_s$~\cite{Monod1979:PRB,Fabian1999:JVST}.
The initially suggested deviation from the Yafet relation for several 
polyvalent 
metals 
(Al, Pd, Be, and Mg) was later resolved by spin-hot-spot theory 
\cite{Silsbee1983:PRB,Fabian1998:PRL,Fabian1999:JVST}, to be described
below.
This work showed that the magnitudes of the spin-mixing probabilities 
$b^2$, taken from 
atomic physics to test Eq.~(\ref{eq:relax:yafet}), should not be used in the
solid-state environment. Various band-structure anomalies (spin hot spots), 
such as crossing of the Brillouin-zone boundaries, 
accidental degeneracy points, 
or symmetry points
on the Fermi surface, can increase the atomic-physics-derived $b^2$ by several 
orders of magnitude, strongly enhancing spin relaxation in polyvalent metals 
as compared to simple 
estimates~\cite{Fabian1998:PRL,Fabian1999:JAP}.

(v) A realistic, first-principles calculation for Al \cite{Fabian1999:PRL} 
(see Sec.~\ref{sec:IIIC}) using
Eq.~(\ref{eq:relax:ey}) shows  
excellent
agreement with experiment. 
 
CESR measurements of $\tau_s$ in various metals are numerous.\footnote{A list
of selected metal includes
Li \cite{Feher1955:PR,Orchard-Webb1970:PSS,Damay1976:PRB};
Na \cite{Feher1955:PR,Vescial1964:PR,Kolbe1971:PRB},
K \cite{Walsh1966:PR};
Rb \cite{Schultz1966:PRL,Walsh1966:PRL};
Cs  \cite{Schultz1966:PRL,Walsh1966:PRL};
Be \cite{Cousins1965:PL,Orchard-Webb1970:PSS}, Mg \cite{Bowring1971:PSS};
Cu \cite{Schultz1965:PRL,Dunifer1976:PRB,Monod1982:JP};
Au \cite{Monod1977:JLTP};
Zn \cite{Stesmans1981:PRB}; 
Al \cite{Lubzens1976:PRL}; 
graphite \cite{Wagoner1960:PR,Matsubara1991:PRB};
Rb$_{3}$C$_{60}$ \cite{Janossy1993:PRL}; MgB$_2$ \cite{simon2001:PRL}.
Various data on CESR $\tau_s$ are collected in 
\cite{Beuneu1978:PRB,Monod1979:PRB}.} 

The spin injection technique (see Sec.~\ref{sec:II}) was also used to measure 
$\tau_s$ for various metals, 
including Al 
\cite{Johnson1985:PRL,Jedema2002:Na,Jedema2002:APL,Jedema2003:PRB}, 
Au \cite{Elezzabi1996:PRL},
Cu \cite{Jedema2001:N,Jedema2003:PRB}, and Nb \cite{Johnson1994:APL}. 
In addition to CESR and spin injection, information about spin-orbit
scattering 
times 
$\tau_{so}$ (see below)
in various (but mostly noble) 
metals at low temperatures has been also 
obtained 
from weak localization magnetoresistance measurements 
on thin films~\cite{Bergmann1982:ZP} and 
tunneling spectroscopy of metallic nanoparticles~\cite{Petta2001:PRL}.
Surface-scattering spin relaxation times in normal metals and 
superconductors 
are
collected by \textcite{Meservey1978:PRL}. Interesting results were
obtained by injecting spin into superconductors. Using YBa$_2$Cu$_3$O$_{7-
\delta}$, for
example, data were interpreted \cite{Fu2002:PRB} to infer that the 
in-plane spin relaxation time is unusually long, 
about 100 $\mu$s at low temperatures to 1 $\mu$s close to the superconducting
transition temperature. For quasiparticles moving along the c-axis, 
$\tau_s$ is more likely to be the usual spin-orbit-induced spin 
relaxation time, having the values of 10-100 ps.
The microscopic origin of quasiparticle spin relaxation in cuprate 
superconductors is not yet known. 

There is one more important time scale, the spin-orbit scattering
time $\tau_{so}$, that is often invoked in mesoscopic transport
as a characteristic of spin relaxation processes. We discuss it briefly
in connection to $\tau_s$.
The spin-orbit scattering time is the scattering time of
Bloch electrons by the spin-orbit potential induced by impurities.
The spin-orbit
part of the Fourier transform of the impurity potential can be written 
as $i c({\bf k}-{\bf k}') ({\bf k} \times {\bf k'})\cdot {\bf \sigma}$,
where $c({\bf q})$ is proportional to the Fourier transform of the 
impurity potential.
The spin-orbit scattering time then is \cite{Werthamer:1969}
\begin{equation}
1/\tau_{so}=\frac{2\pi}{\hbar} N_i \langle |c({\bf k}-{\bf k}')|^2 
|{\bf k}\times {\bf k}' | \rangle^2
{\cal N}(E_F),
\end{equation}
where $N_i$ is the impurity concentration, ${\cal N}(E_F)$   
is the density of states per spin at
the Fermi level, and the angle brackets denote Fermi-surface averaging. 
As a parameter $\tau_{so}$ also includes the spin-orbit coupling of the host 
lattice, 
in the
sense of the EY mechanism. Note, however, that $\tau_s \ne \tau_{so}$, even 
at low
temperatures where the impurity scattering dominates spin relaxation, 
since the spin-orbit
scattering includes both spin-flip and spin-conserving processes, which, 
for
isotropic scattering rates are in the ratio 2:1. In addition, 
the spin relaxation rate is twice the spin-flip scattering rate, since each 
spin flip
equilibrates both spins equally.  For isotropic systems 
$1/\tau_s\approx 4/(3\tau_{so})$.
For a discussion of the effects of the DP processes on weak localization, 
see \textcite{Knap1996:PRB}.}

We illustrate spin relaxation in metals on the case of Al, whose $\tau_s$ was
measured by CESR and spin injection, and numerically calculated
from first principles. The case is instructive since it illustrates 
both the general principles of the EY mechanism as well as the predicting 
power 
of, and the need for realistic band-structure calculations of $\tau_s$. 

Spin relaxation in Al was originally observed in CESR experiments, in which 
$\tau_s$
was measured at low temperatures, from 1 to 100 K \cite{Lubzens1976:PRL}.
The spin relaxation rate $1/\tau_s$ was found to be independent of temperature 
below 10-20 K; at higher temperatures $1/\tau_s$ increases linearly with 
increasing $T$. 
The same behavior was later observed in the original spin injection 
experiment~\cite{Johnson1985:PRL,Johnson1988:PRBb}.
Recently, $\tau_s$  was measured by spin injection at room temperature
~\cite{Jedema2002:Na,Jedema2002:APL,Jedema2003:PRB}. Unlike the CESR 
and the original spin injection experiments, which were performed on 
bulk samples, the room-temperature measurement
used  thin Al films, observing strong spin relaxation due to surface
scattering.

Spin relaxation in Al, as well as in  other polyvalent metals, at first 
appeared \cite{Monod1979:PRB}, in that a simple application
of the Yafet relation, Eq.~(\ref{eq:relax:yafet}), yielded estimates for 
$1/\tau_s$ 
two orders of magnitude smaller than the observed data. Consider Na as a 
reference.
The atomic $\lambda_{so}/\Delta E$ [cf. Eq.~(\ref{eq:relax:lambda})] for 
Na and Al are 
within about 10\% of each other
~\cite{Beuneu1978:PRB,Monod1979:PRB}, yet the corresponding $\tau_s$ for Al is 
about two 
orders of magnitude smaller than that for 
Na~\cite{Feher1955:PR,Vescial1964:PR}. 
This anomaly extends to the $g$ factors as well.
For Na, $\Delta g_{Na} \approx -8 \times 10^{-4}$ and 
for
Al it is six times greater, $\Delta g_{Al}\approx -5\times 10^{-3}$, while
one would expect them to differ also by about 10\%. Note, however, that
the Elliott relation, Eq.~(\ref{eq:relax:elliott}), is unaffected by this
discrepancy, as it predicts that $\tau_s$(Na})/$\tau_s$(Al)$\approx 40$.  
It was later suggested~\cite{Silsbee1983:PRB} that this is due to 
accidental degeneracies in the two-band Fermi surface of Al. 

A full theoretical description, supported by first-principles
calculations, of spin relaxation in Al and other polyvalent metals 
led to the
spin-hot-spots 
theory~\cite{Fabian1998:PRL,Fabian1999:PRL,Fabian1999:JAP,Fabian1999:JVST}. 
Spin hot spots are states on the Fermi surface that have anomalously 
large spin mixing probabilities $|b|^2\approx 
(\lambda_{so}/\Delta E)^2$, 
arising from small energy gaps $\Delta E$. Quite generally, such
states occur near Brillouin-zone boundaries and accidental degeneracy points,
or at high-symmetry points. The condition for a spin hot spot is both a small 
band gap
$\Delta E$ and nonvanishing  
$\lambda_{so}$.\footnote{At some symmetry points 
$|b|$ may be very small. This occurs in the noble metals which 
have Fermi states at 
the Brillouin-zone boundaries, where $\Delta E$ is large,
but the corresponding $\lambda_{so}$ is very small 
due to symmetry.} 

If an electron hops in or out of a spin hot spot, the chance of spin flip 
dramatically increases. Although the total area of spin hot spots on the 
Fermi 
surface
is small, their contribution to $1/\tau_s$ is dominant, due to the large value
of their $|b|^2$ in the Fermi surface average $\langle |b|^2 \rangle$, as
was shown by analytical arguments~\cite{Fabian1998:PRL,Fabian1999:JAP}.
A realistic numerical calculation~\cite{Fabian1999:PRL} for Al, 
also showed that both
the accidental degeneracies considered by~\cite{Silsbee1983:PRB} and states 
close to 
the Brillouin-zone
boundaries dominate spin relaxation. 

A realistic calculation of $\tau_s$ in Al, based on pseudopotentials 
and a realistic phonon description, has been performed by 
\textcite{Fabian1999:PRL} and 
compared to the experimental data available for $T<100$ 
K~\cite{Johnson1985:PRL,Johnson1988:PRBb}. 
Figure \ref{fig:T1_Al} shows both the experiment and the theory. 
In the experimental data only the phonon contribution to $1/\tau_s$ is 
retained
\cite{Johnson1985:PRL}; the constant background impurity scattering is removed.
The figure shows a rapid decrease of $\tau_s$ with increasing $T$ at low $T$, 
where
the agreement between experiment and theory is very good. 
Above 200 K (the Debye temperature $T_D\approx 400$ K)
the calculation predicts a linear dependence $\tau_s [{\rm ns}] 
\approx 24\times
T^{-1}[\rm K^{-1}]$.
In the phonon-dominated linear regime the EY mechanism predicts that 
the ratio $a^{ph}=\tau_p/\tau_s$ does not depend on $T$
(see Sec.~\ref{sec:IIIB1}). 
The calculated value is 
$a^{ph}_{th}=1.2\times 10^{-4}$~\cite{Fabian1999:PRL}, 
showing that $10^4$ phonon scatterings are needed to randomize electron spin.

An important step towards extending spin injection capabilities was undertaken 
recently
by achieving spin injection into Cu and Al at room temperature
~\cite{Jedema2001:N,Jedema2002:Na,Jedema2002:APL,Jedema2003:PRB}; the 
measured data 
are unique
in providing reliable values for spin diffusion lengths and spin relaxation 
times in 
these two important metals at room temperature. 
The measured values for Al are 
somewhat 
sensitive to the experimental procedure/data analysis: $\tau_s= 85$ ps
~\cite{Jedema2002:Na} and $\tau_s=124$ ps ~\cite{Jedema2003:PRB}, as compared 
to $\tau_s=90$ ps predicted by the theory at $T=293$ K. The room temperature 
experimental
data are included in Fig.~\ref{fig:T1_Al} for comparison. They nicely confirm 
the 
theoretical prediction. Less sensitive to data analysis is the ratio $a^{ph}$, 
for
which the experiments give $1.1\times 10^{-4}$~\cite{Jedema2002:Na} and 
$1.3\times 10^{-4}$ \cite{Jedema2003:PRB}, comparing favorably with the
theoretical  $a^{ph}_{th}=1.2\times 10^{-4}$.    
 
\begin{figure}
\centerline{\psfig{file=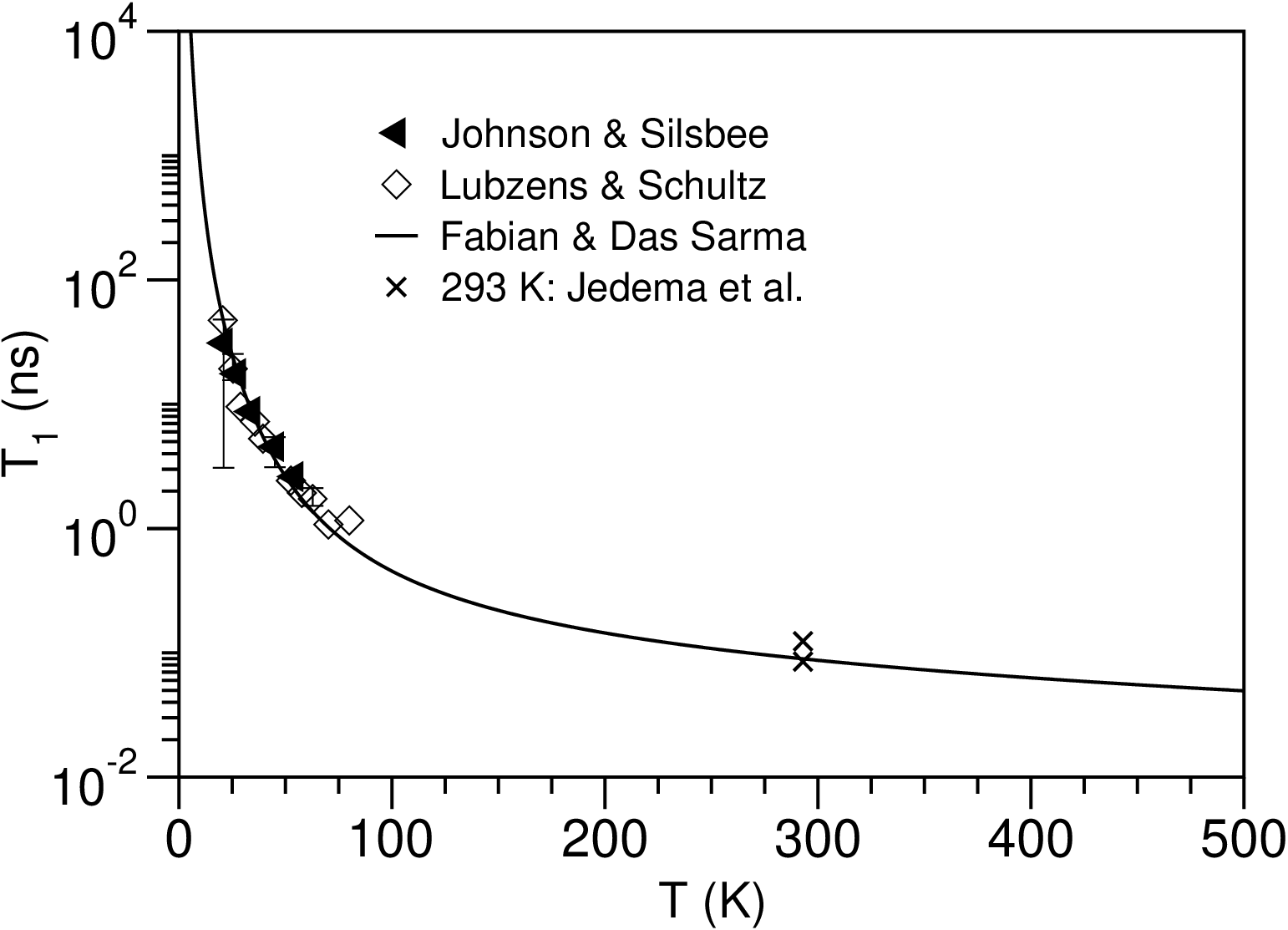,width=1\linewidth,angle=0}}
\caption{Measured and calculated $\tau_s$ in Al. The low-$T$ measurements are 
CESR \cite{Lubzens1976:PRL} and 
spin injection \cite{Johnson1985:PRL}. 
Only the phonon contribution is shown, as adapted from
\textcite{Johnson1985:PRL}. 
The solid line is the first-principles calculation, not a fit to the 
data, \cite{Fabian1998:PRL}.
The data at $T=293$ K are results from room-temperature spin injection 
experiments of \textcite{Jedema2002:Na,Jedema2003:PRB}. 
Adapted from \onlinecite{Fabian1999:PRL}.}
\label{fig:T1_Al}
\end{figure}

Spin relaxation in Al depends rather strongly on magnetic fields at low $T$.
CESR measurements ~\cite{Lubzens1976:PRL,Dunifer1976:PRB} 
show that at temperatures below 100 K, $1/\tau_s$ increases linearly
with increasing $B$. A specific sample~\cite{Lubzens1976:PRL}
showed a decrease of $\tau_s$ from about 
20 ns to 1 ns, upon increase in $B$ from 0.05  to 1.4 T. 
It was proposed that the observed behavior was due to cyclotron motion 
through spin hot spots ~\cite{Silsbee1983:PRB}. The reasoning is as follows.
Assume that there is considerable spread (anisotropy) 
$\delta g \approx \Delta g$ 
of the $g$ factors over the Fermi surface. Such a situation is common in
polyvalent metals, whose spin hot spots have anomalously 
large spin-orbit coupling.
In a magnetic field the electron spins precess correspondingly with rates
varying by $\delta \Omega_L \approx (\delta g/g) \Omega_L$, where 
$\Omega_L$ is the Larmor frequency. 
Motional narrowing leads to $1/\tau_s \approx (\delta\Omega_L)^2 \tau_c$,
where $\tau_c$ is the correlation time for the random changes in $g$. 
At small magnetic fields $\tau_c=\tau_p$ and $1/\tau_s \sim B^2 \tau_p$. 
Such 
a quadratic dependence of $1/\tau_s$ on $B$ is a  typical motional 
narrowing case
and has been observed at low temperatures in Cu \cite{Lubzens1976:PRL}.  
As the field increases $\tau_c$ becomes the time of flight through 
spin hot spots, in which case $\tau_c \sim 1/B$. As a result $1/\tau_s$ 
acquires
a component linear in $B$, in accord with experiment.

In an effort to directly detect phonon-induced spin flips in Al, an interesting 
experiment was devised ~\cite{Lang1996:PRL,Grimaldi1996:PRL} 
using the Zeeman splitting of the energy gap in Al superconducting tunnel 
junctions. 
Although the experiment failed, due to overwhelming spin-flip 
boundary scattering,
it showed the direction for future research in studying spin-flip 
electron-phonon
interactions.

\subsection{ \label{sec:IIID} Spin relaxation in semiconductors}

Although sorting out different spin-relaxation mechanisms of conduction 
electrons in semiconductors
is a difficult task, it has generally been 
observed that the EY mechanism is relevant in small-gap and large-spin-orbit 
coupling semiconductors, while the DP processes are responsible for spin 
dephasing in middle-gap 
materials and at high temperatures. In heavily p-doped samples the BAP  
mechanism 
dominates at lower temperatures, while DP at higher. In low-doped systems the
DP dominates over the whole temperature range 
where electron states are extended.  
Spin relaxation of bound electrons proceeds 
through the hyperfine interaction. Finally, spin relaxation of holes is due to 
the EY processes.
In bulk III-V or II-VI materials, for holes $\tau_s\approx \tau_p$, 
since the valence spin and orbital states are completely
mixed. However, in two-dimensional systems, where the heavy and light hole 
states are split,
hole spin relaxation is much less effective.

\subsubsection{\label{sec:IIID1} Bulk semiconductors}

There is a wealth of useful data on $\tau_s$ in 
semiconductors.\footnote{References for selected semiconductors include
the following:
{p-GaAs}: 
\cite{Fishman1977:PRB,Marushchak1984:SPSS,Seymour1981:PRB,Aronov1983:SPJETP,%
Zerrouati1988:PRB,Sanada2002:APL};
{n-GaAs}: See \ref{sec:IIID3};
{p-Al$_{x}$Ga$_{1-x}$As}: \cite{Garbuzov1971:ZhETF,Clark1975:PRB};
{p-GaSb}: \cite{Aronov1983:SPJETP,Safarov1980:JPSJ,Sakharov1981:SPSS};
{n-GaSb}: \cite{Kauschke1987:PRB};
{n-InSb}: \cite{Chazalviel1975:PRB};
{InAs}: \cite{Boggess2000:APL};
{p-InP}: \cite{Gorelenok1986:SPS};
{n-InP}: \cite{Kauschke1987:PRB};
{n-GaN}: \cite{Fanciulli1993:PRB,Beschoten2001:PRB}.}

A comprehensive theoretical investigation of spin dephasing in bulk 
semiconductors
(both p- and n-types), applied to GaAs, GaSb, 
InAs, and InSb, has been carried out by \textcite{Song2002:PRB} by using the 
EY, DP, and BAP mechanisms.
The calculation uses analytical formulas like 
Eq.~(\ref{eq:relax:chazalviel}), 
while
explicitly evaluating $\tau_p$ for different momentum scattering processes 
at different control parameters (temperature and density), but only 
for nondegenerate 
electron systems (Boltzmann statistics) and zero magnetic field. 
The main results are as follows: in n-type III-V 
semiconductors DP dominates at $T \agt 5$ K. At lower $T$ 
the EY mechanism becomes relevant. This crossover temperature appears to be 
quite 
insensitive to the electron density, being between 1 and 5 K in most 
investigated III-V 
semiconductors for donor densities greater than $10^{14}$ cm$^{-3}$ 
\cite{Song2002:PRB}. 
For p-type materials the dominant mechanisms are DP and BAP, with the 
crossover 
temperature 
sensitive to the acceptor density. For example, in p-GaAs at room temperature, 
the DP mechanism dominates below $10^{18}$ cm$^{-3}$, while at 
large densities 
the BAP 
mechanism dominates. In small-gap InSb the DP mechanism appears to be 
dominant for all 
acceptor densities at temperatures above 50 K. The strong disagreement with 
experiment found at low T (at 5 K, to be specific) points to our still limited  
theoretical understanding of spin relaxation in semiconductors. 
The discrepancy likely arises 
from neglect of
HFI effects. 

Spin relaxation times as long as 300 ns were recently obtained in bulk, 
$\approx 100$ nm wide, GaAs at low 4.2 K, placed in the proximity of QW's
~\cite{Dzhioev2001:JETPL,Dzhioev2002:PRL}. The samples were low doped 
($\approx 10^{14}$ cm$^{-3}$ uncompensated donor density), so that optical 
orientation detected $\tau_s$ 
of
electrons bound on donors. At such low donor concentrations the HFI 
mechanism is 
responsible
for spin relaxation. The unusually large $\tau_s$ is attributed to the 
presence 
of additional
conduction electrons in the structure, coming from the barriers separating 
the 
sample and
the nearest QW. The hyperfine interaction is then motionally narrowed by the 
exchange 
interaction
between the donor-bound and conduction electrons. Upon depletion of the 
conduction electrons
from the sample by resonant excitations in the QW,  
$\tau_s$ decreased to 5 ns~\cite{Dzhioev2002:PRL}, implying that the 
effects of the
static hyperfine fields on bound-electrons spin precessions are not 
reduced by 
motional narrowing.

Spin relaxation of holes in bulk III-V materials is very fast due to a
complete mixing of orbital and spin degrees of freedom in the valence band. 
The 
EY mechanism predicts that hole $\tau_s$ is similar to hole $\tau_p$;
this is a common assumption when considering hole contribution 
to spin-polarized
transport. Hole spin lifetime in undoped GaAs has been measured by optical 
orientation and time-resolved spectroscopy \cite{Hilton2002:PRL}. 
The observed value
at room temperature is $\tau_s \approx 110$ fs, consistent with the 
theoretical 
assumption. 

Spin relaxation of conduction electrons in strained III-V crystals was studied 
experimentally and theoretically by 
\textcite{Dyakonov1986:SPJETP}. Spin relaxation under strain is enhanced 
and becomes
anisotropic due to the 
strain-induced spin splitting of the conduction band, which is linear in $k$, 
similarly to the bulk inversion asymmetry in two-dimensional systems 
(see Sec.~\ref{sec:IIIB2}). 
It was found that $1/\tau_s\sim\sigma^2$, 
where $\sigma$ is the applied stress, and that $\tau_s$ is only weakly 
temperature dependent. Spin relaxation of photoholes in strain crystals has 
been 
studied in 
\cite{Dyakonov1974:SPS}, with the conclusion that the hole spin along 
the strain axis can 
relax (by the EY processes) on time scales much longer than in 
unstrained samples, 
due to 
the lifting of heavy and light hole degeneracy.

Compared to III-V or II-VI, much less effort has been devoted to 
investigation of $\tau_s$ in bulk Si. The reason is that CESR is thus far
the only technique capable of effective detection of spin relaxation 
in Si. Optical orientation is rather weak~\cite{Lampel1968:PRL} due to the 
indirect band-gap structure, while robust 
spin injection in Si is yet to be demonstrated. 
Spin relaxation in Si is slow due to the presence of inversion symmetry 
(the DP mechanism is not applicable) and lack of a nuclear moment for 
the main Si isotope. Earlier experimental studies~\cite{Feher1959b:PR} were
concerned with the hyperfine-interaction-dominated spin dephasing in donor 
states.

A comprehensive experimental study of low-doped Si (P donors were present 
at the 
levels
$7.5\times 10^{14} \le N_d \le 8\times 10^{16}$ cm$^{-3}$), 
at temperatures $20 < T < 300$ K,
was performed by \textcite{Lepine1970:PRB}. Three distinct temperature regimes 
were 
observed:

(a) ($20 < T < 75$ K) Here $\tau_s$ decreases with increasing $T$.
The HFI mechanism dominates: electrons are bound to the ground donor states, 
while
thermal excitations to higher states and the exchange interaction with 
conduction
electrons motionally narrows the hyperfine interaction. 

(b) ($75 < T < 150$ K) In this temperature range 
$\tau_s$ continues to decrease with increasing $T$, the effect caused by the 
spin-orbit interaction in the first excited donor state being motionally 
narrowed by thermal motion. 

(c) ($T> 150$ K) Here
$1/\tau_s$ increases with $T$, in accord with the EY mechanism. 
The observed room-temperature CESR linewidth is
about 8 G, corresponding to the electron spin
lifetime of 7 ns.

\subsubsection{\label{sec:IIID2} Low-dimensional semiconductor structures}

The importance of low-dimensional semiconductor systems (quantum wells, wires, 
and dots)
lies in their great flexibility in manipulating charge and, now, also spin 
properties
of the electronic states. Studies of spin relaxation in those systems are driven
not only by the need for 
fundamental understanding of spin relaxation and decoherence, but also 
by the goal of finding ways to reduce or otherwise control spin 
relaxation and coherence in general. For a survey of spin relaxation 
properties 
of semiconductor 
quantum wells, see \textcite{Sham1993:JPCM}.

Spin relaxation in semiconductor heterostructures is caused by random magnetic
fields originating either from the base material or from the heterostructure 
itself. All four mechanisms of spin relaxation can be important, depending on 
the material, doping, and geometry. The difference from the bulk is 
the localization of the wave function into two, one, or zero dimensions
and the appearance of structure-induced random magnetic fields. 
Of all the mechanisms, the DP and HFI are believed to be most relevant. 

The most studied systems are GaAs/AlGaAs QW's. 
The observed $\tau_s$ varies from nanoseconds
to picoseconds, depending on the range 
of control parameters such as temperature, QW width or confinement 
energy $E_1$, carrier 
concentration, mobility, magnetic field, or bias.\footnote{Here 
is a list of selected references with useful data on $\tau_s$ in 
GaAs/AlGaAs QW's:
confinement energy dependence has been studied by 
\textcite{Tackeuchi1996:APL,Britton1998:APL,Ohno1999:PRL,Malinowski2000:PRB,%
Ohno2000:PE,Endo2000:JJAP};
temperature dependence is treated by 
\textcite{Wagner1993:PRB,Ohno1999:PRL,Malinowski2000:PRB,Ohno2000:PE,%
Adachi2001:PE};
carrier concentration dependence is studied by \textcite{Sandhu2001:PRL};
dependence on mobility is examined by \textcite{Ohno1999:PRL}; 
and dependence on magnetic field 
in studied by \textcite{Zhitomirskii1993:JETPL}.}

Spin relaxation has also been investigated in In/GaAs 
\cite{Paillard2001:PRL,Cortez2002:PE}, in an InAs/GaSb superlattice 
\cite{Olesberg2001:PRB}, in InGaAs \cite{Guettler1998:PRB}, in 
GaAsSb multiple QW's by \cite{Hall1999:APL}.
II-VI QW's (specifically ZnCdSe) were
studied by \textcite{Kikkawa1997:S}, finding $\tau_s\approx 1$ ns, 
weakly dependent
on both mobility and temperature (in the range $5 < T < 270$ K). Electron and 
hole
spin dephasing have also been investigated in dilute magnetic 
semiconductor QW's doped
with Mn ions \cite{Crooker1997:PRB,Camilleri2001:PRB}.

Reduction of spin relaxation by inhibiting the BAP electron-hole exchange 
interaction through spatially
separating the two carriers has been demonstrated in 
$\delta$-doped p-GaAs:Be/AlGaAs 
\cite{Wagner1993:PRB}. The observed $\tau_s$ was $\approx 20$ ns 
at $T < 10$ K, 
which is indeed unusually large.
The exchange interaction was also studied at room temperature, observing
an increase of $\tau_s$ with bias voltage which increases spatial separation 
between electrons
and holes, reducing the BAP effects \cite{Gotoh2000:JAP}. In the fractional 
quantum Hall 
effect regime it was demonstrated \cite{Kuzma1998:S} that nonequilibrium spin 
polarization in 
GaAs QW's can survive for tens of $\mu$s. Spin lifetime was 
also found to be enhanced 
in GaAs QW's
strained by surface acoustic waves \cite{Sogawa2001:PRL}. A theoretical study 
\cite{Kiselev2000:PRB} proposed that spin dephasing in 2DEG can be 
significantly  suppressed by constraining the system to finite stripes, several 
mean free paths wide.

Theoretical studies focusing 
on spin dephasing in III-V and II-VI systems include those of 
\cite{Wu2000:PRB,Wu2001:JS,Wu2002:SSC,Lau2001:PRB,Puller2002:P,Lau2002:JAP,%
Bronold2002:PRB,Krishnamurthy2003:APL}.
Spin relaxation due to the DP mechanism with bulk inversion asymmetry term 
in the important case of GaAs/AlGaAs 
rectangular QW's was investigated by Monte-Carlo simulations 
\cite{Bournel2000:APL} at room 
temperature, including interface roughness scattering. Nice agreement with 
experiment was
found for $\tau_s(E_1)$, where $E_1$ is the confinement energy. 
Interface roughness becomes important at large
values of $E_1$, where scattering increases $\tau_s$ (see also 
\textcite{Sherman2003:APL}). 

Spin relaxation and spin coherence of spin-polarized photoexcited 
electrons and holes in symmetric p- and n-doped 
and undoped GaAs/AlGaAs quantum 
wells was investigated using rate equations 
\cite{Uenoyama1990:PRL,Uenoyama1990:PRB}. 
It was shown that in these heterostructures hole spin relaxation proceeds 
slower 
than electron-hole recombination. Hole relaxation is found to occur mostly 
due to 
acoustic phonon emission. The ratio of the spin-conserving to spin-flip hole 
relaxation
times was found to be 0.46, consistent with the fact that 
luminescence is polarized even in n-doped quantum wells at times greater 
than the
momentum relaxation time. Similar observations hold for  strained 
bulk GaAs, where hole spin relaxation is also reduced. Spin relaxation of 
holes in quantum wells 
was calculated \cite{Ferreira1991:PRB,Bastard1992:SS} using the interaction 
with ionized
impurities and s-d exchange in semimagnetic semiconductors. It was shown that 
size quantization 
significantly reduces 
spin relaxation of holes, due to the 
lifting of
heavy and light hole degeneracy. The observed spin lifetimes for holes at 
low 
temperatures
reached up to 1 ns, while at $T > 50$ K in the same samples $\tau_s$ got
smaller than
5 ps \cite{Baylac1995:SSS}. 

Spin dynamics and spin relaxation of excitons in GaAs 
\cite{Munoz1995:PRB,Vina2001:PB} and ZnSe \cite{Kalt2000:PSS} were
investigated 
experimentally and theoretically \cite{Maialle1993:PRB,Sham1988:PE}.
Coherent spin dynamics in magnetic semiconductors was considered by 
\textcite{Linder1998:PE}.

Spin relaxation in Si heterostructures has been investigated by electron 
spin resonance in modulation doped Si/SiGe QW's.  
Very high mobility (about $10^5$ cm$^2\cdot$V$^{-1}\cdot$s$^{-1}$) samples with 
$n\approx 3\times
10^{11}$ cm$^{-2}$ free electrons forming a 2DEG show, 
at $T=4.2$ K, 
$T_1$ up to 30 $\mu$s \cite{Sandersfeld2000:TSF,Graeff1999:PRB} 
and $T_2$ of the order of 100 ns~\cite{Graeff1999:PRB},
depending on the orientation of $B$ with respect to the QW growth direction. 
Spin relaxation
was attributed to Bychkov-Rashba spin splitting in these asymmetric wells, 
estimating
the corresponding $\hbar\alpha_{BR}$ 
in Eq.~(\ref{eq:relax:BR}) to be around $1\times 
10^{-14}$ eV$\cdot$m~\cite{Wilamowski2002:PE,Wilamowski2002:PRB}. 
Si/Ge heterostructures may have enhanced 
rates of spin relaxation due to 
the leakage of the electron wave function to Ge, 
which is heavier than Si and has 
greater spin-orbit interaction. 
Recent studies \cite{Wilamowski2004:PRB} have confirmed the dominant role
of the DP spin relaxation mechanism, leading to the microsecond spin 
relaxation times. The spin dephasing is argued to be strongly suppressed
by cyclotron motion in high-mobility samples (see Sec.~\ref{sec:IIIB2a}
for a brief discussion of the influence of magnetic field on $\tau_s$).
Spin-orbit coupling in symmetric Si/SiGe quantum wells has been studied 
theoretically by \textcite{Sherman2003:PRB}.

In quantum dots the relevant spin relaxation mechanism is still being 
debated,
as the mechanisms 
(EY and DP) effective for conduction electrons are ineffective for states
localized in QD's~\cite{Khaetskii2000:PRB,Khaetskii2001:PRB,Khaetskii2001:PE}.
It is believed, however, that similar to electrons bound on donors, the dominant
mechanism is a HFI process
\cite{Merkulov2002:PRB,Semenov2002:P,Sousa2003:PRB,Sousa2002:P,%
Khaetskii2002:PRL}.
Unfortunately, experiments on CdSe QD's (of diameter 22-80 \AA) show strong 
inhomogeneous dephasing
($\tau_s\approx 3$ ns at $B=0$, while $\tau_s \approx 100$ ps at 4 T) 
\cite{Gupta1999:PRB}, masking the intrinsic spin dephasing processes.
Recently a lower bound, limited by the signal-to-noise ratio, on $T_1$
of 50 $\mu$s has been measured at 20 mK in a one-electron quantum dot
defined in 2DEG GaAs/AlGaAs heterostructure by 
\textcite{Hanson2003:lanl}. The magnetic field of 7.5 T was oriented 
parallel to the plane heterostructure. While the actual value of
$T_1$ may be orders of magnitude larger, the observed bound suffices
for performing elementary quantum gates (see Sec.~\ref{sec:IVF}).

\subsubsection{\label{sec:IIID3} Example: spin relaxation in GaAs}

We review recent experimental results on spin relaxation in bulk n-GaAs%
\footnote{p-GaAs is extensively discussed by \textcite{Meier:1984}}
and GaAs-based low-dimensional systems.

\paragraph{\label{sec:IIID3a} Bulk n-GaAs.}

The importance of GaAs for spintronics and quantum computing applications 
has been 
recently
underlined by the discovery of rather long spin relaxation times (of the order 
of 100 ns)
in n-doped samples, as well as by the development of experimental 
techniques to 
manipulate spin precession in this semiconductor in a coherent 
manner~\cite{Awschalom2001:PE,Oestreich2002:SST}.

Both optical orientation and time-resolved Faraday rotation spectroscopy have 
been used to measure $\tau_s$ in bulk n-GaAs. 
In the earliest observations of optical spin orientation of
electrons, in 
n-Ga$_{0.7}$Al$_{0.3}$As with $N_d\approx 1\times 10^{16}$ cm$^{-3}$ at 4.2 K,
it was found that $\tau_s\approx 1.2 $ ns~\cite{Ekimov1971:JETPL}.
A much larger spin lifetime was found by optical orientation on n-GaAs 
~\cite{Dzhioev1997:PSS}, 
where for $N_d\approx 1\times 10^{15}$ cm$^{-3}$ the observed
$\tau_s\approx 42$ ns. 
Faraday rotation studies~\cite{Kikkawa1998:PRL,Awschalom2001:PE} 
found even longer  spin lifetimes. At the doping density 
$N_d=1\times 10^{16}$ cm$^{-3}$ of Si donors and  $T=5$ K, the observed 
$\tau_s\approx 130$ ns at zero magnetic field. 
At greater and smaller doping densities, spin relaxation time is significantly 
reduced: for both a nominally undoped sample and for $N_d=1\times 10^{18}$ 
cm$^{-3}$, $\tau_s\approx 0.2$ ns. A comprehensive theoretical investigation
of $\tau_s$ in bulk n-GaAs is reported by \textcite{Wu2000:PSS}, and
\textcite{Wu2001:JS}, who solved numerically kinetic equations in the
presence of magnetic filed. Only the DP mechanism was considered, 
acting with longitudinal phonon and impurity scattering.

A recent comprehensive study of $\tau_s$ based on optical orientation 
revealed a nice, albeit 
complex, picture of spin relaxation in bulk n-GaAs over a large range of doping 
levels~\cite{Dzhioev2002:PRB}. Figure \ref{fig:Dzhioev} summarizes these 
findings. 
The spin relaxation time rises with increasing
$N_d$ at small doping levels, reaching its first maximum (180 ns) at around 
$3\times 
10^{15}$ cm$^{-3}$; $\tau_s$ then decreases until $N_d=N_{dc}=2\times 10^{16}$ 
m$^{-3}$,
where a sudden increase brings $\tau_s$ to another maximum, reaching $\approx 
150$ ns. 
At still higher doping levels $\tau_s$ decreases strongly with increasing 
doping.

\begin{figure}
\centerline{\psfig{file=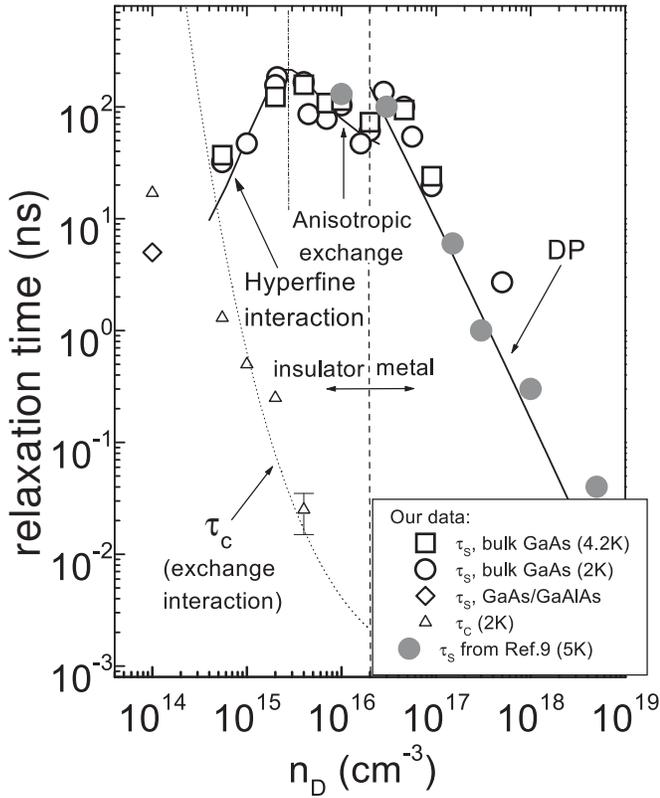,width=1\linewidth,angle=0}}
\caption{Spin relaxation time in n-GaAs as a function of donor density $N_d$ 
(labeled as $n_D$ here) at
low temperatures:  empty symbols, the optical orientation data of 
\textcite{Dzhioev2002a:PRB}; solid circles are the time-resolved 
Faraday 
rotation data
of  \textcite{Kikkawa1998:PRL,Awschalom2001:PE};
open triangles, single-spin decoherence times 
$\tau_{sc}\approx \tau_c$ due to the exchange interaction 
between electron spins on 
neighboring donors;  solid lines, parameter-free theoretical estimates 
with
labels indicating the dominant spin relaxation mechanisms;
dotted line a 
fit to the experimental data on the exchange correlation time 
(triangles) $\tau_c$, 
using a simple model of the exchange coupling between donor states; 
dashed vertical 
line, the metal-insulator transition at 
$N_{dc}=2\times 10^{16}$ cm$^{-3}$. 
From \onlinecite{Dzhioev2002a:PRB}. 
}
\label{fig:Dzhioev}
\end{figure}

The above picture is valid at $T \le 5$ K, where isolated shallow donors 
are not normally ionized, and the sample is a Mott insulator at small dopings. 
Conductivity
is due to hopping between donor states. Beyond the critical density 
$N_{dc}\approx 
2\times 10^{16}$ cm$^{-3}$ (the dashed vertical line in Fig.~\ref{fig:Dzhioev})
the donor states start to overlap and form an impurity 
conduction band---electronic
states delocalize and the sample becomes metallic. Figure~\ref{fig:Dzhioev} 
shows that
it is the rather narrow window around the metal-to-insulator transition where 
the largest $\tau_s$ are found.  

\begin{figure}
\centerline{\psfig{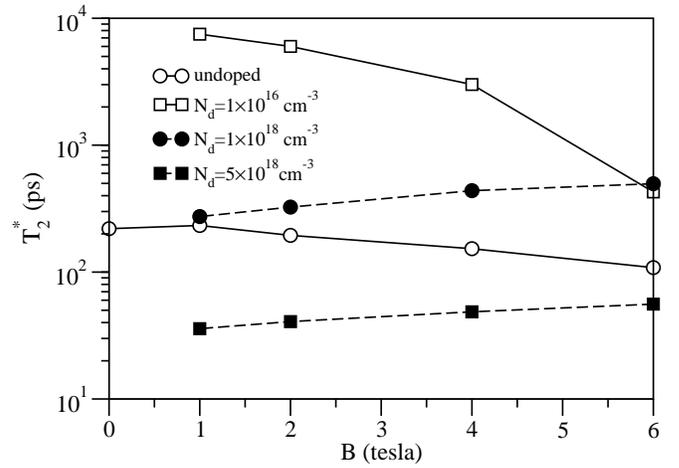}}
\caption{Measured magnetic-field dependence of the spin dephasing time 
(here denoted as $T_2^*$ to indicate the likely presence of inhomogeneous 
broadening; 
see Sec.~\ref{sec:IIIA1}) 
for bulk n-GaAs
at 5 K. Doping levels, varying from insulating 
($N_d < N_{dc}=
2\times 10^{16}$ cm$^{-3}$) to metallic ($N_d > N_{dc}$), are
indicated. Adapted from \onlinecite{Kikkawa1998:PRL}.} 
\label{fig:KA1}
\end{figure}

At $N_d > N_{dc}$ the DP mechanism dominates.  Equation
(\ref{eq:relax:DP}) for degenerate electrons explains the observed data
rather well.  Indeed, considering that $E_F\sim N_d^{2/3}$ and
assuming the Brooks-Herring formula for the impurity scattering
$(1/\tau_p \sim N_d/E_F^{3/2}$), one obtains $\tau_s\sim 1/N_d^2$,
which is observed in Fig.~\ref{fig:Dzhioev}. The EY mechanism, 
Eq.~(\ref{eq:relax:chazalviel}), would give $\tau_s\sim N_d^{-4/3}$.  The
data on the insulating side are consistent with the HFI mechanism: the
precession due to local random magnetic fields from the nuclear moments is
motionally narrowed by the exchange interaction, which increases with
increasing $N_d$ (that is, with increasing overlap between donor
states). The theoretical estimates~\cite{Dzhioev2002:PRB} agree well
with the data.  The behavior of $\tau_s$ in the intermediate regime, 
$3\times 10^{15}$ cm$^{-3} < N_d < N_{dc}$, where $\tau_s$ decreases
with increasing $N_d$, was proposed by~\cite{Kavokin2001:PRB} to be
due to motional narrowing of the antisymmetric exchange interaction
\footnote{The anisotropic exchange interaction of the
Dzyaloshinskii-Moriya form, $\bf{S}_1\times {\bf S}_2$ 
\cite{Dzyaloshinskii1958:PCS,Moriya1960:PR}
appears as a result of spin-orbit coupling in semiconductors 
lacking inversion symmetry.} 
between bound electrons, with the correlation time $\tau_c$
provided by the usual ${\bf S}_1\cdot{\bf S}_2$, direct exchange. This new
mechanism of spin relaxation, which should be generally present for
bound electrons in systems lacking inversion symmetry (such as III-V
and II-VI), although still being
investigated~\cite{Gorkov2003:PRB,Kavokin2002:P}, appears to give
a satisfactory explanation for the experimental data.

In addition to the doping dependence of $\tau_s$, both the temperature and
the magnetic field dependences of spin relaxation in bulk n-GaAs have been
studied by \textcite{Kikkawa1998:PRL}.  Figure \ref{fig:KA1} shows
$\tau_s(B)$ for samples with varying doping levels at $T=5$ K. The spin
relaxation time increases with $B$ in the metallic regime, the
behavior qualitatively consistent with the predictions of the DP
mechanism.  In contrast, $\tau_s$ in the insulating samples decreases
with increasing $B$.  Bound electrons are more susceptible to $g$-factor
anisotropies (due to the distribution of electron energies over donor
states) and local magnetic field variations (due to the hyperfine
interactions).  These anisotropies are amplified by increasing $B$
and motionally narrowed by the exchange interaction. It is thus
likely that $\tau_s\sim B^{-2}\tau_c(B)$, where the exchange
correlation time $\tau_c$ depends on $B$ through magnetic orbital
effects on the bound electron wave functions (magnetic confinement
reduces the extent of the bound orbital, thus reducing the exchange
integrals between neighboring donor states). However, no satisfactory
quantitative explanation for $\tau_s(B)$ in insulating samples
exists.

\begin{figure}
\centerline{\psfig{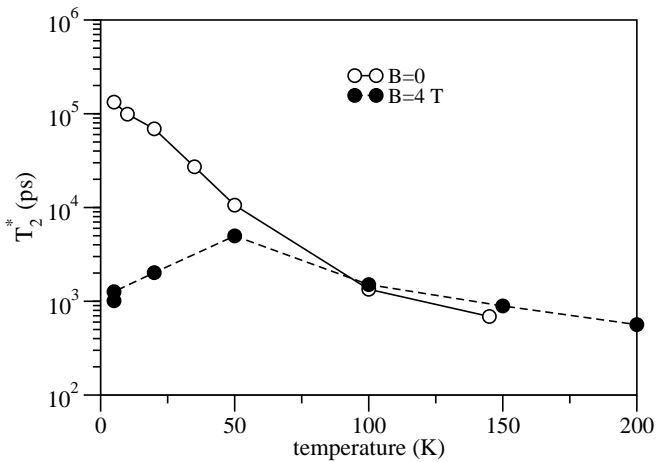}}
\caption{Measured temperature  dependence of the spin dephasing time 
for bulk n-GaAs doped with $N_d=1\times 10^{16}$ cm$^{-3}$ Si donors, at 
$B=0$ and $B=4$ T. 
The sample is insulating at low
$T$ and nondegenerate at high $T$ ($T\agt 50$ K, assuming $\approx 4$ meV for
the donor binding energy), where donors are ionized.
Adapted from \onlinecite{Kikkawa1998:PRL}. 
}
\label{fig:KA2}
\end{figure}

Figure \ref{fig:KA2} plots $\tau_s(T)$ for an insulating sample with
$N_d=1\times 10^{16}$ cm$^{-3}$ at $B=0$ and $B=4$ T.   
For the
zero-field data the initial decrease of $\tau_s$ with $B$ is very
rapid, dropping from 130 ns at 5 K to less than 1 ns at 150 K. However, 
the sample held at $B=4$ T 
shows at first a rapid
increase with increasing $T$, and then a decrease at $T\approx 50$ K.
The decrease of $\tau_s$ with increasing $T$ above 50 K has been found
to be consistent with the DP mechanism~\cite{Kikkawa1998:PRL}, taking
$\tau_s\sim T^{-3}$ in Eq.~(\ref{eq:relax:DPt}), while extracting
the temperature dependence of $\tau_p$ from the measurement of mobility. 
The DP
mechanism for conduction electrons was also observed in p-GaAs in the
regime of nondegenerate hole densities $N_a\approx 10^{17}$ cm$^{-3}$
at temperatures above  100 K~\cite{Aronov1983:SPJETP},
after the contribution from the BAP mechanism was subtracted using a
theoretical prediction. From the observed mobility
it was found that $\tau_p(T)\sim T^{-0.8}$, so that 
according to the
DP mechanism $\tau_s\sim
T^{-2.2}$, which is indeed consistent with the experimental data.  The
origin of $\tau_s(T)$ below 50 K in Fig.~\ref{fig:KA2} is less
obvious.  At low $T$, electrons are localized, so in order to explain
the experimental data the theory should include ionization of donors.
The increase with increasing $T$ of $\tau_s$ at 4 T  
invokes a
picture of motional narrowing in which the correlation time decreases
with increasing $T$ much faster than the dispersion of local Larmor
frequencies. We do not know of a satisfactory quantitative explanation
for these experimental results.\footnote{There is a 
discrepancy in the data presented in Figs. \ref{fig:KA1}
and \ref{fig:KA2}. Take the $N_d=1\times 10^{16}$ cm$^{-3}$ sample.
While Fig.~\ref{fig:KA1} reports $\tau_s\approx 3$ ns at 5 K and 4 T, 
$\tau_s$ is only about 1 ns in Fig.~\ref{fig:KA2}.
The reason for this difference \cite{Kikkawa2003:PC}
turns out to be electronically-induced nuclear polarization 
\cite{Kikkawa2000:S}.
At low temperatures and large magnetic fields, nuclear polarization
develops via the Overhauser effect inhomogeneously throughout the
electron spin excitation region. The inhomogeneous magnetic
field due to polarized nuclei causes inhomogeneous broadening
of the electronic $\tau_s$. The measured spin dephasing time is indeed
$T_2^*$, rather than the intrinsic $T_2$. Furthermore, since nuclear
polarization typically takes minutes to develop, the measured
$T_2^*$ depends on the measurement ``history.'' This is the reason
why two different measurements, reported in Figs. \ref{fig:KA1}
and \ref{fig:KA2}, show different $T_2^*$ under otherwise equivalent
conditions. The nuclear polarization effect is also part of the
reason why the $T_2(T)$ at 4 T 
sharply deviates from that
at zero field at small $T$. The technique should give consistent
results at small fields and large temperatures, as well as 
in heavily doped samples where the nuclear fields are motionally
narrowed by the itinerant nature of electrons.
}
Similar behavior of
$\tau_s(T)$ 
in insulating samples was found in GaN~\cite{Beschoten2001:PRB}.

The temperature dependence of $\tau_s$ for samples with 
$N_d \gg N_{dc}$ has been reported \cite{Kikkawa1998:PRL} to be very weak, 
indicating, for
these degenerate electron densities, that $\tau_p(T)$ is only  weakly
dependent on $T$. What can be expected for $\tau_s$ at room temperature?  The
answer will certainly depend on $N_d$. Recent experiments on time-
resolved Kerr rotation~\cite{Kimel2001:PRB} suggest that 5 ps$ < \tau_s <
10 $ ps for undoped GaAs and 15 ps $< \tau_s < 35$ ps for a heavily doped
n-GaAs with $N_d=2\times 10^{18}$ cm$^{-3}$.

For spintronic applications to make use of the large $\tau_s$
observed in bulk n-GaAs one is limited to both very small temperatures
and small doping levels. Although this may restrict the design of 
room-temperature spintronic devices, such a regime seems acceptable for
spin-based quantum computing (see Sec.~\ref{sec:IVF}), 
where one is interested in the spin lifetime
of single (or a few) electrons, bound to impurities or confined to
quantum dots.  How close is $\tau_s$ to the 
individual spin lifetime $\tau_{sc}$?
There is no clear answer yet.  Ensemble spin dephasing seen for
insulating GaAs samples appears to be due to motional narrowing of the
hyperfine interaction. The randomizing processes are spin flips due to
the direct exchange, leading to the correlation time $\tau_c$, which can
be taken as a measure for the lifetime $\tau_{sc}$ of the individual spins. 
Extracting
this lifetime from the experiment is not easy, but the obvious trend
is the smaller the $\tau_c$, the larger the $\tau_s$.  For a specific
model of spin relaxation in bound electron states $\tau_c$ was
extracted experimentally by \textcite{Dzhioev2002:PRB} by detecting
the changes in the spin polarization due to longitudinal magnetic fields.  The
result is shown in Fig.~\ref{fig:Dzhioev}.  The two times, $\tau_c$
and $\tau_s$ differ by orders of magnitude. For the doping levels where
$\tau_s$ is {\it greater} than 100 ns, $\tau_c$ is {\it smaller} than   
0.1 ns.
Unfortunately, the useful time
for spin quantum computing would be extracted in the limit of very
small dopings, where the data are still sparse. For an informal recent review
of $\tau_s$ in n-GaAs, see \textcite{Kavokin2002:PSS}.

Closely related to spin relaxation is spin diffusion. 
\textcite{Hagele1998:APL} observed the transport of a spin 
population---longitudinal spin drift---in i-GaAs over a length
scale greater than 4 $\mu$m in electric
fields up to 6 kV/cm and at low temperatures. 
This was followed by a remarkable result of \textcite{Kikkawa1999:N},
the observation of the drift of precessing electron 
spins---transverse spin drift---in GaAs with $N_d=1\times 10^{16}$ cm$^{-3}$, 
over 100 $\mu$m in moderate electric fields (tens of V/cm) at $T=1.6$ K,
setting the length scale for the spin dephasing.
By directly analyzing the spreading and drifting of the electron
spin packets in time, 
Kikkawa and Awschalom obtained
the spin diffusion 
(responsible for spreading) and electronic
diffusion
(drift by electric field) 
coefficients.
It was found that
the former is about 20 times as large as the latter. These results are
difficult to interpret, since the sample is just below the metal-to-insulator
transition, where charge is transported via hopping, but they suggest that 
spin diffusion is strongly
enhanced through the exchange interaction. Investigations of this type
in even smaller doping limits may prove important for understanding
single-spin coherence.

\paragraph{GaAs-based quantum wells.}

We discuss selected experimental results on spin relaxation in 
GaAs/Al$_x$Ga$_{1-x}$As QW's,
presenting  the temperature and confinement energy dependence of $\tau_s$.

Figure \ref{fig:QW1} plots the temperature dependence of $1/\tau_s$ in
the interval of $90 < T < 300$ K for QW's of width $L$ ranging from 6
to 20 nm \cite{Malinowski2000:PRB}.  The wells, with $x=0.35$ and
orientation along [001], were grown on a single wafer to minimize
sample-to-sample variations when comparing different wells. The
reported interface roughness was less than the exciton Bohr radius of
13 nm.  In these structures the excitonic effects dominate at $T < 50$ K
(with the reported $\tau_s\approx 50$ ps), while the exciton ionization is
complete roughly at $T > 90$ K, so the data presented are for free
electrons. Spin relaxation was studied  using pump-probe optical
orientation spectroscopy with a 2 ps time resolution and the typical
excitation intensity/pulse of $10^{10}$
cm$^{-2}$.

\begin{figure}
\centerline{\psfig{file=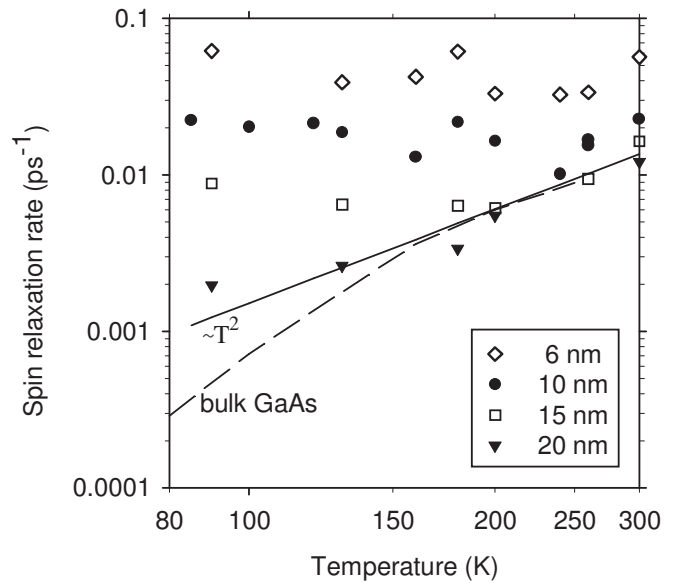,width=1\linewidth,angle=0}}
\caption{Measured temperature dependence of the conduction-electron-spin 
relaxation rate 
$1/\tau_s$
in GaAs/AlGaAs QW's of varying widths: the dashed curve,  data for
a low-doped ($N_a=4\times 10^{16}$ cm$^{-3}$) 
bulk p-GaAs \cite{Meier:1984}; solid line, the 
$\tau_s \sim T^2$ dependence. From \onlinecite{Malinowski2000:PRB}.} 
\label{fig:QW1}
\end{figure}

As Fig.~\ref{fig:QW1} shows, $\tau_s$ depends rather weakly on $T$ for
the narrow wells with $L < 10$ nm.
For the well with $L=15$ nm, after
being approximately constant (or somewhat decreasing) as $T$ increases
to about 200 K, $\tau_s$ 
increases with increasing $T$ at greater temperatures.
The increase is consistent with the $1/\tau_s \sim T^2$ behavior.  The
thickest well increases with the same power law, $1/\tau_s\sim T^2$, over 
the whole temperature range.  In order to make a reliable comparison
with theoretical predictions (the expected mechanism is that of DP in
two-dimensional systems), one needs to know the behavior of $\tau_p(T)$.
The DP mechanism predicts, for the nondegenerate electron densities
employed in the experiment, that $1/\tau_s \sim T^3 \tau_p$ [see 
Eq.~(\ref{eq:relax:DPt})] in the bulk and wide QD's, the condition being that
thermal energy is greater than the subband separation), and $1/\tau_s
\sim T E_1^2 \tau_p$ from Eq.~(\ref{eq:relax:DK}) for the
bulk inversion asymmetry
after making thermal averaging ($E_{\bf k} \rightarrow k_B T$),
when one realizes
that confinement energy $E_1$ is $\sim \langle k_n^2 \rangle$.
When one assumes that momentum relaxation in these elevated temperatures
is due to scattering by phonons, $\tau_p$ should be similar in bulk
and low-dimensional structures. From the observed high-temperature
bulk $\tau_s(T)$ (at low temperatures $\tau_s$ is affected by BAP
processes) one can estimate $\tau_p \sim 1/T$, which is consistent
with the constant $\tau_s$ for the narrow wells, 
and with the quadratic dependence for the wide wells.
At low $T$, in addition
to the BAP mechanism, $\tau_s$ will deviate from that in the bulk due
to impurity scattering. The EY and BAP mechanisms were found not
to be relevant to the observed data \cite{Malinowski2000:PRB}.

\begin{figure}
\centerline{\psfig{file=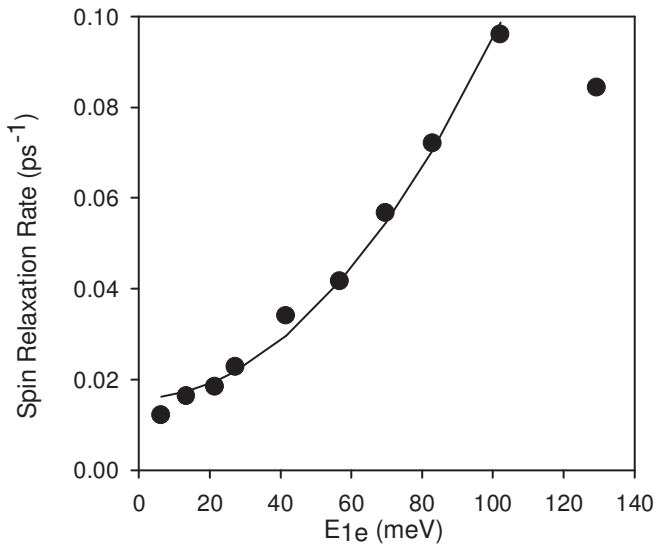,width=1\linewidth,angle=0}}
\caption{Measured room-temperature dependence of $1/\tau_s$ on the
confinement
energy $E_1$ for GaAs/AlGaAs QW's. The solid line is a quadratic fit, 
showing 
behavior
consistent with the DP mechanism. From \onlinecite{Malinowski2000:PRB}.} 
\label{fig:QW2}
\end{figure}

Figure \ref{fig:QW2} shows the dependence of $1/\tau_s$ on the
experimentally determined confinement energy $E_1$ for a variety of
QW's on the same wafer \cite{Malinowski2000:PRB}.  The data are 
at room
temperature. The spin relaxation time 
varies from somewhat less than
100 ps for wide QW's, approximating the bulk data (cf.
\textcite{Kimel2001:PRB} where 15 ps $< \tau_s < 35$ ps 
was found for a heavily
doped n-GaAs), to about 10 ps in most confined structures. The
downturn for the highest-$E_1$ well (of width 3 nm) is most likely due
to the increased importance of interface roughness at such small
widths \cite{Malinowski2000:PRB}.  Confinement strongly enhances spin
relaxation. This is consistent with the DP mechanism for
two-dimensional systems, in which the spin precession about the intrinsic
magnetic fields 
(here induced by bulk inversion asymmetry) 
increases as 
$E_1^2$ with increasing
confinement. The observed data in Fig.~\ref{fig:QW2} are consistent
with the theoretical prediction.

Similarly to bulk GaAs, spin relaxation in GaAs QW's was found to be reduced at 
carrier
concentrations close to the metal-to-insulator transition
($n\approx 5\times 10^{10}$ cm$^{-2}$) \cite{Sandhu2001:PRL}.

\section{\label{sec:IV} Spintronic devices and applications}

In this section we focus primarily on the physical principles 
and materials issues
for various device schemes, which, while not yet commercially viable, are likely
to influence future spintronic research and possible applications.

\subsection{\label{sec:IVA} Spin-polarized transport}

\subsubsection{\label{sec:IVA1} F/I/S tunneling}

Experiments reviewed by \textcite{Tedrow1994:PR} in
ferromagnet/insulator/superconductor (F/I/S) junctions have
established a sensitive technique for measuring the spin polarization $P$
of magnetic thin films and, at the same time, 
has demonstrated that the
current will remain spin-polarized after tunneling through an
insulator. 
These experiments also stimulated more recent
imaging techniques based on the spin-polarized STM  
(see \textcite{Johnson1990:JAP,Wiesendanger1990:PRL}; and a review,
\textcite{Wiesendanger:1998}) 
with the ultimate goal of imaging spin
configurations down to the atomic level.

The degree of spin polarization 
is important for many 
applications such as determining the magnitude of tunneling magnetoresistance
(TMR) in magnetic tunnel junctions (MTJ) [recall Eq.~(\ref{eq:julliere})].
Different probes for spin polarization generally
can measure significantly different values even in experiments
performed on the same homogeneous sample. 
In an actual MTJ, measured polarization is {\it not} an intrinsic
property of the F region and could depend on interfacial
properties and the choice of insulating barrier.  Challenges in 
quantifying $P$, discussed here in the context of F/I/S tunneling, even
when F is a simple ferromagnetic metal, should serve as a caution for
studies of novel, more exotic, spintronic materials.

F/I/S tunneling conductance is shown in 
Fig.~\ref{tunnel}, where for simplicity we assume that the spin-orbit
and spin-flip scattering (see Sec.~\ref{sec:IIIC})  
can be neglected, a good approximation for
Al$_2$O$_3$/Al \cite{Tedrow1971:PRLb,Tedrow1994:PR}, a common choice
for I/S regions.  For each spin the normalized BCS density of states
is ${\tilde {\cal N}}_S(E)=Re(|E|/2\sqrt{E^2-\Delta^2})$, where E is the
quasiparticle excitation energy and $\Delta$ the superconducting
gap.\footnote{Here we focus on a conventional $s$-wave superconductor
with no angular dependence in $\Delta$.} The BCS density of states
is split in a
magnetic field H, applied parallel to the interface, due to a shift in
quasiparticle energy $E \rightarrow E \pm \mu_B H$, for $\uparrow$
($\downarrow$) spin parallel (antiparallel) to the field, where
$\mu_B$ is the Bohr magneton.  The tunneling conductance is normalized
with respect to its normal state value--for an F/I/N junction,
$G(V)\equiv (dI/dV)_S/(dI/dV)_N=G_\uparrow(V)+G_\downarrow(V)$, where
$V$ is the applied bias. This conductance can be expressed 
by generalizing analysis of
\textcite{Giaever1961:PR} as
\begin{eqnarray}
G(V)&=&\int^{\infty}_{-\infty}\frac{1+P}{2} 
\frac{{\tilde {\cal N}}_{S}(E+\mu H) \beta dE}
{4 \cosh^2[\beta(E+qV)/2]}  \\ \nonumber  
&+& \int^{\infty}_{-\infty} \frac{1-P}{2} 
\frac{{\tilde{\cal N}}_{S}(E-\mu H) \beta dE}
{4 \cosh^2[\beta(E+qV)/2]}. 
\label{eq:tg}
\end{eqnarray}

Here $\beta=1/k_BT$, $k_B$ is the Boltzmann constant, $T$ is the temperature, 
and $q$ is the proton charge. The factors $(1\pm P)/2$ represent 
the difference in
tunneling probability 
between for 
$\uparrow$ and $\downarrow$ electrons.
While a rigorous determination of $P$,
in terms of materials parameters,
would require a full calculation
of spin-dependent tunneling, including the appropriate boundary conditions
and a detailed understanding of the interface properties, it is
customary to make some simplifications. Usually $P$ can be identified as 
\cite{Worledge2000:PRB,Maekawa:2002}
\begin{equation}
P \rightarrow P_G=(G_{N\uparrow} -G_{N\downarrow})/
(G_{N\uparrow}+G_{N\downarrow}),
\label{eq:PG}
\end{equation}
the spin polarization of the normal-state conductance (proportional
to the weighted average of the density of states in F and S and 
the square of the  
tunneling matrix element), 
where $\uparrow$ is the electron spin with the magnetic moment parallel 
to the applied field (majority electrons in F). 
With the further simplification of spin-independent and constant 
tunneling matrix element \cite{Tedrow1971:PRLa,Tedrow1994:PR},
Eq.~(\ref{eq:PG}) can be expressed as
\begin{equation}
P \rightarrow P_{\cal N}=({\cal N}_{F\uparrow} -{\cal N}_{F\downarrow})/
({\cal N}_{F\uparrow} + {\cal N}_{F\downarrow}),
\label{eq:PF}
\end{equation}
the spin polarization of the tunneling density of states in the F region 
at the Fermi level.

\begin{figure} 
\centerline{\psfig{file=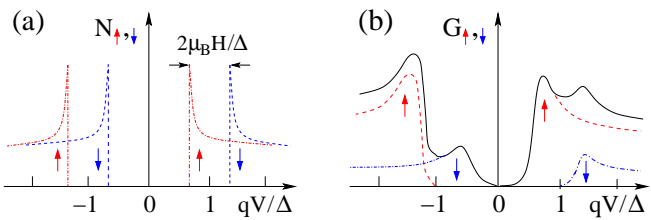,width=\linewidth,angle=0}}
\caption{Ferromagnet/insulator/superconductor tunneling in an applied magnetic
filed: (a) Zeeman splitting of the BCS density of states as a function of 
applied bias; 
(b) normalized spin-resolved conductance (dashed lines) and the total 
conductance (solid line)
at finite temperature.} 
\label{tunnel}
\end{figure}
Spin polarization $P$ of the F electrode can be deduced \cite{Tedrow1994:PR} 
from the asymmetry
of the conductance amplitudes at the four peaks in Fig.~\ref{tunnel} (b) 
[for $P$=0, G(V)=G(-V)].
In CrO$_2$/I/S tunnel junctions, nearly complete
spin polarization $P_G>0.9$ was measured \cite{Parker2002:PRL}. 
Only two of the four peaks sketched
in Fig.~\ref{tunnel}, have been observed, indicating no features due to the
minority spin up to H=2.5 T. \textcite{Parkin2004:P} have shown that
by replacing an aluminum oxide (a typical choice for an insulating 
region) with magnesium oxide, one can significantly increase the 
spin polarization in F/I/S junctions. Correspondingly, extraordinarily large
values of TMR ($>200\%$ at room temperature) can be achieved even
with conventional ferromagnetic CoFe) electrodes.

The assumption of spin-conserving tunneling can be generalized 
\cite{Tedrow1994:PR,Worledge2000:PRB,Monsma2000:APLa,Monsma2000:APLb}  
to extract $P$ in the presence of spin-orbit and spin-flip scattering. 
Theoretical 
analyses \cite{Fulde1973:AP,Maki1964:PTP,Bruno1973:PRB} using 
many-body techniques
show that the  spin-orbit scattering would smear the Zeeman-split 
density of states, eventually merging the
four peaks into two, while the magnetic impurities \cite{Abrikosov1960:ZETP}
act as pair breakers and reduce the value of $\Delta$. 
Neglecting the spin-orbit
scattering was shown to lead to the extraction of higher $P$ 
values \cite{Tedrow1994:PR,Monsma2000:APLa}.

With a few exceptions \cite{Worledge2000:PRL},
F/I/S conductance measurements \cite{Tedrow1994:PR}
have revealed positive $P$--the dominant contribution of 
majority spin electrons for different ferromagnetic films 
(for example, in Fe, Ni, Co and Gd). However, electronic
structure calculations typically give  
that ${\cal N}_{F\uparrow}<{\cal N}_{F\downarrow}$ and $P_{\cal N}<0$
[for Ni and Co  ${\cal N}_{F\uparrow}/{\cal N}_{F\downarrow} \sim 1/10$ 
\cite{Butler2001:PRB}].
Early theoretical work addressed this apparent
difference,\footnote{For a list of references see 
\textcite{Tedrow1973:PRB,Tedrow1994:PR}.}
and efforts to understand precisely what is being experimentally measured 
have continued.

\textcite{Stearns1977:JMMM} suggested that only itinerant, 
freelike electrons
will contribute to tunneling, while nearly localized electrons, 
with a large 
effective mass,
contribute to the total density of states but not to G(V) 
[see also \textcite{Hertz1973:PRB}
and, for spin-unpolarized tunneling, \textcite{Gadzuk1969:PR}].
From the assumed parabolic dispersion of the spin subbands with fixed spin 
splitting, Stearns related the measured polarization 
to the magnetic 
moment, giving positive 
$P \rightarrow P_k=(k_{F\uparrow}-k_{F\downarrow})/(k_{F\uparrow}
+k_{F\downarrow})$,
the spin polarization of the projections of Fermi wave vectors 
perpendicular
to the interface.
Similar arguments, for inequivalent 
density-of-states contributions  
to G(V), were generalized
to more complex electronic structure.
\textcite{Mazin1999:PRL} showed 
the importance of the tunneling matrix elements which have
different Fermi velocities for different bands 
[see also \cite{Yusof1998:PRB},
in the context of tunneling in 
a high temperature superconductor (HTSC)]. 
Consequently, $P_G$ could even have an opposite sign from $P_{\cal N}$--which, 
for example, would be measured by spin-resolved photoemission. 

Good agreement between  
tunneling data and electronic structure calculations was illustrated by 
the example of Ni$_x$Fe$_{1-x}$ \cite{Nadgorny2000:PRB}, showing, however, 
that $P$ is not directly related to the  magnetic moment 
\cite{Meservey1976:PRL}. 
The difference between bulk and the surface densities of states 
of the ferromagnet 
(probed in tunneling measurements) \cite{Oleinik2000:PRB}, 
the choice of tunneling barrier \cite{DeTeresa1999:PRL}, and details of the
interfacial properties, which can change over time \cite{Monsma2000:APLb}, have
all been shown to affect the measured $P$ directly. 

Tedrow-Meservey technique is also considered as a
probe to detect spin injection in Si, where optical methods, due to
the indirect gap, would be ineffective. F/I/S tunneling was also studied
using amorphous Si (a-Si) and Ge (a-Ge) as a barrier. While 
with a-Si  some spin polarization was detected \cite{Meservey1982:JAP} 
no  spin-polarized tunneling was observed using a-Ge  barrier 
\cite{Gibson1985:JAP},
in contrast to the first reports of TMR \cite{Julliere1975:PL}.

Spin-dependent tunneling was also studied using a 
HTSC
electrode as a detector
of spin polarization \cite{Vasko1998:APL,Chen2001:PRB}. 
While this can significantly extend the temperature
range in the tunneling experiments, a lack of  understanding of 
HTSC's makes such structures more a test ground for fundamental physics
than a quantitative tool for 
quantitatively 
determining
$P$.
There are also several important differences between studies using
HTSC's and 
conventional low-temperature superconductors. The superconducting
pairing symmetry no longer yields an isotropic energy gap, and even 
for the BCS-like picture the density of states should be accordingly modified. 
A sign change of the pair potential can result in $G(V=0) >0$
for $T\rightarrow 0$ even for a strong tunneling barrier
and give rise to a zero-bias conductance peak 
\cite{Hu1994:PRL,Kashiwaya1995:PRL,Wei1998:PRL}.
This is explained
by the two-particle process of Andreev reflection (discussed further in
Sec.~\ref{sec:IVA3}),
which, in addition to the usual quasiparticle tunneling, 
contributes to the I-V characteristics of a F/I/S junction 
\cite{Zutic1999:PRBb,Zhu1999:PRB,Zutic2000:PRB,Kashiwaya1999:PRB,Hu1999:PRBb}
[a simpler N/I/S case is reviewed by \textcite{Hu1998:PRB} and
\textcite{Kashiwaya2000:RPP}]. 
The suppression of a zero-bias conductance peak, measured by an STM,
was recently used to detect spin injection into a HTSC \cite{Ngai2003:P}.

\subsubsection{\label{sec:IVA2} F/I/F tunneling}

In the preface to a now classic 
reference on spin-unpolarized tunneling
in solids, \textcite{Duke:1969} concludes that (with only a few exceptions)
the study of tunneling is an art and not a science. 
Perhaps this is also an apt 
description
for the 
present state of experiment on spin-polarized tunneling between 
two ferromagnetic
regions. Even for MTJ's with standard ferromagnetic metals, the bias and the
temperature dependence of the TMR, as identification of the relevant spin 
polarization
remain to be fully understood. In a brief review of current findings we
intend to identify questions that could arise as new materials for
MTJ's are being considered.

A resurgence in interest in the study of MTJ's, following a hiatus after the
early work by \textcite{Julliere1975:PL,Maekawa1982:IEEE}, was 
spurred by the 
observation of large room-temperature TMR 
\cite{Moodera1995:PRL,Miyazaki1995:JMMM}.
This discovery has 
opened the possibility of using MTJ's for fundamental studies
of surface magnetism and room-temperature spin polarization in various 
ferromagnetic electrodes 
as well as to suggesting applications such as highly sensitive 
magnetic-field sensors, magnetic read heads, 
and nonvolatile magnetic memory applications. 

It is instructive to notice the similarity between the schematic geometry 
and the direction of current flow in  an MTJ and that in CPP GMR 
(recall Figs.~\ref{intro:1} and 
\ref{gmr:1}),
which only differ in the middle layer being an insulator and a metal, 
respectively.
By considering the limit of ballistic transport in 
CPP GMR\footnote{Related
applications are usually in a diffusive regime.}
it is possible to give a unified picture of both TMR and CPP GMR by
varying  the strength of the hopping integrals \cite{Mathon1997:PRB} in a
tight-binding representation.

\textcite{Julliere1975:PL} modified Eq.~(\ref{eq:tg}) in the limit
$V \rightarrow 0$, $T \rightarrow 0$,
and applied it to study F/I/F tunneling. 
The two F regions are treated as uncoupled with the spin-conserving 
tunneling across the barrier. This effectively leads to the two-current
model proposed by \textcite{Mott1936:PRCa} and also applied to 
the CPP GMR geometries \cite{Valet1993:PRB,Gijs1997:AP}.
The values for $P$ extracted
from F/I/S measurements are in a good agreement with the observed 
TMR  values (typically positive, as expected from P$_{1,2}>0$). 
However, Julli{\`{e}}re's formula\footnote{For its limitations and extensions
see comprehensive reviews by \textcite{Moodera1999:JMMM,Moodera1999:ARMS}.}
does not provide an explicit TMR dependence on bias and temperature.

Julli{\`{e}}re's result can be obtained as a
limiting case from a more general Kubo/Landauer approach \cite{Mathon1999:PRB}
with the assumption that the component of the wave vector parallel to
the interface ${\bf k}_\|$ is not conserved (incoherent tunneling).
Such a loss of coherence is good approximation for simply capturing the
effects of disorder for amorphous Al$_2$O$_3$, a common choice 
for the I region
with metallic ferromagnets. 
Despite its simplicity, the Julli{\`{e}}re's model for the TMR 
has continued to be used for interpreting
the spin polarization in various MTJ's. 
Recent examples include 
F regions made of manganite perovskites displaying colossal 
magnetoresistance (CMR) \cite{Bowen2002:P} (suggesting $P_{\cal N}$$>$0.95);
magnetite (Fe$_3$O$_4$) \cite{Hu2002:PRL} (with $P$$<$0 and TMR$<$0); 
III-V ferromagnetic semiconductors \cite{Chun2002:PRB}; 
a nonmagnetic semiconductor used as a tunneling barrier 
\cite{Kreuzer2002:APL};
Co/carbon nanotube/Co MTJ \cite{Tsukagoshi1999:N};
and resonant tunneling in F/I/N/F junctions \cite{Yuasa2002:S}.

For novel materials, in which the electronic structure calculations and an
understanding of the interfacial properties are not available, 
Julli{\`{e}}re's 
formula 
still provides useful insights.
A quantitative understanding of MTJ's challenges similar to those 
discussed for F/I/S tunneling, including determining precisely which spin 
polarization is relevant and the related issue of reconciling the 
(typically positive) sign 
of  the observed TMR with the electronic structure 
\cite{Tsymbal1997:JPCM,Oleinik2000:PRB,Bratkovsky1997:PRB,MacLaren1997:PRB,%
Mathon1997:PRB,LeClair2002:PRL}.  

In an approach complementary to Julli{\`{e}}re's, 
\textcite{Slonczewski1989:PRB} 
considered F/I/F as a single quantum-mechanical system in a free-electron 
picture. 
When matching the two-component wave functions at interfaces, 
coherent tunneling was  
assumed, with conserved ${\bf k}_\|$, relevant to epitaxially grown MTJ's 
\cite{Mathon2001:PRB} and the I region was modeled by a square 
barrier.\footnote{A formally analogous problem was considered by 
\textcite{Griffin1971:PRB} in an N/I/S system where the two-component
wave functions represented electron-like and hole-like quasiparticles 
rather then the two spin projections; see Sec.~\ref{sec:IVA3}.}
The resulting TMR can be expressed as in
Eq.~(\ref{eq:julliere}) but with the redefined polarization 
\begin{equation}
P \rightarrow P_k 
(\kappa^2-k_{F\uparrow}k_{F\downarrow})/(\kappa^2
+k_{F\uparrow}k_{F\downarrow}),
\label{eq:slon}
\end{equation}
where $P_k$, as defined by \textcite{Stearns1977:JMMM}, is also
$P_{\cal N}$ (in a free-electron picture) and $i \kappa$ is the usual imaginary
wave vector through a square barrier. Through the dependence of
$\kappa$ on $V$ the resulting polarization in Slonczewski's model can change
sign. A study of a similar geometry using a Boltzmann-like approach shows
\cite{Chui1997:PRB}
that the spin splitting of electrochemical potentials persists in the F region
all the way to the F/I interface, implying 
$\kappa_\uparrow \neq \kappa_\downarrow$
and an additional voltage dependence of the TMR. Variation of  
the density of states [inferred from the
spin-resolved photoemission data \cite{Park1998:PRL,Park1998:N}]  within the
range of applied bias in MTJ's of Co/SrTiO$_3$/La$_{0.7}$Sr$_{0.3}$MnO$_3$ 
(Co/STO/LSMO) 
\cite{DeTeresa1999:PRL}, 
together with Julli{\`{e}}re's model, was used to explain
the large negative TMR (-50\% at 5 K), 
which would even change sign for positive
bias (raising the Co Fermi level above the corresponding one of LSMO). 
The bias dependence of the TMR was also attributed to the density of states
by extending the
model of a trapezoidal tunneling barrier \cite{Brinkman1970:JAP} to the 
spin-polarized
case \cite{Xiang2002:PRB}. 

The decay of TMR with temperature can be attributed to several causes. 
Early theoretical work on N/I/N tunneling 
\cite{Anderson1966:PRL,Appelbaum1966:PRL}
[for a detailed discussion and a review of
related experimental results see \textcite{Duke:1969}]
showed that the presence of magnetic impurities  in the tunneling
barrier produces temperature dependent conductance--referred to as 
zero-bias anomalies. 
These findings, which considered both spin-dependent and 
spin-flip scattering,
were applied to  fit the decay of the TMR with temperature
\cite{Miyazaki:2002,Jansen2000:PRB,Inoue1999:JMMM}.
Hot electrons  localized at F/I interfaces were predicted, 
to create magnons, or collective spin excitations,
near the F/I interfaces, and suppress the TMR \cite{Zhang1997:PRL}.
Magnons were observed \cite{Tsui1971:PRL} in an antiferromagnetic (AFM)
NiO barrier in single crystal Ni/NiO/Pb tunnel junctions and were suggested 
\cite{Moodera1995:PRL} as the cause of
decreasing TMR with $T$ by spin-flip scattering.
Using an $s$-$d$ exchange  (between itinerant $s$ and nearly localized $d$ 
electrons)
Hamiltonian, it was shown \cite{Zhang1997:PRL} that, at  $V\rightarrow 0$,
$G(T)-G(0) \propto T \ln T$, for both 
$\uparrow \uparrow$ and $\uparrow \downarrow$
orientations.
A different temperature dependence of TMR was
suggested by \textcite{Moodera1998:PRL}.
It was related to the decrease of the surface magnetization 
\cite{Pierce1982:PRB,Pierce:1984}
$M(T)/M(0) \propto T^{3/2}$. Such a temperature dependence 
[known as the Bloch's law and reviewed by \textcite{Krey2003:P}],
attributed to magnons, was also obtained for TMR 
\cite{MacDonald1998:PRL}.
An additional decrease of TMR with $T$ was expected due to the spin-independent 
part of G(T)  \cite{Shang1998:PRB}, seen also in N/I/N junctions. 

Systematic studies of MTJ's containing a semiconductor (Sm) region
(used as a tunneling barrier and/or  an F electrode) have begun only 
recently.\footnote{The early F/Ge/F results \cite{Julliere1975:PL} were not 
reproduced and other metallic structures involving Si, Ge, GaAs, and GaN 
as a barrier have shown 
either no \cite{Gibson1985:JAP,Loraine2000:JAP,Boeve2001:JMMM} 
or only a small \cite{Meservey1982:JAP,Jia1996:IEEE,Kreuzer2002:APL} 
spin-dependent signal.}
To improve the performance of MTJ's it is desirable to reduce the junction
resistance. A smaller $RC$ constant would allow faster switching 
times in MRAM (for a detailed discussion see \textcite{DeBoeck2002:SST}). 
Correspondingly, using a semiconducting barrier could prove 
an alternative strategy for difficult fabrication of ultrathin 
($<$1 nm ) oxide barriers 
\cite{Rippard2002:PRL}. Some F/Sm/F MTJ's have been
grown epitaxially, and the amplitude of TMR can be studied as function
of the crystallographic orientation of a F/Sm interface.
For an epitaxially grown Fe/ZnSe/Fe MTJ electronic structure calculations
have predicted  \cite{MacLaren1999:PRB} large TMR (up to $\sim$1000 \%), 
increasing
with ZnSe thickness. However, the observed TMR in Fe/ZnSe/Fe$_{0.85}$Co$_{0.15}$
was limited  below 50 K, reaching 15 \% at 10 K for junctions of higher
resistance and lower defect density\footnote{Interface defects could diminish 
measured TMR. We recall (see Sec.~\ref{sec:IID3}) that at a ZnMnSe/AlGaAs interface 
they limit the spin injection efficiency \cite{Stroud2002:PRL} and from 
Eq.~(\ref{eq:delR}) infer a reduced spin-valve effect.}
\cite{Gustavsson2001:PRB}. 
Results on ZnS, 
another II-VI semiconductor, demonstrated a TMR of $\sim$5 \% at room 
temperature \cite{Guth2001:APL}.

\begin{figure} 
\centerline{\psfig{file=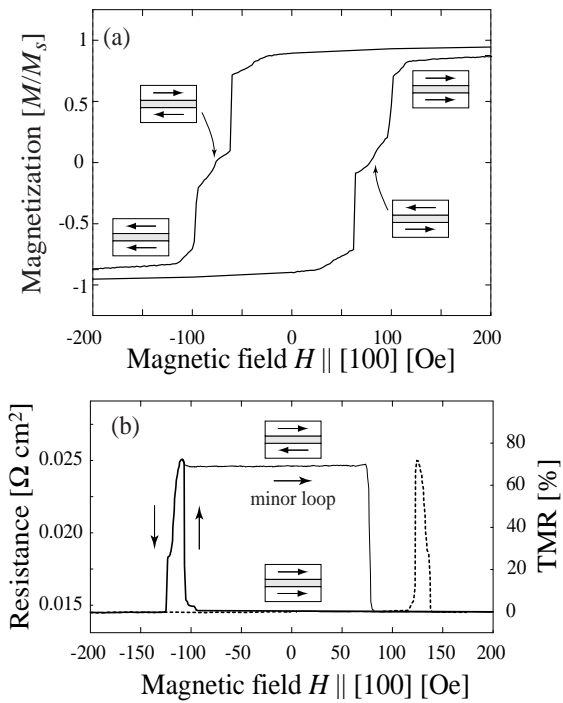,width=0.9\linewidth}}
\caption{All-semiconductor magnetic tunnel junction:
(a) magnetization of Ga$_{1-x}$Mn$_x$As (x=4.0\%,50 nm)/AlAs
(3 nm)/ Ga$_{1-x}$Mn$_x$As (x=3.3\%,50 nm) trilayer measured by a SQUID at 8 K.
The sample size is 3 $\times$ 3 mm$^2$. 
Magnetization shown is normalized with respect to the saturation value M$_s$.
(b) 
TMR curves of a Ga$_{1-x}$Mn$_x$As (x=4.0\%,50 nm)/AlAs
(1.6 nm)/ Ga$_{1-x}$Mn$_x$As (x=3.3\%,50 nm) tunnel junction of 200 $\mu$m
in diameter. Bold solid curve, 
sweep of the magnetic
field from positive to negative;
dashed curve, sweep from
negative to positive; thin solid curve,
a minor loop. From \onlinecite{Tanaka2001:PRL}.}
\label{tmr0}
\end{figure}

There is also a possibility of using 
all-semiconductor F/Sm/F single-crystalline MTJ's
where F is a ferromagnetic semiconductor. These would simplify integration
with the existing conventional semiconductor-based electronics and allow
flexibility of various doping profiles and fabrication of quantum structures,
as compared to the conventional all-metal MTJ's. Large TMR ($>$70 \% at 8 K),
shown in Fig.~\ref{tmr0},
has been measured in an epitaxially grown (Ga,Mn)As/AlAs/(Ga,Mn)As junction 
\cite{Tanaka2001:PRL}. The results are consistent with the $k_\|$ being 
conserved in the tunneling process \cite{Mathon1997:PRB}, with the decrease of 
TMR with $T$ expected from the spin-wave excitations 
\cite{Shang1998:PRB,MacDonald1998:PRL},
discussed above. TMR is nonmonotonic with thickness in AlAs 
(with the peak at
$\sim$ 1.5 nm). For a given AlAs thickness, double MTJ's were also shown
to give similar TMR values and were used to determine electrically
the spin injection in GaAs QW \cite{Mattana2002:PRL}.
However, 
a room-temperature effect 
remains to be demonstrated 
as 
the available
well-characterized ferromagnetic semiconductors do not have 
as high a ferromagnetic
transition temperature. 

A lower barrier in F/Sm/F MTJ's can have important implications in 
determining the actual values of TMR. The standard four-probe technique for
measuring I and V 
has been known to give spurious values when the resistance 
of the F electrodes is non-negligible to the junction resistance.
The tunneling current in that regime has been shown to be highly 
nonuniform\footnote{Nonuniform tunneling current has been studied
in nonmagnetic junctions \cite{Pederson1967:APL}, 
CPP multilayers 
\cite{Lenczowski1994:JAP},    
and conventional MTJ's \cite{Moodera1996:APL,Rzchowski2000:PRB}.} 
and the measured apparent resistance $R_m=V/I$  (different from the
actual junction resistance R$_J$)  can even attain negative values 
\cite{Pederson1967:APL,Moodera1996:APL}. The important implications
for MTJ are the possibility of large overestimates in the TMR amplitude
\cite{Moodera1999:JMMM} and a desirable hysteresis effect--at H=0
the two values of resistance can be used for various nonvolatile 
applications \cite{Moodera1996:APL}. 

A detailed understanding of MTJ's will also require knowing the influence
of the interface and surface roughness \cite{Itoh1999:JPSP}. Even in 
the spin-unpolarized case it is known that the full quantum-mechanical 
approach 
\cite{Tesanovic1986:PRL} can lead to qualitatively different 
results from the 
usual quasiclassical picture and from averaging out the spatial information on the 
length scale of the inverse Fermi wave vector.

A comprehensive review of tunneling phenomena and magnetoresistance in 
granular materials, ferromagnetic single-electron transistors, 
and double tunnel junctions
is given by \textcite{Maekawa:2002}. 
A theoretical study of F/I/F junctions,
in which the I region is a quantum dot,
shows the importance of Coulomb interactions, which could lead to spin 
precession
even in the absence of an applied magnetic field \cite{Konig2003:PRL}.

\subsubsection{\label{sec:IVA3} Andreev reflection}

Andreev reflection \cite{Andreev1964:SPJETP} 
is a scattering process, at an interface with a superconductor, 
responsible for a conversion between a dissipative quasiparticle
current and a dissipationless supercurrent [see also early work
by \textcite{deGennes1963:PL}]. 
For a spin-singlet superconductor
an incident electron (hole) of spin
$\lambda$ is reflected as a
hole (electron) belonging to the opposite spin subband $\overline{\lambda}$, 
back
to the nonsuperconducting region, while a Cooper pair is transferred to the 
superconductor. 
This is a phase-coherent scattering process in which the reflected particle
carries the information about both the phase of 
the incident particle
and the macroscopic phase of the superconductor.\footnote{For instructive
reviews see \textcite{Lambert1998:JPCM,Pannetier2000:JLTP}.} Andreev
reflection  thus is responsible for a proximity effect where the 
phase correlations are introduced to a nonsuperconducting 
material \cite{Demler1997:PRB,Halterman2002:PRB,Bergeret2001:PRL,%
Izyumov2002:PU,Fominov2003:T}.
The probability for Andreev reflection at low bias voltage 
($qV \lesssim \Delta$), which is 
related to the square of the normal-state transmission, 
could
be ignored for  low-transparency junctions with conventional
superconductors, as discussed in Sec.~\ref{sec:IVA1}. In contrast, 
for high-transparency junctions (see the discussion 
of Sharvin conductance in Sec.~\ref{sec:IIC2}),
single-particle tunneling vanishes [recall Eq.~(\ref{eq:tg})]
at low bias and $T=0$ and Andreev
reflection is the dominant process. A convenient description is
provided by the 
Bogoliubov-de Gennes equations \cite{deGennes:1989},
\begin{eqnarray}
\left[\begin{array}{cc} H_\lambda & \Delta \\
\Delta^* & -H^*_{\overline{\lambda}}\end{array} \right]
\left[ \begin{array}{c} u_\lambda \\
v_{\overline{\lambda}} \end{array} \right]=E
\left[ \begin{array}{c} u_\lambda \\
v_{\overline{\lambda}} \end{array} \right],
\label{eq:BdG}
\end{eqnarray} 
and by matching the wave functions at the boundaries (interfaces) between
different regions.
Here $H_\lambda$ is the single-particle Hamiltonian 
for spin $\lambda=\uparrow, \downarrow$ and  $\overline{\lambda}$
denotes a spin opposite to $\lambda$ \cite{deJong1995:PRL,Zutic2000:PRB}.
$\Delta$ is the pair potential \cite{deGennes:1989},  $E$ the excitation energy 
and $u_\lambda$, $v_{\overline{\lambda}}$ are the electronlike quasiparticle 
and holelike quasiparticle amplitudes, 
respectively.\footnote{Equation (\ref{eq:BdG}) 
can be simply modified to include the spin flip and
spin-dependent interfacial scattering \cite{Zutic1999:PRBa}.}   
\textcite{Griffin1971:PRB} have solved the Bogoliubov-de Gennes 
equations with 
square or a $\delta$-function barriers  of  varying strength 
at an N/S interface. They obtained a  
result that interpolates between the clean and the tunneling limits.
\textcite{Blonder1982:PRB} used a similar approach,
known as the Blonder-Tinkham-Klapwijk method, in which 
the two limits correspond to $Z\rightarrow0$ and $Z\rightarrow \infty$,
respectively, and $Z$ is the strength of the  $\delta$-function barrier.
The transparency of this
approach\footnote{A  good agreement \cite{Yan2000:PRB}
was obtained with the more rigorous nonequilibrium 
Keldysh technique \cite{Keldysh1965:SPJETP,Rammer1986:RMP}, for an illustration 
of how such a technique can be used to study spin-polarized transport in a 
wide range of heterojunctions see \textcite{Melin2002:EPJB,Zeng2003:P}.}
makes it suitable for the study of  
ballistic spin-polarized transport and spin injection even in the
absence of a  superconducting region
\cite{Hu2001:PRL,Matsuyama2002:PRB,Hu2001:PRB,Schapers2001b:PRB}.

It is instructive to note a similarity between the
two-component transport in N/S junctions  (for
electronlike and holelike quasiparticles) and F/N junctions 
(for spin $\uparrow$, $\downarrow$), which both lead to current conversion, 
accompanied by the additional boundary resistance
\cite{Blonder1982:PRB,vanSon1987:PRL}.  
In the N/S junction Andreev reflection is responsible for the conversion 
between the 
normal and the 
supercurrent, characterized by the superconducting coherence length,
while in the F/N case a conversion between spin-polarized and unpolarized
current is characterized by the spin diffusion length.

For spin-polarized carriers, with  different populations in two spin
subbands, only a fraction of the incident electrons from a majority subband
will have a  minority subband partner in order to be  Andreev
reflected. This can be simply quantified at zero bias and $Z=0$,
in terms of the total number of scattering channels (for each $k_\|$) 
$N_\lambda=k_{F\lambda}^2 A/4 \pi$ at the Fermi level. Here A is the 
point-contact 
area, and $k_{F\lambda}$ 
is the spin-resolved Fermi wave vector. 
A spherical Fermi surface in the F and S regions, with no (spin-averaged)
Fermi velocity mismatch, is assumed. 
When S is in the normal state, the 
zero-temperature Sharvin conductance is
\begin{equation}
G_{FN}=\frac{e^2}{h}(N_\uparrow+N_\downarrow),
\label{eq:sharvin2} 
\end{equation}
equivalent to $R_{\rm Sharvin}^{-1}$, from Eq.~(\ref{eq:sharvin}).
In the superconducting state all of the $N_\downarrow$ 
and only $(N_\downarrow/N_\uparrow)N_\uparrow$ scattering channels 
contribute to Andreev reflection across the F/S interface and transfer 
charge $2e$, yielding \cite{deJong1995:PRL}
\begin{equation}
G_{FS}=\frac{e^2}{h} \left(2N_\downarrow + \frac{2N_\downarrow}{N_\uparrow} 
N_\uparrow \right)=4\frac{e^2}{h}N_\downarrow.
\label{eq:dejong}
\end{equation}
The suppression of the normalized zero-bias conductance 
at $V=0$ and $Z=0$ \cite{deJong1995:PRL},
\begin{equation}
G_{FS}/G_{FN}=2(1-P_G)
\label{eq:pc}
\end{equation}
with the increase in the spin polarization $P_G \rightarrow 
(N_\uparrow-N_\downarrow)/(N_\uparrow+N_\downarrow)$,
was used as a sensitive transport technique to detect spin polarization 
in a point contact  \cite{Soulen1998:S}. 
Data are 
given in Fig.~\ref{point}. A
similar  study, using a  thin-film nanocontact geometry 
\cite{Uphaday1998:PRL}, emphasized the importance of
fitting the conductance data over a wide range of applied bias, not only
at $V=0$, in order to extract the spin polarization of the F region more 
precisely.

\begin{figure} 
\centerline{\psfig{file=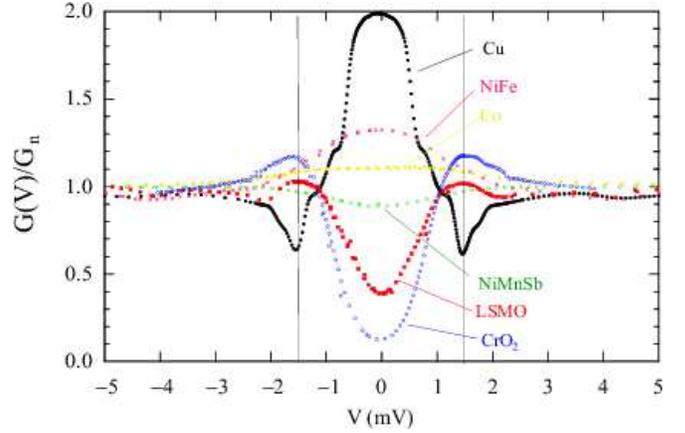,width=\linewidth,angle=0}}
\caption{The differential conductance for several spin-polarized materials,
showing the suppression of Andreev reflection with increasing $P_G$.
The vertical lines denote the bulk superconducting gap for Nb: 
$\Delta$($T=0$)=1.5 meV. 
Note that NiMnSb, one of the Heusler alloys originally proposed 
as half-metallic ferromagnets \cite{deGroot1983:PRL}, shows only
partial spin polarization. From \onlinecite{Soulen1998:S}.} 
\label{point}
\end{figure}
The advantage of such techniques is the detection of polarization
in a much wider range of materials than  those which
can be grown for detection in F/I/S or F/I/F tunnel junctions. 
A large number of
experimental results using spin-polarized Andreev reflection
has since been reported 
\cite{Nadgorny2001:PRB,Ji2001:PRL,Parker2002:PRL,Bourgeois2001:PRB,%
Panguluri2003:Pb},  
including the first direct measurements 
\cite{Braden2003:P,Panguluri2003:P,Panguluri2004:P} of
the spin polarization in (Ga,Mn)As and (In,Mn)Sb.\footnote{Similar 
measurements were
also suggested by \textcite{Zutic1999:PRBa} to yield information 
about the FSm/S interface. A more complete analysis should also quantify
the effects of spin-orbit coupling.}
However,
for a quantitative interpretation of the measured polarization, important
additional factors (similar to the limitations discussed for the application
of Julli{\`{e}}re's formula in Sec.~\ref{sec:IVA2}) need to be incorporated 
in the picture 
provided by Eq.~(\ref{eq:pc}).
For example, the Fermi surface may not be spherical [see the discussion of 
\textcite{Mazin1999:PRL}, specifying what type of spin polarization is
experimentally measured and also that of \textcite{Xia2002:PRL}]. 
The roughness or the size of the  F/S interface 
may lead to a diffusive component of the transport 
\cite{Jedema1999:PRB,Falko1999:PZETF,Mazin2001:JAP}.
As a caution concerning the  possible difficulties in analyzing experimental data,
we mention some subtleties that arise even for the simple model
of a spherical Fermi surface used to describe both F and S regions.
Unlike charge transport in N/S junctions \cite{Blonder1983:PRB} 
in a Griffin-Demers-Blonder-Tinkham-Klapwijk approach, Fermi velocity mismatch
between the F and the S regions, 
does not simply increase the value of effective $Z$.
Specifically, at $Z=V=0$ and normal incidence
it is possible to have perfect transparency even when all the
Fermi velocities differ, satisfying 
$(v_{F\uparrow} v_{F\downarrow})^{1/2}=v_S$,
where $v_S$ is the Fermi velocity in a superconductor 
\cite{Zutic1999:PRBa,Zutic1999:PRBb,Zutic2000:PRB}. In other words, 
unlike in Eq.~(\ref{eq:pc}), the spin polarization 
(nonvanishing exchange energy) 
can {\it increase} the subband conductance, 
for fixed Fermi velocity mismatch. 
Conversely, at a fixed exchange energy, an increase in Fermi velocity mismatch  
could increase the subgap conductance.\footnote{Similar results were 
also obtained when 
F and S region were separated with a quantum dot 
\cite{Zhu2002:PRB,Feng2003:PRB,Zeng2003:P} 
and even in a 1D tight-binding model with  no spin polarization 
\cite{Affleck2000:PRB}.}
In a typical interpretation of a measured conductance, complications can
then arise in trying to disentangle the influence of parameters $Z$, $P_G$,
and Fermi velocity mismatch from the nature of the point contacts 
\cite{Kikuchi2002:PRB} 
and the role of inelastic scattering \cite{Auth2003:PRB}.
Detection of $P$ in HTSC's is even possible with a large barrier or a
vacuum between the F and S regions, 
as proposed by \textcite{Wang2002b:P} 
using resonant Andreev reflection and 
a $d$-wave superconductor.\footnote{Interference effects between the
quasi-electron and quasi-hole scattering trajectories that feel
pair potentials of different sign lead to a large 
conductance near zero bias, even at large interfacial barrier
(referred to as a zero-bias conductance peak in Sec.~\ref{sec:IVA1}).
} 

Large magnetoresistive effects are predicted for crossed Andreev reflection
\cite{Deutscher2000:APL}, when the two F regions, separated within the 
distance of
the superconducting coherence length,\footnote{Recent theoretical
findings suggest that the separation should not exceed the 
Fermi wavelength \cite{Yamashita2003:P}.} are on the same side of the
S region. Such structures have also been theoretically studied to
understand the implications of nonlocal correlations 
\cite{Apinyan2002:EPJB,Melin2002:EPJB}. 

\subsubsection{\label{sec:IVA4} Spin-polarized drift and diffusion}

Traditional semiconductor devices such as field-effect transistors, 
bipolar diodes and transistors, or semiconductor solar cells rely 
in great part on carriers (electrons and holes) whose motion
can be described as drift and diffusion, limited by carrier recombination. In 
inhomogeneous devices where charge buildup is rule, the 
recombination-limited 
drift-diffusion is supplied by Maxwell's equations, to be solved in a 
self-consistent manner. 
Many proposed spintronic devices as well as experimental 
settings for spin injection (see Sec.~\ref{sec:II}) can be described by 
both carrier and spin 
drift and diffusion, limited by carrier recombination
and spin relaxation \cite{Fabian2002:PRB,Zutic2002:PRL}. 
In addition, if spin precession is important for device operation, spin 
dynamics need to be explicitly incorporated into the transport equations 
\cite{Qi2003:PRB}. Drift of the spin-polarized carriers can be due not only 
to the electric field, but also 
to magnetic fields. We illustrate spin-polarized drift and diffusion 
on the transport model of spin-polarized bipolar transport, 
where bipolar refers to the presence of electrons and holes, 
not spin up and down. 
A spin-polarized unipolar transport can be obtained as a limiting case by
setting the electron-hole recombination rate to zero and considering only
one type of carrier (either electrons or holes).

Consider electrons and holes whose density is commonly denoted here as 
$c$ (for carriers), moving in the electrostatic potential $\phi$ which 
comprises both the external bias $V$ and the internal built-in fields 
due to 
charge inhomogeneities. Let the equilibrium spin splitting of the 
carrier band be 
$2q\zeta_{c}$. The spin $\lambda$ resolved carrier charge-current density is
\cite{Zutic2002:PRL}
\begin{equation}
\label{eq:jcl}
{\bf j}_{c\lambda}=
-q\mu_{c\lambda}c_{\lambda}\nabla \phi {\pm}qD_{c\lambda}\nabla c_\lambda
-q\lambda\mu_{c\lambda}c_{\lambda}\nabla \zeta_c,
\end{equation}
where $\mu$ and $D$ stand for mobility and 
diffusion
coefficients, 
the upper sign is for electrons  and the lower sign is for holes. The first term on the
right hand side describes drift caused by the total electric field, 
the second term represents diffusion, while the last term stands for 
magnetic drift---carrier drift in inhomogeneously split 
bands.\footnote{Equation~(\ref{eq:jcl}) can be viewed as the generalization
of the Silsbee-Johnson spin-charge coupling 
\cite{Johnson1987:PRB,Heide2001:PRL,Wegrowe2000:PRB} to bipolar transport 
and to systems with spatially inhomogeneous charge density.}
More transparent are the equations for the total charge, 
$j=j_{\uparrow}+
j_{\downarrow}$, and spin, $j_s=j_{\uparrow}-j_{\downarrow}$, 
current densities:
\begin{eqnarray} \label{eq:jc}
&{\bf j}_c&=-\sigma_c \nabla\phi - 
\sigma_{sc}\nabla \zeta_c \pm qD_c \nabla c \pm qD_{sc}\nabla s_c,   \\
\label{eq:js}
&{\bf j}_{sc}&=-\sigma_{sc} \nabla\phi - 
\sigma_{c}\nabla \zeta_c \pm qD_{sc} \nabla c \pm qD_{c}\nabla s_c,
\end{eqnarray}
where the carrier density $c=c_\uparrow+c_\downarrow$ 
and spin $s_c=c_\uparrow-c_\downarrow$, 
and we introduced the carrier charge and spin conductivities 
$\sigma_c=q(\mu_c c+ \mu_{sc} s_c)$
and $\sigma_{sc}=q(\mu_{sc}c+\mu_cs_c)$, 
where $\mu_c=(\mu_{c\uparrow}+\mu_{c\downarrow})/2$
and $\mu_{cs}=(\mu_{c\uparrow}-\mu_{c\downarrow})/2$ 
are charge and spin mobilities, and similarly
for the diffusion coefficients. Equation (\ref{eq:jc}) 
describes the spin-charge coupling 
in bipolar transport in inhomogeneous magnetic semiconductors. 
Spatial variations in 
spin density can cause charge currents. 
Similarly, it follows from Eq.~(\ref{eq:js})
that spatial variations in carrier densities can lead to spin currents.  

Steady-state carrier recombination and spin relaxation processes 
are described 
by the continuity equations for the spin-resolved carrier densities:
\begin{equation} 
\label{eq:cont}
\nabla\cdot \frac{{\bf j}_{c\lambda}}{q}= 
\pm w_{c\lambda} (c_{\lambda}\bar{c}-
c_{\lambda 0}\bar{c}_0)\pm 
\frac{c_{\lambda}-c_{-\lambda}-\lambda \tilde{s}_c}{2\tau_{sc}}.
\end{equation}
Here $w$ is the spin-dependent recombination rate, the bar denotes
a complementary carrier ($\bar{n}=p$, for example), $\tau_{sc}$ 
is the spin relaxation
time of the carrier $c$ (not to be confused with the single spin
decoherence time discussed in Sec.~\ref{sec:IIIA1}),
and $\tilde{s}_c=P_{c0}c$ is the nonequilibrium spin
density, which appears after realizing that spin relaxation equilibrates 
spin while
preserving carrier density. Finally, the set of equations is 
completed with
Poisson's equation, 
\begin{equation}\label{eq:poisson}
\varepsilon\Delta \phi=-\rho,
\end{equation} 
connecting the electric field and charge $\rho=q(p-n+N_d-N_a)$,
where $N_d$ and $N_a$ are the donor and acceptor densities, respectively, and
$\varepsilon$ is
the dielectric constant. 

In many important cases Eqs.~(\ref{eq:jcl}), (\ref{eq:cont}), 
and (\ref{eq:poisson}) 
need to be solved self-consistently, which usually requires
numerical techniques \cite{Zutic2002:PRL}. 
In some cases it is possible to extract the relevant physics in limiting 
cases analytically,
usually neglecting electric field or magnetic drift. 
In unipolar spin-polarized transport one does not need to consider carrier 
recombination. It also often suffices to study pure spin diffusion, 
if the built-in electric fields are small. Unipolar spin-polarized transport 
in inhomogeneous systems in the presence of electric fields was analyzed by 
\textcite{Fabian2002:PRB,Yu2002:PRBa,Yu2002:PRBb,Martin2003:PRB,Pershin2003:PRL}.
Spin-polarized drift and diffusion in 
model GaAs quantum wires was studied by \textcite{Sogawa2000:PRB}, while 
ramifications of magnetic drift for unipolar transport were 
studied by \textcite{Martin2003:PRB,Fabian2002:PRB}. Bipolar transport 
in the presence of electrical drift and/or 
diffusion has been studied by 
\textcite{Zutic2002:PRL,Fabian2002:PRB,Beck2002:PE,Flatte2000:PRL}.
Transient dynamics of spin drift and diffusion was considered by 
\textcite{Fabian2002b:PRB}.
Recently an interesting study \cite{Saikin2003:JAP} was reported on a 
Monte-Carlo simulation of quantum-mechanical 
spin dynamics limited by spin relaxation, in which quasiclassical 
orbital transport 
was carried out for the in-plane
transport in III-V heterostructures where spin precession is due to  bulk
and structure inversion asymmetry (see Sec.~\ref{sec:IIIB2}).

\subsection{\label{sec:IVB} Materials considerations}

Nominally highly spin-polarized materials, as discussed in the previous
sections, could provide both effective spin injection into nonmagnetic
materials and large MR effects, important for nonvolatile applications.
Examples include  half-metallic oxides such as CrO$_2$, Fe$_3$O$_4$, 
CMR materials, and double perovskites \cite{Kobayashi1998:N} 
[for reviews of half-metallic materials see 
\cite{Pickett2001:PT,Fang2002:JAP}].
Ferromagnetic semiconductors \cite{Nagaev:1983}, known since  CrBr$_3$ 
\cite{Tsubokawa1960:JPSJ}, have been demonstrated to be highly spin polarized.
However, more recent interest in ferromagnetic semiconductors was spurred by the 
fabrication of (III,Mn)V compounds.\footnote{Ferromagnetic order 
with Mn-doping was
obtained previously, for example, in (Sn,Mn)Te \cite{Escorne:1974}, 
(Ge,Mn)Te \cite{Cochrane1974:PRB} and (Pb,Sn,Mn)Te \cite{Story1986:PRL}.}
After the initial discovery of (In,Mn)As 
\cite{Munekata1989:PRL,Ohno1992:PRL,Munekata1991:JCG}, 
most of the research 
has focused on (Ga,Mn)As \cite{Ohno1996:APL,VanEsch1997:PRB,Hayashi1997:JCG}.  
In contrast to (In,Mn)As and
(Ga,Mn)As with high carrier density ($\sim$ 10$^{20}$ cm$^{-3}$),
a much lower carrier density in (Zn,Cr)Te \cite{Saito2002:JAP}, 
a II-VI ferromagnetic 
semiconductor with Curie temperature $T_C$ near room temperature 
\cite{Saito2003:PRL},
suggests that transport properties can be effectively controlled by
carrier doping. Most of the currently studied ferromagnetic semiconductors
are p-doped with holes as spin-polarized carriers, which typically
leads to lower mobilities and shorter spin relaxation times than in
n-doped materials. It is possible to use selective doping to substantially
increase $T_C$, as compared to the uniformly doped bulk ferromagnetic
semiconductors \cite{Nazmul2003:PRB}.

Early work on (Ga,Mn)As \cite{DeBoeck1996:APL} showed the low solubility 
of Mn and the formation of  
magnetic nanoclusters 
characteristic of many 
subsequent compounds and different magnetic impurities. 
The presence of such 
nanoclusters often complicates accurate
determination of  $T_C$ as well as of whether 
the compound is actually in a single phase. Consequently, the reported
room-temperature ferromagnetism  in an increasing number of compounds
reviewed by \textcite{Pearton2003:JAP} is not universally accepted.
Conclusive evidence for intrinsic ferromagnetism in
semiconductors is highly nontrivial. For example, early work reporting 
ferromagnetism even at nearly 900 K in  La-doped CaBa$_6$ 
\cite{Young1999:N,Ott2000:PB,Tromp2001:PRL}, was later revisited
suggesting extrinsic effect \cite{Bennett2003:P}. 
It remains to be understood what the limitations are for using 
extrinsic ferromagnets and, for example,  whether they can be 
effective spin injectors. 

\begin{figure}
\centerline{\psfig{file=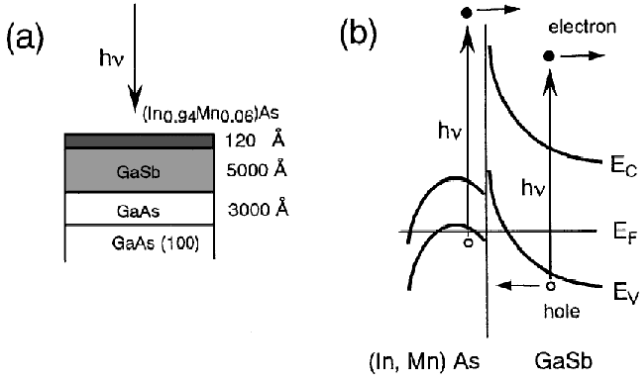,width=1.0\linewidth,angle=0}}
\caption{Photoinduced ferromagnetism in a (In,Mn)As/GaSb heterostructure:
(a) light-irradiated 
sample displaying photoinduced
ferromagnetism--direction of light irradiation is
shown by an arrow; (b) band-edge profile of (In,Mn)As/GaSb heterostructure.
$E_c$, conduction band, $E_v$, valence band; $E_F$,  
Fermi level, respectively. From \onlinecite{Koshihara1997:PRL}.} 
\label{light:1}
\end{figure}

A high $T_C$ 
and almost complete spin polarization in 
bulk samples are alone not sufficient for successful applications. Spintronic 
devices typically rely on inhomogeneous doping, structures of reduced 
dimensionality, 
and/or structures containing different materials. 
Interfacial properties, as discussed in the
previous sections, can significantly influence the magnitude of 
magnetoresistive effects\footnote{In magnetic multilayers GMR is typically
dominated by interfacial scattering \cite{Parkin1993:PRL}, while in MTJ's 
it is the surface rather than the bulk electronic structure which 
influences the relevant spin polarization.}
and the efficiency of spin injection.
Doping properties and possibility of fabricating a wide range of structures
allow spintronic applications beyond MR effects, for example,
spin transistors, spin lasers, and spin-based quantum computers 
(Sec.~\ref{sec:IVF}).
Materials properties of hybrid F/Sm heterostructures, relevant to
device applications, were reviewed by \textcite{Samarth2003:SSC}.

Experiments in which the ferromagnetism is induced 
optically 
\cite{Koshihara1997:PRL,Oiwa2002:PRL,Wang2003:P} and electrically 
\cite{Ohno2000:N,Park2002:S} provide a method for distinguishing the 
carrier-induced
ferromagnetism, based on the exchange interaction between the 
carrier and the magnetic impurity spins,
from ferromagnetism that originates from magnetic nanoclusters. 
Such experiments also suggest a possible nonvolatile multifunctional
devices with tunable, optical, electrical, and magnetic properties. 
Comprehensive surveys of magneto-optical materials and applications, not
limited to semiconductors, are given by \textcite{Zvezdin:2003,Sugamo:2000}.

Photoinduced ferromagnetism was demonstrated by \textcite{Koshihara1997:PRL} 
in p-(In,Mn)As/GaSb 
heterostructure, shown in Figs.~\ref{light:1}, and \ref{light:2}. 
Unpolarized light penetrates through a thin (In,Mn)As layer and is absorbed 
in the GaSb layer. A large band bending across the heterostructures separates,
by a built-in field, electrons and holes. The excess holes generated in a 
GaSb layer are effectively transferred to the p-doped (In,Mn)As layer where 
they enhance the ferromagnetic spin exchange among Mn ions, resulting
in a paramagnetic-ferromagnetic transition.

\begin{figure}
\centerline{\psfig{file=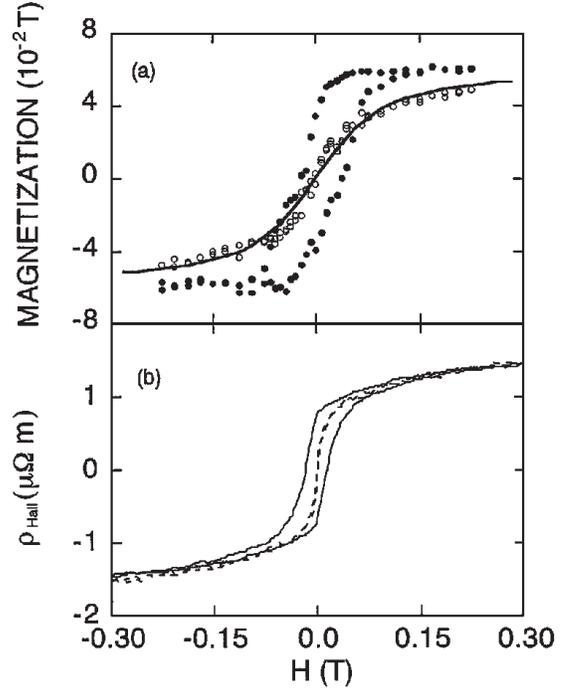,width=0.85\linewidth,angle=0}}
\caption{Magnetization curves for (In,Mn)As/GaSb at 5 K observed before 
(open circles)
and after (solid circles) light irradiation. Solid line show a theoretical 
curve. (b) Hall resistivity at 5 K before (dashed lines) 
and after (solid lines) 
light irradiation. From \onlinecite{Koshihara1997:PRL}.} 
\label{light:2}
\end{figure}
The increase in magnetization, measured by a SQUID, 
is shown in Fig.~\ref{light:2}(a) and in
Fig.~\ref{light:2}(b) the corresponding Hall resistivity 
\begin{equation}
\rho_{\rm Hall}=R_0 B+ R_S M,
\label{eq:hall}
\end{equation}
is shown, where the $R_0$ is the ordinary and $R_S$ 
the anomalous Hall coefficient, 
respectively.
Typical for (III,Mn)V compounds, $\rho_{\rm Hall}$ is dominated by the 
anomalous
contribution, $\rho_{\rm Hall} \propto M$.

\begin{figure}
\centerline{\psfig{file=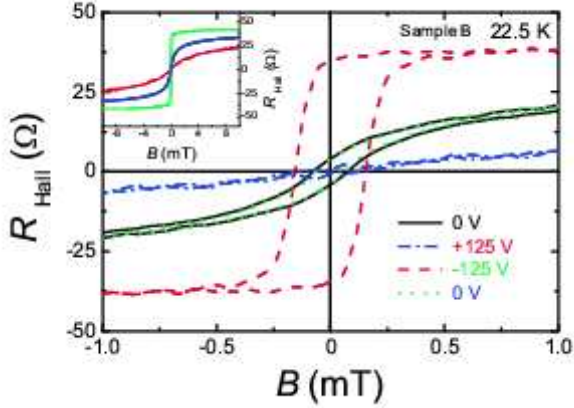,width=1.0\linewidth}}
\caption{Electric-field control of ferromagnetism.
R$_{\rm Hall}$ vs field curves under three different gate biases.
Application of V$_G$=0, +125, and -125 V results in a qualitatively different 
field
dependence of R$_{\rm Hall}$ measured at 22.5 K (sample B):
almost horizontal dash-dotted line, paramagnetic response 
are
partially depleted from the channel (V$_G$=+125 V);
dashed lines, clear hysteresis at low fields
($<$0.7 mT) as holes are accumulated in the channel (V$_G$=-125 V);
solid line, R$_{\rm Hall}$ curve measured at V$_G$=0 V before 
application of $\pm$ 125 V, dotted line, 
R$_{\rm Hall}$ after 
application of $\pm$ 125 V.
Inset, the same curves shown at higher magnetic fields.
From \onlinecite{Ohno2000:N}.} 
\label{ohno}
\end{figure}

A different type of photoinduced magnetization was measured in ferromagnetic 
(Ga,Mn)As.\footnote{Previous studies in paramagnetic (II,Mn)VI materials 
have shown that
nonequilibrium spin-polarized carriers can change the orientation of
magnetic spins in (Hg,Mn)Te \cite{Krenn1985:PRL,Krenn1989:PRB} and in 
(Cd,Mn)Te
\cite{Awschalom1987:PRL}.}
In a Faraday geometry (recall \ref{sec:IID3}), by changing the polarization
of a circularly polarized light, 
one can modulate
the Hall resistance and thus the induced 
magnetization
by up to 15\% of the saturation value \cite{Oiwa2002:PRL}.
Additional experiments on photoinduced magnetization rotation 
\cite{Munekata2003:JS,Oiwa2003:JS} suggest that the 
main contribution of carrier spin to such rotation is realized by generating
an effective magnetic field through the $p$-$d$ exchange interaction, 
rather than by spin-transfer torque as discussed in 
Secs.~\ref{sec:IB1} and \ref{sec:V}
\cite{Moriya2003:P}.
In GaAs-Fe composite films an observation of room temperature photoenhanced 
magnetization was used to demonstrate that a magnetic force can be changed
by light illumination \cite{Shishi2003:APL}.

Electrically induced ferromagnetism was realized  by applying gate
voltage V$_G$ to change the hole concentration in $d$=5nm thick (In,Mn)As
used as a magnetic channel in a metal-insulator semiconductor FET structure.
Below a metal gate and an insulator the (In,Mn)As channel was grown on top
of a InAs/(Al,Ga)Sb/AlSb and GaAs substrate. 
In Fig.~\ref{ohno},
the corresponding data for $R_{\rm Hall}=\rho_{\rm Hall}/d \propto M$ 
[recall Eq.~(\ref{eq:hall}), show that the ferromagnetism can be switched on
and off, as an electric analog of the manipulation of M 
from Fig.~\ref{light:2}.
Subsequent work by \textcite{Park2002:S} showed that in MnGe ferromagnetism 
can
be manipulated at higher temperature and at significantly lower gate voltage
(at $\sim$ 50 K and $\sim$ 1 V). The combination of light and electric-field
control of ferromagnetism was used in modulation-doped p-type
(Cd,Mn)Te QW \cite{Boukari2002:PRL}. It was demonstrated that 
illumination by light in $p-i-n$ diodes would  enhance the spontaneous
magnetization, while illumination in $p-i-p$ structures would 
destroy ferromagnetism.

\begin{figure}
\centerline{\psfig{file=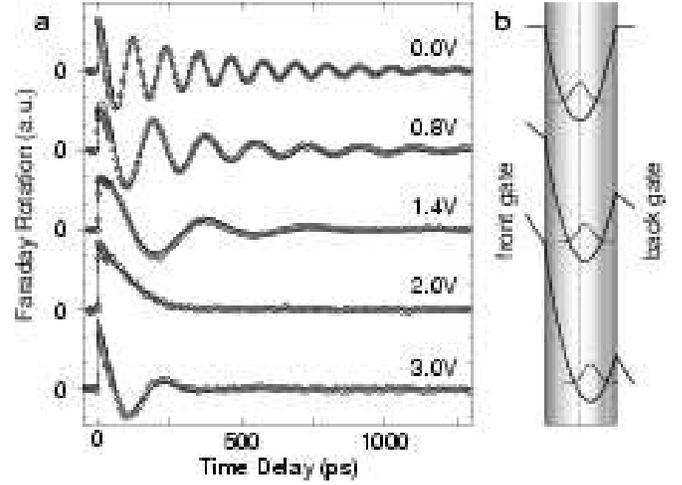,width=1.0\linewidth}}
\caption{Voltage-controlled spin precession: (a) time-resolved Kerr rotation
measurements of electron spin precession in a quantum well at different
gate voltages $V_G$ with Al concentration of 7\% at 5 K and B=6 T;
displacement of the electron wave function
towards the back gate into regions with more Al concentration
as a positive
voltage $V_G$ is applied between back and front gate;
leading to an increase of $g$. 
At $V_G$=2 V, no precession is
observed, corresponding to $g$=0. From \onlinecite{Salis2001:N}.} 
\label{mater:1}
\end{figure}

In semiconductors $g$ factors, which determine the 
spin splitting of carrier bands (and consequently influence the spin dynamics 
and spin resonance), can be very different from the free-electron value. 
With strong spin-orbit coupling in narrow-band III-V's they 
are $\approx$-50 
for InSb and $\approx$-15 for InAs, while, 
as discussed in Sec.~\ref{sec:IID3} the doping 
with magnetic impurities can give even $|g^*|\sim 500$. Manipulation
of the $g$ factor in a GaAs/AlGaAs quantum well (QW)  
in Fig.~\ref{mater:1}, relies on the results for a bulk Al$_x$Ga$_{1-x}$As, 
the variation of Al concentration changes the $g$ factor 
\cite{Chadi1976:PRB,Weisbuch1977:PRB} 
$g$=-0.44 for x=0 and $g$=0.40 for x=0.3. Related experiments on 
modulation-doped 
GaAs/Al$_{0.3}$Ga$_{0.7}$As have shown that by applying V$_G$ 
on can shift the electron
wave function in the QW and produce $\sim$1\% change
in the $g$ factor \cite{Jiang2001:PRB}. Subsequently, 
in an optimized Al$_x$Ga$_{1-x}$As quantum well, where
$x$ varied gradually across the structure, much larger changes were
measured--when $V_G$ is changed, the electron wave function  
efficiently samples 
different regions with different $g$ factors \cite{Salis2001:N}. 
Figure~\ref{mater:1}(a) gives
the time-resolved Kerr rotation data (the technique is discussed 
in Sec.~\ref{sec:III}) 
data which can be described as 
$\propto \exp(-\Delta t/T_2^*)\cos(\Omega \Delta t)$,
where $\Delta t$ is the delay time between the circularly polarized pump
and linearly polarized probe pulses, T$_2^*$ is the
transverse electron spin lifetime with inhomogeneous broadening, and 
the angular precession frequency $\Omega=\mu_B g  B/\hbar$ can be used to 
determine the $g$ factor. 
It is also possible to manipulate $g$ factors dynamically using 
time-dependent $V_G$ \cite{Kato2003:S}. The anisotropy of $g$ factor
($g$ tensor) allows voltage control of both the magnitude and the
direction of the spin precession vector ${\bf \Omega}$.

\subsection{\label{sec:IVC} Spin filters}

Solid state spin filtering (recall the similarity with
spin injection from Sec~\ref{sec:IIC1}) was first realized in N/F/N
tunneling. 
It was shown by \textcite{Esaki1967:PRL} that the magnetic tunneling through
(ferro)magnetic semiconductor Eu chalcogenides 
\cite{Nagaev:1983,Kasuya1968:RMP,vonMolnar1967:JAP}, 
such as EuSe\footnote{At zero magnetic field
EuSe is an antiferromagnet, 
and at moderate fields it becomes a ferromagnet with T$_C$$\approx$ 5K.} 
and EuS,\footnote{At zero magnetic field, exchange
splitting of a conduction band in bulk EuS is $\approx$ 0.36 eV
 \cite{Hao1990:PRB}.}
could be modified by an applied magnetic field. The change in $I-V$ curves 
in the
N/F/N structure, where N is a normal metal and F is a ferromagnet, was 
explained by the influence of the magnetic field on the height of the barrier 
formed at the N/F interface (for EuSe, the barrier height was lowered by
25\% at 2 T). 
The large spin splitting of the Eu chalcogenides was subsequently 
employed in the absence of
applied field with EuS \cite{Moodera1988:PRL} and nearly 100\%
spin polarization was reached  
at $B$=1.2 T with EuSe \cite{Moodera1993:PRL}. 
These spin-filtering properties of the 
Eu chalcogenides, used together with one-electron quantum dots, were proposed 
as the basis for a method to convert 
single spin into single charge 
measurements\footnote{This method could already be 
realized using single-electron transistors or 
quantum point contacts.} and provide an important ingredient in realizing a 
quantum computer \cite{DiVincenzo1999:JAP}, see Sec.~\ref{sec:IVF}.

Zeeman splitting in semiconductor heterostructures and superlattices 
(enhanced by large $g$ factors) \cite{Egues1998:PRL,Guo2001:PRB},
in quantum dots \cite{Recher2000:PRL,Deshmukh2002:PRL,Borda2003:PRL}, 
and nanocrystals  
\cite{Efros2001:PRL} provide effective spin filtering and spin-polarized
currents. Predicted quantum size effects and resonance tunneling 
\cite{Duke:1969}   
also have their spin-dependent counterparts. The structures studied
are typically double-barrier resonant tunneling diodes 
(for an early
spin-unpolarized study see \textcite{Tsu1973:APL}), with either 
Zeeman splitting or using ferromagnetic materials, in which
spin filtering can be tuned by an applied bias.\footnote{See, for example, 
\cite{Brehmer1995:APL,Mendez1986:PRB,Giazotto2003:APL,Ohno1998:S,%
Petukhov2002:PRL,Ting2002:APL,Slobodskyy2003:PRL,petukhov1998:ASS,%
Aleiner1991:JETPL, Vurgaftman2003:PRB}.} 

Several other realizations of spin filtering have been investigated,
relying on  spin-orbit coupling.\footnote{These include the work of 
\cite{Kiselev2001:APL,Governale2002:PRB,Voskoboynikov1998:PRB,%
Koga2002b:PRL,Silva1999:PRB,Voskoboynikov1999:PRB,Perel2003:PRB}.} 
or hot-electron transport across ferromagnetic regions,\footnote{See
\cite{vanDijken2002:PRB,Monsma1995:PRL,Rippard2000:PRL,Upadhyay1999:APL,%
Oberli1998:PRL,Cacho2002:PRL,Filipe1998:PRL}.} discussed in more detail 
in Sec.~\ref{sec:IVE3}.
A choice of particular atomically ordered F/Sm interfaces was suggested
to give a strong spin-filtering effect \cite{Kircenzow2001:PRB},
limited by the spin-orbit coupling and interfacial spin-flip scattering.

\begin{figure} 
\centerline{\psfig{file=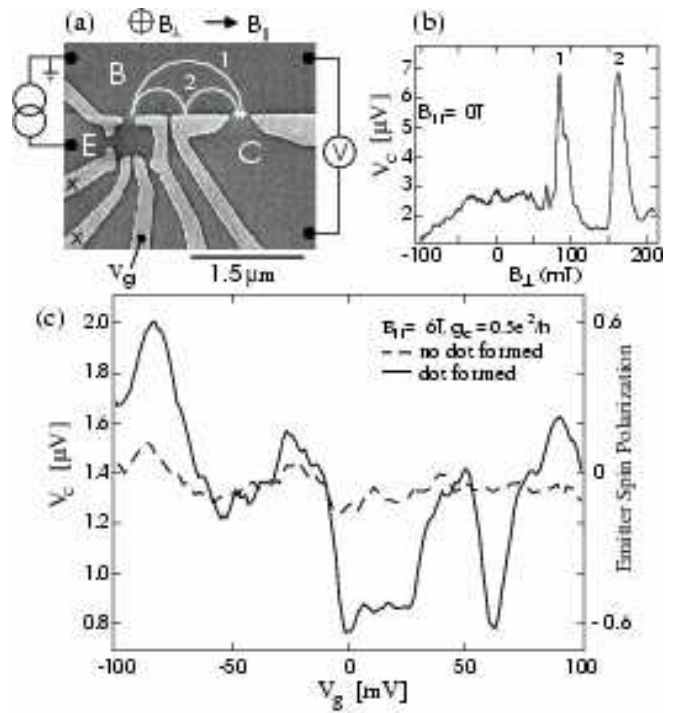,width=1.0\linewidth}}
\caption{Mesoscopic spin filter: 
(a) micrograph and circuit showing the
polarizer-analyzer configuration used in the experiment
of \textcite{Folk2003:S}. The emitter (E)
can be formed into either a quantum dot or a quantum point contact (QPC). 
The collector
(C), is a single point contact. Electrons are focused from E to C through
the base region (B), using a small perpendicular magnetic field. Gates
marked with ``x'' are left undepleted 
when E is operated as a QPC;
(b) base-collector voltage (V$_C$) showing two focusing peaks;
(c) focusing peak height at B$_\|$=6 T with spin-selective collector
QPC conductance ($g_C$=$0.5e^2/h$), comparing E as QPC at $2e^2/h$ (dashed
curve) and E as a quantum dot with both leads at $2e^2/h$ (solid curve). 
Adapted from \onlinecite{Folk2003:S}.} 
\label{filter:1}
\end{figure}

Mesoscopic spin filters have also been suggested 
\cite{Frustaglia2001:PRL,Joshi2001:PRB,Ionicioiu2003:PRB,Avdonin2003:P}
and we discuss here a particular realization.
In an applied magnetic field 
two quantum point contacts (QPC), an emitter (E)
and a collector (C), fabricated on top of a high-mobility 
2DEG in GaAs/AlGaAs,
can act as spin polarizer and analyzer 
\cite{Potok2002:PRL}.\footnote{QPC was also used 
to locally create and probe nonequilibrium nuclear spin
in GaAs/AlGaAs heterostructures 
in the quantum Hall regime \cite{Wald1994:PRL}.} 
In a ballistic regime and at $T=70$ mK 
(mean free path $\approx$ 45 $\mu$m $>>$ sample size $\approx$ 1.5 $\mu$m)
magnetic focusing\footnote{Suggested by \textcite{Sharvin1965:ZETF}
as a technique to study Fermi surfaces; see also \textcite{vanHouten1989:PRB}.} 
with $B_\bot$ results in base-collector voltage peaks, when the separation
of the two QPCs is an even multiple of the cyclotron radius $m^*v_F/eB_\bot$,
where $m^*$ is the effective mass, and v$_F$ the Fermi velocity.
These results are illustrated in Figs.~\ref{filter:1}(a) and (b) 
on a slightly modified structure \cite{Folk2003:S}
where, by applying a gate voltage, one can cause the emitter to form either 
a quantum dot or QPC.  
Effective spin filtering, due to the large in-plane field $B_\|$, can 
be tuned by the gate voltage which changes the conductance of QPC.
The resulting effect of spin filtering modifies the collector voltage
$V_C$  \cite{Potok2002:PRL}, 
\begin{equation}
V_C=\alpha(h/2e^2)I_E(1+P_{I_E}P_{T_C}),
\label{eq:filter}
\end{equation}
where $0< \alpha <1$ parameterizes the imperfections in focusing,
$P_{I_E}$ and $P_{T_C}$ are the spin polarization [recall Eq.~(\ref{eq:polar})] 
of the emitter current $I_E$, and the collector transmission coefficient $T_C$
is related to the collector conductance by $g_C=(2e^2/h)T_C$. 
In Eq.~(\ref{eq:filter}) we note a recurring form for a spin-valve
effect. The measured signal involves the product of two different spin 
polarizations, for example, similar to TMR in Eq.~(\ref{eq:julliere}) 
or to spin-charge coupling due to nonequilibrium spin 
[recall Eqs.~(\ref{eq:DelR}) and (\ref{eq:md})].
Another mesoscopic spin filter with few-electron quantum dot 
(GaAs/AlGaAs-based) was used to demonstrate a nearly complete spin
polarization which could be reversed by adjusting gate voltages 
\cite{Hanson2003:P}.

\subsection{\label{sec:IVD} Spin diodes}

Spin diodes are inhomogeneous two-terminal devices whose electronic or optical 
properties depend on the spin polarization of the carriers. Such devices
were envisaged long before the emergence of 
spintronics. \textcite{Solomon1976:SSC},
for example, proposed and demonstrated a silicon {\it p-n} junction whose 
current was modified by changing the spin polarization of the recombination
centers. In a magnetic field both the mobile carriers and the recombination
centers have an equilibrium spin polarization due to the Zeeman splitting.
The current in a {\it p-n} junction depends on the recombination rate,
which, in turn, depends on the relative orientation of the spin 
of the carriers and the centers \cite{Lepine1972:PRB}. The trick to 
modifying the current is to decrease (even saturate) the spin polarization
of either the electrons or the centers by electron spin
resonance. Indeed, \textcite{Solomon1976:SSC} found a variation of
$\approx 0.01\%$ of the saturation current at small biases where recombination
in the space-charge region dominates. Similar experiments could be used to 
detect nonequilibrium spin due to (potential) spin injection in Si,
where optical methods are ineffective, but also in other semiconductors
where electrical detection would be desirable.\footnote{Spin diodes can 
also probe fundamental properties of electronic
systems. The diode demonstrated by \textcite{Kane1992:PRB} is 
based on a junction
between two coplanar AlGaAs/GaAs 2DEG's, one with $\nu<1$ and 
the other with $\nu>1$, 
where $\nu$ is the Landau-level filling; that is, the two regions have
opposite spins at the Fermi level. The current crossing such a junction,
which has a diode property due to the existence of a built-in field in
the contact, is accompanied by a spin flip. Interestingly, the
current is also time dependent, due to the current-induced dynamic nuclear
polarization.} 

Several spin diodes have recently been proposed or demonstrated 
with the goal of either maximizing the sensitivity of the $I-V$ characteristics
to spin and magnetic field, or to facilitating spin injection and its
detection through semiconductor interfaces comprising a magnetic semiconductor
as the injector. 
Magnetic tunneling diodes have been used for spin injection 
from a ferromagnetic to a nonmagnetic semiconductor, in p-GaMnAs/n-GaAs
{\it p-n} 
junctions~\cite{Kohda2001:JJAP,Johnston-Halperin2002:PRB,vanDorpe2003:P}.
As discussed in Sec.~\ref{sec:IID3},
{\it p-n} 
heterostructures have combined Cr- or Eu-based ferromagnetic
semiconductors and InSb  \cite{Osipov1998:PL,Viglin1997:PLDS}.
Spin light-emitting diodes (recall Figs.~\ref{injexp:3}
and \ref{injexp:4}) 
were employed for injecting and detecting spins 
in semiconductors,
while resonant tunneling diodes have been demonstrated as effective spin 
injectors (Sec.~\ref{sec:IID3}) and spin filters (Sec.~\ref{sec:IVC}).
A magnetic unipolar diode has been proposed by
\textcite{Flatte2001:APL}
to simulate the working of ordinary diodes, but with 
homogeneous monopolar
doping (either donors or acceptors, not both). The role of 
inhomogeneous doping in the {\it p-n} junction is played by the inhomogeneous 
spin splitting
of the carrier band, with the spin up and spin down carriers 
playing roles similar
to those of the electrons and holes in bipolar diodes. 
Si-based {\it p-i-n} diode sandwiched between two ferromagnetic
metals was suggested to allow controlling the device performance
by an externally applied magnetic field \cite{Dugaev2003:PE}.
Finally, 
\textcite{Zutic2002:PRL} 
have proposed the magnetic bipolar diode described below.

The magnetic bipolar diode\footnote{Not to be confused with the usual 
magnetic diodes
which are ordinary diodes in a magnetic field. The $I-V$ characteristics of
such diodes, depend on the magnetic field through small orbital effects 
on diffusion coefficient, 
not through the spin effects described here.} (MBD) is a {\it p-n} 
junction diode with one or both regions magnetic 
\cite{Zutic2002:PRL,Fabian2002:PRB}.
The MBD is the prototypical device of bipolar
spintronics, a subfield of spintronics in which both electrons and holes 
take part 
in carrier transport, while either electrons or holes (or both) are spin 
polarized (see Sec.~\ref{sec:IVA4}). 
Examples of nonmagnetic bipolar spintronic devices are the 
spin-polarized {\it p-n} 
junction \cite{Zutic2001:PRB} and the spin solar cell \cite{Zutic2001:APL}. 
These devices offer 
opportunities for effective spin injection, 
spin amplification (see Sec.~\ref{sec:IIC3}), 
or spin capacity---the effect of 
changing, by voltage, nonequilibrium spin density \cite{Zutic2001:PRB}. 
The advantages of 
magnetic bipolar spintronic devices 
\cite{Zutic2002:PRL,Fabian2002:PRB,Zutic2003:APL,Fabian2002:P}
lie in the combination of equilibrium 
magnetism and nonequilibrium spin and effective methods to manipulate
a minority carrier population.
The most useful effects of the 
spin-charge coupling in MBD's are the 
spin-voltaic
and the giant-magnetoresistive
effects, which are enhanced over those of metallic systems by the 
exponential dependence of the current on bias voltage.

\begin{figure}
\centerline{\psfig{file=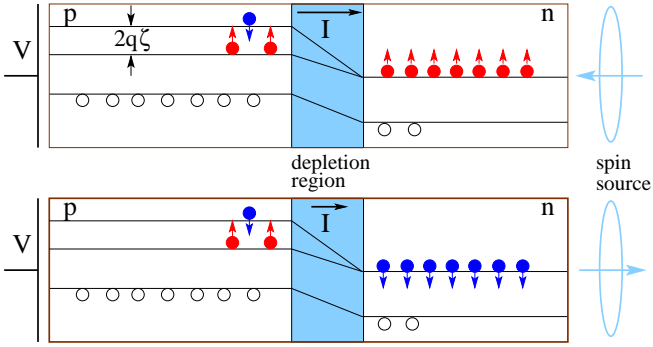,width=1\linewidth,angle=0}}
\caption{Scheme of a magnetic bipolar diode. 
The $p$-region (left) is magnetic, indicated by the spin
splitting $2q\zeta$ of the
conduction band. The $n$-region (right) is nonmagnetic, but spin 
polarized by a spin source: Filled circles, spin-polarized electrons;
empty circles, unpolarized holes.
If the nonequilibrium spin in the 
$n$-region is oriented parallel (top figure) 
to the equilibrium spin in the 
$p$-region, large forward current flows. 
If the relative orientation is antiparallel (bottom), the current drops
significantly. Adapted from \onlinecite{Zutic2002:PRL}. 
}
\label{fig:md}
\end{figure}

A scheme of an MBD is shown in Fig.~\ref{fig:md} 
(also see Fig.~\ref{deplete}). 
The $p$-region is magnetic, by 
which we mean that it has a spin-split conduction band with the spin
splitting (Zeeman or exchange) 
$2q\zeta$ $\sim k_BT$. 
Zeeman splitting can be significantly enhanced by 
the large $g^*$ factors of magnetically doped (Sec.~\ref{sec:IIC3})
or narrow-band-gap semiconductors (Sec.~\ref{sec:IVB}).
Using an MBD with a ferromagnetic semiconductor slightly above its $T_C$ 
is also expected to give large $g^*$ factors.
The $n$-region 
is nonmagnetic, but 
electrons can be 
spin-polarized 
by a spin source (circularly polarized light or 
magnetic electrode). The interplay between the equilibrium spin of polarization 
$P_{n0}=\tanh(q\zeta/k_B T)$ 
in the $p$-region, and the nonequilibrium 
spin source of polarization $\delta P_n$ 
in the $n$-region, at the edge of the depletion layer, 
determines the $I-V$  characteristics of the diodes. 
It is straightforward to generalize these considerations to include 
the spin-polarized holes \cite{Fabian2002:PRB}.

The dependence of the electric current $j$ on $q\zeta$ and $\delta P_n$ was 
obtained by both numerical and analytical methods. 
Numerical calculations \cite{Zutic2002:PRL} 
were performed by self-consistently solving for the drift-diffusion, 
continuity, as well as carrier recombination and 
spin-relaxation equations, 
discussed in  
Sec.~\ref{sec:IVA4}. While the numerical calculations are indispensable in the
high-injection limit,\footnote{The small bias or low-injection limit 
is the regime of applied bias in which the density of the carriers injected
through the depletion layer (the minority carriers) is much smaller than 
the equilibrium density of the majority carriers. 
Here and in Sec.~\ref{sec:IVE2} the terms majority and
minority refer to the relative carrier (electron or hole) population and
not to spin. The large bias or high-injection limit is the regime where 
the injected carrier density
becomes comparable to the equilibrium density. This occurs
at forward biases comparable to the built-in potential, typically  1 V.}
valuable insight and  
analytical formulas can be obtained in the
low-injection limit, where the Shockley theory \cite{Shockley:1950} for 
ordinary {\it p-n}
junctions was generalized by \textcite{Fabian2002:PRB} for the magnetic
case.  

To illustrate the $I-V$ characteristics of MBD's, consider the low-injection 
limit in the configuration of Fig.~\ref{fig:md}. 
The electron contribution to the total electric current is
\cite{Fabian2002:PRB}
\begin{equation}\label{eq:md}
j_n \sim n_{0}(\zeta) 
\left [e^{qV/k_B T} \left (1+\delta P_n P_{n0}\right ) -1 \right ],
\end{equation}
where $V$ is the applied bias (positive for forward bias) 
and 
$n_{0}(\zeta)=(n_i^2/N_a)\cosh(q\zeta/k_B T)$ 
is the equilibrium number of electrons 
in the $p$-region, 
dependent on the splitting, the intrinsic carrier 
density $n_i$, and the acceptor doping $N_a$. Equation (\ref{eq:md}) 
generalizes the 
Silsbee-Johnson spin-charge coupling \cite{Silsbee1980:BMR,Johnson1985:PRL}, 
originally proposed for ferromagnet/paramagnet metal interfaces, to 
the case of magnetic {\it p-n} junctions.  The advantage
of the spin-charge coupling in {\it p-n} junctions, as opposed to metals or 
degenerate systems, is the nonlinear voltage dependence 
of the nonequilibrium carrier and spin densities \cite{Fabian2002:PRB}, 
allowing for 
the exponential enhancement of the effect with increasing $V$. 
Equation (\ref{eq:md}) can be understood qualitatively
from Fig.~\ref{fig:md} \cite{Fabian2002:PRB}. 
In equilibrium, $\delta P_n=0$ 
and $V=0$, no current flows through the
depletion layer, as the electron currents from both sides of the junction 
balance out. The balance is disturbed either by applying bias or 
by selectively populating different spin states,
making the flow of one spin species greater than that of the other. 
In the latter case, 
the effective barriers for crossing
of electrons from the $n$ to the $p$ side is different for spin
up and down electrons (see Fig.~\ref{fig:md}).
Current can flow even at $V=0$ when $\delta P_n\ne 0$. 
This is an example of the spin-voltaic effect (a spin analog
of the photo-voltaic effect), in which  nonequilibrium
spin causes an EMF \cite{Zutic2002:PRL,Zutic2003:P}.
In addition, the direction of the zero-bias current 
is controlled by the relative sign of 
$P_{n0}$ and $\delta P_n$. 

\begin{figure}
\centerline{\psfig{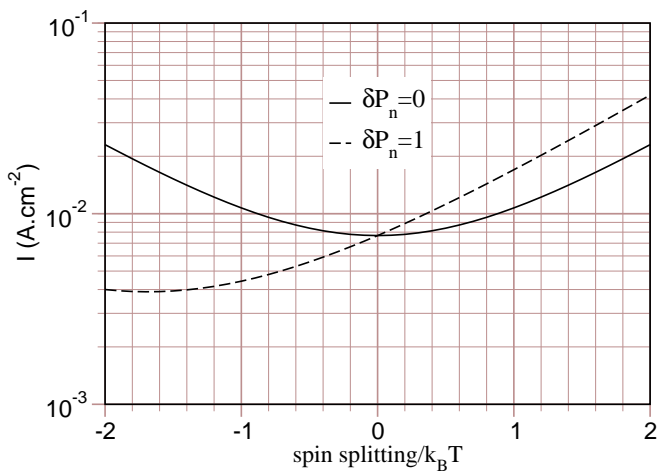}}
\caption{Giant magnetoresistance (GMR) effect in magnetic diodes. 
Current/spin-splitting characteristics ($I$-$\zeta$) 
are calculated self-consistently 
at $V$=0.8 V
for the diode from  Fig.~\ref{fig:md}. 
Spin splitting $2q\zeta$ on the p-side is normalized to $k_B T$. 
The solid curve corresponds to a switched-off spin source. The current is
symmetric in $\zeta$. With spin source on (the extreme case of 100\% spin 
polarization injected into the 
$n$-region 
is shown), the current is a strongly asymmetric 
function of $\zeta$, displaying large GMR, shown by the dashed curve. 
Materials parameters of GaAs were applied.
Adapted from \onlinecite{Zutic2002:PRL}.}
\label{fig:gmrd}
\end{figure}

MBD's can  display an interesting  GMR-like effect, which 
follows from Eq.~(\ref{eq:md}) \cite{Zutic2002:PRL}. The 
current depends strongly on the relative orientation of the 
nonequilibrium spin and the equilibrium magnetization. 
Figure \ref{fig:gmrd} plots $j$,
which also includes the contribution from holes, as a function of 
$2q\zeta/k_B T$ for 
both the unpolarized, $\delta P_n=0$, and fully polarized, $\delta P_n=1$, 
$n$-region. 
In the first case $j$ is
a symmetric function of $\zeta$, increasing exponentially with increasing 
$\zeta$ due to the increase in the 
equilibrium minority carrier density $n_0(\zeta)$. 
In unipolar systems, where transport is due to the majority carriers, 
such a modulation of the current is not likely, as the majority carrier
density is fixed by the density of dopants. 

If $\delta P_n \ne 0$, the current will depend on the sign of 
$P_{n0}\cdot \delta 
P_n$.  For parallel nonequilibrium (in the $n$-region)
and equilibrium spins (in the $p$-region), most electrons cross 
the depletion layer through the 
lower barrier (see Fig.~\ref{fig:md}), increasing the current. 
In the opposite case of antiparallel relative orientation, 
electrons experience a larger barrier and the current is inhibited. 
This is demonstrated in Fig.~\ref{fig:gmrd}
by the strong asymmetry in $j$. The corresponding GMR ratio, the 
difference between  $j$ for parallel and antiparallel orientations,
can also be calculated analytically from Eq.~(\ref{eq:md}) as
$2|\delta P_n P_{n0}|/(1-|\delta P_n P_{n0}|)$ \cite{Fabian2002:PRB}. 
If, for example, $|P_{n0}|=|\delta 
P_n|=0.5$, the relative change is 66\%. 
The GMR effect should be useful for measuring
the spin relaxation rate of bulk semiconductors \cite{Zutic2003:APL},
as well as for detecting nonequilibrium spin in the nonmagnetic
region of the {\it p-n} junction.\footnote{This could be a
way to detect spin injection into Si, where optical detection
is ineffective.}

Although practical MBD's are still to be fabricated and the predicted effects
tested, magnetic {\it p-n} junctions have already been demonstrated. 
Indeed, \textcite{Wen1968:IEEETM}\footnote{We thank M. Field for bringing this
reference to our attention.} were perhaps
the first to show that a ferromagnetic {\it p-n} junction, based on the
ferromagnetic semiconductor CdCr$_2$Se$_4$ doped with Ag acceptors and In 
donors, could act as a diode. Heavily doped p-GaMnAs/n-GaAs junctions
were fabricated 
\cite{Ohno2000:ASS,Kohda2001:JJAP,Johnston-Halperin2002:PRB,Arata2001:PE,%
vanDorpe2003:P} 
to demonstrate tunneling interband spin injection. 
Incorporation of (Ga,Mn)As layer in the intrinsic region of {\it p-i-n}
GaAs diode was shown to lead to an efficient photodiode, in which the
Mn ions function as recombination centers \cite{Teran2003:APL}.
It would be interesting to see such devices combined with a spin
injector in the bulk regions.
Recently, 
\textcite{Tsui2003:APL}
have shown that the current in p-CoMnGe/n-Ge magnetic heterojunction diodes 
can indeed be controlled by magnetic field.
To have functioning MBD's at room temperature, and to
observe the above predicted phenomena, several important challenges
have to be met:

(i) Zeeman or exchange splitting needs to sufficiently large to 
provide equilibrium spin polarization,
$\gtrsim 1-10\%$. This may be difficult at room temperature, unless
the effective $g$ factor is $\sim 100$ at $B \sim 1$ T 
(Sec.~\ref{sec:IID3}). The use of ferromagnetic semiconductors
is limited by their $T_C$ (Sec.~\ref{sec:IVB}).

(ii) For a strong spin-charge coupling [recall the discussion of 
Eq.~(\ref{eq:md})] a nondegenerate carrier density
is desirable, which, while 
likely in (Zn,Cr)Te, is not easily realized in many other ferromagnetic
semiconductors that are typically heavily doped (Sec.~\ref{sec:IVB}).

(iii) An effective integration of magnetic and nonmagnetic structures 
into single devices \cite{Samarth2003:SSC} is needed.

(iv) The samples need to be smaller than the spin diffusion lengths,
requiring high carrier mobility and long spin relaxation (easier to 
realize for spin-polarized electrons).

(v) The effects of actual device structures, such as two- and/or 
three-dimensional spin flow, interface contacts, spin-dependent band offsets 
and band bendings, strong spin relaxation in the depletion layers, etc., 
will need to be understood.

By combining two magnetic {\it p-n} junctions in series one can
obtain a magnetic bipolar transistor (Sec.~\ref{sec:IVE2}), a 
three terminal device which offers spin-dependent amplification.

\subsection{\label{sec:IVE} Spin transistors}

We review several proposals for spin transistors 
that have at least one semiconductor region and 
that aim at integrating spin 
and charge transport within traditional device schemes of the field-effect and 
junction transistors. 
Three important cases are discussed in detail: the Datta-Das  
spin field-effect transistor, the magnetic bipolar transistor, and
the hot-electron spin transistor. 

Various spin transistors that contain metallic (and insulating)
regions have been proposed 
\cite{Johnson1993:S,Bauer2003:APL,Zvezdin2003:TP,You2000:JAP}. 
There is also a large category of
the spin single-electron transistors, first realized by 
\textcite{Ono1996:JPSJ}, and later investigated in 
\cite{Barnas1998:PRL,Korotkov1999:PRB,Ciorga2002:APL,Martinek2002:PRB}. 
Spin single-electron
transistors can be viewed as an extension of magnetic tunneling 
(see Sec.~\ref{sec:IVA2})
to double tunnel junctions, where the Coulomb blockade becomes important 
\cite{Takahashi1998:PRL}. For a review of spin single-electron transistors see 
\cite{Maekawa:2002}. 

\subsubsection{\label{sec:IVE1} Spin field-effect transistors}

\textcite{Datta1990:APL} proposed what became the prototypical spintronic 
device scheme, 
the Datta-Das spin field-effect transistor (SFET) (see Fig.~\ref{fig:DD}). 
The device is based on spin injection and spin detection by a ferromagnetic 
source and 
drain, and on spin precession about the built-in structure inversion asymmetry 
(Bychkov-Rashba) field ${\bf\Omega}$, Eq.~(\ref{eq:relax:BR}), 
in the asymmetric, quasi-one-dimensional channel 
of an ordinary field-effect transistor. The attractive feature of the
Datta-Das SFET is that spin-dependent device operation is controlled
not by external magnetic fields, but by gate bias, which controls the 
spin precession rate.

The structure of the Datta-Das SFET is shown in Fig.~\ref{fig:DD}. 
Consider a 2DEG  confined along the plane of the unit vector $\bf n$. 
The precession axis of $\bf\Omega$
lies always in the channel plane (see Sec.~\ref{sec:IIIB2}), so 
the results 
(unlike those for bulk inversion asymmetry) 
are insensitive 
to the relative orientation of ${\bf n}$ and the principal crystal axes. 
Equation (\ref{eq:relax:BR}) determines the evolution of the 
expectation value for a spin perpendicular to the plane, 
$s_{n}={\bf s}\cdot {\bf n}$, and a spin parallel to the in-plane 
$\bf k$, $s_{\parallel}={\bf s}\cdot{\bf k}/k$:  
\begin{equation}
ds_n/dt=2\alpha_{BR} k s_{\parallel}, \,\,\,\,
ds_\parallel/dt=-2\alpha_{BR} k s_{n},
\end{equation}
where $\alpha_{BR}$ is the structure inversion asymmetry coefficient 
appearing in 
Eq.~(\ref{eq:relax:BR}).
The average spin component along 
$\bf\Omega$, $s_{\perp}={\bf s}\cdot ({\bf k} \times {\bf n})/k$, 
is constant. 
As a result, $s_\parallel=s_{0\parallel}\cos(\omega t)$, where 
$\omega=2\alpha_{BR} k$ 
and the injected spin at the source is labeled with zero. 
If $\varphi$ is the angle between $\bf k$ and the source-drain axis,
the electron will reach the drain at time $t'=L m_c/(\hbar k\cos\varphi)$, 
with the spin 
$s_\parallel$
precessing at
the angle $\phi=2\alpha_{BR} m_c L/\hbar$. The average spin at the drain 
in the direction
of magnetization is $s_{\parallel}(t')\cos\varphi +
s_{0\perp}{\bf m}\cdot ({\bf k}\times {\bf n})$, so the current is 
modulated by 
$1 - \cos^2\varphi\sin^2(\phi/2)$, the probability of finding the spin in 
the direction of magnetization $\bf m$.   

Note that $\phi$ does not depend on the momentum (or energy) of the 
carriers. As 
the spread 
$\varphi$ in the momenta increases, the modulation effect decreases.  
The largest effect is seen for $\varphi=0$, where the current modulation 
factor is $\cos^2(\phi/2)$. It was therefore proposed~\cite{Datta1990:APL} 
that $\varphi$
be limited by further confining the electron motion along $\varphi=0$ using 
a one-dimensional channel as a waveguide. 
Spin modulation of the current becomes ineffective if transport is 
diffusive. 
Taking typical values~\cite{Nitta1997:PRL,Koga2002:PRL}
for 
$\hbar\alpha_{BR}\approx 1\times 10^{-11}$ eV$\cdot$m, and 
$m_c=0.1 m_e$, current modulation should be observable at source-drain 
separations of $L\agt 1$ $\mu$m, setting the scale for ballistic transport.
The device will work best with 
narrow-gap materials ~\cite{Lommer1988:PRL}
like InAs, in which the structure inversion asymmetry dominates the spin 
precession \cite{Luo1988:PRB,Luo1990:PRB,Das1989:PRB}. 
Another option is using Si heterostructures, in which bulk inversion 
asymmetry is
absent. However, the small magnitude of the spin-orbit interaction makes 
$\alpha_{BR}$ in Si probably rather weak.

The Datta-Das SFET is yet to be realized. There are at least four important
difficulties in observing the proposed effects. 

(i) The effective spin 
injection
of spin-polarized carriers from the ferromagnetic source into a 2DEG  
is nontrivial (see Sec.~\ref{sec:IID4}).

(ii) Ballistic spin-polarized transport should be realized through the channel
with $\it uniform$ $\alpha_{BR}$ by eliminating undesirable electric fields 
due to interface inhomogeneities. 

(iii) The parameter $\alpha_{BR}$ should 
be effectively controllable by the gate. 

(iv) 
The structure inversion asymmetry should dominate over the bulk inversion
asymmetry, and the spin precession rate must be 
large enough ($\hbar\alpha_{BR} \agt 10^{-11}$ eV$\cdot$m) to allow
at least a half precession during the ballistic transport. 

These four factors present a great challenge to fabricating a 
Datta-Das SFET at room temperature, limiting the design to 
special materials and very clean interfaces.
However, the modulation of $\alpha_{BR}$ 
by biasing voltage (iii) 
has been already convincingly demonstrated in 
In$_{0.53}$Ga$_{0.47}$As/In$_{0.52}$Al$_{0.48}$As 
QW's~\cite{Nitta1997:PRL,Hu1999:PRB,Grundler2000:PRL} 
[for GaAs/AlGaAs 2DEG see
also \cite{Miller2003:PRL}].  Initial experimental 
investigations of magnetoresistance in the Datta-Das SFET systems 
were performed by \textcite{Gardelis1999:PRB}. Recently spin precession
in the the Datta-Das SFET, including the bulk inversion asymmetry,
was investigated by \cite{Winkler2003:lanl} using $k\cdot p$
model calculations. It is not surprising that the conductance through the
transistor, in the present orientation-dependent bulk inversion asymmetry,
depends rather strongly on the crystallographic orientation of the
two-dimensional channel \cite{Lusakowski2003:PRB}. For more discussion
of the Dresselhaus bulk inversion asymmetry and the Bychkov-Rashba structure
asymmetry see Sec.~\ref{sec:IIIB2b}.

The Datta-Das SFET has generated great interest in mesoscopic spin-polarized 
transport
in the presence of structure inversion asymmetry. 
Model calculations using the tight-binding formulation of $H_{\rm SIA}$
(recall Sec.~II.B.2)
were reported by \textcite{Pareek2002:PRB}.
Further theoretical investigations on the theme of the Datta-Das spin 
transistor
can be found in \textcite{Matsuyama2002:PRB,Nikolic2001:P} 
and in an extensive review by \textcite{Bournel2000:APF}. 
Distinct SFET's have also been suggested, even in the absence of 
ferromagnetic
regions which are replaced by a rotating external magnetic field of 
uniform strength
\cite{Wang2002:P}. \textcite{Ciuti2002:APL} proposed 
a ferromagnetic-oxide-semiconductor 
transistor, with a nonmagnetic source and drain, 
but with two ferromagnetic gates in series above 
the base channel. The relative orientation of the gates' magnetization 
leads to 
magnetoresistance effects. An SFET that can operate in the diffusive regime, 
in the presence of both bulk and structure inversion asymmetry, 
has been considered by \textcite{Schliemann2002:P}.

\subsubsection{\label{sec:IVE2} Magnetic bipolar transistor}

The magnetic bipolar transistor (MBT) is a bipolar transistor
with spin-split carrier bands and, in general, an injected
spin 
\cite{Fabian2002:P,Fabian2003:Pa,Fabian2003:Pb}. 
A related device structure
was already proposed by \textcite{Gregg1997:JMMM} in a push for
silicon-based spintronics. In this proposal (also called
SPICE for spin-polarized injection current emitter)
the semiconductors have no equilibrium spin, while
the 
spin source 
is provided by a ferromagnetic spin
injector attached to the emitter, and another ferromagnetic
metal, a spin detector,
is attached to the base/collector junction to modulate
the current flow. In both configurations the aim is to
control current amplification 
by spin and magnetic field.

A scheme of a particular  MBT is shown in Fig.~\ref{fig:mbt}. 
Such a three-terminal device can be thought of as consisting of 
two magnetic {\it p-n} junctions connected in series. Materials 
considerations discussed in Sec.~\ref{sec:IVD} also apply  
to an MBT in order to provide a sufficient equilibrium polarization
in a magnetic base $P_{B0}$. 
While nonmagnetic, the emitter has a nonequilibrium polarization $\delta P_E$ 
from a spin source, similar to the magnetic diode case in Fig.~\ref{fig:md}.
Only the spin polarization of electrons is assumed. Applying the generalized
Shockley theory to include spin 
effects \cite{Fabian2002:PRB}, a theory of MBT was developed 
by \textcite{Fabian2002:P,Fabian2003:Pb}. 
Later, simplified schemes of MBT [not including the effect of nonequilibrium
spin ($\delta P_E=0$)] were also considered by \textcite{Flatte2003:APL}
and \textcite{Lebedeva2003:JAP}.

\begin{figure}
\centerline{\psfig{file=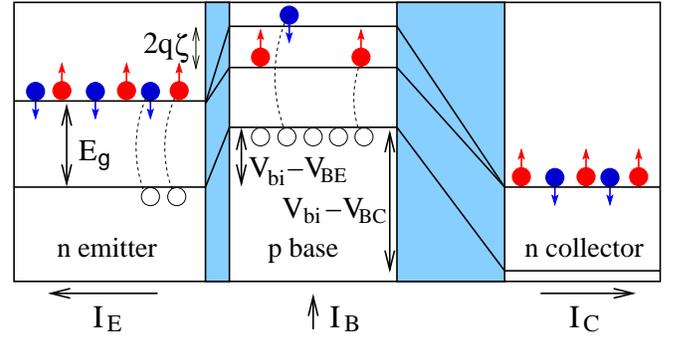,width=1\linewidth,angle=0}}
\caption{
Scheme of an {\it n-p-n} 
magnetic bipolar transistor with magnetic base (B),
nonmagnetic emitter (E), and collector (C). 
Conduction and valence bands are separated by the energy gap $E_g$.
The conduction band has a spin splitting $2q\zeta$, leading
to equilibrium spin polarization $P_{B0}=\tanh(q\zeta/k_B T)$. 
Carriers and depletion regions are represented as in Fig.~\ref{fig:md}.
In the so called forward active regime, where the transistor
can amplify currents, the E-B junction is forward biased (here
with voltage $V_{BE}>0$ lowering the built-in potential $V_{bi}$), while the
B-E junction is reverse biased ($V_{BC}<0$). The directions
of the current flows are indicated. Electrons flow from E to B, 
where they either recombine with holes (dashed 
lines) or continue to be swept by the electric field in the B-E 
depletion layer towards C. Holes form mostly the base current, $I_B$,
flowing to the emitter. The current amplification
$\beta=I_C/I_B$ 
can be controlled by $P_{B0}$ 
as well as by the nonequilibrium spin in E. 
Adapted from \onlinecite{Fabian2003:Pa}.
} 
\label{fig:mbt}
\end{figure}
The current amplification (gain) $\beta=I_C/I_B$ (see Fig.~\ref{fig:mbt})
is typically $\sim 100$ in practical transistors.
This ratio depends on many factors, such as the doping 
densities, carrier lifetimes,  
diffusion coefficients,
and structure geometry. 
In an MBT $\beta$ also depends on the spin splitting 
$2q \zeta$ (see Fig.~\ref{fig:mbt})
and the nonequilibrium polarization $\delta P_E$. 
This additional dependence of $\beta$
in an MBT is called magnetoamplification \cite{Fabian2003:Pb}.
An important prediction is that the nonequilibrium spin can be 
injected at low bias 
all the way from the emitter, through the base, to the collector 
\cite{Fabian2002:P,Fabian2003:Pb}
in order to make possible an
effective control of $\beta$ by $\delta P_E$. 

The calculated dependence of the gain on the spin splitting for 
$\delta P_E=0.9$ is 
shown in Fig.~\ref{fig:gain}, for GaAs and Si materials parameters. 
The gain is
very sensitive to the equilibrium magnetization in Si, while the 
rapid carrier 
recombination in GaAs prevents more effective control of the transport 
across the base.
In Si it is the spin injection at the emitter-collector depletion 
layer which 
controls the current. As the spin-charge coupling is most effective 
across 
the depletion layer (see Sec.~\ref{sec:IVD}), this 
coupling is essential for the current in Si. 
In the limit of slow carrier recombination \cite{Fabian2002:P},
\begin{equation} \label{eq:beta}
\beta \sim \cosh(q\zeta/k_B T) (1 + \delta P_E P_{B0}).  
\end{equation}
Both magnetic field (through $\zeta$) and nonequilibrium spin affect the 
gain, an implication of the spin-voltaic effect
\cite{Zutic2002:PRL,Zutic2003:P}.
The sensitivity of the current to spin can be used to measure  
the injected spin polarization. If no 
spin source 
is present ($\delta P_E=0$),
there is no spin-charge coupling in the space-charge regions, unless
at least two regions are magnetic. The only remaining effects 
on the $I-V$ characteristics come from the sensitivity of the carrier 
densities 
in the equilibrium  
magnetic regions to $\zeta$ [see Eq.~(\ref{eq:beta})
for the case of $\delta P_E=0$]. 

\begin{figure}
\centerline{\psfig{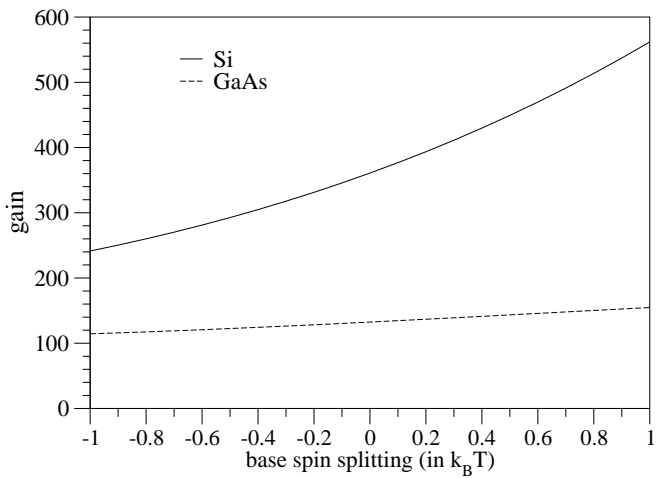}}
\caption{Calculated gain dependence of an MBT as a function of
base spin splitting $2q\zeta$, given in units of thermal energy 
$k_B T$.  The nonequilibrium
spin polarization in the emitter is $\delta P_E=0.9$. 
Si (solid) and GaAs (dashed) 
materials parameters were applied.
Adapted from \onlinecite{Fabian2002:P}. 
} 
\label{fig:gain}
\end{figure}

The MBT is, in effect, a magnetic heterostructure transistor, since its 
functionality depends
on tunability of the band structure of the emitter, base, or collector. The 
advantage 
of MBT, however, is that the band structure is not built-in, but can be tuned 
{\it during} 
the device operation by magnetic field or spin injection signals. 
The challenges to demonstrate the predicted phenomena in MBT are 
similar to those of magnetic bipolar diodes, see Sec.~\ref{sec:IVD}.

\subsubsection{\label{sec:IVE3} Hot-electron spin transistors}

\label{hotspin}
Spin transistors that
rely on transport of hot (nonthermalized) carriers
have the potential to serve of several different purposes. 
On the one hand, they could be used as a diagnostic tool to characterize
spin- and energy-dependent interfacial properties, scattering processes, 
and electronic structure, relevant to spintronic 
devices.\footnote{These efforts 
are motivated in part by the success of (spin-insensitive) 
ballistic-electron-emission microscopy in providing high spatial and
energy resolution of 
properties of metal/semiconductor interfaces 
\cite{Kaiser1988:PRL,Bonnell:2001,Smith2000:PRB}. A subsequent
variation--a ballistic-electron-magnetic microscopy, which also uses an STM 
tip to inject hot carriers, is capable of resolving magnetic features at 
a $\sim10$ nm length scale \cite{Rippard1999:APL,Rippard2000:PRL}.}
On the other hand, hot-electron transistors are also of interest for
their ability to sense magnetic fields, their possible
memory applications, 
and a their potential as a source
of ballistic hot-electron spin injection.
Below we discuss two representative examples,
a spin-valve transistor and a magnetic tunneling transistor.

The spin-valve or Monsma transistor provided an early demonstration of a
hot-electron 
spin transistor and realization of a hybrid spintronic
device that integrates metallic ferromagnets and semiconductors 
\cite{Monsma1995:PRL,Monsma1998:S}.
A three terminal structure 
\footnote{Similar to other hot-electron 
spin devices, the term
transistor characterizes their three-terminal structure rather
than the usual functionality of a conventional semiconductor transistor.
In particular, a semiconductor bipolar transistor,
which also has an emitter/base/collector structure, typically has
a sizable current gain--a small change in the base current 
leads to a large change in the collector current (see Sec.~\ref{sec:IVE2}). 
However, only a small
current gain $\sim2$ (due to large current in a metal base)
was predicted in magnetic tunnel-junction-based devices \cite{Hehn2002:PRB}.}
consisted of a metallic base (B) made
of a ferromagnetic multilayers in a CPP geometry [as depicted in 
Fig.~\ref{gmr:1}(a)] surrounded by a silicon emitter (E) and 
collector (C) with two Schottky contacts, formed at E/B and B/C 
interfaces.\footnote{Another realization of a spin-valve transistor
combines a GaAs emitter with a Si collector \cite{Dessein2000:JAP}.} 
Forward bias $V_{EB}$ controls the emitter current $I_E$
of spin-unpolarized electrons, which are injected
into a base region as hot carriers. 
The scattering processes in the base,
together with the reverse bias $V_{BC}$,
influence how many of the injected electrons can overcome the B/C Schottky
barrier and contribute to the collector current $I_C$. 
Similar to the physics of GMR structures~\cite{Levy:2002,Gijs1997:AP}
scattering in the base region strongly depends on the relative orientation 
of the magnetizations in the ferromagnetic layers.
Electrons with spin which has magnetic moment opposite (antiparallel)
to the magnetization of a ferromagnetic layer typically are scattered
more than electrons with parallel magnetic moments,
resulting
in a spin-filtering effect which can be described in terms of spin-dependent
mean free path 
\cite{Rendell1980:PRL,Pappas1991:PRL,Hong2000:PRB}. 
Generally, both elastic and 
inelastic
scattering processes determine the effective spin-dependent mean free path,
sometimes also referred to as the attenuation length.\footnote{For electrons 
with sufficiently high excess energy, a scattering process
(influencing the mean free path) does not necessarily remove the electron
from the collector current. The attenuation length, which can be determined
by measuring the base layer thickness dependence of the collector current
[see 
\cite{Rippard1999:APL,Rippard2000:PRL,Vlutters2001:PRB,vanDijken2002:PRB}]
can therefore differ from the effective mean free path.}
The magnetorestive response is usually expressed using magneto current (MC),
defined as the change in collector current, normalized to the 
minimum value
\begin{equation}
MC=(I_{C\uparrow \uparrow} -I_{C\uparrow \downarrow})/
I_{C\uparrow \downarrow},
\label{eq:mc}
\end{equation}
analogous to the expression for GMR or TMR structures
[recall Eq.~(\ref{eq:tmr})],
where $\uparrow \uparrow$ (parallel) and $\uparrow \downarrow$ 
(antiparallel) denote the relative orientation of the magnetizations.
The large values of MC\footnote{These values substantially 
exceed the CPP GMR value 
for the same magnetic multilayer used in the base.} ($>$200\%) 
and the sensitivity of $\sim$130\% per G measured  at room temperature 
\cite{AnilKumar2000:JMMM} demonstrate a capability for magnetic-field sensors. 
Several important challenges, raised by the operation of
the spin-valve transistor, need to be addressed to better realize 
the potential of hot-electron 
transistors. These challenges 
include increasing the small collector current and determining 
whether the spin injection of hot carriers into semiconductors is 
feasible. Furthermore, it would be desirable to fabricate
structures in which semiconductor regions played an active
role, not limited to energy selection (via Schottky barriers)
of the carriers injected into the base and collector regions.

\begin{figure} 
\centerline{\psfig{file=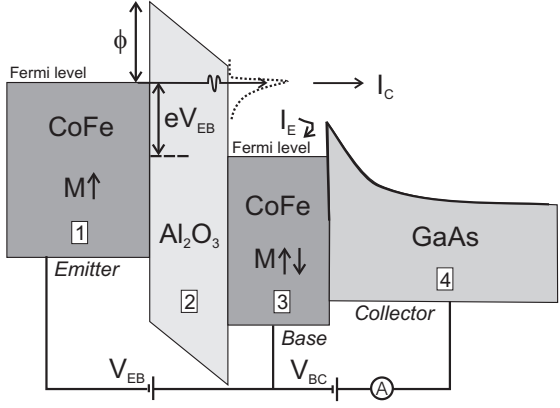,width=0.85\linewidth,angle=0}}
\caption{Schematic energy diagram of a magnetic tunneling transistor.
Region 1 is the emitter, region 2 
the Al$_2$O$_3$ tunnel barrier of height $\phi$, and region 3 the base.
Together they form a magnetic tunnel junction.
Region 4 is a semiconductor collector that has a Schottky barrier
at the interface with the base. 
From \onlinecite{vanDijken2003:PRL}. 
}
\label{mtt}
\end{figure}

An alternative class of hot-electron transistors, often referred to as 
magnetic 
tunneling transistors, has a tunneling junction instead of a Schottky 
barrier 
emitter
\cite{Mizushima1997:IEEETM,Yamauchi1998:PRB,Sato2001:APL,vanDijken2003:PRL,%
vanDijken2002:PRB,Jiang2003:PRL,vanDijken2002:APL}. The
addition of a tunnel junction, combined with a variable $V_{EB}$, allows
exploration of   
hot-electron transport over an energy range
of several eV. At large $V_{EB}$ bias, the ratio $I_C/I_E$, important 
for the device performance, can be substantially increased 
over that of
the 
spin-valve transistor 
\cite{vanDijken2003:APLa,vanDijken2003:APLb,Sato2001:APL}.

A particular realization is depicted in Fig.~\ref{mtt}. Different
coercive fields in the regions 1 and 3 ensure independent switching of
of the corresponding magnetizations in the applied
magnetic field. The magnetocurrent MC, defined in Eq.~(\ref{eq:mc}),
shows a nonmonotonic behavior with $V_{EB}$ \cite{vanDijken2003:PRL}
influenced by the conduction-band structure of a collector. 
In GaAs, in addition to the direct conduction band minimum
at the $\Gamma$ point [recall Fig.~\ref{oo:1} (a)], there
are indirect minima at $L$ points at higher energy \cite{Blakemore1982:JAP}.
After an initial decrease of MC with electron energy, 
at $V_{EB} \approx 0.3$ V 
larger than the base/collector Schottky barrier
there is an onset of hot-electron transport 
into $L$ valleys 
accompanied by an increase in MC \cite{vanDijken2003:PRL}. 

A large magnetocurrent alone, measured in various hot-electron 
spin transistors 
\cite{Monsma1995:PRL,Monsma1998:S,Sato2001:APL,vanDijken2003:APLb},
is not sufficient to demonstrate spin injection in a semiconductor 
collector. 
For conclusive evidence spin detection in a collector region is needed. 
This was first achieved \cite{Jiang2003:PRL} 
using optical detection with a spin LED 
structure\footnote{Analogous to the spin LED from Fig.~\ref{injexp:3},
in which GaAs collector served as an $n$-type spin aligner and InGaAs/GaAs
was used for a quantum well.}
added to the collector in Fig.~\ref{mtt}.
Measurements at $T=1.4$ K and $B=2.5$ T, after a background subtraction, 
showed majority spin injection  with $P_{\rm circ} \approx 10$ \%.

In another realization of a magnetic tunnel transistor, more similar
to the original spin-valve transistor, the emitter was nonmagnetic (Cu) 
while the
base was a magnetic multilayer (F1/N/F2) \cite{vanDijken2003:APLb}.
The resulting strong spin-filtering effect can be inferred  from the 
transmitted hot carriers with a
spin-dependent exponential decay  within the F$_i$, $i=1,2$ layer.
Unpolarized electrons, injected from the emitter, 
after passing an F1 layer of thickness $t$ acquire an effective
transmitted polarization 
\begin{equation}
P_{N1}=\frac{N_\uparrow-N_\downarrow}
{N_\uparrow+N_\downarrow}=
\frac{e^{-t/l_\uparrow}-e^{-t/l_\downarrow}}
{e^{-t/l_\uparrow}+e^{-t/l_\downarrow}},
\end{equation}
where 
$N_\uparrow$ and $N_\downarrow$ 
represent the number of transmitted electrons
with majority or minority spin and $l_\uparrow$ and $l_\downarrow$
are the corresponding attenuation length (the polarization $P_{N2}$ has
an analogous form). 
The resulting magnetocurrent can be
expressed as \cite{vanDijken2003:APLb}
\begin{equation}
MC=2 P_{N1}P_{N2}/(1-P_{N1} P_{N2}),
\label{eq:mc2}
\end{equation}
which is analogous to Eq.~(\ref{eq:julliere}) 
for TMR using Julli{\`{e}}re's model,
but with the redefined definition of spin polarization. 
At $V_{EB}=0.8$ V and at $T=77$ K, the measured
MC exceeds $3400$ \%, while  with Eq.~(\ref{eq:mc2}) the
polarization of the transmitted electrons can be estimated to exceed 90 \%,
even with a ferromagnet only a $\sim 3$ nm thick \cite{vanDijken2003:APLb}.
A theoretical analysis of spin injection and spin filtering in
magnetic tunneling transistors was given by \textcite{Rashba2003:P}
who extended the approach for ballistic spin injection 
\cite{Kravchenko2003:PRB}
(Sec.~\ref{sec:IIC2}) to include the effects of hot-electron transport
and inelastic scattering.

Future studies of hot-electron 
spin transistors are expected to
result in increased spin injection even at room temperatures and
to utilize other semiconductor collectors. It would be particularly 
desirable to demonstrate hot-electron 
spin injection in Si
and facilitate an integration with the CMOS technology.

\subsection{\label{sec:IVF} Spin qubits in semiconductor nanostructures}

A potentially revolutionary idea in spintronics is the possibility
of using the two-level nature of 
electron spin to create a solid-state quantum computer 
\cite{DiVincenzo1995:S,DasSarma2001:SSC,Nielsen:2000}.  
The basic unit in a quantum computer is the quantum bit (or qubit), 
the quantum
analog of the binary bit in a classical digital computer.
A qubit is essentially
a controllable quantum two-level system 
\cite{DasSarma2001:AS,Nielsen:2000}.  
While the dimensionality (2$^n$) of the Hilbert space of $n$ electron spins
is the same as the number of  configurations of a corresponding classical
system, a quantum system can be
in a superposition of all the basis states, effectively performing 
(via a unitary evolution) many classical computations in parallel.  
Several spin-based quantum computer
schemes have been proposed and extensively studied\footnote{See, for example,  
\cite{Loss1998:PRA,Privman1998:PL,Kane1998:N,Hu2000:PRA,Koiller2002:PRL,%
Koiller2003:PRL,Friesen2003:PRB,Piermarocchi2002:PRL,DiVincenzo2000:FP,%
Kane2000:FP,Hu2001:PRA,Burkard1999:PRB,Troiani2003:PRL,%
Skinner2003:PRL,Levy2002:PRL,Meier2003:PRL,Vrijen2000:PRA}.}  
A common theme in these proposals is the idea of
manipulating the dynamics of a single (or a few) electron spin(s) in
semiconductor nanostructures (e.g., quantum dots), with the reasonable
hope that the predicted behavior will extend to many-spin systems,
requisite for practical quantum computation. 

\begin{figure}
\centerline{\psfig{file=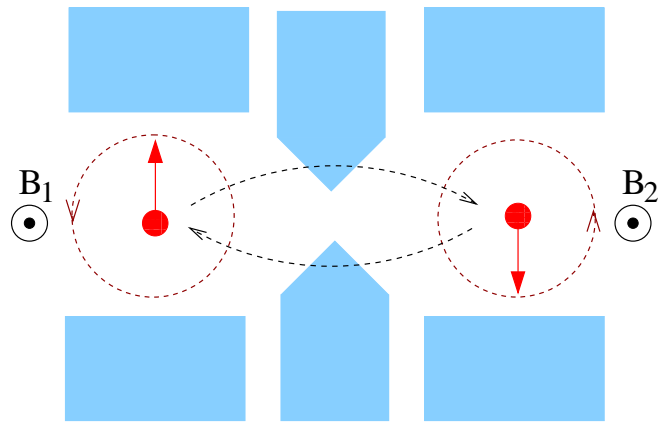,width=1\linewidth,angle=0}}
\caption{The \textcite{Loss1998:PRA} proposal for spin-based solid-state 
quantum computing. Electrons are localized in
electrostatically defined quantum dots, with coupling between 
electron spins---via the exchange interaction---allowed by tunneling 
between the dots. The tunneling is controlled by gate
voltage. The figure shows two electrons localized in the regions defined
by the gates (shaded). Single-qubit
operations are realized by single-spin precessions (circles), 
performed by applying 
local magnetic fields (here perpendicular to the page) to each dot. 
Two-qubit operations are done 
through the exchange interaction indicated by the dashed curves. 
The scheme works according to the time-dependent Hamiltonian 
$H(t)=\sum_{i,j}' J_{i,j}(t) {\bf S}_i\cdot {\bf S}_j 
+\mu_B g \sum {\bf B}_i(t) {\bf S}_i$, 
where the first summation, over all neighboring spin pairs, 
describes the local
exchange interaction ($J$ is the exchange coupling), 
while the second describes
the single spin operations by local magnetic fields. Variations to 
this scheme are described by \textcite{Burkard2000:FP}.  
}
\label{fig:QC}
\end{figure}

The control of spin dynamics and entanglement [many-spin quantum 
correlations \cite{Nielsen:2000}]
at the single-spin level in a semiconductor quantum dot structure is a 
formidable task,
which has not been achieved even at mK temperatures, 
although impressive experimental advances have recently been made
\cite{Fujisawa2002:N,Elzerman2003:PRB,Hanson2003:lanl}.
The current architectures for spin-based quantum computing 
employ GaAs quantum dots \cite{Loss1998:PRA} or
Si (or Si-Ge) systems \cite{Kane1998:N}, with different 
variations. 
The basic idea (see Fig.~\ref{fig:QC}) is to manipulate
the spin states of a single electron using external magnetic fields (or
microwaves) for single-qubit operations and to utilize the quantum exchange
coupling between two neighboring electrons to carry out two-qubit
operations.  

State-of-the-art 
techniques, to measure a single spin in a solid,
such as magnetic resonance force microscopy 
\cite{Sidles1995:RMP,Mamin2003:P,Barbic2002:JAP} 
or the spin-selective single-electron-transistor spectroscopy
\cite{Ono2002:S}, are 
still not sensitive enough 
for quantum computing operations.  
However, recently a single shot readout of the spin of an individual electron
has been demonstrated using an
electrical pump-probe measurement \cite{Kouwenhoven2004:PC}.
A single electron with an unknown spin was trapped in a quantum dot
for a few miliseconds. At the end of the trapping time the spin was
measured by quickly shifting the Zeeman resolved spin states towards
the Fermi energy. A spin to charge conversion allowed for an
electrical readout of the spin.

The real motivation for using the spin
as a qubit is its long coherence time, microseconds or longer
in operational experimental conditions \cite{Sousa2003:PRB}, to be
contrasted with typical picosecond coherence times for charge or orbital
states in solids. 
Interest in spin-based quantum computing
will only increase as we understand
more about spin coherence and relaxation from other spintronic studies. 
The broad
subject of spin-based quantum computing, 
which is related to the areas of quantum measurement and quantum 
decoherence \cite{Zurek2003:RMP}
is beyond the scope of 
this review.

\section{\label{sec:V} Outlook}

We have reviewed selected topics on spintronics, emphasizing both the
fundamental aspects of spin dynamics, transport,
and relaxation, and the potential applications.
While the current push for spintronics
is driven by the prospect of technological applications, the fundamental
spin physics, which has a longstanding tradition in the solid-state
community, is by itself exciting and worth pursuing. Furthermore,
even though many proposed spintronic device schemes may turn out
to be impractical in the end, their importance lies in stimulating
interesting experimental and theoretical research.

There are many challenges and open questions to be tackled by future
research, in particular a robust
spin injection into silicon.\footnote{Small signals attributed to spin 
injection have already been reported \cite{Jia1996:IEEE}.}
While GaAs is of great technological
importance, the control of spin in silicon would raise hopes for
seamless integration of spintronics with the current
information technology. In addition, the small magnitude of the
spin-orbit interaction and the absence of inversion symmetry
lead to relatively long room-temperature spin lifetimes (of about
10 ns; see Sec.~\ref{sec:IIID1}), relaxing some constraints on the
operational length and time scales. Important materials advances
have been realized in improving the compatibility of Si/III-V
structures \cite{Sieg1998:APL}
suggesting a possibility that the existing control
of spin in GaAs or in III-V ferromagnetic semiconductors might be
extended to Si.

Future progress in spin-polarized transport will be largely driven
by the materials advances. In the context of semiconductors, considering
all-semiconductor structures rather than the hybrid structures with
metallic ferromagnets will depend on the improvements in ferromagnetic
semiconductors, for example, whether they 
can achieve higher mobility, 
higher Curie temperature,\footnote{There still remain many challenges in
accurately predicting Curie temperature. First principles results 
suggest that dominant models of ferromagnetism in semiconductors
cannot be used to explain a variation of Curie temperature across
different materials \cite{Erwin2004:NM}. For reviews of ferromagnetic
semiconductor theories outside the scope of this article, see,
for example, \textcite{Nagaev:1983}; \textcite{Bhat2002:JS};
\textcite{Dietl2002:SST}; \textcite{Sanvito2002:JS}, 
\textcite{DasSarma2003:PRB}; \textcite{Koning:2003}; 
\textcite{Timm2003:JPCM}.}
and a simple fabrication of high quality
interfaces with nonmagnetic materials. What is missing, even in the
currently available materials, is a systematic understanding of the
effects of magnetic interfaces and materials inhomogeneities on 
spin-polarized transport. A comprehensive transport calculation
in the actual devices
with realistic electronic structure of
the studied materials would provide valuable insights into 
both the  spin polarization being measured and
how it is reduced from the moment it was generated.

Spin relaxation and spin dephasing of conduction electrons
is a rather field, with the basic principles well understood.
What is needed are accurate band-structure-derived calculations
of spin relaxation times, in both metals and semiconductors.
The same can be said of
$g$ factor, calculation of which from first
principles is a nontrivial task that has not been accomplished even
for the elemental metals. An important and still debated issue
is spin relaxation and decoherence of localized or
confined electrons, where the hyperfine-interaction mechanism
dominates. Furthermore, single-spin relaxation and
decoherence, and their relation to the ensemble spin dephasing,
need to be 
pursued further 
in the context of quantum computing. 
A first step towards understanding
single-spin relaxation  is the recent experiment of
\textcite{Hanson2003:lanl} in a one-electron quantum dot.

While dynamic nuclear polarization induced by electron spin can often
be a nuisance for detecting intrinsic spin dynamics 
(see Sec.~\ref{sec:IIID3}),
the interaction between electron and nuclear spins 
\cite{Paget:1984,Fleisher:1984,Vagner2003:P,Smet2002:N} 
is of fundamental
importance for spintronics. An NMR of the nuclear spin polarized
by spin-polarized photoexcited electrons has already been used to detect
the nonequilibrium electron spin in Si \cite{Lampel1968:PRL}. On the
other hand, an NMR signal can be detected optically through measuring
changes in the circular polarization of photoluminescence
due to resonant variations of the nuclear field 
\cite{Dyakonov:1984}, as shown first in
p-doped Ga$_{0.7}$Al$_{0.3}$As \cite{Ekimov1972:JETPL}. 
The early work of \textcite{Lampel1968:PRL}, 
and \textcite{Ekimov1972:JETPL} established the basic
principles for a series of experiments that demonstrated various 
realizations of an all-optical NMR. The role of the resonant 
radio waves is played by periodically optically excited electron spins
\cite{Salis2001:PRL,Kalevich1986:FTT,Kalevich1981:FTT,Kalevich1980:FTT,%
Kikkawa2000:S,Eickhoff2002:PRB,Fleisher:1984b}.
Electron-nuclear spintronics is likely to become relevant for
quantum computation and for few-spin manipulations, which can benefit
from long nuclear spin coherence times (even lasting minutes).

The range of potential spintronic applications 
goes beyond the use of
large magnetoresistive
effects.
\textcite{Rudolph2003:APL}, for example, have demonstrated the operation of
a spin laser. The laser is a vertical-cavity surface-emitting laser 
(VCSEL), optically pumped in the gain medium, here two InGaAs quantum
wells, with 50\% spin-polarized electrons. The electrons recombine with 
heavy holes, which are effectively unpolarized, emitting circularly
polarized light (see Sec.~\ref{sec:IIB}). The threshold electrical current, 
extracted from the pump power for the lasing operation, was found to be 
0.5 A$\cdot$cm$^{-2}$,
which is 23\% below the threshold current of the spin-unpolarized VCSEL.
Furthermore, for a fixed pump power, the emission power of the laser
changed by 400\% upon changing the degree of circular polarization of
the pump laser. The reason for the decrease in threshold is the selective
coupling of spin-polarized electrons to photons with one helicity.
While the experiment was conducted at 6 K, a room-temperature
operation and an electrically pumped laser should be viable as 
well.\footnote{The requirement is that the spin relaxation time be longer 
than
the carrier recombination time, and that the spin injection spot and the
gain medium be within the spin transport length.}

\begin{figure}
\centerline{\psfig{file=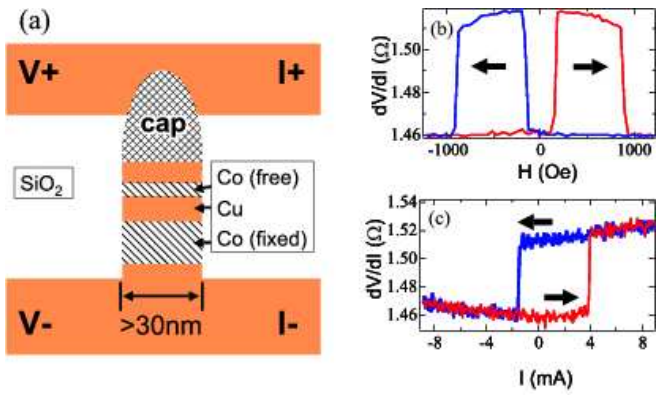,width=1.0\linewidth}}
\caption{(a) Nanopilar device: schematic diagram of
a nanopilar device
operating at room temperature. 
The direction of magnetization
is fixed (pinned) in the thick Co film and free in the thin Co film;
(b) differential resistance $dV/dI$ of a nanopilar device as
a function of applied field; (c) $dV/dI$ of the device as a
function of applied current. 
The arrows in panels (b) and (c) represent the direction of
magnetic field and current sweeps, respectively. 
For positive current, 
electrons flow from the thin to the thick Co film.
Adapted from \onlinecite{Albert2002:PRL}.} 
\label{ralph}
\end{figure}

The demonstration that the flow of spin-polarized carriers, rather
than applied magnetic field, can also be used to manipulate
magnetization of ferromagnetic materials brings the exciting
prospect of a novel class of spintronic devices. In addition
to reversal of magnetization, which is a key element in realizing
various
magnetoresitive applications, the driving of a spin-polarized current
can lead to coherent 
microwave oscillations in nanomagnets \cite{Kiselev2003:P}. 
Spin-transfer
torque (Sec.~\ref{sec:IB1}) 
has already been realized in several 
experimental geometries. These include nanowires 
\cite{Wegrowe1999:EL,Kelly2003:PRB}, 
point contacts 
\cite{Tsoi1998:PRLa,Yi2003:PRL,Tsoi2000:N,Tsoi2002:PRL}, 
nanoconstrictions \cite{Myers1999:S,Rippard2003:P}, 
and nanopilars \cite{Katine2000:PRL,Albert2002:PRL,Urazhdin2003:PRL} 
(see Fig.~\ref{ralph}), all involving metallic ferromagnets.
The common feature of all these geometries is a need for very 
large current densities ($\sim10^7$ Acm$^{-2}$).  
Ongoing
experiments \cite{Munekata2003:PC,Chiba2004:P,Yamanouchi2004:N}
to demonstrate 
spin-transfer torque 
(together with other cooperative phenomena) 
in ferromagnetic semiconductors, which have
much smaller magnetization than their metallic counterparts, are expected
to also require much smaller switching currents. Based on the findings
in electric-field controlled ferromagnetism (see Fig.~\ref{ohno}),
it has been demonstrated that the reversal of magnetization in (In,Mn)As can
be manipulated by modifying the carrier density, using a gate voltage
in a FET structure \cite{Chiba2003:S}.

\acknowledgments{We thank R. H. Buhrman, K. Bussmann, R. de Sousa, 
M. I. D'yakonov,  
S. C. Erwin, M. E. Fisher, A. M. Goldman, K. Halterman, X. Hu, S. V. Iordanskii,
M. Johnson, B. T. Jonker, K. Kavokin, J. M. Kikkawa,  
S. Maekawa, C. M. Marcus, I. I. Mazin, B. D. McCombe, J. S. Moodera, 
H. Munekata, 
B. E. Nadgorny, H. Ohno, 
S. S. P. Parkin,  
D. C. Ralph, E. I. Rashba, W. H. Rippard, V. I. Safarov, G. Schmidt, 
E. Sherman, 
R. H. Silsbee, D. D. Smith, 
M. D. Stiles, O. T. Valls, S. van Dijken, 
H. M. van Driel, T. Venkatesan, S. von Moln\'{a}r, and J. Y. T. Wei 
for useful discussions.  
We thank D. D. Awschalom, N. C. Emley, R. T. Harley, M. Johnson, B. T. Jonker, 
K. Kavokin, J. M. Kikkawa, C. M. Marcus, H. Munekata, Y. Ohno, A. Oiwa, 
S. S. P. Parkin, R. J. Soulen, Jr.,  and M. Tanaka for providing us with 
illustrative
figures from their published works, and A. Kaminski and D. J. Priour
for help with preparation of the manuscript. This work was supported by 
DARPA, the US ONR, and the NSF-ECS.} 

\begin{table}
\begin{tabular}{ll}
\hline
\hline
2DEG  & two dimensional electron gas \\
BAP   & Bir-Aronov-Pikus \\
BCS   & Bardeen Cooper Schrieffer \\
BIA   & bulk inversion asymmetry \\
BTK   & Blonder-Tinkham-Klapwijk \\
CESR  & conduction electron spin resonance \\
CIP   & current in plane \\
CMOS  & complementary metal oxide semiconductor \\
CMR   & colossal magnetoresistance \\
CPP   & current perpendicular to plane \\
DNA   & deoxyribonucleic acid \\
DP    & D'yakonov-Perel' \\
EDSR  & electron dipole spin resonance \\
EMF   & electromotive force \\
ESR   & electron spin resonance \\
EY    & Elliot-Yafet \\
F     & ferromagnet \\
FSm   & ferromagnetic semiconductor \\
GMR   & giant magnetoresistance \\
FET   & field effect transistor  \\
HFI   & hyperfine interaction \\
I     & insulator \\
HTSC  & high temperature superconductor \\
LED   & light emitting diode \\
MBD   & magnetic bipolar diode \\
MBE   & molecular beam epitaxy \\
MBT   & magnetic bipolar transistor \\
MC    & magnetocurrent \\
MR    & magnetoresistance \\
MRAM  & magnetic random access memory \\
MTJ   & magnetic tunnel junction \\
N     & normal (paramagnetic) metal \\
NMR   & nuclear magnetic resonance \\
QD    & quantum dot \\
QPC   & quantum point contact \\
QW    & quantum well \\
S     & superconductor \\
SET   & single electron transistor  \\
SFET  & spin field effect transistor  \\
SIA   & structure inversion asymmetry \\
Sm    & semiconductor \\
SQUID & superconducting interference quantum device \\
STM   & scanning tunneling microscope \\
TESR  & transmission electron spin resonance \\
\hline
\hline
\end{tabular}
\caption{List of acronyms used in the text.}
\label{tab:2}
\end{table}

\bibliographystyle{apsrmp}
\bibliography{references}

\begin{thebibliography}{916}
\expandafter\ifx\csname natexlab\endcsname\relax\def\natexlab#1{#1}\fi
\expandafter\ifx\csname bibnamefont\endcsname\relax
  \def\bibnamefont#1{#1}\fi
\expandafter\ifx\csname bibfnamefont\endcsname\relax
  \def\bibfnamefont#1{#1}\fi
\expandafter\ifx\csname citenamefont\endcsname\relax
  \def\citenamefont#1{#1}\fi
\expandafter\ifx\csname url\endcsname\relax
  \def\url#1{\texttt{#1}}\fi
\expandafter\ifx\csname urlprefix\endcsname\relax\def\urlprefix{URL }\fi
\providecommand{\bibinfo}[2]{#2}
\providecommand{\eprint}[2][]{\url{#2}}

\bibitem[{\citenamefont{Abragam}(1961)}]{Abragam:1961}
\bibinfo{author}{\bibnamefont{Abragam}, \bibfnamefont{A.}},
  \bibinfo{year}{1961}, \emph{\bibinfo{title}{The Principles of Nucelar
  Magnetism}} (\bibinfo{publisher}{Oxford University Press, London}).

\bibitem[{\citenamefont{Abrikosov and Gorkov}(1960)}]{Abrikosov1960:ZETP}
\bibinfo{author}{\bibnamefont{Abrikosov}, \bibfnamefont{A.~A.}}, and
  \bibinfo{author}{\bibfnamefont{L.~P.} \bibnamefont{Gorkov}},
  \bibinfo{year}{1960}, {``}\bibinfo{title}{Contribution to the theory of
  superconducting alloys with paramagnetic impurities},{''}
  \bibinfo{journal}{Zh. Eksp. Teor. Fiz.} \textbf{\bibinfo{volume}{39}},
  \bibinfo{pages}{1781--1796} \bibinfo{note}{[Sov. Phys. JETP {\bf 12},
  1243-1253 (1961)]}.

\bibitem[{\citenamefont{Adachi} \emph{et~al.}(2001)\citenamefont{Adachi, Ohno,
  Matsukura, and Ohno}}]{Adachi2001:PE}
\bibinfo{author}{\bibnamefont{Adachi}, \bibfnamefont{T.}},
  \bibinfo{author}{\bibfnamefont{Y.}~\bibnamefont{Ohno}},
  \bibinfo{author}{\bibfnamefont{F.}~\bibnamefont{Matsukura}}, and
  \bibinfo{author}{\bibfnamefont{H.}~\bibnamefont{Ohno}}, \bibinfo{year}{2001},
  {``}\bibinfo{title}{Spin relaxation in n-modulation doped {GaAs/AlGaAs}(110)
  quantum wells},{''} \bibinfo{journal}{Physica E}
  \textbf{\bibinfo{volume}{10}},  \bibinfo{pages}{36--39}.

\bibitem[{\citenamefont{Affleck} \emph{et~al.}(2000)\citenamefont{Affleck,
  Caux, and Zagoskin}}]{Affleck2000:PRB}
\bibinfo{author}{\bibnamefont{Affleck}, \bibfnamefont{I.}},
  \bibinfo{author}{\bibfnamefont{J.-S.} \bibnamefont{Caux}}, and
  \bibinfo{author}{\bibfnamefont{A.~M.} \bibnamefont{Zagoskin}},
  \bibinfo{year}{2000}, {``}\bibinfo{title}{Andreev scattering and {Josephson}
  current in a one-dimensional electron liquid},{''} \bibinfo{journal}{Phys.
  Rev. B} \textbf{\bibinfo{volume}{62}},  \bibinfo{pages}{1433--1445}.

\bibitem[{\citenamefont{Akinaga}(2002)}]{Akinaga2002:SST}
\bibinfo{author}{\bibnamefont{Akinaga}, \bibfnamefont{H.}},
  \bibinfo{year}{2002}, {``}\bibinfo{title}{Magnetoresistive switch effect in
  metal/semiconductor hybrid granular films: extremely huge magnetoresistance
  effect at room temperature},{''} \bibinfo{journal}{Semicond. Sci. Technol.}
  \textbf{\bibinfo{volume}{17}},  \bibinfo{pages}{322--326}.

\bibitem[{\citenamefont{Albert} \emph{et~al.}(2002)\citenamefont{Albert, Emley,
  Myers, Ralph, and Buhrman}}]{Albert2002:PRL}
\bibinfo{author}{\bibnamefont{Albert}, \bibfnamefont{F.~J.}},
  \bibinfo{author}{\bibfnamefont{N.~C.} \bibnamefont{Emley}},
  \bibinfo{author}{\bibfnamefont{E.~B.} \bibnamefont{Myers}},
  \bibinfo{author}{\bibfnamefont{D.~C.} \bibnamefont{Ralph}}, and
  \bibinfo{author}{\bibfnamefont{R.~A.} \bibnamefont{Buhrman}},
  \bibinfo{year}{2002}, {``}\bibinfo{title}{Quantitative study of magnetization
  reversal by spin-polarized current in magnetic multilayer nanopillars},{''}
  \bibinfo{journal}{Phys. Rev. Lett.} \textbf{\bibinfo{volume}{89}},
  \bibinfo{pages}{226802}.

\bibitem[{\citenamefont{Albrecht and Smith}(2002)}]{Albrecht2002:PRB}
\bibinfo{author}{\bibnamefont{Albrecht}, \bibfnamefont{J.~D.}}, and
  \bibinfo{author}{\bibfnamefont{D.~L.} \bibnamefont{Smith}},
  \bibinfo{year}{2002}, {``}\bibinfo{title}{Electron spin injection at a
  {Schottky} contact},{''} \bibinfo{journal}{Phys. Rev. B}
  \textbf{\bibinfo{volume}{66}},  \bibinfo{pages}{113303}.

\bibitem[{\citenamefont{Albrecht and Smith}(2003)}]{Albrecht2003:PRB}
\bibinfo{author}{\bibnamefont{Albrecht}, \bibfnamefont{J.~D.}}, and
  \bibinfo{author}{\bibfnamefont{D.~L.} \bibnamefont{Smith}},
  \bibinfo{year}{2003}, {``}\bibinfo{title}{Spin-polarized electron transport
  at ferromagnet/semiconductor {Schottky} contacts},{''}
  \bibinfo{journal}{Phys. Rev. B} \textbf{\bibinfo{volume}{68}},
  \bibinfo{pages}{035340}.

\bibitem[{\citenamefont{Aleiner and {Lyanda-Geller}}(1991)}]{Aleiner1991:JETPL}
\bibinfo{author}{\bibnamefont{Aleiner}, \bibfnamefont{I.~L.}}, and
  \bibinfo{author}{\bibfnamefont{Y.~B.} \bibnamefont{{Lyanda-Geller}}},
  \bibinfo{year}{1991}, {``}\bibinfo{title}{Conductance and electron spin
  polarization in resonant tunneling},{''} \bibinfo{journal}{Zh. Eksp. Teor.
  Fiz. Pisma Red.} \textbf{\bibinfo{volume}{53}},  \bibinfo{pages}{89--92}
  \bibinfo{note}{[JETP Lett. {\bf 53}, 91-95 (1991)]}.

\bibitem[{\citenamefont{Alvarado}(1995)}]{Alvarado1995:PRL}
\bibinfo{author}{\bibnamefont{Alvarado}, \bibfnamefont{S.~F.}},
  \bibinfo{year}{1995}, {``}\bibinfo{title}{Tunneling potential barrier
  dependence of electron spin polarization},{''} \bibinfo{journal}{Phys. Rev.
  Lett.} \textbf{\bibinfo{volume}{75}},  \bibinfo{pages}{513--516}.

\bibitem[{\citenamefont{Alvarado and Renaud}(1992)}]{Alvarado1992:PRL}
\bibinfo{author}{\bibnamefont{Alvarado}, \bibfnamefont{S.~F.}}, and
  \bibinfo{author}{\bibfnamefont{P.}~\bibnamefont{Renaud}},
  \bibinfo{year}{1992}, {``}\bibinfo{title}{Observation of
  spin-polarized-electron tunneling from a ferromagnet into {GaAs}},{''}
  \bibinfo{journal}{Phys. Rev. Lett.} \textbf{\bibinfo{volume}{68}},
  \bibinfo{pages}{1387--1390}.

\bibitem[{\citenamefont{Anderson}(1966)}]{Anderson1966:PRL}
\bibinfo{author}{\bibnamefont{Anderson}, \bibfnamefont{P.~W.}},
  \bibinfo{year}{1966}, {``}\bibinfo{title}{Localized magnetic states and
  {Fermi}-surface anomalies in tunneling},{''} \bibinfo{journal}{Phys. Rev.
  Lett.} \textbf{\bibinfo{volume}{17}},  \bibinfo{pages}{95--97}.

\bibitem[{\citenamefont{Ando} \emph{et~al.}(1982)\citenamefont{Ando, Fowler,
  and Stern}}]{Ando1982:RMP}
\bibinfo{author}{\bibnamefont{Ando}, \bibfnamefont{T.}},
  \bibinfo{author}{\bibfnamefont{A.~B.} \bibnamefont{Fowler}}, and
  \bibinfo{author}{\bibfnamefont{F.}~\bibnamefont{Stern}},
  \bibinfo{year}{1982}, {``}\bibinfo{title}{Electronic properties of
  two-dimensional systems},{''} \bibinfo{journal}{Rev. Mod. Phys.}
  \textbf{\bibinfo{volume}{54}},  \bibinfo{pages}{437--672}.

\bibitem[{\citenamefont{Andreev}(1964)}]{Andreev1964:SPJETP}
\bibinfo{author}{\bibnamefont{Andreev}, \bibfnamefont{A.~F.}},
  \bibinfo{year}{1964}, {``}\bibinfo{title}{The thermal conductivity of the
  intermediate state in superconductors},{''} \bibinfo{journal}{Zh. Eksp. Teor.
  Fiz.} \textbf{\bibinfo{volume}{46}},  \bibinfo{pages}{1823--1825}
  \bibinfo{note}{[Sov. Phys. JETP {\bf 19}, 1228-1231 (1964)]}.

\bibitem[{\citenamefont{Andreev and Gerasimenko}(1958)}]{Andreev1958:JETP}
\bibinfo{author}{\bibnamefont{Andreev}, \bibfnamefont{V.~V.}}, and
  \bibinfo{author}{\bibfnamefont{V.~I.} \bibnamefont{Gerasimenko}},
  \bibinfo{year}{1958}, {``}\bibinfo{title}{On the theory of paramagnetic
  resonance and paramagnetic relaxation in metals},{''} \bibinfo{journal}{Zh.
  Exp. Teor. Fiz.} \textbf{\bibinfo{volume}{35}},  \bibinfo{pages}{1209--1215}
  \bibinfo{note}{[Sov. Phys. JETP {\bf 35}, 846-850 (1959)]}.

\bibitem[{\citenamefont{{Anil Kumar}} \emph{et~al.}(2000)\citenamefont{{Anil
  Kumar}, Jansen, {van `t Erve}, Vlutters, {de Haan}, and
  Lodder}}]{AnilKumar2000:JMMM}
\bibinfo{author}{\bibnamefont{{Anil Kumar}}, \bibfnamefont{P.~S.}},
  \bibinfo{author}{\bibfnamefont{R.}~\bibnamefont{Jansen}},
  \bibinfo{author}{\bibfnamefont{O.~M.~J.} \bibnamefont{{van `t Erve}}},
  \bibinfo{author}{\bibfnamefont{R.}~\bibnamefont{Vlutters}},
  \bibinfo{author}{\bibfnamefont{P.}~\bibnamefont{{de Haan}}}, and
  \bibinfo{author}{\bibfnamefont{J.~C.} \bibnamefont{Lodder}},
  \bibinfo{year}{2000}, {``}\bibinfo{title}{Low-field magnetocurrent above
  200\% in a spin-valve transistor at room temperature},{''}
  \bibinfo{journal}{J. Magn. Magn. Mater.} \textbf{\bibinfo{volume}{214}},
  \bibinfo{pages}{L1--L6}.

\bibitem[{\citenamefont{Ansermet}(1998)}]{Ansermet1998:JPCM}
\bibinfo{author}{\bibnamefont{Ansermet}, \bibfnamefont{J.-P.}},
  \bibinfo{year}{1998}, {``}\bibinfo{title}{Perpendicular transport of
  spin-polarized electrons through magnetic nanostructures},{''}
  \bibinfo{journal}{J. Phys.: Condens. Matter} \textbf{\bibinfo{volume}{10}},
  \bibinfo{pages}{6027--6050}.

\bibitem[{\citenamefont{Apinyan and M{\'e}lin}(2002)}]{Apinyan2002:EPJB}
\bibinfo{author}{\bibnamefont{Apinyan}, \bibfnamefont{V.}}, and
  \bibinfo{author}{\bibfnamefont{R.}~\bibnamefont{M{\'e}lin}},
  \bibinfo{year}{2002}, {``}\bibinfo{title}{Microscopic theory of non local
  pair correlations in metallic {F/S/F} trilayers},{''} \bibinfo{journal}{Eur.
  Phys. J. B} \textbf{\bibinfo{volume}{25}},  \bibinfo{pages}{373--389}.

\bibitem[{\citenamefont{Appelbaum}(1966)}]{Appelbaum1966:PRL}
\bibinfo{author}{\bibnamefont{Appelbaum}, \bibfnamefont{J.}},
  \bibinfo{year}{1966}, {``}\bibinfo{title}{`s-d' exchange model of zero-bias
  tunneling anomalies},{''} \bibinfo{journal}{Phys. Rev. Lett.}
  \textbf{\bibinfo{volume}{17}},  \bibinfo{pages}{91--95}.

\bibitem[{\citenamefont{Arata} \emph{et~al.}(2001)\citenamefont{Arata, Ohno,
  Matsukura, and Ohno}}]{Arata2001:PE}
\bibinfo{author}{\bibnamefont{Arata}, \bibfnamefont{I.}},
  \bibinfo{author}{\bibfnamefont{Y.}~\bibnamefont{Ohno}},
  \bibinfo{author}{\bibfnamefont{F.}~\bibnamefont{Matsukura}}, and
  \bibinfo{author}{\bibfnamefont{H.}~\bibnamefont{Ohno}}, \bibinfo{year}{2001},
  {``}\bibinfo{title}{Temperature dependence of electroluminescence and {I-V}
  characteristics of ferromagnetic/non-magnetic semiconductor pn
  junctions},{''} \bibinfo{journal}{Physica E} \textbf{\bibinfo{volume}{10}},
  \bibinfo{pages}{288--291}.

\bibitem[{\citenamefont{Aronov}(1976{\natexlab{a}})}]{Aronov1976:SPJETP}
\bibinfo{author}{\bibnamefont{Aronov}, \bibfnamefont{A.~G.}},
  \bibinfo{year}{1976}{\natexlab{a}}, {``}\bibinfo{title}{Spin injection and
  polarization of excitations and nuclei in superoncductors},{''}
  \bibinfo{journal}{Zh. Eksp. Teor. Fiz.} \textbf{\bibinfo{volume}{71}},
  \bibinfo{pages}{370--376} \bibinfo{note}{[Sov. Phys. JETP {\bf 44}, 193-196
  (1976)]}.

\bibitem[{\citenamefont{Aronov}(1976{\natexlab{b}})}]{Aronov1976:JETPL}
\bibinfo{author}{\bibnamefont{Aronov}, \bibfnamefont{A.~G.}},
  \bibinfo{year}{1976}{\natexlab{b}}, {``}\bibinfo{title}{Spin injection in
  metals and polarization of nuclei},{''} \bibinfo{journal}{Zh. Eksp. Teor.
  Fiz. Pisma Red.} \textbf{\bibinfo{volume}{24}},  \bibinfo{pages}{37--39}
  \bibinfo{note}{[JETP Lett. {\bf 24}, 32-34 (1976)]}.

\bibitem[{\citenamefont{Aronov and Pikus}(1976)}]{Aronov1976:SPS}
\bibinfo{author}{\bibnamefont{Aronov}, \bibfnamefont{A.~G.}}, and
  \bibinfo{author}{\bibfnamefont{G.~E.} \bibnamefont{Pikus}},
  \bibinfo{year}{1976}, {``}\bibinfo{title}{Spin injection into
  semiconductors},{''} \bibinfo{journal}{Fiz. Tekh. Poluprovodn.}
  \textbf{\bibinfo{volume}{10}},  \bibinfo{pages}{1177--1180}
  \bibinfo{note}{[Sov. Phys. Semicond. {\bf 10}, 698-700 (1976)]}.

\bibitem[{\citenamefont{Aronov} \emph{et~al.}(1983)\citenamefont{Aronov, Pikus,
  and Titkov}}]{Aronov1983:SPJETP}
\bibinfo{author}{\bibnamefont{Aronov}, \bibfnamefont{A.~G.}},
  \bibinfo{author}{\bibfnamefont{G.~E.} \bibnamefont{Pikus}}, and
  \bibinfo{author}{\bibfnamefont{A.~N.} \bibnamefont{Titkov}},
  \bibinfo{year}{1983}, {``}\bibinfo{title}{Spin relaxation of conduction
  electrons in p-type {III-V} compounds},{''} \bibinfo{journal}{Zh. Eksp. Teor.
  Fiz.} \textbf{\bibinfo{volume}{84}},  \bibinfo{pages}{1170--1184}
  \bibinfo{note}{[Sov. Phys. JETP {\bf 57}, 680-687 (1983)]}.

\bibitem[{\citenamefont{Ashcroft and Mermin}(1976)}]{Ashcroft:1976}
\bibinfo{author}{\bibnamefont{Ashcroft}, \bibfnamefont{N.~W.}}, and
  \bibinfo{author}{\bibfnamefont{N.~D.} \bibnamefont{Mermin}},
  \bibinfo{year}{1976}, \emph{\bibinfo{title}{Solid State Physics}}
  (\bibinfo{publisher}{Sounders, Philadelphia}).

\bibitem[{\citenamefont{Asnin} \emph{et~al.}(1979)\citenamefont{Asnin, Bakun,
  Danishevskii, Ivchenko, Pikus, and Rogachev}}]{Asnin1979:SSC}
\bibinfo{author}{\bibnamefont{Asnin}, \bibfnamefont{V.~M.}},
  \bibinfo{author}{\bibfnamefont{A.~A.} \bibnamefont{Bakun}},
  \bibinfo{author}{\bibfnamefont{A.~M.} \bibnamefont{Danishevskii}},
  \bibinfo{author}{\bibfnamefont{E.~L.} \bibnamefont{Ivchenko}},
  \bibinfo{author}{\bibfnamefont{G.~E.} \bibnamefont{Pikus}}, and
  \bibinfo{author}{\bibfnamefont{A.~A.} \bibnamefont{Rogachev}},
  \bibinfo{year}{1979}, {``}\bibinfo{title}{Circular photogalvanic effect in
  optically active crystals},{''} \bibinfo{journal}{Solid State Commun.}
  \textbf{\bibinfo{volume}{30}},  \bibinfo{pages}{565--570}.

\bibitem[{\citenamefont{Atanasov} \emph{et~al.}(1996)\citenamefont{Atanasov,
  Hach{\'e}, Hughes, {van Driel}, and Sipe}}]{Atanasov1996:PRL}
\bibinfo{author}{\bibnamefont{Atanasov}, \bibfnamefont{R.}},
  \bibinfo{author}{\bibfnamefont{A.}~\bibnamefont{Hach{\'e}}},
  \bibinfo{author}{\bibfnamefont{J.~L.~P.} \bibnamefont{Hughes}},
  \bibinfo{author}{\bibfnamefont{H.~M.} \bibnamefont{{van Driel}}}, and
  \bibinfo{author}{\bibfnamefont{J.~E.} \bibnamefont{Sipe}},
  \bibinfo{year}{1996}, {``}\bibinfo{title}{Coherent control of photocurrent
  generation in bulk semiconductors},{''} \bibinfo{journal}{Phys. Rev. Lett.}
  \textbf{\bibinfo{volume}{76}},  \bibinfo{pages}{1703--1706}.

\bibitem[{\citenamefont{Auth} \emph{et~al.}(2003)\citenamefont{Auth, Jakob,
  Block, and Felser}}]{Auth2003:PRB}
\bibinfo{author}{\bibnamefont{Auth}, \bibfnamefont{N.}},
  \bibinfo{author}{\bibfnamefont{G.}~\bibnamefont{Jakob}},
  \bibinfo{author}{\bibfnamefont{T.}~\bibnamefont{Block}}, and
  \bibinfo{author}{\bibfnamefont{C.}~\bibnamefont{Felser}},
  \bibinfo{year}{2003}, {``}\bibinfo{title}{Spin polarization of
  magnetoresistive materials by point contact spectroscopy},{''}
  \bibinfo{journal}{Phys. Rev. B} \textbf{\bibinfo{volume}{68}},
  \bibinfo{pages}{024403}.

\bibitem[{\citenamefont{Avdonin} \emph{et~al.}(2003)\citenamefont{Avdonin,
  Dmitrieva, Kuperin, and Sartan}}]{Avdonin2003:P}
\bibinfo{author}{\bibnamefont{Avdonin}, \bibfnamefont{S.~A.}},
  \bibinfo{author}{\bibfnamefont{L.~A.} \bibnamefont{Dmitrieva}},
  \bibinfo{author}{\bibfnamefont{Y.~A.} \bibnamefont{Kuperin}}, and
  \bibinfo{author}{\bibfnamefont{V.~V.} \bibnamefont{Sartan}},
  \bibinfo{year}{2003}, {``}\bibinfo{title}{Spin-dependent transport through
  the finite array of quantum dots: Spin gun},{''} \eprint{cond-mat/0316302}.

\bibitem[{\citenamefont{Averkiev and Golub}(1999)}]{Averkiev1999:PRB}
\bibinfo{author}{\bibnamefont{Averkiev}, \bibfnamefont{N.~S.}}, and
  \bibinfo{author}{\bibfnamefont{L.~E.} \bibnamefont{Golub}},
  \bibinfo{year}{1999}, {``}\bibinfo{title}{Giant spin relaxation anisotropy in
  zinc-blende heterostructures},{''} \bibinfo{journal}{Phys. Rev. B}
  \textbf{\bibinfo{volume}{60}},  \bibinfo{pages}{15582--15584}.

\bibitem[{\citenamefont{Awschalom}(2001)}]{Awschalom2001:PE}
\bibinfo{author}{\bibnamefont{Awschalom}, \bibfnamefont{D.~D.}},
  \bibinfo{year}{2001}, {``}\bibinfo{title}{Manipulating and storing spin
  coherence in semiconductors},{''} \bibinfo{journal}{Physica E}
  \textbf{\bibinfo{volume}{10}},  \bibinfo{pages}{1--6}.

\bibitem[{\citenamefont{Awschalom and Kikkawa}(1999)}]{Awschalom1999:PT}
\bibinfo{author}{\bibnamefont{Awschalom}, \bibfnamefont{D.~D.}}, and
  \bibinfo{author}{\bibfnamefont{J.~M.} \bibnamefont{Kikkawa}},
  \bibinfo{year}{1999}, {``}\bibinfo{title}{Electron spin and optical coherence
  in semiconductors},{''} \bibinfo{journal}{Phys. Today}
  \textbf{\bibinfo{volume}{52 (6)}},  \bibinfo{pages}{33--38}.

\bibitem[{\citenamefont{Awschalom} \emph{et~al.}(1987)\citenamefont{Awschalom,
  Warnock, and {von Moln{\'a}r}}}]{Awschalom1987:PRL}
\bibinfo{author}{\bibnamefont{Awschalom}, \bibfnamefont{D.~D.}},
  \bibinfo{author}{\bibfnamefont{J.}~\bibnamefont{Warnock}}, and
  \bibinfo{author}{\bibfnamefont{S.}~\bibnamefont{{von Moln{\'a}r}}},
  \bibinfo{year}{1987}, {``}\bibinfo{title}{Low-temperature magnetic
  spectroscopy of a dilute magnetic semiconductor},{''} \bibinfo{journal}{Phys.
  Rev. Lett.} \textbf{\bibinfo{volume}{58}},  \bibinfo{pages}{812--815}.

\bibitem[{\citenamefont{Baibich} \emph{et~al.}(1988)\citenamefont{Baibich,
  Broto, Fert, {Nguyen Van Dau}, Petroff, Eitenne, Creuzet, Friederich, and
  Chazelas}}]{Baibich1988:PRL}
\bibinfo{author}{\bibnamefont{Baibich}, \bibfnamefont{M.~N.}},
  \bibinfo{author}{\bibfnamefont{J.~M.} \bibnamefont{Broto}},
  \bibinfo{author}{\bibfnamefont{A.}~\bibnamefont{Fert}},
  \bibinfo{author}{\bibfnamefont{F.}~\bibnamefont{{Nguyen Van Dau}}},
  \bibinfo{author}{\bibfnamefont{F.}~\bibnamefont{Petroff}},
  \bibinfo{author}{\bibfnamefont{P.}~\bibnamefont{Eitenne}},
  \bibinfo{author}{\bibfnamefont{G.}~\bibnamefont{Creuzet}},
  \bibinfo{author}{\bibfnamefont{A.}~\bibnamefont{Friederich}}, and
  \bibinfo{author}{\bibfnamefont{J.}~\bibnamefont{Chazelas}},
  \bibinfo{year}{1988}, {``}\bibinfo{title}{Giant magnetoresistance of
  {(001)Fe/(001)Cr} magnetic superlattices},{''} \bibinfo{journal}{Phys. Rev.
  Lett.} \textbf{\bibinfo{volume}{61}},  \bibinfo{pages}{2472--2475}.

\bibitem[{\citenamefont{Balents and Egger}(2000)}]{Balents2000:PRL}
\bibinfo{author}{\bibnamefont{Balents}, \bibfnamefont{L.}}, and
  \bibinfo{author}{\bibfnamefont{R.}~\bibnamefont{Egger}},
  \bibinfo{year}{2000}, {``}\bibinfo{title}{Spin transport in interacting
  quantum wires and carbon nanotubes},{''} \bibinfo{journal}{Phys. Rev. Lett.}
  \textbf{\bibinfo{volume}{85}},  \bibinfo{pages}{3464--3467}.

\bibitem[{\citenamefont{Balents and Egger}(2001)}]{Balents2001:PRB}
\bibinfo{author}{\bibnamefont{Balents}, \bibfnamefont{L.}}, and
  \bibinfo{author}{\bibfnamefont{R.}~\bibnamefont{Egger}},
  \bibinfo{year}{2001}, {``}\bibinfo{title}{Spin-dependent transport in a
  {Luttinger} liquid},{''} \bibinfo{journal}{Phys. Rev. B}
  \textbf{\bibinfo{volume}{64}},  \bibinfo{pages}{035310}.

\bibitem[{\citenamefont{Barbic}(2002)}]{Barbic2002:JAP}
\bibinfo{author}{\bibnamefont{Barbic}, \bibfnamefont{M.}},
  \bibinfo{year}{2002}, {``}\bibinfo{title}{Magnetic resonance diffraction
  using the magnetic field from ferromagnetic sphere},{''} \bibinfo{journal}{J.
  Appl. Phys.} \textbf{\bibinfo{volume}{91}},  \bibinfo{pages}{9987--9994}.

\bibitem[{\citenamefont{Barna{\'s} and Fert}(1998)}]{Barnas1998:PRL}
\bibinfo{author}{\bibnamefont{Barna{\'s}}, \bibfnamefont{J.}}, and
  \bibinfo{author}{\bibfnamefont{A.}~\bibnamefont{Fert}}, \bibinfo{year}{1998},
  {``}\bibinfo{title}{Magnetoresistance oscillations due to charging effects in
  double ferromagnetic tunnel junctions},{''} \bibinfo{journal}{Phys. Rev.
  Lett.} \textbf{\bibinfo{volume}{80}},  \bibinfo{pages}{1058--1061}.

\bibitem[{\citenamefont{Bass and {Pratt, Jr.}}(1999)}]{Bass1999:JMMM}
\bibinfo{author}{\bibnamefont{Bass}, \bibfnamefont{J.}}, and
  \bibinfo{author}{\bibfnamefont{W.~P.} \bibnamefont{{Pratt, Jr.}}},
  \bibinfo{year}{1999}, {``}\bibinfo{title}{Current-perpendicular {(CPP)}
  magnetoresistance in magnetic metallic multilayers},{''} \bibinfo{journal}{J.
  Magn. Magn. Mater.} \textbf{\bibinfo{volume}{200}},
  \bibinfo{pages}{274--289}.

\bibitem[{\citenamefont{Bastard and Ferreira}(1992)}]{Bastard1992:SS}
\bibinfo{author}{\bibnamefont{Bastard}, \bibfnamefont{G.}}, and
  \bibinfo{author}{\bibfnamefont{R.}~\bibnamefont{Ferreira}},
  \bibinfo{year}{1992}, {``}\bibinfo{title}{Spin-flip scattering times in
  semiconductor quantum wells},{''} \bibinfo{journal}{Surf. Sci.}
  \textbf{\bibinfo{volume}{267}},  \bibinfo{pages}{335--341}.

\bibitem[{\citenamefont{Bauer} \emph{et~al.}(2003)\citenamefont{Bauer, Brataas,
  and {van Wees}}}]{Bauer2003:APL}
\bibinfo{author}{\bibnamefont{Bauer}, \bibfnamefont{G.~E.~W.}},
  \bibinfo{author}{\bibfnamefont{A.}~\bibnamefont{Brataas}}, and
  \bibinfo{author}{\bibfnamefont{Y.~T. B.~J.} \bibnamefont{{van Wees}}},
  \bibinfo{year}{2003}, {``}\bibinfo{title}{Spin-torque transistor},{''}
  \bibinfo{journal}{Appl. Phys. Lett.} \textbf{\bibinfo{volume}{82}},
  \bibinfo{pages}{3928--3930}.

\bibitem[{\citenamefont{Baylac} \emph{et~al.}(1995)\citenamefont{Baylac, Amand,
  Marie, Dareys, Brousseau, Bacquet, and Thierry-Mieg}}]{Baylac1995:SSS}
\bibinfo{author}{\bibnamefont{Baylac}, \bibfnamefont{B.}},
  \bibinfo{author}{\bibfnamefont{T.}~\bibnamefont{Amand}},
  \bibinfo{author}{\bibfnamefont{X.}~\bibnamefont{Marie}},
  \bibinfo{author}{\bibfnamefont{B.}~\bibnamefont{Dareys}},
  \bibinfo{author}{\bibfnamefont{M.}~\bibnamefont{Brousseau}},
  \bibinfo{author}{\bibfnamefont{G.}~\bibnamefont{Bacquet}}, and
  \bibinfo{author}{\bibfnamefont{V.}~\bibnamefont{Thierry-Mieg}},
  \bibinfo{year}{1995}, {``}\bibinfo{title}{Hole spin relaxation in
  n-modulation doped quantum wells},{''} \bibinfo{journal}{Solid State Commun.}
  \textbf{\bibinfo{volume}{93}},  \bibinfo{pages}{57--60}.

\bibitem[{\citenamefont{Bazaliy} \emph{et~al.}(1998)\citenamefont{Bazaliy,
  Jones, and Zhang}}]{Bazily1998:PRB}
\bibinfo{author}{\bibnamefont{Bazaliy}, \bibfnamefont{Y.~B.}},
  \bibinfo{author}{\bibfnamefont{B.~A.} \bibnamefont{Jones}}, and
  \bibinfo{author}{\bibfnamefont{S.-C.} \bibnamefont{Zhang}},
  \bibinfo{year}{1998}, {``}\bibinfo{title}{Modification of the
  {Landau-Lifshitz} equation in the presence of a spin-polarized current in
  colossal- and giant-magnetoresistive materials},{''} \bibinfo{journal}{Phys.
  Rev. B.} \textbf{\bibinfo{volume}{57}},  \bibinfo{pages}{R3213--R3216}.

\bibitem[{\citenamefont{Beck} \emph{et~al.}(2002)\citenamefont{Beck, Kiesel,
  Malzer, and {D\"{o}hler}}}]{Beck2002:PE}
\bibinfo{author}{\bibnamefont{Beck}, \bibfnamefont{M.}},
  \bibinfo{author}{\bibfnamefont{P.}~\bibnamefont{Kiesel}},
  \bibinfo{author}{\bibfnamefont{S.}~\bibnamefont{Malzer}}, and
  \bibinfo{author}{\bibfnamefont{G.~H.} \bibnamefont{{D\"{o}hler}}},
  \bibinfo{year}{2002}, {``}\bibinfo{title}{Spin transport by giant ambipolar
  diffusion},{''} \bibinfo{journal}{Physica E} \textbf{\bibinfo{volume}{12}},
  \bibinfo{pages}{407--411}.

\bibitem[{\citenamefont{Bennett} \emph{et~al.}(2003)\citenamefont{Bennett, {van
  Lierop}, Berkeley, Mansfield, Henderson, Aronson, Young, Bianchi, Fisk,
  Balakirev, and Lacerda}}]{Bennett2003:P}
\bibinfo{author}{\bibnamefont{Bennett}, \bibfnamefont{M.~C.}},
  \bibinfo{author}{\bibfnamefont{J.}~\bibnamefont{{van Lierop}}},
  \bibinfo{author}{\bibfnamefont{E.~M.} \bibnamefont{Berkeley}},
  \bibinfo{author}{\bibfnamefont{J.~F.} \bibnamefont{Mansfield}},
  \bibinfo{author}{\bibfnamefont{C.}~\bibnamefont{Henderson}},
  \bibinfo{author}{\bibfnamefont{M.~C.} \bibnamefont{Aronson}},
  \bibinfo{author}{\bibfnamefont{D.~P.} \bibnamefont{Young}},
  \bibinfo{author}{\bibfnamefont{A.}~\bibnamefont{Bianchi}},
  \bibinfo{author}{\bibfnamefont{Z.}~\bibnamefont{Fisk}},
  \bibinfo{author}{\bibfnamefont{F.}~\bibnamefont{Balakirev}}, and
  \bibinfo{author}{\bibfnamefont{A.}~\bibnamefont{Lacerda}},
  \bibinfo{year}{2003}, {``}\bibinfo{title}{The origin of weak ferromagnetism
  in {CaB$_6$}},{''} \eprint{cond-mat/0306709}.

\bibitem[{\citenamefont{Berciu and Jank{\'o}}(2003)}]{Berciu2003:PRL}
\bibinfo{author}{\bibnamefont{Berciu}, \bibfnamefont{M.}}, and
  \bibinfo{author}{\bibfnamefont{B.}~\bibnamefont{Jank{\'o}}},
  \bibinfo{year}{2003}, {``}\bibinfo{title}{{Zeeman} localization of charge
  carriers in diluted magnetic semiconductor--permalloy hybrids},{''}
  \bibinfo{journal}{Phys. Rev. Lett.} \textbf{\bibinfo{volume}{90}},
  \bibinfo{pages}{246804}.

\bibitem[{\citenamefont{Berger}(1996)}]{Berger1996:PRB}
\bibinfo{author}{\bibnamefont{Berger}, \bibfnamefont{L.}},
  \bibinfo{year}{1996}, {``}\bibinfo{title}{Emission of spin waves by a
  magnetic multilayer transversed by a current},{''} \bibinfo{journal}{Phys.
  Rev. B} \textbf{\bibinfo{volume}{54}},  \bibinfo{pages}{9353--9358}.

\bibitem[{\citenamefont{Bergeret} \emph{et~al.}(2001)\citenamefont{Bergeret,
  Volkov, and Efetov}}]{Bergeret2001:PRL}
\bibinfo{author}{\bibnamefont{Bergeret}, \bibfnamefont{F.~S.}},
  \bibinfo{author}{\bibfnamefont{A.~F.} \bibnamefont{Volkov}}, and
  \bibinfo{author}{\bibfnamefont{K.~B.} \bibnamefont{Efetov}},
  \bibinfo{year}{2001}, {``}\bibinfo{title}{Long-range proximity effects in
  superconductor-ferromagnet structures},{''} \bibinfo{journal}{Phys. Rev.
  Lett.} \textbf{\bibinfo{volume}{86}},  \bibinfo{pages}{4096--4099}.

\bibitem[{\citenamefont{Bergmann}(1982)}]{Bergmann1982:ZP}
\bibinfo{author}{\bibnamefont{Bergmann}, \bibfnamefont{G.}},
  \bibinfo{year}{1982}, {``}\bibinfo{title}{Studies of spin-orbit scattering in
  noble-metal nanoparticles using energy-level tunneling spectroscopy},{''}
  \bibinfo{journal}{Z. Phys. B} \textbf{\bibinfo{volume}{48}},
  \bibinfo{pages}{5--16}.

\bibitem[{\citenamefont{Beschoten} \emph{et~al.}(2001)\citenamefont{Beschoten,
  Johnston-Halperin, Young, Poggio, Grimaldi, Keller, DenBaars, Mishra, Hu, and
  Awschalom}}]{Beschoten2001:PRB}
\bibinfo{author}{\bibnamefont{Beschoten}, \bibfnamefont{B.}},
  \bibinfo{author}{\bibfnamefont{E.}~\bibnamefont{Johnston-Halperin}},
  \bibinfo{author}{\bibfnamefont{D.~K.} \bibnamefont{Young}},
  \bibinfo{author}{\bibfnamefont{M.}~\bibnamefont{Poggio}},
  \bibinfo{author}{\bibfnamefont{J.~E.} \bibnamefont{Grimaldi}},
  \bibinfo{author}{\bibfnamefont{S.}~\bibnamefont{Keller}},
  \bibinfo{author}{\bibfnamefont{S.~P.} \bibnamefont{DenBaars}},
  \bibinfo{author}{\bibfnamefont{U.~K.} \bibnamefont{Mishra}},
  \bibinfo{author}{\bibfnamefont{E.~L.} \bibnamefont{Hu}}, and
  \bibinfo{author}{\bibfnamefont{D.~D.} \bibnamefont{Awschalom}},
  \bibinfo{year}{2001}, {``}\bibinfo{title}{Spin coherence and dephasing in
  {GaN}},{''} \bibinfo{journal}{Phys. Rev. B} \textbf{\bibinfo{volume}{63}},
  \bibinfo{pages}{121202}.

\bibitem[{\citenamefont{Beuneu and Monod}(1978)}]{Beuneu1978:PRB}
\bibinfo{author}{\bibnamefont{Beuneu}, \bibfnamefont{F.}}, and
  \bibinfo{author}{\bibfnamefont{P.}~\bibnamefont{Monod}},
  \bibinfo{year}{1978}, {``}\bibinfo{title}{The {Elliott} relation in pure
  metals},{''} \bibinfo{journal}{Phys. Rev. B} \textbf{\bibinfo{volume}{18}},
  \bibinfo{pages}{2422--2425}.

\bibitem[{\citenamefont{Bhat and Sipe}(2000)}]{Bhat2000:PRL}
\bibinfo{author}{\bibnamefont{Bhat}, \bibfnamefont{R.~D.~R.}}, and
  \bibinfo{author}{\bibfnamefont{J.~E.} \bibnamefont{Sipe}},
  \bibinfo{year}{2000}, {``}\bibinfo{title}{Optically injected spin currents in
  semiconductors},{''} \bibinfo{journal}{Phys. Rev. Lett.}
  \textbf{\bibinfo{volume}{85}},  \bibinfo{pages}{5432--5435}.

\bibitem[{\citenamefont{Bhat} \emph{et~al.}(2002)\citenamefont{Bhat, Berciu,
  Kennett, and Wan}}]{Bhat2002:JS}
\bibinfo{author}{\bibnamefont{Bhat}, \bibfnamefont{R.~N.}},
  \bibinfo{author}{\bibfnamefont{M.}~\bibnamefont{Berciu}},
  \bibinfo{author}{\bibfnamefont{M.~P.} \bibnamefont{Kennett}}, and
  \bibinfo{author}{\bibfnamefont{X.}~\bibnamefont{Wan}}, \bibinfo{year}{2002},
  {``}\bibinfo{title}{Dilute magnetic semiconductors in the low density
  regime},{''} \bibinfo{journal}{J. Supercond.} \textbf{\bibinfo{volume}{15}},
  \bibinfo{pages}{71--83}.

\bibitem[{\citenamefont{Binasch} \emph{et~al.}(1989)\citenamefont{Binasch,
  Gr{\"u}nberg, Saurenbach, and Zinn}}]{Binasch1989:PRB}
\bibinfo{author}{\bibnamefont{Binasch}, \bibfnamefont{G.}},
  \bibinfo{author}{\bibfnamefont{P.}~\bibnamefont{Gr{\"u}nberg}},
  \bibinfo{author}{\bibfnamefont{F.}~\bibnamefont{Saurenbach}}, and
  \bibinfo{author}{\bibfnamefont{W.}~\bibnamefont{Zinn}}, \bibinfo{year}{1989},
  {``}\bibinfo{title}{Enhanced magnetoresistance in layered magnetic structures
  with antiferromagnetic interlayer exchange},{''} \bibinfo{journal}{Phys. Rev.
  B} \textbf{\bibinfo{volume}{39}},  \bibinfo{pages}{4828--4830}.

\bibitem[{\citenamefont{Bir} \emph{et~al.}(1975)\citenamefont{Bir, Aronov, and
  Pikus}}]{Bir1976:SPJETP}
\bibinfo{author}{\bibnamefont{Bir}, \bibfnamefont{G.~L.}},
  \bibinfo{author}{\bibfnamefont{A.~G.} \bibnamefont{Aronov}}, and
  \bibinfo{author}{\bibfnamefont{G.~E.} \bibnamefont{Pikus}},
  \bibinfo{year}{1975}, {``}\bibinfo{title}{Spin relaxation of electrons due to
  scattering by holes},{''} \bibinfo{journal}{Zh. Eksp. Teor. Fiz.}
  \textbf{\bibinfo{volume}{69}},  \bibinfo{pages}{1382--1397}
  \bibinfo{note}{[Sov. Phys. JETP {\bf 42}, 705-712 (1976)]}.

\bibitem[{\citenamefont{Blakemore}(1982)}]{Blakemore1982:JAP}
\bibinfo{author}{\bibnamefont{Blakemore}, \bibfnamefont{J.~S.}},
  \bibinfo{year}{1982}, {``}\bibinfo{title}{Semiconducting and other major
  properties of gallium arsenide},{''} \bibinfo{journal}{J. Appl. Phys.}
  \textbf{\bibinfo{volume}{53}},  \bibinfo{pages}{R123--R181}.

\bibitem[{\citenamefont{Bloch}(1946)}]{Bloch1946:PR}
\bibinfo{author}{\bibnamefont{Bloch}, \bibfnamefont{F.}}, \bibinfo{year}{1946},
  {``}\bibinfo{title}{Nuclear induction},{''} \bibinfo{journal}{Phys. Rev.}
  \textbf{\bibinfo{volume}{70}},  \bibinfo{pages}{460--474}.

\bibitem[{\citenamefont{Blonder and Tinkham}(1983)}]{Blonder1983:PRB}
\bibinfo{author}{\bibnamefont{Blonder}, \bibfnamefont{G.~E.}}, and
  \bibinfo{author}{\bibfnamefont{M.}~\bibnamefont{Tinkham}},
  \bibinfo{year}{1983}, {``}\bibinfo{title}{Metallic to tunneling transition in
  {Cu-Nb} point contacts},{''} \bibinfo{journal}{Phys. Rev. B}
  \textbf{\bibinfo{volume}{27}},  \bibinfo{pages}{112--118}.

\bibitem[{\citenamefont{Blonder} \emph{et~al.}(1982)\citenamefont{Blonder,
  Tinkham, and Klapwijk}}]{Blonder1982:PRB}
\bibinfo{author}{\bibnamefont{Blonder}, \bibfnamefont{G.~E.}},
  \bibinfo{author}{\bibfnamefont{M.}~\bibnamefont{Tinkham}}, and
  \bibinfo{author}{\bibfnamefont{T.~M.} \bibnamefont{Klapwijk}},
  \bibinfo{year}{1982}, {``}\bibinfo{title}{Transition from metallic to
  tunneling regimes in superconducting microconstrictions: {Excess} current,
  charge imbalance, and supercurrent conversion},{''} \bibinfo{journal}{Phys.
  Rev. B} \textbf{\bibinfo{volume}{25}},  \bibinfo{pages}{4515--4532}.

\bibitem[{\citenamefont{Boeve} \emph{et~al.}(2001)\citenamefont{Boeve,
  N{\'{e}}meth, Liu, {De Boeck}, and Borghs}}]{Boeve2001:JMMM}
\bibinfo{author}{\bibnamefont{Boeve}, \bibfnamefont{H.}},
  \bibinfo{author}{\bibfnamefont{{\v{S}}.}~\bibnamefont{N{\'{e}}meth}},
  \bibinfo{author}{\bibfnamefont{Z.}~\bibnamefont{Liu}},
  \bibinfo{author}{\bibfnamefont{J.}~\bibnamefont{{De Boeck}}}, and
  \bibinfo{author}{\bibfnamefont{G.}~\bibnamefont{Borghs}},
  \bibinfo{year}{2001}, {``}\bibinfo{title}{Transport properties of epitaxial
  {Fe-GaN-Fe} tunnel junctions on {GaAs}},{''} \bibinfo{journal}{J. Magn. Magn.
  Mater.} \textbf{\bibinfo{volume}{226-230}},  \bibinfo{pages}{933--935}.

\bibitem[{\citenamefont{Boggess} \emph{et~al.}(2000)\citenamefont{Boggess,
  Olesberg, Yu, Flatt{\'e}, and Lau}}]{Boggess2000:APL}
\bibinfo{author}{\bibnamefont{Boggess}, \bibfnamefont{T.}},
  \bibinfo{author}{\bibfnamefont{J.~T.} \bibnamefont{Olesberg}},
  \bibinfo{author}{\bibfnamefont{C.}~\bibnamefont{Yu}},
  \bibinfo{author}{\bibfnamefont{M.~E.} \bibnamefont{Flatt{\'e}}}, and
  \bibinfo{author}{\bibfnamefont{W.~H.} \bibnamefont{Lau}},
  \bibinfo{year}{2000}, {``}\bibinfo{title}{Room-temperature electron spin
  relaxation in bulk {InAs}},{''} \bibinfo{journal}{Appl. Phys. Lett.}
  \textbf{\bibinfo{volume}{77}},  \bibinfo{pages}{1333--1335}.

\bibitem[{\citenamefont{Boguslawski}(1980)}]{Boguslawski1980:SSC}
\bibinfo{author}{\bibnamefont{Boguslawski}, \bibfnamefont{P.}},
  \bibinfo{year}{1980}, {``}\bibinfo{title}{Electron-electron spin-flip
  scattering and spin relaxation in {III-V} and {II-VI} semiconductors},{''}
  \bibinfo{journal}{Solid State Commun.} \textbf{\bibinfo{volume}{33}},
  \bibinfo{pages}{389--391}.

\bibitem[{\citenamefont{{Bonnell Ed.}}(2001)}]{Bonnell:2001}
\bibinfo{author}{\bibnamefont{{Bonnell Ed.}}, \bibfnamefont{D.~A.}},
  \bibinfo{year}{2001}, \emph{\bibinfo{title}{Scanning Probe Microscopy and
  Spectroscopy}} (\bibinfo{publisher}{Wiley, New York}).

\bibitem[{\citenamefont{Borda} \emph{et~al.}(2003)\citenamefont{Borda,
  Zar{\'a}nd, Hofstetter, Halperin, and {von Delft}}}]{Borda2003:PRL}
\bibinfo{author}{\bibnamefont{Borda}, \bibfnamefont{L.}},
  \bibinfo{author}{\bibfnamefont{G.}~\bibnamefont{Zar{\'a}nd}},
  \bibinfo{author}{\bibfnamefont{W.}~\bibnamefont{Hofstetter}},
  \bibinfo{author}{\bibfnamefont{B.~I.} \bibnamefont{Halperin}}, and
  \bibinfo{author}{\bibfnamefont{J.}~\bibnamefont{{von Delft}}},
  \bibinfo{year}{2003}, {``}\bibinfo{title}{{SU(4)} {Fermi} liquid state and
  spin filtering in a double quantum dot system},{''} \bibinfo{journal}{Phys.
  Rev. Lett.} \textbf{\bibinfo{volume}{90}},  \bibinfo{pages}{026602}.

\bibitem[{\citenamefont{Boukari} \emph{et~al.}(2002)\citenamefont{Boukari,
  Kossacki, Bertolini, Ferrand, Cibert, Tatarenko, Wasiela, Gaj, and
  Dietl}}]{Boukari2002:PRL}
\bibinfo{author}{\bibnamefont{Boukari}, \bibfnamefont{H.}},
  \bibinfo{author}{\bibfnamefont{P.}~\bibnamefont{Kossacki}},
  \bibinfo{author}{\bibfnamefont{M.}~\bibnamefont{Bertolini}},
  \bibinfo{author}{\bibfnamefont{D.}~\bibnamefont{Ferrand}},
  \bibinfo{author}{\bibfnamefont{J.}~\bibnamefont{Cibert}},
  \bibinfo{author}{\bibfnamefont{S.}~\bibnamefont{Tatarenko}},
  \bibinfo{author}{\bibfnamefont{A.}~\bibnamefont{Wasiela}},
  \bibinfo{author}{\bibfnamefont{J.~A.} \bibnamefont{Gaj}}, and
  \bibinfo{author}{\bibfnamefont{T.}~\bibnamefont{Dietl}},
  \bibinfo{year}{2002}, {``}\bibinfo{title}{Light and electric field control of
  ferromagnetism in magnetic quantum structures},{''} \bibinfo{journal}{Phys.
  Rev. Lett.} \textbf{\bibinfo{volume}{88}},  \bibinfo{pages}{207204}.

\bibitem[{\citenamefont{Bourgeois} \emph{et~al.}(2001)\citenamefont{Bourgeois,
  Gandit, Sulpice, J.~Chauss, and Grison}}]{Bourgeois2001:PRB}
\bibinfo{author}{\bibnamefont{Bourgeois}, \bibfnamefont{O.}},
  \bibinfo{author}{\bibfnamefont{P.}~\bibnamefont{Gandit}},
  \bibinfo{author}{\bibfnamefont{A.}~\bibnamefont{Sulpice}},
  \bibinfo{author}{\bibfnamefont{J.~L.} \bibnamefont{J.~Chauss}}, and
  \bibinfo{author}{\bibfnamefont{X.}~\bibnamefont{Grison}},
  \bibinfo{year}{2001}, {``}\bibinfo{title}{Transport in
  superconductor/ferromagnet/superconductor junctions dominated by interface
  resistance},{''} \bibinfo{journal}{Phys. Rev. B}
  \textbf{\bibinfo{volume}{63}},  \bibinfo{pages}{064517}.

\bibitem[{\citenamefont{Bournel}(2000)}]{Bournel2000:APF}
\bibinfo{author}{\bibnamefont{Bournel}, \bibfnamefont{A.}},
  \bibinfo{year}{2000}, {``}\bibinfo{title}{Magn{\'e}to-{\'e}lectronique dans
  des dispositifs {\'a} semiconducteurs},{''} \bibinfo{journal}{Ann. Phys.
  (Paris)} \textbf{\bibinfo{volume}{25}},  \bibinfo{pages}{1--167}.

\bibitem[{\citenamefont{Bournel} \emph{et~al.}(2000)\citenamefont{Bournel,
  Dollfus, Cassan, and Hesto}}]{Bournel2000:APL}
\bibinfo{author}{\bibnamefont{Bournel}, \bibfnamefont{A.}},
  \bibinfo{author}{\bibfnamefont{P.}~\bibnamefont{Dollfus}},
  \bibinfo{author}{\bibfnamefont{E.}~\bibnamefont{Cassan}}, and
  \bibinfo{author}{\bibfnamefont{P.}~\bibnamefont{Hesto}},
  \bibinfo{year}{2000}, {``}\bibinfo{title}{Monte {Carlo} study of spin
  relaxation in {AlGaAs/GaAs} quantum wells},{''} \bibinfo{journal}{Appl. Phys.
  Lett.} \textbf{\bibinfo{volume}{77}},  \bibinfo{pages}{2346--2348}.

\bibitem[{\citenamefont{Bowen} \emph{et~al.}(2003)\citenamefont{Bowen, Bibes,
  Barthelemy, Contour, Anane, Lemaitre, and Fert}}]{Bowen2002:P}
\bibinfo{author}{\bibnamefont{Bowen}, \bibfnamefont{M.}},
  \bibinfo{author}{\bibfnamefont{M.}~\bibnamefont{Bibes}},
  \bibinfo{author}{\bibfnamefont{A.}~\bibnamefont{Barthelemy}},
  \bibinfo{author}{\bibfnamefont{J.-P.} \bibnamefont{Contour}},
  \bibinfo{author}{\bibfnamefont{A.}~\bibnamefont{Anane}},
  \bibinfo{author}{\bibfnamefont{Y.}~\bibnamefont{Lemaitre}}, and
  \bibinfo{author}{\bibfnamefont{A.}~\bibnamefont{Fert}}, \bibinfo{year}{2003},
  {``}\bibinfo{title}{Nearly total spin polarization in
  {La$_{2/3}$Sr$_{1/3}$MnO$_3$} from tunneling experiments},{''}
  \bibinfo{journal}{Appl. Phys. Lett.} \textbf{\bibinfo{volume}{82}},
  \bibinfo{pages}{233--235}.

\bibitem[{\citenamefont{Bowring} \emph{et~al.}(1971)\citenamefont{Bowring,
  Smithard, and Cousins}}]{Bowring1971:PSS}
\bibinfo{author}{\bibnamefont{Bowring}, \bibfnamefont{C.~S.}},
  \bibinfo{author}{\bibfnamefont{M.~A.} \bibnamefont{Smithard}}, and
  \bibinfo{author}{\bibfnamefont{J.~E.} \bibnamefont{Cousins}},
  \bibinfo{year}{1971}, {``}\bibinfo{title}{Conduction electron spin resonance
  in magnesium},{''} \bibinfo{journal}{Phys. Status Solidi B}
  \textbf{\bibinfo{volume}{43}},  \bibinfo{pages}{625--630}.

\bibitem[{\citenamefont{Braden} \emph{et~al.}(2003)\citenamefont{Braden,
  Parker, Xiong, Chun, and Samarth}}]{Braden2003:P}
\bibinfo{author}{\bibnamefont{Braden}, \bibfnamefont{J.~G.}},
  \bibinfo{author}{\bibfnamefont{J.~S.} \bibnamefont{Parker}},
  \bibinfo{author}{\bibfnamefont{P.}~\bibnamefont{Xiong}},
  \bibinfo{author}{\bibfnamefont{S.~H.} \bibnamefont{Chun}}, and
  \bibinfo{author}{\bibfnamefont{N.}~\bibnamefont{Samarth}},
  \bibinfo{year}{2003}, {``}\bibinfo{title}{Direct measurement of the spin
  polarization of the magnetic semiconductor {(Ga,Mn)As}},{''}
  \bibinfo{journal}{Phys. Rev. Lett.} \textbf{\bibinfo{volume}{91}},
  \bibinfo{pages}{056602}.

\bibitem[{\citenamefont{Brand} \emph{et~al.}(2002)\citenamefont{Brand,
  Malinowski, Karimov, Marsden, Harley, Shields, Sanvitto, Ritchie, and
  Simmons}}]{Brand2002:PRL}
\bibinfo{author}{\bibnamefont{Brand}, \bibfnamefont{M.~A.}},
  \bibinfo{author}{\bibfnamefont{A.}~\bibnamefont{Malinowski}},
  \bibinfo{author}{\bibfnamefont{O.~Z.} \bibnamefont{Karimov}},
  \bibinfo{author}{\bibfnamefont{P.~A.} \bibnamefont{Marsden}},
  \bibinfo{author}{\bibfnamefont{R.~T.} \bibnamefont{Harley}},
  \bibinfo{author}{\bibfnamefont{A.~J.} \bibnamefont{Shields}},
  \bibinfo{author}{\bibfnamefont{D.}~\bibnamefont{Sanvitto}},
  \bibinfo{author}{\bibfnamefont{D.~A.} \bibnamefont{Ritchie}}, and
  \bibinfo{author}{\bibfnamefont{M.~Y.} \bibnamefont{Simmons}},
  \bibinfo{year}{2002}, {``}\bibinfo{title}{Precession and motional slowing of
  spin evolution in a high mobility two-dimensional electron gas},{''}
  \bibinfo{journal}{Phys. Rev. Lett.} \textbf{\bibinfo{volume}{89}},
  \bibinfo{pages}{236601}.

\bibitem[{\citenamefont{Brandt and Moshchalkov}(1984)}]{Brandt1984:AP}
\bibinfo{author}{\bibnamefont{Brandt}, \bibfnamefont{N.~B.}}, and
  \bibinfo{author}{\bibfnamefont{V.~V.} \bibnamefont{Moshchalkov}},
  \bibinfo{year}{1984}, {``}\bibinfo{title}{Semimagnetic semiconductors},{''}
  \bibinfo{journal}{Adv. Phys.} \textbf{\bibinfo{volume}{33}},
  \bibinfo{pages}{193--256}.

\bibitem[{\citenamefont{Brataas} \emph{et~al.}(2002)\citenamefont{Brataas,
  Tserkovnyak, Bauer, and Halperin}}]{Brataas2002:PRB}
\bibinfo{author}{\bibnamefont{Brataas}, \bibfnamefont{A.}},
  \bibinfo{author}{\bibfnamefont{Y.}~\bibnamefont{Tserkovnyak}},
  \bibinfo{author}{\bibfnamefont{G.~E.~W.} \bibnamefont{Bauer}}, and
  \bibinfo{author}{\bibfnamefont{B.~I.} \bibnamefont{Halperin}},
  \bibinfo{year}{2002}, {``}\bibinfo{title}{Spin battery operated by
  ferromagnetic resonance},{''} \bibinfo{journal}{Phys. Rev. B}
  \textbf{\bibinfo{volume}{66}},  \bibinfo{pages}{060404}.

\bibitem[{\citenamefont{Bratkovsky}(1997)}]{Bratkovsky1997:PRB}
\bibinfo{author}{\bibnamefont{Bratkovsky}, \bibfnamefont{A.~M.}},
  \bibinfo{year}{1997}, {``}\bibinfo{title}{Tunneling of electrons in
  conventional and half-metallic systems: Towards very large
  magnetoresistance},{''} \bibinfo{journal}{Phys. Rev. B}
  \textbf{\bibinfo{volume}{56}},  \bibinfo{pages}{2344--2347}.

\bibitem[{\citenamefont{Bratkovsky and Osipov}(2003)}]{Bratkovsky2003:P}
\bibinfo{author}{\bibnamefont{Bratkovsky}, \bibfnamefont{A.~M.}}, and
  \bibinfo{author}{\bibfnamefont{V.~V.} \bibnamefont{Osipov}},
  \bibinfo{year}{2003}, {``}\bibinfo{title}{Spin extraction from a non-magnetic
  semiconductor},{''} \eprint{cond-mat/0307656}.

\bibitem[{\citenamefont{Brehmer} \emph{et~al.}(1995)\citenamefont{Brehmer,
  Zhang, Schwarz, Chau, Allen, Ibbetson, Zhang, Palmstr$\o$m, and
  Wilkens}}]{Brehmer1995:APL}
\bibinfo{author}{\bibnamefont{Brehmer}, \bibfnamefont{D.~E.}},
  \bibinfo{author}{\bibfnamefont{K.}~\bibnamefont{Zhang}},
  \bibinfo{author}{\bibfnamefont{C.~J.} \bibnamefont{Schwarz}},
  \bibinfo{author}{\bibfnamefont{S.}~\bibnamefont{Chau}},
  \bibinfo{author}{\bibfnamefont{S.~J.} \bibnamefont{Allen}},
  \bibinfo{author}{\bibfnamefont{J.~P.} \bibnamefont{Ibbetson}},
  \bibinfo{author}{\bibfnamefont{J.~P.} \bibnamefont{Zhang}},
  \bibinfo{author}{\bibfnamefont{C.~J.} \bibnamefont{Palmstr$\o$m}}, and
  \bibinfo{author}{\bibfnamefont{B.}~\bibnamefont{Wilkens}},
  \bibinfo{year}{1995}, {``}\bibinfo{title}{Resonant tunneling through {ErAs}
  semimetal quantum wells},{''} \bibinfo{journal}{Appl. Phys. Lett.}
  \textbf{\bibinfo{volume}{67}},  \bibinfo{pages}{1268--1270}.

\bibitem[{\citenamefont{Brinkman} \emph{et~al.}(1970)\citenamefont{Brinkman,
  Dynes, and Rowell}}]{Brinkman1970:JAP}
\bibinfo{author}{\bibnamefont{Brinkman}, \bibfnamefont{W.~F.}},
  \bibinfo{author}{\bibfnamefont{R.~C.} \bibnamefont{Dynes}}, and
  \bibinfo{author}{\bibfnamefont{J.~M.} \bibnamefont{Rowell}},
  \bibinfo{year}{1970}, {``}\bibinfo{title}{Tunneling conductance of
  asymmetrical barriers},{''} \bibinfo{journal}{J. Appl. Phys.}
  \textbf{\bibinfo{volume}{41}},  \bibinfo{pages}{1915--1921}.

\bibitem[{\citenamefont{Britton} \emph{et~al.}(1998)\citenamefont{Britton,
  Grevatt, Malinowski, Harley, Perozzo, Cameron, and Miller}}]{Britton1998:APL}
\bibinfo{author}{\bibnamefont{Britton}, \bibfnamefont{R.~S.}},
  \bibinfo{author}{\bibfnamefont{T.}~\bibnamefont{Grevatt}},
  \bibinfo{author}{\bibfnamefont{A.}~\bibnamefont{Malinowski}},
  \bibinfo{author}{\bibfnamefont{R.~T.} \bibnamefont{Harley}},
  \bibinfo{author}{\bibfnamefont{P.}~\bibnamefont{Perozzo}},
  \bibinfo{author}{\bibfnamefont{A.~R.} \bibnamefont{Cameron}}, and
  \bibinfo{author}{\bibfnamefont{A.}~\bibnamefont{Miller}},
  \bibinfo{year}{1998}, {``}\bibinfo{title}{Room temperature spin relaxation in
  {GaAs/AlGaAs} multiple quantum wells},{''} \bibinfo{journal}{Appl. Phys.
  Lett.} \textbf{\bibinfo{volume}{73}},  \bibinfo{pages}{2140--2142}.

\bibitem[{\citenamefont{Bronold} \emph{et~al.}(2002)\citenamefont{Bronold,
  Martin, Saxena, and Smith}}]{Bronold2002:PRB}
\bibinfo{author}{\bibnamefont{Bronold}, \bibfnamefont{F.~X.}},
  \bibinfo{author}{\bibfnamefont{I.}~\bibnamefont{Martin}},
  \bibinfo{author}{\bibfnamefont{A.}~\bibnamefont{Saxena}}, and
  \bibinfo{author}{\bibfnamefont{D.~L.} \bibnamefont{Smith}},
  \bibinfo{year}{2002}, {``}\bibinfo{title}{Magnetic-field dependence of
  electron spin relaxation in n-type semiconductors},{''}
  \bibinfo{journal}{Phys. Rev. B} \textbf{\bibinfo{volume}{66}},
  \bibinfo{pages}{233206}.

\bibitem[{\citenamefont{Brosig} \emph{et~al.}(1999)\citenamefont{Brosig,
  Ensslin, Warburton, Nguyen, Brar, Thomas, and Kroemer}}]{Brosig1999:PRB}
\bibinfo{author}{\bibnamefont{Brosig}, \bibfnamefont{K.}},
  \bibinfo{author}{\bibfnamefont{K.}~\bibnamefont{Ensslin}},
  \bibinfo{author}{\bibfnamefont{R.~J.} \bibnamefont{Warburton}},
  \bibinfo{author}{\bibfnamefont{C.}~\bibnamefont{Nguyen}},
  \bibinfo{author}{\bibfnamefont{B.}~\bibnamefont{Brar}},
  \bibinfo{author}{\bibfnamefont{M.}~\bibnamefont{Thomas}}, and
  \bibinfo{author}{\bibfnamefont{H.}~\bibnamefont{Kroemer}},
  \bibinfo{year}{1999}, {``}\bibinfo{title}{Zero-field spin splitting in
  {InAs-AlSb} quantum wells revisited},{''} \bibinfo{journal}{Phys. Rev. B}
  \textbf{\bibinfo{volume}{60}},  \bibinfo{pages}{R13989--R13992}.

\bibitem[{\citenamefont{Brouwer}(1998)}]{Brouwer1998:PRB}
\bibinfo{author}{\bibnamefont{Brouwer}, \bibfnamefont{P.~W.}},
  \bibinfo{year}{1998}, {``}\bibinfo{title}{Scattering approach to parametric
  pumping},{''} \bibinfo{journal}{Phys. Rev. B} \textbf{\bibinfo{volume}{58}},
  \bibinfo{pages}{R10135--R10138}.

\bibitem[{\citenamefont{Bruno}(1999)}]{Bruno1999:PRL}
\bibinfo{author}{\bibnamefont{Bruno}, \bibfnamefont{P.}}, \bibinfo{year}{1999},
  {``}\bibinfo{title}{Geometrically constrained magnetic wall},{''}
  \bibinfo{journal}{Phys. Rev. Lett.} \textbf{\bibinfo{volume}{83}},
  \bibinfo{pages}{2425--2428}.

\bibitem[{\citenamefont{Bruno and Schwartz}(1973)}]{Bruno1973:PRB}
\bibinfo{author}{\bibnamefont{Bruno}, \bibfnamefont{R.~C.}}, and
  \bibinfo{author}{\bibfnamefont{B.~B.} \bibnamefont{Schwartz}},
  \bibinfo{year}{1973}, {``}\bibinfo{title}{Magnetic field splitting of the
  density of states of thin superconductors},{''} \bibinfo{journal}{Phys. Rev.
  B} \textbf{\bibinfo{volume}{8}},  \bibinfo{pages}{3161--3178}.

\bibitem[{\citenamefont{Burkard} \emph{et~al.}(2000)\citenamefont{Burkard,
  Engel, and Loss}}]{Burkard2000:FP}
\bibinfo{author}{\bibnamefont{Burkard}, \bibfnamefont{G.}},
  \bibinfo{author}{\bibfnamefont{H.~A.} \bibnamefont{Engel}}, and
  \bibinfo{author}{\bibfnamefont{D.}~\bibnamefont{Loss}}, \bibinfo{year}{2000},
  {``}\bibinfo{title}{Spintronics and quantum dots for quantum computing and
  quantum communication},{''} \bibinfo{journal}{Fortschr. Phys.}
  \textbf{\bibinfo{volume}{48}},  \bibinfo{pages}{965--986}.

\bibitem[{\citenamefont{Burkard} \emph{et~al.}(1999)\citenamefont{Burkard,
  Loss, and DiVincenzo}}]{Burkard1999:PRB}
\bibinfo{author}{\bibnamefont{Burkard}, \bibfnamefont{G.}},
  \bibinfo{author}{\bibfnamefont{D.}~\bibnamefont{Loss}}, and
  \bibinfo{author}{\bibfnamefont{D.~P.} \bibnamefont{DiVincenzo}},
  \bibinfo{year}{1999}, {``}\bibinfo{title}{Coupled quantum dots as quantum
  gates},{''} \bibinfo{journal}{Phys. Rev. B} \textbf{\bibinfo{volume}{59}},
  \bibinfo{pages}{2070--2078}.

\bibitem[{\citenamefont{Busch} \emph{et~al.}(1969)\citenamefont{Busch,
  Campagna, Cotti, and {Ch. Siegmann}}}]{Busch1969:PRL}
\bibinfo{author}{\bibnamefont{Busch}, \bibfnamefont{G.}},
  \bibinfo{author}{\bibfnamefont{M.}~\bibnamefont{Campagna}},
  \bibinfo{author}{\bibfnamefont{P.}~\bibnamefont{Cotti}}, and
  \bibinfo{author}{\bibfnamefont{H.}~\bibnamefont{{Ch. Siegmann}}},
  \bibinfo{year}{1969}, {``}\bibinfo{title}{Observation of electron
  polarization in photoemission},{''} \bibinfo{journal}{Phys. Rev. Lett.}
  \textbf{\bibinfo{volume}{22}},  \bibinfo{pages}{597--599}.

\bibitem[{\citenamefont{Bussmann} \emph{et~al.}(1998)\citenamefont{Bussmann,
  Cheng, Prinz, Hu, Gutmann, Wang, Beech, and Zhu}}]{Bussmann1998:IEEETM}
\bibinfo{author}{\bibnamefont{Bussmann}, \bibfnamefont{K.}},
  \bibinfo{author}{\bibfnamefont{S.~F.} \bibnamefont{Cheng}},
  \bibinfo{author}{\bibfnamefont{G.~A.} \bibnamefont{Prinz}},
  \bibinfo{author}{\bibfnamefont{Y.}~\bibnamefont{Hu}},
  \bibinfo{author}{\bibfnamefont{R.}~\bibnamefont{Gutmann}},
  \bibinfo{author}{\bibfnamefont{D.}~\bibnamefont{Wang}},
  \bibinfo{author}{\bibfnamefont{R.}~\bibnamefont{Beech}}, and
  \bibinfo{author}{\bibfnamefont{J.}~\bibnamefont{Zhu}}, \bibinfo{year}{1998},
  {``}\bibinfo{title}{{CPP} giant magnetoresistance of {NiFeCo/Cu/CoFe/Cu}
  multilayers},{''} \bibinfo{journal}{IEEE Trans. Magn.}
  \textbf{\bibinfo{volume}{34}},  \bibinfo{pages}{924--926}.

\bibitem[{\citenamefont{Bussmann} \emph{et~al.}(1999)\citenamefont{Bussmann,
  Prinz, Cheng, and Wang}}]{Bussman1999:APL}
\bibinfo{author}{\bibnamefont{Bussmann}, \bibfnamefont{K.}},
  \bibinfo{author}{\bibfnamefont{G.~A.} \bibnamefont{Prinz}},
  \bibinfo{author}{\bibfnamefont{S.-F.} \bibnamefont{Cheng}}, and
  \bibinfo{author}{\bibfnamefont{D.}~\bibnamefont{Wang}}, \bibinfo{year}{1999},
  {``}\bibinfo{title}{Switching of vertical giant magnetoresistance devices by
  current through the device},{''} \bibinfo{journal}{Appl. Phys. Lett.}
  \textbf{\bibinfo{volume}{75}},  \bibinfo{pages}{2476--2478}.

\bibitem[{\citenamefont{Butler} \emph{et~al.}(2001)\citenamefont{Butler, Zhang,
  Schulthess, and MacLaren}}]{Butler2001:PRB}
\bibinfo{author}{\bibnamefont{Butler}, \bibfnamefont{W.~H.}},
  \bibinfo{author}{\bibfnamefont{X.-G.} \bibnamefont{Zhang}},
  \bibinfo{author}{\bibfnamefont{T.~C.} \bibnamefont{Schulthess}}, and
  \bibinfo{author}{\bibfnamefont{J.~M.} \bibnamefont{MacLaren}},
  \bibinfo{year}{2001}, {``}\bibinfo{title}{Reduction of electron tunneling
  current due to lateral variation of the wave function},{''}
  \bibinfo{journal}{Phys. Rev. B} \textbf{\bibinfo{volume}{63}},
  \bibinfo{pages}{092402}.

\bibitem[{\citenamefont{Bychkov and
  Rashba}(1984{\natexlab{a}})}]{Bychkov1984:JPC}
\bibinfo{author}{\bibnamefont{Bychkov}, \bibfnamefont{Y.~A.}}, and
  \bibinfo{author}{\bibfnamefont{E.~I.} \bibnamefont{Rashba}},
  \bibinfo{year}{1984}{\natexlab{a}}, {``}\bibinfo{title}{Oscillatory effects
  and the magnetic-susceptibility of carriers in inversion-layers},{''}
  \bibinfo{journal}{J. Phys. C} \textbf{\bibinfo{volume}{17}},
  \bibinfo{pages}{6039--6045}.

\bibitem[{\citenamefont{Bychkov and
  Rashba}(1984{\natexlab{b}})}]{Bychkov1984:JETPL}
\bibinfo{author}{\bibnamefont{Bychkov}, \bibfnamefont{Y.~A.}}, and
  \bibinfo{author}{\bibfnamefont{E.~I.} \bibnamefont{Rashba}},
  \bibinfo{year}{1984}{\natexlab{b}}, {``}\bibinfo{title}{Properties of a {2D}
  with lifted spectral degeneracy},{''} \bibinfo{journal}{Zh. Eksp. Teor. Fiz.
  Pisma Red.} \textbf{\bibinfo{volume}{39}},  \bibinfo{pages}{66--69}
  \bibinfo{note}{[JETP Lett. {\bf 39}, 78-81 (1984)]}.

\bibitem[{\citenamefont{Cacho} \emph{et~al.}(2002)\citenamefont{Cacho,
  Lassailly, Drouhin, Lampel, and Peretti}}]{Cacho2002:PRL}
\bibinfo{author}{\bibnamefont{Cacho}, \bibfnamefont{C.}},
  \bibinfo{author}{\bibfnamefont{Y.}~\bibnamefont{Lassailly}},
  \bibinfo{author}{\bibfnamefont{H.}~\bibnamefont{Drouhin}},
  \bibinfo{author}{\bibfnamefont{G.}~\bibnamefont{Lampel}}, and
  \bibinfo{author}{\bibfnamefont{J.}~\bibnamefont{Peretti}},
  \bibinfo{year}{2002}, {``}\bibinfo{title}{Spin filtering of free electrons by
  magnetic multilayers: towards an efficient self-calibrated spin
  polarimeter},{''} \bibinfo{journal}{Phys. Rev. Lett.}
  \textbf{\bibinfo{volume}{88}},  \bibinfo{pages}{066601}.

\bibitem[{\citenamefont{Camilleri} \emph{et~al.}(2001)\citenamefont{Camilleri,
  Teppe, Scalbert, Semenov, Nawrocki, D'yakonov, Cibert, Tatarenko, and
  Wojtowicz}}]{Camilleri2001:PRB}
\bibinfo{author}{\bibnamefont{Camilleri}, \bibfnamefont{C.}},
  \bibinfo{author}{\bibfnamefont{F.}~\bibnamefont{Teppe}},
  \bibinfo{author}{\bibfnamefont{D.}~\bibnamefont{Scalbert}},
  \bibinfo{author}{\bibfnamefont{Y.~G.} \bibnamefont{Semenov}},
  \bibinfo{author}{\bibfnamefont{M.}~\bibnamefont{Nawrocki}},
  \bibinfo{author}{\bibfnamefont{M.~I.} \bibnamefont{D'yakonov}},
  \bibinfo{author}{\bibfnamefont{J.}~\bibnamefont{Cibert}},
  \bibinfo{author}{\bibfnamefont{S.}~\bibnamefont{Tatarenko}}, and
  \bibinfo{author}{\bibfnamefont{T.}~\bibnamefont{Wojtowicz}},
  \bibinfo{year}{2001}, {``}\bibinfo{title}{Electron and hole spin relaxation
  in modulation-doped {CdMnTe} quantum wells},{''} \bibinfo{journal}{Phys. Rev.
  B} \textbf{\bibinfo{volume}{64}},  \bibinfo{pages}{085331}.

\bibitem[{\citenamefont{Campbell} \emph{et~al.}(1967)\citenamefont{Campbell,
  Fert, and Pomeroy}}]{Campbell1967:PM}
\bibinfo{author}{\bibnamefont{Campbell}, \bibfnamefont{I.~A.}},
  \bibinfo{author}{\bibfnamefont{A.}~\bibnamefont{Fert}}, and
  \bibinfo{author}{\bibfnamefont{A.~R.} \bibnamefont{Pomeroy}},
  \bibinfo{year}{1967}, {``}\bibinfo{title}{Evidence for two current conduction
  iron},{''} \bibinfo{journal}{Phil. Mag.} \textbf{\bibinfo{volume}{15}},
  \bibinfo{pages}{977--983}.

\bibitem[{\citenamefont{Chadi} \emph{et~al.}(1976)\citenamefont{Chadi, Clark,
  and Burnham}}]{Chadi1976:PRB}
\bibinfo{author}{\bibnamefont{Chadi}, \bibfnamefont{D.~J.}},
  \bibinfo{author}{\bibfnamefont{A.~H.} \bibnamefont{Clark}}, and
  \bibinfo{author}{\bibfnamefont{R.~D.} \bibnamefont{Burnham}},
  \bibinfo{year}{1976}, {``}\bibinfo{title}{{$\Gamma_1$} conduction electron g
  factor and matrix elements in {GaAs} and {Al$_x$Ga$_{1-x}$As} alloys},{''}
  \bibinfo{journal}{Phys. Rev. B} \textbf{\bibinfo{volume}{13}},
  \bibinfo{pages}{4466--4469}.

\bibitem[{\citenamefont{Chan} \emph{et~al.}(1999)\citenamefont{Chan, Ashoori,
  Pfeiffer, and West}}]{Chan1999:PRL}
\bibinfo{author}{\bibnamefont{Chan}, \bibfnamefont{H.~B.}},
  \bibinfo{author}{\bibfnamefont{R.~C.} \bibnamefont{Ashoori}},
  \bibinfo{author}{\bibfnamefont{L.~N.} \bibnamefont{Pfeiffer}}, and
  \bibinfo{author}{\bibfnamefont{K.~W.} \bibnamefont{West}},
  \bibinfo{year}{1999}, {``}\bibinfo{title}{Tunneling into ferromagnetic
  quantum {Hall} states: {Observation} of a spin bottleneck},{''}
  \bibinfo{journal}{Phys. Rev. Lett.} \textbf{\bibinfo{volume}{83}},
  \bibinfo{pages}{3258--3261}.

\bibitem[{\citenamefont{Chazalviel}(1975)}]{Chazalviel1975:PRB}
\bibinfo{author}{\bibnamefont{Chazalviel}, \bibfnamefont{J.-N.}},
  \bibinfo{year}{1975}, {``}\bibinfo{title}{Spin relaxation of conduction
  electrons in n-type indium antimonide at low temperature},{''}
  \bibinfo{journal}{Phys. Rev. B} \textbf{\bibinfo{volume}{11}},
  \bibinfo{pages}{1555--1562}.

\bibitem[{\citenamefont{Chen} \emph{et~al.}(2001)\citenamefont{Chen, Biswas,
  \v{Z}uti\'c, Wu, Ogale, Greene, and Venkatesan}}]{Chen2001:PRB}
\bibinfo{author}{\bibnamefont{Chen}, \bibfnamefont{Z.~Y.}},
  \bibinfo{author}{\bibfnamefont{A.}~\bibnamefont{Biswas}},
  \bibinfo{author}{\bibfnamefont{I.}~\bibnamefont{\v{Z}uti\'c}},
  \bibinfo{author}{\bibfnamefont{T.}~\bibnamefont{Wu}},
  \bibinfo{author}{\bibfnamefont{S.~B.} \bibnamefont{Ogale}},
  \bibinfo{author}{\bibfnamefont{R.~L.} \bibnamefont{Greene}}, and
  \bibinfo{author}{\bibfnamefont{T.}~\bibnamefont{Venkatesan}},
  \bibinfo{year}{2001}, {``}\bibinfo{title}{Spin-polarized transport across a
  {La$_{0.7}$Sr$_{0.3}$MnO$_3$/YBa$_2$Cu$_3$O$_{7-x}$} interface: Role of
  {Andreev} bound states},{''} \bibinfo{journal}{Phys. Rev. B}
  \textbf{\bibinfo{volume}{63}},  \bibinfo{pages}{212508}.

\bibitem[{\citenamefont{Chiba} \emph{et~al.}(2004)\citenamefont{Chiba, Sato,
  Kita, Matsukura, and Ohno}}]{Chiba2004:P}
\bibinfo{author}{\bibnamefont{Chiba}, \bibfnamefont{D.}},
  \bibinfo{author}{\bibfnamefont{Y.}~\bibnamefont{Sato}},
  \bibinfo{author}{\bibfnamefont{T.}~\bibnamefont{Kita}},
  \bibinfo{author}{\bibfnamefont{F.}~\bibnamefont{Matsukura}}, and
  \bibinfo{author}{\bibfnamefont{H.}~\bibnamefont{Ohno}}, \bibinfo{year}{2004},
  {``}\bibinfo{title}{Current-driven magnetization reversal in a ferromagnetic
  semiconductor {(Ga,Mn)As/GaAs/(Ga,Mn)As} tunnel junction},{''}
  \eprint{cond-mat/0403500}.

\bibitem[{\citenamefont{Chiba} \emph{et~al.}(2003)\citenamefont{Chiba,
  Yamanouchi, Matsukura, and Ohno}}]{Chiba2003:S}
\bibinfo{author}{\bibnamefont{Chiba}, \bibfnamefont{D.}},
  \bibinfo{author}{\bibfnamefont{M.}~\bibnamefont{Yamanouchi}},
  \bibinfo{author}{\bibfnamefont{F.}~\bibnamefont{Matsukura}}, and
  \bibinfo{author}{\bibfnamefont{H.}~\bibnamefont{Ohno}}, \bibinfo{year}{2003},
  {``}\bibinfo{title}{Electrical manipulation of magnetization reversal in a
  ferromagnetic semiconductor},{''} \bibinfo{journal}{{\sl Science}}
  \textbf{\bibinfo{volume}{301}},  \bibinfo{pages}{943--945}.

\bibitem[{\citenamefont{Chtchelkanova}
  \emph{et~al.}(2003)\citenamefont{Chtchelkanova, Wolf, and {Idzerda
  (Eds.)}}}]{Chtchelkanova:2003}
\bibinfo{author}{\bibnamefont{Chtchelkanova}, \bibfnamefont{A.}},
  \bibinfo{author}{\bibfnamefont{S.}~\bibnamefont{Wolf}}, and
  \bibinfo{author}{\bibfnamefont{Y.}~\bibnamefont{{Idzerda (Eds.)}}},
  \bibinfo{year}{2003}, \emph{\bibinfo{title}{Magnetic Interactions and Spin
  Transport}} (\bibinfo{publisher}{Kluwer Academic Dordrecht/Plenum, New
  York}).

\bibitem[{\citenamefont{Chui}(1995)}]{Chui1995:PRB}
\bibinfo{author}{\bibnamefont{Chui}, \bibfnamefont{S.~T.}},
  \bibinfo{year}{1995}, {``}\bibinfo{title}{Electron interaction on the giant
  magnetoresistance in the perpendicular geometry},{''} \bibinfo{journal}{Phys.
  Rev. B} \textbf{\bibinfo{volume}{52}},  \bibinfo{pages}{R3832--R3835}.

\bibitem[{\citenamefont{Chui}(1997)}]{Chui1997:PRB}
\bibinfo{author}{\bibnamefont{Chui}, \bibfnamefont{S.~T.}},
  \bibinfo{year}{1997}, {``}\bibinfo{title}{Bias dependence in spin-polarized
  tunneling},{''} \bibinfo{journal}{Phys. Rev. B}
  \textbf{\bibinfo{volume}{55}},  \bibinfo{pages}{5600--5603}.

\bibitem[{\citenamefont{Chui and Cullen}(1995)}]{Chui1995:PRL}
\bibinfo{author}{\bibnamefont{Chui}, \bibfnamefont{S.~T.}}, and
  \bibinfo{author}{\bibfnamefont{J.~R.} \bibnamefont{Cullen}},
  \bibinfo{year}{1995}, {``}\bibinfo{title}{Spin transmission in metallic
  trilayers},{''} \bibinfo{journal}{Phys. Rev. Lett.}
  \textbf{\bibinfo{volume}{74}},  \bibinfo{pages}{2118--2121}.

\bibitem[{\citenamefont{Chun} \emph{et~al.}(2002)\citenamefont{Chun, Potashnik,
  Ku, Schiffer, and Samarth}}]{Chun2002:PRB}
\bibinfo{author}{\bibnamefont{Chun}, \bibfnamefont{S.~H.}},
  \bibinfo{author}{\bibfnamefont{S.~J.} \bibnamefont{Potashnik}},
  \bibinfo{author}{\bibfnamefont{K.~C.} \bibnamefont{Ku}},
  \bibinfo{author}{\bibfnamefont{P.}~\bibnamefont{Schiffer}}, and
  \bibinfo{author}{\bibfnamefont{N.}~\bibnamefont{Samarth}},
  \bibinfo{year}{2002}, {``}\bibinfo{title}{Spin-polarized tunneling in hybrid
  metal-semiconductor magnetic tunnel junctions},{''} \bibinfo{journal}{Phys.
  Rev. B} \textbf{\bibinfo{volume}{66}},  \bibinfo{pages}{100408}.

\bibitem[{\citenamefont{Chung} \emph{et~al.}(2002)\citenamefont{Chung, Munoz,
  Garcia, Egelhoff, and Gomez}}]{Chung2002:PRL}
\bibinfo{author}{\bibnamefont{Chung}, \bibfnamefont{S.~H.}},
  \bibinfo{author}{\bibfnamefont{M.}~\bibnamefont{Munoz}},
  \bibinfo{author}{\bibfnamefont{N.}~\bibnamefont{Garcia}},
  \bibinfo{author}{\bibfnamefont{W.~F.} \bibnamefont{Egelhoff}}, and
  \bibinfo{author}{\bibfnamefont{R.~D.} \bibnamefont{Gomez}},
  \bibinfo{year}{2002}, {``}\bibinfo{title}{Universal scaling of ballistic
  magnetoresistance in magnetic nanocontacts},{''} \bibinfo{journal}{Phys. Rev.
  Lett.} \textbf{\bibinfo{volume}{89}},  \bibinfo{pages}{287203}.

\bibitem[{\citenamefont{Chye} \emph{et~al.}(2002)\citenamefont{Chye, White,
  {Johnston-Halperin}, Gerardot, Awschalom, and Petroff}}]{Chye2002:PRB}
\bibinfo{author}{\bibnamefont{Chye}, \bibfnamefont{Y.}},
  \bibinfo{author}{\bibfnamefont{M.~E.} \bibnamefont{White}},
  \bibinfo{author}{\bibfnamefont{E.}~\bibnamefont{{Johnston-Halperin}}},
  \bibinfo{author}{\bibfnamefont{B.~D.} \bibnamefont{Gerardot}},
  \bibinfo{author}{\bibfnamefont{D.~D.} \bibnamefont{Awschalom}}, and
  \bibinfo{author}{\bibfnamefont{P.~M.} \bibnamefont{Petroff}},
  \bibinfo{year}{2002}, {``}\bibinfo{title}{Spin injection from {(Ga,Mn)As}
  into {InAs} quantum dots},{''} \bibinfo{journal}{Phys. Rev. B}
  \textbf{\bibinfo{volume}{66}},  \bibinfo{pages}{201301}.

\bibitem[{\citenamefont{Ciorga} \emph{et~al.}(2002)\citenamefont{Ciorga,
  Pioro-Ladriere, Zawadzki, Hawrylak, and Sachrajda}}]{Ciorga2002:APL}
\bibinfo{author}{\bibnamefont{Ciorga}, \bibfnamefont{M.}},
  \bibinfo{author}{\bibfnamefont{M.}~\bibnamefont{Pioro-Ladriere}},
  \bibinfo{author}{\bibfnamefont{P.}~\bibnamefont{Zawadzki}},
  \bibinfo{author}{\bibfnamefont{P.}~\bibnamefont{Hawrylak}}, and
  \bibinfo{author}{\bibfnamefont{A.~S.} \bibnamefont{Sachrajda}},
  \bibinfo{year}{2002}, {``}\bibinfo{title}{Tunable negative differential
  resistance controlled by spin blockade in single-electron transistors},{''}
  \bibinfo{journal}{Appl. Phys. Lett.} \textbf{\bibinfo{volume}{80}},
  \bibinfo{pages}{2177--2179}.

\bibitem[{\citenamefont{Ciuti}
  \emph{et~al.}(2002{\natexlab{a}})\citenamefont{Ciuti, McGuire, and
  Sham}}]{Ciuti2002:PRL}
\bibinfo{author}{\bibnamefont{Ciuti}, \bibfnamefont{C.}},
  \bibinfo{author}{\bibfnamefont{J.~P.} \bibnamefont{McGuire}}, and
  \bibinfo{author}{\bibfnamefont{L.~J.} \bibnamefont{Sham}},
  \bibinfo{year}{2002}{\natexlab{a}}, {``}\bibinfo{title}{Ferromagnetic
  imprinting of spin polarization in a semiconductor},{''}
  \bibinfo{journal}{Phys. Rev. Lett.} \textbf{\bibinfo{volume}{89}},
  \bibinfo{pages}{156601}.

\bibitem[{\citenamefont{Ciuti}
  \emph{et~al.}(2002{\natexlab{b}})\citenamefont{Ciuti, McGuire, and
  Sham}}]{Ciuti2002:APL}
\bibinfo{author}{\bibnamefont{Ciuti}, \bibfnamefont{C.}},
  \bibinfo{author}{\bibfnamefont{J.~P.} \bibnamefont{McGuire}}, and
  \bibinfo{author}{\bibfnamefont{L.~J.} \bibnamefont{Sham}},
  \bibinfo{year}{2002}{\natexlab{b}}, {``}\bibinfo{title}{Spin-dependent
  properties of a two-dimensional electron gas with ferromagnetic gates},{''}
  \bibinfo{journal}{Appl. Phys. Lett.} \textbf{\bibinfo{volume}{81}},
  \bibinfo{pages}{4781--4783}.

\bibitem[{\citenamefont{Clark} \emph{et~al.}(1975)\citenamefont{Clark, Burnham,
  Chadi, and White}}]{Clark1975:PRB}
\bibinfo{author}{\bibnamefont{Clark}, \bibfnamefont{A.~H.}},
  \bibinfo{author}{\bibfnamefont{R.~D.} \bibnamefont{Burnham}},
  \bibinfo{author}{\bibfnamefont{D.~J.} \bibnamefont{Chadi}}, and
  \bibinfo{author}{\bibfnamefont{R.~M.} \bibnamefont{White}},
  \bibinfo{year}{1975}, {``}\bibinfo{title}{Spin relaxation of conduction
  electrons in {Al$_x$Ga$_{1-x}$As}},{''} \bibinfo{journal}{Phys. Rev. B}
  \textbf{\bibinfo{volume}{12}},  \bibinfo{pages}{5758--5765}.

\bibitem[{\citenamefont{Clark and Feher}(1963)}]{Clark1963:PRL}
\bibinfo{author}{\bibnamefont{Clark}, \bibfnamefont{W.~G.}}, and
  \bibinfo{author}{\bibfnamefont{G.}~\bibnamefont{Feher}},
  \bibinfo{year}{1963}, {``}\bibinfo{title}{Nuclear polarization in {InSb} by a
  dc current},{''} \bibinfo{journal}{Phys. Rev. Lett.}
  \textbf{\bibinfo{volume}{10}},  \bibinfo{pages}{134--138}.

\bibitem[{\citenamefont{Cochrane} \emph{et~al.}(1974)\citenamefont{Cochrane,
  Plischke, and {Str\"{o}m-Olsen}}}]{Cochrane1974:PRB}
\bibinfo{author}{\bibnamefont{Cochrane}, \bibfnamefont{R.~W.}},
  \bibinfo{author}{\bibfnamefont{M.}~\bibnamefont{Plischke}}, and
  \bibinfo{author}{\bibfnamefont{J.~O.} \bibnamefont{{Str\"{o}m-Olsen}}},
  \bibinfo{year}{1974}, {``}\bibinfo{title}{Magnetization studies of
  {(GeTe)$_{1-x}$(MnTe)$_x$} pseudobinary alloys},{''} \bibinfo{journal}{Phys.
  Rev. B} \textbf{\bibinfo{volume}{9}},  \bibinfo{pages}{3013--3021}.

\bibitem[{\citenamefont{Cohen-Tannoudji and
  Kostler}(1966)}]{Cohen-Tannoudji:1966}
\bibinfo{author}{\bibnamefont{Cohen-Tannoudji}, \bibfnamefont{C.}}, and
  \bibinfo{author}{\bibfnamefont{A.}~\bibnamefont{Kostler}},
  \bibinfo{year}{1966}, {``}\bibinfo{title}{Optical Pumping},{''} in
  \emph{\bibinfo{booktitle}{Progress in Optics, Vol.25}}, edited by
  \bibinfo{editor}{\bibfnamefont{E.}~\bibnamefont{Wolf}}
  (\bibinfo{publisher}{North-Holland, Amsterdam}), ~\bibinfo{pages}{33}.

\bibitem[{\citenamefont{Cortez} \emph{et~al.}(2002)\citenamefont{Cortez, Jbeli,
  Marie, Krebs, Ferreira, Amand, Voisin, and Gerard}}]{Cortez2002:PE}
\bibinfo{author}{\bibnamefont{Cortez}, \bibfnamefont{S.}},
  \bibinfo{author}{\bibfnamefont{A.}~\bibnamefont{Jbeli}},
  \bibinfo{author}{\bibfnamefont{X.}~\bibnamefont{Marie}},
  \bibinfo{author}{\bibfnamefont{O.}~\bibnamefont{Krebs}},
  \bibinfo{author}{\bibfnamefont{R.}~\bibnamefont{Ferreira}},
  \bibinfo{author}{\bibfnamefont{T.}~\bibnamefont{Amand}},
  \bibinfo{author}{\bibfnamefont{P.}~\bibnamefont{Voisin}}, and
  \bibinfo{author}{\bibfnamefont{J.~M.} \bibnamefont{Gerard}},
  \bibinfo{year}{2002}, {``}\bibinfo{title}{Spin polarization dynamics in
  n-doped {InAs/GaAs} quantum dots},{''} \bibinfo{journal}{Physica E}
  \textbf{\bibinfo{volume}{13}},  \bibinfo{pages}{508--511}.

\bibitem[{\citenamefont{Cousins and Dupree}(1965)}]{Cousins1965:PL}
\bibinfo{author}{\bibnamefont{Cousins}, \bibfnamefont{J.~E.}}, and
  \bibinfo{author}{\bibfnamefont{R.}~\bibnamefont{Dupree}},
  \bibinfo{year}{1965}, {``}\bibinfo{title}{Electron spin resonance of
  conduction electrons in beryllium},{''} \bibinfo{journal}{Phys. Lett.}
  \textbf{\bibinfo{volume}{19}},  \bibinfo{pages}{464--465}.

\bibitem[{\citenamefont{Crooker} \emph{et~al.}(1997)\citenamefont{Crooker,
  Awschalom, Baumberg, Flack, and Samarth}}]{Crooker1997:PRB}
\bibinfo{author}{\bibnamefont{Crooker}, \bibfnamefont{S.~A.}},
  \bibinfo{author}{\bibfnamefont{D.~D.} \bibnamefont{Awschalom}},
  \bibinfo{author}{\bibfnamefont{J.~J.} \bibnamefont{Baumberg}},
  \bibinfo{author}{\bibfnamefont{F.}~\bibnamefont{Flack}}, and
  \bibinfo{author}{\bibfnamefont{N.}~\bibnamefont{Samarth}},
  \bibinfo{year}{1997}, {``}\bibinfo{title}{Optical spin resonance and
  transverse spin relaxation in magnetic semiconductor quantum wells},{''}
  \bibinfo{journal}{Phys. Rev. Lett.} \textbf{\bibinfo{volume}{56}},
  \bibinfo{pages}{7574--7588}.

\bibitem[{\citenamefont{Damay and Sienko}(1976)}]{Damay1976:PRB}
\bibinfo{author}{\bibnamefont{Damay}, \bibfnamefont{P.}}, and
  \bibinfo{author}{\bibfnamefont{M.~J.} \bibnamefont{Sienko}},
  \bibinfo{year}{1976}, {``}\bibinfo{title}{Conduction-electron spin resonance
  in metallic lithium},{''} \bibinfo{journal}{Phys. Rev. B}
  \textbf{\bibinfo{volume}{13}},  \bibinfo{pages}{603--606}.

\bibitem[{\citenamefont{Dargys}(2002)}]{Dargys2002:PRB}
\bibinfo{author}{\bibnamefont{Dargys}, \bibfnamefont{A.}},
  \bibinfo{year}{2002}, {``}\bibinfo{title}{{Luttinger}-{Kohn} Hamiltonian and
  coherent excitation of the valence-band holes},{''} \bibinfo{journal}{Phys.
  Rev. B} \textbf{\bibinfo{volume}{66}},  \bibinfo{pages}{165216}.

\bibitem[{\citenamefont{Das} \emph{et~al.}(1989)\citenamefont{Das, Miller,
  Datta, Reifenberger, Hong, Bhattacharya, Singh, and Jaffe}}]{Das1989:PRB}
\bibinfo{author}{\bibnamefont{Das}, \bibfnamefont{B.}},
  \bibinfo{author}{\bibfnamefont{D.~C.} \bibnamefont{Miller}},
  \bibinfo{author}{\bibfnamefont{S.}~\bibnamefont{Datta}},
  \bibinfo{author}{\bibfnamefont{R.}~\bibnamefont{Reifenberger}},
  \bibinfo{author}{\bibfnamefont{W.~P.} \bibnamefont{Hong}},
  \bibinfo{author}{\bibfnamefont{P.~K.} \bibnamefont{Bhattacharya}},
  \bibinfo{author}{\bibfnamefont{J.}~\bibnamefont{Singh}}, and
  \bibinfo{author}{\bibfnamefont{M.}~\bibnamefont{Jaffe}},
  \bibinfo{year}{1989}, {``}\bibinfo{title}{Evidence for spin splitting in
  {In$_x$Ga$_{1-x}$As/In$_{0.52}$Al$_{0.48}$As} heterostructures as {$B
  \rightarrow 0$}},{''} \bibinfo{journal}{Phys. Rev. B}
  \textbf{\bibinfo{volume}{39}},  \bibinfo{pages}{1411--1414}.

\bibitem[{\citenamefont{{Das Sarma}}(2001)}]{DasSarma2001:AS}
\bibinfo{author}{\bibnamefont{{Das Sarma}}, \bibfnamefont{S.}},
  \bibinfo{year}{2001}, {``}\bibinfo{title}{Spintronics},{''}
  \bibinfo{journal}{Am. Sci.} \textbf{\bibinfo{volume}{89}},
  \bibinfo{pages}{516--523}.

\bibitem[{\citenamefont{{Das Sarma}}
  \emph{et~al.}(2000{\natexlab{a}})\citenamefont{{Das Sarma}, Fabian, Hu, and
  \v{Z}uti\'{c}}}]{DasSarma2000:DRC}
\bibinfo{author}{\bibnamefont{{Das Sarma}}, \bibfnamefont{S.}},
  \bibinfo{author}{\bibfnamefont{J.}~\bibnamefont{Fabian}},
  \bibinfo{author}{\bibfnamefont{X.}~\bibnamefont{Hu}}, and
  \bibinfo{author}{\bibfnamefont{I.}~\bibnamefont{\v{Z}uti\'{c}}},
  \bibinfo{year}{2000}{\natexlab{a}}, {``}\bibinfo{title}{Issues, concepts, and
  challenges in spintronics},{''} \bibinfo{journal}{58th DRC (Device Research
  Conference) Conference Digest (IEEE, Piscataway)}  \bibinfo{pages}{{95 }--98}
  \eprint{{ cond-mat/0006369}}.

\bibitem[{\citenamefont{{Das Sarma}}
  \emph{et~al.}(2000{\natexlab{b}})\citenamefont{{Das Sarma}, Fabian, Hu, and
  \v{Z}uti\'{c}}}]{DasSarma2000:IEEE}
\bibinfo{author}{\bibnamefont{{Das Sarma}}, \bibfnamefont{S.}},
  \bibinfo{author}{\bibfnamefont{J.}~\bibnamefont{Fabian}},
  \bibinfo{author}{\bibfnamefont{X.}~\bibnamefont{Hu}}, and
  \bibinfo{author}{\bibfnamefont{I.}~\bibnamefont{\v{Z}uti\'{c}}},
  \bibinfo{year}{2000}{\natexlab{b}}, {``}\bibinfo{title}{Theoretical
  perspectives on spintronics and spin-polarized transport},{''}
  \bibinfo{journal}{IEEE Trans. Magn.} \textbf{\bibinfo{volume}{36}},
  \bibinfo{pages}{2821--2826}.

\bibitem[{\citenamefont{{Das Sarma}} \emph{et~al.}(2001)\citenamefont{{Das
  Sarma}, Fabian, Hu, and \v{Z}uti\'{c}}}]{DasSarma2001:SSC}
\bibinfo{author}{\bibnamefont{{Das Sarma}}, \bibfnamefont{S.}},
  \bibinfo{author}{\bibfnamefont{J.}~\bibnamefont{Fabian}},
  \bibinfo{author}{\bibfnamefont{X.}~\bibnamefont{Hu}}, and
  \bibinfo{author}{\bibfnamefont{I.}~\bibnamefont{\v{Z}uti\'{c}}},
  \bibinfo{year}{2001}, {``}\bibinfo{title}{Spin electronics and spin
  computation},{''} \bibinfo{journal}{Solid State Commun.}
  \textbf{\bibinfo{volume}{119}},  \bibinfo{pages}{207--215}.

\bibitem[{\citenamefont{{Das Sarma}}
  \emph{et~al.}(2000{\natexlab{c}})\citenamefont{{Das Sarma}, Fabian, Hu, and
  {\v{Z}uti\'{c}}}}]{DasSarma2000:SM}
\bibinfo{author}{\bibnamefont{{Das Sarma}}, \bibfnamefont{S.}},
  \bibinfo{author}{\bibfnamefont{J.}~\bibnamefont{Fabian}},
  \bibinfo{author}{\bibfnamefont{X.}~\bibnamefont{Hu}}, and
  \bibinfo{author}{\bibfnamefont{I.}~\bibnamefont{{\v{Z}uti\'{c}}}},
  \bibinfo{year}{2000}{\natexlab{c}}, {``}\bibinfo{title}{Spintronics: electron
  spin coherence, entanglement, and transport},{''}
  \bibinfo{journal}{Superlattices Microstruct.} \textbf{\bibinfo{volume}{27}},
  \bibinfo{pages}{289--295}.

\bibitem[{\citenamefont{{Das Sarma}} \emph{et~al.}(2003)\citenamefont{{Das
  Sarma}, Hwang, and Kaminski}}]{DasSarma2003:PRB}
\bibinfo{author}{\bibnamefont{{Das Sarma}}, \bibfnamefont{S.}},
  \bibinfo{author}{\bibfnamefont{E.~H.} \bibnamefont{Hwang}}, and
  \bibinfo{author}{\bibfnamefont{A.}~\bibnamefont{Kaminski}},
  \bibinfo{year}{2003}, {``}\bibinfo{title}{Temperature-dependent magnetization
  in diluted magnetic semiconductors},{''} \bibinfo{journal}{Phys. Rev. B}
  \textbf{\bibinfo{volume}{67}},  \bibinfo{pages}{155201}.

\bibitem[{\citenamefont{Datta and Das}(1990)}]{Datta1990:APL}
\bibinfo{author}{\bibnamefont{Datta}, \bibfnamefont{S.}}, and
  \bibinfo{author}{\bibfnamefont{B.}~\bibnamefont{Das}}, \bibinfo{year}{1990},
  {``}\bibinfo{title}{Electronic analog of the electro-optic modulator},{''}
  \bibinfo{journal}{Appl. Phys. Lett.} \textbf{\bibinfo{volume}{56}},
  \bibinfo{pages}{665--667}.

\bibitem[{\citenamefont{Daughton} \emph{et~al.}(1999)\citenamefont{Daughton,
  Pohm, Fayfield, and Smith}}]{Daughton1999:JPDAP}
\bibinfo{author}{\bibnamefont{Daughton}, \bibfnamefont{J.~M.}},
  \bibinfo{author}{\bibfnamefont{A.~V.} \bibnamefont{Pohm}},
  \bibinfo{author}{\bibfnamefont{R.~T.} \bibnamefont{Fayfield}}, and
  \bibinfo{author}{\bibfnamefont{C.~H.} \bibnamefont{Smith}},
  \bibinfo{year}{1999}, {``}\bibinfo{title}{Applications of spin dependent
  transport materials},{''} \bibinfo{journal}{J. Phys. D}
  \textbf{\bibinfo{volume}{32}},  \bibinfo{pages}{R169--R177}.

\bibitem[{\citenamefont{Davis and Bussmann}(2003)}]{Davis2003:JAP}
\bibinfo{author}{\bibnamefont{Davis}, \bibfnamefont{A.~H.}}, and
  \bibinfo{author}{\bibfnamefont{K.}~\bibnamefont{Bussmann}},
  \bibinfo{year}{2003}, {``}\bibinfo{title}{Organic luminescent devices and
  magnetoelectronics},{''} \bibinfo{journal}{J. Appl. Phys.}
  \textbf{\bibinfo{volume}{93}},  \bibinfo{pages}{7358--7360}.

\bibitem[{\citenamefont{{de Andrada e Silva}}
  \emph{et~al.}(1997)\citenamefont{{de Andrada e Silva}, {G. C. La Rocca}, and
  Bassani}}]{Silva1997:PRB}
\bibinfo{author}{\bibnamefont{{de Andrada e Silva}}, \bibfnamefont{E.~A.}},
  \bibinfo{author}{\bibnamefont{{G. C. La Rocca}}}, and
  \bibinfo{author}{\bibfnamefont{F.}~\bibnamefont{Bassani}},
  \bibinfo{year}{1997}, {``}\bibinfo{title}{Spin-orbit splitting of electronic
  states in semiconductor asymmetric quantum wells},{''}
  \bibinfo{journal}{Phys. Rev. B} \textbf{\bibinfo{volume}{55}},
  \bibinfo{pages}{16293--16299}.

\bibitem[{\citenamefont{{de Andrada e Silva} and {La
  Rocca}}(1999)}]{Silva1999:PRB}
\bibinfo{author}{\bibnamefont{{de Andrada e Silva}}, \bibfnamefont{E.~A.}}, and
  \bibinfo{author}{\bibfnamefont{G.~C.} \bibnamefont{{La Rocca}}},
  \bibinfo{year}{1999}, {``}\bibinfo{title}{Electron-spin polarization by
  resonant tunneling},{''} \bibinfo{journal}{Phys. Rev. B}
  \textbf{\bibinfo{volume}{59}},  \bibinfo{pages}{R15583--R15585}.

\bibitem[{\citenamefont{{De Boeck}} \emph{et~al.}(1996)\citenamefont{{De
  Boeck}, Oesterholt, {Van Esch}, Bender, Bruynseraede, {Van Hoof}, and
  Borghs}}]{DeBoeck1996:APL}
\bibinfo{author}{\bibnamefont{{De Boeck}}, \bibfnamefont{J.}},
  \bibinfo{author}{\bibfnamefont{R.}~\bibnamefont{Oesterholt}},
  \bibinfo{author}{\bibfnamefont{A.}~\bibnamefont{{Van Esch}}},
  \bibinfo{author}{\bibfnamefont{H.}~\bibnamefont{Bender}},
  \bibinfo{author}{\bibfnamefont{C.}~\bibnamefont{Bruynseraede}},
  \bibinfo{author}{\bibfnamefont{C.}~\bibnamefont{{Van Hoof}}}, and
  \bibinfo{author}{\bibfnamefont{G.}~\bibnamefont{Borghs}},
  \bibinfo{year}{1996}, {``}\bibinfo{title}{Nanometer-scale magnetic {MnAs}
  particles in {GaAs} grown by molecular beam epitaxy},{''}
  \bibinfo{journal}{App. Phys. Lett.} \textbf{\bibinfo{volume}{68}},
  \bibinfo{pages}{2744--2746}.

\bibitem[{\citenamefont{{De Boeck}} \emph{et~al.}(2002)\citenamefont{{De
  Boeck}, {Van Roy}, Das, Motsnyi, Liu, Lagae, Boeve, Dessein, and
  Borghs}}]{DeBoeck2002:SST}
\bibinfo{author}{\bibnamefont{{De Boeck}}, \bibfnamefont{J.}},
  \bibinfo{author}{\bibfnamefont{W.}~\bibnamefont{{Van Roy}}},
  \bibinfo{author}{\bibfnamefont{J.}~\bibnamefont{Das}},
  \bibinfo{author}{\bibfnamefont{V.~F.} \bibnamefont{Motsnyi}},
  \bibinfo{author}{\bibfnamefont{Z.}~\bibnamefont{Liu}},
  \bibinfo{author}{\bibfnamefont{L.}~\bibnamefont{Lagae}},
  \bibinfo{author}{\bibfnamefont{H.}~\bibnamefont{Boeve}},
  \bibinfo{author}{\bibfnamefont{K.}~\bibnamefont{Dessein}}, and
  \bibinfo{author}{\bibfnamefont{G.}~\bibnamefont{Borghs}},
  \bibinfo{year}{2002}, {``}\bibinfo{title}{Technology and materials issues in
  semiconductor-based magnetoelectronics},{''} \bibinfo{journal}{Semicond. Sci.
  Technol.} \textbf{\bibinfo{volume}{17}},  \bibinfo{pages}{342--354}.

\bibitem[{\citenamefont{{de Gennes}}(1989)}]{deGennes:1989}
\bibinfo{author}{\bibnamefont{{de Gennes}}, \bibfnamefont{P.~G.}},
  \bibinfo{year}{1989}, \emph{\bibinfo{title}{Superconductivty of Metals and
  Alloys}} (\bibinfo{publisher}{Addison-Wesley, Reading MA}).

\bibitem[{\citenamefont{{de Gennes} and {Saint James}}(1963)}]{deGennes1963:PL}
\bibinfo{author}{\bibnamefont{{de Gennes}}, \bibfnamefont{P.~G.}}, and
  \bibinfo{author}{\bibfnamefont{D.}~\bibnamefont{{Saint James}}},
  \bibinfo{year}{1963}, {``}\bibinfo{title}{Elementary excitations in the
  vicinity of a normal metal-superconducting metal contact},{''}
  \bibinfo{journal}{Phys. Lett.} \textbf{\bibinfo{volume}{4}},
  \bibinfo{pages}{151--152}.

\bibitem[{\citenamefont{{de Groot}}
  \emph{et~al.}(1983{\natexlab{a}})\citenamefont{{de Groot}, Janner, and
  Mueller}}]{deGroot1983:PA}
\bibinfo{author}{\bibnamefont{{de Groot}}, \bibfnamefont{R.~A.}},
  \bibinfo{author}{\bibfnamefont{A.~G.} \bibnamefont{Janner}}, and
  \bibinfo{author}{\bibfnamefont{F.~M.} \bibnamefont{Mueller}},
  \bibinfo{year}{1983}{\natexlab{a}} \bibinfo{journal}{{Dutch patent 8, 300,
  602, Canadian patent 1,216,375}} .

\bibitem[{\citenamefont{{de Groot}}
  \emph{et~al.}(1983{\natexlab{b}})\citenamefont{{de Groot}, Mueller, van
  Engen, and Buschow}}]{deGroot1983:PRL}
\bibinfo{author}{\bibnamefont{{de Groot}}, \bibfnamefont{R.~A.}},
  \bibinfo{author}{\bibfnamefont{F.~M.} \bibnamefont{Mueller}},
  \bibinfo{author}{\bibfnamefont{P.~G.} \bibnamefont{van Engen}}, and
  \bibinfo{author}{\bibfnamefont{K.~H.~J.} \bibnamefont{Buschow}},
  \bibinfo{year}{1983}{\natexlab{b}}, {``}\bibinfo{title}{New class of
  materials: {Half}-metallic ferromagnets},{''} \bibinfo{journal}{Phys. Rev.
  Lett.} \textbf{\bibinfo{volume}{50}},  \bibinfo{pages}{2024--2027}.

\bibitem[{\citenamefont{{de Jong}}(1994)}]{deJong1994:PRB}
\bibinfo{author}{\bibnamefont{{de Jong}}, \bibfnamefont{M.~J.~M.}},
  \bibinfo{year}{1994}, {``}\bibinfo{title}{Transition from {Sharvin to Drude}
  resistance in high-mobility wires},{''} \bibinfo{journal}{Phys. Rev. B}
  \textbf{\bibinfo{volume}{49}},  \bibinfo{pages}{7778--7781}.

\bibitem[{\citenamefont{{de Sousa} and {Das
  Sarma}}(2003{\natexlab{a}})}]{Sousa2003:PRB}
\bibinfo{author}{\bibnamefont{{de Sousa}}, \bibfnamefont{R.}}, and
  \bibinfo{author}{\bibfnamefont{S.}~\bibnamefont{{Das Sarma}}},
  \bibinfo{year}{2003}{\natexlab{a}}, {``}\bibinfo{title}{Electron spin
  coherence in semiconductors: {Considerations} for a spin-based solid-state
  quantum computer architecture},{''} \bibinfo{journal}{Phys. Rev. B}
  \textbf{\bibinfo{volume}{67}},  \bibinfo{pages}{033301}.

\bibitem[{\citenamefont{{de Sousa} and {Das
  Sarma}}(2003{\natexlab{b}})}]{deSousa2003:PRBc}
\bibinfo{author}{\bibnamefont{{de Sousa}}, \bibfnamefont{R.}}, and
  \bibinfo{author}{\bibfnamefont{S.}~\bibnamefont{{Das Sarma}}},
  \bibinfo{year}{2003}{\natexlab{b}}, {``}\bibinfo{title}{Gate control of spin
  dynamics in {III-V} semiconductor quantum dots},{''} \bibinfo{journal}{Phys.
  Rev. B} \textbf{\bibinfo{volume}{68}},  \bibinfo{pages}{155330}.

\bibitem[{\citenamefont{{de Sousa} and {Das
  Sarma}}(2003{\natexlab{c}})}]{Sousa2002:P}
\bibinfo{author}{\bibnamefont{{de Sousa}}, \bibfnamefont{R.}}, and
  \bibinfo{author}{\bibfnamefont{S.}~\bibnamefont{{Das Sarma}}},
  \bibinfo{year}{2003}{\natexlab{c}}, {``}\bibinfo{title}{Theory of
  nuclear-induced spectral diffusion: Spin decoherence of phosphorous donors in
  {Si} and {GaAs} quantum dots},{''} \bibinfo{journal}{Phys. Rev. B}
  \textbf{\bibinfo{volume}{68}},  \bibinfo{pages}{115322}.

\bibitem[{\citenamefont{{De Teresa}} \emph{et~al.}(1999)\citenamefont{{De
  Teresa}, Barthelemy, Fert, Contour, Lyonnet, Montaigne, Seneor, and
  Vaur{\`{e}}s}}]{DeTeresa1999:PRL}
\bibinfo{author}{\bibnamefont{{De Teresa}}, \bibfnamefont{J.}},
  \bibinfo{author}{\bibfnamefont{A.}~\bibnamefont{Barthelemy}},
  \bibinfo{author}{\bibfnamefont{A.}~\bibnamefont{Fert}},
  \bibinfo{author}{\bibfnamefont{J.}~\bibnamefont{Contour}},
  \bibinfo{author}{\bibfnamefont{R.}~\bibnamefont{Lyonnet}},
  \bibinfo{author}{\bibfnamefont{F.}~\bibnamefont{Montaigne}},
  \bibinfo{author}{\bibfnamefont{P.}~\bibnamefont{Seneor}}, and
  \bibinfo{author}{\bibfnamefont{A.}~\bibnamefont{Vaur{\`{e}}s}},
  \bibinfo{year}{1999}, {``}\bibinfo{title}{Inverse tunnel magnetoresistance in
  {Co/SrTiO$_3$/La$_{0.7}$Sr$_{0.3}$MnO$_3$}: {New} ideas on spin-polarized
  tunneling},{''} \bibinfo{journal}{Phys. Rev. Lett.}
  \textbf{\bibinfo{volume}{82}},  \bibinfo{pages}{4288--4291}.

\bibitem[{\citenamefont{Dediu} \emph{et~al.}(2002)\citenamefont{Dediu, Murgia,
  Matacotta, Taliani, and Barbanera}}]{Dediu2002:SSC}
\bibinfo{author}{\bibnamefont{Dediu}, \bibfnamefont{V.}},
  \bibinfo{author}{\bibfnamefont{M.}~\bibnamefont{Murgia}},
  \bibinfo{author}{\bibfnamefont{F.~C.} \bibnamefont{Matacotta}},
  \bibinfo{author}{\bibfnamefont{C.}~\bibnamefont{Taliani}}, and
  \bibinfo{author}{\bibfnamefont{S.}~\bibnamefont{Barbanera}},
  \bibinfo{year}{2002}, {``}\bibinfo{title}{Room temperature spin polarized
  injection in organic semiconductor},{''} \bibinfo{journal}{Solid State
  Commun.} \textbf{\bibinfo{volume}{122}},  \bibinfo{pages}{181--184}.

\bibitem[{\citenamefont{Demler} \emph{et~al.}(1997)\citenamefont{Demler,
  Arnold, and Beasley}}]{Demler1997:PRB}
\bibinfo{author}{\bibnamefont{Demler}, \bibfnamefont{E.~A.}},
  \bibinfo{author}{\bibfnamefont{G.~B.} \bibnamefont{Arnold}}, and
  \bibinfo{author}{\bibfnamefont{M.~R.} \bibnamefont{Beasley}},
  \bibinfo{year}{1997}, {``}\bibinfo{title}{Superconducting proximity effects
  in magnetic metals},{''} \bibinfo{journal}{Phys. Rev. B}
  \textbf{\bibinfo{volume}{55}},  \bibinfo{pages}{15174--15182}.

\bibitem[{\citenamefont{Deshmukh and Ralph}(2002)}]{Deshmukh2002:PRL}
\bibinfo{author}{\bibnamefont{Deshmukh}, \bibfnamefont{M.~M.}}, and
  \bibinfo{author}{\bibfnamefont{D.~C.} \bibnamefont{Ralph}},
  \bibinfo{year}{2002}, {``}\bibinfo{title}{Using single quantum states as spin
  filters to study spin polarization in ferromagnets},{''}
  \bibinfo{journal}{Phys. Rev. Lett.} \textbf{\bibinfo{volume}{89}},
  \bibinfo{pages}{266803}.

\bibitem[{\citenamefont{Dessein} \emph{et~al.}(2000)\citenamefont{Dessein,
  Boeve, Kumar, Boeck, Lodder, Delaey, and Borghs}}]{Dessein2000:JAP}
\bibinfo{author}{\bibnamefont{Dessein}, \bibfnamefont{K.}},
  \bibinfo{author}{\bibfnamefont{H.}~\bibnamefont{Boeve}},
  \bibinfo{author}{\bibfnamefont{P.~S.~A.} \bibnamefont{Kumar}},
  \bibinfo{author}{\bibfnamefont{J.~D.} \bibnamefont{Boeck}},
  \bibinfo{author}{\bibfnamefont{J.~C.} \bibnamefont{Lodder}},
  \bibinfo{author}{\bibfnamefont{L.}~\bibnamefont{Delaey}}, and
  \bibinfo{author}{\bibfnamefont{G.}~\bibnamefont{Borghs}},
  \bibinfo{year}{2000}, {``}\bibinfo{title}{Evaluation of vacuum bonded
  {GaAs/Si} spin-valve transistors},{''} \bibinfo{journal}{J. Appl. Phys.}
  \textbf{\bibinfo{volume}{87}},  \bibinfo{pages}{5155--5157}.

\bibitem[{\citenamefont{Deutscher and Feinberg}(2000)}]{Deutscher2000:APL}
\bibinfo{author}{\bibnamefont{Deutscher}, \bibfnamefont{G.}}, and
  \bibinfo{author}{\bibfnamefont{D.}~\bibnamefont{Feinberg}},
  \bibinfo{year}{2000}, {``}\bibinfo{title}{Coupling
  superconducting-ferromagnetic point contacts by {Andreev} reflections},{''}
  \bibinfo{journal}{Appl. Phys. Lett.} \textbf{\bibinfo{volume}{76}},
  \bibinfo{pages}{487--489}.

\bibitem[{\citenamefont{Dieny} \emph{et~al.}(1991)\citenamefont{Dieny,
  Speriosu, Parkin, Gurney, Wilhoit, and Maur}}]{Dieny1991:PRB}
\bibinfo{author}{\bibnamefont{Dieny}, \bibfnamefont{B.}},
  \bibinfo{author}{\bibfnamefont{V.~S.} \bibnamefont{Speriosu}},
  \bibinfo{author}{\bibfnamefont{S.~S.~P.} \bibnamefont{Parkin}},
  \bibinfo{author}{\bibfnamefont{B.~A.} \bibnamefont{Gurney}},
  \bibinfo{author}{\bibfnamefont{D.~R.} \bibnamefont{Wilhoit}}, and
  \bibinfo{author}{\bibfnamefont{D.}~\bibnamefont{Maur}}, \bibinfo{year}{1991},
  {``}\bibinfo{title}{Giant magnetoresistive in soft ferromagnetic
  multilayers},{''} \bibinfo{journal}{Phys. Rev. B}
  \textbf{\bibinfo{volume}{43}},  \bibinfo{pages}{1297--1300}.

\bibitem[{\citenamefont{Dietl}(1994)}]{Dietl:1994}
\bibinfo{author}{\bibnamefont{Dietl}, \bibfnamefont{T.}}, \bibinfo{year}{1994},
  {``}\bibinfo{title}{Diluted Magnetic Semiconductors},{''} in
  \emph{\bibinfo{booktitle}{Handbook of Semiconductors, {Vol. 3}}}, edited by
  \bibinfo{editor}{\bibfnamefont{T.~S.} \bibnamefont{Moss}} and
  \bibinfo{editor}{\bibfnamefont{S.}~\bibnamefont{Mahajan}}
  (\bibinfo{publisher}{{North-Holland}, New York}),  \bibinfo{pages}{1251}.

\bibitem[{\citenamefont{Dietl}(2002)}]{Dietl2002:SST}
\bibinfo{author}{\bibnamefont{Dietl}, \bibfnamefont{T.}}, \bibinfo{year}{2002},
  {``}\bibinfo{title}{Ferromagnetic semiconductors},{''}
  \bibinfo{journal}{Semicond. Sci. Technol.} \textbf{\bibinfo{volume}{17}},
  \bibinfo{pages}{377--392}.

\bibitem[{\citenamefont{DiVincenzo}(1995)}]{DiVincenzo1995:S}
\bibinfo{author}{\bibnamefont{DiVincenzo}, \bibfnamefont{D.~P.}},
  \bibinfo{year}{1995}, {``}\bibinfo{title}{Quantum computation},{''}
  \bibinfo{journal}{{\sl Science}} \textbf{\bibinfo{volume}{270}},
  \bibinfo{pages}{255--261}.

\bibitem[{\citenamefont{{DiVincenzo}}(1999)}]{DiVincenzo1999:JAP}
\bibinfo{author}{\bibnamefont{{DiVincenzo}}, \bibfnamefont{D.~P.}},
  \bibinfo{year}{1999}, {``}\bibinfo{title}{Quantum computing and single-qubit
  measurements using the spin-filter effect},{''} \bibinfo{journal}{J. Appl.
  Phys.} \textbf{\bibinfo{volume}{85}},  \bibinfo{pages}{4785--4787}.

\bibitem[{\citenamefont{DiVincenzo}(2000)}]{DiVincenzo2000:FP}
\bibinfo{author}{\bibnamefont{DiVincenzo}, \bibfnamefont{D.~P.}},
  \bibinfo{year}{2000}, {``}\bibinfo{title}{The physical implementation of
  quantum computation},{''} \bibinfo{journal}{Fortschr. Phys.}
  \textbf{\bibinfo{volume}{48}},  \bibinfo{pages}{771--783}.

\bibitem[{\citenamefont{Dobers} \emph{et~al.}(1988)\citenamefont{Dobers, {von
  Klitzing}, and Weimann}}]{Dobers1988:PRB}
\bibinfo{author}{\bibnamefont{Dobers}, \bibfnamefont{M.}},
  \bibinfo{author}{\bibfnamefont{K.}~\bibnamefont{{von Klitzing}}}, and
  \bibinfo{author}{\bibfnamefont{G.}~\bibnamefont{Weimann}},
  \bibinfo{year}{1988}, {``}\bibinfo{title}{Electron-spin resonance in the
  two-dimensional electron gas of {GaAs-Al$_x$Ga$_{1-x}$As}
  heterostructures},{''} \bibinfo{journal}{Phys. Rev. B}
  \textbf{\bibinfo{volume}{38}},  \bibinfo{pages}{5453--5456}.

\bibitem[{\citenamefont{Dong} \emph{et~al.}(1997)\citenamefont{Dong, Ramesh,
  Venkatesan, Johnson, Chen, Pai, Talyansky, Sharma, Shreekala, Lobb, and
  Greene}}]{Dong1997:APL}
\bibinfo{author}{\bibnamefont{Dong}, \bibfnamefont{Z.~W.}},
  \bibinfo{author}{\bibfnamefont{R.}~\bibnamefont{Ramesh}},
  \bibinfo{author}{\bibfnamefont{T.}~\bibnamefont{Venkatesan}},
  \bibinfo{author}{\bibfnamefont{M.}~\bibnamefont{Johnson}},
  \bibinfo{author}{\bibfnamefont{Z.~Y.} \bibnamefont{Chen}},
  \bibinfo{author}{\bibfnamefont{S.~P.} \bibnamefont{Pai}},
  \bibinfo{author}{\bibfnamefont{V.}~\bibnamefont{Talyansky}},
  \bibinfo{author}{\bibfnamefont{R.~P.} \bibnamefont{Sharma}},
  \bibinfo{author}{\bibfnamefont{R.}~\bibnamefont{Shreekala}},
  \bibinfo{author}{\bibfnamefont{C.~J.} \bibnamefont{Lobb}}, and
  \bibinfo{author}{\bibfnamefont{R.~L.} \bibnamefont{Greene}},
  \bibinfo{year}{1997}, {``}\bibinfo{title}{Spin-polarized quasiparticle
  injection devices using
  {Au/YBa$_2$Cu$_3$O$_7$/LaAlO$_3$/Nd$_{0.7}$Sr$_{0.3}$MnO$_3$}
  heterostructures},{''} \bibinfo{journal}{Appl. Phys. Lett.}
  \textbf{\bibinfo{volume}{71}},  \bibinfo{pages}{1718--1720}.

\bibitem[{\citenamefont{Dresselhaus}(1955)}]{Dresselhaus1955:PR}
\bibinfo{author}{\bibnamefont{Dresselhaus}, \bibfnamefont{G.}},
  \bibinfo{year}{1955}, {``}\bibinfo{title}{Spin-orbit coupling effects in zinc
  blende structures},{''} \bibinfo{journal}{Phys. Rev.}
  \textbf{\bibinfo{volume}{100}},  \bibinfo{pages}{580--586}.

\bibitem[{\citenamefont{Dugaev} \emph{et~al.}(2003)\citenamefont{Dugaev,
  Vygranenko, Vieira, Litvinov, and Barnas}}]{Dugaev2003:PE}
\bibinfo{author}{\bibnamefont{Dugaev}, \bibfnamefont{V.~K.}},
  \bibinfo{author}{\bibfnamefont{Y.}~\bibnamefont{Vygranenko}},
  \bibinfo{author}{\bibfnamefont{M.}~\bibnamefont{Vieira}},
  \bibinfo{author}{\bibfnamefont{V.~I.} \bibnamefont{Litvinov}}, and
  \bibinfo{author}{\bibfnamefont{J.}~\bibnamefont{Barnas}},
  \bibinfo{year}{2003}, {``}\bibinfo{title}{Modeling of magnetically controlled
  {Si-based} optoelectronic devices},{''} \bibinfo{journal}{Physica E}
  \textbf{\bibinfo{volume}{16}},  \bibinfo{pages}{558--562}.

\bibitem[{\citenamefont{Duke}(1969)}]{Duke:1969}
\bibinfo{author}{\bibnamefont{Duke}, \bibfnamefont{C.~B.}},
  \bibinfo{year}{1969}, {``}\bibinfo{title}{Tunneling in Solids},{''} in
  \emph{\bibinfo{booktitle}{Solid State Physics, Supplement 10}}, edited by
  \bibinfo{editor}{\bibfnamefont{F.}~\bibnamefont{Seitz}},
  \bibinfo{editor}{\bibfnamefont{D.}~\bibnamefont{Turnbull}}, and
  \bibinfo{editor}{\bibfnamefont{H.}~\bibnamefont{Ehrenreich}}
  (\bibinfo{publisher}{Academic, New York}).

\bibitem[{\citenamefont{Dupree and Holland}(1967)}]{Dupree1967:PSS}
\bibinfo{author}{\bibnamefont{Dupree}, \bibfnamefont{R.}}, and
  \bibinfo{author}{\bibfnamefont{B.~W.} \bibnamefont{Holland}},
  \bibinfo{year}{1967}, {``}\bibinfo{title}{The range of g factors and the
  breakdown of motional narrowing in conduction electron spin resonance},{''}
  \bibinfo{journal}{Phys. Status Solidi} \textbf{\bibinfo{volume}{24}},
  \bibinfo{pages}{275--279}.

\bibitem[{\citenamefont{D'yakonov and Kachorovskii}(1986)}]{Dyakonov1986:SPS}
\bibinfo{author}{\bibnamefont{D'yakonov}, \bibfnamefont{M.~I.}}, and
  \bibinfo{author}{\bibfnamefont{V.~Y.} \bibnamefont{Kachorovskii}},
  \bibinfo{year}{1986}, {``}\bibinfo{title}{Spin relaxation of two-dimensional
  electrons in noncentrosymmetric semiconductors},{''} \bibinfo{journal}{Fiz.
  Tekh. Poluprovodn.} \textbf{\bibinfo{volume}{20}},  \bibinfo{pages}{178--181}
  \bibinfo{note}{[Sov. Phys. Semicond. {\bf 20}, 110-112 (1986)]}.

\bibitem[{\citenamefont{D'yakonov} \emph{et~al.}(1986)\citenamefont{D'yakonov,
  Marushchak, Perel', and Titkov}}]{Dyakonov1986:SPJETP}
\bibinfo{author}{\bibnamefont{D'yakonov}, \bibfnamefont{M.~I.}},
  \bibinfo{author}{\bibfnamefont{V.~A.} \bibnamefont{Marushchak}},
  \bibinfo{author}{\bibfnamefont{V.~I.} \bibnamefont{Perel'}}, and
  \bibinfo{author}{\bibfnamefont{A.~N.} \bibnamefont{Titkov}},
  \bibinfo{year}{1986}, {``}\bibinfo{title}{The effect of strain on the spin
  relaxation of conduction electrons in {III-V} semiconductors},{''}
  \bibinfo{journal}{Zh. Eksp. Teor. Fiz.} \textbf{\bibinfo{volume}{90}},
  \bibinfo{pages}{1123--1133} \bibinfo{note}{[Sov. Phys. JETP {\bf 63}, 655-661
  (1986)]}.

\bibitem[{\citenamefont{D'yakonov and
  Perel'}(1971{\natexlab{a}})}]{Dyakonov1971:PL}
\bibinfo{author}{\bibnamefont{D'yakonov}, \bibfnamefont{M.~I.}}, and
  \bibinfo{author}{\bibfnamefont{V.~I.} \bibnamefont{Perel'}},
  \bibinfo{year}{1971}{\natexlab{a}}, {``}\bibinfo{title}{Current-induced spin
  orientation of electrons in semiconductors},{''} \bibinfo{journal}{Phys.
  Lett. A} \textbf{\bibinfo{volume}{35}},  \bibinfo{pages}{459--460}.

\bibitem[{\citenamefont{D'yakonov and
  Perel'}(1971{\natexlab{b}})}]{Dyakonov1971:JETPL}
\bibinfo{author}{\bibnamefont{D'yakonov}, \bibfnamefont{M.~I.}}, and
  \bibinfo{author}{\bibfnamefont{V.~I.} \bibnamefont{Perel'}},
  \bibinfo{year}{1971}{\natexlab{b}}, {``}\bibinfo{title}{Feasibility of
  optical orientation of equilibrium electrons in semiconductors},{''},
  \textbf{\bibinfo{volume}{13}},  \bibinfo{pages}{206--208}
  \bibinfo{note}{[JETP Lett. {\bf 13}, 144-146 (1971)]}.

\bibitem[{\citenamefont{D'yakonov and
  Perel'}(1971{\natexlab{c}})}]{Dyakonov1971:JETPLa}
\bibinfo{author}{\bibnamefont{D'yakonov}, \bibfnamefont{M.~I.}}, and
  \bibinfo{author}{\bibfnamefont{V.~I.} \bibnamefont{Perel'}},
  \bibinfo{year}{1971}{\natexlab{c}}, {``}\bibinfo{title}{Possibility of
  orienting electron spins with current},{''} \bibinfo{journal}{Zh. Eksp. Teor.
  Fiz. Pisma Red.} \textbf{\bibinfo{volume}{13}},  \bibinfo{pages}{657--660}
  \bibinfo{note}{[JETP Lett. {\bf 13}, 467-469 (1971)]}.

\bibitem[{\citenamefont{D'yakonov and
  Perel'}(1971{\natexlab{d}})}]{Dyakonov1971:SPJETP}
\bibinfo{author}{\bibnamefont{D'yakonov}, \bibfnamefont{M.~I.}}, and
  \bibinfo{author}{\bibfnamefont{V.~I.} \bibnamefont{Perel'}},
  \bibinfo{year}{1971}{\natexlab{d}}, {``}\bibinfo{title}{Spin orientation of
  electrons associated with the interband absorption of light in
  semiconductors},{''} \bibinfo{journal}{Zh. Eksp. Teor. Fiz.}
  \textbf{\bibinfo{volume}{60}},  \bibinfo{pages}{1954--1965}
  \bibinfo{note}{[Sov. Phys. JETP {\bf 33}, 1053-1059 (1971)]}.

\bibitem[{\citenamefont{D'yakonov and
  Perel'}(1971{\natexlab{e}})}]{Dyakonov1972:SPSS}
\bibinfo{author}{\bibnamefont{D'yakonov}, \bibfnamefont{M.~I.}}, and
  \bibinfo{author}{\bibfnamefont{V.~I.} \bibnamefont{Perel'}},
  \bibinfo{year}{1971}{\natexlab{e}}, {``}\bibinfo{title}{Spin relaxation of
  conduction electrons in noncentrosymmetric semiconductors},{''}
  \bibinfo{journal}{Fiz. Tverd. Tela} \textbf{\bibinfo{volume}{13}},
  \bibinfo{pages}{3581--3585} \bibinfo{note}{[Sov. Phys. Solid State {\bf 13},
  3023-3026 (1971)]}.

\bibitem[{\citenamefont{D'yakonov and
  Perel'}(1973{\natexlab{a}})}]{Dyakonov1974:SPS}
\bibinfo{author}{\bibnamefont{D'yakonov}, \bibfnamefont{M.~I.}}, and
  \bibinfo{author}{\bibfnamefont{V.~I.} \bibnamefont{Perel'}},
  \bibinfo{year}{1973}{\natexlab{a}}, {``}\bibinfo{title}{Influence of an
  electric field and deformation on the optical orientation in
  semiconductors},{''} \bibinfo{journal}{Fiz. Tekh. Poluprovodn.}
  \textbf{\bibinfo{volume}{7}},  \bibinfo{pages}{2335--2339}
  \bibinfo{note}{[Sov. Phys. Semicond. {\bf 7}, 1551-1553 (1974)]}.

\bibitem[{\citenamefont{D'yakonov and
  Perel'}(1973{\natexlab{b}})}]{Dyakonov1974:SPJETP}
\bibinfo{author}{\bibnamefont{D'yakonov}, \bibfnamefont{M.~I.}}, and
  \bibinfo{author}{\bibfnamefont{V.~I.} \bibnamefont{Perel'}},
  \bibinfo{year}{1973}{\natexlab{b}}, {``}\bibinfo{title}{Optical orientation
  in a system of electrons and lattice nuclei in semiconductors.
  {Theory.}},{''} \bibinfo{journal}{Zh. Eksp. Teor. Fiz.}
  \textbf{\bibinfo{volume}{38}},  \bibinfo{pages}{362--376}
  \bibinfo{note}{[Sov. Phys. JETP {\bf 38}, 177-183 (1973)]}.

\bibitem[{\citenamefont{D'yakonov and Perel'}(1976)}]{D'yakonov1976:FTP}
\bibinfo{author}{\bibnamefont{D'yakonov}, \bibfnamefont{M.~I.}}, and
  \bibinfo{author}{\bibfnamefont{V.~I.} \bibnamefont{Perel'}},
  \bibinfo{year}{1976}, {``}\bibinfo{title}{Theory of the {Hanle} effect in
  optical orientation of electrons in n-type semiconductors},{''}
  \bibinfo{journal}{Fiz. Tekh. Poluprovodn.} \textbf{\bibinfo{volume}{10}},
  \bibinfo{pages}{350--353} \bibinfo{note}{[Sov. Phys. Semicond. {\bf 10},
  208-210 (1976)]}.

\bibitem[{\citenamefont{D'yakonov and
  Perel'}(1984{\natexlab{a}})}]{Dyakonov:1984}
\bibinfo{author}{\bibnamefont{D'yakonov}, \bibfnamefont{M.~I.}}, and
  \bibinfo{author}{\bibfnamefont{V.~I.} \bibnamefont{Perel'}},
  \bibinfo{year}{1984}{\natexlab{a}}, {``}\bibinfo{title}{Theory of Optical
  Spin Orientation of Electrons and Nuclei in Semiconductors},{''} in
  \emph{\bibinfo{booktitle}{Optical Orientation, Modern Problems in Condensed
  Matter Science, Vol. 8}}, edited by
  \bibinfo{editor}{\bibfnamefont{F.}~\bibnamefont{Meier}} and
  \bibinfo{editor}{\bibfnamefont{B.~P.} \bibnamefont{Zakharchenya}}
  (\bibinfo{publisher}{North-Holland, Amsterdam}),  \bibinfo{pages}{11--71}.

\bibitem[{\citenamefont{D'yakonov and
  Perel'}(1984{\natexlab{b}})}]{Dyakonov:1984b}
\bibinfo{author}{\bibnamefont{D'yakonov}, \bibfnamefont{M.~I.}}, and
  \bibinfo{author}{\bibfnamefont{V.~I.} \bibnamefont{Perel'}},
  \bibinfo{year}{1984}{\natexlab{b}}, {``}\bibinfo{title}{Theory of optical
  spin orientation of electrons and nuclei in semiconductors},{''} in
  \emph{\bibinfo{booktitle}{Optical Orientation, Modern Problems in Condensed
  Matter Science, Vol. 8}}, edited by
  \bibinfo{editor}{\bibfnamefont{F.}~\bibnamefont{Meier}} and
  \bibinfo{editor}{\bibfnamefont{B.~P.} \bibnamefont{Zakharchenya}}
  (\bibinfo{publisher}{North-Holland, Amsterdam}), ~\bibinfo{pages}{40}.

\bibitem[{\citenamefont{D'yakonov} \emph{et~al.}(1974)\citenamefont{D'yakonov,
  Perel', Berkovits, and Safarov}}]{D'yakonov1975:SPJETP}
\bibinfo{author}{\bibnamefont{D'yakonov}, \bibfnamefont{M.~I.}},
  \bibinfo{author}{\bibfnamefont{V.~I.} \bibnamefont{Perel'}},
  \bibinfo{author}{\bibfnamefont{V.~L.} \bibnamefont{Berkovits}}, and
  \bibinfo{author}{\bibfnamefont{V.~I.} \bibnamefont{Safarov}},
  \bibinfo{year}{1974}, {``}\bibinfo{title}{Optical effects due to polarization
  of nuclei in semiconductors},{''} \bibinfo{journal}{Zh. Eksp. Teor. Fiz.}
  \textbf{\bibinfo{volume}{67}},  \bibinfo{pages}{1912--1924}
  \bibinfo{note}{[Sov. Phys. JETP {\bf 40}, 950-955 (1975)]}.

\bibitem[{\citenamefont{Dyson}(1955)}]{Dyson1955:PR}
\bibinfo{author}{\bibnamefont{Dyson}, \bibfnamefont{F.~J.}},
  \bibinfo{year}{1955}, {``}\bibinfo{title}{Electron spin resonance absorption
  in metals. {II.} {Theory} of electron diffusion and the skin effect},{''}
  \bibinfo{journal}{Phys. Rev.} \textbf{\bibinfo{volume}{98}},
  \bibinfo{pages}{349--359}.

\bibitem[{\citenamefont{Dzero} \emph{et~al.}(2003)\citenamefont{Dzero,
  {Gor'kov}, Zvezdin, and Zvezdin}}]{Dzero2003:PRB}
\bibinfo{author}{\bibnamefont{Dzero}, \bibfnamefont{M.}},
  \bibinfo{author}{\bibfnamefont{L.~P.} \bibnamefont{{Gor'kov}}},
  \bibinfo{author}{\bibfnamefont{A.~K.} \bibnamefont{Zvezdin}}, and
  \bibinfo{author}{\bibfnamefont{K.~A.} \bibnamefont{Zvezdin}},
  \bibinfo{year}{2003}, {``}\bibinfo{title}{Even-odd effects in
  magnetoresistance of ferromagnetic domain walls},{''} \bibinfo{journal}{Phys.
  Rev. B} \textbf{\bibinfo{volume}{67}},  \bibinfo{pages}{100402}.

\bibitem[{\citenamefont{Dzhioev}
  \emph{et~al.}(2002{\natexlab{a}})\citenamefont{Dzhioev, Kavokin, Korenev,
  Lazarev, Meltser, Stepanova, Zakharchenya, Gammon, and
  Katzer}}]{Dzhioev2002a:PRB}
\bibinfo{author}{\bibnamefont{Dzhioev}, \bibfnamefont{R.~I.}},
  \bibinfo{author}{\bibfnamefont{K.~V.} \bibnamefont{Kavokin}},
  \bibinfo{author}{\bibfnamefont{V.~L.} \bibnamefont{Korenev}},
  \bibinfo{author}{\bibfnamefont{M.~V.} \bibnamefont{Lazarev}},
  \bibinfo{author}{\bibfnamefont{B.~Y.} \bibnamefont{Meltser}},
  \bibinfo{author}{\bibfnamefont{M.~N.} \bibnamefont{Stepanova}},
  \bibinfo{author}{\bibfnamefont{B.~P.} \bibnamefont{Zakharchenya}},
  \bibinfo{author}{\bibfnamefont{D.}~\bibnamefont{Gammon}}, and
  \bibinfo{author}{\bibfnamefont{D.~S.} \bibnamefont{Katzer}},
  \bibinfo{year}{2002}{\natexlab{a}}, {``}\bibinfo{title}{Low-temperature spin
  relaxation in n-type {GaAs}},{''} \bibinfo{journal}{Phys. Rev. B}
  \textbf{\bibinfo{volume}{66}},  \bibinfo{pages}{245204}.

\bibitem[{\citenamefont{Dzhioev}
  \emph{et~al.}(2002{\natexlab{b}})\citenamefont{Dzhioev, Korenev, Merkulov,
  Zakharchenya, Gammon, Efros, and Katzer}}]{Dzhioev2002:PRL}
\bibinfo{author}{\bibnamefont{Dzhioev}, \bibfnamefont{R.~I.}},
  \bibinfo{author}{\bibfnamefont{V.~L.} \bibnamefont{Korenev}},
  \bibinfo{author}{\bibfnamefont{I.~A.} \bibnamefont{Merkulov}},
  \bibinfo{author}{\bibfnamefont{B.~P.} \bibnamefont{Zakharchenya}},
  \bibinfo{author}{\bibfnamefont{D.}~\bibnamefont{Gammon}},
  \bibinfo{author}{\bibfnamefont{A.~L.} \bibnamefont{Efros}}, and
  \bibinfo{author}{\bibfnamefont{D.~S.} \bibnamefont{Katzer}},
  \bibinfo{year}{2002}{\natexlab{b}}, {``}\bibinfo{title}{Manipulation of the
  spin memory of electrons in n-{GaAs}},{''} \bibinfo{journal}{Phys. Rev.
  Lett.} \textbf{\bibinfo{volume}{88}},  \bibinfo{pages}{256801}.

\bibitem[{\citenamefont{Dzhioev}
  \emph{et~al.}(2002{\natexlab{c}})\citenamefont{Dzhioev, Korenev,
  Zakharchenya, Gammon, Bracker, Tischler, and Katzer}}]{Dzhioev2002:PRB}
\bibinfo{author}{\bibnamefont{Dzhioev}, \bibfnamefont{R.~I.}},
  \bibinfo{author}{\bibfnamefont{V.~L.} \bibnamefont{Korenev}},
  \bibinfo{author}{\bibfnamefont{B.~P.} \bibnamefont{Zakharchenya}},
  \bibinfo{author}{\bibfnamefont{D.}~\bibnamefont{Gammon}},
  \bibinfo{author}{\bibfnamefont{A.~S.} \bibnamefont{Bracker}},
  \bibinfo{author}{\bibfnamefont{J.~G.} \bibnamefont{Tischler}}, and
  \bibinfo{author}{\bibfnamefont{D.~S.} \bibnamefont{Katzer}},
  \bibinfo{year}{2002}{\natexlab{c}}, {``}\bibinfo{title}{Optical orientation
  and the {Hanle} effect of neutral and negatively charged excitons in
  {GaAs/Al$_x$Ga$_{1-x}$As} quantum wells},{''} \bibinfo{journal}{Phys. Rev. B}
  \textbf{\bibinfo{volume}{66}},  \bibinfo{pages}{153409}.

\bibitem[{\citenamefont{Dzhioev} \emph{et~al.}(2001)\citenamefont{Dzhioev,
  Zakharchenya, Korenev, Gammon, and Katzer}}]{Dzhioev2001:JETPL}
\bibinfo{author}{\bibnamefont{Dzhioev}, \bibfnamefont{R.~I.}},
  \bibinfo{author}{\bibfnamefont{B.~P.} \bibnamefont{Zakharchenya}},
  \bibinfo{author}{\bibfnamefont{V.~L.} \bibnamefont{Korenev}},
  \bibinfo{author}{\bibfnamefont{D.}~\bibnamefont{Gammon}}, and
  \bibinfo{author}{\bibfnamefont{D.~S.} \bibnamefont{Katzer}},
  \bibinfo{year}{2001}, {``}\bibinfo{title}{Long electron spin memory times in
  gallium arsenide},{''} \bibinfo{journal}{Zh. Eksp. Teor. Fiz. Pisma Red.}
  \textbf{\bibinfo{volume}{74}},  \bibinfo{pages}{182--185}
  \bibinfo{note}{[JETP Lett. {\bf 74}, 200-203 (2001)]}.

\bibitem[{\citenamefont{Dzhioev} \emph{et~al.}(1997)\citenamefont{Dzhioev,
  Zakharchenya, Korenev, and Stepanova}}]{Dzhioev1997:PSS}
\bibinfo{author}{\bibnamefont{Dzhioev}, \bibfnamefont{R.~I.}},
  \bibinfo{author}{\bibfnamefont{B.~P.} \bibnamefont{Zakharchenya}},
  \bibinfo{author}{\bibfnamefont{V.~L.} \bibnamefont{Korenev}}, and
  \bibinfo{author}{\bibfnamefont{M.~N.} \bibnamefont{Stepanova}},
  \bibinfo{year}{1997}, {``}\bibinfo{title}{Spin diffusion of optically
  oriented electrons and photon entrainment in n-gallium arsenide},{''}
  \bibinfo{journal}{Fiz. Tverd. Tela} \textbf{\bibinfo{volume}{39}},
  \bibinfo{pages}{1975--1979} \bibinfo{note}{[Phys. Solid State {\bf 39},
  1765-1768 (1997)]}.

\bibitem[{\citenamefont{Dzihoev} \emph{et~al.}(2003)\citenamefont{Dzihoev,
  Zakharchenya, Kavokin, and Lazarev}}]{Dzihoev2003:FTT}
\bibinfo{author}{\bibnamefont{Dzihoev}, \bibfnamefont{R.~I.}},
  \bibinfo{author}{\bibfnamefont{B.~P.} \bibnamefont{Zakharchenya}},
  \bibinfo{author}{\bibfnamefont{K.~V.} \bibnamefont{Kavokin}}, and
  \bibinfo{author}{\bibfnamefont{M.~V.} \bibnamefont{Lazarev}},
  \bibinfo{year}{2003}, {``}\bibinfo{title}{The {Hanle} effect in nonuniformly
  doped {GaAs}},{''} \bibinfo{journal}{Fiz. Tverd. Tela}
  \textbf{\bibinfo{volume}{45}},  \bibinfo{pages}{2153--2160}
  \bibinfo{note}{[Phys. Solid State {\bf 45}, 2225-2263 (2003)]}.

\bibitem[{\citenamefont{Dzyaloshinskii}(1958)}]{Dzyaloshinskii1958:PCS}
\bibinfo{author}{\bibnamefont{Dzyaloshinskii}, \bibfnamefont{I.}},
  \bibinfo{year}{1958}, {``}\bibinfo{title}{A thermodynamic theory of {"weak"}
  ferromagnetism of antiferromagnetics},{''} \bibinfo{journal}{Phys. Chem.
  Solids} \textbf{\bibinfo{volume}{4}},  \bibinfo{pages}{241--255}.

\bibitem[{\citenamefont{Efros} \emph{et~al.}(2001)\citenamefont{Efros, Rashba,
  and Rosen}}]{Efros2001:PRL}
\bibinfo{author}{\bibnamefont{Efros}, \bibfnamefont{A.~L.}},
  \bibinfo{author}{\bibfnamefont{E.~I.} \bibnamefont{Rashba}}, and
  \bibinfo{author}{\bibfnamefont{M.}~\bibnamefont{Rosen}},
  \bibinfo{year}{2001}, {``}\bibinfo{title}{Paramagnetic Ion-Doped Nanocrystal
  as a Voltage-Controlled Spin Filter},{''} \bibinfo{journal}{Phys. Rev. Lett.}
  \textbf{\bibinfo{volume}{87}},  \bibinfo{pages}{206601}.

\bibitem[{\citenamefont{{Egelhoff, Jr.}}
  \emph{et~al.}(2002)\citenamefont{{Egelhoff, Jr.}, Stiles, Pappas, Pierce,
  Byers, Johnson, Jonker, Alvarado, Gregg, Bland, and
  Buhrman}}]{Egelhoff2002:S}
\bibinfo{author}{\bibnamefont{{Egelhoff, Jr.}}, \bibfnamefont{W.~F.}},
  \bibinfo{author}{\bibfnamefont{M.~D.} \bibnamefont{Stiles}},
  \bibinfo{author}{\bibfnamefont{D.~P.} \bibnamefont{Pappas}},
  \bibinfo{author}{\bibfnamefont{D.~T.} \bibnamefont{Pierce}},
  \bibinfo{author}{\bibfnamefont{J.~M.} \bibnamefont{Byers}},
  \bibinfo{author}{\bibfnamefont{M.~B.} \bibnamefont{Johnson}},
  \bibinfo{author}{\bibfnamefont{B.~T.} \bibnamefont{Jonker}},
  \bibinfo{author}{\bibfnamefont{S.~F.} \bibnamefont{Alvarado}},
  \bibinfo{author}{\bibfnamefont{J.~F.} \bibnamefont{Gregg}},
  \bibinfo{author}{\bibfnamefont{J.~A.~C.} \bibnamefont{Bland}}, and
  \bibinfo{author}{\bibfnamefont{R.~A.} \bibnamefont{Buhrman}},
  \bibinfo{year}{2002}, {``}\bibinfo{title}{Spin polarization of injected
  electrons},{''} \bibinfo{journal}{{\sl Science}}
  \textbf{\bibinfo{volume}{296}},  \bibinfo{pages}{1195}.

\bibitem[{\citenamefont{Egues}(1998)}]{Egues1998:PRL}
\bibinfo{author}{\bibnamefont{Egues}, \bibfnamefont{J.~C.}},
  \bibinfo{year}{1998}, {``}\bibinfo{title}{Spin-dependent perpendicular
  magnetotransport through a tunable {ZnSe/Zn$_{1-x}$Mn$_x$Se} heterostructure:
  a possible spin filter?},{''} \bibinfo{journal}{Phys. Rev. Lett.}
  \textbf{\bibinfo{volume}{80}},  \bibinfo{pages}{4578--4581}.

\bibitem[{\citenamefont{Eickhoff} \emph{et~al.}(2002)\citenamefont{Eickhoff,
  Lenzmann, Flinn, and Suter}}]{Eickhoff2002:PRB}
\bibinfo{author}{\bibnamefont{Eickhoff}, \bibfnamefont{M.}},
  \bibinfo{author}{\bibfnamefont{B.}~\bibnamefont{Lenzmann}},
  \bibinfo{author}{\bibfnamefont{G.}~\bibnamefont{Flinn}}, and
  \bibinfo{author}{\bibfnamefont{D.}~\bibnamefont{Suter}},
  \bibinfo{year}{2002}, {``}\bibinfo{title}{Coupling mechanisms for optically
  induced {NMR} in {GaAs} quantum wells},{''} \bibinfo{journal}{Phys. Rev. B}
  \textbf{\bibinfo{volume}{65}},  \bibinfo{pages}{125301}.

\bibitem[{\citenamefont{Ekimov and Safarov}(1970)}]{Ekimov1970:ZhETF}
\bibinfo{author}{\bibnamefont{Ekimov}, \bibfnamefont{A.~I.}}, and
  \bibinfo{author}{\bibfnamefont{V.~I.} \bibnamefont{Safarov}},
  \bibinfo{year}{1970}, {``}\bibinfo{title}{Optical orientation of carriers in
  interband transitions in semiconductors},{''} \bibinfo{journal}{Zh. Eksp.
  Teor. Fiz. Pisma Red.} \textbf{\bibinfo{volume}{12}},
  \bibinfo{pages}{293--297} \bibinfo{note}{[JETP Lett. {\bf 12}, 198-201
  (1970)]}.

\bibitem[{\citenamefont{Ekimov and Safarov}(1971)}]{Ekimov1971:JETPL}
\bibinfo{author}{\bibnamefont{Ekimov}, \bibfnamefont{A.~I.}}, and
  \bibinfo{author}{\bibfnamefont{V.~I.} \bibnamefont{Safarov}},
  \bibinfo{year}{1971}, {``}\bibinfo{title}{Observation of optical orientation
  of equilibrium electrons in n-type semiconductors},{''} \bibinfo{journal}{Zh.
  Eksp. Teor. Fiz. Pisma Red.} \textbf{\bibinfo{volume}{13}},
  \bibinfo{pages}{251--254} \bibinfo{note}{[JETP Lett. {\bf 13}, 177-179
  (1971)]}.

\bibitem[{\citenamefont{Ekimov and Safarov}(1972)}]{Ekimov1972:JETPL}
\bibinfo{author}{\bibnamefont{Ekimov}, \bibfnamefont{A.~I.}}, and
  \bibinfo{author}{\bibfnamefont{V.~I.} \bibnamefont{Safarov}},
  \bibinfo{year}{1972}, {``}\bibinfo{title}{Optical detection of dynamic of
  nuclei in semiconductors},{''} \bibinfo{journal}{Zh. Eksp. Teor. Fiz. Pisma
  Red.} \textbf{\bibinfo{volume}{15}},  \bibinfo{pages}{257--261}
  \bibinfo{note}{[JETP Lett. {\bf 15}, 179-181 (1972)]}.

\bibitem[{\citenamefont{Elezzabi} \emph{et~al.}(1996)\citenamefont{Elezzabi,
  Freeman, and Johnson}}]{Elezzabi1996:PRL}
\bibinfo{author}{\bibnamefont{Elezzabi}, \bibfnamefont{Y.}},
  \bibinfo{author}{\bibfnamefont{M.~R.} \bibnamefont{Freeman}}, and
  \bibinfo{author}{\bibfnamefont{M.}~\bibnamefont{Johnson}},
  \bibinfo{year}{1996}, {``}\bibinfo{title}{Direct measurement of the
  conduction electron spin-lattice relaxation time {$T_1$} in gold},{''}
  \bibinfo{journal}{Phys. Rev. Lett.} \textbf{\bibinfo{volume}{77}},
  \bibinfo{pages}{3220--3223}.

\bibitem[{\citenamefont{Elliott}(1954)}]{Elliott1954:PR}
\bibinfo{author}{\bibnamefont{Elliott}, \bibfnamefont{R.~J.}},
  \bibinfo{year}{1954}, {``}\bibinfo{title}{Theory of the effect of spin-orbit
  coupling on magnetic resonance in some semiconductors},{''}
  \bibinfo{journal}{Phys. Rev.} \textbf{\bibinfo{volume}{96}},
  \bibinfo{pages}{266--279}.

\bibitem[{\citenamefont{Elzerman} \emph{et~al.}(2003)\citenamefont{Elzerman,
  Hanson, Greidanus, {Willems van Beveren}, Franceschi, Vandersypen, Tarucha,
  and Kouwenhoven}}]{Elzerman2003:PRB}
\bibinfo{author}{\bibnamefont{Elzerman}, \bibfnamefont{J.~M.}},
  \bibinfo{author}{\bibfnamefont{R.}~\bibnamefont{Hanson}},
  \bibinfo{author}{\bibfnamefont{J.~S.} \bibnamefont{Greidanus}},
  \bibinfo{author}{\bibfnamefont{L.~H.} \bibnamefont{{Willems van Beveren}}},
  \bibinfo{author}{\bibfnamefont{S.~D.} \bibnamefont{Franceschi}},
  \bibinfo{author}{\bibfnamefont{L.~M.~K.} \bibnamefont{Vandersypen}},
  \bibinfo{author}{\bibfnamefont{S.}~\bibnamefont{Tarucha}}, and
  \bibinfo{author}{\bibfnamefont{L.~P.} \bibnamefont{Kouwenhoven}},
  \bibinfo{year}{2003}, {``}\bibinfo{title}{Few-electron quantum dot circuit
  with integrated charge read out},{''} \bibinfo{journal}{Phys. Rev. B}
  \textbf{\bibinfo{volume}{67}},  \bibinfo{pages}{161308}.

\bibitem[{\citenamefont{Endo} \emph{et~al.}(2000)\citenamefont{Endo, Sueoka,
  and Mukasa}}]{Endo2000:JJAP}
\bibinfo{author}{\bibnamefont{Endo}, \bibfnamefont{T.}},
  \bibinfo{author}{\bibfnamefont{K.}~\bibnamefont{Sueoka}}, and
  \bibinfo{author}{\bibfnamefont{K.}~\bibnamefont{Mukasa}},
  \bibinfo{year}{2000}, {``}\bibinfo{title}{Electron spin-relaxation times in
  p-type $\delta$-doped {GaAs/AlGaAs} double heterostructures},{''}
  \bibinfo{journal}{Jpn. J. Appl. Phys.} \textbf{\bibinfo{volume}{39}},
  \bibinfo{pages}{397--401}.

\bibitem[{\citenamefont{Engels} \emph{et~al.}(1997)\citenamefont{Engels, Lange,
  Sch{\"{a}}pers, and L{\"{u}}th}}]{Engels1997:PRB}
\bibinfo{author}{\bibnamefont{Engels}, \bibfnamefont{G.}},
  \bibinfo{author}{\bibfnamefont{J.}~\bibnamefont{Lange}},
  \bibinfo{author}{\bibfnamefont{T.}~\bibnamefont{Sch{\"{a}}pers}}, and
  \bibinfo{author}{\bibfnamefont{H.}~\bibnamefont{L{\"{u}}th}},
  \bibinfo{year}{1997}, {``}\bibinfo{title}{Experimental and theoretical
  approach to spin splitting in modulation-doped {In$_x$Ga$_{1-x}$As/InP}
  quantum wells for {$B\rightarrow 0$}},{''} \bibinfo{journal}{Phys. Rev. B}
  \textbf{\bibinfo{volume}{55}},  \bibinfo{pages}{R1958--R1961}.

\bibitem[{\citenamefont{Epstein}(2003)}]{Epstein2003:MRS}
\bibinfo{author}{\bibnamefont{Epstein}, \bibfnamefont{A.~J.}},
  \bibinfo{year}{2003}, {``}\bibinfo{title}{Organic-based magnets:
  opportunities in photoinduced magnetism, spintronics, fractal magnetism and
  beyond},{''} \bibinfo{journal}{MRS Bull.} \textbf{\bibinfo{volume}{28}},
  \bibinfo{pages}{492--499}.

\bibitem[{\citenamefont{Erlingsson}
  \emph{et~al.}(2001)\citenamefont{Erlingsson, Nazarov, and
  Falko}}]{Erlingsson2001:PRB}
\bibinfo{author}{\bibnamefont{Erlingsson}, \bibfnamefont{S.~I.}},
  \bibinfo{author}{\bibfnamefont{Y.~V.} \bibnamefont{Nazarov}}, and
  \bibinfo{author}{\bibfnamefont{V.~I.} \bibnamefont{Falko}},
  \bibinfo{year}{2001}, {``}\bibinfo{title}{Nucleus-mediated spin-flip
  transitions in {GaAs} quantum dots},{''} \bibinfo{journal}{Phys. Rev. B}
  \textbf{\bibinfo{volume}{64}},  \bibinfo{pages}{195306}.

\bibitem[{\citenamefont{Erwin} \emph{et~al.}(2002)\citenamefont{Erwin, Lee, and
  Scheffler}}]{Erwin2002:PRB}
\bibinfo{author}{\bibnamefont{Erwin}, \bibfnamefont{S.~C.}},
  \bibinfo{author}{\bibfnamefont{S.}~\bibnamefont{Lee}}, and
  \bibinfo{author}{\bibfnamefont{M.}~\bibnamefont{Scheffler}},
  \bibinfo{year}{2002}, {``}\bibinfo{title}{First-principles study of
  nucleation, growth, and interface structure of {Fe/GaAs}},{''}
  \bibinfo{journal}{Phys. Rev. B} \textbf{\bibinfo{volume}{65}},
  \bibinfo{pages}{205422}.

\bibitem[{\citenamefont{Erwin and {\v{Z}uti\'c}}(2004)}]{Erwin2004:NM}
\bibinfo{author}{\bibnamefont{Erwin}, \bibfnamefont{S.~C.}}, and
  \bibinfo{author}{\bibfnamefont{I.}~\bibnamefont{{\v{Z}uti\'c}}},
  \bibinfo{year}{2004}, {``}\bibinfo{title}{Tailoring ferromagnetic
  chalcopyrites},{''} \bibinfo{note}{{\sl Nature Mater.}, published online: 16
  May 2004.} \eprint{cond-mat/0401157}.

\bibitem[{\citenamefont{Esaki} \emph{et~al.}(1967)\citenamefont{Esaki, Stiles,
  and {von Moln\'{a}r}}}]{Esaki1967:PRL}
\bibinfo{author}{\bibnamefont{Esaki}, \bibfnamefont{L.}},
  \bibinfo{author}{\bibfnamefont{P.}~\bibnamefont{Stiles}}, and
  \bibinfo{author}{\bibfnamefont{S.}~\bibnamefont{{von Moln\'{a}r}}},
  \bibinfo{year}{1967}, {``}\bibinfo{title}{Magnetointernal field emission in
  junctions of magnetic insulators},{''} \bibinfo{journal}{Phys. Rev. Lett.}
  \textbf{\bibinfo{volume}{19}},  \bibinfo{pages}{852--854}.

\bibitem[{\citenamefont{Escorne} \emph{et~al.}(1974)\citenamefont{Escorne,
  Ghazalli, and {Leroux-Hugon}}}]{Escorne:1974}
\bibinfo{author}{\bibnamefont{Escorne}, \bibfnamefont{M.}},
  \bibinfo{author}{\bibfnamefont{A.}~\bibnamefont{Ghazalli}}, and
  \bibinfo{author}{\bibfnamefont{P.}~\bibnamefont{{Leroux-Hugon}}},
  \bibinfo{year}{1974} in \emph{\bibinfo{booktitle}{Proceedings of the 12th
  International Conference on the Physics of Semiconductors}}, edited by
  \bibinfo{editor}{\bibfnamefont{M.~H.} \bibnamefont{Pilkuhn}}
  (\bibinfo{publisher}{Teubner, Stuttgart}),  \bibinfo{pages}{915}.

\bibitem[{\citenamefont{Fabian and {Das Sarma}}(1998)}]{Fabian1998:PRL}
\bibinfo{author}{\bibnamefont{Fabian}, \bibfnamefont{J.}}, and
  \bibinfo{author}{\bibfnamefont{S.}~\bibnamefont{{Das Sarma}}},
  \bibinfo{year}{1998}, {``}\bibinfo{title}{Spin relaxation of conduction
  electrons in polyvalent metals: theory and a realistic calculation},{''}
  \bibinfo{journal}{Phys. Rev. Lett.} \textbf{\bibinfo{volume}{81}},
  \bibinfo{pages}{5624--5627}.

\bibitem[{\citenamefont{Fabian and {Das
  Sarma}}(1999{\natexlab{a}})}]{Fabian1999:JAP}
\bibinfo{author}{\bibnamefont{Fabian}, \bibfnamefont{J.}}, and
  \bibinfo{author}{\bibfnamefont{S.}~\bibnamefont{{Das Sarma}}},
  \bibinfo{year}{1999}{\natexlab{a}}, {``}\bibinfo{title}{Band-structure
  effects in the spin relaxation of conduction electrons},{''}
  \bibinfo{journal}{J. Appl. Phys.} \textbf{\bibinfo{volume}{85}},
  \bibinfo{pages}{5075--5079}.

\bibitem[{\citenamefont{Fabian and {Das
  Sarma}}(1999{\natexlab{b}})}]{Fabian1999:PRL}
\bibinfo{author}{\bibnamefont{Fabian}, \bibfnamefont{J.}}, and
  \bibinfo{author}{\bibfnamefont{S.}~\bibnamefont{{Das Sarma}}},
  \bibinfo{year}{1999}{\natexlab{b}}, {``}\bibinfo{title}{Phonon-induced spin
  relaxation of conduction electrons in aluminum},{''} \bibinfo{journal}{Phys.
  Rev. Lett.} \textbf{\bibinfo{volume}{83}},  \bibinfo{pages}{1211--1214}.

\bibitem[{\citenamefont{Fabian and {Das
  Sarma}}(1999{\natexlab{c}})}]{Fabian1999:JVST}
\bibinfo{author}{\bibnamefont{Fabian}, \bibfnamefont{J.}}, and
  \bibinfo{author}{\bibfnamefont{S.}~\bibnamefont{{Das Sarma}}},
  \bibinfo{year}{1999}{\natexlab{c}}, {``}\bibinfo{title}{Spin relaxation of
  conduction electrons},{''} \bibinfo{journal}{J. Vac. Sci. Technol. B}
  \textbf{\bibinfo{volume}{17}},  \bibinfo{pages}{1708--1715}.

\bibitem[{\citenamefont{Fabian and {Das Sarma}}(2002)}]{Fabian2002b:PRB}
\bibinfo{author}{\bibnamefont{Fabian}, \bibfnamefont{J.}}, and
  \bibinfo{author}{\bibfnamefont{S.}~\bibnamefont{{Das Sarma}}},
  \bibinfo{year}{2002}, {``}\bibinfo{title}{Spin transport in inhomogeneous
  magnetic fields: a proposal for {Stern}-{Gerlach}-like experiments with
  conduction electrons},{''} \bibinfo{journal}{Phys. Rev. B}
  \textbf{\bibinfo{volume}{66}},  \bibinfo{pages}{024436}.

\bibitem[{\citenamefont{Fabian}
  \emph{et~al.}(2002{\natexlab{a}})\citenamefont{Fabian, \v{Z}uti\'{c}, and
  {Das Sarma}}}]{Fabian2002:P}
\bibinfo{author}{\bibnamefont{Fabian}, \bibfnamefont{J.}},
  \bibinfo{author}{\bibfnamefont{I.}~\bibnamefont{\v{Z}uti\'{c}}}, and
  \bibinfo{author}{\bibfnamefont{S.}~\bibnamefont{{Das Sarma}}},
  \bibinfo{year}{2002}{\natexlab{a}}, {``}\bibinfo{title}{Theory of magnetic
  bipolar transistor},{''} \eprint{cond-mat/0211639}.

\bibitem[{\citenamefont{Fabian and {\v{Z}uti\'{c}}}(2004)}]{Fabian2003:Pb}
\bibinfo{author}{\bibnamefont{Fabian}, \bibfnamefont{J.}}, and
  \bibinfo{author}{\bibfnamefont{I.}~\bibnamefont{{\v{Z}uti\'{c}}}},
  \bibinfo{year}{2004}, {``}\bibinfo{title}{Spin-polarized current
  amplification and spin injection in magnetic bipolar transistors},{''}
  \bibinfo{journal}{Phys. Rev. B} \textbf{\bibinfo{volume}{69}},
  \bibinfo{pages}{115314}.

\bibitem[{\citenamefont{Fabian}
  \emph{et~al.}(2002{\natexlab{b}})\citenamefont{Fabian, {\v{Z}uti\'{c}}, and
  {Das Sarma}}}]{Fabian2002:PRB}
\bibinfo{author}{\bibnamefont{Fabian}, \bibfnamefont{J.}},
  \bibinfo{author}{\bibfnamefont{I.}~\bibnamefont{{\v{Z}uti\'{c}}}}, and
  \bibinfo{author}{\bibfnamefont{S.}~\bibnamefont{{Das Sarma}}},
  \bibinfo{year}{2002}{\natexlab{b}}, {``}\bibinfo{title}{Theory of
  spin-polarized bipolar transport in magnetic p-n junctions},{''}
  \bibinfo{journal}{Phys. Rev. B} \textbf{\bibinfo{volume}{66}},
  \bibinfo{pages}{165301}.

\bibitem[{\citenamefont{Fabian} \emph{et~al.}(2004)\citenamefont{Fabian,
  {\v{Z}uti\'{c}}, and Sarma}}]{Fabian2003:Pa}
\bibinfo{author}{\bibnamefont{Fabian}, \bibfnamefont{J.}},
  \bibinfo{author}{\bibfnamefont{I.}~\bibnamefont{{\v{Z}uti\'{c}}}}, and
  \bibinfo{author}{\bibfnamefont{S.~D.} \bibnamefont{Sarma}},
  \bibinfo{year}{2004}, {``}\bibinfo{title}{Magnetic bipolar transistor},{''}
  \bibinfo{journal}{Appl. Phys. Lett.} \textbf{\bibinfo{volume}{84}},
  \bibinfo{pages}{85--87}.

\bibitem[{\citenamefont{Falicov} \emph{et~al.}(1990)\citenamefont{Falicov,
  Pierce, Bader, Gronsky, Hathaway, Hopster, Lambeth, Parkin, Prinz, Salamon,
  Schuller, and Victora}}]{Falicov1990:JMR}
\bibinfo{author}{\bibnamefont{Falicov}, \bibfnamefont{L.~M.}},
  \bibinfo{author}{\bibfnamefont{D.~T.} \bibnamefont{Pierce}},
  \bibinfo{author}{\bibfnamefont{S.~D.} \bibnamefont{Bader}},
  \bibinfo{author}{\bibfnamefont{R.}~\bibnamefont{Gronsky}},
  \bibinfo{author}{\bibfnamefont{K.~B.} \bibnamefont{Hathaway}},
  \bibinfo{author}{\bibfnamefont{H.~J.} \bibnamefont{Hopster}},
  \bibinfo{author}{\bibfnamefont{D.~N.} \bibnamefont{Lambeth}},
  \bibinfo{author}{\bibfnamefont{S.~S.~P.} \bibnamefont{Parkin}},
  \bibinfo{author}{\bibfnamefont{G.}~\bibnamefont{Prinz}},
  \bibinfo{author}{\bibfnamefont{M.}~\bibnamefont{Salamon}},
  \bibinfo{author}{\bibfnamefont{I.~K.} \bibnamefont{Schuller}}, and
  \bibinfo{author}{\bibfnamefont{R.~H.} \bibnamefont{Victora}},
  \bibinfo{year}{1990}, {``}\bibinfo{title}{Surface, interface, and thin-film
  magnetism},{''} \bibinfo{journal}{J. Mater. Res.}
  \textbf{\bibinfo{volume}{5}},  \bibinfo{pages}{1299--1340}.

\bibitem[{\citenamefont{Fal'ko} \emph{et~al.}(1999)\citenamefont{Fal'ko,
  Lambert, and Volkov}}]{Falko1999:PZETF}
\bibinfo{author}{\bibnamefont{Fal'ko}, \bibfnamefont{V.~I.}},
  \bibinfo{author}{\bibfnamefont{C.~J.} \bibnamefont{Lambert}}, and
  \bibinfo{author}{\bibfnamefont{A.~F.} \bibnamefont{Volkov}},
  \bibinfo{year}{1999}, {``}\bibinfo{title}{Andreev reflections and
  magnetoresistance in ferromagnet-superconductor mesoscopic structures},{''}
  \bibinfo{journal}{Zh. Eksp. Teor. Fiz. Pisma Red.}
  \textbf{\bibinfo{volume}{69}},  \bibinfo{pages}{497--503}
  \bibinfo{note}{[JETP Lett. {\bf 69}, 532-538 (1999)]}.

\bibitem[{\citenamefont{Fanciulli} \emph{et~al.}(1993)\citenamefont{Fanciulli,
  Lei, and Moustakas}}]{Fanciulli1993:PRB}
\bibinfo{author}{\bibnamefont{Fanciulli}, \bibfnamefont{M.}},
  \bibinfo{author}{\bibfnamefont{T.}~\bibnamefont{Lei}}, and
  \bibinfo{author}{\bibfnamefont{T.~D.} \bibnamefont{Moustakas}},
  \bibinfo{year}{1993}, {``}\bibinfo{title}{Conduction-electron spin resonance
  in zinc-blende {GaN} thin films},{''} \bibinfo{journal}{Phys. Rev. B}
  \textbf{\bibinfo{volume}{48}},  \bibinfo{pages}{15144--15147}.

\bibitem[{\citenamefont{Fang} \emph{et~al.}(2002)\citenamefont{Fang, de~Wijs,
  and de~Groot}}]{Fang2002:JAP}
\bibinfo{author}{\bibnamefont{Fang}, \bibfnamefont{C.~M.}},
  \bibinfo{author}{\bibfnamefont{G.~A.} \bibnamefont{de~Wijs}}, and
  \bibinfo{author}{\bibfnamefont{R.~A.} \bibnamefont{de~Groot}},
  \bibinfo{year}{2002}, {``}\bibinfo{title}{Spin-polarization in
  half-metals},{''} \bibinfo{journal}{J. Appl. Phys.}
  \textbf{\bibinfo{volume}{91}},  \bibinfo{pages}{8340--8344}.

\bibitem[{\citenamefont{Feher}(1959)}]{Feher1959:PRL}
\bibinfo{author}{\bibnamefont{Feher}, \bibfnamefont{G.}}, \bibinfo{year}{1959},
  {``}\bibinfo{title}{Nuclear polarization via `hot' conduction Electrons},{''}
  \bibinfo{journal}{Phys. Rev. Lett.} \textbf{\bibinfo{volume}{3}},
  \bibinfo{pages}{135--137}.

\bibitem[{\citenamefont{Feher and Gere}(1959)}]{Feher1959b:PR}
\bibinfo{author}{\bibnamefont{Feher}, \bibfnamefont{G.}}, and
  \bibinfo{author}{\bibfnamefont{E.~A.} \bibnamefont{Gere}},
  \bibinfo{year}{1959}, {``}\bibinfo{title}{Electron spin resonance experiments
  on donors in silicon: {II.} electron spin relaxation effects},{''}
  \bibinfo{journal}{Phys. Rev.} \textbf{\bibinfo{volume}{114}},
  \bibinfo{pages}{1245--1256}.

\bibitem[{\citenamefont{Feher and Kip}(1955)}]{Feher1955:PR}
\bibinfo{author}{\bibnamefont{Feher}, \bibfnamefont{G.}}, and
  \bibinfo{author}{\bibfnamefont{A.~F.} \bibnamefont{Kip}},
  \bibinfo{year}{1955}, {``}\bibinfo{title}{Electron spin resonance absorption
  in metals. {I.} {Experimental}},{''} \bibinfo{journal}{Phys. Rev.}
  \textbf{\bibinfo{volume}{98}},  \bibinfo{pages}{337--348}.

\bibitem[{\citenamefont{Feng and Xiong}(2003)}]{Feng2003:PRB}
\bibinfo{author}{\bibnamefont{Feng}, \bibfnamefont{J.-F.}}, and
  \bibinfo{author}{\bibfnamefont{S.-J.} \bibnamefont{Xiong}},
  \bibinfo{year}{2003}, {``}\bibinfo{title}{Tunneling resonances and {Andreev}
  reflection in transport of electrons through a ferromagnetic metal/quantum
  dot/superconductor system},{''} \bibinfo{journal}{Phys. Rev. B}
  \textbf{\bibinfo{volume}{67}},  \bibinfo{pages}{045316}.

\bibitem[{\citenamefont{Ferreira and Bastard}(1991)}]{Ferreira1991:PRB}
\bibinfo{author}{\bibnamefont{Ferreira}, \bibfnamefont{R.}}, and
  \bibinfo{author}{\bibfnamefont{G.}~\bibnamefont{Bastard}},
  \bibinfo{year}{1991}, {``}\bibinfo{title}{Spin-flip scattering of holes in
  semiconductor quantum wells},{''} \bibinfo{journal}{Phys. Rev. B}
  \textbf{\bibinfo{volume}{43}},  \bibinfo{pages}{9687--9691}.

\bibitem[{\citenamefont{Fert and Campbell}(1968)}]{Fert1968:PRL}
\bibinfo{author}{\bibnamefont{Fert}, \bibfnamefont{A.}}, and
  \bibinfo{author}{\bibfnamefont{I.~A.} \bibnamefont{Campbell}},
  \bibinfo{year}{1968}, {``}\bibinfo{title}{Two-current conduction in
  nickel},{''} \bibinfo{journal}{Phys. Rev. Lett.}
  \textbf{\bibinfo{volume}{21}},  \bibinfo{pages}{1190--1192}.

\bibitem[{\citenamefont{Fert and Jaffres}(2001)}]{Fert2001:PRB}
\bibinfo{author}{\bibnamefont{Fert}, \bibfnamefont{A.}}, and
  \bibinfo{author}{\bibfnamefont{H.}~\bibnamefont{Jaffres}},
  \bibinfo{year}{2001}, {``}\bibinfo{title}{Conditions for efficient spin
  injection from a ferromagnetic metal into a semiconductor},{''}
  \bibinfo{journal}{Phys. Rev. B} \textbf{\bibinfo{volume}{64}},
  \bibinfo{pages}{184420}.

\bibitem[{\citenamefont{Fert and Lee}(1996)}]{Fert1996:PRB}
\bibinfo{author}{\bibnamefont{Fert}, \bibfnamefont{A.}}, and
  \bibinfo{author}{\bibfnamefont{S.-F.} \bibnamefont{Lee}},
  \bibinfo{year}{1996}, {``}\bibinfo{title}{Theory of the bipolar spin
  switch},{''} \bibinfo{journal}{Phys. Rev. B} \textbf{\bibinfo{volume}{53}},
  \bibinfo{pages}{6554--6565}.

\bibitem[{\citenamefont{Fiederling}
  \emph{et~al.}(2003)\citenamefont{Fiederling, Grabs, Ossau, Schmidt, and
  Molenkamp}}]{Fiederling2003:APL}
\bibinfo{author}{\bibnamefont{Fiederling}, \bibfnamefont{R.}},
  \bibinfo{author}{\bibfnamefont{P.}~\bibnamefont{Grabs}},
  \bibinfo{author}{\bibfnamefont{W.}~\bibnamefont{Ossau}},
  \bibinfo{author}{\bibfnamefont{G.}~\bibnamefont{Schmidt}}, and
  \bibinfo{author}{\bibfnamefont{L.~W.} \bibnamefont{Molenkamp}},
  \bibinfo{year}{2003}, {``}\bibinfo{title}{Detection of electrical spin
  injection by light-emitting diodes in top- and side-emission
  configurations},{''} \bibinfo{journal}{Appl. Phys. Lett.}
  \textbf{\bibinfo{volume}{82}},  \bibinfo{pages}{2160--2162}.

\bibitem[{\citenamefont{Fiederling}
  \emph{et~al.}(1999)\citenamefont{Fiederling, Kleim, Reuscher, Ossau, Schmidt,
  Waag, and Molenkamp}}]{Fiederling1999:N}
\bibinfo{author}{\bibnamefont{Fiederling}, \bibfnamefont{R.}},
  \bibinfo{author}{\bibfnamefont{M.}~\bibnamefont{Kleim}},
  \bibinfo{author}{\bibfnamefont{G.}~\bibnamefont{Reuscher}},
  \bibinfo{author}{\bibfnamefont{W.}~\bibnamefont{Ossau}},
  \bibinfo{author}{\bibfnamefont{G.}~\bibnamefont{Schmidt}},
  \bibinfo{author}{\bibfnamefont{A.}~\bibnamefont{Waag}}, and
  \bibinfo{author}{\bibfnamefont{L.~W.} \bibnamefont{Molenkamp}},
  \bibinfo{year}{1999}, {``}\bibinfo{title}{Injection and detection of a
  spin-polarized current in a light-emitting diode},{''} \bibinfo{journal}{{\sl
  Nature}} \textbf{\bibinfo{volume}{402}},  \bibinfo{pages}{787--790}.

\bibitem[{\citenamefont{Filip} \emph{et~al.}(2000)\citenamefont{Filip, Hoving,
  Jedema, {B. J. van Wees}, Dutta, and Borghs}}]{Filip2000:PRB}
\bibinfo{author}{\bibnamefont{Filip}, \bibfnamefont{A.~T.}},
  \bibinfo{author}{\bibfnamefont{B.~H.} \bibnamefont{Hoving}},
  \bibinfo{author}{\bibfnamefont{F.~J.} \bibnamefont{Jedema}},
  \bibinfo{author}{\bibnamefont{{B. J. van Wees}}},
  \bibinfo{author}{\bibfnamefont{B.}~\bibnamefont{Dutta}}, and
  \bibinfo{author}{\bibfnamefont{S.}~\bibnamefont{Borghs}},
  \bibinfo{year}{2000}, {``}\bibinfo{title}{Experimental search for the
  electrical spin injection in a semiconductor},{''} \bibinfo{journal}{Phys.
  Rev. B} \textbf{\bibinfo{volume}{62}},  \bibinfo{pages}{9996--9999}.

\bibitem[{\citenamefont{Filip} \emph{et~al.}(2002)\citenamefont{Filip, LeClair,
  Smits, Kohlhepp, Swagten, Koopmans, and {de Jonge}}}]{Filip2002:APL}
\bibinfo{author}{\bibnamefont{Filip}, \bibfnamefont{A.~T.}},
  \bibinfo{author}{\bibfnamefont{P.}~\bibnamefont{LeClair}},
  \bibinfo{author}{\bibfnamefont{C.~J.~P.} \bibnamefont{Smits}},
  \bibinfo{author}{\bibfnamefont{J.~T.} \bibnamefont{Kohlhepp}},
  \bibinfo{author}{\bibfnamefont{H.~J.~M.} \bibnamefont{Swagten}},
  \bibinfo{author}{\bibfnamefont{B.}~\bibnamefont{Koopmans}}, and
  \bibinfo{author}{\bibfnamefont{W.~J.~M.} \bibnamefont{{de Jonge}}},
  \bibinfo{year}{2002}, {``}\bibinfo{title}{Spin-injection device based on
  {EuS} magnetic tunnel barriers},{''} \bibinfo{journal}{Appl. Phys. Lett.}
  \textbf{\bibinfo{volume}{81}},  \bibinfo{pages}{1815--1817}.

\bibitem[{\citenamefont{Filipe} \emph{et~al.}(1998)\citenamefont{Filipe,
  Drouhin, Lampel, Lassailly, Nagle, Peretti, Safarov, and
  Schuhl}}]{Filipe1998:PRL}
\bibinfo{author}{\bibnamefont{Filipe}, \bibfnamefont{A.}},
  \bibinfo{author}{\bibfnamefont{H.}~\bibnamefont{Drouhin}},
  \bibinfo{author}{\bibfnamefont{G.}~\bibnamefont{Lampel}},
  \bibinfo{author}{\bibfnamefont{Y.}~\bibnamefont{Lassailly}},
  \bibinfo{author}{\bibfnamefont{J.}~\bibnamefont{Nagle}},
  \bibinfo{author}{\bibfnamefont{J.}~\bibnamefont{Peretti}},
  \bibinfo{author}{\bibfnamefont{V.~I.} \bibnamefont{Safarov}}, and
  \bibinfo{author}{\bibfnamefont{A.}~\bibnamefont{Schuhl}},
  \bibinfo{year}{1998}, {``}\bibinfo{title}{Spin-dependent transmission of
  electrons through the ferromagnetic metal base of a hot-electron
  transistorlike system},{''} \bibinfo{journal}{Phys. Rev. Lett.}
  \textbf{\bibinfo{volume}{80}},  \bibinfo{pages}{2425--2428}.

\bibitem[{\citenamefont{Fisher}(1967)}]{Fisher1967:PRL}
\bibinfo{author}{\bibnamefont{Fisher}, \bibfnamefont{M.~E.}},
  \bibinfo{year}{1967}, {``}\bibinfo{title}{Interfacial, boundary, and size
  effects at critical points},{''} \bibinfo{journal}{Phys. Rev. Lett.}
  \textbf{\bibinfo{volume}{19}},  \bibinfo{pages}{169--172}.

\bibitem[{\citenamefont{Fishman and Lampel}(1977)}]{Fishman1977:PRB}
\bibinfo{author}{\bibnamefont{Fishman}, \bibfnamefont{G.}}, and
  \bibinfo{author}{\bibfnamefont{G.}~\bibnamefont{Lampel}},
  \bibinfo{year}{1977}, {``}\bibinfo{title}{Spin relaxation of photoelectrons
  in p-type gallium arsenide},{''} \bibinfo{journal}{Phys. Rev. B}
  \textbf{\bibinfo{volume}{16}},  \bibinfo{pages}{820--831}.

\bibitem[{\citenamefont{Flatt{\'e} and Byers}(2000)}]{Flatte2000:PRL}
\bibinfo{author}{\bibnamefont{Flatt{\'e}}, \bibfnamefont{M.~E.}}, and
  \bibinfo{author}{\bibfnamefont{J.~M.} \bibnamefont{Byers}},
  \bibinfo{year}{2000}, {``}\bibinfo{title}{Spin diffusion in
  semiconductors},{''} \bibinfo{journal}{Phys. Rev. Lett.}
  \textbf{\bibinfo{volume}{84}},  \bibinfo{pages}{4220--4223}.

\bibitem[{\citenamefont{Flatt{\'e} and Vignale}(2001)}]{Flatte2001:APL}
\bibinfo{author}{\bibnamefont{Flatt{\'e}}, \bibfnamefont{M.~E.}}, and
  \bibinfo{author}{\bibfnamefont{G.}~\bibnamefont{Vignale}},
  \bibinfo{year}{2001}, {``}\bibinfo{title}{Unipolar spin diodes and
  transistors},{''} \bibinfo{journal}{Appl. Phys. Lett.}
  \textbf{\bibinfo{volume}{78}},  \bibinfo{pages}{1273--1275}.

\bibitem[{\citenamefont{Flatt{\'{e}}}
  \emph{et~al.}(2003)\citenamefont{Flatt{\'{e}}, Yu, Johnston-Halperin, and
  Awschalom}}]{Flatte2003:APL}
\bibinfo{author}{\bibnamefont{Flatt{\'{e}}}, \bibfnamefont{M.~E.}},
  \bibinfo{author}{\bibfnamefont{Z.~G.} \bibnamefont{Yu}},
  \bibinfo{author}{\bibfnamefont{E.}~\bibnamefont{Johnston-Halperin}}, and
  \bibinfo{author}{\bibfnamefont{D.~D.} \bibnamefont{Awschalom}},
  \bibinfo{year}{2003}, {``}\bibinfo{title}{Theory of semiconductor magnetic
  bipolar transistors},{''} \bibinfo{journal}{Appl. Phys. Lett.}
  \textbf{\bibinfo{volume}{82}},  \bibinfo{pages}{4740--4742}.

\bibitem[{\citenamefont{Fleisher and
  Merkulov}(1984{\natexlab{a}})}]{Fleisher:1984}
\bibinfo{author}{\bibnamefont{Fleisher}, \bibfnamefont{V.~G.}}, and
  \bibinfo{author}{\bibfnamefont{I.~A.} \bibnamefont{Merkulov}},
  \bibinfo{year}{1984}{\natexlab{a}}, {``}\bibinfo{title}{Optical Orientation
  of the Coupled Electron-Nuclear Spin System of a Semiconductor},{''} in
  \emph{\bibinfo{booktitle}{Optical Orientation, Modern Problems in Condensed
  Matter Science, Vol. 8}}, edited by
  \bibinfo{editor}{\bibfnamefont{F.}~\bibnamefont{Meier}} and
  \bibinfo{editor}{\bibfnamefont{B.~P.} \bibnamefont{Zakharchenya}}
  (\bibinfo{publisher}{North-Holland, Amsterdam}),  \bibinfo{pages}{173--258}.

\bibitem[{\citenamefont{Fleisher and
  Merkulov}(1984{\natexlab{b}})}]{Fleisher:1984b}
\bibinfo{author}{\bibnamefont{Fleisher}, \bibfnamefont{V.~G.}}, and
  \bibinfo{author}{\bibfnamefont{I.~A.} \bibnamefont{Merkulov}},
  \bibinfo{year}{1984}{\natexlab{b}}, {``}\bibinfo{title}{Optical Orientation
  of the Coupled Electron-Nuclear Spin System of a Semiconductor},{''} in
  \emph{\bibinfo{booktitle}{Optical Orientation, Modern Problems in Condensed
  Matter Science, Vol. 8}}, edited by
  \bibinfo{editor}{\bibfnamefont{F.}~\bibnamefont{Meier}} and
  \bibinfo{editor}{\bibfnamefont{B.~P.} \bibnamefont{Zakharchenya}}
  (\bibinfo{publisher}{North-Holland, Amsterdam}),  \bibinfo{pages}{198--208}.

\bibitem[{\citenamefont{Flensberg} \emph{et~al.}(2001)\citenamefont{Flensberg,
  Jensen, and Mortensen}}]{Flensberg2001:PRB}
\bibinfo{author}{\bibnamefont{Flensberg}, \bibfnamefont{K.}},
  \bibinfo{author}{\bibfnamefont{T.~S.} \bibnamefont{Jensen}}, and
  \bibinfo{author}{\bibfnamefont{N.~A.} \bibnamefont{Mortensen}},
  \bibinfo{year}{2001}, {``}\bibinfo{title}{Diffusion equation and spin drag in
  spin-polarized transport},{''} \bibinfo{journal}{Phys. Rev. B}
  \textbf{\bibinfo{volume}{64}},  \bibinfo{pages}{245308}.

\bibitem[{\citenamefont{Folk} \emph{et~al.}(2003)\citenamefont{Folk, Potok,
  Marcus, and Umansky}}]{Folk2003:S}
\bibinfo{author}{\bibnamefont{Folk}, \bibfnamefont{J.~A.}},
  \bibinfo{author}{\bibfnamefont{R.~M.} \bibnamefont{Potok}},
  \bibinfo{author}{\bibfnamefont{C.~M.} \bibnamefont{Marcus}}, and
  \bibinfo{author}{\bibfnamefont{V.}~\bibnamefont{Umansky}},
  \bibinfo{year}{2003}, {``}\bibinfo{title}{A gate-controlled bidirectional
  spin filter using quantum coherence},{''} \bibinfo{journal}{{\sl Science}}
  \textbf{\bibinfo{volume}{299}},  \bibinfo{pages}{679--682}.

\bibitem[{\citenamefont{Fominov}(2003)}]{Fominov2003:T}
\bibinfo{author}{\bibnamefont{Fominov}, \bibfnamefont{Y.~V.}},
  \bibinfo{year}{2003}, {``}\bibinfo{title}{Proximity and {Josephson} effects
  in superconducting hybrid structures},{''} \bibinfo{journal}{Ph.D. Thesis
  (Universiteit Twente)} .

\bibitem[{\citenamefont{Fraser} \emph{et~al.}(1999)\citenamefont{Fraser,
  Shkrebtii, Sipe, and {van Driel}}}]{Fraser1999:PRL}
\bibinfo{author}{\bibnamefont{Fraser}, \bibfnamefont{J.~M.}},
  \bibinfo{author}{\bibfnamefont{A.~I.} \bibnamefont{Shkrebtii}},
  \bibinfo{author}{\bibfnamefont{J.~E.} \bibnamefont{Sipe}}, and
  \bibinfo{author}{\bibfnamefont{H.~M.} \bibnamefont{{van Driel}}},
  \bibinfo{year}{1999}, {``}\bibinfo{title}{Quantum interference in
  electron-hole generation in noncentrosymmetric semiconductors},{''}
  \bibinfo{journal}{Phys. Rev. Lett.} \textbf{\bibinfo{volume}{83}},
  \bibinfo{pages}{4192--4195}.

\bibitem[{\citenamefont{Friesen} \emph{et~al.}(2003)\citenamefont{Friesen,
  Rugheimer, Savage, Lagally, {van der Weide}, Joynt, and
  Eriksson}}]{Friesen2003:PRB}
\bibinfo{author}{\bibnamefont{Friesen}, \bibfnamefont{M.}},
  \bibinfo{author}{\bibfnamefont{P.}~\bibnamefont{Rugheimer}},
  \bibinfo{author}{\bibfnamefont{D.~E.} \bibnamefont{Savage}},
  \bibinfo{author}{\bibfnamefont{M.~G.} \bibnamefont{Lagally}},
  \bibinfo{author}{\bibfnamefont{D.~W.} \bibnamefont{{van der Weide}}},
  \bibinfo{author}{\bibfnamefont{R.}~\bibnamefont{Joynt}}, and
  \bibinfo{author}{\bibfnamefont{M.~A.} \bibnamefont{Eriksson}},
  \bibinfo{year}{2003}, {``}\bibinfo{title}{Practical design and simulation of
  silicon-based quantum-dot qubits},{''} \bibinfo{journal}{Phys. Rev. B}
  \textbf{\bibinfo{volume}{67}},  \bibinfo{pages}{121301}.

\bibitem[{\citenamefont{Frustaglia}
  \emph{et~al.}(2001)\citenamefont{Frustaglia, Hentschel, and
  Richter}}]{Frustaglia2001:PRL}
\bibinfo{author}{\bibnamefont{Frustaglia}, \bibfnamefont{D.}},
  \bibinfo{author}{\bibfnamefont{M.}~\bibnamefont{Hentschel}}, and
  \bibinfo{author}{\bibfnamefont{K.}~\bibnamefont{Richter}},
  \bibinfo{year}{2001}, {``}\bibinfo{title}{Quantum transport in nonuniform
  magnetic fields: {Aharonov}-{Bohm} ring as a spin switch},{''}
  \bibinfo{journal}{Phys. Rev. Lett.} \textbf{\bibinfo{volume}{87}},
  \bibinfo{pages}{256602}.

\bibitem[{\citenamefont{Fu} \emph{et~al.}(2002)\citenamefont{Fu, Huang, and
  Yeh}}]{Fu2002:PRB}
\bibinfo{author}{\bibnamefont{Fu}, \bibfnamefont{C.~C.}},
  \bibinfo{author}{\bibfnamefont{Z.}~\bibnamefont{Huang}}, and
  \bibinfo{author}{\bibfnamefont{N.~C.} \bibnamefont{Yeh}},
  \bibinfo{year}{2002}, {``}\bibinfo{title}{Spin-polarized quasiparticle
  transport in cuprate superconductors},{''} \bibinfo{journal}{Phys. Rev. B}
  \textbf{\bibinfo{volume}{65}},  \bibinfo{pages}{224516}.

\bibitem[{\citenamefont{Fujisawa} \emph{et~al.}(2002)\citenamefont{Fujisawa,
  Austing, Tokura, Hirayama, and Tarucha}}]{Fujisawa2002:N}
\bibinfo{author}{\bibnamefont{Fujisawa}, \bibfnamefont{T.}},
  \bibinfo{author}{\bibfnamefont{D.~G.} \bibnamefont{Austing}},
  \bibinfo{author}{\bibfnamefont{Y.}~\bibnamefont{Tokura}},
  \bibinfo{author}{\bibfnamefont{Y.}~\bibnamefont{Hirayama}}, and
  \bibinfo{author}{\bibfnamefont{S.}~\bibnamefont{Tarucha}},
  \bibinfo{year}{2002}, {``}\bibinfo{title}{Allowed and forbidden transitions
  in artificial hydrogen and helium atoms},{''} \bibinfo{journal}{{\sl Nature}}
  \textbf{\bibinfo{volume}{419}},  \bibinfo{pages}{278--281}.

\bibitem[{\citenamefont{Fulde}(1973)}]{Fulde1973:AP}
\bibinfo{author}{\bibnamefont{Fulde}, \bibfnamefont{P.}}, \bibinfo{year}{1973},
  {``}\bibinfo{title}{High field superconductivity in thin films},{''}
  \bibinfo{journal}{Adv. Phys.} \textbf{\bibinfo{volume}{22}},
  \bibinfo{pages}{667--719}.

\bibitem[{\citenamefont{Furdyna}(1988)}]{Furdyna1988:JAP}
\bibinfo{author}{\bibnamefont{Furdyna}, \bibfnamefont{J.~K.}},
  \bibinfo{year}{1988}, {``}\bibinfo{title}{Diluted magnetic
  semiconductors},{''} \bibinfo{journal}{J. Appl. Phys.}
  \textbf{\bibinfo{volume}{64}},  \bibinfo{pages}{R29--R64}.

\bibitem[{\citenamefont{Gadzuk}(1969)}]{Gadzuk1969:PR}
\bibinfo{author}{\bibnamefont{Gadzuk}, \bibfnamefont{J.~W.}},
  \bibinfo{year}{1969}, {``}\bibinfo{title}{Band-structure effects in the
  field-induced tunneling of electrons from metals},{''}
  \bibinfo{journal}{Phys. Rev.} \textbf{\bibinfo{volume}{182}},
  \bibinfo{pages}{416--426}.

\bibitem[{\citenamefont{Gaj}(1988)}]{Gaj:1988}
\bibinfo{author}{\bibnamefont{Gaj}, \bibfnamefont{J.~A.}},
  \bibinfo{year}{1988}, {``}\bibinfo{title}{Diluted Magnetic
  Semiconductors},{''} in \emph{\bibinfo{booktitle}{Semiconductors and
  Semimetals, Vol. 25}}, edited by \bibinfo{editor}{\bibfnamefont{J.~K.}
  \bibnamefont{Furdyna}} and
  \bibinfo{editor}{\bibfnamefont{J.}~\bibnamefont{Kossut}}
  (\bibinfo{publisher}{Academic, New York}),  \bibinfo{pages}{286}.

\bibitem[{\citenamefont{Ganichev}
  \emph{et~al.}(2002{\natexlab{a}})\citenamefont{Ganichev, Danilov, Belkov,
  Ivchenko, Bichler, Wegscheider, Weiss, and Prettl}}]{Ganichev2002:PRL}
\bibinfo{author}{\bibnamefont{Ganichev}, \bibfnamefont{S.~D.}},
  \bibinfo{author}{\bibfnamefont{S.~N.} \bibnamefont{Danilov}},
  \bibinfo{author}{\bibfnamefont{V.~V.} \bibnamefont{Belkov}},
  \bibinfo{author}{\bibfnamefont{E.~L.} \bibnamefont{Ivchenko}},
  \bibinfo{author}{\bibfnamefont{M.}~\bibnamefont{Bichler}},
  \bibinfo{author}{\bibfnamefont{W.}~\bibnamefont{Wegscheider}},
  \bibinfo{author}{\bibfnamefont{D.}~\bibnamefont{Weiss}}, and
  \bibinfo{author}{\bibfnamefont{W.}~\bibnamefont{Prettl}},
  \bibinfo{year}{2002}{\natexlab{a}}, {``}\bibinfo{title}{Spin-sensitive
  bleaching and monopolar spin orientation in quantum wells},{''}
  \bibinfo{journal}{Phys. Rev. Lett.} \textbf{\bibinfo{volume}{88}},
  \bibinfo{pages}{057401}.

\bibitem[{\citenamefont{Ganichev}
  \emph{et~al.}(2002{\natexlab{b}})\citenamefont{Ganichev, Ivchenko, Belkov,
  Tarasenko, Sollinger, Weiss, Wegschelder, and Prettl}}]{Ganichev2002:N}
\bibinfo{author}{\bibnamefont{Ganichev}, \bibfnamefont{S.~D.}},
  \bibinfo{author}{\bibfnamefont{E.~L.} \bibnamefont{Ivchenko}},
  \bibinfo{author}{\bibfnamefont{V.~V.} \bibnamefont{Belkov}},
  \bibinfo{author}{\bibfnamefont{S.~A.} \bibnamefont{Tarasenko}},
  \bibinfo{author}{\bibfnamefont{M.}~\bibnamefont{Sollinger}},
  \bibinfo{author}{\bibfnamefont{D.}~\bibnamefont{Weiss}},
  \bibinfo{author}{\bibfnamefont{W.}~\bibnamefont{Wegschelder}}, and
  \bibinfo{author}{\bibfnamefont{W.}~\bibnamefont{Prettl}},
  \bibinfo{year}{2002}{\natexlab{b}}, {``}\bibinfo{title}{Spin-galvanic
  effect},{''} \bibinfo{journal}{{\sl Nature}} \textbf{\bibinfo{volume}{417}},
  \bibinfo{pages}{153--156}.

\bibitem[{\citenamefont{Ganichev} \emph{et~al.}(2001)\citenamefont{Ganichev,
  Ivchenko, Danilov, Eroms, Wegscheider, Weiss, and Prettl}}]{Ganichev2001:PRL}
\bibinfo{author}{\bibnamefont{Ganichev}, \bibfnamefont{S.~D.}},
  \bibinfo{author}{\bibfnamefont{E.~L.} \bibnamefont{Ivchenko}},
  \bibinfo{author}{\bibfnamefont{S.~N.} \bibnamefont{Danilov}},
  \bibinfo{author}{\bibfnamefont{J.}~\bibnamefont{Eroms}},
  \bibinfo{author}{\bibfnamefont{W.}~\bibnamefont{Wegscheider}},
  \bibinfo{author}{\bibfnamefont{D.}~\bibnamefont{Weiss}}, and
  \bibinfo{author}{\bibfnamefont{W.}~\bibnamefont{Prettl}},
  \bibinfo{year}{2001}, {``}\bibinfo{title}{Conversion of spin into directed
  electric current in quantum wells},{''} \bibinfo{journal}{Phys. Rev. Lett.}
  \textbf{\bibinfo{volume}{86}},  \bibinfo{pages}{4358--4561}.

\bibitem[{\citenamefont{Ganichev and Prettl}(2003)}]{Ganichev2003:JPCM}
\bibinfo{author}{\bibnamefont{Ganichev}, \bibfnamefont{S.~D.}}, and
  \bibinfo{author}{\bibfnamefont{W.}~\bibnamefont{Prettl}},
  \bibinfo{year}{2003}, {``}\bibinfo{title}{Spin photocurrents in quantum
  wells},{''} \bibinfo{journal}{J. Phys.: Condens. Matter.}
  \textbf{\bibinfo{volume}{15}},  \bibinfo{pages}{R935--R983}.

\bibitem[{\citenamefont{Ganichev} \emph{et~al.}(2003)\citenamefont{Ganichev,
  Schneider, Bel'kov, Ivchenko, Tarasenko, Wegscheider, Weiss, Schuh, Murdin,
  Pidgeon, Clarke, Merrick} \emph{et~al.}}]{Ganichev2003:PRB}
\bibinfo{author}{\bibnamefont{Ganichev}, \bibfnamefont{S.~D.}},
  \bibinfo{author}{\bibfnamefont{P.}~\bibnamefont{Schneider}},
  \bibinfo{author}{\bibfnamefont{V.~V.} \bibnamefont{Bel'kov}},
  \bibinfo{author}{\bibfnamefont{E.~L.} \bibnamefont{Ivchenko}},
  \bibinfo{author}{\bibfnamefont{S.~A.} \bibnamefont{Tarasenko}},
  \bibinfo{author}{\bibfnamefont{W.}~\bibnamefont{Wegscheider}},
  \bibinfo{author}{\bibfnamefont{D.}~\bibnamefont{Weiss}},
  \bibinfo{author}{\bibfnamefont{D.}~\bibnamefont{Schuh}},
  \bibinfo{author}{\bibfnamefont{B.~N.} \bibnamefont{Murdin}},
  \bibinfo{author}{\bibfnamefont{P.~J. P. C.~R.} \bibnamefont{Pidgeon}},
  \bibinfo{author}{\bibfnamefont{D.~G.} \bibnamefont{Clarke}},
  \bibinfo{author}{\bibfnamefont{M.}~\bibnamefont{Merrick}}, \emph{et~al.},
  \bibinfo{year}{2003}, {``}\bibinfo{title}{Spin-galvanic effect due to optical
  spin orientation in n-type {GaAs} quantum well structures},{''}
  \bibinfo{journal}{Phys. Rev. B} \textbf{\bibinfo{volume}{68}},
  \bibinfo{pages}{081302}.

\bibitem[{\citenamefont{Garbuzov} \emph{et~al.}(1971)\citenamefont{Garbuzov,
  Ekimov, and Safarov}}]{Garbuzov1971:ZhETF}
\bibinfo{author}{\bibnamefont{Garbuzov}, \bibfnamefont{D.~Z.}},
  \bibinfo{author}{\bibfnamefont{A.~I.} \bibnamefont{Ekimov}}, and
  \bibinfo{author}{\bibfnamefont{V.~I.} \bibnamefont{Safarov}},
  \bibinfo{year}{1971}, {``}\bibinfo{title}{Measurement of the lifetime and of
  the spin-relaxation time of electrons in semiconductors by the
  optical-orientation method},{''} \bibinfo{journal}{Zh. Eksp. Teor. Fiz. Pisma
  Red.} \textbf{\bibinfo{volume}{13}},  \bibinfo{pages}{36--40}
  \bibinfo{note}{[JETP Lett. {\bf 13}, 24-26 (1971)]}.

\bibitem[{\citenamefont{Garcia} \emph{et~al.}(1999)\citenamefont{Garcia, Munoz,
  and Zhao}}]{Garcia1999:PRL}
\bibinfo{author}{\bibnamefont{Garcia}, \bibfnamefont{N.}},
  \bibinfo{author}{\bibfnamefont{M.}~\bibnamefont{Munoz}}, and
  \bibinfo{author}{\bibfnamefont{Y.-W.} \bibnamefont{Zhao}},
  \bibinfo{year}{1999}, {``}\bibinfo{title}{Magnetoresistance in excess of
  200\% in Ballistic {Ni} nanocontacts at room temperature and 100 {Oe}},{''}
  \bibinfo{journal}{Phys. Rev. Lett.} \textbf{\bibinfo{volume}{82}},
  \bibinfo{pages}{2923--2926}.

\bibitem[{\citenamefont{Gardelis} \emph{et~al.}(1999)\citenamefont{Gardelis,
  Smith, Barnes, Linfield, and Ritchie}}]{Gardelis1999:PRB}
\bibinfo{author}{\bibnamefont{Gardelis}, \bibfnamefont{S.}},
  \bibinfo{author}{\bibfnamefont{C.~G.} \bibnamefont{Smith}},
  \bibinfo{author}{\bibfnamefont{C.~H.~W.} \bibnamefont{Barnes}},
  \bibinfo{author}{\bibfnamefont{E.~H.} \bibnamefont{Linfield}}, and
  \bibinfo{author}{\bibfnamefont{D.~A.} \bibnamefont{Ritchie}},
  \bibinfo{year}{1999}, {``}\bibinfo{title}{Spin-valve effects in a
  semiconductor field-effect transistor: a spintronic device},{''}
  \bibinfo{journal}{Phys. Rev. B} \textbf{\bibinfo{volume}{60}},
  \bibinfo{pages}{7764--7767}.

\bibitem[{\citenamefont{Geux} \emph{et~al.}(2000)\citenamefont{Geux, Brataas,
  and Bauer}}]{Geux2000:APPA}
\bibinfo{author}{\bibnamefont{Geux}, \bibfnamefont{L.~S.}},
  \bibinfo{author}{\bibfnamefont{A.}~\bibnamefont{Brataas}}, and
  \bibinfo{author}{\bibfnamefont{G.~E.~W.} \bibnamefont{Bauer}},
  \bibinfo{year}{2000}, {``}\bibinfo{title}{Scattering theory of the {Johnson}
  spin transistor},{''} \bibinfo{journal}{Acta Phys. Pol. A}
  \textbf{\bibinfo{volume}{97}},  \bibinfo{pages}{119--128}.

\bibitem[{\citenamefont{Giaever and Megerle}(1961)}]{Giaever1961:PR}
\bibinfo{author}{\bibnamefont{Giaever}, \bibfnamefont{I.}}, and
  \bibinfo{author}{\bibfnamefont{K.}~\bibnamefont{Megerle}},
  \bibinfo{year}{1961}, {``}\bibinfo{title}{Study of superconductors by
  electron tunneling},{''} \bibinfo{journal}{Phys. Rev.}
  \textbf{\bibinfo{volume}{122}},  \bibinfo{pages}{1101--1111}.

\bibitem[{\citenamefont{Giazotto} \emph{et~al.}(2003)\citenamefont{Giazotto,
  Taddei, Fazio, and Beltram}}]{Giazotto2003:APL}
\bibinfo{author}{\bibnamefont{Giazotto}, \bibfnamefont{F.}},
  \bibinfo{author}{\bibfnamefont{F.}~\bibnamefont{Taddei}},
  \bibinfo{author}{\bibfnamefont{R.}~\bibnamefont{Fazio}}, and
  \bibinfo{author}{\bibfnamefont{F.}~\bibnamefont{Beltram}},
  \bibinfo{year}{2003}, {``}\bibinfo{title}{Ferromagnetic resonant tunneling
  diodes as spin polarimeters},{''} \bibinfo{journal}{Appl. Phys. Lett.}
  \textbf{\bibinfo{volume}{82}},  \bibinfo{pages}{2449--2451}.

\bibitem[{\citenamefont{Gibson and Meservey}(1985)}]{Gibson1985:JAP}
\bibinfo{author}{\bibnamefont{Gibson}, \bibfnamefont{G.~A.}}, and
  \bibinfo{author}{\bibfnamefont{R.}~\bibnamefont{Meservey}},
  \bibinfo{year}{1985}, {``}\bibinfo{title}{Properties of amorphous germanium
  tunnel barriers},{''} \bibinfo{journal}{J. Appl. Phys.}
  \textbf{\bibinfo{volume}{58}},  \bibinfo{pages}{1584--1596}.

\bibitem[{\citenamefont{Gijs and Bauer}(1997)}]{Gijs1997:AP}
\bibinfo{author}{\bibnamefont{Gijs}, \bibfnamefont{M.~A.~M.}}, and
  \bibinfo{author}{\bibfnamefont{G.~E.~W.} \bibnamefont{Bauer}},
  \bibinfo{year}{1997}, {``}\bibinfo{title}{Perpendicular giant
  magnetoresistance of magnetic multilayers},{''} \bibinfo{journal}{Adv. Phys.}
  \textbf{\bibinfo{volume}{46}},  \bibinfo{pages}{285--445}.

\bibitem[{\citenamefont{Gim} \emph{et~al.}(2001)\citenamefont{Gim, Kleinsasser,
  and Barner}}]{Gim2001:JAP}
\bibinfo{author}{\bibnamefont{Gim}, \bibfnamefont{Y.}},
  \bibinfo{author}{\bibfnamefont{A.~W.} \bibnamefont{Kleinsasser}}, and
  \bibinfo{author}{\bibfnamefont{J.~B.} \bibnamefont{Barner}},
  \bibinfo{year}{2001}, {``}\bibinfo{title}{Current injection into
  high-temperature superconductors: Does spin matter?},{''}
  \bibinfo{journal}{J. Appl. Phys.} \textbf{\bibinfo{volume}{90}},
  \bibinfo{pages}{4063--4077}.

\bibitem[{\citenamefont{Glazov and Ivchenko}(2002)}]{Glazov2002:JETPL}
\bibinfo{author}{\bibnamefont{Glazov}, \bibfnamefont{M.~M.}}, and
  \bibinfo{author}{\bibfnamefont{E.~L.} \bibnamefont{Ivchenko}},
  \bibinfo{year}{2002}, {``}\bibinfo{title}{Precession spin relaxation
  mechanism caused by frequent electron-electron collisions},{''}
  \bibinfo{journal}{Zh. Eksp. Teor. Fiz. Pisma Red.}
  \textbf{\bibinfo{volume}{75}},  \bibinfo{pages}{476--478}
  \bibinfo{note}{[JETP Lett. {\bf 75}, 403-405 (2002)]}.

\bibitem[{\citenamefont{Glazov and Ivchenko}(2003)}]{Glazov2003:P}
\bibinfo{author}{\bibnamefont{Glazov}, \bibfnamefont{M.~M.}}, and
  \bibinfo{author}{\bibfnamefont{E.~L.} \bibnamefont{Ivchenko}},
  \bibinfo{year}{2003}, {``}\bibinfo{title}{{D'yakonov-Perel'} spin relaxation
  under electron-electron collisions in {QWs}},{''} \eprint{cond-mat/0301519}.

\bibitem[{\citenamefont{Goldman} \emph{et~al.}(2001)\citenamefont{Goldman,
  Nikolaev, Kraus, {Vas'ko}, Bhattacharya, and Cooley}}]{Goldman2001:JS}
\bibinfo{author}{\bibnamefont{Goldman}, \bibfnamefont{A.~M.}},
  \bibinfo{author}{\bibfnamefont{K.}~\bibnamefont{Nikolaev}},
  \bibinfo{author}{\bibfnamefont{P.}~\bibnamefont{Kraus}},
  \bibinfo{author}{\bibfnamefont{V.}~\bibnamefont{{Vas'ko}}},
  \bibinfo{author}{\bibfnamefont{A.}~\bibnamefont{Bhattacharya}}, and
  \bibinfo{author}{\bibfnamefont{W.}~\bibnamefont{Cooley}},
  \bibinfo{year}{2001}, {``}\bibinfo{title}{Spin injection and transport in
  magnetic-superconducting oxide heterostructures},{''} \bibinfo{journal}{J.
  Supercond.} \textbf{\bibinfo{volume}{14}},  \bibinfo{pages}{283--290}.

\bibitem[{\citenamefont{Goldman} \emph{et~al.}(1999)\citenamefont{Goldman,
  {Vas'ko}, Kraus, Nikolaev, and Larkin}}]{Goldman1999:JMMM}
\bibinfo{author}{\bibnamefont{Goldman}, \bibfnamefont{A.~M.}},
  \bibinfo{author}{\bibfnamefont{V.}~\bibnamefont{{Vas'ko}}},
  \bibinfo{author}{\bibfnamefont{P.}~\bibnamefont{Kraus}},
  \bibinfo{author}{\bibfnamefont{K.}~\bibnamefont{Nikolaev}}, and
  \bibinfo{author}{\bibfnamefont{V.~A.} \bibnamefont{Larkin}},
  \bibinfo{year}{1999}, {``}\bibinfo{title}{Cuprate/manganite
  heterostructures},{''} \bibinfo{journal}{J. Mag. Magn. Mater.}
  \textbf{\bibinfo{volume}{200}},  \bibinfo{pages}{69--82}.

\bibitem[{\citenamefont{Gordon and Browers}(1958)}]{Gordon1958:PR}
\bibinfo{author}{\bibnamefont{Gordon}, \bibfnamefont{J.~P.}}, and
  \bibinfo{author}{\bibfnamefont{K.~D.} \bibnamefont{Browers}},
  \bibinfo{year}{1958}, {``}\bibinfo{title}{Microwave spin echoes from donor
  electrons in silicon},{''} \bibinfo{journal}{Phys. Rev. Lett.}
  \textbf{\bibinfo{volume}{1}},  \bibinfo{pages}{368--370}.

\bibitem[{\citenamefont{Gorelenok} \emph{et~al.}(1986)\citenamefont{Gorelenok,
  Gruzdov, Marushchak, and Titkov}}]{Gorelenok1986:SPS}
\bibinfo{author}{\bibnamefont{Gorelenok}, \bibfnamefont{A.~T.}},
  \bibinfo{author}{\bibfnamefont{V.~G.} \bibnamefont{Gruzdov}},
  \bibinfo{author}{\bibfnamefont{V.~A.} \bibnamefont{Marushchak}}, and
  \bibinfo{author}{\bibfnamefont{A.~N.} \bibnamefont{Titkov}},
  \bibinfo{year}{1986}, {``}\bibinfo{title}{Spin splitting of the conduction
  band of {InP}},{''} \bibinfo{journal}{Sov. Phys. Semicond.}
  \textbf{\bibinfo{volume}{20}},  \bibinfo{pages}{347--350}
  \bibinfo{note}{[Sov. Phys. Semicond. {\bf 20}, 216-218 (1986)]}.

\bibitem[{\citenamefont{Gorkov and Krotkov}(2003)}]{Gorkov2003:PRB}
\bibinfo{author}{\bibnamefont{Gorkov}, \bibfnamefont{L.~P.}}, and
  \bibinfo{author}{\bibfnamefont{P.~L.} \bibnamefont{Krotkov}},
  \bibinfo{year}{2003}, {``}\bibinfo{title}{Spin relaxation and antisymmetric
  exchange in n-doped {III-V} semiconductors},{''} \bibinfo{journal}{Phys. Rev.
  B} \textbf{\bibinfo{volume}{67}},  \bibinfo{pages}{033203}.

\bibitem[{\citenamefont{Gotoh} \emph{et~al.}(2000)\citenamefont{Gotoh, Ando,
  Sogawa, Kamada, Kagawa, and Iwamura}}]{Gotoh2000:JAP}
\bibinfo{author}{\bibnamefont{Gotoh}, \bibfnamefont{H.}},
  \bibinfo{author}{\bibfnamefont{H.}~\bibnamefont{Ando}},
  \bibinfo{author}{\bibfnamefont{T.}~\bibnamefont{Sogawa}},
  \bibinfo{author}{\bibfnamefont{H.}~\bibnamefont{Kamada}},
  \bibinfo{author}{\bibfnamefont{T.}~\bibnamefont{Kagawa}}, and
  \bibinfo{author}{\bibfnamefont{H.}~\bibnamefont{Iwamura}},
  \bibinfo{year}{2000}, {``}\bibinfo{title}{Effect of electron-hole interaction
  on electron spin relaxation in {GaAs/AlGaAs} quantum wells at room
  temperature},{''} \bibinfo{journal}{J. Appl. Phys.}
  \textbf{\bibinfo{volume}{87}},  \bibinfo{pages}{3394--3398}.

\bibitem[{\citenamefont{Governale} \emph{et~al.}(2002)\citenamefont{Governale,
  Boese, Z{\"u}licke, and Schroll}}]{Governale2002:PRB}
\bibinfo{author}{\bibnamefont{Governale}, \bibfnamefont{M.}},
  \bibinfo{author}{\bibfnamefont{D.}~\bibnamefont{Boese}},
  \bibinfo{author}{\bibfnamefont{U.}~\bibnamefont{Z{\"u}licke}}, and
  \bibinfo{author}{\bibfnamefont{C.}~\bibnamefont{Schroll}},
  \bibinfo{year}{2002}, {``}\bibinfo{title}{Filtering spin with tunnel-coupled
  electron wave guides},{''} \bibinfo{journal}{Phys. Rev. B}
  \textbf{\bibinfo{volume}{65}},  \bibinfo{pages}{140403}.

\bibitem[{\citenamefont{Governale} \emph{et~al.}(2003)\citenamefont{Governale,
  Taddei, and Fazio}}]{Governale2003:PRB}
\bibinfo{author}{\bibnamefont{Governale}, \bibfnamefont{M.}},
  \bibinfo{author}{\bibfnamefont{F.}~\bibnamefont{Taddei}}, and
  \bibinfo{author}{\bibfnamefont{R.}~\bibnamefont{Fazio}},
  \bibinfo{year}{2003}, {``}\bibinfo{title}{Pumping spin with electrical
  fields},{''} \bibinfo{journal}{Phys. Rev. B} \textbf{\bibinfo{volume}{68}},
  \bibinfo{pages}{155324}.

\bibitem[{\citenamefont{Graeff} \emph{et~al.}(1999)\citenamefont{Graeff,
  Brandt, Stutzmann, Holzmann, Abstreiter, and
  {Sch\"{a}ffler}}}]{Graeff1999:PRB}
\bibinfo{author}{\bibnamefont{Graeff}, \bibfnamefont{C.~F.~O.}},
  \bibinfo{author}{\bibfnamefont{M.~S.} \bibnamefont{Brandt}},
  \bibinfo{author}{\bibfnamefont{M.}~\bibnamefont{Stutzmann}},
  \bibinfo{author}{\bibfnamefont{M.}~\bibnamefont{Holzmann}},
  \bibinfo{author}{\bibfnamefont{G.}~\bibnamefont{Abstreiter}}, and
  \bibinfo{author}{\bibfnamefont{F.}~\bibnamefont{{Sch\"{a}ffler}}},
  \bibinfo{year}{1999}, {``}\bibinfo{title}{Electrically detected magnetic
  resonance of two-dimensional electron gases in {Si/SeGe}
  heterostructures},{''} \bibinfo{journal}{Phys. Rev. B}
  \textbf{\bibinfo{volume}{59}},  \bibinfo{pages}{13242--13250}.

\bibitem[{\citenamefont{Gregg} \emph{et~al.}(1997)\citenamefont{Gregg, Allen,
  Viart, Kirschman, Schille, Gester, Thompson, Sparks, {Da Costa}, Ounadjela,
  and Skvarla.}}]{Gregg1997:JMMM}
\bibinfo{author}{\bibnamefont{Gregg}, \bibfnamefont{J.}},
  \bibinfo{author}{\bibfnamefont{W.}~\bibnamefont{Allen}},
  \bibinfo{author}{\bibfnamefont{N.}~\bibnamefont{Viart}},
  \bibinfo{author}{\bibfnamefont{R.}~\bibnamefont{Kirschman}},
  \bibinfo{author}{\bibfnamefont{C.~S. J.-P.} \bibnamefont{Schille}},
  \bibinfo{author}{\bibfnamefont{M.}~\bibnamefont{Gester}},
  \bibinfo{author}{\bibfnamefont{S.}~\bibnamefont{Thompson}},
  \bibinfo{author}{\bibfnamefont{P.}~\bibnamefont{Sparks}},
  \bibinfo{author}{\bibfnamefont{V.}~\bibnamefont{{Da Costa}}},
  \bibinfo{author}{\bibfnamefont{K.}~\bibnamefont{Ounadjela}}, and
  \bibinfo{author}{\bibfnamefont{M.}~\bibnamefont{Skvarla.}},
  \bibinfo{year}{1997}, {``}\bibinfo{title}{The art of spin electronics},{''}
  \bibinfo{journal}{J. Magn. Magn. Mater.} \textbf{\bibinfo{volume}{175}},
  \bibinfo{pages}{1--9}.

\bibitem[{\citenamefont{Griffin and Demers}(1971)}]{Griffin1971:PRB}
\bibinfo{author}{\bibnamefont{Griffin}, \bibfnamefont{A.}}, and
  \bibinfo{author}{\bibfnamefont{J.}~\bibnamefont{Demers}},
  \bibinfo{year}{1971}, {``}\bibinfo{title}{Tunneling in the
  normal-metal-insulator-superconductor geometry using the {Bogoliubov}
  equations of motion},{''} \bibinfo{journal}{Phys. Rev. B}
  \textbf{\bibinfo{volume}{4}},  \bibinfo{pages}{2202--2208}.

\bibitem[{\citenamefont{Grimaldi and Fulde}(1996)}]{Grimaldi1996:PRL}
\bibinfo{author}{\bibnamefont{Grimaldi}, \bibfnamefont{C.}}, and
  \bibinfo{author}{\bibfnamefont{P.}~\bibnamefont{Fulde}},
  \bibinfo{year}{1996}, {``}\bibinfo{title}{Spin-orbit scattering effects on
  the phonon emission and absorption in superconducting tunneling
  junctions},{''} \bibinfo{journal}{Phys. Rev. Lett.}
  \textbf{\bibinfo{volume}{77}},  \bibinfo{pages}{2550--2553}.

\bibitem[{\citenamefont{Grimaldi and Fulde}(1997)}]{Grimaldi1997:PRB}
\bibinfo{author}{\bibnamefont{Grimaldi}, \bibfnamefont{C.}}, and
  \bibinfo{author}{\bibfnamefont{P.}~\bibnamefont{Fulde}},
  \bibinfo{year}{1997}, {``}\bibinfo{title}{Theory of screening of the
  phonon-modulated spin-orbit interaction in metals},{''}
  \bibinfo{journal}{Phys. Rev. B} \textbf{\bibinfo{volume}{55}},
  \bibinfo{pages}{15523--15530}.

\bibitem[{\citenamefont{Griswold} \emph{et~al.}(1952)\citenamefont{Griswold,
  Kip, and Kittel}}]{Griswold1952:PR}
\bibinfo{author}{\bibnamefont{Griswold}, \bibfnamefont{T.~W.}},
  \bibinfo{author}{\bibfnamefont{A.~F.} \bibnamefont{Kip}}, and
  \bibinfo{author}{\bibfnamefont{C.}~\bibnamefont{Kittel}},
  \bibinfo{year}{1952}, {``}\bibinfo{title}{Microwave spin resonance absorption
  by conduction electrons in metallic sodium},{''} \bibinfo{journal}{Phys.
  Rev.} \textbf{\bibinfo{volume}{88}},  \bibinfo{pages}{951--952}.

\bibitem[{\citenamefont{Gruber} \emph{et~al.}(2001)\citenamefont{Gruber, Kein,
  Fiederling, Reuscher, Ossau, Schmidt, and Molenkamp}}]{Gruber2001:APL}
\bibinfo{author}{\bibnamefont{Gruber}, \bibfnamefont{T.}},
  \bibinfo{author}{\bibfnamefont{M.}~\bibnamefont{Kein}},
  \bibinfo{author}{\bibfnamefont{R.}~\bibnamefont{Fiederling}},
  \bibinfo{author}{\bibfnamefont{G.}~\bibnamefont{Reuscher}},
  \bibinfo{author}{\bibfnamefont{W.}~\bibnamefont{Ossau}},
  \bibinfo{author}{\bibfnamefont{G.}~\bibnamefont{Schmidt}}, and
  \bibinfo{author}{\bibfnamefont{L.~W.} \bibnamefont{Molenkamp}},
  \bibinfo{year}{2001}, {``}\bibinfo{title}{Electron spin manipulation using
  semimagnetic resonant tunneling diodes},{''} \bibinfo{journal}{Appl. Phys.
  Lett.} \textbf{\bibinfo{volume}{78}},  \bibinfo{pages}{1101--1103}.

\bibitem[{\citenamefont{Grundler}(2000)}]{Grundler2000:PRL}
\bibinfo{author}{\bibnamefont{Grundler}, \bibfnamefont{D.}},
  \bibinfo{year}{2000}, {``}\bibinfo{title}{Large {Rashba} splitting in {InAs}
  quantum wells due to electron wave function penetration into the barrier
  layers},{''} \bibinfo{journal}{Phys. Rev. Lett.}
  \textbf{\bibinfo{volume}{84}},  \bibinfo{pages}{6074--6077}.

\bibitem[{\citenamefont{Gu} \emph{et~al.}(2002)\citenamefont{Gu, Caballero,
  Slater, Loloee, and Pratt}}]{Gu2002:PRB}
\bibinfo{author}{\bibnamefont{Gu}, \bibfnamefont{J.~Y.}},
  \bibinfo{author}{\bibfnamefont{J.~A.} \bibnamefont{Caballero}},
  \bibinfo{author}{\bibfnamefont{R.~D.} \bibnamefont{Slater}},
  \bibinfo{author}{\bibfnamefont{R.}~\bibnamefont{Loloee}}, and
  \bibinfo{author}{\bibfnamefont{W.~P.} \bibnamefont{Pratt}},
  \bibinfo{year}{2002}, {``}\bibinfo{title}{Direct measurement of quasiparticle
  evanescent waves in a dirty superconductor},{''} \bibinfo{journal}{Phys. Rev.
  B} \textbf{\bibinfo{volume}{66}},  \bibinfo{pages}{140507}.

\bibitem[{\citenamefont{Guettler} \emph{et~al.}(1998)\citenamefont{Guettler,
  Triques, Vervoort, Roussignol, Voisin, Rondi, and
  Harmand}}]{Guettler1998:PRB}
\bibinfo{author}{\bibnamefont{Guettler}, \bibfnamefont{T.}},
  \bibinfo{author}{\bibfnamefont{A.~L.} \bibnamefont{Triques}},
  \bibinfo{author}{\bibfnamefont{L.}~\bibnamefont{Vervoort}},
  \bibinfo{author}{\bibfnamefont{R.~F.~P.} \bibnamefont{Roussignol}},
  \bibinfo{author}{\bibfnamefont{P.}~\bibnamefont{Voisin}},
  \bibinfo{author}{\bibfnamefont{D.}~\bibnamefont{Rondi}}, and
  \bibinfo{author}{\bibfnamefont{J.~C.} \bibnamefont{Harmand}},
  \bibinfo{year}{1998}, {``}\bibinfo{title}{Optical polarization in
  {In$_x$Ga$_{1-x}$As}-based quantum wells: {E}vidence of the interface
  symmetry-reduction effect},{''} \bibinfo{journal}{Phys. Rev. B}
  \textbf{\bibinfo{volume}{58}},  \bibinfo{pages}{R10179--R10182}.

\bibitem[{\citenamefont{Guo} \emph{et~al.}(2001)\citenamefont{Guo, Lu, Gu, and
  Kawazoe}}]{Guo2001:PRB}
\bibinfo{author}{\bibnamefont{Guo}, \bibfnamefont{Y.}},
  \bibinfo{author}{\bibfnamefont{J.-Q.} \bibnamefont{Lu}},
  \bibinfo{author}{\bibfnamefont{B.-L.} \bibnamefont{Gu}}, and
  \bibinfo{author}{\bibfnamefont{Y.}~\bibnamefont{Kawazoe}},
  \bibinfo{year}{2001}, {``}\bibinfo{title}{Spin-resonant splitting in
  magnetically modulated semimagnetic semiconductor superlattices},{''}
  \bibinfo{journal}{Phys. Rev. B} \textbf{\bibinfo{volume}{64}},
  \bibinfo{pages}{155312}.

\bibitem[{\citenamefont{Gupta} \emph{et~al.}(2002)\citenamefont{Gupta,
  Awschalom, Efros, and Rodina}}]{Gupta2002:PRB}
\bibinfo{author}{\bibnamefont{Gupta}, \bibfnamefont{J.~A.}},
  \bibinfo{author}{\bibfnamefont{D.~D.} \bibnamefont{Awschalom}},
  \bibinfo{author}{\bibfnamefont{A.~L.} \bibnamefont{Efros}}, and
  \bibinfo{author}{\bibfnamefont{A.~V.} \bibnamefont{Rodina}},
  \bibinfo{year}{2002}, {``}\bibinfo{title}{Spin dynamics in semiconductor
  nanocrystals},{''} \bibinfo{journal}{Phys. Rev. B}
  \textbf{\bibinfo{volume}{66}},  \bibinfo{pages}{125307}.

\bibitem[{\citenamefont{Gupta} \emph{et~al.}(1999)\citenamefont{Gupta,
  Awschalom, Peng, and Alivisatos}}]{Gupta1999:PRB}
\bibinfo{author}{\bibnamefont{Gupta}, \bibfnamefont{J.~A.}},
  \bibinfo{author}{\bibfnamefont{D.~D.} \bibnamefont{Awschalom}},
  \bibinfo{author}{\bibfnamefont{X.}~\bibnamefont{Peng}}, and
  \bibinfo{author}{\bibfnamefont{A.~P.} \bibnamefont{Alivisatos}},
  \bibinfo{year}{1999}, {``}\bibinfo{title}{Spin coherence in semiconductor
  quantum dots},{''} \bibinfo{journal}{Phys. Rev. B}
  \textbf{\bibinfo{volume}{59}},  \bibinfo{pages}{10421--10424}.

\bibitem[{\citenamefont{Gurzhi} \emph{et~al.}(2001)\citenamefont{Gurzhi,
  Kalinenko, Kopeliovich, and Yanovskii}}]{Gurzhi2001:FNT}
\bibinfo{author}{\bibnamefont{Gurzhi}, \bibfnamefont{R.~N.}},
  \bibinfo{author}{\bibfnamefont{A.~N.} \bibnamefont{Kalinenko}},
  \bibinfo{author}{\bibfnamefont{A.~I.} \bibnamefont{Kopeliovich}}, and
  \bibinfo{author}{\bibfnamefont{A.~V.} \bibnamefont{Yanovskii}},
  \bibinfo{year}{2001}, {``}\bibinfo{title}{Nonmagnetic spinguides and spin
  transport in semiconductors},{''} \bibinfo{journal}{Fiz. Nizk. Temp.}
  \textbf{\bibinfo{volume}{27}},  \bibinfo{pages}{1332--1334}
  \bibinfo{note}{[Low Temp. Phys. {\bf 27}, 985-986 (2001)]}.

\bibitem[{\citenamefont{Gurzhi} \emph{et~al.}(2003)\citenamefont{Gurzhi,
  Kalinenko, Kopeliovich, Yanovsky, Bogachek, and Landman}}]{Gurzhi2003:P}
\bibinfo{author}{\bibnamefont{Gurzhi}, \bibfnamefont{R.~N.}},
  \bibinfo{author}{\bibfnamefont{A.~N.} \bibnamefont{Kalinenko}},
  \bibinfo{author}{\bibfnamefont{A.~I.} \bibnamefont{Kopeliovich}},
  \bibinfo{author}{\bibfnamefont{A.~V.} \bibnamefont{Yanovsky}},
  \bibinfo{author}{\bibfnamefont{E.~N.} \bibnamefont{Bogachek}}, and
  \bibinfo{author}{\bibfnamefont{U.}~\bibnamefont{Landman}},
  \bibinfo{year}{2003}, {``}\bibinfo{title}{Spin-guide source for the
  generation of the high spin-polarized current},{''} \bibinfo{journal}{Phys.
  Rev. B} \textbf{\bibinfo{volume}{68}},  \bibinfo{pages}{125115}.

\bibitem[{\citenamefont{Gustavsson}
  \emph{et~al.}(2001)\citenamefont{Gustavsson, George, Etgens, and
  Eddrief}}]{Gustavsson2001:PRB}
\bibinfo{author}{\bibnamefont{Gustavsson}, \bibfnamefont{F.}},
  \bibinfo{author}{\bibfnamefont{J.-M.} \bibnamefont{George}},
  \bibinfo{author}{\bibfnamefont{V.~H.} \bibnamefont{Etgens}}, and
  \bibinfo{author}{\bibfnamefont{M.}~\bibnamefont{Eddrief}},
  \bibinfo{year}{2001}, {``}\bibinfo{title}{Structural and transport properties
  of epitaxial {Fe/ZnSe/FeCo} magnetic tunnel junctions},{''}
  \bibinfo{journal}{Phys. Rev. B} \textbf{\bibinfo{volume}{64}},
  \bibinfo{pages}{184422}.

\bibitem[{\citenamefont{Guth} \emph{et~al.}(2001)\citenamefont{Guth, Dinia,
  Schmerber, and {van den Berg}}}]{Guth2001:APL}
\bibinfo{author}{\bibnamefont{Guth}, \bibfnamefont{M.}},
  \bibinfo{author}{\bibfnamefont{A.}~\bibnamefont{Dinia}},
  \bibinfo{author}{\bibfnamefont{G.}~\bibnamefont{Schmerber}}, and
  \bibinfo{author}{\bibfnamefont{H.~A.~M.} \bibnamefont{{van den Berg}}},
  \bibinfo{year}{2001}, {``}\bibinfo{title}{Tunnel magnetoresistance in
  magnetic tunnel junctions with a {ZnS} barrier},{''} \bibinfo{journal}{Appl.
  Phys. Lett.} \textbf{\bibinfo{volume}{78}},  \bibinfo{pages}{3487--3489}.

\bibitem[{\citenamefont{Hach{\'e}} \emph{et~al.}(1997)\citenamefont{Hach{\'e},
  Kostoulas, Atanasov, Hughes, Sipe, and {van Driel}}}]{Hache1997:PRL}
\bibinfo{author}{\bibnamefont{Hach{\'e}}, \bibfnamefont{A.}},
  \bibinfo{author}{\bibfnamefont{Y.}~\bibnamefont{Kostoulas}},
  \bibinfo{author}{\bibfnamefont{R.}~\bibnamefont{Atanasov}},
  \bibinfo{author}{\bibfnamefont{J.~L.~P.} \bibnamefont{Hughes}},
  \bibinfo{author}{\bibfnamefont{J.~E.} \bibnamefont{Sipe}}, and
  \bibinfo{author}{\bibfnamefont{H.~M.} \bibnamefont{{van Driel}}},
  \bibinfo{year}{1997}, {``}\bibinfo{title}{Observation of coherently
  controlled photocurrent in unbiased, bulk {GaAs}},{''}
  \bibinfo{journal}{Phys. Rev. Lett.} \textbf{\bibinfo{volume}{78}},
  \bibinfo{pages}{306--309}.

\bibitem[{\citenamefont{{H\"{a}gele}}
  \emph{et~al.}(1998)\citenamefont{{H\"{a}gele}, Ostreich, {R\"{u}hle}, Nestle,
  and Eberl}}]{Hagele1998:APL}
\bibinfo{author}{\bibnamefont{{H\"{a}gele}}, \bibfnamefont{D.}},
  \bibinfo{author}{\bibfnamefont{M.}~\bibnamefont{Ostreich}},
  \bibinfo{author}{\bibfnamefont{W.~W.} \bibnamefont{{R\"{u}hle}}},
  \bibinfo{author}{\bibfnamefont{N.}~\bibnamefont{Nestle}}, and
  \bibinfo{author}{\bibfnamefont{K.}~\bibnamefont{Eberl}},
  \bibinfo{year}{1998}, {``}\bibinfo{title}{Spin transport in {GaAs}},{''}
  \bibinfo{journal}{Appl. Phys. Lett.} \textbf{\bibinfo{volume}{73}},
  \bibinfo{pages}{1580--1582}.

\bibitem[{\citenamefont{Hall} \emph{et~al.}(1999)\citenamefont{Hall, Leonard,
  {van Driel}, Kost, Selvig, and Chow}}]{Hall1999:APL}
\bibinfo{author}{\bibnamefont{Hall}, \bibfnamefont{K.~C.}},
  \bibinfo{author}{\bibfnamefont{S.~W.} \bibnamefont{Leonard}},
  \bibinfo{author}{\bibfnamefont{H.~M.} \bibnamefont{{van Driel}}},
  \bibinfo{author}{\bibfnamefont{A.~R.} \bibnamefont{Kost}},
  \bibinfo{author}{\bibfnamefont{E.}~\bibnamefont{Selvig}}, and
  \bibinfo{author}{\bibfnamefont{D.~H.} \bibnamefont{Chow}},
  \bibinfo{year}{1999}, {``}\bibinfo{title}{Subpicosecond spin relaxation in
  {GaAsSb} multiple quantum wells},{''} \bibinfo{journal}{Appl. Phys. Lett.}
  \textbf{\bibinfo{volume}{75}},  \bibinfo{pages}{4156--4158}.

\bibitem[{\citenamefont{Halterman and Valls}(2002)}]{Halterman2002:PRB}
\bibinfo{author}{\bibnamefont{Halterman}, \bibfnamefont{K.}}, and
  \bibinfo{author}{\bibfnamefont{O.~T.} \bibnamefont{Valls}},
  \bibinfo{year}{2002}, {``}\bibinfo{title}{Proximity effects at
  ferromagnet-superconductor interfaces},{''} \bibinfo{journal}{Phys. Rev. B}
  \textbf{\bibinfo{volume}{65}},  \bibinfo{pages}{014509}.

\bibitem[{\citenamefont{Hammar} \emph{et~al.}(1999)\citenamefont{Hammar,
  Bennett, Yang, and Johnson}}]{Hammar1999:PRL}
\bibinfo{author}{\bibnamefont{Hammar}, \bibfnamefont{P.~R.}},
  \bibinfo{author}{\bibfnamefont{B.~R.} \bibnamefont{Bennett}},
  \bibinfo{author}{\bibfnamefont{M.~J.} \bibnamefont{Yang}}, and
  \bibinfo{author}{\bibfnamefont{M.}~\bibnamefont{Johnson}},
  \bibinfo{year}{1999}, {``}\bibinfo{title}{Observation of spin injection at a
  ferromagnet-semiconductor interface},{''} \bibinfo{journal}{Phys. Rev. Lett.}
  \textbf{\bibinfo{volume}{83}},  \bibinfo{pages}{203--206}.

\bibitem[{\citenamefont{Hammar} \emph{et~al.}(2000)\citenamefont{Hammar,
  Bennett, Yang, and Johnson}}]{Hammar2000:PRL}
\bibinfo{author}{\bibnamefont{Hammar}, \bibfnamefont{P.~R.}},
  \bibinfo{author}{\bibfnamefont{B.~R.} \bibnamefont{Bennett}},
  \bibinfo{author}{\bibfnamefont{M.~J.} \bibnamefont{Yang}}, and
  \bibinfo{author}{\bibfnamefont{M.}~\bibnamefont{Johnson}},
  \bibinfo{year}{2000}, {``}\bibinfo{title}{{A reply to the comment by F. G.
  Monzon and H. X. Tang and M. L. Roukes}},{''} \bibinfo{journal}{Phys. Rev.
  Lett.} \textbf{\bibinfo{volume}{84}},  \bibinfo{pages}{5024--5025}.

\bibitem[{\citenamefont{Hammar and Johnson}(2000)}]{Hammar2000:PRB}
\bibinfo{author}{\bibnamefont{Hammar}, \bibfnamefont{P.~R.}}, and
  \bibinfo{author}{\bibfnamefont{M.}~\bibnamefont{Johnson}},
  \bibinfo{year}{2000}, {``}\bibinfo{title}{Potentiometric measurements of the
  spin-split subbands in a two-dimensional electron gas},{''}
  \bibinfo{journal}{Phys. Rev. B} \textbf{\bibinfo{volume}{61}},
  \bibinfo{pages}{7207--7210}.

\bibitem[{\citenamefont{Hammar and Johnson}(2001)}]{Hammar2001:APL}
\bibinfo{author}{\bibnamefont{Hammar}, \bibfnamefont{P.~R.}}, and
  \bibinfo{author}{\bibfnamefont{M.}~\bibnamefont{Johnson}},
  \bibinfo{year}{2001}, {``}\bibinfo{title}{Spin-dependent current transmission
  across a ferromagnet-insulator-two-dimensional electron gas junction},{''}
  \bibinfo{journal}{Appl. Phys. Lett.} \textbf{\bibinfo{volume}{79}},
  \bibinfo{pages}{2591--2593}.

\bibitem[{\citenamefont{Hammar and Johnson}(2002)}]{Hammar2002:PRL}
\bibinfo{author}{\bibnamefont{Hammar}, \bibfnamefont{P.~R.}}, and
  \bibinfo{author}{\bibfnamefont{M.}~\bibnamefont{Johnson}},
  \bibinfo{year}{2002}, {``}\bibinfo{title}{Detection of spin-polarized
  electrons injected into a two-dimensional electron gas},{''}
  \bibinfo{journal}{Phys. Rev. Lett.} \textbf{\bibinfo{volume}{88}},
  \bibinfo{pages}{066806}.

\bibitem[{\citenamefont{Hanbicki} \emph{et~al.}(2003)\citenamefont{Hanbicki,
  {van t Erve}, Magno, Kioseoglou, Li, Jonker, Itskos, Mallory, Yasar, and
  Petrou}}]{Hanbicki2003:P}
\bibinfo{author}{\bibnamefont{Hanbicki}, \bibfnamefont{.~A.}},
  \bibinfo{author}{\bibfnamefont{O.~M.~J.} \bibnamefont{{van t Erve}}},
  \bibinfo{author}{\bibfnamefont{R.}~\bibnamefont{Magno}},
  \bibinfo{author}{\bibfnamefont{G.}~\bibnamefont{Kioseoglou}},
  \bibinfo{author}{\bibfnamefont{C.~H.} \bibnamefont{Li}},
  \bibinfo{author}{\bibfnamefont{B.~T.} \bibnamefont{Jonker}},
  \bibinfo{author}{\bibfnamefont{G.}~\bibnamefont{Itskos}},
  \bibinfo{author}{\bibfnamefont{R.}~\bibnamefont{Mallory}},
  \bibinfo{author}{\bibfnamefont{M.}~\bibnamefont{Yasar}}, and
  \bibinfo{author}{\bibfnamefont{A.}~\bibnamefont{Petrou}},
  \bibinfo{year}{2003}, {``}\bibinfo{title}{Analysis of the transport process
  providing spin injection through an {Fe/AlGaAs} {Schottky} barrier},{''}
  \bibinfo{journal}{Appl. Phys. Lett.} \textbf{\bibinfo{volume}{82}},
  \bibinfo{pages}{4092--4094}.

\bibitem[{\citenamefont{Hanbicki and Jonker}(2002)}]{Hanbicki2002:APLb}
\bibinfo{author}{\bibnamefont{Hanbicki}, \bibfnamefont{A.~T.}}, and
  \bibinfo{author}{\bibfnamefont{B.~T.} \bibnamefont{Jonker}},
  \bibinfo{year}{2002}, {``}\bibinfo{title}{Response to Comment on `{Efficient}
  electrical spin injection from a magnetic metal/tunnel barrier contact into a
  semiconductor'},{''} \bibinfo{journal}{Appl. Phys. Lett.}
  \textbf{\bibinfo{volume}{81}},  \bibinfo{pages}{2131}.

\bibitem[{\citenamefont{Hanbicki} \emph{et~al.}(2002)\citenamefont{Hanbicki,
  Jonker, Itskos, Kioseoglou, and Petrou}}]{Hanbicki2002:APLa}
\bibinfo{author}{\bibnamefont{Hanbicki}, \bibfnamefont{A.~T.}},
  \bibinfo{author}{\bibfnamefont{B.~T.} \bibnamefont{Jonker}},
  \bibinfo{author}{\bibfnamefont{G.}~\bibnamefont{Itskos}},
  \bibinfo{author}{\bibfnamefont{G.}~\bibnamefont{Kioseoglou}}, and
  \bibinfo{author}{\bibfnamefont{A.}~\bibnamefont{Petrou}},
  \bibinfo{year}{2002}, {``}\bibinfo{title}{Efficient electrical spin injection
  from a magnetic metal/tunnel barrier contact into a semiconductor},{''}
  \bibinfo{journal}{Appl. Phys. Lett.} \textbf{\bibinfo{volume}{80}},
  \bibinfo{pages}{1240--1242}.

\bibitem[{\citenamefont{Hanle}(1924)}]{Hanle1924:ZP}
\bibinfo{author}{\bibnamefont{Hanle}, \bibfnamefont{W.}}, \bibinfo{year}{1924},
  {``}\bibinfo{title}{{\"U}ber magnetische beeinflussung der polarisation der
  resonanzfluoreszenz},{''} \bibinfo{journal}{Z. Phys.}
  \textbf{\bibinfo{volume}{30}},  \bibinfo{pages}{93--105}.

\bibitem[{\citenamefont{Hanson}
  \emph{et~al.}(2003{\natexlab{a}})\citenamefont{Hanson, Vandersypen, {van
  Beveren}, Elzerman, Vink, and Kouwenhoven}}]{Hanson2003:P}
\bibinfo{author}{\bibnamefont{Hanson}, \bibfnamefont{R.}},
  \bibinfo{author}{\bibfnamefont{L.~M.~K.} \bibnamefont{Vandersypen}},
  \bibinfo{author}{\bibfnamefont{L.~H.~W.} \bibnamefont{{van Beveren}}},
  \bibinfo{author}{\bibfnamefont{J.~M.} \bibnamefont{Elzerman}},
  \bibinfo{author}{\bibfnamefont{I.~T.} \bibnamefont{Vink}}, and
  \bibinfo{author}{\bibfnamefont{L.~P.} \bibnamefont{Kouwenhoven}},
  \bibinfo{year}{2003}{\natexlab{a}}, {``}\bibinfo{title}{Semiconductor
  few-electron quantum dot operated as a bipolar spin filter},{''}
  \eprint{cond-mat/0311414}.

\bibitem[{\citenamefont{Hanson}
  \emph{et~al.}(2003{\natexlab{b}})\citenamefont{Hanson, Witkamp, Vandersypen,
  {Willems van Beveren}, Elzerman, and Kouwenhoven}}]{Hanson2003:lanl}
\bibinfo{author}{\bibnamefont{Hanson}, \bibfnamefont{R.}},
  \bibinfo{author}{\bibfnamefont{B.}~\bibnamefont{Witkamp}},
  \bibinfo{author}{\bibfnamefont{L.~M.~K.} \bibnamefont{Vandersypen}},
  \bibinfo{author}{\bibfnamefont{L.~H.} \bibnamefont{{Willems van Beveren}}},
  \bibinfo{author}{\bibfnamefont{J.~M.} \bibnamefont{Elzerman}}, and
  \bibinfo{author}{\bibfnamefont{L.~P.} \bibnamefont{Kouwenhoven}},
  \bibinfo{year}{2003}{\natexlab{b}}, {``}\bibinfo{title}{Zeeman energy and
  spin relaxation in a one-electron quantum dot},{''} \bibinfo{journal}{Phys.
  Rev. Lett.} \textbf{\bibinfo{volume}{91}},  \bibinfo{pages}{196802}.

\bibitem[{\citenamefont{Hao} \emph{et~al.}(1990)\citenamefont{Hao, Moodera, and
  Meservey}}]{Hao1990:PRB}
\bibinfo{author}{\bibnamefont{Hao}, \bibfnamefont{X.}},
  \bibinfo{author}{\bibfnamefont{J.~S.} \bibnamefont{Moodera}}, and
  \bibinfo{author}{\bibfnamefont{R.}~\bibnamefont{Meservey}},
  \bibinfo{year}{1990}, {``}\bibinfo{title}{Spin-filter effect of ferromagnetic
  europium sulfide tunnel barriers},{''} \bibinfo{journal}{Phys. Rev. B}
  \textbf{\bibinfo{volume}{42}},  \bibinfo{pages}{8235--8243}.

\bibitem[{\citenamefont{Happer}(1972)}]{Happer1972:RMP}
\bibinfo{author}{\bibnamefont{Happer}, \bibfnamefont{W.}},
  \bibinfo{year}{1972}, {``}\bibinfo{title}{Optical Pumping},{''}
  \bibinfo{journal}{Rev. Mod. Phys.} \textbf{\bibinfo{volume}{44}},
  \bibinfo{pages}{169--249}.

\bibitem[{\citenamefont{{Hartman (Ed.)}}(2000)}]{Hartmann:2000}
\bibinfo{author}{\bibnamefont{{Hartman (Ed.)}}, \bibfnamefont{U.}},
  \bibinfo{year}{2000}, \emph{\bibinfo{title}{Magnetic Multilayers and Giant
  Magnetoresistance}} (\bibinfo{publisher}{Springer, Berlin}).

\bibitem[{\citenamefont{Hass} \emph{et~al.}(1994)\citenamefont{Hass, Covington,
  Feldmann, Greene, and Johnson}}]{Hass1994:PC}
\bibinfo{author}{\bibnamefont{Hass}, \bibfnamefont{N.}},
  \bibinfo{author}{\bibfnamefont{M.}~\bibnamefont{Covington}},
  \bibinfo{author}{\bibfnamefont{W.~L.} \bibnamefont{Feldmann}},
  \bibinfo{author}{\bibfnamefont{L.~H.} \bibnamefont{Greene}}, and
  \bibinfo{author}{\bibfnamefont{M.}~\bibnamefont{Johnson}},
  \bibinfo{year}{1994}, {``}\bibinfo{title}{Transport properties of
  {YBa$_2$Cu$_3$O$_{7-\delta}$}/ferromagnetic interfaces},{''}
  \bibinfo{journal}{Physica C} \textbf{\bibinfo{volume}{235-240}},
  \bibinfo{pages}{1905--1906}.

\bibitem[{\citenamefont{Hayashi} \emph{et~al.}(1997)\citenamefont{Hayashi,
  Tanaka, Nishinaga, Shimada, Tsuchiya, and Otuka}}]{Hayashi1997:JCG}
\bibinfo{author}{\bibnamefont{Hayashi}, \bibfnamefont{T.}},
  \bibinfo{author}{\bibfnamefont{M.}~\bibnamefont{Tanaka}},
  \bibinfo{author}{\bibfnamefont{T.}~\bibnamefont{Nishinaga}},
  \bibinfo{author}{\bibfnamefont{H.}~\bibnamefont{Shimada}},
  \bibinfo{author}{\bibfnamefont{T.}~\bibnamefont{Tsuchiya}}, and
  \bibinfo{author}{\bibfnamefont{Y.}~\bibnamefont{Otuka}},
  \bibinfo{year}{1997}, {``}\bibinfo{title}{{(GaMn)As}: {GaAs}-based {III-V}
  diluted magnetic semiconductors grown by molecular beam epitaxy},{''}
  \bibinfo{journal}{J. Cryst. Growth} \textbf{\bibinfo{volume}{175/176}},
  \bibinfo{pages}{1063--1068}.

\bibitem[{\citenamefont{Heberle} \emph{et~al.}(1996)\citenamefont{Heberle,
  Blaumberg, Binder, Kuhn, {K\"{o}hler}, and Ploog}}]{Heberle1996:IEEE}
\bibinfo{author}{\bibnamefont{Heberle}, \bibfnamefont{A.~P.}},
  \bibinfo{author}{\bibfnamefont{J.~J.} \bibnamefont{Blaumberg}},
  \bibinfo{author}{\bibfnamefont{E.}~\bibnamefont{Binder}},
  \bibinfo{author}{\bibfnamefont{T.}~\bibnamefont{Kuhn}},
  \bibinfo{author}{\bibfnamefont{K.}~\bibnamefont{{K\"{o}hler}}}, and
  \bibinfo{author}{\bibfnamefont{K.~H.} \bibnamefont{Ploog}},
  \bibinfo{year}{1996}, {``}\bibinfo{title}{Coherent control of exciton density
  and spin},{''} \bibinfo{journal}{IEEE J. Sel. Top. Quant.}
  \textbf{\bibinfo{volume}{2}},  \bibinfo{pages}{769--775}.

\bibitem[{\citenamefont{Heberle} \emph{et~al.}(1994)\citenamefont{Heberle,
  {R\"{u}hle}, and Ploog}}]{Heberle1994:PRL}
\bibinfo{author}{\bibnamefont{Heberle}, \bibfnamefont{A.~P.}},
  \bibinfo{author}{\bibfnamefont{W.~W.} \bibnamefont{{R\"{u}hle}}}, and
  \bibinfo{author}{\bibfnamefont{K.}~\bibnamefont{Ploog}},
  \bibinfo{year}{1994}, {``}\bibinfo{title}{Quantum beats of electron {Larmor}
  precession in {GaAs} wells},{''} \bibinfo{journal}{Phys. Rev. Lett.}
  \textbf{\bibinfo{volume}{72}},  \bibinfo{pages}{3887--3890}.

\bibitem[{\citenamefont{Heersche} \emph{et~al.}(2001)\citenamefont{Heersche,
  {Sch\"{a}pers}, Nitta, and Takayanagi}}]{Schapers2001b:PRB}
\bibinfo{author}{\bibnamefont{Heersche}, \bibfnamefont{H.~B.}},
  \bibinfo{author}{\bibfnamefont{T.}~\bibnamefont{{Sch\"{a}pers}}},
  \bibinfo{author}{\bibfnamefont{J.}~\bibnamefont{Nitta}}, and
  \bibinfo{author}{\bibfnamefont{H.}~\bibnamefont{Takayanagi}},
  \bibinfo{year}{2001}, {``}\bibinfo{title}{Enhancement of spin injection from
  ferromagnetic metal into a two-dimensional electron gas using a tunnel
  barrier},{''} \bibinfo{journal}{Phys. Rev. B} \textbf{\bibinfo{volume}{64}},
  \bibinfo{pages}{161307}.

\bibitem[{\citenamefont{Hehn} \emph{et~al.}(2002)\citenamefont{Hehn, Montaigne,
  and Schuhl}}]{Hehn2002:PRB}
\bibinfo{author}{\bibnamefont{Hehn}, \bibfnamefont{M.}},
  \bibinfo{author}{\bibfnamefont{F.}~\bibnamefont{Montaigne}}, and
  \bibinfo{author}{\bibfnamefont{A.}~\bibnamefont{Schuhl}},
  \bibinfo{year}{2002}, {``}\bibinfo{title}{Hot-electron three-terminal devices
  based on magnetic tunnel junction stacks},{''} \bibinfo{journal}{Phys. Rev.
  B} \textbf{\bibinfo{volume}{66}},  \bibinfo{pages}{144411}.

\bibitem[{\citenamefont{Heida} \emph{et~al.}(1998)\citenamefont{Heida, {van
  Wees}, Kuipers, Klapwijk, and Borgh}}]{Heida1998:PRB}
\bibinfo{author}{\bibnamefont{Heida}, \bibfnamefont{J.~P.}},
  \bibinfo{author}{\bibfnamefont{B.~J.} \bibnamefont{{van Wees}}},
  \bibinfo{author}{\bibfnamefont{J.~J.} \bibnamefont{Kuipers}},
  \bibinfo{author}{\bibfnamefont{T.~M.} \bibnamefont{Klapwijk}}, and
  \bibinfo{author}{\bibfnamefont{G.}~\bibnamefont{Borgh}},
  \bibinfo{year}{1998}, {``}\bibinfo{title}{Spin-orbit interaction in a
  two-dimensional electron gas in a {InAs/AlSb} quantum well with
  gate-controlled electron density},{''} \bibinfo{journal}{Phys. Rev. B}
  \textbf{\bibinfo{volume}{57}},  \bibinfo{pages}{11911--11914}.

\bibitem[{\citenamefont{Heide}(2001)}]{Heide2001:PRL}
\bibinfo{author}{\bibnamefont{Heide}, \bibfnamefont{C.}}, \bibinfo{year}{2001},
  {``}\bibinfo{title}{Spin currents in magnetic films},{''}
  \bibinfo{journal}{Phys. Rev. Lett.} \textbf{\bibinfo{volume}{87}},
  \bibinfo{pages}{197201}.

\bibitem[{\citenamefont{Hermann} \emph{et~al.}(1985)\citenamefont{Hermann,
  Lampel, and Safarov}}]{Hermann1985:APF}
\bibinfo{author}{\bibnamefont{Hermann}, \bibfnamefont{C.}},
  \bibinfo{author}{\bibfnamefont{G.}~\bibnamefont{Lampel}}, and
  \bibinfo{author}{\bibfnamefont{V.~I.} \bibnamefont{Safarov}},
  \bibinfo{year}{1985}, {``}\bibinfo{title}{Optical pumping in
  semiconductors},{''} \bibinfo{journal}{Ann. Phys. (Paris)}
  \textbf{\bibinfo{volume}{10}},  \bibinfo{pages}{1117--1138}.

\bibitem[{\citenamefont{Hershfield and Zhao}(1997)}]{Hershfield1997:PRB}
\bibinfo{author}{\bibnamefont{Hershfield}, \bibfnamefont{S.}}, and
  \bibinfo{author}{\bibfnamefont{H.~L.} \bibnamefont{Zhao}},
  \bibinfo{year}{1997}, {``}\bibinfo{title}{Charge and spin transport through a
  metallic ferromagnetic-paramagnetic-ferromagnetic junction},{''}
  \bibinfo{journal}{Phys. Rev. B} \textbf{\bibinfo{volume}{56}},
  \bibinfo{pages}{3296--3305}.

\bibitem[{\citenamefont{Hertz and Aoi}(1973)}]{Hertz1973:PRB}
\bibinfo{author}{\bibnamefont{Hertz}, \bibfnamefont{J.~A.}}, and
  \bibinfo{author}{\bibfnamefont{K.}~\bibnamefont{Aoi}}, \bibinfo{year}{1973},
  {``}\bibinfo{title}{Spin-Dependent Tunneling from Transition-Metal
  Ferromagnets},{''} \bibinfo{journal}{Phys. Rev. B}
  \textbf{\bibinfo{volume}{8}},  \bibinfo{pages}{3252--3256}.

\bibitem[{\citenamefont{Hilton and Tang}(2002)}]{Hilton2002:PRL}
\bibinfo{author}{\bibnamefont{Hilton}, \bibfnamefont{D.~J.}}, and
  \bibinfo{author}{\bibfnamefont{C.~L.} \bibnamefont{Tang}},
  \bibinfo{year}{2002}, {``}\bibinfo{title}{Optical orientation and femtosecond
  relaxation of spin-polarized holes in {GaAs}},{''} \bibinfo{journal}{Phys.
  Rev. Lett.} \textbf{\bibinfo{volume}{89}},  \bibinfo{pages}{146601}.

\bibitem[{\citenamefont{Hirohata} \emph{et~al.}(2001)\citenamefont{Hirohata,
  Xu, Guertler, Bland, and Holmes}}]{Hirohata2001:PRB}
\bibinfo{author}{\bibnamefont{Hirohata}, \bibfnamefont{A.}},
  \bibinfo{author}{\bibfnamefont{Y.~B.} \bibnamefont{Xu}},
  \bibinfo{author}{\bibfnamefont{C.~M.} \bibnamefont{Guertler}},
  \bibinfo{author}{\bibfnamefont{J.~A.~C.} \bibnamefont{Bland}}, and
  \bibinfo{author}{\bibfnamefont{S.~N.} \bibnamefont{Holmes}},
  \bibinfo{year}{2001}, {``}\bibinfo{title}{Spin-polarized electron transport
  in ferromagnet/semiconductor hybrid structures induced by photon
  excitation},{''} \bibinfo{journal}{Phys. Rev. B}
  \textbf{\bibinfo{volume}{63}},  \bibinfo{pages}{104425}.

\bibitem[{\citenamefont{Hirota} \emph{et~al.}(2002)\citenamefont{Hirota,
  Sakakima, and Inomata}}]{Hirota:2002}
\bibinfo{author}{\bibnamefont{Hirota}, \bibfnamefont{E.}},
  \bibinfo{author}{\bibfnamefont{H.}~\bibnamefont{Sakakima}}, and
  \bibinfo{author}{\bibfnamefont{K.}~\bibnamefont{Inomata}},
  \bibinfo{year}{2002}, \emph{\bibinfo{title}{Giant Magneto-Resistance
  Devices}} (\bibinfo{publisher}{Springer, Berlin}).

\bibitem[{\citenamefont{Hirsch}(1999)}]{Hirsch1999:PRL}
\bibinfo{author}{\bibnamefont{Hirsch}, \bibfnamefont{J.~E.}},
  \bibinfo{year}{1999}, {``}\bibinfo{title}{Spin {Hall} effect},{''}
  \bibinfo{journal}{Phys. Rev. Lett.} \textbf{\bibinfo{volume}{83}},
  \bibinfo{pages}{1834--1837}.

\bibitem[{\citenamefont{Hong and Mills}(2000)}]{Hong2000:PRB}
\bibinfo{author}{\bibnamefont{Hong}, \bibfnamefont{J.}}, and
  \bibinfo{author}{\bibfnamefont{D.~L.} \bibnamefont{Mills}},
  \bibinfo{year}{2000}, {``}\bibinfo{title}{Spin dependence of the inelastic
  electron mean free path in {Fe} and {Ni}: Explicit calculations and
  implications},{''} \bibinfo{journal}{Phys. Rev. B}
  \textbf{\bibinfo{volume}{62}},  \bibinfo{pages}{5589--5600}.

\bibitem[{\citenamefont{Hu} \emph{et~al.}(1999)\citenamefont{Hu, Nitta,
  Akazaki, Takayanagi, Osaka, Pfeffer, and Zawadzki}}]{Hu1999:PRB}
\bibinfo{author}{\bibnamefont{Hu}, \bibfnamefont{C.}},
  \bibinfo{author}{\bibfnamefont{J.}~\bibnamefont{Nitta}},
  \bibinfo{author}{\bibfnamefont{T.}~\bibnamefont{Akazaki}},
  \bibinfo{author}{\bibfnamefont{H.}~\bibnamefont{Takayanagi}},
  \bibinfo{author}{\bibfnamefont{J.}~\bibnamefont{Osaka}},
  \bibinfo{author}{\bibfnamefont{P.}~\bibnamefont{Pfeffer}}, and
  \bibinfo{author}{\bibfnamefont{W.}~\bibnamefont{Zawadzki}},
  \bibinfo{year}{1999}, {``}\bibinfo{title}{Zero-field spin splitting in an
  inverted {In$_{0.53}$Ga$_{0.47}$As/In$_{0.52}$Al$_{0.48}$As} heterostructure:
  Band nonparabolicity influence and the subband dependence},{''}
  \bibinfo{journal}{Phys. Rev. B} \textbf{\bibinfo{volume}{60}},
  \bibinfo{pages}{7736--7739}.

\bibitem[{\citenamefont{Hu and Matsuyama}(2001)}]{Hu2001:PRL}
\bibinfo{author}{\bibnamefont{Hu}, \bibfnamefont{C.-M.}}, and
  \bibinfo{author}{\bibfnamefont{T.}~\bibnamefont{Matsuyama}},
  \bibinfo{year}{2001}, {``}\bibinfo{title}{Spin injection across a
  heterojunction: {A} ballistic picture},{''} \bibinfo{journal}{Phys. Rev.
  Lett.} \textbf{\bibinfo{volume}{87}},  \bibinfo{pages}{066803}.

\bibitem[{\citenamefont{Hu} \emph{et~al.}(2001{\natexlab{a}})\citenamefont{Hu,
  Nitta, Jensen, Hansen, and Takayanagi}}]{Hu2001:PRB}
\bibinfo{author}{\bibnamefont{Hu}, \bibfnamefont{C.-M.}},
  \bibinfo{author}{\bibfnamefont{J.}~\bibnamefont{Nitta}},
  \bibinfo{author}{\bibfnamefont{A.}~\bibnamefont{Jensen}},
  \bibinfo{author}{\bibfnamefont{J.~B.} \bibnamefont{Hansen}}, and
  \bibinfo{author}{\bibfnamefont{H.}~\bibnamefont{Takayanagi}},
  \bibinfo{year}{2001}{\natexlab{a}}, {``}\bibinfo{title}{Spin-polarized
  transport in a two-dimensional electron gas with interdigital-ferromagnetic
  contacts},{''} \bibinfo{journal}{Phys. Rev. B} \textbf{\bibinfo{volume}{63}},
   \bibinfo{pages}{125333}.

\bibitem[{\citenamefont{Hu}(1994)}]{Hu1994:PRL}
\bibinfo{author}{\bibnamefont{Hu}, \bibfnamefont{C.~R.}}, \bibinfo{year}{1994},
  {``}\bibinfo{title}{Midgap surface states as a novel signature for
  $d_{{x^2_a}-{x^2_b}}$-wave superconductivity},{''} \bibinfo{journal}{Phys.
  Rev. Lett.} \textbf{\bibinfo{volume}{72}},  \bibinfo{pages}{1526--1529}.

\bibitem[{\citenamefont{Hu}(1998)}]{Hu1998:PRB}
\bibinfo{author}{\bibnamefont{Hu}, \bibfnamefont{C.-R.}}, \bibinfo{year}{1998},
  {``}\bibinfo{title}{Origin of the zero-bias conductance peaks observed
  ubiquitously in high-{$T_c$} superconductors},{''} \bibinfo{journal}{Phys.
  Rev. B} \textbf{\bibinfo{volume}{57}},  \bibinfo{pages}{1266--1276}.

\bibitem[{\citenamefont{Hu and Yan}(1999)}]{Hu1999:PRBb}
\bibinfo{author}{\bibnamefont{Hu}, \bibfnamefont{C.-R.}}, and
  \bibinfo{author}{\bibfnamefont{X.-Z.} \bibnamefont{Yan}},
  \bibinfo{year}{1999}, {``}\bibinfo{title}{Predicted giant magnetic moment on
  non-\{n0m\} surfaces of d-wave superconductors},{''} \bibinfo{journal}{Phys.
  Rev. B} \textbf{\bibinfo{volume}{60}},  \bibinfo{pages}{R12573--R12576}.

\bibitem[{\citenamefont{Hu and Suzuki}(2002)}]{Hu2002:PRL}
\bibinfo{author}{\bibnamefont{Hu}, \bibfnamefont{G.}}, and
  \bibinfo{author}{\bibfnamefont{Y.}~\bibnamefont{Suzuki}},
  \bibinfo{year}{2002}, {``}\bibinfo{title}{Negative spin polarization of
  {Fe$_3$O$_4$} in magnetite/manganite-based junctions},{''}
  \bibinfo{journal}{Phys. Rev. Lett.} \textbf{\bibinfo{volume}{89}},
  \bibinfo{pages}{276601}.

\bibitem[{\citenamefont{Hu and {Das Sarma}}(2000)}]{Hu2000:PRA}
\bibinfo{author}{\bibnamefont{Hu}, \bibfnamefont{X.}}, and
  \bibinfo{author}{\bibfnamefont{S.}~\bibnamefont{{Das Sarma}}},
  \bibinfo{year}{2000}, {``}\bibinfo{title}{Hilbert-space structure of a
  solid-state quantum computer: two-electron states of a double-quantum-dot
  artificial molecule},{''} \bibinfo{journal}{Phys. Rev. A}
  \textbf{\bibinfo{volume}{61}},  \bibinfo{pages}{062301}.

\bibitem[{\citenamefont{Hu and {Das Sarma}}(2001)}]{Hu2001:PRA}
\bibinfo{author}{\bibnamefont{Hu}, \bibfnamefont{X.}}, and
  \bibinfo{author}{\bibfnamefont{S.}~\bibnamefont{{Das Sarma}}},
  \bibinfo{year}{2001}, {``}\bibinfo{title}{Spin-based quantum computation in
  multielectron quantum dots},{''} \bibinfo{journal}{Phys. Rev. A}
  \textbf{\bibinfo{volume}{64}},  \bibinfo{pages}{042312}.

\bibitem[{\citenamefont{Hu} \emph{et~al.}(2001{\natexlab{b}})\citenamefont{Hu,
  de~Sousa, and {Das Sarma}}}]{Hu2001:lanl}
\bibinfo{author}{\bibnamefont{Hu}, \bibfnamefont{X.}},
  \bibinfo{author}{\bibfnamefont{R.}~\bibnamefont{de~Sousa}}, and
  \bibinfo{author}{\bibfnamefont{S.}~\bibnamefont{{Das Sarma}}},
  \bibinfo{year}{2001}{\natexlab{b}}, {``}\bibinfo{title}{Decoherence and
  dephasing in spin-based solid state quantum computers},{''}
  \eprint{cond-mat/0108339}.

\bibitem[{\citenamefont{{H\"{u}bner}}
  \emph{et~al.}(2003)\citenamefont{{H\"{u}bner}, {R\"{u}hle}, Klude, Hommel,
  {R. Bhat}, Sipe, and {van Driel}}}]{Hubner2003:PRL}
\bibinfo{author}{\bibnamefont{{H\"{u}bner}}, \bibfnamefont{J.}},
  \bibinfo{author}{\bibfnamefont{W.~W.} \bibnamefont{{R\"{u}hle}}},
  \bibinfo{author}{\bibfnamefont{M.}~\bibnamefont{Klude}},
  \bibinfo{author}{\bibfnamefont{D.}~\bibnamefont{Hommel}},
  \bibinfo{author}{\bibfnamefont{R.~D.} \bibnamefont{{R. Bhat}}},
  \bibinfo{author}{\bibfnamefont{J.~E.} \bibnamefont{Sipe}}, and
  \bibinfo{author}{\bibfnamefont{H.~M.} \bibnamefont{{van Driel}}},
  \bibinfo{year}{2003}, {``}\bibinfo{title}{Direct observation of optically
  injected spin-polarized currents in semiconductors},{''}
  \bibinfo{journal}{Phys. Rev. Lett.} \textbf{\bibinfo{volume}{90}},
  \bibinfo{pages}{216601}.

\bibitem[{\citenamefont{Imamura} \emph{et~al.}(2000)\citenamefont{Imamura,
  Kobayashi, Takahashi, and Maekawa}}]{Imamura2000:PRL}
\bibinfo{author}{\bibnamefont{Imamura}, \bibfnamefont{H.}},
  \bibinfo{author}{\bibfnamefont{N.}~\bibnamefont{Kobayashi}},
  \bibinfo{author}{\bibfnamefont{S.}~\bibnamefont{Takahashi}}, and
  \bibinfo{author}{\bibfnamefont{S.}~\bibnamefont{Maekawa}},
  \bibinfo{year}{2000}, {``}\bibinfo{title}{Conductance quantization and
  magnetoresistance in magnetic point contacts},{''} \bibinfo{journal}{Phys.
  Rev. Lett.} \textbf{\bibinfo{volume}{84}},  \bibinfo{pages}{1003--1006}.

\bibitem[{\citenamefont{Inoue and Makeawa}(1999)}]{Inoue1999:JMMM}
\bibinfo{author}{\bibnamefont{Inoue}, \bibfnamefont{J.}}, and
  \bibinfo{author}{\bibfnamefont{S.}~\bibnamefont{Makeawa}},
  \bibinfo{year}{1999}, {``}\bibinfo{title}{Effects of spin-flip and
  magnon-inelastic scattering on tunnel magnetoresistance},{''}
  \bibinfo{journal}{J. Magn. Magn. Mater.} \textbf{\bibinfo{volume}{198}},
  \bibinfo{pages}{167--169}.

\bibitem[{\citenamefont{Ionicioiu and {D'Amico}}(2003)}]{Ionicioiu2003:PRB}
\bibinfo{author}{\bibnamefont{Ionicioiu}, \bibfnamefont{R.}}, and
  \bibinfo{author}{\bibfnamefont{I.}~\bibnamefont{{D'Amico}}},
  \bibinfo{year}{2003}, {``}\bibinfo{title}{Mesoscopic {Stern-Gerlach} device
  to polarize spin currents},{''} \bibinfo{journal}{Phys. Rev. B}
  \textbf{\bibinfo{volume}{67}},  \bibinfo{pages}{041307}.

\bibitem[{\citenamefont{Isakovi\'c}
  \emph{et~al.}(2001)\citenamefont{Isakovi\'c, Carr, Strand, Schultz,
  Palmstr$\o$m, and Crowell}}]{Isakovic2001:PRB}
\bibinfo{author}{\bibnamefont{Isakovi\'c}, \bibfnamefont{A.~F.}},
  \bibinfo{author}{\bibfnamefont{D.~M.} \bibnamefont{Carr}},
  \bibinfo{author}{\bibfnamefont{J.}~\bibnamefont{Strand}},
  \bibinfo{author}{\bibfnamefont{B.~D.} \bibnamefont{Schultz}},
  \bibinfo{author}{\bibfnamefont{C.~J.} \bibnamefont{Palmstr$\o$m}}, and
  \bibinfo{author}{\bibfnamefont{P.~A.} \bibnamefont{Crowell}},
  \bibinfo{year}{2001}, {``}\bibinfo{title}{Optical pumping in
  ferromagnet-semiconductor heterostructures: Magneto-optics and spin
  transport},{''} \bibinfo{journal}{Phys. Rev. B}
  \textbf{\bibinfo{volume}{64}},  \bibinfo{pages}{161304}.

\bibitem[{\citenamefont{Itoh} \emph{et~al.}(1999)\citenamefont{Itoh, Shibata,
  Kumazaki, Inoue, and Maekawa}}]{Itoh1999:JPSP}
\bibinfo{author}{\bibnamefont{Itoh}, \bibfnamefont{H.}},
  \bibinfo{author}{\bibfnamefont{A.}~\bibnamefont{Shibata}},
  \bibinfo{author}{\bibfnamefont{T.}~\bibnamefont{Kumazaki}},
  \bibinfo{author}{\bibfnamefont{J.}~\bibnamefont{Inoue}}, and
  \bibinfo{author}{\bibfnamefont{S.}~\bibnamefont{Maekawa}},
  \bibinfo{year}{1999}, {``}\bibinfo{title}{Effects of randomness on tunnel
  conductance and magnetoresistance in ferromagnetic tunnel junctions},{''}
  \bibinfo{journal}{J. Phys. Soc. Jpn.} \textbf{\bibinfo{volume}{68}},
  \bibinfo{pages}{1632--1639}.

\bibitem[{\citenamefont{Ivchenko and Pikus}(1997)}]{Ivchenko:1997}
\bibinfo{author}{\bibnamefont{Ivchenko}, \bibfnamefont{E.~L.}}, and
  \bibinfo{author}{\bibfnamefont{G.~E.} \bibnamefont{Pikus}},
  \bibinfo{year}{1997}, \emph{\bibinfo{title}{Superlattices and Other
  Heterostructures, Symmetry and Optical Phenomena, {\rm 2nd {Ed.}}}}
  (\bibinfo{publisher}{Springer, New York}).

\bibitem[{\citenamefont{Izyumov} \emph{et~al.}(2002)\citenamefont{Izyumov,
  Proshin, and Khusainov}}]{Izyumov2002:PU}
\bibinfo{author}{\bibnamefont{Izyumov}, \bibfnamefont{Y.~A.}},
  \bibinfo{author}{\bibfnamefont{Y.~N.} \bibnamefont{Proshin}}, and
  \bibinfo{author}{\bibfnamefont{M.~G.} \bibnamefont{Khusainov}},
  \bibinfo{year}{2002}, {``}\bibinfo{title}{Competition between
  superconductivty and magnetism in ferromagnet/superconductor
  heterostructures},{''} \bibinfo{journal}{Usph. Fiz. Nauk}
  \textbf{\bibinfo{volume}{172}},  \bibinfo{pages}{113--154}
  \bibinfo{note}{[Phys. Usp. {\bf 45}, 109-148 (2002)]}.

\bibitem[{\citenamefont{{J\'{a}nossy}}(1980)}]{Janossy1980:PRB}
\bibinfo{author}{\bibnamefont{{J\'{a}nossy}}, \bibfnamefont{A.}},
  \bibinfo{year}{1980}, {``}\bibinfo{title}{Resonant and nonresonant
  conduction-electron-spin transmission in normal metals},{''}
  \bibinfo{journal}{Phys. Rev. B} \textbf{\bibinfo{volume}{21}},
  \bibinfo{pages}{3793--3810}.

\bibitem[{\citenamefont{{J\'{a}nossy}}
  \emph{et~al.}(1993)\citenamefont{{J\'{a}nossy}, Chauvet, Pekker, Cooper, and
  {Forr\'{o}}}}]{Janossy1993:PRL}
\bibinfo{author}{\bibnamefont{{J\'{a}nossy}}, \bibfnamefont{A.}},
  \bibinfo{author}{\bibfnamefont{O.}~\bibnamefont{Chauvet}},
  \bibinfo{author}{\bibfnamefont{S.}~\bibnamefont{Pekker}},
  \bibinfo{author}{\bibfnamefont{J.~R.} \bibnamefont{Cooper}}, and
  \bibinfo{author}{\bibfnamefont{L.}~\bibnamefont{{Forr\'{o}}}},
  \bibinfo{year}{1993}, {``}\bibinfo{title}{Conduction electron spin resonance
  in {Rb$_3$C$_{60}$}},{''} \bibinfo{journal}{Phys. Rev. Lett.}
  \textbf{\bibinfo{volume}{71}},  \bibinfo{pages}{1091--1094}.

\bibitem[{\citenamefont{Jansen} \emph{et~al.}(1980)\citenamefont{Jansen, {van
  Gelder}, and Wyder}}]{Jansen1980:JPCSSP}
\bibinfo{author}{\bibnamefont{Jansen}, \bibfnamefont{A.~G.~M.}},
  \bibinfo{author}{\bibfnamefont{A.~P.} \bibnamefont{{van Gelder}}}, and
  \bibinfo{author}{\bibfnamefont{P.}~\bibnamefont{Wyder}},
  \bibinfo{year}{1980}, {``}\bibinfo{title}{Point-contact spectroscopy in
  metals},{''} \bibinfo{journal}{J. Phys. C} \textbf{\bibinfo{volume}{13}},
  \bibinfo{pages}{6073--6118}.

\bibitem[{\citenamefont{Jansen}(2002)}]{Jansen2002:APL}
\bibinfo{author}{\bibnamefont{Jansen}, \bibfnamefont{R.}},
  \bibinfo{year}{2002}, {``}\bibinfo{title}{Comment on `{Efficient} electrical
  spin injection from a magnetic metal/tunnel barrier contact into a
  semiconductor'},{''} \bibinfo{journal}{Appl. Phys. Lett.}
  \textbf{\bibinfo{volume}{81}},  \bibinfo{pages}{2130}.

\bibitem[{\citenamefont{Jansen and Moodera}(2000)}]{Jansen2000:PRB}
\bibinfo{author}{\bibnamefont{Jansen}, \bibfnamefont{R.}}, and
  \bibinfo{author}{\bibfnamefont{J.~S.} \bibnamefont{Moodera}},
  \bibinfo{year}{2000}, {``}\bibinfo{title}{Magnetoresistance in doped magnetic
  tunnel junctions: Effect of spin scattering and impurity-assisted
  transport},{''} \bibinfo{journal}{Phys. Rev. B}
  \textbf{\bibinfo{volume}{61}},  \bibinfo{pages}{9047--9050}.

\bibitem[{\citenamefont{Jedema}(2002)}]{Jedema2002:T}
\bibinfo{author}{\bibnamefont{Jedema}, \bibfnamefont{F.~J.}},
  \bibinfo{year}{2002}, {``}\bibinfo{title}{Electrical spin injection in
  metallic mesoscopic spin valves},{''} \bibinfo{journal}{Ph.D. Thesis
  (Rijksuniversiteit Groningen)} .

\bibitem[{\citenamefont{Jedema}
  \emph{et~al.}(2002{\natexlab{a}})\citenamefont{Jedema, Costache, Heersche,
  Baselmans, and {van Wees}}}]{Jedema2002:APL}
\bibinfo{author}{\bibnamefont{Jedema}, \bibfnamefont{F.~J.}},
  \bibinfo{author}{\bibfnamefont{M.~V.} \bibnamefont{Costache}},
  \bibinfo{author}{\bibfnamefont{H.~B.} \bibnamefont{Heersche}},
  \bibinfo{author}{\bibfnamefont{J.~J.~A.} \bibnamefont{Baselmans}}, and
  \bibinfo{author}{\bibfnamefont{B.~J.} \bibnamefont{{van Wees}}},
  \bibinfo{year}{2002}{\natexlab{a}}, {``}\bibinfo{title}{Electrical detection
  of spin accumulation and spin precession at room temperature in metallic spin
  valves},{''} \bibinfo{journal}{Appl. Phys. Lett.}
  \textbf{\bibinfo{volume}{81}},  \bibinfo{pages}{5162--5164}.

\bibitem[{\citenamefont{Jedema} \emph{et~al.}(2001)\citenamefont{Jedema, Filip,
  and {van Wees}}}]{Jedema2001:N}
\bibinfo{author}{\bibnamefont{Jedema}, \bibfnamefont{F.~J.}},
  \bibinfo{author}{\bibfnamefont{A.~T.} \bibnamefont{Filip}}, and
  \bibinfo{author}{\bibfnamefont{B.~J.} \bibnamefont{{van Wees}}},
  \bibinfo{year}{2001}, {``}\bibinfo{title}{Electrical spin injection and
  accumulation at room temperature in an all-metal mesoscopic spin valve},{''}
  \bibinfo{journal}{{\sl Nature}} \textbf{\bibinfo{volume}{410}},
  \bibinfo{pages}{345--348}.

\bibitem[{\citenamefont{Jedema}
  \emph{et~al.}(2002{\natexlab{b}})\citenamefont{Jedema, Heersche, Filip,
  Baselmans, and {van Wees}}}]{Jedema2002:Na}
\bibinfo{author}{\bibnamefont{Jedema}, \bibfnamefont{F.~J.}},
  \bibinfo{author}{\bibfnamefont{H.~B.} \bibnamefont{Heersche}},
  \bibinfo{author}{\bibfnamefont{A.~T.} \bibnamefont{Filip}},
  \bibinfo{author}{\bibfnamefont{J.~J.~A.} \bibnamefont{Baselmans}}, and
  \bibinfo{author}{\bibfnamefont{B.~J.} \bibnamefont{{van Wees}}},
  \bibinfo{year}{2002}{\natexlab{b}}, {``}\bibinfo{title}{Electrical detection
  of spin precession in a metallic mesoscopic spin valve},{''}
  \bibinfo{journal}{{\sl Nature}} \textbf{\bibinfo{volume}{416}},
  \bibinfo{pages}{713--716}.

\bibitem[{\citenamefont{Jedema}
  \emph{et~al.}(2002{\natexlab{c}})\citenamefont{Jedema, Heersche, Filip,
  Baselmans, and {van Wees}}}]{Jedema2002:Nb}
\bibinfo{author}{\bibnamefont{Jedema}, \bibfnamefont{F.~J.}},
  \bibinfo{author}{\bibfnamefont{H.~B.} \bibnamefont{Heersche}},
  \bibinfo{author}{\bibfnamefont{A.~T.} \bibnamefont{Filip}},
  \bibinfo{author}{\bibfnamefont{J.~J.~A.} \bibnamefont{Baselmans}}, and
  \bibinfo{author}{\bibfnamefont{B.~J.} \bibnamefont{{van Wees}}},
  \bibinfo{year}{2002}{\natexlab{c}}, {``}\bibinfo{title}{Spintronics
  (Communication arising): {Spin} accumulation in mesoscopic systems},{''}
  \bibinfo{journal}{{\sl Nature}} \textbf{\bibinfo{volume}{416}},
  \bibinfo{pages}{810}.

\bibitem[{\citenamefont{Jedema}
  \emph{et~al.}(2002{\natexlab{d}})\citenamefont{Jedema, Nijboer, Filip, and
  {van Wees}}}]{Jedema2002:JS}
\bibinfo{author}{\bibnamefont{Jedema}, \bibfnamefont{F.~J.}},
  \bibinfo{author}{\bibfnamefont{M.~S.} \bibnamefont{Nijboer}},
  \bibinfo{author}{\bibfnamefont{A.~T.} \bibnamefont{Filip}}, and
  \bibinfo{author}{\bibfnamefont{B.~J.} \bibnamefont{{van Wees}}},
  \bibinfo{year}{2002}{\natexlab{d}}, {``}\bibinfo{title}{Spin Injection and
  spin accumulation in permalloy-cooper mesoscopic spin valves},{''}
  \bibinfo{journal}{J. Supercond.} \textbf{\bibinfo{volume}{15}},
  \bibinfo{pages}{27--35}.

\bibitem[{\citenamefont{Jedema} \emph{et~al.}(2003)\citenamefont{Jedema,
  Nijboer, Filip, and {van Wees}}}]{Jedema2003:PRB}
\bibinfo{author}{\bibnamefont{Jedema}, \bibfnamefont{F.~J.}},
  \bibinfo{author}{\bibfnamefont{M.~S.} \bibnamefont{Nijboer}},
  \bibinfo{author}{\bibfnamefont{A.~T.} \bibnamefont{Filip}}, and
  \bibinfo{author}{\bibfnamefont{B.~J.} \bibnamefont{{van Wees}}},
  \bibinfo{year}{2003}, {``}\bibinfo{title}{Spin injection and spin
  accumulation in all-metal mesoscopic spin valves},{''}
  \bibinfo{journal}{Phys. Rev. B} \textbf{\bibinfo{volume}{67}},
  \bibinfo{pages}{085319}.

\bibitem[{\citenamefont{Jedema} \emph{et~al.}(1999)\citenamefont{Jedema, {van
  Wees}, Hoving, Filip, and Klapwijk}}]{Jedema1999:PRB}
\bibinfo{author}{\bibnamefont{Jedema}, \bibfnamefont{F.~J.}},
  \bibinfo{author}{\bibfnamefont{B.~J.} \bibnamefont{{van Wees}}},
  \bibinfo{author}{\bibfnamefont{B.~H.} \bibnamefont{Hoving}},
  \bibinfo{author}{\bibfnamefont{A.~T.} \bibnamefont{Filip}}, and
  \bibinfo{author}{\bibfnamefont{T.~M.} \bibnamefont{Klapwijk}},
  \bibinfo{year}{1999}, {``}\bibinfo{title}{Spin-accumulation-induced
  resistance in mesoscopic ferromagnet-superconductor junctions},{''}
  \bibinfo{journal}{Phys. Rev. B} \textbf{\bibinfo{volume}{60}},
  \bibinfo{pages}{16549--16552}.

\bibitem[{\citenamefont{Ji} \emph{et~al.}(2003)\citenamefont{Ji, Chien, and
  Stiles}}]{Yi2003:PRL}
\bibinfo{author}{\bibnamefont{Ji}, \bibfnamefont{Y.}},
  \bibinfo{author}{\bibfnamefont{C.~L.} \bibnamefont{Chien}}, and
  \bibinfo{author}{\bibfnamefont{M.~D.} \bibnamefont{Stiles}},
  \bibinfo{year}{2003}, {``}\bibinfo{title}{Current-induced spin-wave
  excitations in a single ferromagnetic layer},{''} \bibinfo{journal}{Phys.
  Rev. Lett.} \textbf{\bibinfo{volume}{90}},  \bibinfo{pages}{106601}.

\bibitem[{\citenamefont{Ji} \emph{et~al.}(2001)\citenamefont{Ji, Strijkers,
  Yang, Chien, Byers, Anguelouch, Xiao, and Gupta}}]{Ji2001:PRL}
\bibinfo{author}{\bibnamefont{Ji}, \bibfnamefont{Y.}},
  \bibinfo{author}{\bibfnamefont{G.~J.} \bibnamefont{Strijkers}},
  \bibinfo{author}{\bibfnamefont{F.~Y.} \bibnamefont{Yang}},
  \bibinfo{author}{\bibfnamefont{C.~L.} \bibnamefont{Chien}},
  \bibinfo{author}{\bibfnamefont{J.~M.} \bibnamefont{Byers}},
  \bibinfo{author}{\bibfnamefont{A.}~\bibnamefont{Anguelouch}},
  \bibinfo{author}{\bibfnamefont{G.}~\bibnamefont{Xiao}}, and
  \bibinfo{author}{\bibfnamefont{A.}~\bibnamefont{Gupta}},
  \bibinfo{year}{2001}, {``}\bibinfo{title}{Determination of the spin
  polarization of half-metallic {CrO$_2$} by point-contact {Andreev}
  reflection},{''} \bibinfo{journal}{Phys. Rev. Lett.}
  \textbf{\bibinfo{volume}{86}},  \bibinfo{pages}{5585--5588}.

\bibitem[{\citenamefont{Jia} \emph{et~al.}(1996)\citenamefont{Jia, Shi, and
  Chou}}]{Jia1996:IEEE}
\bibinfo{author}{\bibnamefont{Jia}, \bibfnamefont{Y.~Q.}},
  \bibinfo{author}{\bibfnamefont{R.~C.} \bibnamefont{Shi}}, and
  \bibinfo{author}{\bibfnamefont{S.~Y.} \bibnamefont{Chou}},
  \bibinfo{year}{1996}, {``}\bibinfo{title}{Spin-valve effects in
  nickel/silicon/nickel junctions},{''} \bibinfo{journal}{IEEE Trans. Magn.}
  \textbf{\bibinfo{volume}{32}},  \bibinfo{pages}{4707--4709}.

\bibitem[{\citenamefont{Jiang and Yablonovitch}(2001)}]{Jiang2001:PRB}
\bibinfo{author}{\bibnamefont{Jiang}, \bibfnamefont{H.~W.}}, and
  \bibinfo{author}{\bibfnamefont{E.}~\bibnamefont{Yablonovitch}},
  \bibinfo{year}{2001}, {``}\bibinfo{title}{Gate-controlled electron spin
  resonance in {GaAs/Al$_x$Ga$_{1-x}$As} heterostructures},{''}
  \bibinfo{journal}{Phys. Rev. B} \textbf{\bibinfo{volume}{64}},
  \bibinfo{pages}{041307}.

\bibitem[{\citenamefont{Jiang} \emph{et~al.}(2004)\citenamefont{Jiang, Shelby,
  Wang, Macfarlane, and Bank}}]{Jiang2004:P}
\bibinfo{author}{\bibnamefont{Jiang}, \bibfnamefont{X.}},
  \bibinfo{author}{\bibfnamefont{R.}~\bibnamefont{Shelby}},
  \bibinfo{author}{\bibfnamefont{R.}~\bibnamefont{Wang}},
  \bibinfo{author}{\bibfnamefont{R.~M.} \bibnamefont{Macfarlane}}, and
  \bibinfo{author}{\bibfnamefont{S.}~\bibnamefont{Bank}}, \bibinfo{year}{2004},
  {``}\bibinfo{title}{Highly spin polarized tunnel injector for semiconductor
  spintronics using MgO(100)},{''} \bibinfo{note}{preprint}.

\bibitem[{\citenamefont{Jiang} \emph{et~al.}(2003)\citenamefont{Jiang, Wang,
  van Dijken, Shelby, Macfarlane, Solomon, Harris, and Parkin}}]{Jiang2003:PRL}
\bibinfo{author}{\bibnamefont{Jiang}, \bibfnamefont{X.}},
  \bibinfo{author}{\bibfnamefont{R.}~\bibnamefont{Wang}},
  \bibinfo{author}{\bibfnamefont{S.}~\bibnamefont{van Dijken}},
  \bibinfo{author}{\bibfnamefont{R.}~\bibnamefont{Shelby}},
  \bibinfo{author}{\bibfnamefont{R.}~\bibnamefont{Macfarlane}},
  \bibinfo{author}{\bibfnamefont{G.~S.} \bibnamefont{Solomon}},
  \bibinfo{author}{\bibfnamefont{J.}~\bibnamefont{Harris}}, and
  \bibinfo{author}{\bibfnamefont{S.~S.~P.} \bibnamefont{Parkin}},
  \bibinfo{year}{2003}, {``}\bibinfo{title}{Optical detection of hot-electron
  spin injection into {GaAs} from a magnetic tunnel transistor source},{''}
  \bibinfo{journal}{Phys. Rev. Lett.} \textbf{\bibinfo{volume}{90}},
  \bibinfo{pages}{256603}.

\bibitem[{\citenamefont{Johnson}(1991)}]{Johnson1991:PRL}
\bibinfo{author}{\bibnamefont{Johnson}, \bibfnamefont{M.}},
  \bibinfo{year}{1991}, {``}\bibinfo{title}{Analysis of anomalous multilayer
  magnetoresistance within the thermomagnetic system},{''}
  \bibinfo{journal}{Phys. Rev. Lett.} \textbf{\bibinfo{volume}{67}},
  \bibinfo{pages}{3594--3597}.

\bibitem[{\citenamefont{Johnson}(1993{\natexlab{a}})}]{Johnson1993:S}
\bibinfo{author}{\bibnamefont{Johnson}, \bibfnamefont{M.}},
  \bibinfo{year}{1993}{\natexlab{a}}, {``}\bibinfo{title}{Bipolar spin
  switch},{''} \bibinfo{journal}{{\sl Science}} \textbf{\bibinfo{volume}{260}},
   \bibinfo{pages}{320--323}.

\bibitem[{\citenamefont{Johnson}(1993{\natexlab{b}})}]{Johnson1993:PRL}
\bibinfo{author}{\bibnamefont{Johnson}, \bibfnamefont{M.}},
  \bibinfo{year}{1993}{\natexlab{b}}, {``}\bibinfo{title}{Spin accumulation in
  gold films},{''} \bibinfo{journal}{Phys. Rev. Lett.}
  \textbf{\bibinfo{volume}{70}},  \bibinfo{pages}{2142--2145}.

\bibitem[{\citenamefont{Johnson}(1994)}]{Johnson1994:APL}
\bibinfo{author}{\bibnamefont{Johnson}, \bibfnamefont{M.}},
  \bibinfo{year}{1994}, {``}\bibinfo{title}{Spin coupled resistance observed in
  ferromagnet-superconductor-ferromagnet trilayers},{''}
  \bibinfo{journal}{Appl. Phys. Lett.} \textbf{\bibinfo{volume}{65}},
  \bibinfo{pages}{1460--1462}.

\bibitem[{\citenamefont{Johnson}(2000)}]{Johnson2000:APL}
\bibinfo{author}{\bibnamefont{Johnson}, \bibfnamefont{M.}},
  \bibinfo{year}{2000}, {``}\bibinfo{title}{Dynamic nuclear polarization by
  spin injection},{''} \bibinfo{journal}{Appl. Phys. Lett.}
  \textbf{\bibinfo{volume}{77}},  \bibinfo{pages}{1680--1682}.

\bibitem[{\citenamefont{Johnson}(2001)}]{Johnson2001:JS}
\bibinfo{author}{\bibnamefont{Johnson}, \bibfnamefont{M.}},
  \bibinfo{year}{2001}, {``}\bibinfo{title}{Spin Injection: {A} survey and
  review},{''} \bibinfo{journal}{J. Supercond.} \textbf{\bibinfo{volume}{14}},
  \bibinfo{pages}{273--281}.

\bibitem[{\citenamefont{Johnson}(2002{\natexlab{a}})}]{Johnson2002:SST}
\bibinfo{author}{\bibnamefont{Johnson}, \bibfnamefont{M.}},
  \bibinfo{year}{2002}{\natexlab{a}}, {``}\bibinfo{title}{Spin injection into
  metals and semiconductors},{''} \bibinfo{journal}{Semicond. Sci. Technol.}
  \textbf{\bibinfo{volume}{17}},  \bibinfo{pages}{298--309}.

\bibitem[{\citenamefont{Johnson}(2002{\natexlab{b}})}]{Johnson2002:N}
\bibinfo{author}{\bibnamefont{Johnson}, \bibfnamefont{M.}},
  \bibinfo{year}{2002}{\natexlab{b}}, {``}\bibinfo{title}{Spintronics
  (Communication arising): {Spin} accumulation in mesoscopic systems},{''}
  \bibinfo{journal}{{\sl Nature}} \textbf{\bibinfo{volume}{416}},
  \bibinfo{pages}{809--810}.

\bibitem[{\citenamefont{Johnson}(2003)}]{Johnson:2003}
\bibinfo{author}{\bibnamefont{Johnson}, \bibfnamefont{M.}},
  \bibinfo{year}{2003}, {``}\bibinfo{title}{Hybrid Devices},{''} in
  \emph{\bibinfo{booktitle}{Magnetic Interactions and Spin Transport}}, edited
  by \bibinfo{editor}{\bibfnamefont{A.}~\bibnamefont{Chtchelkanova}},
  \bibinfo{editor}{\bibfnamefont{S.}~\bibnamefont{Wolf}}, and
  \bibinfo{editor}{\bibfnamefont{Y.}~\bibnamefont{Idzerda}}
  (\bibinfo{publisher}{Kluwer Academic Dordrecht/Plenum, New York}),
  \bibinfo{pages}{515--564}.

\bibitem[{\citenamefont{Johnson and Byers}(2003)}]{Johnson2003:PRB}
\bibinfo{author}{\bibnamefont{Johnson}, \bibfnamefont{M.}}, and
  \bibinfo{author}{\bibfnamefont{J.}~\bibnamefont{Byers}},
  \bibinfo{year}{2003}, {``}\bibinfo{title}{Charge and spin diffusion in
  mesoscopic metal wires and at ferromagnet/nonmagnet interfaces},{''}
  \bibinfo{journal}{Phys. Rev. B} \textbf{\bibinfo{volume}{67}},
  \bibinfo{pages}{125112}.

\bibitem[{\citenamefont{Johnson and Clarke}(1990)}]{Johnson1990:JAP}
\bibinfo{author}{\bibnamefont{Johnson}, \bibfnamefont{M.}}, and
  \bibinfo{author}{\bibfnamefont{J.}~\bibnamefont{Clarke}},
  \bibinfo{year}{1990}, {``}\bibinfo{title}{Spin-polarized scanning tunneling
  microscope: Concept, design, and preliminary results from a prototype
  operated in air},{''} \bibinfo{journal}{J. Appl. Phys.}
  \textbf{\bibinfo{volume}{67}},  \bibinfo{pages}{6141--6152}.

\bibitem[{\citenamefont{Johnson and Silsbee}(1985)}]{Johnson1985:PRL}
\bibinfo{author}{\bibnamefont{Johnson}, \bibfnamefont{M.}}, and
  \bibinfo{author}{\bibfnamefont{R.~H.} \bibnamefont{Silsbee}},
  \bibinfo{year}{1985}, {``}\bibinfo{title}{Interfacial charge-spin coupling:
  {Injection} and detection of spin magnetization in metals},{''}
  \bibinfo{journal}{Phys. Rev. Lett.} \textbf{\bibinfo{volume}{55}},
  \bibinfo{pages}{1790--1793}.

\bibitem[{\citenamefont{Johnson and Silsbee}(1987)}]{Johnson1987:PRB}
\bibinfo{author}{\bibnamefont{Johnson}, \bibfnamefont{M.}}, and
  \bibinfo{author}{\bibfnamefont{R.~H.} \bibnamefont{Silsbee}},
  \bibinfo{year}{1987}, {``}\bibinfo{title}{Thermodynamic analysis of
  interfacial transport and of the thermomagnetoelectric system},{''}
  \bibinfo{journal}{Phys. Rev. B} \textbf{\bibinfo{volume}{35}},
  \bibinfo{pages}{4959--4972}.

\bibitem[{\citenamefont{Johnson and
  Silsbee}(1988{\natexlab{a}})}]{Johnson1988:PRBa}
\bibinfo{author}{\bibnamefont{Johnson}, \bibfnamefont{M.}}, and
  \bibinfo{author}{\bibfnamefont{R.~H.} \bibnamefont{Silsbee}},
  \bibinfo{year}{1988}{\natexlab{a}}, {``}\bibinfo{title}{Coupling of
  electronic charge and spin at a ferromagnetic-paramagnetic metal
  interface},{''} \bibinfo{journal}{Phys. Rev. B}
  \textbf{\bibinfo{volume}{37}},  \bibinfo{pages}{5312--5325}.

\bibitem[{\citenamefont{Johnson and
  Silsbee}(1988{\natexlab{b}})}]{Johnson1988:JAP}
\bibinfo{author}{\bibnamefont{Johnson}, \bibfnamefont{M.}}, and
  \bibinfo{author}{\bibfnamefont{R.~H.} \bibnamefont{Silsbee}},
  \bibinfo{year}{1988}{\natexlab{b}}, {``}\bibinfo{title}{Electron spin
  injection and detection at a ferromagnetic-paramagnetic interface},{''}
  \bibinfo{journal}{J. Appl. Phys.} \textbf{\bibinfo{volume}{63}},
  \bibinfo{pages}{3934--3939}.

\bibitem[{\citenamefont{Johnson and
  Silsbee}(1988{\natexlab{c}})}]{Johnson1988:PRL}
\bibinfo{author}{\bibnamefont{Johnson}, \bibfnamefont{M.}}, and
  \bibinfo{author}{\bibfnamefont{R.~H.} \bibnamefont{Silsbee}},
  \bibinfo{year}{1988}{\natexlab{c}},
  {``}\bibinfo{title}{Ferromagnet-nonferromagnet interface resistance},{''}
  \bibinfo{journal}{Phys. Rev. Lett.} \textbf{\bibinfo{volume}{60}},
  \bibinfo{pages}{377}.

\bibitem[{\citenamefont{Johnson and
  Silsbee}(1988{\natexlab{d}})}]{Johnson1988:PRBb}
\bibinfo{author}{\bibnamefont{Johnson}, \bibfnamefont{M.}}, and
  \bibinfo{author}{\bibfnamefont{R.~H.} \bibnamefont{Silsbee}},
  \bibinfo{year}{1988}{\natexlab{d}}, {``}\bibinfo{title}{Spin injection
  experiment},{''} \bibinfo{journal}{Phys. Rev. B}
  \textbf{\bibinfo{volume}{37}},  \bibinfo{pages}{5326--5335}.

\bibitem[{\citenamefont{Johnston-Halperin}
  \emph{et~al.}(2002)\citenamefont{Johnston-Halperin, Lofgreen, Kawakami,
  Young, Coldren, Gossard, and Awschalom}}]{Johnston-Halperin2002:PRB}
\bibinfo{author}{\bibnamefont{Johnston-Halperin}, \bibfnamefont{E.}},
  \bibinfo{author}{\bibfnamefont{D.}~\bibnamefont{Lofgreen}},
  \bibinfo{author}{\bibfnamefont{R.~K.} \bibnamefont{Kawakami}},
  \bibinfo{author}{\bibfnamefont{D.~K.} \bibnamefont{Young}},
  \bibinfo{author}{\bibfnamefont{L.}~\bibnamefont{Coldren}},
  \bibinfo{author}{\bibfnamefont{A.~C.} \bibnamefont{Gossard}}, and
  \bibinfo{author}{\bibfnamefont{D.~D.} \bibnamefont{Awschalom}},
  \bibinfo{year}{2002}, {``}\bibinfo{title}{Spin-polarized {Zener} tunneling in
  {(Ga,Mn)As}},{''} \bibinfo{journal}{Phys. Rev. B}
  \textbf{\bibinfo{volume}{65}},  \bibinfo{pages}{041306}.

\bibitem[{\citenamefont{de~Jong and Beenakker}(1995)}]{deJong1995:PRL}
\bibinfo{author}{\bibnamefont{de~Jong}, \bibfnamefont{M.~J.~M.}}, and
  \bibinfo{author}{\bibfnamefont{C.~W.~J.} \bibnamefont{Beenakker}},
  \bibinfo{year}{1995}, {``}\bibinfo{title}{Andreev reflection in
  ferromagnet-superconductor junctions},{''} \bibinfo{journal}{Phys. Rev.
  Lett.} \textbf{\bibinfo{volume}{74}},  \bibinfo{pages}{1657--1660}.

\bibitem[{\citenamefont{Jonker}(1999)}]{Jonker1999:PA}
\bibinfo{author}{\bibnamefont{Jonker}, \bibfnamefont{B.~T.}},
  \bibinfo{year}{1999} \bibinfo{journal}{{U.S.} patent No. 5,874,749 (filed 23
  {June} 1993, awarded 23 {February} 1999 to {U.S. Navy})} .

\bibitem[{\citenamefont{Jonker}
  \emph{et~al.}(2003{\natexlab{a}})\citenamefont{Jonker, Erwin, Petrou, and
  Petukhov}}]{Jonker2003:MRS}
\bibinfo{author}{\bibnamefont{Jonker}, \bibfnamefont{B.~T.}},
  \bibinfo{author}{\bibfnamefont{S.~C.} \bibnamefont{Erwin}},
  \bibinfo{author}{\bibfnamefont{A.}~\bibnamefont{Petrou}}, and
  \bibinfo{author}{\bibfnamefont{A.~G.} \bibnamefont{Petukhov}},
  \bibinfo{year}{2003}{\natexlab{a}}, {``}\bibinfo{title}{Electrical spin
  injection and transport in semiconductor spintronic devices},{''}
  \bibinfo{journal}{MRS Bull.} \textbf{\bibinfo{volume}{28}},
  \bibinfo{pages}{740--748}.

\bibitem[{\citenamefont{Jonker} \emph{et~al.}(1997)\citenamefont{Jonker,
  Glembocki, Holm, and Wagner}}]{Jonker1997:PRL}
\bibinfo{author}{\bibnamefont{Jonker}, \bibfnamefont{B.~T.}},
  \bibinfo{author}{\bibfnamefont{O.~J.} \bibnamefont{Glembocki}},
  \bibinfo{author}{\bibfnamefont{R.~T.} \bibnamefont{Holm}}, and
  \bibinfo{author}{\bibfnamefont{R.~J.} \bibnamefont{Wagner}},
  \bibinfo{year}{1997}, {``}\bibinfo{title}{Enhanced carrier lifetimes and
  suppression of midgap states in {GaAs} at a magnetic metal interface},{''}
  \bibinfo{journal}{Phys. Rev. Lett.} \textbf{\bibinfo{volume}{79}},
  \bibinfo{pages}{4886--4889}.

\bibitem[{\citenamefont{Jonker} \emph{et~al.}(2001)\citenamefont{Jonker,
  Hanbicki, Park, Itskos, Furis, Kioseoglou, Petrou, and Wei}}]{Jonker2001:APL}
\bibinfo{author}{\bibnamefont{Jonker}, \bibfnamefont{B.~T.}},
  \bibinfo{author}{\bibfnamefont{A.~T.} \bibnamefont{Hanbicki}},
  \bibinfo{author}{\bibfnamefont{Y.~D.} \bibnamefont{Park}},
  \bibinfo{author}{\bibfnamefont{G.}~\bibnamefont{Itskos}},
  \bibinfo{author}{\bibfnamefont{M.}~\bibnamefont{Furis}},
  \bibinfo{author}{\bibfnamefont{G.}~\bibnamefont{Kioseoglou}},
  \bibinfo{author}{\bibfnamefont{A.}~\bibnamefont{Petrou}}, and
  \bibinfo{author}{\bibfnamefont{X.}~\bibnamefont{Wei}}, \bibinfo{year}{2001},
  {``}\bibinfo{title}{Quantifying electrical spin injection: Component-resolved
  electroluminescence from spin-polarized light-emitting diodes},{''}
  \bibinfo{journal}{Appl. Phys. Lett.} \textbf{\bibinfo{volume}{79}},
  \bibinfo{pages}{3098--3100}.

\bibitem[{\citenamefont{Jonker}
  \emph{et~al.}(2003{\natexlab{b}})\citenamefont{Jonker, Hanbicki, Pierece, and
  Stiles}}]{Jonker2003:P}
\bibinfo{author}{\bibnamefont{Jonker}, \bibfnamefont{B.~T.}},
  \bibinfo{author}{\bibfnamefont{A.~T.} \bibnamefont{Hanbicki}},
  \bibinfo{author}{\bibfnamefont{D.~T.} \bibnamefont{Pierece}}, and
  \bibinfo{author}{\bibfnamefont{M.~D.} \bibnamefont{Stiles}},
  \bibinfo{year}{2003}{\natexlab{b}}, {``}\bibinfo{title}{Spin nomenclature for
  semiconductors and magnetic metals},{''} \bibinfo{note}{{J. Magn. Magn.
  Mater.}, in press.} \eprint{cond-mat/030445}.

\bibitem[{\citenamefont{Jonker} \emph{et~al.}(2000)\citenamefont{Jonker, Park,
  Bennett, Cheong, Kioseoglou, and Petrou}}]{Jonker2000:PRB}
\bibinfo{author}{\bibnamefont{Jonker}, \bibfnamefont{B.~T.}},
  \bibinfo{author}{\bibfnamefont{Y.~D.} \bibnamefont{Park}},
  \bibinfo{author}{\bibfnamefont{B.~R.} \bibnamefont{Bennett}},
  \bibinfo{author}{\bibfnamefont{H.~D.} \bibnamefont{Cheong}},
  \bibinfo{author}{\bibfnamefont{G.}~\bibnamefont{Kioseoglou}}, and
  \bibinfo{author}{\bibfnamefont{A.}~\bibnamefont{Petrou}},
  \bibinfo{year}{2000}, {``}\bibinfo{title}{Robust electrical spin injection
  into a semiconductor heterostructure},{''} \bibinfo{journal}{Phys. Rev. B}
  \textbf{\bibinfo{volume}{62}},  \bibinfo{pages}{8180--8183}.

\bibitem[{\citenamefont{Joshi} \emph{et~al.}(2001)\citenamefont{Joshi, Sahoo,
  and Jayannavar}}]{Joshi2001:PRB}
\bibinfo{author}{\bibnamefont{Joshi}, \bibfnamefont{S.~K.}},
  \bibinfo{author}{\bibfnamefont{D.}~\bibnamefont{Sahoo}}, and
  \bibinfo{author}{\bibfnamefont{A.~M.} \bibnamefont{Jayannavar}},
  \bibinfo{year}{2001}, {``}\bibinfo{title}{Aharonov-{Bohm} oscillations and
  spin-polarized transport in a mesoscopic ring with a magnetic impurity},{''}
  \bibinfo{journal}{Phys. Rev. B} \textbf{\bibinfo{volume}{64}},
  \bibinfo{pages}{075320}.

\bibitem[{\citenamefont{Jullier{\`{e}}}(1975)}]{Julliere1975:PL}
\bibinfo{author}{\bibnamefont{Jullier{\`{e}}}, \bibfnamefont{M.}},
  \bibinfo{year}{1975}, {``}\bibinfo{title}{Tunneling between ferromagnetic
  films},{''} \bibinfo{journal}{Phys. Lett.} \textbf{\bibinfo{volume}{54 A}},
  \bibinfo{pages}{225--226}.

\bibitem[{\citenamefont{Kainz} \emph{et~al.}(2003)\citenamefont{Kainz,
  {R\"{o}ssler}, and Winkler}}]{Kainz2003:PRB}
\bibinfo{author}{\bibnamefont{Kainz}, \bibfnamefont{J.}},
  \bibinfo{author}{\bibfnamefont{U.}~\bibnamefont{{R\"{o}ssler}}}, and
  \bibinfo{author}{\bibfnamefont{R.}~\bibnamefont{Winkler}},
  \bibinfo{year}{2003}, {``}\bibinfo{title}{Anisotropic spin splitting and spin
  relaxation in asymmetric zinc-blende semiconductor quantum structures},{''}
  \bibinfo{journal}{Phys. Rev. B} \textbf{\bibinfo{volume}{68}},
  \bibinfo{pages}{075322}.

\bibitem[{\citenamefont{Kaiser and Bell}(1988)}]{Kaiser1988:PRL}
\bibinfo{author}{\bibnamefont{Kaiser}, \bibfnamefont{W.~J.}}, and
  \bibinfo{author}{\bibfnamefont{L.~D.} \bibnamefont{Bell}},
  \bibinfo{year}{1988}, {``}\bibinfo{title}{Direct investigation of subsurface
  interface electronic structure by ballistic-electron-emission
  microscopy},{''} \bibinfo{journal}{Phys. Rev. Lett.}
  \textbf{\bibinfo{volume}{60}},  \bibinfo{pages}{1406--1409}.

\bibitem[{\citenamefont{Kalevich}(1986)}]{Kalevich1986:FTT}
\bibinfo{author}{\bibnamefont{Kalevich}, \bibfnamefont{V.~K.}},
  \bibinfo{year}{1986}, {``}\bibinfo{title}{Optically induced nuclear magnetic
  resonance in semiconductors},{''} \bibinfo{journal}{Fiz. Tverd. Tela}
  \textbf{\bibinfo{volume}{28}},  \bibinfo{pages}{3462--3465}
  \bibinfo{note}{[Sov. Phys. Solid State {\bf 28}, 1947-1949 (1986)]}.

\bibitem[{\citenamefont{Kalevich} \emph{et~al.}(1980)\citenamefont{Kalevich,
  Kul'kov, and Fleisher}}]{Kalevich1980:FTT}
\bibinfo{author}{\bibnamefont{Kalevich}, \bibfnamefont{V.~K.}},
  \bibinfo{author}{\bibfnamefont{V.~D.} \bibnamefont{Kul'kov}}, and
  \bibinfo{author}{\bibfnamefont{V.~G.} \bibnamefont{Fleisher}},
  \bibinfo{year}{1980}, {``}\bibinfo{title}{Manifestation of the sign of the g
  factor of conduction electrons in resonant cooling of the nuclear spin system
  of a semiconductor},{''} \bibinfo{journal}{Fiz. Tverd. Tela}
  \textbf{\bibinfo{volume}{22}},  \bibinfo{pages}{1208--1211}
  \bibinfo{note}{[Sov. Phys. Solid State {\bf 22}, 703-705 (1980)]}.

\bibitem[{\citenamefont{Kalevich} \emph{et~al.}(1981)\citenamefont{Kalevich,
  Kul'kov, and Fleisher}}]{Kalevich1981:FTT}
\bibinfo{author}{\bibnamefont{Kalevich}, \bibfnamefont{V.~K.}},
  \bibinfo{author}{\bibfnamefont{V.~D.} \bibnamefont{Kul'kov}}, and
  \bibinfo{author}{\bibfnamefont{V.~G.} \bibnamefont{Fleisher}},
  \bibinfo{year}{1981}, {``}\bibinfo{title}{Manifestation of the sign of the g
  factor of conduction electrons in resonant cooling of the nuclear spin system
  of a semiconductor},{''} \bibinfo{journal}{Fiz. Tverd. Tela}
  \textbf{\bibinfo{volume}{23}},  \bibinfo{pages}{1524--1526}
  \bibinfo{note}{[Sov. Phys. Solid State {\bf 23}, 892-893 (1981)]}.

\bibitem[{\citenamefont{Kalt} \emph{et~al.}(2000)\citenamefont{Kalt,
  {Tr\"{o}ndle}, Wachter, Blewett, Brown, and Galbraith}}]{Kalt2000:PSS}
\bibinfo{author}{\bibnamefont{Kalt}, \bibfnamefont{H.}},
  \bibinfo{author}{\bibfnamefont{D.}~\bibnamefont{{Tr\"{o}ndle}}},
  \bibinfo{author}{\bibfnamefont{S.}~\bibnamefont{Wachter}},
  \bibinfo{author}{\bibfnamefont{I.~J.} \bibnamefont{Blewett}},
  \bibinfo{author}{\bibfnamefont{G.}~\bibnamefont{Brown}}, and
  \bibinfo{author}{\bibfnamefont{I.}~\bibnamefont{Galbraith}},
  \bibinfo{year}{2000}, {``}\bibinfo{title}{The role of spin orientation and
  relaxation in exciton-exciton scattering},{''} \bibinfo{journal}{Phys. Status
  Solidi B} \textbf{\bibinfo{volume}{221}},  \bibinfo{pages}{477--480}.

\bibitem[{\citenamefont{Kane}(1998)}]{Kane1998:N}
\bibinfo{author}{\bibnamefont{Kane}, \bibfnamefont{B.~E.}},
  \bibinfo{year}{1998}, {``}\bibinfo{title}{A silicon based nuclear spin
  quantum computer},{''} \bibinfo{journal}{{\sl Nature}}
  \textbf{\bibinfo{volume}{393}},  \bibinfo{pages}{133--137}.

\bibitem[{\citenamefont{Kane}(2000)}]{Kane2000:FP}
\bibinfo{author}{\bibnamefont{Kane}, \bibfnamefont{B.~E.}},
  \bibinfo{year}{2000}, {``}\bibinfo{title}{Silicon-based quantum
  computation},{''} \bibinfo{journal}{Fortschr. Phys.}
  \textbf{\bibinfo{volume}{48}},  \bibinfo{pages}{1023--1041}.

\bibitem[{\citenamefont{Kane} \emph{et~al.}(1992)\citenamefont{Kane, Pfeiffer,
  and West}}]{Kane1992:PRB}
\bibinfo{author}{\bibnamefont{Kane}, \bibfnamefont{B.~E.}},
  \bibinfo{author}{\bibfnamefont{L.~N.} \bibnamefont{Pfeiffer}}, and
  \bibinfo{author}{\bibfnamefont{K.~W.} \bibnamefont{West}},
  \bibinfo{year}{1992}, {``}\bibinfo{title}{Evidence for an
  electric-field-induced phase transition in a spin-polarized two-dimensional
  electron gas},{''} \bibinfo{journal}{Phys. Rev. B}
  \textbf{\bibinfo{volume}{46}},  \bibinfo{pages}{7264--7267}.

\bibitem[{\citenamefont{Kaplan}(1959)}]{Kaplan1959:PR}
\bibinfo{author}{\bibnamefont{Kaplan}, \bibfnamefont{J.~I.}},
  \bibinfo{year}{1959}, {``}\bibinfo{title}{Application of the
  diffusion-modified {Bloch} equation to electron spin resonance in ordinary
  and ferromagnetic metals},{''} \bibinfo{journal}{Phys. Rev.}
  \textbf{\bibinfo{volume}{115}},  \bibinfo{pages}{575--577}.

\bibitem[{\citenamefont{Kashiwaya and Tanaka}(2000)}]{Kashiwaya2000:RPP}
\bibinfo{author}{\bibnamefont{Kashiwaya}, \bibfnamefont{S.}}, and
  \bibinfo{author}{\bibfnamefont{Y.}~\bibnamefont{Tanaka}},
  \bibinfo{year}{2000}, {``}\bibinfo{title}{Tunnelling effects on surface bound
  states in unconventional superconductors},{''} \bibinfo{journal}{Rep. Prog.
  Phys.} \textbf{\bibinfo{volume}{63}},  \bibinfo{pages}{1641--1724}.

\bibitem[{\citenamefont{Kashiwaya} \emph{et~al.}(1999)\citenamefont{Kashiwaya,
  Tanaka, Yoshida, and Beasley}}]{Kashiwaya1999:PRB}
\bibinfo{author}{\bibnamefont{Kashiwaya}, \bibfnamefont{S.}},
  \bibinfo{author}{\bibfnamefont{Y.}~\bibnamefont{Tanaka}},
  \bibinfo{author}{\bibfnamefont{N.}~\bibnamefont{Yoshida}}, and
  \bibinfo{author}{\bibfnamefont{M.~R.} \bibnamefont{Beasley}},
  \bibinfo{year}{1999}, {``}\bibinfo{title}{Spin current in
  ferromagnet-insulator-superconductor junctions},{''} \bibinfo{journal}{Phys.
  Rev. B} \textbf{\bibinfo{volume}{60}},  \bibinfo{pages}{3572--3580}.

\bibitem[{\citenamefont{Kastler}(1950)}]{Kastler1950:JDP}
\bibinfo{author}{\bibnamefont{Kastler}, \bibfnamefont{A.}},
  \bibinfo{year}{1950}, {``}\bibinfo{title}{Quelques suggestions concernant la
  production optique et la d{\'e}tection optique d'une inegalit{\'e} de
  population des niveaux de quantification spatiale des atomes - application
  {\'a} l'experience de {Stern} et {Gerlach} et a la resonance magnetique},{''}
  \bibinfo{journal}{J. Phys. (Paris)} \textbf{\bibinfo{volume}{11}},
  \bibinfo{pages}{255--265}.

\bibitem[{\citenamefont{Kasuya and Yanase}(1968)}]{Kasuya1968:RMP}
\bibinfo{author}{\bibnamefont{Kasuya}, \bibfnamefont{T.}}, and
  \bibinfo{author}{\bibfnamefont{A.}~\bibnamefont{Yanase}},
  \bibinfo{year}{1968}, {``}\bibinfo{title}{Anomalous transport phenomena in
  {Eu}-chalcogenide alloys},{''} \bibinfo{journal}{Rev. Mod. Phys.}
  \textbf{\bibinfo{volume}{40}},  \bibinfo{pages}{684--696}.

\bibitem[{\citenamefont{Katine} \emph{et~al.}(2000)\citenamefont{Katine,
  Albert, Buhrman, Myers, and Ralph}}]{Katine2000:PRL}
\bibinfo{author}{\bibnamefont{Katine}, \bibfnamefont{J.~A.}},
  \bibinfo{author}{\bibfnamefont{F.~J.} \bibnamefont{Albert}},
  \bibinfo{author}{\bibfnamefont{R.~A.} \bibnamefont{Buhrman}},
  \bibinfo{author}{\bibfnamefont{E.~B.} \bibnamefont{Myers}}, and
  \bibinfo{author}{\bibfnamefont{D.}~\bibnamefont{Ralph}},
  \bibinfo{year}{2000}, {``}\bibinfo{title}{Current-driven magnetization
  reversal and spin-wave excitations in {Co/Cu/Co} pillars},{''}
  \bibinfo{journal}{Phys. Rev. Lett.} \textbf{\bibinfo{volume}{84}},
  \bibinfo{pages}{3149--3152}.

\bibitem[{\citenamefont{Kato} \emph{et~al.}(2003)\citenamefont{Kato, Myers,
  Driscoll, Gossard, Levy, and Awschalom}}]{Kato2003:S}
\bibinfo{author}{\bibnamefont{Kato}, \bibfnamefont{Y.}},
  \bibinfo{author}{\bibfnamefont{R.~C.} \bibnamefont{Myers}},
  \bibinfo{author}{\bibfnamefont{D.~C.} \bibnamefont{Driscoll}},
  \bibinfo{author}{\bibfnamefont{A.~C.} \bibnamefont{Gossard}},
  \bibinfo{author}{\bibfnamefont{J.}~\bibnamefont{Levy}}, and
  \bibinfo{author}{\bibfnamefont{D.~D.} \bibnamefont{Awschalom}},
  \bibinfo{year}{2003}, {``}\bibinfo{title}{Gigahertz electron spin
  manipulation using voltage-controlled g-tensor modulation},{''}
  \bibinfo{journal}{{\sl Science}} \textbf{\bibinfo{volume}{299}},
  \bibinfo{pages}{1201--1204}.

\bibitem[{\citenamefont{Kauschke} \emph{et~al.}(1987)\citenamefont{Kauschke,
  Mestres, and Cardona}}]{Kauschke1987:PRB}
\bibinfo{author}{\bibnamefont{Kauschke}, \bibfnamefont{W.}},
  \bibinfo{author}{\bibfnamefont{N.}~\bibnamefont{Mestres}}, and
  \bibinfo{author}{\bibfnamefont{M.}~\bibnamefont{Cardona}},
  \bibinfo{year}{1987}, {``}\bibinfo{title}{Spin relaxation of holes in the
  split-hole band of {InP} and {GaSb}},{''} \bibinfo{journal}{Phys. Rev. B}
  \textbf{\bibinfo{volume}{35}},  \bibinfo{pages}{3843--3853}.

\bibitem[{\citenamefont{Kavokin}(2001)}]{Kavokin2001:PRB}
\bibinfo{author}{\bibnamefont{Kavokin}, \bibfnamefont{K.~V.}},
  \bibinfo{year}{2001}, {``}\bibinfo{title}{Anisotropic exchange interaction of
  localized conduction-band electrons in semiconductors},{''}
  \bibinfo{journal}{Phys. Rev. B} \textbf{\bibinfo{volume}{64}},
  \bibinfo{pages}{075305}.

\bibitem[{\citenamefont{Kavokin}(2002{\natexlab{a}})}]{Kavokin2002:PSS}
\bibinfo{author}{\bibnamefont{Kavokin}, \bibfnamefont{K.~V.}},
  \bibinfo{year}{2002}{\natexlab{a}}, {``}\bibinfo{title}{Optical
  manifestations of electron spin transport and relaxation in
  semiconductors},{''} \bibinfo{journal}{Phys. Status Solidi A}
  \textbf{\bibinfo{volume}{190}},  \bibinfo{pages}{221--227}.

\bibitem[{\citenamefont{Kavokin}(2002{\natexlab{b}})}]{Kavokin2002:P}
\bibinfo{author}{\bibnamefont{Kavokin}, \bibfnamefont{K.~V.}},
  \bibinfo{year}{2002}{\natexlab{b}}, {``}\bibinfo{title}{Symmetry of
  anisotropic exchange interactions in semiconductor nanostructures},{''}
  \eprint{cond-mat/0212347}.

\bibitem[{\citenamefont{Keldysh}(1964)}]{Keldysh1965:SPJETP}
\bibinfo{author}{\bibnamefont{Keldysh}, \bibfnamefont{L.~V.}},
  \bibinfo{year}{1964}, {``}\bibinfo{title}{Diagram technique for
  nonequilibrium processes},{''} \bibinfo{journal}{Zh. Eksp. Teor. Fiz.}
  \textbf{\bibinfo{volume}{47}},  \bibinfo{pages}{1515--1527}
  \bibinfo{note}{[Sov. Phys. JETP {\bf 20}, 1018-1026 (1965)]}.

\bibitem[{\citenamefont{Kelly} \emph{et~al.}(2003)\citenamefont{Kelly, Wegrowe,
  k.~Truong, Hoffer, and Ansermet}}]{Kelly2003:PRB}
\bibinfo{author}{\bibnamefont{Kelly}, \bibfnamefont{D.}},
  \bibinfo{author}{\bibfnamefont{J.-E.} \bibnamefont{Wegrowe}},
  \bibinfo{author}{\bibfnamefont{T.}~\bibnamefont{k.~Truong}},
  \bibinfo{author}{\bibfnamefont{X.}~\bibnamefont{Hoffer}}, and
  \bibinfo{author}{\bibfnamefont{J.-P.} \bibnamefont{Ansermet}},
  \bibinfo{year}{2003}, {``}\bibinfo{title}{Spin-polarized current-induced
  magnetization reversal in single nanowires},{''} \bibinfo{journal}{Phys. Rev.
  B} \textbf{\bibinfo{volume}{68}},  \bibinfo{pages}{134425}.

\bibitem[{\citenamefont{Kessler}(1976)}]{Kessler:1976}
\bibinfo{author}{\bibnamefont{Kessler}, \bibfnamefont{J.}},
  \bibinfo{year}{1976}, \emph{\bibinfo{title}{Polarized Electrons}}
  (\bibinfo{publisher}{Springer, New York}).

\bibitem[{\citenamefont{Khaetskii}(2001)}]{Khaetskii2001:PE}
\bibinfo{author}{\bibnamefont{Khaetskii}, \bibfnamefont{A.~V.}},
  \bibinfo{year}{2001}, {``}\bibinfo{title}{Spin relaxation in semiconductor
  mesoscopic systems},{''} \bibinfo{journal}{Physica E}
  \textbf{\bibinfo{volume}{10}},  \bibinfo{pages}{27--31}.

\bibitem[{\citenamefont{Khaetskii} \emph{et~al.}(2002)\citenamefont{Khaetskii,
  Loss, and Glazman}}]{Khaetskii2002:PRL}
\bibinfo{author}{\bibnamefont{Khaetskii}, \bibfnamefont{A.~V.}},
  \bibinfo{author}{\bibfnamefont{D.}~\bibnamefont{Loss}}, and
  \bibinfo{author}{\bibfnamefont{L.}~\bibnamefont{Glazman}},
  \bibinfo{year}{2002}, {``}\bibinfo{title}{Electron spin decoherence in
  quantum dots due to interaction with nuclei},{''} \bibinfo{journal}{Phys.
  Rev. Lett.} \textbf{\bibinfo{volume}{88}},  \bibinfo{pages}{186802}.

\bibitem[{\citenamefont{Khaetskii} \emph{et~al.}(2003)\citenamefont{Khaetskii,
  Loss, and Glazman}}]{Khaetskii2002:P}
\bibinfo{author}{\bibnamefont{Khaetskii}, \bibfnamefont{A.~V.}},
  \bibinfo{author}{\bibfnamefont{D.}~\bibnamefont{Loss}}, and
  \bibinfo{author}{\bibfnamefont{L.}~\bibnamefont{Glazman}},
  \bibinfo{year}{2003}, {``}\bibinfo{title}{Electron spin evolution by
  interaction with nuclei in a quantum dot},{''} \bibinfo{journal}{Phys. Rev.
  B} \textbf{\bibinfo{volume}{67}},  \bibinfo{pages}{195329}.

\bibitem[{\citenamefont{Khaetskii and Nazarov}(2000)}]{Khaetskii2000:PRB}
\bibinfo{author}{\bibnamefont{Khaetskii}, \bibfnamefont{A.~V.}}, and
  \bibinfo{author}{\bibfnamefont{Y.~V.} \bibnamefont{Nazarov}},
  \bibinfo{year}{2000}, {``}\bibinfo{title}{Spin relaxation in semiconductor
  quantum dots},{''} \bibinfo{journal}{Phys. Rev. B}
  \textbf{\bibinfo{volume}{61}},  \bibinfo{pages}{12639--12642}.

\bibitem[{\citenamefont{Khaetskii and Nazarov}(2001)}]{Khaetskii2001:PRB}
\bibinfo{author}{\bibnamefont{Khaetskii}, \bibfnamefont{A.~V.}}, and
  \bibinfo{author}{\bibfnamefont{Y.~V.} \bibnamefont{Nazarov}},
  \bibinfo{year}{2001}, {``}\bibinfo{title}{Spin-flip transitions between
  {Zeeman} sublevels in semiconductor quantum dots},{''}
  \bibinfo{journal}{Phys. Rev. B} \textbf{\bibinfo{volume}{64}},
  \bibinfo{pages}{125316}.

\bibitem[{\citenamefont{Kikkawa}(2003)}]{Kikkawa2003:PC}
\bibinfo{author}{\bibnamefont{Kikkawa}, \bibfnamefont{J.~M.}},
  \bibinfo{year}{2003} \bibinfo{journal}{private communication} .

\bibitem[{\citenamefont{Kikkawa and Awschalom}(1998)}]{Kikkawa1998:PRL}
\bibinfo{author}{\bibnamefont{Kikkawa}, \bibfnamefont{J.~M.}}, and
  \bibinfo{author}{\bibfnamefont{D.~D.} \bibnamefont{Awschalom}},
  \bibinfo{year}{1998}, {``}\bibinfo{title}{Resonant spin amplification in
  n-type {GaAs}},{''} \bibinfo{journal}{Phys. Rev. Lett.}
  \textbf{\bibinfo{volume}{80}},  \bibinfo{pages}{4313--4316}.

\bibitem[{\citenamefont{Kikkawa and Awschalom}(1999)}]{Kikkawa1999:N}
\bibinfo{author}{\bibnamefont{Kikkawa}, \bibfnamefont{J.~M.}}, and
  \bibinfo{author}{\bibfnamefont{D.~D.} \bibnamefont{Awschalom}},
  \bibinfo{year}{1999}, {``}\bibinfo{title}{Lateral drag of spin coherence in
  gallium arsenide},{''} \bibinfo{journal}{{\sl Nature}}
  \textbf{\bibinfo{volume}{397}},  \bibinfo{pages}{139--141}.

\bibitem[{\citenamefont{Kikkawa and Awschalom}(2000)}]{Kikkawa2000:S}
\bibinfo{author}{\bibnamefont{Kikkawa}, \bibfnamefont{J.~M.}}, and
  \bibinfo{author}{\bibfnamefont{D.~D.} \bibnamefont{Awschalom}},
  \bibinfo{year}{2000}, {``}\bibinfo{title}{All-optical magnetic resonance in
  semiconductors},{''} \bibinfo{journal}{{\sl Science}}
  \textbf{\bibinfo{volume}{287}},  \bibinfo{pages}{473--476}.

\bibitem[{\citenamefont{Kikkawa} \emph{et~al.}(2001)\citenamefont{Kikkawa,
  Gupta, Malajovich, and Awschalom}}]{Kikkawa2001:PE}
\bibinfo{author}{\bibnamefont{Kikkawa}, \bibfnamefont{J.~M.}},
  \bibinfo{author}{\bibfnamefont{J.~A.} \bibnamefont{Gupta}},
  \bibinfo{author}{\bibfnamefont{I.}~\bibnamefont{Malajovich}}, and
  \bibinfo{author}{\bibfnamefont{D.~D.} \bibnamefont{Awschalom}},
  \bibinfo{year}{2001}, {``}\bibinfo{title}{Spin coherence in semiconductors:
  storage, transport, and reduced dimensionality},{''}
  \bibinfo{journal}{Physica E} \textbf{\bibinfo{volume}{9}},
  \bibinfo{pages}{194--201}.

\bibitem[{\citenamefont{Kikkawa} \emph{et~al.}(1997)\citenamefont{Kikkawa,
  Smorchkova, Samarth, and Awschalom}}]{Kikkawa1997:S}
\bibinfo{author}{\bibnamefont{Kikkawa}, \bibfnamefont{J.~M.}},
  \bibinfo{author}{\bibfnamefont{I.~P.} \bibnamefont{Smorchkova}},
  \bibinfo{author}{\bibfnamefont{N.}~\bibnamefont{Samarth}}, and
  \bibinfo{author}{\bibfnamefont{D.~D.} \bibnamefont{Awschalom}},
  \bibinfo{year}{1997}, {``}\bibinfo{title}{Room-temperature spin memory in
  two-dimensional electron gases},{''} \bibinfo{journal}{{\sl Science}}
  \textbf{\bibinfo{volume}{277}},  \bibinfo{pages}{1284--1287}.

\bibitem[{\citenamefont{Kikuchi} \emph{et~al.}(2002)\citenamefont{Kikuchi,
  Immamura, Takahashi, and Maekawa}}]{Kikuchi2002:PRB}
\bibinfo{author}{\bibnamefont{Kikuchi}, \bibfnamefont{K.}},
  \bibinfo{author}{\bibfnamefont{H.}~\bibnamefont{Immamura}},
  \bibinfo{author}{\bibfnamefont{S.}~\bibnamefont{Takahashi}}, and
  \bibinfo{author}{\bibfnamefont{S.}~\bibnamefont{Maekawa}},
  \bibinfo{year}{2002}, {``}\bibinfo{title}{Conductance quantization and
  {Andreev} reflection in narrow ferromagnet/superconductor point
  contacts},{''} \bibinfo{journal}{Phys. Rev. B} \textbf{\bibinfo{volume}{65}},
   \bibinfo{pages}{020508}.

\bibitem[{\citenamefont{Kimel} \emph{et~al.}(2001)\citenamefont{Kimel,
  Bentivegna, Gridnev, Pavlov, Pisarev, and Rasing}}]{Kimel2001:PRB}
\bibinfo{author}{\bibnamefont{Kimel}, \bibfnamefont{A.~V.}},
  \bibinfo{author}{\bibfnamefont{F.}~\bibnamefont{Bentivegna}},
  \bibinfo{author}{\bibfnamefont{V.~N.} \bibnamefont{Gridnev}},
  \bibinfo{author}{\bibfnamefont{V.~V.} \bibnamefont{Pavlov}},
  \bibinfo{author}{\bibfnamefont{R.~V.} \bibnamefont{Pisarev}}, and
  \bibinfo{author}{\bibfnamefont{T.}~\bibnamefont{Rasing}},
  \bibinfo{year}{2001}, {``}\bibinfo{title}{Room-temperature ultrafast carrier
  and spin dynamics in {GaAs} probed by the photoinduced magneto-optical {Kerr}
  effect},{''} \bibinfo{journal}{Phys. Rev. B} \textbf{\bibinfo{volume}{63}},
  \bibinfo{pages}{235201}.

\bibitem[{\citenamefont{Kimel} \emph{et~al.}(2000)\citenamefont{Kimel, Pavlov,
  Pisarev, Gridnev, Bentivegna, and Rasing}}]{Kimel2000:PRB}
\bibinfo{author}{\bibnamefont{Kimel}, \bibfnamefont{A.~V.}},
  \bibinfo{author}{\bibfnamefont{V.~V.} \bibnamefont{Pavlov}},
  \bibinfo{author}{\bibfnamefont{R.~V.} \bibnamefont{Pisarev}},
  \bibinfo{author}{\bibfnamefont{V.~N.} \bibnamefont{Gridnev}},
  \bibinfo{author}{\bibfnamefont{F.}~\bibnamefont{Bentivegna}}, and
  \bibinfo{author}{\bibfnamefont{T.}~\bibnamefont{Rasing}},
  \bibinfo{year}{2000}, {``}\bibinfo{title}{Ultrafast dynamics of the
  photo-induced magneto-optical {Kerr} effect in {CdTe} at room
  temperature},{''} \bibinfo{journal}{Phys. Rev. B}
  \textbf{\bibinfo{volume}{62}},  \bibinfo{pages}{R10610--R10613}.

\bibitem[{\citenamefont{Kirczenow}(2001)}]{Kircenzow2001:PRB}
\bibinfo{author}{\bibnamefont{Kirczenow}, \bibfnamefont{G.}},
  \bibinfo{year}{2001}, {``}\bibinfo{title}{Ideal spin filters: A theoretical
  study of electron transmission through ordered and disordered interfaces
  between ferromagnetic metals and semiconductors},{''} \bibinfo{journal}{Phys.
  Rev. B} \textbf{\bibinfo{volume}{63}},  \bibinfo{pages}{054422}.

\bibitem[{\citenamefont{Kiselev and Kim}(2000)}]{Kiselev2000:PRB}
\bibinfo{author}{\bibnamefont{Kiselev}, \bibfnamefont{A.~A.}}, and
  \bibinfo{author}{\bibfnamefont{K.~W.} \bibnamefont{Kim}},
  \bibinfo{year}{2000}, {``}\bibinfo{title}{Progressive suppression of spin
  relaxation in two-dimensional channels of finite width},{''}
  \bibinfo{journal}{Phys. Rev. B} \textbf{\bibinfo{volume}{61}},
  \bibinfo{pages}{13115--13120}.

\bibitem[{\citenamefont{Kiselev and Kim}(2001)}]{Kiselev2001:APL}
\bibinfo{author}{\bibnamefont{Kiselev}, \bibfnamefont{A.~A.}}, and
  \bibinfo{author}{\bibfnamefont{K.~W.} \bibnamefont{Kim}},
  \bibinfo{year}{2001}, {``}\bibinfo{title}{T-shaped ballistic spin
  filter},{''} \bibinfo{journal}{Appl. Phys. Lett.}
  \textbf{\bibinfo{volume}{78}},  \bibinfo{pages}{775--777}.

\bibitem[{\citenamefont{Kiselev} \emph{et~al.}(2003)\citenamefont{Kiselev,
  Sankey, Krivorotov, Emley, Schoelkopf, Buhrman, and Ralph}}]{Kiselev2003:P}
\bibinfo{author}{\bibnamefont{Kiselev}, \bibfnamefont{S.~I.}},
  \bibinfo{author}{\bibfnamefont{J.~C.} \bibnamefont{Sankey}},
  \bibinfo{author}{\bibfnamefont{I.~N.} \bibnamefont{Krivorotov}},
  \bibinfo{author}{\bibfnamefont{N.~C.} \bibnamefont{Emley}},
  \bibinfo{author}{\bibfnamefont{R.~J.} \bibnamefont{Schoelkopf}},
  \bibinfo{author}{\bibfnamefont{R.~A.} \bibnamefont{Buhrman}}, and
  \bibinfo{author}{\bibfnamefont{D.~C.} \bibnamefont{Ralph}},
  \bibinfo{year}{2003}, {``}\bibinfo{title}{Microwave oscillations of a
  nanomagnet driven by a spin-polarized current},{''} \bibinfo{journal}{{\sl
  Nature}} \textbf{\bibinfo{volume}{425}},  \bibinfo{pages}{380--383}.

\bibitem[{\citenamefont{Kittel}(1963)}]{Kittel:1963}
\bibinfo{author}{\bibnamefont{Kittel}, \bibfnamefont{C.}},
  \bibinfo{year}{1963}, \emph{\bibinfo{title}{Quantum Theory of Solids}}
  (\bibinfo{publisher}{Wiley, New York}).

\bibitem[{\citenamefont{Kittel}(1996)}]{Kittel:1996}
\bibinfo{author}{\bibnamefont{Kittel}, \bibfnamefont{C.}},
  \bibinfo{year}{1996}, \emph{\bibinfo{title}{Introduction to Solid State
  Physics, 7th {Ed.}}} (\bibinfo{publisher}{Wiley, New York}).

\bibitem[{\citenamefont{Kivelson and Rokhsar}(1990)}]{Kivelson1990:PRB}
\bibinfo{author}{\bibnamefont{Kivelson}, \bibfnamefont{S.~A.}}, and
  \bibinfo{author}{\bibfnamefont{D.~S.} \bibnamefont{Rokhsar}},
  \bibinfo{year}{1990}, {``}\bibinfo{title}{Bogoliubov quasiparticles, spinons,
  and spin-charge decoupling in superconductors},{''} \bibinfo{journal}{Phys.
  Rev. B} \textbf{\bibinfo{volume}{41}},  \bibinfo{pages}{11693--11696}.

\bibitem[{\citenamefont{Knap} \emph{et~al.}(1996)\citenamefont{Knap,
  Skierbiszewski, Zduniak, {E. Litwin-Staszewska}, Bertho, Kobbi, Robert,
  Pikus, Pikus, Iordanskii, Mosser, Zekentes} \emph{et~al.}}]{Knap1996:PRB}
\bibinfo{author}{\bibnamefont{Knap}, \bibfnamefont{W.}},
  \bibinfo{author}{\bibfnamefont{C.}~\bibnamefont{Skierbiszewski}},
  \bibinfo{author}{\bibfnamefont{A.}~\bibnamefont{Zduniak}},
  \bibinfo{author}{\bibnamefont{{E. Litwin-Staszewska}}},
  \bibinfo{author}{\bibfnamefont{D.}~\bibnamefont{Bertho}},
  \bibinfo{author}{\bibfnamefont{F.}~\bibnamefont{Kobbi}},
  \bibinfo{author}{\bibfnamefont{J.~L.} \bibnamefont{Robert}},
  \bibinfo{author}{\bibfnamefont{G.~E.} \bibnamefont{Pikus}},
  \bibinfo{author}{\bibfnamefont{F.~G.} \bibnamefont{Pikus}},
  \bibinfo{author}{\bibfnamefont{S.~V.} \bibnamefont{Iordanskii}},
  \bibinfo{author}{\bibfnamefont{V.}~\bibnamefont{Mosser}},
  \bibinfo{author}{\bibfnamefont{K.}~\bibnamefont{Zekentes}}, \emph{et~al.},
  \bibinfo{year}{1996}, {``}\bibinfo{title}{Weak antilocalization and spin
  precession in quantum wells},{''} \bibinfo{journal}{Phys. Rev. B}
  \textbf{\bibinfo{volume}{53}},  \bibinfo{pages}{3912--3924}.

\bibitem[{\citenamefont{Kobayashi} \emph{et~al.}(1998)\citenamefont{Kobayashi,
  Kimura, Saweda, Terakura, and Tokura}}]{Kobayashi1998:N}
\bibinfo{author}{\bibnamefont{Kobayashi}, \bibfnamefont{K.~L.}},
  \bibinfo{author}{\bibfnamefont{T.}~\bibnamefont{Kimura}},
  \bibinfo{author}{\bibfnamefont{H.}~\bibnamefont{Saweda}},
  \bibinfo{author}{\bibfnamefont{K.}~\bibnamefont{Terakura}}, and
  \bibinfo{author}{\bibfnamefont{Y.}~\bibnamefont{Tokura}},
  \bibinfo{year}{1998}, {``}\bibinfo{title}{Room-temperature magnetoresistance
  in an oxide material with an ordered double-perovskite structure},{''}
  \bibinfo{journal}{{\sl Nature}} \textbf{\bibinfo{volume}{395}},
  \bibinfo{pages}{677--680}.

\bibitem[{\citenamefont{Koga}
  \emph{et~al.}(2002{\natexlab{a}})\citenamefont{Koga, Nitta, Akazaki, and
  Takayanagi}}]{Koga2002:PRL}
\bibinfo{author}{\bibnamefont{Koga}, \bibfnamefont{T.}},
  \bibinfo{author}{\bibfnamefont{J.}~\bibnamefont{Nitta}},
  \bibinfo{author}{\bibfnamefont{T.}~\bibnamefont{Akazaki}}, and
  \bibinfo{author}{\bibfnamefont{H.}~\bibnamefont{Takayanagi}},
  \bibinfo{year}{2002}{\natexlab{a}}, {``}\bibinfo{title}{Rashba spin-orbit
  coupling probed by the weak antilocalization analysis in
  {InAlAs}/{InGaAs}/{InAlAs} quantum wells as a function of quantum well
  asymmetry},{''} \bibinfo{journal}{Phys. Rev. Lett.}
  \textbf{\bibinfo{volume}{89}},  \bibinfo{pages}{046801}.

\bibitem[{\citenamefont{Koga}
  \emph{et~al.}(2002{\natexlab{b}})\citenamefont{Koga, Nitta, Takayanagi, and
  Datta}}]{Koga2002b:PRL}
\bibinfo{author}{\bibnamefont{Koga}, \bibfnamefont{T.}},
  \bibinfo{author}{\bibfnamefont{J.}~\bibnamefont{Nitta}},
  \bibinfo{author}{\bibfnamefont{H.}~\bibnamefont{Takayanagi}}, and
  \bibinfo{author}{\bibfnamefont{S.}~\bibnamefont{Datta}},
  \bibinfo{year}{2002}{\natexlab{b}}, {``}\bibinfo{title}{Spin-filter device
  based on the {Rashba} effect using a nonmagnetic resonant tunneling
  diode},{''} \bibinfo{journal}{Phys. Rev. Lett.}
  \textbf{\bibinfo{volume}{88}},  \bibinfo{pages}{126601}.

\bibitem[{\citenamefont{Kohda} \emph{et~al.}(2001)\citenamefont{Kohda, Ohno,
  Takamura, Matsukura, and Ohno}}]{Kohda2001:JJAP}
\bibinfo{author}{\bibnamefont{Kohda}, \bibfnamefont{M.}},
  \bibinfo{author}{\bibfnamefont{Y.}~\bibnamefont{Ohno}},
  \bibinfo{author}{\bibfnamefont{K.}~\bibnamefont{Takamura}},
  \bibinfo{author}{\bibfnamefont{F.}~\bibnamefont{Matsukura}}, and
  \bibinfo{author}{\bibfnamefont{H.}~\bibnamefont{Ohno}}, \bibinfo{year}{2001},
  {``}\bibinfo{title}{A spin {Esaki} diode},{''} \bibinfo{journal}{Jpn. J.
  Appl. Phys.} \textbf{\bibinfo{volume}{40}},  \bibinfo{pages}{L1274--L1276}.

\bibitem[{\citenamefont{Koiller} \emph{et~al.}(2002)\citenamefont{Koiller, Hu,
  and {Das Sarma}}}]{Koiller2002:PRL}
\bibinfo{author}{\bibnamefont{Koiller}, \bibfnamefont{B.}},
  \bibinfo{author}{\bibfnamefont{X.}~\bibnamefont{Hu}}, and
  \bibinfo{author}{\bibfnamefont{S.}~\bibnamefont{{Das Sarma}}},
  \bibinfo{year}{2002}, {``}\bibinfo{title}{Exchange in silicon-based quantum
  computer architecture},{''} \bibinfo{journal}{Phys. Rev. Lett.}
  \textbf{\bibinfo{volume}{88}},  \bibinfo{pages}{027903}.

\bibitem[{\citenamefont{Koiller} \emph{et~al.}(2003)\citenamefont{Koiller, Hu,
  and {Das Sarma}}}]{Koiller2003:PRL}
\bibinfo{author}{\bibnamefont{Koiller}, \bibfnamefont{B.}},
  \bibinfo{author}{\bibfnamefont{X.}~\bibnamefont{Hu}}, and
  \bibinfo{author}{\bibfnamefont{S.}~\bibnamefont{{Das Sarma}}},
  \bibinfo{year}{2003}, {``}\bibinfo{title}{Disentangling the exchange coupling
  of entangled donors in the {Si} quantum computer architecture},{''}
  \bibinfo{journal}{Phys. Rev. Lett.} \textbf{\bibinfo{volume}{90}},
  \bibinfo{pages}{067401}.

\bibitem[{\citenamefont{Kolbe}(1971)}]{Kolbe1971:PRB}
\bibinfo{author}{\bibnamefont{Kolbe}, \bibfnamefont{W.}}, \bibinfo{year}{1971},
  {``}\bibinfo{title}{Spin relaxation time of conduction electrons in bulk
  sodium metal},{''} \bibinfo{journal}{Phys. Rev. B}
  \textbf{\bibinfo{volume}{3}},  \bibinfo{pages}{320--323}.

\bibitem[{\citenamefont{K{\"o}nig} \emph{et~al.}(2001)\citenamefont{K{\"o}nig,
  B$\o$nsager, and MacDonald}}]{Konig2001:PRL}
\bibinfo{author}{\bibnamefont{K{\"o}nig}, \bibfnamefont{J.}},
  \bibinfo{author}{\bibfnamefont{M.~C.} \bibnamefont{B$\o$nsager}}, and
  \bibinfo{author}{\bibfnamefont{A.~H.} \bibnamefont{MacDonald}},
  \bibinfo{year}{2001}, {``}\bibinfo{title}{Dissipationless spin transport in
  thin film ferromagnets},{''} \bibinfo{journal}{Phys. Rev. Lett.}
  \textbf{\bibinfo{volume}{87}},  \bibinfo{pages}{187202}.

\bibitem[{\citenamefont{K{\"o}nig and Martinek}(2003)}]{Konig2003:PRL}
\bibinfo{author}{\bibnamefont{K{\"o}nig}, \bibfnamefont{J.}}, and
  \bibinfo{author}{\bibfnamefont{J.}~\bibnamefont{Martinek}},
  \bibinfo{year}{2003}, {``}\bibinfo{title}{Interaction-driven spin precession
  in quantum-dot spin valves},{''} \bibinfo{journal}{Phys. Rev. Lett.}
  \textbf{\bibinfo{volume}{90}},  \bibinfo{pages}{166602}.

\bibitem[{\citenamefont{K{\"{o}}nig}
  \emph{et~al.}(2003)\citenamefont{K{\"{o}}nig, Schliemann, Jungwirth, and
  MacDonald}}]{Koning:2003}
\bibinfo{author}{\bibnamefont{K{\"{o}}nig}, \bibfnamefont{J.}},
  \bibinfo{author}{\bibfnamefont{J.}~\bibnamefont{Schliemann}},
  \bibinfo{author}{\bibfnamefont{T.}~\bibnamefont{Jungwirth}}, and
  \bibinfo{author}{\bibfnamefont{A.~H.} \bibnamefont{MacDonald}},
  \bibinfo{year}{2003}, {``}\bibinfo{title}{Ferromagnetism in (III,Mn)V
  semiconductors},{''} in \emph{\bibinfo{booktitle}{Electronic Structure and
  Magnetism of Complex Materials}}, edited by
  \bibinfo{editor}{\bibfnamefont{D.~J.} \bibnamefont{Singh}} and
  \bibinfo{editor}{\bibfnamefont{D.~A.} \bibnamefont{Papaconstantopoulos}}
  (\bibinfo{publisher}{Academic, New York}),  \bibinfo{pages}{163--211}.

\bibitem[{\citenamefont{Korenblum and Rashba}(2002)}]{Korenblum2002:PRL}
\bibinfo{author}{\bibnamefont{Korenblum}, \bibfnamefont{B.}}, and
  \bibinfo{author}{\bibfnamefont{E.~I.} \bibnamefont{Rashba}},
  \bibinfo{year}{2002}, {``}\bibinfo{title}{Classical Properties of
  Low-Dimensional Conductors: Giant capacitance and non-{Ohmic} potential
  drop},{''} \bibinfo{journal}{Phys. Rev. Lett.} \textbf{\bibinfo{volume}{89}},
   \bibinfo{pages}{096803}.

\bibitem[{\citenamefont{Korotkov and Safarov}(1999)}]{Korotkov1999:PRB}
\bibinfo{author}{\bibnamefont{Korotkov}, \bibfnamefont{A.~N.}}, and
  \bibinfo{author}{\bibfnamefont{V.~I.} \bibnamefont{Safarov}},
  \bibinfo{year}{1999}, {``}\bibinfo{title}{Nonequilibrium spin distribution in
  a single-electron transistor},{''} \bibinfo{journal}{Phys. Rev. B}
  \textbf{\bibinfo{volume}{59}},  \bibinfo{pages}{89--92}.

\bibitem[{\citenamefont{Koshihara} \emph{et~al.}(1997)\citenamefont{Koshihara,
  Oiwa, Hirasawa, Katsumoto, Iye, Urano, Takagi, and
  Munekata}}]{Koshihara1997:PRL}
\bibinfo{author}{\bibnamefont{Koshihara}, \bibfnamefont{S.}},
  \bibinfo{author}{\bibfnamefont{A.}~\bibnamefont{Oiwa}},
  \bibinfo{author}{\bibfnamefont{M.}~\bibnamefont{Hirasawa}},
  \bibinfo{author}{\bibfnamefont{S.}~\bibnamefont{Katsumoto}},
  \bibinfo{author}{\bibfnamefont{Y.}~\bibnamefont{Iye}},
  \bibinfo{author}{\bibfnamefont{S.}~\bibnamefont{Urano}},
  \bibinfo{author}{\bibfnamefont{H.}~\bibnamefont{Takagi}}, and
  \bibinfo{author}{\bibfnamefont{H.}~\bibnamefont{Munekata}},
  \bibinfo{year}{1997}, {``}\bibinfo{title}{Ferromagnetic order induced by
  photogenerated carriers in magnetic {III-V} semiconductor heterostructures of
  {(In,Mn)As/GaSb}},{''} \bibinfo{journal}{Phys. Rev. Lett.}
  \textbf{\bibinfo{volume}{78}},  \bibinfo{pages}{4617--4620}.

\bibitem[{\citenamefont{Kouwenhoven}(2004)}]{Kouwenhoven2004:PC}
\bibinfo{author}{\bibnamefont{Kouwenhoven}, \bibfnamefont{L.~P.}},
  \bibinfo{year}{2004} \bibinfo{journal}{private communication} .

\bibitem[{\citenamefont{Kravchenko}(2002)}]{Kravchenko2002:JETP}
\bibinfo{author}{\bibnamefont{Kravchenko}, \bibfnamefont{V.~Y.}},
  \bibinfo{year}{2002}, {``}\bibinfo{title}{The electric conductivity of a
  laminated metal system (Alternating magnetic and nonmagnetic layers)},{''}
  \bibinfo{journal}{Zh. Eksp. Teor. Fiz.} \textbf{\bibinfo{volume}{121}},
  \bibinfo{pages}{703--727} \bibinfo{note}{[JETP {\bf 94}, 603-626 (2002)]}.

\bibitem[{\citenamefont{Kravchenko and Rashba}(2003)}]{Kravchenko2003:PRB}
\bibinfo{author}{\bibnamefont{Kravchenko}, \bibfnamefont{V.~Y.}}, and
  \bibinfo{author}{\bibfnamefont{E.~I.} \bibnamefont{Rashba}},
  \bibinfo{year}{2003}, {``}\bibinfo{title}{Spin injection into a ballistic
  semiconductor microstructure},{''} \bibinfo{journal}{Phys. Rev. B}
  \textbf{\bibinfo{volume}{67}},  \bibinfo{pages}{121310}.

\bibitem[{\citenamefont{Krenn} \emph{et~al.}(1989)\citenamefont{Krenn,
  Kaltenegger, Dietl, Spalek, and Bauer}}]{Krenn1989:PRB}
\bibinfo{author}{\bibnamefont{Krenn}, \bibfnamefont{H.}},
  \bibinfo{author}{\bibfnamefont{K.}~\bibnamefont{Kaltenegger}},
  \bibinfo{author}{\bibfnamefont{T.}~\bibnamefont{Dietl}},
  \bibinfo{author}{\bibfnamefont{J.}~\bibnamefont{Spalek}}, and
  \bibinfo{author}{\bibfnamefont{G.}~\bibnamefont{Bauer}},
  \bibinfo{year}{1989}, {``}\bibinfo{title}{Photoinduced magnetization in
  dilute magnetic (semimagnetic) semiconductors},{''} \bibinfo{journal}{Phys.
  Rev. B} \textbf{\bibinfo{volume}{39}},  \bibinfo{pages}{10918--10934}.

\bibitem[{\citenamefont{Krenn} \emph{et~al.}(1985)\citenamefont{Krenn,
  Zawadzki, and Bauer}}]{Krenn1985:PRL}
\bibinfo{author}{\bibnamefont{Krenn}, \bibfnamefont{H.}},
  \bibinfo{author}{\bibfnamefont{W.}~\bibnamefont{Zawadzki}}, and
  \bibinfo{author}{\bibfnamefont{G.}~\bibnamefont{Bauer}},
  \bibinfo{year}{1985}, {``}\bibinfo{title}{Optically induced magnetization in
  a dilute magnetic semiconductor: {Hg$_{1-x}$Mn$_x$Te}},{''}
  \bibinfo{journal}{Phys. Rev. Lett.} \textbf{\bibinfo{volume}{55}},
  \bibinfo{pages}{1510--1513}.

\bibitem[{\citenamefont{Kreuzer} \emph{et~al.}(2002)\citenamefont{Kreuzer,
  Moser, Wegscheider, Weiss, Bichler, and Schuh}}]{Kreuzer2002:APL}
\bibinfo{author}{\bibnamefont{Kreuzer}, \bibfnamefont{S.}},
  \bibinfo{author}{\bibfnamefont{J.}~\bibnamefont{Moser}},
  \bibinfo{author}{\bibfnamefont{W.}~\bibnamefont{Wegscheider}},
  \bibinfo{author}{\bibfnamefont{D.}~\bibnamefont{Weiss}},
  \bibinfo{author}{\bibfnamefont{M.}~\bibnamefont{Bichler}}, and
  \bibinfo{author}{\bibfnamefont{D.}~\bibnamefont{Schuh}},
  \bibinfo{year}{2002}, {``}\bibinfo{title}{Spin polarized tunneling through
  single-crystal {GaAs(001)} barriers},{''} \bibinfo{journal}{Appl. Phys.
  Lett.} \textbf{\bibinfo{volume}{80}},  \bibinfo{pages}{4582--4584}.

\bibitem[{\citenamefont{Krey}(2004)}]{Krey2003:P}
\bibinfo{author}{\bibnamefont{Krey}, \bibfnamefont{U.}}, \bibinfo{year}{2004},
  {``}\bibinfo{title}{On the significance of quantum effects and interactions
  for the apparent universality of {Bloch} laws for {$M_s(T)$}},{''}
  \bibinfo{journal}{J. Magn. Magn. Mater.} \textbf{\bibinfo{volume}{268}},
  \bibinfo{pages}{277--291}.

\bibitem[{\citenamefont{Krinichnyi}(2000)}]{Krinichnyi2000:SM}
\bibinfo{author}{\bibnamefont{Krinichnyi}, \bibfnamefont{V.~I.}},
  \bibinfo{year}{2000}, {``}\bibinfo{title}{2-mm waveband electron paramagnetic
  resonance spectroscopy of conducting polymers},{''} \bibinfo{journal}{Synth.
  Metals} \textbf{\bibinfo{volume}{108}},  \bibinfo{pages}{173--221}.

\bibitem[{\citenamefont{Krishnamurthy}
  \emph{et~al.}(2003)\citenamefont{Krishnamurthy, {van Schilfgaarde}, and
  Newman}}]{Krishnamurthy2003:APL}
\bibinfo{author}{\bibnamefont{Krishnamurthy}, \bibfnamefont{S.}},
  \bibinfo{author}{\bibfnamefont{M.}~\bibnamefont{{van Schilfgaarde}}}, and
  \bibinfo{author}{\bibfnamefont{N.}~\bibnamefont{Newman}},
  \bibinfo{year}{2003}, {``}\bibinfo{title}{Spin lifetimes of electrons
  injected into {GaAs} and {GaN}},{''} \bibinfo{journal}{Appl. Phys. Lett.}
  \textbf{\bibinfo{volume}{83}},  \bibinfo{pages}{1761--1763}.

\bibitem[{\citenamefont{Kuzma} \emph{et~al.}(1998)\citenamefont{Kuzma,
  Khandelwal, Barrett, Pfeiffer, and West}}]{Kuzma1998:S}
\bibinfo{author}{\bibnamefont{Kuzma}, \bibfnamefont{N.~N.}},
  \bibinfo{author}{\bibfnamefont{P.}~\bibnamefont{Khandelwal}},
  \bibinfo{author}{\bibfnamefont{S.~E.} \bibnamefont{Barrett}},
  \bibinfo{author}{\bibfnamefont{L.~N.} \bibnamefont{Pfeiffer}}, and
  \bibinfo{author}{\bibfnamefont{K.~W.} \bibnamefont{West}},
  \bibinfo{year}{1998}, {``}\bibinfo{title}{Ultraslow electron spin dynamics in
  {GaAs} quantum wells probed by optically pumped {NMR}},{''}
  \bibinfo{journal}{{\sl Science}} \textbf{\bibinfo{volume}{281}},
  \bibinfo{pages}{686--690}.

\bibitem[{\citenamefont{LaBella} \emph{et~al.}(2001)\citenamefont{LaBella,
  Bullock, Ding, Emery, Venkatesan, Oliver, Salamo, Thibado, and
  Mortazavi}}]{LaBella2001:S}
\bibinfo{author}{\bibnamefont{LaBella}, \bibfnamefont{V.~P.}},
  \bibinfo{author}{\bibfnamefont{D.~W.} \bibnamefont{Bullock}},
  \bibinfo{author}{\bibfnamefont{Z.}~\bibnamefont{Ding}},
  \bibinfo{author}{\bibfnamefont{C.}~\bibnamefont{Emery}},
  \bibinfo{author}{\bibfnamefont{A.}~\bibnamefont{Venkatesan}},
  \bibinfo{author}{\bibfnamefont{W.~F.} \bibnamefont{Oliver}},
  \bibinfo{author}{\bibfnamefont{G.~J.} \bibnamefont{Salamo}},
  \bibinfo{author}{\bibfnamefont{P.~M.} \bibnamefont{Thibado}}, and
  \bibinfo{author}{\bibfnamefont{M.}~\bibnamefont{Mortazavi}},
  \bibinfo{year}{2001}, {``}\bibinfo{title}{Spatially resolved spin-injection
  probability for gallium arsenide},{''} \bibinfo{journal}{{\sl Science}}
  \textbf{\bibinfo{volume}{292}},  \bibinfo{pages}{1518--1521}.

\bibitem[{\citenamefont{LaBella} \emph{et~al.}(2002)\citenamefont{LaBella,
  Bullock, Ding, Emery, Venkatesan, Oliver, Salamo, Thibado, and
  Mortazavi}}]{LaBella2002:S}
\bibinfo{author}{\bibnamefont{LaBella}, \bibfnamefont{V.~P.}},
  \bibinfo{author}{\bibfnamefont{D.~W.} \bibnamefont{Bullock}},
  \bibinfo{author}{\bibfnamefont{Z.}~\bibnamefont{Ding}},
  \bibinfo{author}{\bibfnamefont{C.}~\bibnamefont{Emery}},
  \bibinfo{author}{\bibfnamefont{A.}~\bibnamefont{Venkatesan}},
  \bibinfo{author}{\bibfnamefont{W.~F.} \bibnamefont{Oliver}},
  \bibinfo{author}{\bibfnamefont{G.~J.} \bibnamefont{Salamo}},
  \bibinfo{author}{\bibfnamefont{P.~M.} \bibnamefont{Thibado}}, and
  \bibinfo{author}{\bibfnamefont{M.}~\bibnamefont{Mortazavi}},
  \bibinfo{year}{2002}, {``}\bibinfo{title}{Spin polarization of injected
  electrons (Reply)},{''} \bibinfo{journal}{{\sl Science}}
  \textbf{\bibinfo{volume}{292}},  \bibinfo{pages}{1195}.

\bibitem[{\citenamefont{Lamari}(2003)}]{Lamari2003:PRB}
\bibinfo{author}{\bibnamefont{Lamari}, \bibfnamefont{S.}},
  \bibinfo{year}{2003}, {``}\bibinfo{title}{Effect of the doping concentration
  on the zero-field splitting and {Rashba} parameter in a {p-InAs}
  {MOSFET}},{''} \bibinfo{journal}{Phys. Rev. B} \textbf{\bibinfo{volume}{67}},
   \bibinfo{pages}{165329}.

\bibitem[{\citenamefont{Lambert and Raimondi}(1998)}]{Lambert1998:JPCM}
\bibinfo{author}{\bibnamefont{Lambert}, \bibfnamefont{C.~J.}}, and
  \bibinfo{author}{\bibfnamefont{R.}~\bibnamefont{Raimondi}},
  \bibinfo{year}{1998}, {``}\bibinfo{title}{Phase-coherent transport in hybrid
  superconducting nanostructures},{''} \bibinfo{journal}{J. Phys.: Condens.
  Matter} \textbf{\bibinfo{volume}{10}},  \bibinfo{pages}{901--941}.

\bibitem[{\citenamefont{Lampel}(1968)}]{Lampel1968:PRL}
\bibinfo{author}{\bibnamefont{Lampel}, \bibfnamefont{G.}},
  \bibinfo{year}{1968}, {``}\bibinfo{title}{Nuclear dynamic polarization by
  optical electronic saturation and optical pumping in semiconductors},{''}
  \bibinfo{journal}{Phys. Rev. Lett.} \textbf{\bibinfo{volume}{20}},
  \bibinfo{pages}{491--493}.

\bibitem[{\citenamefont{Lang} \emph{et~al.}(1996)\citenamefont{Lang,
  Eisenmenger, and Fulde}}]{Lang1996:PRL}
\bibinfo{author}{\bibnamefont{Lang}, \bibfnamefont{J.}},
  \bibinfo{author}{\bibfnamefont{W.}~\bibnamefont{Eisenmenger}}, and
  \bibinfo{author}{\bibfnamefont{P.}~\bibnamefont{Fulde}},
  \bibinfo{year}{1996}, {``}\bibinfo{title}{Phonon-induced spin flip in
  extremely thin superconducting {Al} tunneling junctions in high magnetic
  fields},{''} \bibinfo{journal}{Phys. Rev. Lett.}
  \textbf{\bibinfo{volume}{77}},  \bibinfo{pages}{2546--2549}.

\bibitem[{\citenamefont{Lau and {Flatt\'{e}}}(2002)}]{Lau2002:JAP}
\bibinfo{author}{\bibnamefont{Lau}, \bibfnamefont{W.~H.}}, and
  \bibinfo{author}{\bibfnamefont{M.~E.} \bibnamefont{{Flatt\'{e}}}},
  \bibinfo{year}{2002}, {``}\bibinfo{title}{Tunability of electron spin
  coherence in {III-V} quantum wells},{''} \bibinfo{journal}{J. Appl. Phys.}
  \textbf{\bibinfo{volume}{91}},  \bibinfo{pages}{8682--8684}.

\bibitem[{\citenamefont{Lau} \emph{et~al.}(2001)\citenamefont{Lau, Olesberg,
  and {Flatt\'{e}}}}]{Lau2001:PRB}
\bibinfo{author}{\bibnamefont{Lau}, \bibfnamefont{W.~H.}},
  \bibinfo{author}{\bibfnamefont{J.~T.} \bibnamefont{Olesberg}}, and
  \bibinfo{author}{\bibfnamefont{M.~E.} \bibnamefont{{Flatt\'{e}}}},
  \bibinfo{year}{2001}, {``}\bibinfo{title}{Electron-spin decoherence in bulk
  and quantum-well zinc-blende semiconductors},{''} \bibinfo{journal}{Phys.
  Rev. B} \textbf{\bibinfo{volume}{64}},  \bibinfo{pages}{161301}.

\bibitem[{\citenamefont{Lebedeva and Kuivalainen}(2003)}]{Lebedeva2003:JAP}
\bibinfo{author}{\bibnamefont{Lebedeva}, \bibfnamefont{N.}}, and
  \bibinfo{author}{\bibfnamefont{P.}~\bibnamefont{Kuivalainen}},
  \bibinfo{year}{2003}, {``}\bibinfo{title}{Modeling of ferromagnetic
  semiconductor devices for spintronics},{''} \bibinfo{journal}{J. Appl. Phys.}
  \textbf{\bibinfo{volume}{93}},  \bibinfo{pages}{9845--9864}.

\bibitem[{\citenamefont{LeClair} \emph{et~al.}(2002)\citenamefont{LeClair,
  Kohlhepp, {van de Vin}, Wieldraaijer, Swagten, {de Jonge}, Davis, MacLaren,
  Moodera, and Jansen}}]{LeClair2002:PRL}
\bibinfo{author}{\bibnamefont{LeClair}, \bibfnamefont{P.}},
  \bibinfo{author}{\bibfnamefont{J.~T.} \bibnamefont{Kohlhepp}},
  \bibinfo{author}{\bibfnamefont{C.~H.} \bibnamefont{{van de Vin}}},
  \bibinfo{author}{\bibfnamefont{H.}~\bibnamefont{Wieldraaijer}},
  \bibinfo{author}{\bibfnamefont{H.~J.~M.} \bibnamefont{Swagten}},
  \bibinfo{author}{\bibfnamefont{W.~J.~M.} \bibnamefont{{de Jonge}}},
  \bibinfo{author}{\bibfnamefont{A.~H.} \bibnamefont{Davis}},
  \bibinfo{author}{\bibfnamefont{J.~M.} \bibnamefont{MacLaren}},
  \bibinfo{author}{\bibfnamefont{J.~S.} \bibnamefont{Moodera}}, and
  \bibinfo{author}{\bibfnamefont{R.}~\bibnamefont{Jansen}},
  \bibinfo{year}{2002}, {``}\bibinfo{title}{Band structure and density of
  states effects in {Co}-based magnetic tunnel junctions},{''}
  \bibinfo{journal}{Phys. Rev. Lett.} \textbf{\bibinfo{volume}{88}},
  \bibinfo{pages}{107201}.

\bibitem[{\citenamefont{Lee} \emph{et~al.}(1999)\citenamefont{Lee, Gardelis,
  Choi, Xu, Smith, Barnes, Ritchie, Linfield, and Bland}}]{Lee1999:JAP}
\bibinfo{author}{\bibnamefont{Lee}, \bibfnamefont{W.~Y.}},
  \bibinfo{author}{\bibfnamefont{S.}~\bibnamefont{Gardelis}},
  \bibinfo{author}{\bibfnamefont{B.-C.} \bibnamefont{Choi}},
  \bibinfo{author}{\bibfnamefont{Y.~B.} \bibnamefont{Xu}},
  \bibinfo{author}{\bibfnamefont{C.~G.} \bibnamefont{Smith}},
  \bibinfo{author}{\bibfnamefont{C.~H.~W.} \bibnamefont{Barnes}},
  \bibinfo{author}{\bibfnamefont{D.~A.} \bibnamefont{Ritchie}},
  \bibinfo{author}{\bibfnamefont{E.~H.} \bibnamefont{Linfield}}, and
  \bibinfo{author}{\bibfnamefont{J.~A.~C.} \bibnamefont{Bland}},
  \bibinfo{year}{1999}, {``}\bibinfo{title}{Magnetization reversal and
  magnetoresistance in a lateral spin-injection device},{''}
  \bibinfo{journal}{J. Appl. Phys.} \textbf{\bibinfo{volume}{85}},
  \bibinfo{pages}{6682--6685}.

\bibitem[{\citenamefont{Lenczowski}
  \emph{et~al.}(1994)\citenamefont{Lenczowski, {van de Veerdonk}, Gijs,
  Giesbers, and Janssen}}]{Lenczowski1994:JAP}
\bibinfo{author}{\bibnamefont{Lenczowski}, \bibfnamefont{S.~K.~J.}},
  \bibinfo{author}{\bibfnamefont{R.~J.~M.} \bibnamefont{{van de Veerdonk}}},
  \bibinfo{author}{\bibfnamefont{M.~A.~M.} \bibnamefont{Gijs}},
  \bibinfo{author}{\bibfnamefont{J.~B.} \bibnamefont{Giesbers}}, and
  \bibinfo{author}{\bibfnamefont{H.~H. J.~M.} \bibnamefont{Janssen}},
  \bibinfo{year}{1994}, {``}\bibinfo{title}{Current-distribution effects in
  microstructures for perpendicular magnetoresistance experiments},{''}
  \bibinfo{journal}{J. Appl. Phys.} \textbf{\bibinfo{volume}{75}},
  \bibinfo{pages}{5154--5159}.

\bibitem[{\citenamefont{Lepine}(1970)}]{Lepine1970:PRB}
\bibinfo{author}{\bibnamefont{Lepine}, \bibfnamefont{D.~J.}},
  \bibinfo{year}{1970}, {``}\bibinfo{title}{Spin resonance of localized and
  delocalized electrons in phosphorus-doped silicon between 20 and 30 {K}},{''}
  \bibinfo{journal}{Phys. Rev. B} \textbf{\bibinfo{volume}{2}},
  \bibinfo{pages}{2429--2439}.

\bibitem[{\citenamefont{Lepine}(1972)}]{Lepine1972:PRB}
\bibinfo{author}{\bibnamefont{Lepine}, \bibfnamefont{D.~J.}},
  \bibinfo{year}{1972}, {``}\bibinfo{title}{Spin-dependent recombination on
  silicon surface},{''} \bibinfo{journal}{Phys. Rev. B}
  \textbf{\bibinfo{volume}{6}},  \bibinfo{pages}{436--441}.

\bibitem[{\citenamefont{Levitov} \emph{et~al.}(1984)\citenamefont{Levitov,
  Nazarov, and \'{E}liashberg}}]{Levitov1985:SPJETP}
\bibinfo{author}{\bibnamefont{Levitov}, \bibfnamefont{L.~S.}},
  \bibinfo{author}{\bibfnamefont{Y.~V.} \bibnamefont{Nazarov}}, and
  \bibinfo{author}{\bibfnamefont{G.~M.} \bibnamefont{\'{E}liashberg}},
  \bibinfo{year}{1984}, {``}\bibinfo{title}{Magnetoelectric effects in
  conductors with mirror isomer symmetry},{''} \bibinfo{journal}{Zh. Eksp.
  Teor. Fiz.} \textbf{\bibinfo{volume}{88}},  \bibinfo{pages}{229--236}
  \bibinfo{note}{[Sov. Phys. JETP {\bf 61}, 133-137 (1973)]}.

\bibitem[{\citenamefont{Levitov and Rashba}(2003)}]{Levitov2003:PRB}
\bibinfo{author}{\bibnamefont{Levitov}, \bibfnamefont{L.~S.}}, and
  \bibinfo{author}{\bibfnamefont{E.~I.} \bibnamefont{Rashba}},
  \bibinfo{year}{2003}, {``}\bibinfo{title}{Dynamical spin-electric coupling in
  a quantum dot},{''} \bibinfo{journal}{Phys. Rev. B}
  \textbf{\bibinfo{volume}{67}},  \bibinfo{pages}{115324}.

\bibitem[{\citenamefont{Levy}(2002)}]{Levy2002:PRL}
\bibinfo{author}{\bibnamefont{Levy}, \bibfnamefont{J.}}, \bibinfo{year}{2002},
  {``}\bibinfo{title}{Universal quantum computation with spin-1/2 pairs and
  {Heisenberg} exchange},{''} \bibinfo{journal}{Phys. Rev. Lett.}
  \textbf{\bibinfo{volume}{89}},  \bibinfo{pages}{147902}.

\bibitem[{\citenamefont{Levy and Mertig}(2002)}]{Levy:2002}
\bibinfo{author}{\bibnamefont{Levy}, \bibfnamefont{P.~M.}}, and
  \bibinfo{author}{\bibfnamefont{I.}~\bibnamefont{Mertig}},
  \bibinfo{year}{2002}, {``}\bibinfo{title}{Theory of Giant
  Magnetoresistance},{''} in \emph{\bibinfo{booktitle}{Spin Dependent Transport
  in Magnetic Nanostructures}}, edited by
  \bibinfo{editor}{\bibfnamefont{S.}~\bibnamefont{Maekawa}} and
  \bibinfo{editor}{\bibfnamefont{T.}~\bibnamefont{Shinjo}}
  (\bibinfo{publisher}{Taylor and Francis, New York}),
  \bibinfo{pages}{47--111}.

\bibitem[{\citenamefont{Linder and Sham}(1998)}]{Linder1998:PE}
\bibinfo{author}{\bibnamefont{Linder}, \bibfnamefont{N.}}, and
  \bibinfo{author}{\bibfnamefont{L.~J.} \bibnamefont{Sham}},
  \bibinfo{year}{1998}, {``}\bibinfo{title}{Theory of the coherent spin
  dynamics in magnetic semiconductor quantum wells},{''}
  \bibinfo{journal}{Physica E} \textbf{\bibinfo{volume}{2}},
  \bibinfo{pages}{412--416}.

\bibitem[{\citenamefont{Lommer} \emph{et~al.}(1988)\citenamefont{Lommer,
  Malcher, and R{\"{o}ssler}}}]{Lommer1988:PRL}
\bibinfo{author}{\bibnamefont{Lommer}, \bibfnamefont{G.}},
  \bibinfo{author}{\bibfnamefont{F.}~\bibnamefont{Malcher}}, and
  \bibinfo{author}{\bibfnamefont{U.}~\bibnamefont{R{\"{o}ssler}}},
  \bibinfo{year}{1988}, {``}\bibinfo{title}{Spin splitting in semiconductor
  heterostructures for {$B \rightarrow 0$}},{''} \bibinfo{journal}{Phys. Rev.
  Lett.} \textbf{\bibinfo{volume}{60}},  \bibinfo{pages}{728--731}.

\bibitem[{\citenamefont{Long} \emph{et~al.}(2003)\citenamefont{Long, Sun, Guo,
  and Wang}}]{Long2002:APL}
\bibinfo{author}{\bibnamefont{Long}, \bibfnamefont{W.}},
  \bibinfo{author}{\bibfnamefont{Q.-F.} \bibnamefont{Sun}},
  \bibinfo{author}{\bibfnamefont{H.}~\bibnamefont{Guo}}, and
  \bibinfo{author}{\bibfnamefont{J.}~\bibnamefont{Wang}}, \bibinfo{year}{2003},
  {``}\bibinfo{title}{Gate-controllable spin battery},{''}
  \bibinfo{journal}{Appl. Phys. Lett.} \textbf{\bibinfo{volume}{83}},
  \bibinfo{pages}{1397--1399}.

\bibitem[{\citenamefont{Loraine} \emph{et~al.}(2000)\citenamefont{Loraine,
  Pugh, Jenniches, Kirschman, Thompson, Allen, Sirisathikul, and
  Gregg}}]{Loraine2000:JAP}
\bibinfo{author}{\bibnamefont{Loraine}, \bibfnamefont{D.~R.}},
  \bibinfo{author}{\bibfnamefont{D.~I.} \bibnamefont{Pugh}},
  \bibinfo{author}{\bibfnamefont{H.}~\bibnamefont{Jenniches}},
  \bibinfo{author}{\bibfnamefont{R.}~\bibnamefont{Kirschman}},
  \bibinfo{author}{\bibfnamefont{S.~M.} \bibnamefont{Thompson}},
  \bibinfo{author}{\bibfnamefont{W.}~\bibnamefont{Allen}},
  \bibinfo{author}{\bibfnamefont{C.}~\bibnamefont{Sirisathikul}}, and
  \bibinfo{author}{\bibfnamefont{J.~F.} \bibnamefont{Gregg}},
  \bibinfo{year}{2000}, {``}\bibinfo{title}{Effect of silicon crystal structure
  on spin transmission through spin electronic devices},{''}
  \bibinfo{journal}{J. Appl. Phys.} \textbf{\bibinfo{volume}{87}},
  \bibinfo{pages}{5161--5163}.

\bibitem[{\citenamefont{Loss and DiVincenzo}(1998)}]{Loss1998:PRA}
\bibinfo{author}{\bibnamefont{Loss}, \bibfnamefont{D.}}, and
  \bibinfo{author}{\bibfnamefont{D.~P.} \bibnamefont{DiVincenzo}},
  \bibinfo{year}{1998}, {``}\bibinfo{title}{Quantum computation with quantum
  dots},{''} \bibinfo{journal}{Phys. Rev. A} \textbf{\bibinfo{volume}{57}},
  \bibinfo{pages}{120--122}.

\bibitem[{\citenamefont{Lubzens and
  Schultz}(1976{\natexlab{a}})}]{Dunifer1976:PRB}
\bibinfo{author}{\bibnamefont{Lubzens}, \bibfnamefont{D.}}, and
  \bibinfo{author}{\bibfnamefont{S.}~\bibnamefont{Schultz}},
  \bibinfo{year}{1976}{\natexlab{a}}, {``}\bibinfo{title}{Conduction-electron
  spin resonance in aluminum at 79 {GHz}},{''} \bibinfo{journal}{Phys. Rev. B}
  \textbf{\bibinfo{volume}{14}},  \bibinfo{pages}{945--950}.

\bibitem[{\citenamefont{Lubzens and
  Schultz}(1976{\natexlab{b}})}]{Lubzens1976:PRL}
\bibinfo{author}{\bibnamefont{Lubzens}, \bibfnamefont{D.}}, and
  \bibinfo{author}{\bibfnamefont{S.}~\bibnamefont{Schultz}},
  \bibinfo{year}{1976}{\natexlab{b}}, {``}\bibinfo{title}{Observation of an
  anomalous frequency dependence of the conduction-electron spin resonance in
  {Al}},{''} \bibinfo{journal}{Phys. Rev. Lett.} \textbf{\bibinfo{volume}{36}},
   \bibinfo{pages}{1104--1106}.

\bibitem[{\citenamefont{Luo} \emph{et~al.}(1988)\citenamefont{Luo, Munekata,
  Fang, and Stiles}}]{Luo1988:PRB}
\bibinfo{author}{\bibnamefont{Luo}, \bibfnamefont{J.}},
  \bibinfo{author}{\bibfnamefont{H.}~\bibnamefont{Munekata}},
  \bibinfo{author}{\bibfnamefont{F.~F.} \bibnamefont{Fang}}, and
  \bibinfo{author}{\bibfnamefont{P.~J.} \bibnamefont{Stiles}},
  \bibinfo{year}{1988}, {``}\bibinfo{title}{Observation of the zero-field spin
  splitting of the ground electron subband in {GaSb-InAs-GaSb} quantum
  wells},{''} \bibinfo{journal}{Phys. Rev. B} \textbf{\bibinfo{volume}{38}},
  \bibinfo{pages}{10142--10145}.

\bibitem[{\citenamefont{Luo} \emph{et~al.}(1990)\citenamefont{Luo, Munekata,
  Fang, and Stiles}}]{Luo1990:PRB}
\bibinfo{author}{\bibnamefont{Luo}, \bibfnamefont{J.}},
  \bibinfo{author}{\bibfnamefont{H.}~\bibnamefont{Munekata}},
  \bibinfo{author}{\bibfnamefont{F.~F.} \bibnamefont{Fang}}, and
  \bibinfo{author}{\bibfnamefont{P.~J.} \bibnamefont{Stiles}},
  \bibinfo{year}{1990}, {``}\bibinfo{title}{Effects of inversion asymmetry on
  electron energy band structures in {GaSb/InAs/GaSb} quantum wells},{''}
  \bibinfo{journal}{Phys. Rev. B} \textbf{\bibinfo{volume}{41}},
  \bibinfo{pages}{7685--7693}.

\bibitem[{\citenamefont{Lusakowski}
  \emph{et~al.}(2003)\citenamefont{Lusakowski, {Wr\'{o}bel}, and
  Dietl}}]{Lusakowski2003:PRB}
\bibinfo{author}{\bibnamefont{Lusakowski}, \bibfnamefont{A.}},
  \bibinfo{author}{\bibfnamefont{J.}~\bibnamefont{{Wr\'{o}bel}}}, and
  \bibinfo{author}{\bibfnamefont{T.}~\bibnamefont{Dietl}},
  \bibinfo{year}{2003}, {``}\bibinfo{title}{Effect of bulk inversion asymmetry
  on the {Datta-Das} transistor},{''} \bibinfo{journal}{Phys. Rev. B}
  \textbf{\bibinfo{volume}{68}},  \bibinfo{pages}{115324}.

\bibitem[{\citenamefont{MacDonald}(1999)}]{MacDonald1999:PRL}
\bibinfo{author}{\bibnamefont{MacDonald}, \bibfnamefont{A.~H.}},
  \bibinfo{year}{1999}, {``}\bibinfo{title}{Spin bottlenecks in the quantum
  {Hall} regime},{''} \bibinfo{journal}{Phys. Rev. Lett.}
  \textbf{\bibinfo{volume}{83}},  \bibinfo{pages}{3262--3265}.

\bibitem[{\citenamefont{MacDonald} \emph{et~al.}(1998)\citenamefont{MacDonald,
  Jungwirth, and Kasner}}]{MacDonald1998:PRL}
\bibinfo{author}{\bibnamefont{MacDonald}, \bibfnamefont{A.~H.}},
  \bibinfo{author}{\bibfnamefont{T.}~\bibnamefont{Jungwirth}}, and
  \bibinfo{author}{\bibfnamefont{M.}~\bibnamefont{Kasner}},
  \bibinfo{year}{1998}, {``}\bibinfo{title}{Temperature dependence of itinerant
  electron junction magnetoresistance},{''} \bibinfo{journal}{Phys. Rev. Lett.}
  \textbf{\bibinfo{volume}{81}},  \bibinfo{pages}{705--708}.

\bibitem[{\citenamefont{MacLaren} \emph{et~al.}(1997)\citenamefont{MacLaren,
  Zhang, and Butler}}]{MacLaren1997:PRB}
\bibinfo{author}{\bibnamefont{MacLaren}, \bibfnamefont{J.~M.}},
  \bibinfo{author}{\bibfnamefont{X.-G.} \bibnamefont{Zhang}}, and
  \bibinfo{author}{\bibfnamefont{W.~H.} \bibnamefont{Butler}},
  \bibinfo{year}{1997}, {``}\bibinfo{title}{Validity of the {Jullier{\`{e}}}
  model of spin-dependent tunneling},{''} \bibinfo{journal}{Phys. Rev. Lett.}
  \textbf{\bibinfo{volume}{56}},  \bibinfo{pages}{11827--11832}.

\bibitem[{\citenamefont{MacLaren} \emph{et~al.}(1999)\citenamefont{MacLaren,
  Zhang, Butler, and Wang}}]{MacLaren1999:PRB}
\bibinfo{author}{\bibnamefont{MacLaren}, \bibfnamefont{J.~M.}},
  \bibinfo{author}{\bibfnamefont{X.-G.} \bibnamefont{Zhang}},
  \bibinfo{author}{\bibfnamefont{W.~H.} \bibnamefont{Butler}}, and
  \bibinfo{author}{\bibfnamefont{X.}~\bibnamefont{Wang}}, \bibinfo{year}{1999},
  {``}\bibinfo{title}{Layer {KKR} approach to {Bloch}-wave transmission and
  reflection: {Application} to spin-dependent tunneling},{''}
  \bibinfo{journal}{Phys. Rev. B} \textbf{\bibinfo{volume}{59}},
  \bibinfo{pages}{5470--5478}.

\bibitem[{\citenamefont{Maekawa and G{\"a}fvert}(1982)}]{Maekawa1982:IEEE}
\bibinfo{author}{\bibnamefont{Maekawa}, \bibfnamefont{S.}}, and
  \bibinfo{author}{\bibfnamefont{U.}~\bibnamefont{G{\"a}fvert}},
  \bibinfo{year}{1982}, {``}\bibinfo{title}{Electron tunneling between
  ferromagnetic films},{''} \bibinfo{journal}{IEEE Trans. Magn.}
  \textbf{\bibinfo{volume}{18}},  \bibinfo{pages}{707--708}.

\bibitem[{\citenamefont{Maekawa} \emph{et~al.}(2001)\citenamefont{Maekawa,
  Takahashi, and Imamura}}]{Maekawa2001:MSE}
\bibinfo{author}{\bibnamefont{Maekawa}, \bibfnamefont{S.}},
  \bibinfo{author}{\bibfnamefont{S.}~\bibnamefont{Takahashi}}, and
  \bibinfo{author}{\bibfnamefont{H.}~\bibnamefont{Imamura}},
  \bibinfo{year}{2001}, {``}\bibinfo{title}{Spin-polarized tunneling and spin
  injection in superconductor-ferromagnet junctions},{''}
  \bibinfo{journal}{Mater. Sci. Eng. B} \textbf{\bibinfo{volume}{84}},
  \bibinfo{pages}{44--48}.

\bibitem[{\citenamefont{Maekawa} \emph{et~al.}(2002)\citenamefont{Maekawa,
  Takahashi, and Imamura}}]{Maekawa:2002}
\bibinfo{author}{\bibnamefont{Maekawa}, \bibfnamefont{S.}},
  \bibinfo{author}{\bibfnamefont{S.}~\bibnamefont{Takahashi}}, and
  \bibinfo{author}{\bibfnamefont{H.}~\bibnamefont{Imamura}},
  \bibinfo{year}{2002}, {``}\bibinfo{title}{Theory of Tunnel
  Magnetoresistance},{''} in \emph{\bibinfo{booktitle}{Spin Dependent Transport
  in Magnetic Nanostructures}}, edited by
  \bibinfo{editor}{\bibfnamefont{S.}~\bibnamefont{Maekawa}} and
  \bibinfo{editor}{\bibfnamefont{T.}~\bibnamefont{Shinjo}}
  (\bibinfo{publisher}{Taylor and Francis, New York}),
  \bibinfo{pages}{143--236}.

\bibitem[{\citenamefont{Maialle}(1996)}]{Maialle1996:PRB}
\bibinfo{author}{\bibnamefont{Maialle}, \bibfnamefont{M.~Z.}},
  \bibinfo{year}{1996}, {``}\bibinfo{title}{Spin relaxation of electrons in
  p-doped quantum wells via the electron-hole exchange interaction},{''}
  \bibinfo{journal}{Phys. Rev. B} \textbf{\bibinfo{volume}{54}},
  \bibinfo{pages}{1967--1974}.

\bibitem[{\citenamefont{Maialle and Degani}(1997)}]{Maialle1997:PRB}
\bibinfo{author}{\bibnamefont{Maialle}, \bibfnamefont{M.~Z.}}, and
  \bibinfo{author}{\bibfnamefont{M.~H.} \bibnamefont{Degani}},
  \bibinfo{year}{1997}, {``}\bibinfo{title}{Electron-spin relaxation in p-type
  quantum wells via the electron-hole exchange interaction: the effects of the
  valence-band spin mixing and of an applied longitudinal electric field},{''}
  \bibinfo{journal}{Phys. Rev. B} \textbf{\bibinfo{volume}{55}},
  \bibinfo{pages}{13771--13777}.

\bibitem[{\citenamefont{Maialle} \emph{et~al.}(1993)\citenamefont{Maialle, {E.
  A. de Andrada e Silva}, and Sham}}]{Maialle1993:PRB}
\bibinfo{author}{\bibnamefont{Maialle}, \bibfnamefont{M.~Z.}},
  \bibinfo{author}{\bibnamefont{{E. A. de Andrada e Silva}}}, and
  \bibinfo{author}{\bibfnamefont{L.~J.} \bibnamefont{Sham}},
  \bibinfo{year}{1993}, {``}\bibinfo{title}{Exciton spin dynamics in quantum
  wells},{''} \bibinfo{journal}{Phys. Rev. B} \textbf{\bibinfo{volume}{47}},
  \bibinfo{pages}{15776--15788}.

\bibitem[{\citenamefont{Majewski and Vogl}(2003)}]{Majewski:2003}
\bibinfo{author}{\bibnamefont{Majewski}, \bibfnamefont{J.~A.}}, and
  \bibinfo{author}{\bibfnamefont{P.}~\bibnamefont{Vogl}}, \bibinfo{year}{2003},
  {``}\bibinfo{title}{Resonant spin-orbit interactions and phonon relaxation
  rates in superlattices},{''} in \emph{\bibinfo{booktitle}{Physics of
  Semiconductors 2002}}, edited by \bibinfo{editor}{\bibfnamefont{A.~R.}
  \bibnamefont{Long}} and \bibinfo{editor}{\bibfnamefont{J.~H.}
  \bibnamefont{Davies}} (\bibinfo{publisher}{IOP, Bristol}),
  \bibinfo{pages}{305}.

\bibitem[{\citenamefont{Maki}(1964)}]{Maki1964:PTP}
\bibinfo{author}{\bibnamefont{Maki}, \bibfnamefont{K.}}, \bibinfo{year}{1964},
  {``}\bibinfo{title}{Pauli paramagnetism and superconducting state. {II}},{''}
  \bibinfo{journal}{Prog. Theor. Phys.} \textbf{\bibinfo{volume}{32}},
  \bibinfo{pages}{29--36}.

\bibitem[{\citenamefont{Malajovich}
  \emph{et~al.}(2001)\citenamefont{Malajovich, Berry, Samarth, and
  Awschalom}}]{Malajovich2001:N}
\bibinfo{author}{\bibnamefont{Malajovich}, \bibfnamefont{I.}},
  \bibinfo{author}{\bibfnamefont{J.~J.} \bibnamefont{Berry}},
  \bibinfo{author}{\bibfnamefont{N.}~\bibnamefont{Samarth}}, and
  \bibinfo{author}{\bibfnamefont{D.~D.} \bibnamefont{Awschalom}},
  \bibinfo{year}{2001}, {``}\bibinfo{title}{Persistent sourcing of coherent
  spins for multifunctional semiconductor spintronics},{''}
  \bibinfo{journal}{{\sl Nature}} \textbf{\bibinfo{volume}{411}},
  \bibinfo{pages}{770--772}.

\bibitem[{\citenamefont{Malinowski}
  \emph{et~al.}(2000)\citenamefont{Malinowski, Britton, Grevatt, Harley,
  Ritschie, and Simmons}}]{Malinowski2000:PRB}
\bibinfo{author}{\bibnamefont{Malinowski}, \bibfnamefont{A.}},
  \bibinfo{author}{\bibfnamefont{R.~S.} \bibnamefont{Britton}},
  \bibinfo{author}{\bibfnamefont{T.}~\bibnamefont{Grevatt}},
  \bibinfo{author}{\bibfnamefont{R.~T.} \bibnamefont{Harley}},
  \bibinfo{author}{\bibfnamefont{D.~A.} \bibnamefont{Ritschie}}, and
  \bibinfo{author}{\bibfnamefont{M.~Y.} \bibnamefont{Simmons}},
  \bibinfo{year}{2000}, {``}\bibinfo{title}{Spin relaxation in
  {GaAs/Al$_x$Ga$_{1-x}$As} quantum wells},{''} \bibinfo{journal}{Phys. Rev. B}
  \textbf{\bibinfo{volume}{62}},  \bibinfo{pages}{13034--13039}.

\bibitem[{\citenamefont{Mal'shukov}
  \emph{et~al.}(2003)\citenamefont{Mal'shukov, Tang, Chu, and
  Chao}}]{Malshukov2003:PRB}
\bibinfo{author}{\bibnamefont{Mal'shukov}, \bibfnamefont{A.~G.}},
  \bibinfo{author}{\bibfnamefont{C.~S.} \bibnamefont{Tang}},
  \bibinfo{author}{\bibfnamefont{C.~S.} \bibnamefont{Chu}}, and
  \bibinfo{author}{\bibfnamefont{K.~A.} \bibnamefont{Chao}},
  \bibinfo{year}{2003}, {``}\bibinfo{title}{Spin-current generation and
  detection in the presence of an ac gate},{''} \bibinfo{journal}{Phys. Rev. B}
  \textbf{\bibinfo{volume}{68}},  \bibinfo{pages}{233307}.

\bibitem[{\citenamefont{Mamin} \emph{et~al.}(2003)\citenamefont{Mamin,
  Budakian, Chui, and Rugar}}]{Mamin2003:P}
\bibinfo{author}{\bibnamefont{Mamin}, \bibfnamefont{H.~J.}},
  \bibinfo{author}{\bibfnamefont{R.}~\bibnamefont{Budakian}},
  \bibinfo{author}{\bibfnamefont{B.~W.} \bibnamefont{Chui}}, and
  \bibinfo{author}{\bibfnamefont{D.}~\bibnamefont{Rugar}},
  \bibinfo{year}{2003}, {``}\bibinfo{title}{Detection and manipulation of
  statistical polarization in small spin ensembles},{''}
  \bibinfo{journal}{Phys. Rev. Lett.} \textbf{\bibinfo{volume}{91}},
  \bibinfo{pages}{207604}.

\bibitem[{\citenamefont{Margulis and Margulis}(1994)}]{Margulis1994:PB}
\bibinfo{author}{\bibnamefont{Margulis}, \bibfnamefont{A.~D.}}, and
  \bibinfo{author}{\bibfnamefont{V.~A.} \bibnamefont{Margulis}},
  \bibinfo{year}{1994}, {``}\bibinfo{title}{Effect of spin injection on the
  {CESR} at a ferromagnet-semiconductor contact},{''} \bibinfo{journal}{Physica
  B} \textbf{\bibinfo{volume}{193}},  \bibinfo{pages}{179--187}.

\bibitem[{\citenamefont{Martin}(2003)}]{Martin2003:PRB}
\bibinfo{author}{\bibnamefont{Martin}, \bibfnamefont{I.}},
  \bibinfo{year}{2003}, {``}\bibinfo{title}{Spin-drift transport and its
  applications},{''} \bibinfo{journal}{Phys. Rev. B}
  \textbf{\bibinfo{volume}{67}},  \bibinfo{pages}{014421}.

\bibitem[{\citenamefont{Martinek} \emph{et~al.}(2002)\citenamefont{Martinek,
  Barna{\'{s}}, Maekawa, Schoeller, and {Sch\"{o}n}}}]{Martinek2002:PRB}
\bibinfo{author}{\bibnamefont{Martinek}, \bibfnamefont{J.}},
  \bibinfo{author}{\bibfnamefont{J.}~\bibnamefont{Barna{\'{s}}}},
  \bibinfo{author}{\bibfnamefont{S.}~\bibnamefont{Maekawa}},
  \bibinfo{author}{\bibfnamefont{H.}~\bibnamefont{Schoeller}}, and
  \bibinfo{author}{\bibfnamefont{G.}~\bibnamefont{{Sch\"{o}n}}},
  \bibinfo{year}{2002}, {``}\bibinfo{title}{Spin accumulation in ferromagnetic
  single-electron transistors in the cotunneling regime},{''}
  \bibinfo{journal}{Phys. Rev. B} \textbf{\bibinfo{volume}{66}},
  \bibinfo{pages}{014402}.

\bibitem[{\citenamefont{Marushchak}
  \emph{et~al.}(1984)\citenamefont{Marushchak, Stepanova, and
  Titkov}}]{Marushchak1984:SPSS}
\bibinfo{author}{\bibnamefont{Marushchak}, \bibfnamefont{V.~A.}},
  \bibinfo{author}{\bibfnamefont{M.~N.} \bibnamefont{Stepanova}}, and
  \bibinfo{author}{\bibfnamefont{A.~N.} \bibnamefont{Titkov}},
  \bibinfo{year}{1984}, {``}\bibinfo{title}{Spin relaxation of conduction
  electrons in moderately doped gallium arsenide crystals},{''}
  \bibinfo{journal}{Fiz. Tverd. Tela} \textbf{\bibinfo{volume}{25}},
  \bibinfo{pages}{3537--3542} \bibinfo{note}{[Sov. Phys. Solid State {\bf 25},
  2035-2038 (1983)]}.

\bibitem[{\citenamefont{Masterov and Makovskii}(1979)}]{Masterov1979:SPS}
\bibinfo{author}{\bibnamefont{Masterov}, \bibfnamefont{V.~F.}}, and
  \bibinfo{author}{\bibfnamefont{L.~L.} \bibnamefont{Makovskii}},
  \bibinfo{year}{1979}, {``}\bibinfo{title}{Spin-dependent recombination on
  injection of electrons from a ferromagnet into a semiconductor},{''}
  \bibinfo{journal}{Fiz. Tekh. Poluprovodn.} \textbf{\bibinfo{volume}{13}},
  \bibinfo{pages}{2042--2044} \bibinfo{note}{[Sov. Phys. Semicond. {\bf 13},
  1192-1193 (1979)]}.

\bibitem[{\citenamefont{Mathon and Umerski}(1997)}]{Mathon1997:PRB}
\bibinfo{author}{\bibnamefont{Mathon}, \bibfnamefont{J.}}, and
  \bibinfo{author}{\bibfnamefont{A.}~\bibnamefont{Umerski}},
  \bibinfo{year}{1997}, {``}\bibinfo{title}{Tight-binding theory of tunneling
  giant magnetoresistance},{''} \bibinfo{journal}{Phys. Rev. B}
  \textbf{\bibinfo{volume}{56}},  \bibinfo{pages}{11810--11819}.

\bibitem[{\citenamefont{Mathon and Umerski}(1999)}]{Mathon1999:PRB}
\bibinfo{author}{\bibnamefont{Mathon}, \bibfnamefont{J.}}, and
  \bibinfo{author}{\bibfnamefont{A.}~\bibnamefont{Umerski}},
  \bibinfo{year}{1999}, {``}\bibinfo{title}{Theory of tunneling
  magnetoresistance in a junction with a nonmagnetic metallic interlayer},{''}
  \bibinfo{journal}{Phys. Rev. B} \textbf{\bibinfo{volume}{60}},
  \bibinfo{pages}{1117--1121}.

\bibitem[{\citenamefont{Mathon and Umerski}(2001)}]{Mathon2001:PRB}
\bibinfo{author}{\bibnamefont{Mathon}, \bibfnamefont{J.}}, and
  \bibinfo{author}{\bibfnamefont{A.}~\bibnamefont{Umerski}},
  \bibinfo{year}{2001}, {``}\bibinfo{title}{Theory of tunneling
  magnetoresistance of an epitaxial {Fe/MgO/Fe(001)} junction},{''}
  \bibinfo{journal}{Phys. Rev. B} \textbf{\bibinfo{volume}{63}},
  \bibinfo{pages}{220403}.

\bibitem[{\citenamefont{Matsubara} \emph{et~al.}(1991)\citenamefont{Matsubara,
  Tsuzuku, and Sugihara}}]{Matsubara1991:PRB}
\bibinfo{author}{\bibnamefont{Matsubara}, \bibfnamefont{K.}},
  \bibinfo{author}{\bibfnamefont{T.}~\bibnamefont{Tsuzuku}}, and
  \bibinfo{author}{\bibfnamefont{K.}~\bibnamefont{Sugihara}},
  \bibinfo{year}{1991}, {``}\bibinfo{title}{Electron spin resonance in
  graphite},{''} \bibinfo{journal}{Phys. Rev. B} \textbf{\bibinfo{volume}{44}},
   \bibinfo{pages}{11845--11851}.

\bibitem[{\citenamefont{Matsuyama} \emph{et~al.}(2002)\citenamefont{Matsuyama,
  Hu, Grundler, Meier, and Merkt}}]{Matsuyama2002:PRB}
\bibinfo{author}{\bibnamefont{Matsuyama}, \bibfnamefont{T.}},
  \bibinfo{author}{\bibfnamefont{C.-M.} \bibnamefont{Hu}},
  \bibinfo{author}{\bibfnamefont{D.}~\bibnamefont{Grundler}},
  \bibinfo{author}{\bibfnamefont{G.}~\bibnamefont{Meier}}, and
  \bibinfo{author}{\bibfnamefont{U.}~\bibnamefont{Merkt}},
  \bibinfo{year}{2002}, {``}\bibinfo{title}{Ballistic spin transport and spin
  interference in {ferromagnetic/InAs(2DES)/ferromagnetic} devices},{''}
  \bibinfo{journal}{Phys. Rev. B} \textbf{\bibinfo{volume}{65}},
  \bibinfo{pages}{155322}.

\bibitem[{\citenamefont{Mattana} \emph{et~al.}(2003)\citenamefont{Mattana,
  George, Jaffres, {Nguyen Van Dau}, Fert, Lepine, Guivarc'h, and
  Jezequel}}]{Mattana2002:PRL}
\bibinfo{author}{\bibnamefont{Mattana}, \bibfnamefont{R.}},
  \bibinfo{author}{\bibfnamefont{J.-M.} \bibnamefont{George}},
  \bibinfo{author}{\bibfnamefont{H.}~\bibnamefont{Jaffres}},
  \bibinfo{author}{\bibfnamefont{F.}~\bibnamefont{{Nguyen Van Dau}}},
  \bibinfo{author}{\bibfnamefont{A.}~\bibnamefont{Fert}},
  \bibinfo{author}{\bibfnamefont{B.}~\bibnamefont{Lepine}},
  \bibinfo{author}{\bibfnamefont{A.}~\bibnamefont{Guivarc'h}}, and
  \bibinfo{author}{\bibfnamefont{G.}~\bibnamefont{Jezequel}},
  \bibinfo{year}{2003}, {``}\bibinfo{title}{Electrical detection of spin
  accumulation in a p-type {GaAs} quantum well},{''} \bibinfo{journal}{Phys.
  Rev. Lett.} \textbf{\bibinfo{volume}{90}},  \bibinfo{pages}{166601}.

\bibitem[{\citenamefont{Mavropoulos}
  \emph{et~al.}(2002)\citenamefont{Mavropoulos, Wunnicke, and
  Dederichs}}]{Mavropoulos2002:PRB}
\bibinfo{author}{\bibnamefont{Mavropoulos}, \bibfnamefont{P.}},
  \bibinfo{author}{\bibfnamefont{O.}~\bibnamefont{Wunnicke}}, and
  \bibinfo{author}{\bibfnamefont{P.~H.} \bibnamefont{Dederichs}},
  \bibinfo{year}{2002}, {``}\bibinfo{title}{Ballistic spin injection and
  detection in {Fe/semiconductor/Fe} junctions},{''} \bibinfo{journal}{Phys.
  Rev. B} \textbf{\bibinfo{volume}{66}},  \bibinfo{pages}{024416}.

\bibitem[{\citenamefont{Mazin}(1999)}]{Mazin1999:PRL}
\bibinfo{author}{\bibnamefont{Mazin}, \bibfnamefont{I.~I.}},
  \bibinfo{year}{1999}, {``}\bibinfo{title}{How to define and calculate the
  degree of spin polarization in ferromagnets},{''} \bibinfo{journal}{Phys.
  Rev. Lett.} \textbf{\bibinfo{volume}{83}},  \bibinfo{pages}{1427--1430}.

\bibitem[{\citenamefont{Mazin} \emph{et~al.}(2001)\citenamefont{Mazin, Golubov,
  and Nadgorny}}]{Mazin2001:JAP}
\bibinfo{author}{\bibnamefont{Mazin}, \bibfnamefont{I.~I.}},
  \bibinfo{author}{\bibfnamefont{A.~A.} \bibnamefont{Golubov}}, and
  \bibinfo{author}{\bibfnamefont{B.}~\bibnamefont{Nadgorny}},
  \bibinfo{year}{2001}, {``}\bibinfo{title}{Probing spin polarization with
  {Andreev} reflection: A theoretical basis},{''} \bibinfo{journal}{J. Appl.
  Phys.} \textbf{\bibinfo{volume}{89}},  \bibinfo{pages}{7576--7578}.

\bibitem[{\citenamefont{McCombe and Wagner}(1971)}]{McCombe1971:PRB}
\bibinfo{author}{\bibnamefont{McCombe}, \bibfnamefont{B.~D.}}, and
  \bibinfo{author}{\bibfnamefont{R.~J.} \bibnamefont{Wagner}},
  \bibinfo{year}{1971}, {``}\bibinfo{title}{Electric-dipole-excited electron
  spin resonance in {InSb}},{''} \bibinfo{journal}{Phys. Rev. B}
  \textbf{\bibinfo{volume}{4}},  \bibinfo{pages}{1285--1288}.

\bibitem[{\citenamefont{Meier} \emph{et~al.}(2003)\citenamefont{Meier, Levy,
  and Loss}}]{Meier2003:PRL}
\bibinfo{author}{\bibnamefont{Meier}, \bibfnamefont{F.}},
  \bibinfo{author}{\bibfnamefont{J.}~\bibnamefont{Levy}}, and
  \bibinfo{author}{\bibfnamefont{D.}~\bibnamefont{Loss}}, \bibinfo{year}{2003},
  {``}\bibinfo{title}{Quantum computing with spin cluster qubits},{''}
  \bibinfo{journal}{Phys. Rev. Lett.} \textbf{\bibinfo{volume}{90}},
  \bibinfo{pages}{047901}.

\bibitem[{\citenamefont{Meier and {Zakharchenya (Eds.)}}(1984)}]{Meier:1984}
\bibinfo{author}{\bibnamefont{Meier}, \bibfnamefont{F.}}, and
  \bibinfo{author}{\bibfnamefont{B.~P.} \bibnamefont{{Zakharchenya (Eds.)}}},
  \bibinfo{year}{1984}, \emph{\bibinfo{title}{Optical Orientation}}
  (\bibinfo{publisher}{North-Holand, New York}).

\bibitem[{\citenamefont{M{\'e}lin and Feinberg}(2002)}]{Melin2002:EPJB}
\bibinfo{author}{\bibnamefont{M{\'e}lin}, \bibfnamefont{R.}}, and
  \bibinfo{author}{\bibfnamefont{D.}~\bibnamefont{Feinberg}},
  \bibinfo{year}{2002}, {``}\bibinfo{title}{Transport theory of multiterminal
  hybrid structures},{''} \bibinfo{journal}{Eur. Phys. J. B}
  \textbf{\bibinfo{volume}{26}},  \bibinfo{pages}{101--114}.

\bibitem[{\citenamefont{Mendez} \emph{et~al.}(1998)\citenamefont{Mendez, Esaki,
  and Wang}}]{Mendez1986:PRB}
\bibinfo{author}{\bibnamefont{Mendez}, \bibfnamefont{E.~E.}},
  \bibinfo{author}{\bibfnamefont{L.}~\bibnamefont{Esaki}}, and
  \bibinfo{author}{\bibfnamefont{W.~I.} \bibnamefont{Wang}},
  \bibinfo{year}{1998}, {``}\bibinfo{title}{Resonant magnetotunneling in
  {GaAlAs-GaAs-GaAlAs} heterostructures},{''} \bibinfo{journal}{Phys. Rev. B}
  \textbf{\bibinfo{volume}{33}},  \bibinfo{pages}{R2893--R2896}.

\bibitem[{\citenamefont{Merkulov} \emph{et~al.}(2002)\citenamefont{Merkulov,
  Efros, and Rosen}}]{Merkulov2002:PRB}
\bibinfo{author}{\bibnamefont{Merkulov}, \bibfnamefont{I.~A.}},
  \bibinfo{author}{\bibfnamefont{A.~L.} \bibnamefont{Efros}}, and
  \bibinfo{author}{\bibfnamefont{M.}~\bibnamefont{Rosen}},
  \bibinfo{year}{2002}, {``}\bibinfo{title}{Electron spin relaxation by nuclei
  in semiconductor quantum dots},{''} \bibinfo{journal}{Phys. Rev. B}
  \textbf{\bibinfo{volume}{65}},  \bibinfo{pages}{205309}.

\bibitem[{\citenamefont{Meservey} \emph{et~al.}(1976)\citenamefont{Meservey,
  Paraskevopoulos, and Tedrow}}]{Meservey1976:PRL}
\bibinfo{author}{\bibnamefont{Meservey}, \bibfnamefont{R.}},
  \bibinfo{author}{\bibfnamefont{D.}~\bibnamefont{Paraskevopoulos}}, and
  \bibinfo{author}{\bibfnamefont{P.~M.} \bibnamefont{Tedrow}},
  \bibinfo{year}{1976}, {``}\bibinfo{title}{Correlation between spin
  polarization of tunnel currents from 3d ferromagnets and their magnetic
  moments},{''} \bibinfo{journal}{Phys. Rev. Lett.}
  \textbf{\bibinfo{volume}{37}},  \bibinfo{pages}{858--860}.

\bibitem[{\citenamefont{Meservey and Tedrow}(1978)}]{Meservey1978:PRL}
\bibinfo{author}{\bibnamefont{Meservey}, \bibfnamefont{R.}}, and
  \bibinfo{author}{\bibfnamefont{P.~M.} \bibnamefont{Tedrow}},
  \bibinfo{year}{1978}, {``}\bibinfo{title}{Surface relaxation times of
  conduction-electron spins in superconductors and normal metals},{''}
  \bibinfo{journal}{Phys. Rev. Lett.} \textbf{\bibinfo{volume}{41}},
  \bibinfo{pages}{805--808}.

\bibitem[{\citenamefont{Meservey} \emph{et~al.}(1982)\citenamefont{Meservey,
  Tedrow, and Brooks}}]{Meservey1982:JAP}
\bibinfo{author}{\bibnamefont{Meservey}, \bibfnamefont{R.}},
  \bibinfo{author}{\bibfnamefont{P.~M.} \bibnamefont{Tedrow}}, and
  \bibinfo{author}{\bibfnamefont{J.~S.} \bibnamefont{Brooks}},
  \bibinfo{year}{1982}, {``}\bibinfo{title}{Tunneling characteristics of
  amorphous {Si} barriers},{''} \bibinfo{journal}{J. Appl. Phys.}
  \textbf{\bibinfo{volume}{53}},  \bibinfo{pages}{1563--1570}.

\bibitem[{\citenamefont{Meservey} \emph{et~al.}(1983)\citenamefont{Meservey,
  Tedrow, and Moodera}}]{Meservey1983:JMMM}
\bibinfo{author}{\bibnamefont{Meservey}, \bibfnamefont{R.}},
  \bibinfo{author}{\bibfnamefont{P.~M.} \bibnamefont{Tedrow}}, and
  \bibinfo{author}{\bibfnamefont{J.~S.} \bibnamefont{Moodera}},
  \bibinfo{year}{1983}, {``}\bibinfo{title}{Electron spin polarized tunneling
  study of ferromagnetic thin films},{''} \bibinfo{journal}{J. Magn. Magn.
  Mater.} \textbf{\bibinfo{volume}{35}},  \bibinfo{pages}{1--6}.

\bibitem[{\citenamefont{Miller} \emph{et~al.}(2003)\citenamefont{Miller,
  Zumb{\"u}hl, Marcus, {Lyanda-Geller}, {Goldhaber-Gordon}, Campman, and
  Gossard}}]{Miller2003:PRL}
\bibinfo{author}{\bibnamefont{Miller}, \bibfnamefont{J.~B.}},
  \bibinfo{author}{\bibfnamefont{D.~M.} \bibnamefont{Zumb{\"u}hl}},
  \bibinfo{author}{\bibfnamefont{C.~M.} \bibnamefont{Marcus}},
  \bibinfo{author}{\bibfnamefont{Y.~B.} \bibnamefont{{Lyanda-Geller}}},
  \bibinfo{author}{\bibfnamefont{D.}~\bibnamefont{{Goldhaber-Gordon}}},
  \bibinfo{author}{\bibfnamefont{K.}~\bibnamefont{Campman}}, and
  \bibinfo{author}{\bibfnamefont{A.~C.} \bibnamefont{Gossard}},
  \bibinfo{year}{2003}, {``}\bibinfo{title}{Gate-controlled spin-orbit quantum
  interference eEffects in lateral transport},{''} \bibinfo{journal}{Phys. Rev.
  Lett.} \textbf{\bibinfo{volume}{90}},  \bibinfo{pages}{076807}.

\bibitem[{\citenamefont{Mills}(1971)}]{Mills1971:PRB}
\bibinfo{author}{\bibnamefont{Mills}, \bibfnamefont{D.~L.}},
  \bibinfo{year}{1971}, {``}\bibinfo{title}{Surface effects in magnetic
  crystals near the ordering temperature},{''} \bibinfo{journal}{Phys. Rev. B}
  \textbf{\bibinfo{volume}{3}},  \bibinfo{pages}{3887}.

\bibitem[{\citenamefont{Mireles and Kirczenow}(2001)}]{Mireles2001:PRB}
\bibinfo{author}{\bibnamefont{Mireles}, \bibfnamefont{F.}}, and
  \bibinfo{author}{\bibfnamefont{G.}~\bibnamefont{Kirczenow}},
  \bibinfo{year}{2001}, {``}\bibinfo{title}{Ballistic spin-polarized transport
  and {Rashba} spin precession in semiconductor nanowires},{''}
  \bibinfo{journal}{Phys. Rev. B} \textbf{\bibinfo{volume}{64}},
  \bibinfo{pages}{024426}.

\bibitem[{\citenamefont{Miyazaki}(2002)}]{Miyazaki:2002}
\bibinfo{author}{\bibnamefont{Miyazaki}, \bibfnamefont{T.}},
  \bibinfo{year}{2002}, {``}\bibinfo{title}{Experiments of Tunnel
  Magnetoresistance},{''} in \emph{\bibinfo{booktitle}{Spin Dependent Transport
  in Magnetic Nanostructures}}, edited by
  \bibinfo{editor}{\bibfnamefont{S.}~\bibnamefont{Maekawa}} and
  \bibinfo{editor}{\bibfnamefont{T.}~\bibnamefont{Shinjo}}
  (\bibinfo{publisher}{Taylor and Francis, New York}),
  \bibinfo{pages}{143--236}.

\bibitem[{\citenamefont{Miyazaki and Tezuka}(1995)}]{Miyazaki1995:JMMM}
\bibinfo{author}{\bibnamefont{Miyazaki}, \bibfnamefont{T.}}, and
  \bibinfo{author}{\bibfnamefont{N.}~\bibnamefont{Tezuka}},
  \bibinfo{year}{1995}, {``}\bibinfo{title}{Giant magnetic tunneling effect in
  {Fe/Al$_2$O$_3$/Fe} junction},{''} \bibinfo{journal}{J. Mag. Magn. Mater.}
  \textbf{\bibinfo{volume}{139}},  \bibinfo{pages}{L231--L234}.

\bibitem[{\citenamefont{Mizushima} \emph{et~al.}(1997)\citenamefont{Mizushima,
  Kinno, Yamauchi, and Tanaka}}]{Mizushima1997:IEEETM}
\bibinfo{author}{\bibnamefont{Mizushima}, \bibfnamefont{K.}},
  \bibinfo{author}{\bibfnamefont{T.}~\bibnamefont{Kinno}},
  \bibinfo{author}{\bibfnamefont{T.}~\bibnamefont{Yamauchi}}, and
  \bibinfo{author}{\bibfnamefont{K.}~\bibnamefont{Tanaka}},
  \bibinfo{year}{1997}, {``}\bibinfo{title}{Energy-dependent hot electron
  transport across a spin-valve},{''} \bibinfo{journal}{IEEE Trans. Magn.}
  \textbf{\bibinfo{volume}{33}},  \bibinfo{pages}{3500--3504}.

\bibitem[{\citenamefont{Monod and Beuneu}(1979)}]{Monod1979:PRB}
\bibinfo{author}{\bibnamefont{Monod}, \bibfnamefont{P.}}, and
  \bibinfo{author}{\bibfnamefont{F.}~\bibnamefont{Beuneu}},
  \bibinfo{year}{1979}, {``}\bibinfo{title}{Conduction-electron spin flip by
  phonons in metals: {Analysis} of experimental data},{''}
  \bibinfo{journal}{Phys. Rev. B} \textbf{\bibinfo{volume}{19}},
  \bibinfo{pages}{911--915}.

\bibitem[{\citenamefont{Monod and {J\'{a}nossy}}(1977)}]{Monod1977:JLTP}
\bibinfo{author}{\bibnamefont{Monod}, \bibfnamefont{P.}}, and
  \bibinfo{author}{\bibfnamefont{A.}~\bibnamefont{{J\'{a}nossy}}},
  \bibinfo{year}{1977}, {``}\bibinfo{title}{Conduction electron spin resonance
  in gold},{''} \bibinfo{journal}{J. Low Temp. Phys.}
  \textbf{\bibinfo{volume}{26}},  \bibinfo{pages}{311--316}.

\bibitem[{\citenamefont{Monod and Schultz}(1982)}]{Monod1982:JP}
\bibinfo{author}{\bibnamefont{Monod}, \bibfnamefont{P.}}, and
  \bibinfo{author}{\bibfnamefont{S.}~\bibnamefont{Schultz}},
  \bibinfo{year}{1982}, {``}\bibinfo{title}{Conduction electron spin-flip
  scattering by impurities in copper},{''} \bibinfo{journal}{J. Phys. (Paris)}
  \textbf{\bibinfo{volume}{43}},  \bibinfo{pages}{393--401}.

\bibitem[{\citenamefont{Monsma} \emph{et~al.}(1995)\citenamefont{Monsma,
  Lodder, Popma, and Dieny}}]{Monsma1995:PRL}
\bibinfo{author}{\bibnamefont{Monsma}, \bibfnamefont{D.~J.}},
  \bibinfo{author}{\bibfnamefont{J.~C.} \bibnamefont{Lodder}},
  \bibinfo{author}{\bibfnamefont{T.~J.~A.} \bibnamefont{Popma}}, and
  \bibinfo{author}{\bibfnamefont{B.}~\bibnamefont{Dieny}},
  \bibinfo{year}{1995}, {``}\bibinfo{title}{Perpendicular hot electron
  spin-valve effect in a new magnetic field sensor: {The} spin-valve
  transistor},{''} \bibinfo{journal}{Phys. Rev. Lett.}
  \textbf{\bibinfo{volume}{74}},  \bibinfo{pages}{5260--5263}.

\bibitem[{\citenamefont{Monsma and
  Parkin}(2000{\natexlab{a}})}]{Monsma2000:APLa}
\bibinfo{author}{\bibnamefont{Monsma}, \bibfnamefont{D.~J.}}, and
  \bibinfo{author}{\bibfnamefont{S.~S.~P.} \bibnamefont{Parkin}},
  \bibinfo{year}{2000}{\natexlab{a}}, {``}\bibinfo{title}{Spin polarization of
  tunneling current from ferromagnet/{Al$_2$O$_3$} interfaces using
  copper-doped aluminum superconducting films},{''} \bibinfo{journal}{Appl.
  Phys. Lett.} \textbf{\bibinfo{volume}{77}},  \bibinfo{pages}{720--722}.

\bibitem[{\citenamefont{Monsma and
  Parkin}(2000{\natexlab{b}})}]{Monsma2000:APLb}
\bibinfo{author}{\bibnamefont{Monsma}, \bibfnamefont{D.~J.}}, and
  \bibinfo{author}{\bibfnamefont{S.~S.~P.} \bibnamefont{Parkin}},
  \bibinfo{year}{2000}{\natexlab{b}}, {``}\bibinfo{title}{Temporal evolution of
  spin-polarization in ferromagnetic tunnel junctions},{''}
  \bibinfo{journal}{Appl. Phys. Lett.} \textbf{\bibinfo{volume}{77}},
  \bibinfo{pages}{883--885}.

\bibitem[{\citenamefont{Monsma} \emph{et~al.}(1998)\citenamefont{Monsma,
  Vlutters, and Lodder}}]{Monsma1998:S}
\bibinfo{author}{\bibnamefont{Monsma}, \bibfnamefont{D.~J.}},
  \bibinfo{author}{\bibfnamefont{R.}~\bibnamefont{Vlutters}}, and
  \bibinfo{author}{\bibfnamefont{J.~C.} \bibnamefont{Lodder}},
  \bibinfo{year}{1998}, {``}\bibinfo{title}{Room temperature-operating
  spin-valve transistors formed by vacuum bonding},{''} \bibinfo{journal}{{\sl
  Science}} \textbf{\bibinfo{volume}{281}},  \bibinfo{pages}{407--409}.

\bibitem[{\citenamefont{Monzon} \emph{et~al.}(1997)\citenamefont{Monzon,
  Johnson, and Roukes}}]{Monzon1997:APL}
\bibinfo{author}{\bibnamefont{Monzon}, \bibfnamefont{F.~G.}},
  \bibinfo{author}{\bibfnamefont{M.}~\bibnamefont{Johnson}}, and
  \bibinfo{author}{\bibfnamefont{M.~L.} \bibnamefont{Roukes}},
  \bibinfo{year}{1997}, {``}\bibinfo{title}{Strong {Hall} voltage modulation in
  hybrid ferromagnet/semiconductor microstructures},{''}
  \bibinfo{journal}{Appl. Phys. Lett.} \textbf{\bibinfo{volume}{71}},
  \bibinfo{pages}{3087--3089}.

\bibitem[{\citenamefont{Monzon and Roukes}(1999)}]{Monzon1999:JMMM}
\bibinfo{author}{\bibnamefont{Monzon}, \bibfnamefont{F.~G.}}, and
  \bibinfo{author}{\bibfnamefont{M.~L.} \bibnamefont{Roukes}},
  \bibinfo{year}{1999}, {``}\bibinfo{title}{Spin injection and the local {Hall}
  effect in {InAs} quantum wells},{''} \bibinfo{journal}{J. Magn. Magn. Mater.}
  \textbf{\bibinfo{volume}{199}},  \bibinfo{pages}{632--635}.

\bibitem[{\citenamefont{Monzon} \emph{et~al.}(2000)\citenamefont{Monzon, Tang,
  and Roukes}}]{Monzon2000:PRL}
\bibinfo{author}{\bibnamefont{Monzon}, \bibfnamefont{F.~G.}},
  \bibinfo{author}{\bibfnamefont{H.~X.} \bibnamefont{Tang}}, and
  \bibinfo{author}{\bibfnamefont{M.~L.} \bibnamefont{Roukes}},
  \bibinfo{year}{2000}, {``}\bibinfo{title}{Magnetoelectronic phenomena at a
  ferromagnet-semiconductor interface},{''} \bibinfo{journal}{Phys. Rev. Lett.}
  \textbf{\bibinfo{volume}{84}},  \bibinfo{pages}{5022}.

\bibitem[{\citenamefont{Moodera} \emph{et~al.}(1988)\citenamefont{Moodera, Hao,
  Gibson, and Meservey}}]{Moodera1988:PRL}
\bibinfo{author}{\bibnamefont{Moodera}, \bibfnamefont{J.~S.}},
  \bibinfo{author}{\bibfnamefont{X.}~\bibnamefont{Hao}},
  \bibinfo{author}{\bibfnamefont{G.~A.} \bibnamefont{Gibson}}, and
  \bibinfo{author}{\bibfnamefont{R.}~\bibnamefont{Meservey}},
  \bibinfo{year}{1988}, {``}\bibinfo{title}{Electron-spin polarization in
  tunnel junctions in zero applied field with ferromagnetic {EuS}
  barriers},{''} \bibinfo{journal}{Phys. Rev. B} \textbf{\bibinfo{volume}{42}},
   \bibinfo{pages}{8235--8243}.

\bibitem[{\citenamefont{Moodera} \emph{et~al.}(1996)\citenamefont{Moodera,
  Kinder, Nowak, LeClair, and Meservey}}]{Moodera1996:APL}
\bibinfo{author}{\bibnamefont{Moodera}, \bibfnamefont{J.~S.}},
  \bibinfo{author}{\bibfnamefont{L.~R.} \bibnamefont{Kinder}},
  \bibinfo{author}{\bibfnamefont{J.}~\bibnamefont{Nowak}},
  \bibinfo{author}{\bibfnamefont{P.}~\bibnamefont{LeClair}}, and
  \bibinfo{author}{\bibfnamefont{R.}~\bibnamefont{Meservey}},
  \bibinfo{year}{1996}, {``}\bibinfo{title}{Geometrically enhanced
  magnetoresistance in ferromagnet-insulator-ferromagnet tunnel junctions},{''}
  \bibinfo{journal}{Appl. Phys. Lett.} \textbf{\bibinfo{volume}{69}},
  \bibinfo{pages}{708--710}.

\bibitem[{\citenamefont{Moodera} \emph{et~al.}(1995)\citenamefont{Moodera,
  Kinder, Wong, and Meservey}}]{Moodera1995:PRL}
\bibinfo{author}{\bibnamefont{Moodera}, \bibfnamefont{J.~S.}},
  \bibinfo{author}{\bibfnamefont{L.~R.} \bibnamefont{Kinder}},
  \bibinfo{author}{\bibfnamefont{T.~M.} \bibnamefont{Wong}}, and
  \bibinfo{author}{\bibfnamefont{R.}~\bibnamefont{Meservey}},
  \bibinfo{year}{1995}, {``}\bibinfo{title}{Large magnetoresistance at room
  temperature in ferromagnetic thin film tunnel junctions},{''}
  \bibinfo{journal}{Phys. Rev. Lett.} \textbf{\bibinfo{volume}{74}},
  \bibinfo{pages}{3273--3276}.

\bibitem[{\citenamefont{Moodera and Mathon}(1999)}]{Moodera1999:JMMM}
\bibinfo{author}{\bibnamefont{Moodera}, \bibfnamefont{J.~S.}}, and
  \bibinfo{author}{\bibfnamefont{G.}~\bibnamefont{Mathon}},
  \bibinfo{year}{1999}, {``}\bibinfo{title}{Spin polarized tunneling in
  ferromagnetic junctions},{''} \bibinfo{journal}{J. Mag. Magn. Mater.}
  \textbf{\bibinfo{volume}{200}},  \bibinfo{pages}{248--273}.

\bibitem[{\citenamefont{Moodera} \emph{et~al.}(1993)\citenamefont{Moodera,
  Meservey, and Hao}}]{Moodera1993:PRL}
\bibinfo{author}{\bibnamefont{Moodera}, \bibfnamefont{J.~S.}},
  \bibinfo{author}{\bibfnamefont{R.}~\bibnamefont{Meservey}}, and
  \bibinfo{author}{\bibfnamefont{X.}~\bibnamefont{Hao}}, \bibinfo{year}{1993},
  {``}\bibinfo{title}{Variation of the electron-spin polarization in {EuSe}
  tunnel junctions from zero to near 100\% in a magnetic field},{''}
  \bibinfo{journal}{Phys. Rev. Lett.} \textbf{\bibinfo{volume}{70}},
  \bibinfo{pages}{853--856}.

\bibitem[{\citenamefont{Moodera} \emph{et~al.}(1999)\citenamefont{Moodera,
  Nassar, and Mathon}}]{Moodera1999:ARMS}
\bibinfo{author}{\bibnamefont{Moodera}, \bibfnamefont{J.~S.}},
  \bibinfo{author}{\bibfnamefont{J.}~\bibnamefont{Nassar}}, and
  \bibinfo{author}{\bibfnamefont{G.}~\bibnamefont{Mathon}},
  \bibinfo{year}{1999}, {``}\bibinfo{title}{Spin-tunneling in ferromagnetic
  junctions},{''} \bibinfo{journal}{Annu. Rev. Mater. Sci.}
  \textbf{\bibinfo{volume}{29}},  \bibinfo{pages}{381--432}.

\bibitem[{\citenamefont{Moodera} \emph{et~al.}(1998)\citenamefont{Moodera,
  Nowak, and {van de Veerdonk}}}]{Moodera1998:PRL}
\bibinfo{author}{\bibnamefont{Moodera}, \bibfnamefont{J.~S.}},
  \bibinfo{author}{\bibfnamefont{J.}~\bibnamefont{Nowak}}, and
  \bibinfo{author}{\bibfnamefont{R.~J.~M.} \bibnamefont{{van de Veerdonk}}},
  \bibinfo{year}{1998}, {``}\bibinfo{title}{Interface magnetism and spin wave
  scattering in ferromagnet-insulator-ferromagnet tunnel junctions},{''}
  \bibinfo{journal}{Phys. Rev. Lett.} \textbf{\bibinfo{volume}{80}},
  \bibinfo{pages}{2941--2944}.

\bibitem[{\citenamefont{Moriya} \emph{et~al.}(2003)\citenamefont{Moriya,
  Hamaya, Oiwa, and Munekata}}]{Moriya2003:P}
\bibinfo{author}{\bibnamefont{Moriya}, \bibfnamefont{R.}},
  \bibinfo{author}{\bibfnamefont{K.}~\bibnamefont{Hamaya}},
  \bibinfo{author}{\bibfnamefont{A.}~\bibnamefont{Oiwa}}, and
  \bibinfo{author}{\bibfnamefont{H.}~\bibnamefont{Munekata}},
  \bibinfo{year}{2003}, {``}\bibinfo{title}{Effect of electrical spin injection
  on magnetization reversal in {(Ga,Mn)As}/{AlAs}/{(Ga,Mn)As} trilayer
  structures},{''} \bibinfo{note}{preprint, {H. Munekata}, private
  communication}.

\bibitem[{\citenamefont{Moriya}(1960)}]{Moriya1960:PR}
\bibinfo{author}{\bibnamefont{Moriya}, \bibfnamefont{T.}},
  \bibinfo{year}{1960}, {``}\bibinfo{title}{Anisotropic superexchange
  interaction and weak ferromagnetism},{''} \bibinfo{journal}{Phys. Rev.}
  \textbf{\bibinfo{volume}{120}},  \bibinfo{pages}{91--98}.

\bibitem[{\citenamefont{Motsnyi} \emph{et~al.}(2002)\citenamefont{Motsnyi,
  Safarov, {De Boeck}, Das, {Van Roy}, Goovaerts, and
  Borghs}}]{Motsnyi2002:APL}
\bibinfo{author}{\bibnamefont{Motsnyi}, \bibfnamefont{V.~F.}},
  \bibinfo{author}{\bibfnamefont{V.~I.} \bibnamefont{Safarov}},
  \bibinfo{author}{\bibfnamefont{J.}~\bibnamefont{{De Boeck}}},
  \bibinfo{author}{\bibfnamefont{J.}~\bibnamefont{Das}},
  \bibinfo{author}{\bibfnamefont{W.}~\bibnamefont{{Van Roy}}},
  \bibinfo{author}{\bibfnamefont{E.}~\bibnamefont{Goovaerts}}, and
  \bibinfo{author}{\bibfnamefont{G.}~\bibnamefont{Borghs}},
  \bibinfo{year}{2002}, {``}\bibinfo{title}{Electrical spin injection in a
  ferromagnet/tunnel barrier/semiconductor heterostructure},{''}
  \bibinfo{journal}{Appl. Phys. Lett.} \textbf{\bibinfo{volume}{81}},
  \bibinfo{pages}{265--267}.

\bibitem[{\citenamefont{Motsnyi} \emph{et~al.}(2003)\citenamefont{Motsnyi, {Van
  Dorpe}, {Van Roy}, Goovaerts, Safarov, Borghs, and Boeck}}]{Motsnyi2003:PRB}
\bibinfo{author}{\bibnamefont{Motsnyi}, \bibfnamefont{V.~F.}},
  \bibinfo{author}{\bibfnamefont{P.}~\bibnamefont{{Van Dorpe}}},
  \bibinfo{author}{\bibfnamefont{W.}~\bibnamefont{{Van Roy}}},
  \bibinfo{author}{\bibfnamefont{E.}~\bibnamefont{Goovaerts}},
  \bibinfo{author}{\bibfnamefont{V.~I.} \bibnamefont{Safarov}},
  \bibinfo{author}{\bibfnamefont{G.}~\bibnamefont{Borghs}}, and
  \bibinfo{author}{\bibfnamefont{J.~D.} \bibnamefont{Boeck}},
  \bibinfo{year}{2003}, {``}\bibinfo{title}{Optical investigation of electrical
  spin injection into semiconductor},{''} \bibinfo{journal}{Phys. Rev. B}
  \textbf{\bibinfo{volume}{68}},  \bibinfo{pages}{245319}.

\bibitem[{\citenamefont{Mott}(1936{\natexlab{a}})}]{Mott1936:PRCa}
\bibinfo{author}{\bibnamefont{Mott}, \bibfnamefont{N.~F.}},
  \bibinfo{year}{1936}{\natexlab{a}}, {``}\bibinfo{title}{The electrical
  conductivity of transition metals},{''} \bibinfo{journal}{Proc. R. Soc.
  London, Ser. A} \textbf{\bibinfo{volume}{153}},  \bibinfo{pages}{699--717}.

\bibitem[{\citenamefont{Mott}(1936{\natexlab{b}})}]{Mott1936:PRCb}
\bibinfo{author}{\bibnamefont{Mott}, \bibfnamefont{N.~F.}},
  \bibinfo{year}{1936}{\natexlab{b}}, {``}\bibinfo{title}{The resistance and
  thermoelectric properties of the transition metals},{''}
  \bibinfo{journal}{Proc. R. Soc. London, Ser. A}
  \textbf{\bibinfo{volume}{156}},  \bibinfo{pages}{368--382}.

\bibitem[{\citenamefont{Mott and Massey}(1965)}]{Mott:1965}
\bibinfo{author}{\bibnamefont{Mott}, \bibfnamefont{N.~F.}}, and
  \bibinfo{author}{\bibfnamefont{H.~S.~W.} \bibnamefont{Massey}},
  \bibinfo{year}{1965}, \emph{\bibinfo{title}{The Theory of Atomic Collisions,
  {\rm 3rd {Ed.}}}} (\bibinfo{publisher}{Clarendon, Oxford}).

\bibitem[{\citenamefont{Moussa} \emph{et~al.}(2003)\citenamefont{Moussa,
  {Ram-Mohan}, Rowe, and Solin}}]{Moussa2003:JAP}
\bibinfo{author}{\bibnamefont{Moussa}, \bibfnamefont{J.}},
  \bibinfo{author}{\bibfnamefont{L.~R.} \bibnamefont{{Ram-Mohan}}},
  \bibinfo{author}{\bibfnamefont{A.~C.~H.} \bibnamefont{Rowe}}, and
  \bibinfo{author}{\bibfnamefont{S.~A.} \bibnamefont{Solin}},
  \bibinfo{year}{2003}, {``}\bibinfo{title}{Response of an extraordinary
  magnetoresistance read head to a magnetic bit},{''} \bibinfo{journal}{J.
  Appl. Phys.} \textbf{\bibinfo{volume}{94}},  \bibinfo{pages}{1110--1114}.

\bibitem[{\citenamefont{Mucciolo} \emph{et~al.}(2002)\citenamefont{Mucciolo,
  Chamon, and Marcus}}]{Mucciolo2002:PRL}
\bibinfo{author}{\bibnamefont{Mucciolo}, \bibfnamefont{E.~R.}},
  \bibinfo{author}{\bibfnamefont{C.}~\bibnamefont{Chamon}}, and
  \bibinfo{author}{\bibfnamefont{C.~M.} \bibnamefont{Marcus}},
  \bibinfo{year}{2002}, {``}\bibinfo{title}{An adiabatic quantum pump of spin
  polarized current},{''} \bibinfo{journal}{Phys. Rev. Lett.}
  \textbf{\bibinfo{volume}{89}},  \bibinfo{pages}{146802}.

\bibitem[{\citenamefont{Munekata}(2003)}]{Munekata2003:PC}
\bibinfo{author}{\bibnamefont{Munekata}, \bibfnamefont{H.}},
  \bibinfo{year}{2003} \bibinfo{journal}{private communications} .

\bibitem[{\citenamefont{Munekata} \emph{et~al.}(1989)\citenamefont{Munekata,
  Ohno, von Moln\'{a}r, Segm{\"u}ller, Chang, and Esaki}}]{Munekata1989:PRL}
\bibinfo{author}{\bibnamefont{Munekata}, \bibfnamefont{H.}},
  \bibinfo{author}{\bibfnamefont{H.}~\bibnamefont{Ohno}},
  \bibinfo{author}{\bibfnamefont{S.}~\bibnamefont{von Moln\'{a}r}},
  \bibinfo{author}{\bibfnamefont{A.}~\bibnamefont{Segm{\"u}ller}},
  \bibinfo{author}{\bibfnamefont{L.~L.} \bibnamefont{Chang}}, and
  \bibinfo{author}{\bibfnamefont{L.}~\bibnamefont{Esaki}},
  \bibinfo{year}{1989}, {``}\bibinfo{title}{Diluted magnetic {III-V}
  semiconductors},{''} \bibinfo{journal}{Phys. Rev. Lett.}
  \textbf{\bibinfo{volume}{63}},  \bibinfo{pages}{1849--1852}.

\bibitem[{\citenamefont{Munekata} \emph{et~al.}(1991)\citenamefont{Munekata,
  Ohno, Ruf, Gambino, and Chang}}]{Munekata1991:JCG}
\bibinfo{author}{\bibnamefont{Munekata}, \bibfnamefont{H.}},
  \bibinfo{author}{\bibfnamefont{H.}~\bibnamefont{Ohno}},
  \bibinfo{author}{\bibfnamefont{R.~R.} \bibnamefont{Ruf}},
  \bibinfo{author}{\bibfnamefont{R.~J.} \bibnamefont{Gambino}}, and
  \bibinfo{author}{\bibfnamefont{L.~L.} \bibnamefont{Chang}},
  \bibinfo{year}{1991}, {``}\bibinfo{title}{P-type diluted magnetic {III-V}
  semiconductors},{''} \bibinfo{journal}{J. Cryst. Growth}
  \textbf{\bibinfo{volume}{111}},  \bibinfo{pages}{1011--1015}.

\bibitem[{\citenamefont{Munekata} \emph{et~al.}(2003)\citenamefont{Munekata,
  Oiwa, Mitsumori, Moriya, and Slupinski}}]{Munekata2003:JS}
\bibinfo{author}{\bibnamefont{Munekata}, \bibfnamefont{H.}},
  \bibinfo{author}{\bibfnamefont{A.}~\bibnamefont{Oiwa}},
  \bibinfo{author}{\bibfnamefont{Y.}~\bibnamefont{Mitsumori}},
  \bibinfo{author}{\bibfnamefont{R.}~\bibnamefont{Moriya}}, and
  \bibinfo{author}{\bibfnamefont{T.}~\bibnamefont{Slupinski}},
  \bibinfo{year}{2003}, {``}\bibinfo{title}{Rotation of ferromagnetically
  coupled {Mn} spins in {(Ga,Mn)As} by hole spins},{''} \bibinfo{journal}{J.
  Supercond.} \textbf{\bibinfo{volume}{16}},  \bibinfo{pages}{411--414}.

\bibitem[{\citenamefont{{Munoz}} \emph{et~al.}(1995)\citenamefont{{Munoz},
  Perez, {Vina}, and Ploog}}]{Munoz1995:PRB}
\bibinfo{author}{\bibnamefont{{Munoz}}, \bibfnamefont{L.}},
  \bibinfo{author}{\bibfnamefont{E.}~\bibnamefont{Perez}},
  \bibinfo{author}{\bibfnamefont{L.}~\bibnamefont{{Vina}}}, and
  \bibinfo{author}{\bibfnamefont{K.}~\bibnamefont{Ploog}},
  \bibinfo{year}{1995}, {``}\bibinfo{title}{Spin relaxation in intrinsic {GaAs}
  quantum wells: influence of excitonic localization},{''}
  \bibinfo{journal}{Phys. Rev. B} \textbf{\bibinfo{volume}{51}},
  \bibinfo{pages}{4247--4257}.

\bibitem[{\citenamefont{Murakami} \emph{et~al.}(2003)\citenamefont{Murakami,
  Nagosa, and Zhang}}]{Murakami2003:P}
\bibinfo{author}{\bibnamefont{Murakami}, \bibfnamefont{S.}},
  \bibinfo{author}{\bibfnamefont{N.}~\bibnamefont{Nagosa}}, and
  \bibinfo{author}{\bibfnamefont{S.-C.} \bibnamefont{Zhang}},
  \bibinfo{year}{2003}, {``}\bibinfo{title}{Dissipationless quantum spin
  current at room temperature},{''} \bibinfo{journal}{{\sl Science}}
  \textbf{\bibinfo{volume}{301}},  \bibinfo{pages}{1348--1351}.

\bibitem[{\citenamefont{Myers} \emph{et~al.}(1999)\citenamefont{Myers, Ralph,
  Katine, Louie, and Buhrman}}]{Myers1999:S}
\bibinfo{author}{\bibnamefont{Myers}, \bibfnamefont{E.~B.}},
  \bibinfo{author}{\bibfnamefont{D.~C.} \bibnamefont{Ralph}},
  \bibinfo{author}{\bibfnamefont{J.~A.} \bibnamefont{Katine}},
  \bibinfo{author}{\bibfnamefont{R.~N.} \bibnamefont{Louie}}, and
  \bibinfo{author}{\bibfnamefont{R.}~\bibnamefont{Buhrman}},
  \bibinfo{year}{1999}, {``}\bibinfo{title}{Current-induced switching of
  domains in magnetic multilayer devices},{''} \bibinfo{journal}{{\sl Science}}
  \textbf{\bibinfo{volume}{285}},  \bibinfo{pages}{867--870}.

\bibitem[{\citenamefont{Nadgorny} \emph{et~al.}(2001)\citenamefont{Nadgorny,
  Mazin, Osofsky, R.~J.~Soulen, Broussard, Stroud, Singh, Harris, Arsenov, and
  Mukovskii}}]{Nadgorny2001:PRB}
\bibinfo{author}{\bibnamefont{Nadgorny}, \bibfnamefont{B.}},
  \bibinfo{author}{\bibfnamefont{I.~I.} \bibnamefont{Mazin}},
  \bibinfo{author}{\bibfnamefont{M.}~\bibnamefont{Osofsky}},
  \bibinfo{author}{\bibfnamefont{J.}~\bibnamefont{R.~J.~Soulen}},
  \bibinfo{author}{\bibfnamefont{P.}~\bibnamefont{Broussard}},
  \bibinfo{author}{\bibfnamefont{R.~M.} \bibnamefont{Stroud}},
  \bibinfo{author}{\bibfnamefont{D.~J.} \bibnamefont{Singh}},
  \bibinfo{author}{\bibfnamefont{V.~G.} \bibnamefont{Harris}},
  \bibinfo{author}{\bibfnamefont{A.}~\bibnamefont{Arsenov}}, and
  \bibinfo{author}{\bibfnamefont{Y.}~\bibnamefont{Mukovskii}},
  \bibinfo{year}{2001}, {``}\bibinfo{title}{Origin of high transport spin
  polarization in {La$_{0.7}$Sr$_{0.3}$MnO$_3$}: Direct evidence for minority
  spin states},{''} \bibinfo{journal}{Phys. Rev. B}
  \textbf{\bibinfo{volume}{63}},  \bibinfo{pages}{184433}.

\bibitem[{\citenamefont{Nadgorny} \emph{et~al.}(2000)\citenamefont{Nadgorny,
  {Soulen, Jr.}, Osofsky, Mazin, Laprade, {van de Veerdonk}, Smits, Cheng,
  Skelton, and Qadri}}]{Nadgorny2000:PRB}
\bibinfo{author}{\bibnamefont{Nadgorny}, \bibfnamefont{B.}},
  \bibinfo{author}{\bibfnamefont{R.~J.} \bibnamefont{{Soulen, Jr.}}},
  \bibinfo{author}{\bibfnamefont{M.~S.} \bibnamefont{Osofsky}},
  \bibinfo{author}{\bibfnamefont{I.~I.} \bibnamefont{Mazin}},
  \bibinfo{author}{\bibfnamefont{G.}~\bibnamefont{Laprade}},
  \bibinfo{author}{\bibfnamefont{R.~J.~M.} \bibnamefont{{van de Veerdonk}}},
  \bibinfo{author}{\bibfnamefont{A.~A.} \bibnamefont{Smits}},
  \bibinfo{author}{\bibfnamefont{S.~F.} \bibnamefont{Cheng}},
  \bibinfo{author}{\bibfnamefont{E.~F.} \bibnamefont{Skelton}}, and
  \bibinfo{author}{\bibfnamefont{S.~B.} \bibnamefont{Qadri}},
  \bibinfo{year}{2000}, {``}\bibinfo{title}{Transport spin polarization of
  {Ni$_x$Fe$_{1-x}$}: Electronic kinematics and band structure},{''}
  \bibinfo{journal}{Phys. Rev. B} \textbf{\bibinfo{volume}{61}},
  \bibinfo{pages}{R3788--R3791}.

\bibitem[{\citenamefont{Nagaev}(1983)}]{Nagaev:1983}
\bibinfo{author}{\bibnamefont{Nagaev}, \bibfnamefont{E.~L.}},
  \bibinfo{year}{1983}, \emph{\bibinfo{title}{Physics of Magnetic
  Semiconductors}} (\bibinfo{publisher}{Mir, Moscow}).

\bibitem[{\citenamefont{Nazmul} \emph{et~al.}(2003)\citenamefont{Nazmul,
  Sugahara, and Tanaka}}]{Nazmul2003:PRB}
\bibinfo{author}{\bibnamefont{Nazmul}, \bibfnamefont{A.~M.}},
  \bibinfo{author}{\bibfnamefont{S.}~\bibnamefont{Sugahara}}, and
  \bibinfo{author}{\bibfnamefont{M.}~\bibnamefont{Tanaka}},
  \bibinfo{year}{2003}, {``}\bibinfo{title}{Ferromagnetism and high {Curie}
  temperature in semiconductor heterostructureswith {Mn} delta-doped {GaAs} and
  p-type selective doping},{''} \bibinfo{journal}{Phys. Rev. B}
  \textbf{\bibinfo{volume}{67}},  \bibinfo{pages}{241308}.

\bibitem[{\citenamefont{Ngai} \emph{et~al.}(2004)\citenamefont{Ngai, Tseng,
  Morales, Wei, Chen, and Perovic}}]{Ngai2003:P}
\bibinfo{author}{\bibnamefont{Ngai}, \bibfnamefont{J.}},
  \bibinfo{author}{\bibfnamefont{Y.~C.} \bibnamefont{Tseng}},
  \bibinfo{author}{\bibfnamefont{P.}~\bibnamefont{Morales}},
  \bibinfo{author}{\bibfnamefont{J.~Y.~T.} \bibnamefont{Wei}},
  \bibinfo{author}{\bibfnamefont{F.}~\bibnamefont{Chen}}, and
  \bibinfo{author}{\bibfnamefont{D.~D.} \bibnamefont{Perovic}},
  \bibinfo{year}{2004}, {``}\bibinfo{title}{Scanning tunneling spectroscopy
  under pulsed spin injection},{''} \bibinfo{journal}{Appl. Phys. Lett.}
  \textbf{\bibinfo{volume}{84}},  \bibinfo{pages}{1907--1909}.

\bibitem[{\citenamefont{Nielsen and Chuang}(2000)}]{Nielsen:2000}
\bibinfo{author}{\bibnamefont{Nielsen}, \bibfnamefont{M.~A.}}, and
  \bibinfo{author}{\bibfnamefont{I.~L.} \bibnamefont{Chuang}},
  \bibinfo{year}{2000}, \emph{\bibinfo{title}{Quantum Computation and Quantum
  Information}} (\bibinfo{publisher}{Cambridge University, Cambridege/New
  York}).

\bibitem[{\citenamefont{Nikoli\'{c} and Allen}(1999)}]{Nikolic1999:PRB}
\bibinfo{author}{\bibnamefont{Nikoli\'{c}}, \bibfnamefont{B.}}, and
  \bibinfo{author}{\bibfnamefont{P.~B.} \bibnamefont{Allen}},
  \bibinfo{year}{1999}, {``}\bibinfo{title}{Electron transport through a
  circular constriction},{''} \bibinfo{journal}{Phys. Rev. B}
  \textbf{\bibinfo{volume}{60}},  \bibinfo{pages}{3963--3969}.

\bibitem[{\citenamefont{Nikoli\'c and Freericks}(2001)}]{Nikolic2001:P}
\bibinfo{author}{\bibnamefont{Nikoli\'c}, \bibfnamefont{B.~K.}}, and
  \bibinfo{author}{\bibfnamefont{J.~K.} \bibnamefont{Freericks}},
  \bibinfo{year}{2001}, {``}\bibinfo{title}{Mesoscopics in Spintronics:
  Conductance fluctuations of spin-polarized electrons},{''}
  \eprint{cond-mat/0111144}.

\bibitem[{\citenamefont{Nishikawa} \emph{et~al.}(1995)\citenamefont{Nishikawa,
  Tackeuchi, Nakamura, Muto, and Yokoyama}}]{Nishikawa1995:APL}
\bibinfo{author}{\bibnamefont{Nishikawa}, \bibfnamefont{Y.}},
  \bibinfo{author}{\bibfnamefont{A.}~\bibnamefont{Tackeuchi}},
  \bibinfo{author}{\bibfnamefont{S.}~\bibnamefont{Nakamura}},
  \bibinfo{author}{\bibfnamefont{S.}~\bibnamefont{Muto}}, and
  \bibinfo{author}{\bibfnamefont{N.}~\bibnamefont{Yokoyama}},
  \bibinfo{year}{1995}, {``}\bibinfo{title}{All-optical picosecond switching of
  a quantum well etalon using spin-polarization relaxation},{''}
  \bibinfo{journal}{Appl. Phys. Lett.} \textbf{\bibinfo{volume}{66}},
  \bibinfo{pages}{839--841}.

\bibitem[{\citenamefont{Nitta} \emph{et~al.}(1997)\citenamefont{Nitta, Akazaki,
  Takayanagi, and Enoki}}]{Nitta1997:PRL}
\bibinfo{author}{\bibnamefont{Nitta}, \bibfnamefont{J.}},
  \bibinfo{author}{\bibfnamefont{T.}~\bibnamefont{Akazaki}},
  \bibinfo{author}{\bibfnamefont{H.}~\bibnamefont{Takayanagi}}, and
  \bibinfo{author}{\bibfnamefont{T.}~\bibnamefont{Enoki}},
  \bibinfo{year}{1997}, {``}\bibinfo{title}{Gate control of spin-orbit
  interaction in an inverted
  {In$_{0.53}$Ga$_{0.47}$As/In$_{0.52}$Al$_{0.48}$As} heterostructure},{''}
  \bibinfo{journal}{Phys. Rev. Lett.} \textbf{\bibinfo{volume}{78}},
  \bibinfo{pages}{1335--1338}.

\bibitem[{\citenamefont{Oberli} \emph{et~al.}(1998)\citenamefont{Oberli,
  Burgermeister, Riesen, Weber, and Siegmann}}]{Oberli1998:PRL}
\bibinfo{author}{\bibnamefont{Oberli}, \bibfnamefont{D.}},
  \bibinfo{author}{\bibfnamefont{R.}~\bibnamefont{Burgermeister}},
  \bibinfo{author}{\bibfnamefont{S.}~\bibnamefont{Riesen}},
  \bibinfo{author}{\bibfnamefont{W.}~\bibnamefont{Weber}}, and
  \bibinfo{author}{\bibfnamefont{H.~C.} \bibnamefont{Siegmann}},
  \bibinfo{year}{1998}, {``}\bibinfo{title}{Total scattering cross section and
  spin motion of low energy electrons passing through a ferromagnet},{''}
  \bibinfo{journal}{Phys. Rev. Lett.} \textbf{\bibinfo{volume}{88}},
  \bibinfo{pages}{4228--4231}.

\bibitem[{\citenamefont{Oestreich} \emph{et~al.}(2002)\citenamefont{Oestreich,
  Brender, H{\"u}bner, R{\"u}hle, Klar, Heimbrodt, Lampalzer, Voltz, and
  Stolz}}]{Oestreich2002:SST}
\bibinfo{author}{\bibnamefont{Oestreich}, \bibfnamefont{M.}},
  \bibinfo{author}{\bibfnamefont{M.}~\bibnamefont{Brender}},
  \bibinfo{author}{\bibfnamefont{J.}~\bibnamefont{H{\"u}bner}},
  \bibinfo{author}{\bibfnamefont{D.~H. W.~W.} \bibnamefont{R{\"u}hle}},
  \bibinfo{author}{\bibfnamefont{T.~H. P.~J.} \bibnamefont{Klar}},
  \bibinfo{author}{\bibfnamefont{W.}~\bibnamefont{Heimbrodt}},
  \bibinfo{author}{\bibfnamefont{M.}~\bibnamefont{Lampalzer}},
  \bibinfo{author}{\bibfnamefont{K.}~\bibnamefont{Voltz}}, and
  \bibinfo{author}{\bibfnamefont{W.}~\bibnamefont{Stolz}},
  \bibinfo{year}{2002}, {``}\bibinfo{title}{Spin injection, spin transport and
  spin coherence},{''} \bibinfo{journal}{Semicond. Sci. Technol.}
  \textbf{\bibinfo{volume}{17}},  \bibinfo{pages}{285--297}.

\bibitem[{\citenamefont{Oestreich} \emph{et~al.}(1999)\citenamefont{Oestreich,
  {H\"{u}bner}, {H\"{a}gele}, Klar, Heimbrodt, {R\"{u}hle}, Ashenford, and
  Lunn}}]{Oestreich1999:APL}
\bibinfo{author}{\bibnamefont{Oestreich}, \bibfnamefont{M.}},
  \bibinfo{author}{\bibfnamefont{J.}~\bibnamefont{{H\"{u}bner}}},
  \bibinfo{author}{\bibfnamefont{D.}~\bibnamefont{{H\"{a}gele}}},
  \bibinfo{author}{\bibfnamefont{P.~J.} \bibnamefont{Klar}},
  \bibinfo{author}{\bibfnamefont{W.}~\bibnamefont{Heimbrodt}},
  \bibinfo{author}{\bibfnamefont{W.~W.} \bibnamefont{{R\"{u}hle}}},
  \bibinfo{author}{\bibfnamefont{D.~E.} \bibnamefont{Ashenford}}, and
  \bibinfo{author}{\bibfnamefont{B.}~\bibnamefont{Lunn}}, \bibinfo{year}{1999},
  {``}\bibinfo{title}{Spin injection into semiconductors},{''}
  \bibinfo{journal}{Appl. Phys. Lett.} \textbf{\bibinfo{volume}{74}},
  \bibinfo{pages}{1251--1253}.

\bibitem[{\citenamefont{Ohno}(1998)}]{Ohno1998:S}
\bibinfo{author}{\bibnamefont{Ohno}, \bibfnamefont{H.}}, \bibinfo{year}{1998},
  {``}\bibinfo{title}{Making Nonmagnetic Semiconductors Ferromagnetic},{''}
  \bibinfo{journal}{{\sl Science}} \textbf{\bibinfo{volume}{281}},
  \bibinfo{pages}{951--956}.

\bibitem[{\citenamefont{Ohno}
  \emph{et~al.}(2000{\natexlab{a}})\citenamefont{Ohno, Chiba, Matsukura, Abe,
  Dietl, Ohno, and Ohtani}}]{Ohno2000:N}
\bibinfo{author}{\bibnamefont{Ohno}, \bibfnamefont{H.}},
  \bibinfo{author}{\bibfnamefont{D.}~\bibnamefont{Chiba}},
  \bibinfo{author}{\bibfnamefont{F.}~\bibnamefont{Matsukura}},
  \bibinfo{author}{\bibfnamefont{T.~O.~E.} \bibnamefont{Abe}},
  \bibinfo{author}{\bibfnamefont{T.}~\bibnamefont{Dietl}},
  \bibinfo{author}{\bibfnamefont{Y.}~\bibnamefont{Ohno}}, and
  \bibinfo{author}{\bibfnamefont{K.}~\bibnamefont{Ohtani}},
  \bibinfo{year}{2000}{\natexlab{a}}, {``}\bibinfo{title}{Electric-field
  control of ferromagnetism},{''} \bibinfo{journal}{{\sl Nature}}
  \textbf{\bibinfo{volume}{408}},  \bibinfo{pages}{944--946}.

\bibitem[{\citenamefont{Ohno} \emph{et~al.}(1992)\citenamefont{Ohno, Munekata,
  Penney, von Moln\'{a}r, and Chang}}]{Ohno1992:PRL}
\bibinfo{author}{\bibnamefont{Ohno}, \bibfnamefont{H.}},
  \bibinfo{author}{\bibfnamefont{H.}~\bibnamefont{Munekata}},
  \bibinfo{author}{\bibfnamefont{T.}~\bibnamefont{Penney}},
  \bibinfo{author}{\bibfnamefont{S.}~\bibnamefont{von Moln\'{a}r}}, and
  \bibinfo{author}{\bibfnamefont{L.~L.} \bibnamefont{Chang}},
  \bibinfo{year}{1992}, {``}\bibinfo{title}{Magnetotransport properties of
  p-type {(In,Mn)As} diluted magnetic {III-V} semiconductors},{''}
  \bibinfo{journal}{Phys. Rev. Lett.} \textbf{\bibinfo{volume}{68}},
  \bibinfo{pages}{2664--2667}.

\bibitem[{\citenamefont{Ohno} \emph{et~al.}(1996)\citenamefont{Ohno, Shen,
  Matsukura, Oiwa, End, Katsumoto, and Iye}}]{Ohno1996:APL}
\bibinfo{author}{\bibnamefont{Ohno}, \bibfnamefont{H.}},
  \bibinfo{author}{\bibfnamefont{A.}~\bibnamefont{Shen}},
  \bibinfo{author}{\bibfnamefont{F.}~\bibnamefont{Matsukura}},
  \bibinfo{author}{\bibfnamefont{A.}~\bibnamefont{Oiwa}},
  \bibinfo{author}{\bibfnamefont{A.}~\bibnamefont{End}},
  \bibinfo{author}{\bibfnamefont{S.}~\bibnamefont{Katsumoto}}, and
  \bibinfo{author}{\bibfnamefont{Y.}~\bibnamefont{Iye}}, \bibinfo{year}{1996},
  {``}\bibinfo{title}{{(Ga,Mn)As}: A new diluted magnetic semiconductor based
  on {GaAs}},{''} \bibinfo{journal}{Appl. Phys. Lett.}
  \textbf{\bibinfo{volume}{69}},  \bibinfo{pages}{363--365}.

\bibitem[{\citenamefont{Ohno}
  \emph{et~al.}(2000{\natexlab{b}})\citenamefont{Ohno, Arata, Matsukura,
  Ohtani, Wang, and Ohno}}]{Ohno2000:ASS}
\bibinfo{author}{\bibnamefont{Ohno}, \bibfnamefont{Y.}},
  \bibinfo{author}{\bibfnamefont{I.}~\bibnamefont{Arata}},
  \bibinfo{author}{\bibfnamefont{F.}~\bibnamefont{Matsukura}},
  \bibinfo{author}{\bibfnamefont{K.}~\bibnamefont{Ohtani}},
  \bibinfo{author}{\bibfnamefont{S.}~\bibnamefont{Wang}}, and
  \bibinfo{author}{\bibfnamefont{H.}~\bibnamefont{Ohno}},
  \bibinfo{year}{2000}{\natexlab{b}}, {``}\bibinfo{title}{{MBE} growth and
  electroluminescence of ferromagnetic/non-magnetic semiconductor pn junctions
  based on {(Ga,Mn)As}},{''} \bibinfo{journal}{Appl. Surface Sc.}
  \textbf{\bibinfo{volume}{159-160}},  \bibinfo{pages}{308--312}.

\bibitem[{\citenamefont{Ohno}
  \emph{et~al.}(2000{\natexlab{c}})\citenamefont{Ohno, Terauchi, Adachi,
  Matsukura, and Ohno}}]{Ohno2000:PE}
\bibinfo{author}{\bibnamefont{Ohno}, \bibfnamefont{Y.}},
  \bibinfo{author}{\bibfnamefont{R.}~\bibnamefont{Terauchi}},
  \bibinfo{author}{\bibfnamefont{T.}~\bibnamefont{Adachi}},
  \bibinfo{author}{\bibfnamefont{F.}~\bibnamefont{Matsukura}}, and
  \bibinfo{author}{\bibfnamefont{H.}~\bibnamefont{Ohno}},
  \bibinfo{year}{2000}{\natexlab{c}}, {``}\bibinfo{title}{Electron spin
  relaxation beyond {D'yakonov-Perel'} interaction in {GaAs/AlGaAs} quantum
  wells},{''} \bibinfo{journal}{Physica E} \textbf{\bibinfo{volume}{6}},
  \bibinfo{pages}{817--820}.

\bibitem[{\citenamefont{Ohno}
  \emph{et~al.}(1999{\natexlab{a}})\citenamefont{Ohno, Terauchi, Matsukura, and
  Ohno}}]{Ohno1999:PRL}
\bibinfo{author}{\bibnamefont{Ohno}, \bibfnamefont{Y.}},
  \bibinfo{author}{\bibfnamefont{R.}~\bibnamefont{Terauchi}},
  \bibinfo{author}{\bibfnamefont{F.}~\bibnamefont{Matsukura}}, and
  \bibinfo{author}{\bibfnamefont{H.}~\bibnamefont{Ohno}},
  \bibinfo{year}{1999}{\natexlab{a}}, {``}\bibinfo{title}{Spin relaxation in
  {GaAs(110)} quantum wells},{''} \bibinfo{journal}{Phys. Rev. Lett.}
  \textbf{\bibinfo{volume}{83}},  \bibinfo{pages}{4196--4199}.

\bibitem[{\citenamefont{Ohno}
  \emph{et~al.}(1999{\natexlab{b}})\citenamefont{Ohno, Young, Beschoten,
  Matsukura, Ohno, and Awschalom}}]{Ohno1999:N}
\bibinfo{author}{\bibnamefont{Ohno}, \bibfnamefont{Y.}},
  \bibinfo{author}{\bibfnamefont{D.~K.} \bibnamefont{Young}},
  \bibinfo{author}{\bibfnamefont{B.}~\bibnamefont{Beschoten}},
  \bibinfo{author}{\bibfnamefont{F.}~\bibnamefont{Matsukura}},
  \bibinfo{author}{\bibfnamefont{H.}~\bibnamefont{Ohno}}, and
  \bibinfo{author}{\bibfnamefont{D.~D.} \bibnamefont{Awschalom}},
  \bibinfo{year}{1999}{\natexlab{b}}, {``}\bibinfo{title}{Electrical spin
  injection in a ferromagnetic semiconductor heterostructure},{''}
  \bibinfo{journal}{{\sl Nature}} \textbf{\bibinfo{volume}{402}},
  \bibinfo{pages}{790--792}.

\bibitem[{\citenamefont{Oiwa} \emph{et~al.}(2002)\citenamefont{Oiwa, Mitsumori,
  Moriya, Supinski, and Munekata}}]{Oiwa2002:PRL}
\bibinfo{author}{\bibnamefont{Oiwa}, \bibfnamefont{A.}},
  \bibinfo{author}{\bibfnamefont{Y.}~\bibnamefont{Mitsumori}},
  \bibinfo{author}{\bibfnamefont{R.}~\bibnamefont{Moriya}},
  \bibinfo{author}{\bibfnamefont{T.}~\bibnamefont{Supinski}}, and
  \bibinfo{author}{\bibfnamefont{H.}~\bibnamefont{Munekata}},
  \bibinfo{year}{2002}, {``}\bibinfo{title}{Effect of optical spin injection on
  ferromagnetically coupled {Mn} spins in the {III-V} magnetic alloy
  semiconductor {(Ga,Mn)As}},{''} \bibinfo{journal}{Phys. Rev. Lett.}
  \textbf{\bibinfo{volume}{88}},  \bibinfo{pages}{137202}.

\bibitem[{\citenamefont{Oiwa} \emph{et~al.}(2003)\citenamefont{Oiwa, Moriya,
  Mitsumori, Slupinski, and Munekata}}]{Oiwa2003:JS}
\bibinfo{author}{\bibnamefont{Oiwa}, \bibfnamefont{A.}},
  \bibinfo{author}{\bibfnamefont{R.}~\bibnamefont{Moriya}},
  \bibinfo{author}{\bibfnamefont{Y.}~\bibnamefont{Mitsumori}},
  \bibinfo{author}{\bibfnamefont{T.}~\bibnamefont{Slupinski}}, and
  \bibinfo{author}{\bibfnamefont{H.}~\bibnamefont{Munekata}},
  \bibinfo{year}{2003}, {``}\bibinfo{title}{Manifestation of local magnetic
  domain reversal by spin-polarized carrier injection in {(Ga,Mn)As} thin
  films},{''} \bibinfo{journal}{J. Supercond.} \textbf{\bibinfo{volume}{16}},
  \bibinfo{pages}{439--442}.

\bibitem[{\citenamefont{Oleinik} \emph{et~al.}(2000)\citenamefont{Oleinik, {Yu.
  Tsymbal}, and Pettifor}}]{Oleinik2000:PRB}
\bibinfo{author}{\bibnamefont{Oleinik}, \bibfnamefont{I.~I.}},
  \bibinfo{author}{\bibfnamefont{E.}~\bibnamefont{{Yu. Tsymbal}}}, and
  \bibinfo{author}{\bibfnamefont{D.~G.} \bibnamefont{Pettifor}},
  \bibinfo{year}{2000}, {``}\bibinfo{title}{Structural and electronic
  properties of {Co/Al$_2$O$_3$/Co} magnetic tunnel junction from first
  principles},{''} \bibinfo{journal}{Phys. Rev. B}
  \textbf{\bibinfo{volume}{62}},  \bibinfo{pages}{3952--3959}.

\bibitem[{\citenamefont{Olesberg} \emph{et~al.}(2001)\citenamefont{Olesberg,
  Lau, Flatt{\'e}, Yu, Altunkaya, Shaw, Hasenberg, and
  Boggess}}]{Olesberg2001:PRB}
\bibinfo{author}{\bibnamefont{Olesberg}, \bibfnamefont{J.~T.}},
  \bibinfo{author}{\bibfnamefont{W.~H.} \bibnamefont{Lau}},
  \bibinfo{author}{\bibfnamefont{M.~E.} \bibnamefont{Flatt{\'e}}},
  \bibinfo{author}{\bibfnamefont{C.}~\bibnamefont{Yu}},
  \bibinfo{author}{\bibfnamefont{E.}~\bibnamefont{Altunkaya}},
  \bibinfo{author}{\bibfnamefont{E.~M.} \bibnamefont{Shaw}},
  \bibinfo{author}{\bibfnamefont{T.~C.} \bibnamefont{Hasenberg}}, and
  \bibinfo{author}{\bibfnamefont{T.~F.} \bibnamefont{Boggess}},
  \bibinfo{year}{2001}, {``}\bibinfo{title}{Interface contributions to spin
  relaxation in a short-period {InAs/GaSb} superlattice},{''}
  \bibinfo{journal}{Phys. Rev. B} \textbf{\bibinfo{volume}{64}},
  \bibinfo{pages}{201301}.

\bibitem[{\citenamefont{Ono} \emph{et~al.}(2002)\citenamefont{Ono, Austin,
  Tokura, and Tarucha}}]{Ono2002:S}
\bibinfo{author}{\bibnamefont{Ono}, \bibfnamefont{K.}},
  \bibinfo{author}{\bibfnamefont{D.~G.} \bibnamefont{Austin}},
  \bibinfo{author}{\bibfnamefont{Y.}~\bibnamefont{Tokura}}, and
  \bibinfo{author}{\bibfnamefont{S.}~\bibnamefont{Tarucha}},
  \bibinfo{year}{2002}, {``}\bibinfo{title}{Current rectification by {Pauli}
  exclusion in a weakly coupled double quantum dot system},{''}
  \bibinfo{journal}{{\sl Science}} \textbf{\bibinfo{volume}{297}},
  \bibinfo{pages}{1313--1317}.

\bibitem[{\citenamefont{Ono} \emph{et~al.}(1996)\citenamefont{Ono, Shimada,
  Kobayashi, and Outuka}}]{Ono1996:JPSJ}
\bibinfo{author}{\bibnamefont{Ono}, \bibfnamefont{K.}},
  \bibinfo{author}{\bibfnamefont{H.}~\bibnamefont{Shimada}},
  \bibinfo{author}{\bibfnamefont{S.}~\bibnamefont{Kobayashi}}, and
  \bibinfo{author}{\bibfnamefont{Y.}~\bibnamefont{Outuka}},
  \bibinfo{year}{1996}, {``}\bibinfo{title}{Magnetoresistance of {Ni/NiO/Co}
  small tunnel junctions in {Coulomb} blockade regime},{''}
  \bibinfo{journal}{J. Phys. Soc. Jpn.} \textbf{\bibinfo{volume}{65}},
  \bibinfo{pages}{3449--3451}.

\bibitem[{\citenamefont{Orchard-Webb}
  \emph{et~al.}(1970)\citenamefont{Orchard-Webb, Watts, Smithard, and
  Cousins}}]{Orchard-Webb1970:PSS}
\bibinfo{author}{\bibnamefont{Orchard-Webb}, \bibfnamefont{J.~H.}},
  \bibinfo{author}{\bibfnamefont{A.~J.} \bibnamefont{Watts}},
  \bibinfo{author}{\bibfnamefont{M.~A.} \bibnamefont{Smithard}}, and
  \bibinfo{author}{\bibfnamefont{J.~E.} \bibnamefont{Cousins}},
  \bibinfo{year}{1970}, {``}\bibinfo{title}{The effect of concentration
  gradients of impurities on the temperature dependence of the spin-lattice
  relaxation time in lithium and beryllium},{''} \bibinfo{journal}{Phys. Stat.
  Sol.} \textbf{\bibinfo{volume}{41}},  \bibinfo{pages}{325--332}.

\bibitem[{\citenamefont{Osipov and Bratkovsky}(2003)}]{Osipov2003:P}
\bibinfo{author}{\bibnamefont{Osipov}, \bibfnamefont{V.~V.}}, and
  \bibinfo{author}{\bibfnamefont{A.~M.} \bibnamefont{Bratkovsky}},
  \bibinfo{year}{2003}, {``}\bibinfo{title}{Efficient nonlinear
  room-temperature spin tunneling-emission in ferromagnet-semiconductor
  heterostructures with extended penetration depth},{''}
  \eprint{cond-mat/0307030}.

\bibitem[{\citenamefont{Osipov} \emph{et~al.}(1990)\citenamefont{Osipov,
  Viglin, Kochev, and Samokhvalov}}]{Osipov1990:PZETF}
\bibinfo{author}{\bibnamefont{Osipov}, \bibfnamefont{V.~V.}},
  \bibinfo{author}{\bibfnamefont{N.~A.} \bibnamefont{Viglin}},
  \bibinfo{author}{\bibfnamefont{I.~V.} \bibnamefont{Kochev}}, and
  \bibinfo{author}{\bibfnamefont{A.~A.} \bibnamefont{Samokhvalov}},
  \bibinfo{year}{1990}, {``}\bibinfo{title}{Microwave-absorption at a junction
  between the ferromagnetic semiconductor {HgCr$_2$Se$_4$} and the
  semiconductor {InSb}},{''} \bibinfo{journal}{Zh. Eksp. Teor. Fiz. Pisma Red.}
  \textbf{\bibinfo{volume}{52}},  \bibinfo{pages}{996--998}
  \bibinfo{note}{[JETP Lett. {\bf 52}, 386-389 (1990)]}.

\bibitem[{\citenamefont{Osipov} \emph{et~al.}(1998)\citenamefont{Osipov,
  Viglin, and Samokhvalov}}]{Osipov1998:PL}
\bibinfo{author}{\bibnamefont{Osipov}, \bibfnamefont{V.~V.}},
  \bibinfo{author}{\bibfnamefont{N.~A.} \bibnamefont{Viglin}}, and
  \bibinfo{author}{\bibfnamefont{A.~A.} \bibnamefont{Samokhvalov}},
  \bibinfo{year}{1998}, {``}\bibinfo{title}{Investigation of heterostructure
  `ferromagnetic semiconductor-semiconductor' in the milimeter and submilimeter
  microwave range},{''} \bibinfo{journal}{Phys. Lett. A}
  \textbf{\bibinfo{volume}{247}},  \bibinfo{pages}{353--359}.

\bibitem[{\citenamefont{Oskotskij} \emph{et~al.}(1997)\citenamefont{Oskotskij,
  Subashiev, and Mamaev}}]{Oskotski1997:PLDS}
\bibinfo{author}{\bibnamefont{Oskotskij}, \bibfnamefont{B.~D.}},
  \bibinfo{author}{\bibfnamefont{A.~V.} \bibnamefont{Subashiev}}, and
  \bibinfo{author}{\bibfnamefont{Y.~A.} \bibnamefont{Mamaev}},
  \bibinfo{year}{1997}, {``}\bibinfo{title}{Polarized photoemission spectra of
  the strained semiconductor layers},{''} \bibinfo{journal}{Phys. Low-Dimens.
  Semicond. Struct.} \textbf{\bibinfo{volume}{1/2}},  \bibinfo{pages}{77--87}.

\bibitem[{\citenamefont{Osofsky}(2000)}]{Osofsky2000:JS}
\bibinfo{author}{\bibnamefont{Osofsky}, \bibfnamefont{M.}},
  \bibinfo{year}{2000}, {``}\bibinfo{title}{Spin-injection},{''}
  \bibinfo{journal}{J. Supercond.} \textbf{\bibinfo{volume}{13}},
  \bibinfo{pages}{209--219}.

\bibitem[{\citenamefont{Ott} \emph{et~al.}(2000)\citenamefont{Ott, Gavilano,
  Ambrosini, Vonlanthen, Felder, Degiorgi, Young, Fisk, and
  Zysler}}]{Ott2000:PB}
\bibinfo{author}{\bibnamefont{Ott}, \bibfnamefont{H.~R.}},
  \bibinfo{author}{\bibfnamefont{J.~L.} \bibnamefont{Gavilano}},
  \bibinfo{author}{\bibfnamefont{B.}~\bibnamefont{Ambrosini}},
  \bibinfo{author}{\bibfnamefont{P.}~\bibnamefont{Vonlanthen}},
  \bibinfo{author}{\bibfnamefont{E.}~\bibnamefont{Felder}},
  \bibinfo{author}{\bibfnamefont{L.}~\bibnamefont{Degiorgi}},
  \bibinfo{author}{\bibfnamefont{D.~P.} \bibnamefont{Young}},
  \bibinfo{author}{\bibfnamefont{Z.}~\bibnamefont{Fisk}}, and
  \bibinfo{author}{\bibfnamefont{R.}~\bibnamefont{Zysler}},
  \bibinfo{year}{2000}, {``}\bibinfo{title}{Unusual magnetism of
  hexaborides},{''} \bibinfo{journal}{Physica B}
  \textbf{\bibinfo{volume}{281/282}},  \bibinfo{pages}{423--427}.

\bibitem[{\citenamefont{Overhauser}(1953{\natexlab{a}})}]{Overhauser1953:PR}
\bibinfo{author}{\bibnamefont{Overhauser}, \bibfnamefont{A.~W.}},
  \bibinfo{year}{1953}{\natexlab{a}}, {``}\bibinfo{title}{Paramagnetic
  relaxation in metals},{''} \bibinfo{journal}{Phys. Rev.}
  \textbf{\bibinfo{volume}{89}},  \bibinfo{pages}{689--700}.

\bibitem[{\citenamefont{Overhauser}(1953{\natexlab{b}})}]{Overhauser1953b:PR}
\bibinfo{author}{\bibnamefont{Overhauser}, \bibfnamefont{A.~W.}},
  \bibinfo{year}{1953}{\natexlab{b}}, {``}\bibinfo{title}{Polarization of
  nuclei in metals},{''} \bibinfo{journal}{Phys. Rev.}
  \textbf{\bibinfo{volume}{92}},  \bibinfo{pages}{411--415}.

\bibitem[{\citenamefont{Paget and Berkovits}(1984)}]{Paget:1984}
\bibinfo{author}{\bibnamefont{Paget}, \bibfnamefont{D.}}, and
  \bibinfo{author}{\bibfnamefont{V.~L.} \bibnamefont{Berkovits}},
  \bibinfo{year}{1984}, {``}\bibinfo{title}{Optical Investigation of Hypefine
  Couplng between Electronic and Nuclear Spins},{''} in
  \emph{\bibinfo{booktitle}{Optical Orientation, Modern Problems in Condensed
  Matter Science, Vol. 8}}, edited by
  \bibinfo{editor}{\bibfnamefont{F.}~\bibnamefont{Meier}} and
  \bibinfo{editor}{\bibfnamefont{B.~P.} \bibnamefont{Zakharchenya}}
  (\bibinfo{publisher}{North-Holland, Amsterdam}),  \bibinfo{pages}{381--421}.

\bibitem[{\citenamefont{Paget} \emph{et~al.}(1977)\citenamefont{Paget, Lampel,
  Sapoval, and Safarov}}]{Paget1977:PRB}
\bibinfo{author}{\bibnamefont{Paget}, \bibfnamefont{D.}},
  \bibinfo{author}{\bibfnamefont{G.}~\bibnamefont{Lampel}},
  \bibinfo{author}{\bibfnamefont{B.}~\bibnamefont{Sapoval}}, and
  \bibinfo{author}{\bibfnamefont{V.~I.} \bibnamefont{Safarov}},
  \bibinfo{year}{1977}, {``}\bibinfo{title}{Low field electron-nuclear spin
  coupling in gallium arsenide under optical pumping conditions},{''}
  \bibinfo{journal}{Phys. Rev. B} \textbf{\bibinfo{volume}{15}},
  \bibinfo{pages}{5780--5796}.

\bibitem[{\citenamefont{Paillard} \emph{et~al.}(2001)\citenamefont{Paillard,
  Marie, Renucci, Amand, Jbeli, and Gerard}}]{Paillard2001:PRL}
\bibinfo{author}{\bibnamefont{Paillard}, \bibfnamefont{M.}},
  \bibinfo{author}{\bibfnamefont{X.}~\bibnamefont{Marie}},
  \bibinfo{author}{\bibfnamefont{P.}~\bibnamefont{Renucci}},
  \bibinfo{author}{\bibfnamefont{T.}~\bibnamefont{Amand}},
  \bibinfo{author}{\bibfnamefont{A.}~\bibnamefont{Jbeli}}, and
  \bibinfo{author}{\bibfnamefont{J.~M.} \bibnamefont{Gerard}},
  \bibinfo{year}{2001}, {``}\bibinfo{title}{Spin relaxation quenching in
  semiconductor quantum dots},{''} \bibinfo{journal}{Phys. Rev. Lett.}
  \textbf{\bibinfo{volume}{86}},  \bibinfo{pages}{1634--1637}.

\bibitem[{\citenamefont{Panguluri}
  \emph{et~al.}(2003{\natexlab{a}})\citenamefont{Panguluri, Ku, Samarth,
  Wojtowicz, Liu, Furdyna, Mazin, and Nadgorny}}]{Panguluri2003:P}
\bibinfo{author}{\bibnamefont{Panguluri}, \bibfnamefont{R.~P.}},
  \bibinfo{author}{\bibfnamefont{K.~C.} \bibnamefont{Ku}},
  \bibinfo{author}{\bibfnamefont{N.}~\bibnamefont{Samarth}},
  \bibinfo{author}{\bibfnamefont{T.}~\bibnamefont{Wojtowicz}},
  \bibinfo{author}{\bibfnamefont{X.}~\bibnamefont{Liu}},
  \bibinfo{author}{\bibfnamefont{J.~K.} \bibnamefont{Furdyna}},
  \bibinfo{author}{\bibfnamefont{I.~I.} \bibnamefont{Mazin}}, and
  \bibinfo{author}{\bibfnamefont{B.}~\bibnamefont{Nadgorny}},
  \bibinfo{year}{2003}{\natexlab{a}}, {``}\bibinfo{title}{Andreev reflection
  and pairbreaking effects at the superconductor/ferromagnetic (Ga,Mn)As
  interface},{''} \bibinfo{note}{preprint}.

\bibitem[{\citenamefont{Panguluri} \emph{et~al.}(2004)\citenamefont{Panguluri,
  Nadgorny, Wojtowicz, Lim, Liu, and Furdyna}}]{Panguluri2004:P}
\bibinfo{author}{\bibnamefont{Panguluri}, \bibfnamefont{R.~P.}},
  \bibinfo{author}{\bibfnamefont{B.}~\bibnamefont{Nadgorny}},
  \bibinfo{author}{\bibfnamefont{T.}~\bibnamefont{Wojtowicz}},
  \bibinfo{author}{\bibfnamefont{W.~L.} \bibnamefont{Lim}},
  \bibinfo{author}{\bibfnamefont{X.}~\bibnamefont{Liu}}, and
  \bibinfo{author}{\bibfnamefont{J.~K.} \bibnamefont{Furdyna}},
  \bibinfo{year}{2004}, {``}\bibinfo{title}{Measurement of spin polarization by
  {Andreev} reflection in ferromagnetic $In_{1-x}Mn_{x}Sb$ epilayers},{''}
  \eprint{cond-mat/0403451}.

\bibitem[{\citenamefont{Panguluri}
  \emph{et~al.}(2003{\natexlab{b}})\citenamefont{Panguluri, Tsoi, Nadgorny,
  Chun, Samarth, and Mazin}}]{Panguluri2003:Pb}
\bibinfo{author}{\bibnamefont{Panguluri}, \bibfnamefont{R.~P.}},
  \bibinfo{author}{\bibfnamefont{G.}~\bibnamefont{Tsoi}},
  \bibinfo{author}{\bibfnamefont{B.}~\bibnamefont{Nadgorny}},
  \bibinfo{author}{\bibfnamefont{S.~H.} \bibnamefont{Chun}},
  \bibinfo{author}{\bibfnamefont{N.}~\bibnamefont{Samarth}}, and
  \bibinfo{author}{\bibfnamefont{I.~I.} \bibnamefont{Mazin}},
  \bibinfo{year}{2003}{\natexlab{b}}, {``}\bibinfo{title}{Point contact spin
  spectroscopy of ferromagnetic {MnAs} epitaxial films},{''}
  \bibinfo{journal}{Phys. Rev. B} \textbf{\bibinfo{volume}{68}},
  \bibinfo{pages}{201307}.

\bibitem[{\citenamefont{Pankove}(1971)}]{Pankove:1971}
\bibinfo{author}{\bibnamefont{Pankove}, \bibfnamefont{J.~I.}},
  \bibinfo{year}{1971}, \emph{\bibinfo{title}{Optical Processes in
  Semiconductors}} (\bibinfo{publisher}{Prentice-Hall, Inc. Englewood Cliffs,
  N. J.}).

\bibitem[{\citenamefont{Pannetier and Courtois}(2000)}]{Pannetier2000:JLTP}
\bibinfo{author}{\bibnamefont{Pannetier}, \bibfnamefont{B.}}, and
  \bibinfo{author}{\bibfnamefont{H.}~\bibnamefont{Courtois}},
  \bibinfo{year}{2000}, {``}\bibinfo{title}{Andreev Reflection and Proximity
  Effect},{''} \bibinfo{journal}{J. Low Temp. Phys.}
  \textbf{\bibinfo{volume}{118}},  \bibinfo{pages}{599--615}.

\bibitem[{\citenamefont{Pappas} \emph{et~al.}(1991)\citenamefont{Pappas,
  K{\"a}mper, Miller, Hopster, Fowler, Brundle, and Luntz}}]{Pappas1991:PRL}
\bibinfo{author}{\bibnamefont{Pappas}, \bibfnamefont{D.~P.}},
  \bibinfo{author}{\bibfnamefont{K.}~\bibnamefont{K{\"a}mper}},
  \bibinfo{author}{\bibfnamefont{B.~P.} \bibnamefont{Miller}},
  \bibinfo{author}{\bibfnamefont{H.}~\bibnamefont{Hopster}},
  \bibinfo{author}{\bibfnamefont{D.~E.} \bibnamefont{Fowler}},
  \bibinfo{author}{\bibfnamefont{C.~R.} \bibnamefont{Brundle}}, and
  \bibinfo{author}{\bibfnamefont{A.~C.} \bibnamefont{Luntz}},
  \bibinfo{year}{1991}, {``}\bibinfo{title}{Spin-dependent electron attenuation
  by transmission through thin ferromagnetic film},{''} \bibinfo{journal}{Phys.
  Rev. Lett.} \textbf{\bibinfo{volume}{66}},  \bibinfo{pages}{504--507}.

\bibitem[{\citenamefont{Pareek and Bruno}(2002)}]{Pareek2002:PRB}
\bibinfo{author}{\bibnamefont{Pareek}, \bibfnamefont{T.~P.}}, and
  \bibinfo{author}{\bibfnamefont{P.}~\bibnamefont{Bruno}},
  \bibinfo{year}{2002}, {``}\bibinfo{title}{Spin coherence in a two-dimensional
  electron gas with {Rashba} spin-orbit interaction},{''}
  \bibinfo{journal}{Phys. Rev. B} \textbf{\bibinfo{volume}{65}},
  \bibinfo{pages}{241305}.

\bibitem[{\citenamefont{Park}
  \emph{et~al.}(1998{\natexlab{a}})\citenamefont{Park, Vescovo, Kim, Kwon,
  Ramesh, and Venkatesan}}]{Park1998:N}
\bibinfo{author}{\bibnamefont{Park}, \bibfnamefont{J.-H.}},
  \bibinfo{author}{\bibfnamefont{E.}~\bibnamefont{Vescovo}},
  \bibinfo{author}{\bibfnamefont{H.-J.} \bibnamefont{Kim}},
  \bibinfo{author}{\bibfnamefont{C.}~\bibnamefont{Kwon}},
  \bibinfo{author}{\bibfnamefont{R.}~\bibnamefont{Ramesh}}, and
  \bibinfo{author}{\bibfnamefont{T.}~\bibnamefont{Venkatesan}},
  \bibinfo{year}{1998}{\natexlab{a}}, {``}\bibinfo{title}{Direct evidence for a
  half-metallic ferromagnet},{''} \bibinfo{journal}{{\sl Nature}}
  \textbf{\bibinfo{volume}{392}},  \bibinfo{pages}{794--796}.

\bibitem[{\citenamefont{Park}
  \emph{et~al.}(1998{\natexlab{b}})\citenamefont{Park, Vescovo, Kim, Kwon,
  Ramesh, and Venkatesan}}]{Park1998:PRL}
\bibinfo{author}{\bibnamefont{Park}, \bibfnamefont{J.-H.}},
  \bibinfo{author}{\bibfnamefont{E.}~\bibnamefont{Vescovo}},
  \bibinfo{author}{\bibfnamefont{H.-J.} \bibnamefont{Kim}},
  \bibinfo{author}{\bibfnamefont{C.}~\bibnamefont{Kwon}},
  \bibinfo{author}{\bibfnamefont{R.}~\bibnamefont{Ramesh}}, and
  \bibinfo{author}{\bibfnamefont{T.}~\bibnamefont{Venkatesan}},
  \bibinfo{year}{1998}{\natexlab{b}}, {``}\bibinfo{title}{Magnetic properties
  at surface boundary of a half-metallic ferromagnet
  {La$_{0.7}$Sr$_{0.3}$MnO$_3$}},{''} \bibinfo{journal}{Phys. Rev. Lett.}
  \textbf{\bibinfo{volume}{81}},  \bibinfo{pages}{1953--1956}.

\bibitem[{\citenamefont{Park} \emph{et~al.}(2002)\citenamefont{Park, Hanbicki,
  Erwin, Hellberg, Sullivan, Mattson, Ambrose, Wilson, Spanos, and
  Jonker}}]{Park2002:S}
\bibinfo{author}{\bibnamefont{Park}, \bibfnamefont{Y.~D.}},
  \bibinfo{author}{\bibfnamefont{A.~T.} \bibnamefont{Hanbicki}},
  \bibinfo{author}{\bibfnamefont{S.~C.} \bibnamefont{Erwin}},
  \bibinfo{author}{\bibfnamefont{C.~S.} \bibnamefont{Hellberg}},
  \bibinfo{author}{\bibfnamefont{J.~M.} \bibnamefont{Sullivan}},
  \bibinfo{author}{\bibfnamefont{J.~E.} \bibnamefont{Mattson}},
  \bibinfo{author}{\bibfnamefont{T.~F.} \bibnamefont{Ambrose}},
  \bibinfo{author}{\bibfnamefont{A.}~\bibnamefont{Wilson}},
  \bibinfo{author}{\bibfnamefont{G.}~\bibnamefont{Spanos}}, and
  \bibinfo{author}{\bibfnamefont{B.~T.} \bibnamefont{Jonker}},
  \bibinfo{year}{2002}, {``}\bibinfo{title}{A group-{IV} ferromagnetic
  semiconductor: {Mn$_x$Ge$_{1-x}$}},{''} \bibinfo{journal}{{\sl Science}}
  \textbf{\bibinfo{volume}{295}},  \bibinfo{pages}{651--654}.

\bibitem[{\citenamefont{Park} \emph{et~al.}(2000)\citenamefont{Park, Jonker,
  Bennett, Itskos, Furis, Kioseoglou, and Petrou}}]{Park2000:APL}
\bibinfo{author}{\bibnamefont{Park}, \bibfnamefont{Y.~D.}},
  \bibinfo{author}{\bibfnamefont{B.~T.} \bibnamefont{Jonker}},
  \bibinfo{author}{\bibfnamefont{B.~R.} \bibnamefont{Bennett}},
  \bibinfo{author}{\bibfnamefont{G.}~\bibnamefont{Itskos}},
  \bibinfo{author}{\bibfnamefont{M.}~\bibnamefont{Furis}},
  \bibinfo{author}{\bibfnamefont{G.}~\bibnamefont{Kioseoglou}}, and
  \bibinfo{author}{\bibfnamefont{A.}~\bibnamefont{Petrou}},
  \bibinfo{year}{2000}, {``}\bibinfo{title}{Electrical spin injection across
  air-exposed epitaxially regrown semiconductor interfaces},{''}
  \bibinfo{journal}{Appl. Phys. Lett.} \textbf{\bibinfo{volume}{77}},
  \bibinfo{pages}{3989--3991}.

\bibitem[{\citenamefont{Parker} \emph{et~al.}(2002)\citenamefont{Parker, Watts,
  Ivanov, and Xiong}}]{Parker2002:PRL}
\bibinfo{author}{\bibnamefont{Parker}, \bibfnamefont{J.~S.}},
  \bibinfo{author}{\bibfnamefont{S.~M.} \bibnamefont{Watts}},
  \bibinfo{author}{\bibfnamefont{P.~G.} \bibnamefont{Ivanov}}, and
  \bibinfo{author}{\bibfnamefont{P.}~\bibnamefont{Xiong}},
  \bibinfo{year}{2002}, {``}\bibinfo{title}{Spin polarization of {CrO$_2$} at
  and across an artificial barrier},{''} \bibinfo{journal}{Phys. Rev. Lett.}
  \textbf{\bibinfo{volume}{88}},  \bibinfo{pages}{196601}.

\bibitem[{\citenamefont{Parkin}(1993)}]{Parkin1993:PRL}
\bibinfo{author}{\bibnamefont{Parkin}, \bibfnamefont{S.~S.~P.}},
  \bibinfo{year}{1993}, {``}\bibinfo{title}{Origin of enhanced
  magnetoresistance of magnetic multilayers: Spin-dependent scattering from
  magnetic interface states},{''} \bibinfo{journal}{Phys. Rev. Lett.}
  \textbf{\bibinfo{volume}{71}},  \bibinfo{pages}{1641--1644}.

\bibitem[{\citenamefont{Parkin}(2002)}]{Parkin:2002}
\bibinfo{author}{\bibnamefont{Parkin}, \bibfnamefont{S.~S.~P.}},
  \bibinfo{year}{2002}, {``}\bibinfo{title}{Applications of Magnetic
  Nanostructures},{''} in \emph{\bibinfo{booktitle}{Spin Dependent Transport in
  Magnetic Nanostructures}}, edited by
  \bibinfo{editor}{\bibfnamefont{S.}~\bibnamefont{Maekawa}} and
  \bibinfo{editor}{\bibfnamefont{T.}~\bibnamefont{Shinjo}}
  (\bibinfo{publisher}{Taylor and Francis, New York}),
  \bibinfo{pages}{237--271}.

\bibitem[{\citenamefont{Parkin}
  \emph{et~al.}(1991{\natexlab{a}})\citenamefont{Parkin, Bhadra, and
  Roche}}]{Parkin1991:PRL}
\bibinfo{author}{\bibnamefont{Parkin}, \bibfnamefont{S.~S.~P.}},
  \bibinfo{author}{\bibfnamefont{R.}~\bibnamefont{Bhadra}}, and
  \bibinfo{author}{\bibfnamefont{K.~P.} \bibnamefont{Roche}},
  \bibinfo{year}{1991}{\natexlab{a}}, {``}\bibinfo{title}{Oscillatory magnetic
  exchange coupling through thin copper layers},{''} \bibinfo{journal}{Phys.
  Rev. Lett.} \textbf{\bibinfo{volume}{66}},  \bibinfo{pages}{2152--2155}.

\bibitem[{\citenamefont{Parkin} \emph{et~al.}(2003)\citenamefont{Parkin, Jiang,
  Kaiser, Panchula, Roche, and Samant}}]{Parkin2003:PIEEE}
\bibinfo{author}{\bibnamefont{Parkin}, \bibfnamefont{S.~S.~P.}},
  \bibinfo{author}{\bibfnamefont{X.}~\bibnamefont{Jiang}},
  \bibinfo{author}{\bibfnamefont{C.}~\bibnamefont{Kaiser}},
  \bibinfo{author}{\bibfnamefont{A.}~\bibnamefont{Panchula}},
  \bibinfo{author}{\bibfnamefont{K.}~\bibnamefont{Roche}}, and
  \bibinfo{author}{\bibfnamefont{M.}~\bibnamefont{Samant}},
  \bibinfo{year}{2003}, {``}\bibinfo{title}{Magnetically engineered spintronic
  sensors and memory},{''} \bibinfo{journal}{Proc. IEEE}
  \textbf{\bibinfo{volume}{91}},  \bibinfo{pages}{661--680}.

\bibitem[{\citenamefont{Parkin} \emph{et~al.}(2004)\citenamefont{Parkin,
  Kaiser, Panchula, Rice, Samant, and Yang}}]{Parkin2004:P}
\bibinfo{author}{\bibnamefont{Parkin}, \bibfnamefont{S.~S.~P.}},
  \bibinfo{author}{\bibfnamefont{C.}~\bibnamefont{Kaiser}},
  \bibinfo{author}{\bibfnamefont{A.}~\bibnamefont{Panchula}},
  \bibinfo{author}{\bibfnamefont{P.}~\bibnamefont{Rice}},
  \bibinfo{author}{\bibfnamefont{M.}~\bibnamefont{Samant}}, and
  \bibinfo{author}{\bibfnamefont{S.-H.} \bibnamefont{Yang}},
  \bibinfo{year}{2004}, {``}\bibinfo{title}{Giant room temperature tunneling
  magnetoresistance in magnetic tunnel junctions with {MgO}(100) tunnel
  barriers},{''} \bibinfo{note}{preprint}.

\bibitem[{\citenamefont{Parkin}
  \emph{et~al.}(1991{\natexlab{b}})\citenamefont{Parkin, Li, and
  Smith}}]{Parkin1991:APL}
\bibinfo{author}{\bibnamefont{Parkin}, \bibfnamefont{S.~S.~P.}},
  \bibinfo{author}{\bibfnamefont{Z.~G.} \bibnamefont{Li}}, and
  \bibinfo{author}{\bibfnamefont{D.~J.} \bibnamefont{Smith}},
  \bibinfo{year}{1991}{\natexlab{b}}, {``}\bibinfo{title}{Giant
  magnetoresistance in antiferromagnetic {Co/Cu} multilayers},{''}
  \bibinfo{journal}{Appl. Phys. Lett.} \textbf{\bibinfo{volume}{58}},
  \bibinfo{pages}{2710--2712}.

\bibitem[{\citenamefont{Parkin} \emph{et~al.}(1999)\citenamefont{Parkin, Roche,
  Samant, Rice, Beyers, Scheuerlein, {O'Sullivan}, Brown, Bucchigano, Abraham,
  Lu, Rooks} \emph{et~al.}}]{Parkin1999:JAP}
\bibinfo{author}{\bibnamefont{Parkin}, \bibfnamefont{S.~S.~P.}},
  \bibinfo{author}{\bibfnamefont{K.~P.} \bibnamefont{Roche}},
  \bibinfo{author}{\bibfnamefont{M.~G.} \bibnamefont{Samant}},
  \bibinfo{author}{\bibfnamefont{P.~M.} \bibnamefont{Rice}},
  \bibinfo{author}{\bibfnamefont{R.~B.} \bibnamefont{Beyers}},
  \bibinfo{author}{\bibfnamefont{R.~E.} \bibnamefont{Scheuerlein}},
  \bibinfo{author}{\bibfnamefont{E.~J.} \bibnamefont{{O'Sullivan}}},
  \bibinfo{author}{\bibfnamefont{S.~L.} \bibnamefont{Brown}},
  \bibinfo{author}{\bibfnamefont{J.}~\bibnamefont{Bucchigano}},
  \bibinfo{author}{\bibfnamefont{D.~W.} \bibnamefont{Abraham}},
  \bibinfo{author}{\bibfnamefont{Y.}~\bibnamefont{Lu}},
  \bibinfo{author}{\bibfnamefont{M.}~\bibnamefont{Rooks}}, \emph{et~al.},
  \bibinfo{year}{1999}, {``}\bibinfo{title}{Exchange-biased magnetic tunnel
  junctions and application to nonvolatile magnetic random access memory},{''}
  \bibinfo{journal}{J. Appl. Phys.} \textbf{\bibinfo{volume}{85}},
  \bibinfo{pages}{5828--5833}.

\bibitem[{\citenamefont{Parsons}(1969)}]{Parsons1969:PRL}
\bibinfo{author}{\bibnamefont{Parsons}, \bibfnamefont{R.~R.}},
  \bibinfo{year}{1969}, {``}\bibinfo{title}{Band-to-band optical pumping in
  solids and polarized photoluminescence},{''} \bibinfo{journal}{Phys. Rev.
  Lett.} \textbf{\bibinfo{volume}{23}},  \bibinfo{pages}{1152--1154}.

\bibitem[{\citenamefont{Pearton} \emph{et~al.}(2003)\citenamefont{Pearton,
  Abernathy, Overberg, Thaler, Norton, Theodoropoulou, Hebard, Park, Ren, Kim,
  and Boatner}}]{Pearton2003:JAP}
\bibinfo{author}{\bibnamefont{Pearton}, \bibfnamefont{S.~J.}},
  \bibinfo{author}{\bibfnamefont{C.~R.} \bibnamefont{Abernathy}},
  \bibinfo{author}{\bibfnamefont{M.~E.} \bibnamefont{Overberg}},
  \bibinfo{author}{\bibfnamefont{G.~T.} \bibnamefont{Thaler}},
  \bibinfo{author}{\bibfnamefont{D.~P.} \bibnamefont{Norton}},
  \bibinfo{author}{\bibfnamefont{N.}~\bibnamefont{Theodoropoulou}},
  \bibinfo{author}{\bibfnamefont{A.~F.} \bibnamefont{Hebard}},
  \bibinfo{author}{\bibfnamefont{Y.~D.} \bibnamefont{Park}},
  \bibinfo{author}{\bibfnamefont{F.}~\bibnamefont{Ren}},
  \bibinfo{author}{\bibfnamefont{J.}~\bibnamefont{Kim}}, and
  \bibinfo{author}{\bibfnamefont{L.~A.} \bibnamefont{Boatner}},
  \bibinfo{year}{2003}, {``}\bibinfo{title}{Wide band gap ferromagnetic
  semiconductors and oxides},{''} \bibinfo{journal}{J. Appl. Phys.}
  \textbf{\bibinfo{volume}{93}},  \bibinfo{pages}{1--13}.

\bibitem[{\citenamefont{Pederson and {Vernon, Jr.}}(1967)}]{Pederson1967:APL}
\bibinfo{author}{\bibnamefont{Pederson}, \bibfnamefont{R.~J.}}, and
  \bibinfo{author}{\bibfnamefont{F.~L.} \bibnamefont{{Vernon, Jr.}}},
  \bibinfo{year}{1967}, {``}\bibinfo{title}{Effect of film resistance on
  low-impedance tunneling measurements},{''} \bibinfo{journal}{Appl. Phys.
  Lett.} \textbf{\bibinfo{volume}{10}},  \bibinfo{pages}{29--31}.

\bibitem[{\citenamefont{Pejakovi{\'{c}}}
  \emph{et~al.}(2002)\citenamefont{Pejakovi{\'{c}}, Kitamura, Miller, and
  Epstein}}]{Pejakovic2002:PRL}
\bibinfo{author}{\bibnamefont{Pejakovi{\'{c}}}, \bibfnamefont{D.~A.}},
  \bibinfo{author}{\bibfnamefont{C.}~\bibnamefont{Kitamura}},
  \bibinfo{author}{\bibfnamefont{J.~S.} \bibnamefont{Miller}}, and
  \bibinfo{author}{\bibfnamefont{A.~J.} \bibnamefont{Epstein}},
  \bibinfo{year}{2002}, {``}\bibinfo{title}{Photoinduced magnetization in the
  organic-based magnet {Mn(TCNE)$_x$ $\cdot$ y(CH$_2$Cl$_2$)}},{''}
  \bibinfo{journal}{Phys. Rev. Lett.} \textbf{\bibinfo{volume}{88}},
  \bibinfo{pages}{057202}.

\bibitem[{\citenamefont{Perel'} \emph{et~al.}(2003)\citenamefont{Perel',
  Tarasenko, Yassievich, Ganichev, Bel'kov, and Prettl}}]{Perel2003:PRB}
\bibinfo{author}{\bibnamefont{Perel'}, \bibfnamefont{V.~I.}},
  \bibinfo{author}{\bibfnamefont{S.~A.} \bibnamefont{Tarasenko}},
  \bibinfo{author}{\bibfnamefont{I.~N.} \bibnamefont{Yassievich}},
  \bibinfo{author}{\bibfnamefont{S.~D.} \bibnamefont{Ganichev}},
  \bibinfo{author}{\bibfnamefont{V.~V.} \bibnamefont{Bel'kov}}, and
  \bibinfo{author}{\bibfnamefont{W.}~\bibnamefont{Prettl}},
  \bibinfo{year}{2003}, {``}\bibinfo{title}{Spin-dependent tunneling through a
  symmetric semiconductor barrier},{''} \bibinfo{journal}{Phys. Rev. B}
  \textbf{\bibinfo{volume}{67}},  \bibinfo{pages}{201304}.

\bibitem[{\citenamefont{Pershin and
  Privman}(2003{\natexlab{a}})}]{Pershin2003:PRL}
\bibinfo{author}{\bibnamefont{Pershin}, \bibfnamefont{Y.~V.}}, and
  \bibinfo{author}{\bibfnamefont{V.}~\bibnamefont{Privman}},
  \bibinfo{year}{2003}{\natexlab{a}}, {``}\bibinfo{title}{Focusing of spin
  polarization in semiconductors by inhomogeneous doping},{''}
  \bibinfo{journal}{Phys. Rev. Lett.} \textbf{\bibinfo{volume}{90}},
  \bibinfo{pages}{256603}.

\bibitem[{\citenamefont{Pershin and
  Privman}(2003{\natexlab{b}})}]{Pershin2003:Pb}
\bibinfo{author}{\bibnamefont{Pershin}, \bibfnamefont{Y.~V.}}, and
  \bibinfo{author}{\bibfnamefont{V.}~\bibnamefont{Privman}},
  \bibinfo{year}{2003}{\natexlab{b}}, {``}\bibinfo{title}{Propagation of
  spin-polarized electrons through interfaces separating differently Doped
  semiconductor regions},{''} \eprint{cond-mat/0306041}.

\bibitem[{\citenamefont{Pershin and
  Privman}(2003{\natexlab{c}})}]{Pershin2003:P}
\bibinfo{author}{\bibnamefont{Pershin}, \bibfnamefont{Y.~V.}}, and
  \bibinfo{author}{\bibfnamefont{V.}~\bibnamefont{Privman}},
  \bibinfo{year}{2003}{\natexlab{c}}, {``}\bibinfo{title}{Spin relaxation of
  conduction electrons in semiconductors due to interaction with nuclear
  spins},{''} \bibinfo{journal}{Nano Lett.} \textbf{\bibinfo{volume}{3}},
  \bibinfo{pages}{695--700}.

\bibitem[{\citenamefont{Petta and Ralph}(2001)}]{Petta2001:PRL}
\bibinfo{author}{\bibnamefont{Petta}, \bibfnamefont{J.~R.}}, and
  \bibinfo{author}{\bibfnamefont{D.~C.} \bibnamefont{Ralph}},
  \bibinfo{year}{2001}, {``}\bibinfo{title}{Studies of spin-orbit scattering in
  noble-metal nanoparticles using energy-level tunneling spectroscopy},{''}
  \bibinfo{journal}{Phys. Rev. Lett.} \textbf{\bibinfo{volume}{87}},
  \bibinfo{pages}{266801}.

\bibitem[{\citenamefont{Petukhov}(1998)}]{petukhov1998:ASS}
\bibinfo{author}{\bibnamefont{Petukhov}, \bibfnamefont{A.~G.}},
  \bibinfo{year}{1998}, {``}\bibinfo{title}{Transmission coefficients of
  {GaAs/ErAs} resonant tunneling diodes in strong magnetic fields},{''}
  \bibinfo{journal}{Appl. Surf. Sci.} \textbf{\bibinfo{volume}{123}},
  \bibinfo{pages}{385--390}.

\bibitem[{\citenamefont{Petukhov} \emph{et~al.}(2002)\citenamefont{Petukhov,
  Chantis, and Demchenko}}]{Petukhov2002:PRL}
\bibinfo{author}{\bibnamefont{Petukhov}, \bibfnamefont{A.~G.}},
  \bibinfo{author}{\bibfnamefont{A.~N.} \bibnamefont{Chantis}}, and
  \bibinfo{author}{\bibfnamefont{D.~O.} \bibnamefont{Demchenko}},
  \bibinfo{year}{2002}, {``}\bibinfo{title}{Resonant enhancement of tunneling
  magnetoresistance in double-barrier Magnetic Heterostructures},{''}
  \bibinfo{journal}{Phys. Rev. Lett.} \textbf{\bibinfo{volume}{89}},
  \bibinfo{pages}{107205}.

\bibitem[{\citenamefont{Pfeffer}(1997)}]{Pfeffer1997:PRB}
\bibinfo{author}{\bibnamefont{Pfeffer}, \bibfnamefont{P.}},
  \bibinfo{year}{1997}, {``}\bibinfo{title}{Spin splitting of conduction
  energies in GaAs-Ga$_{0.7}$Al$_{0.3}$As heterostructures at $B=0$ and $B\ne
  0$ due to inversion asymmetry},{''} \bibinfo{journal}{Phys. Rev. B}
  \textbf{\bibinfo{volume}{55}},  \bibinfo{pages}{R7359--R7362}.

\bibitem[{\citenamefont{Pfeffer and Zawadzki}(1995)}]{Pfeffer1995:PRB}
\bibinfo{author}{\bibnamefont{Pfeffer}, \bibfnamefont{P.}}, and
  \bibinfo{author}{\bibfnamefont{W.}~\bibnamefont{Zawadzki}},
  \bibinfo{year}{1995}, {``}\bibinfo{title}{Spin splitting of conduction
  subbands in GaAs-Ga$_{0.7}$Al$_{0.3}$As heterostructures},{''}
  \bibinfo{journal}{Phys. Rev. B} \textbf{\bibinfo{volume}{52}},
  \bibinfo{pages}{R14332--R14335}.

\bibitem[{\citenamefont{Pfeffer and Zawadzki}(1999)}]{Pfeffer1999:PRB}
\bibinfo{author}{\bibnamefont{Pfeffer}, \bibfnamefont{P.}}, and
  \bibinfo{author}{\bibfnamefont{W.}~\bibnamefont{Zawadzki}},
  \bibinfo{year}{1999}, {``}\bibinfo{title}{Spin splitting of conduction
  subbands in III-V heterostructures due to inversion asymmetry},{''}
  \bibinfo{journal}{Phys. Rev. B} \textbf{\bibinfo{volume}{59}},
  \bibinfo{pages}{R5312--R5315}.

\bibitem[{\citenamefont{Pfeffer and Zawadzki}(2003)}]{Pfeffer2003:PRB}
\bibinfo{author}{\bibnamefont{Pfeffer}, \bibfnamefont{P.}}, and
  \bibinfo{author}{\bibfnamefont{W.}~\bibnamefont{Zawadzki}},
  \bibinfo{year}{2003}, {``}\bibinfo{title}{{Bychkov-Rashba} spin splitting and
  its dependence on magnetic field in {InSb/In$_{0.91}$Al$_{0.09}$Sb}
  asymmetric quantum wells},{''} \bibinfo{journal}{Phys. Rev. B}
  \textbf{\bibinfo{volume}{68}},  \bibinfo{pages}{035315}.

\bibitem[{\citenamefont{Pickett and Moodera}(2001)}]{Pickett2001:PT}
\bibinfo{author}{\bibnamefont{Pickett}, \bibfnamefont{W.~E.}}, and
  \bibinfo{author}{\bibfnamefont{J.~S.} \bibnamefont{Moodera}},
  \bibinfo{year}{2001}, {``}\bibinfo{title}{Half metallic magnets},{''}
  \bibinfo{journal}{Phys. Today} \textbf{\bibinfo{volume}{54 (5)}},
  \bibinfo{pages}{39--44}.

\bibitem[{\citenamefont{Pierce and Celotta}(1984)}]{Pierce:1984}
\bibinfo{author}{\bibnamefont{Pierce}, \bibfnamefont{D.~T.}}, and
  \bibinfo{author}{\bibfnamefont{R.~J.} \bibnamefont{Celotta}},
  \bibinfo{year}{1984}, {``}\bibinfo{title}{Applications of Polarized Electron
  Sources Utilizing Optical Orinetation in Solids},{''} in
  \emph{\bibinfo{booktitle}{Optical Orientation, Modern Problems in Condensed
  Matter Science, Vol. 8}}, edited by
  \bibinfo{editor}{\bibfnamefont{F.}~\bibnamefont{Meier}} and
  \bibinfo{editor}{\bibfnamefont{B.~P.} \bibnamefont{Zakharchenya}}
  (\bibinfo{publisher}{North-Holland, Amsterdam}),  \bibinfo{pages}{259--294}.

\bibitem[{\citenamefont{Pierce} \emph{et~al.}(1982)\citenamefont{Pierce,
  Celotta, Unguris, and Siegmann}}]{Pierce1982:PRB}
\bibinfo{author}{\bibnamefont{Pierce}, \bibfnamefont{D.~T.}},
  \bibinfo{author}{\bibfnamefont{R.~J.} \bibnamefont{Celotta}},
  \bibinfo{author}{\bibfnamefont{J.}~\bibnamefont{Unguris}}, and
  \bibinfo{author}{\bibfnamefont{H.~C.} \bibnamefont{Siegmann}},
  \bibinfo{year}{1982}, {``}\bibinfo{title}{Spin-dependent elastic scattering
  of electrons from a ferromagnetic glass, {Ni$_{40}$Fe$_{40}$B$_{20}$}},{''}
  \bibinfo{journal}{Phys. Rev. B} \textbf{\bibinfo{volume}{26}},
  \bibinfo{pages}{2566--2574}.

\bibitem[{\citenamefont{Pierce and Meier}(1976)}]{Pierce1976:PRB}
\bibinfo{author}{\bibnamefont{Pierce}, \bibfnamefont{D.~T.}}, and
  \bibinfo{author}{\bibfnamefont{F.}~\bibnamefont{Meier}},
  \bibinfo{year}{1976}, {``}\bibinfo{title}{Photoemission of spin-polarized
  electrons from {GaAs}},{''} \bibinfo{journal}{Phys. Rev. B}
  \textbf{\bibinfo{volume}{13}},  \bibinfo{pages}{5484--5500}.

\bibitem[{\citenamefont{Piermarocchi}
  \emph{et~al.}(2002)\citenamefont{Piermarocchi, Chen, Sham, and
  Steel}}]{Piermarocchi2002:PRL}
\bibinfo{author}{\bibnamefont{Piermarocchi}, \bibfnamefont{C.}},
  \bibinfo{author}{\bibfnamefont{P.}~\bibnamefont{Chen}},
  \bibinfo{author}{\bibfnamefont{L.~J.} \bibnamefont{Sham}}, and
  \bibinfo{author}{\bibfnamefont{D.~G.} \bibnamefont{Steel}},
  \bibinfo{year}{2002}, {``}\bibinfo{title}{Optical {RKKY} interaction between
  charged semiconductor quantum dots},{''} \bibinfo{journal}{Phys. Rev. Lett.}
  \textbf{\bibinfo{volume}{89}},  \bibinfo{pages}{167402}.

\bibitem[{\citenamefont{Pikus and Pikus}(1995)}]{Pikus1995:PRB}
\bibinfo{author}{\bibnamefont{Pikus}, \bibfnamefont{F.~G.}}, and
  \bibinfo{author}{\bibfnamefont{G.~E.} \bibnamefont{Pikus}},
  \bibinfo{year}{1995}, {``}\bibinfo{title}{Conduction-band spin splitting and
  negative magnetoresistance in {A$_3$B$_5$} heterostructures},{''}
  \bibinfo{journal}{Phys. Rev. B} \textbf{\bibinfo{volume}{51}},
  \bibinfo{pages}{16928--16935}.

\bibitem[{\citenamefont{Pikus and Titkov}(1984)}]{Pikus:1984}
\bibinfo{author}{\bibnamefont{Pikus}, \bibfnamefont{G.~E.}}, and
  \bibinfo{author}{\bibfnamefont{A.~N.} \bibnamefont{Titkov}},
  \bibinfo{year}{1984}, {``}\bibinfo{title}{Spin Relaxation under Optical
  Orientation in Semiconductors},{''} in \emph{\bibinfo{booktitle}{Optical
  Orientation, Modern Problems in Condensed Matter Science, Vol. 8}}, edited by
  \bibinfo{editor}{\bibfnamefont{F.}~\bibnamefont{Meier}} and
  \bibinfo{editor}{\bibfnamefont{B.~P.} \bibnamefont{Zakharchenya}}
  (\bibinfo{publisher}{North-Holland, Amsterdam}),  \bibinfo{pages}{109}.

\bibitem[{\citenamefont{Pines and Slichter}(1955)}]{Pines1955:PR}
\bibinfo{author}{\bibnamefont{Pines}, \bibfnamefont{D.}}, and
  \bibinfo{author}{\bibfnamefont{C.~P.} \bibnamefont{Slichter}},
  \bibinfo{year}{1955}, {``}\bibinfo{title}{Relaxation times in magnetic
  resonance},{''} \bibinfo{journal}{Phys. Rev.} \textbf{\bibinfo{volume}{100}},
   \bibinfo{pages}{1014--1020}.

\bibitem[{\citenamefont{Pohm} \emph{et~al.}(1987)\citenamefont{Pohm, Daughton,
  Comstock, Yoo, and Hur}}]{Pohm1987:IEEETM}
\bibinfo{author}{\bibnamefont{Pohm}, \bibfnamefont{A.~V.}},
  \bibinfo{author}{\bibfnamefont{J.~M.} \bibnamefont{Daughton}},
  \bibinfo{author}{\bibfnamefont{C.~S.} \bibnamefont{Comstock}},
  \bibinfo{author}{\bibfnamefont{H.~Y.} \bibnamefont{Yoo}}, and
  \bibinfo{author}{\bibfnamefont{J.}~\bibnamefont{Hur}}, \bibinfo{year}{1987},
  {``}\bibinfo{title}{Threshold properties of 1, 2 and 4 $\mu$m multilayer
  magento-resistive memory cells},{''} \bibinfo{journal}{IEEE Trans. Magn.}
  \textbf{\bibinfo{volume}{23}},  \bibinfo{pages}{2575--2577}.

\bibitem[{\citenamefont{Pohm} \emph{et~al.}(1988)\citenamefont{Pohm, Huang,
  Daughton, Krahn, and Mehra}}]{Pohm1988:IEEETM}
\bibinfo{author}{\bibnamefont{Pohm}, \bibfnamefont{A.~V.}},
  \bibinfo{author}{\bibfnamefont{J.~S.~T.} \bibnamefont{Huang}},
  \bibinfo{author}{\bibfnamefont{J.~M.} \bibnamefont{Daughton}},
  \bibinfo{author}{\bibfnamefont{D.~R.} \bibnamefont{Krahn}}, and
  \bibinfo{author}{\bibfnamefont{V.}~\bibnamefont{Mehra}},
  \bibinfo{year}{1988}, {``}\bibinfo{title}{The design of a one megabit
  non-nolatile {M-R} memory chip using 1.5 $\times$ 5 $\mu$m cells},{''}
  \bibinfo{journal}{IEEE Trans. Magn.} \textbf{\bibinfo{volume}{24}},
  \bibinfo{pages}{3117--3119}.

\bibitem[{\citenamefont{Potok} \emph{et~al.}(2002)\citenamefont{Potok, Folk,
  Marcus, and Umansky}}]{Potok2002:PRL}
\bibinfo{author}{\bibnamefont{Potok}, \bibfnamefont{R.~M.}},
  \bibinfo{author}{\bibfnamefont{J.~A.} \bibnamefont{Folk}},
  \bibinfo{author}{\bibfnamefont{C.~M.} \bibnamefont{Marcus}}, and
  \bibinfo{author}{\bibfnamefont{V.}~\bibnamefont{Umansky}},
  \bibinfo{year}{2002}, {``}\bibinfo{title}{Detecting spin-polarized currents
  in ballistic nanostructures},{''} \bibinfo{journal}{Phys. Rev. Lett.}
  \textbf{\bibinfo{volume}{89}},  \bibinfo{pages}{266602}.

\bibitem[{\citenamefont{{Pratt, Jr.}} \emph{et~al.}(1991)\citenamefont{{Pratt,
  Jr.}, Lee, Slaughter, Loloee, Schroeder, and Bass}}]{Pratt1991:PRL}
\bibinfo{author}{\bibnamefont{{Pratt, Jr.}}, \bibfnamefont{W.~P.}},
  \bibinfo{author}{\bibfnamefont{S.-F.} \bibnamefont{Lee}},
  \bibinfo{author}{\bibfnamefont{J.~M.} \bibnamefont{Slaughter}},
  \bibinfo{author}{\bibfnamefont{R.}~\bibnamefont{Loloee}},
  \bibinfo{author}{\bibfnamefont{P.~A.} \bibnamefont{Schroeder}}, and
  \bibinfo{author}{\bibfnamefont{J.}~\bibnamefont{Bass}}, \bibinfo{year}{1991},
  {``}\bibinfo{title}{Perpendicular giant magnetoresistances of {Ag/Co}
  multilayers},{''} \bibinfo{journal}{Phys. Rev. Lett.}
  \textbf{\bibinfo{volume}{66}},  \bibinfo{pages}{3060--3063}.

\bibitem[{\citenamefont{Preskill}(1998)}]{Preskill1998:PRSL}
\bibinfo{author}{\bibnamefont{Preskill}, \bibfnamefont{J.}},
  \bibinfo{year}{1998}, {``}\bibinfo{title}{Reliable quantum computers},{''}
  \bibinfo{journal}{Proc. R. Soc. London, Ser. A}
  \textbf{\bibinfo{volume}{454}},  \bibinfo{pages}{385--410}.

\bibitem[{\citenamefont{Prins} \emph{et~al.}(1995)\citenamefont{Prins, {van
  Kempen}, {van Leuken}, {de Groot}, {van Roy}, and {De
  Boeck}}}]{Prins1995:JPCM}
\bibinfo{author}{\bibnamefont{Prins}, \bibfnamefont{M.~W.~J.}},
  \bibinfo{author}{\bibfnamefont{H.}~\bibnamefont{{van Kempen}}},
  \bibinfo{author}{\bibfnamefont{H.}~\bibnamefont{{van Leuken}}},
  \bibinfo{author}{\bibfnamefont{R.~A.} \bibnamefont{{de Groot}}},
  \bibinfo{author}{\bibfnamefont{W.}~\bibnamefont{{van Roy}}}, and
  \bibinfo{author}{\bibfnamefont{J.}~\bibnamefont{{De Boeck}}},
  \bibinfo{year}{1995}, {``}\bibinfo{title}{Spin-dependent transport in
  metal/semiconductor tunnel junctions},{''} \bibinfo{journal}{J. Phys.:
  Condens. Matter} \textbf{\bibinfo{volume}{7}},  \bibinfo{pages}{9447--9464}.

\bibitem[{\citenamefont{Prinz}(1995)}]{Prinz1995:PT}
\bibinfo{author}{\bibnamefont{Prinz}, \bibfnamefont{G.}}, \bibinfo{year}{1995},
  {``}\bibinfo{title}{Spin-polarized transport},{''} \bibinfo{journal}{Phys.
  Today} \textbf{\bibinfo{volume}{48 (4)}},  \bibinfo{pages}{58--63}.

\bibitem[{\citenamefont{Prinz}(1998)}]{Prinz1998:S}
\bibinfo{author}{\bibnamefont{Prinz}, \bibfnamefont{G.}}, \bibinfo{year}{1998},
  {``}\bibinfo{title}{Magnetoelectronics},{''} \bibinfo{journal}{{\sl Science}}
  \textbf{\bibinfo{volume}{282}},  \bibinfo{pages}{1660--1663}.

\bibitem[{\citenamefont{Privman} \emph{et~al.}(1998)\citenamefont{Privman,
  Vagner, and Kventsel}}]{Privman1998:PL}
\bibinfo{author}{\bibnamefont{Privman}, \bibfnamefont{V.}},
  \bibinfo{author}{\bibfnamefont{I.~D.} \bibnamefont{Vagner}}, and
  \bibinfo{author}{\bibfnamefont{G.}~\bibnamefont{Kventsel}},
  \bibinfo{year}{1998}, {``}\bibinfo{title}{Quantum computation in
  quantum-{Hall} systems},{''} \bibinfo{journal}{Phys. Lett. A}
  \textbf{\bibinfo{volume}{239}},  \bibinfo{pages}{141--146}.

\bibitem[{\citenamefont{Puller} \emph{et~al.}(2003)\citenamefont{Puller,
  Mourokh, and Horing}}]{Puller2002:P}
\bibinfo{author}{\bibnamefont{Puller}, \bibfnamefont{V.~I.}},
  \bibinfo{author}{\bibfnamefont{L.~G.} \bibnamefont{Mourokh}}, and
  \bibinfo{author}{\bibfnamefont{N.~J.~M.} \bibnamefont{Horing}},
  \bibinfo{year}{2003}, {``}\bibinfo{title}{Electron spin relaxation in a
  semiconductor quantum well},{''} \bibinfo{journal}{Phys. Rev. B}
  \textbf{\bibinfo{volume}{67}},  \bibinfo{pages}{155309}.

\bibitem[{\citenamefont{Qi and Zhang}(2003)}]{Qi2003:PRB}
\bibinfo{author}{\bibnamefont{Qi}, \bibfnamefont{Y.}}, and
  \bibinfo{author}{\bibfnamefont{S.}~\bibnamefont{Zhang}},
  \bibinfo{year}{2003}, {``}\bibinfo{title}{Spin diffusion at finite electric
  and magnetic fields},{''} \bibinfo{journal}{Phys. Rev. B}
  \textbf{\bibinfo{volume}{67}},  \bibinfo{pages}{052407}.

\bibitem[{\citenamefont{Rammer and Smith}(1986)}]{Rammer1986:RMP}
\bibinfo{author}{\bibnamefont{Rammer}, \bibfnamefont{J.}}, and
  \bibinfo{author}{\bibfnamefont{H.}~\bibnamefont{Smith}},
  \bibinfo{year}{1986}, {``}\bibinfo{title}{Quantum field-theoretical methods
  in transport theory of metals},{''} \bibinfo{journal}{Rev. Mod. Phys.}
  \textbf{\bibinfo{volume}{58}},  \bibinfo{pages}{323--359}.

\bibitem[{\citenamefont{Ramsteiner}
  \emph{et~al.}(2002)\citenamefont{Ramsteiner, Hao, Kawaharazuka, Zhu,
  K{\"a}stner, Hey, D{\"a}weritz, Grahn, and Ploog}}]{Ramsteiner2002:PRB}
\bibinfo{author}{\bibnamefont{Ramsteiner}, \bibfnamefont{M.}},
  \bibinfo{author}{\bibfnamefont{H.~Y.} \bibnamefont{Hao}},
  \bibinfo{author}{\bibfnamefont{A.}~\bibnamefont{Kawaharazuka}},
  \bibinfo{author}{\bibfnamefont{H.~J.} \bibnamefont{Zhu}},
  \bibinfo{author}{\bibfnamefont{M.}~\bibnamefont{K{\"a}stner}},
  \bibinfo{author}{\bibfnamefont{R.}~\bibnamefont{Hey}},
  \bibinfo{author}{\bibfnamefont{L.}~\bibnamefont{D{\"a}weritz}},
  \bibinfo{author}{\bibfnamefont{H.~T.} \bibnamefont{Grahn}}, and
  \bibinfo{author}{\bibfnamefont{K.~H.} \bibnamefont{Ploog}},
  \bibinfo{year}{2002}, {``}\bibinfo{title}{Electrical spin injection from
  ferromagnetic {MnAs} metal layers into {GaAs}},{''} \bibinfo{journal}{Phys.
  Rev. B} \textbf{\bibinfo{volume}{66}},  \bibinfo{pages}{081304}.

\bibitem[{\citenamefont{Rashba}(1960)}]{Rashba1960:SPSS}
\bibinfo{author}{\bibnamefont{Rashba}, \bibfnamefont{E.~I.}},
  \bibinfo{year}{1960}, {``}\bibinfo{title}{Properties of semiconductors with
  an extremum loop {I}. {Cyclotron} and combinational resonance in a magnetic
  field perpendicular to the plane of the loop},{''} \bibinfo{journal}{Fiz.
  Tverd. Tela} \textbf{\bibinfo{volume}{2}},  \bibinfo{pages}{1109--1122}
  \bibinfo{note}{[Sov. Phys. Solid State {\bf 2}, 1224-1238 (1960)]}.

\bibitem[{\citenamefont{Rashba}(2000)}]{Rashba2000:PRB}
\bibinfo{author}{\bibnamefont{Rashba}, \bibfnamefont{E.~I.}},
  \bibinfo{year}{2000}, {``}\bibinfo{title}{Theory of electrical spin
  injection: {Tunnel} contacts as a solution of the conductivity mismatch
  problem},{''} \bibinfo{journal}{Phys. Rev. B} \textbf{\bibinfo{volume}{62}},
  \bibinfo{pages}{R16267--R16270}.

\bibitem[{\citenamefont{Rashba}(2002{\natexlab{a}})}]{Rashba2002:PC}
\bibinfo{author}{\bibnamefont{Rashba}, \bibfnamefont{E.~I.}},
  \bibinfo{year}{2002}{\natexlab{a}} \bibinfo{journal}{private communication} .

\bibitem[{\citenamefont{Rashba}(2002{\natexlab{b}})}]{Rashba2002:APL}
\bibinfo{author}{\bibnamefont{Rashba}, \bibfnamefont{E.~I.}},
  \bibinfo{year}{2002}{\natexlab{b}}, {``}\bibinfo{title}{Complex impedance of
  a spin injecting junction},{''} \bibinfo{journal}{Appl. Phys. Lett.}
  \textbf{\bibinfo{volume}{80}},  \bibinfo{pages}{2329--2331}.

\bibitem[{\citenamefont{Rashba}(2002{\natexlab{c}})}]{Rashba2002:EPJ}
\bibinfo{author}{\bibnamefont{Rashba}, \bibfnamefont{E.~I.}},
  \bibinfo{year}{2002}{\natexlab{c}}, {``}\bibinfo{title}{Diffusion theory of
  spin injection through resisitive contacts},{''} \bibinfo{journal}{Eur. Phys.
  J. B} \textbf{\bibinfo{volume}{29}},  \bibinfo{pages}{513--527}.

\bibitem[{\citenamefont{Rashba}(2002{\natexlab{d}})}]{Rashba2002:JS}
\bibinfo{author}{\bibnamefont{Rashba}, \bibfnamefont{E.~I.}},
  \bibinfo{year}{2002}{\natexlab{d}}, {``}\bibinfo{title}{Spintronics:
  {Sources} and challenge},{''} \bibinfo{journal}{J. Supercond.}
  \textbf{\bibinfo{volume}{15}},  \bibinfo{pages}{13--17}.

\bibitem[{\citenamefont{Rashba}(2003{\natexlab{a}})}]{Rashba2003:P}
\bibinfo{author}{\bibnamefont{Rashba}, \bibfnamefont{E.~I.}},
  \bibinfo{year}{2003}{\natexlab{a}}, {``}\bibinfo{title}{Inelastic scattering
  approach to the theory of a tunnel magnetic transistor source},{''}
  \bibinfo{journal}{Phys. Rev. B} \textbf{\bibinfo{volume}{68}},
  \bibinfo{pages}{241310}.

\bibitem[{\citenamefont{Rashba}(2003{\natexlab{b}})}]{Rashba2003:PRB}
\bibinfo{author}{\bibnamefont{Rashba}, \bibfnamefont{E.~I.}},
  \bibinfo{year}{2003}{\natexlab{b}}, {``}\bibinfo{title}{Spin currents in
  theromdynamic equilibrium: {The} challenge of discerning transport
  currents},{''} \bibinfo{journal}{Phys. Rev. B} \textbf{\bibinfo{volume}{68}},
   \bibinfo{pages}{241315}.

\bibitem[{\citenamefont{Rashba and Efros}(2003)}]{Rashba2003:PRL}
\bibinfo{author}{\bibnamefont{Rashba}, \bibfnamefont{E.~I.}}, and
  \bibinfo{author}{\bibfnamefont{A.~L.} \bibnamefont{Efros}},
  \bibinfo{year}{2003}, {``}\bibinfo{title}{Orbital mechanisms of electron-spin
  manipulation by an electric field},{''} \bibinfo{journal}{Phys. Rev. Lett.}
  \textbf{\bibinfo{volume}{91}},  \bibinfo{pages}{126405}.

\bibitem[{\citenamefont{Rashba and Sheka}(1961)}]{Rashba1961:SPSS}
\bibinfo{author}{\bibnamefont{Rashba}, \bibfnamefont{E.~I.}}, and
  \bibinfo{author}{\bibfnamefont{V.~I.} \bibnamefont{Sheka}},
  \bibinfo{year}{1961}, {``}\bibinfo{title}{Combinatorial resonance of zonal
  electrons in crystals having a zinc blende lattice},{''}
  \bibinfo{journal}{Fiz. Tverd. Tela} \textbf{\bibinfo{volume}{3}},
  \bibinfo{pages}{1735--1749} \bibinfo{note}{[Sov. Phys. Solid State {\bf 3},
  1257-1267 (1961)]}.

\bibitem[{\citenamefont{Rashba and Sheka}(1991)}]{Rashba:1991}
\bibinfo{author}{\bibnamefont{Rashba}, \bibfnamefont{E.~I.}}, and
  \bibinfo{author}{\bibfnamefont{V.~I.} \bibnamefont{Sheka}},
  \bibinfo{year}{1991}, {``}\bibinfo{title}{Electric-Dipole Spin
  Resonances},{''} in \emph{\bibinfo{booktitle}{Landau Level Spectroscopy,
  Modern Problems in Condensed Matter Science, Vol. 27}}, edited by
  \bibinfo{editor}{\bibfnamefont{G.}~\bibnamefont{Landwehr}} and
  \bibinfo{editor}{\bibfnamefont{E.~I.} \bibnamefont{Rashba}}
  (\bibinfo{publisher}{North-Holland, Amsterdam}),  \bibinfo{pages}{131--206}.

\bibitem[{\citenamefont{Rashba and Sherman}(1988)}]{Rashba1988:PLA}
\bibinfo{author}{\bibnamefont{Rashba}, \bibfnamefont{E.~I.}}, and
  \bibinfo{author}{\bibfnamefont{E.~Y.} \bibnamefont{Sherman}},
  \bibinfo{year}{1988}, {``}\bibinfo{title}{Spin-orbital band splitting in
  symmetric quantum wells},{''} \bibinfo{journal}{Phys. Lett. A}
  \textbf{\bibinfo{volume}{129}},  \bibinfo{pages}{175--179}.

\bibitem[{\citenamefont{Recher} \emph{et~al.}(2000)\citenamefont{Recher,
  Sukhorukov, and Loss}}]{Recher2000:PRL}
\bibinfo{author}{\bibnamefont{Recher}, \bibfnamefont{P.}},
  \bibinfo{author}{\bibfnamefont{E.~V.} \bibnamefont{Sukhorukov}}, and
  \bibinfo{author}{\bibfnamefont{D.}~\bibnamefont{Loss}}, \bibinfo{year}{2000},
  {``}\bibinfo{title}{Quantum dot as spin filter and spin memory},{''}
  \bibinfo{journal}{Phys. Rev. Lett.} \textbf{\bibinfo{volume}{85}},
  \bibinfo{pages}{1962--1965}.

\bibitem[{\citenamefont{Rendell and Penn}(1980)}]{Rendell1980:PRL}
\bibinfo{author}{\bibnamefont{Rendell}, \bibfnamefont{R.~W.}}, and
  \bibinfo{author}{\bibfnamefont{D.~R.} \bibnamefont{Penn}},
  \bibinfo{year}{1980}, {``}\bibinfo{title}{Spin dependence of the electron
  mean free path in {Fe}, {Co}, and {Ni}},{''} \bibinfo{journal}{Phys. Rev.
  Lett.} \textbf{\bibinfo{volume}{45}},  \bibinfo{pages}{2057--2060}.

\bibitem[{\citenamefont{Rippard and Buhrman}(1999)}]{Rippard1999:APL}
\bibinfo{author}{\bibnamefont{Rippard}, \bibfnamefont{W.~H.}}, and
  \bibinfo{author}{\bibfnamefont{R.~A.} \bibnamefont{Buhrman}},
  \bibinfo{year}{1999}, {``}\bibinfo{title}{Ballistic electron magnetic
  microscopy: Imaging magnetic domains with nanometer resolution},{''}
  \bibinfo{journal}{Appl. Phys. Lett.} \textbf{\bibinfo{volume}{84}},
  \bibinfo{pages}{1001--1003}.

\bibitem[{\citenamefont{Rippard and Buhrman}(2000)}]{Rippard2000:PRL}
\bibinfo{author}{\bibnamefont{Rippard}, \bibfnamefont{W.~H.}}, and
  \bibinfo{author}{\bibfnamefont{R.~A.} \bibnamefont{Buhrman}},
  \bibinfo{year}{2000}, {``}\bibinfo{title}{Spin-dependent hot electron
  transport in {Co/Cu} thin films},{''} \bibinfo{journal}{Phys. Rev. Lett.}
  \textbf{\bibinfo{volume}{84}},  \bibinfo{pages}{971--974}.

\bibitem[{\citenamefont{Rippard} \emph{et~al.}(2002)\citenamefont{Rippard,
  Perrella, Albert, and Buhrman}}]{Rippard2002:PRL}
\bibinfo{author}{\bibnamefont{Rippard}, \bibfnamefont{W.~H.}},
  \bibinfo{author}{\bibfnamefont{A.~C.} \bibnamefont{Perrella}},
  \bibinfo{author}{\bibfnamefont{F.~J.} \bibnamefont{Albert}}, and
  \bibinfo{author}{\bibfnamefont{R.~A.} \bibnamefont{Buhrman}},
  \bibinfo{year}{2002}, {``}\bibinfo{title}{Ultrathin aluminum oxide tunnel
  barriers},{''} \bibinfo{journal}{Phys. Rev. Lett.}
  \textbf{\bibinfo{volume}{88}},  \bibinfo{pages}{046805}.

\bibitem[{\citenamefont{Rippard} \emph{et~al.}(2004)\citenamefont{Rippard,
  Pufall, Kaka, Russek, and Silva}}]{Rippard2003:P}
\bibinfo{author}{\bibnamefont{Rippard}, \bibfnamefont{W.~H.}},
  \bibinfo{author}{\bibfnamefont{M.~R.} \bibnamefont{Pufall}},
  \bibinfo{author}{\bibfnamefont{S.}~\bibnamefont{Kaka}},
  \bibinfo{author}{\bibfnamefont{S.~E.} \bibnamefont{Russek}}, and
  \bibinfo{author}{\bibfnamefont{T.~J.} \bibnamefont{Silva}},
  \bibinfo{year}{2004}, {``}\bibinfo{title}{Direct-current induced yynamics in
  {Co$_{90}$Fe$_{10}$/Ni$_{80}$Fe$_{20}$} point contacts},{''}
  \bibinfo{journal}{Phys. Rev. Lett.} \textbf{\bibinfo{volume}{92}},
  \bibinfo{pages}{027201}.

\bibitem[{\citenamefont{Rowe} \emph{et~al.}(2001)\citenamefont{Rowe, Nehls,
  Stradling, and Ferguson}}]{Rowe2001:PRB}
\bibinfo{author}{\bibnamefont{Rowe}, \bibfnamefont{A.~C.}},
  \bibinfo{author}{\bibfnamefont{J.}~\bibnamefont{Nehls}},
  \bibinfo{author}{\bibfnamefont{R.~A.} \bibnamefont{Stradling}}, and
  \bibinfo{author}{\bibfnamefont{R.~S.} \bibnamefont{Ferguson}},
  \bibinfo{year}{2001}, {``}\bibinfo{title}{Origin of beat patterns in the
  quantum magnetoresistance of gated {InAs/GaSb} and {InAs/AlSb} quantum
  wells},{''} \bibinfo{journal}{Phys. Rev. B} \textbf{\bibinfo{volume}{63}},
  \bibinfo{pages}{201307}.

\bibitem[{\citenamefont{Rudolph} \emph{et~al.}(2003)\citenamefont{Rudolph,
  {H\"{a}gele}, Gibbs, Khitrova, and Oestreich}}]{Rudolph2003:APL}
\bibinfo{author}{\bibnamefont{Rudolph}, \bibfnamefont{J.}},
  \bibinfo{author}{\bibfnamefont{D.}~\bibnamefont{{H\"{a}gele}}},
  \bibinfo{author}{\bibfnamefont{H.~M.} \bibnamefont{Gibbs}},
  \bibinfo{author}{\bibfnamefont{G.}~\bibnamefont{Khitrova}}, and
  \bibinfo{author}{\bibfnamefont{M.}~\bibnamefont{Oestreich}},
  \bibinfo{year}{2003}, {``}\bibinfo{title}{Laser threshold reduction in a
  spintronic device},{''} \bibinfo{journal}{Appl. Phys. Lett.}
  \textbf{\bibinfo{volume}{82}},  \bibinfo{pages}{4516--4518}.

\bibitem[{\citenamefont{Rzchowski and Wu}(2000)}]{Rzchowski2000:PRB}
\bibinfo{author}{\bibnamefont{Rzchowski}, \bibfnamefont{M.~S.}}, and
  \bibinfo{author}{\bibfnamefont{X.~W.} \bibnamefont{Wu}},
  \bibinfo{year}{2000}, {``}\bibinfo{title}{Bias dependence of magnetic tunnel
  junctions},{''} \bibinfo{journal}{Phys. Rev. B}
  \textbf{\bibinfo{volume}{61}},  \bibinfo{pages}{5884--5887}.

\bibitem[{\citenamefont{Safarov and Titkov}(1980)}]{Safarov1980:JPSJ}
\bibinfo{author}{\bibnamefont{Safarov}, \bibfnamefont{V.~I.}}, and
  \bibinfo{author}{\bibfnamefont{A.~N.} \bibnamefont{Titkov}},
  \bibinfo{year}{1980}, {``}\bibinfo{title}{Spin relaxation of conduction
  electrons in p-type semiconductors},{''} \bibinfo{journal}{J. Phys. Soc.
  Jpn.} \textbf{\bibinfo{volume}{49}},  \bibinfo{pages}{623--626}.

\bibitem[{\citenamefont{Saikin} \emph{et~al.}(2002)\citenamefont{Saikin,
  Mozyrsky, and Privman}}]{Saikin2002:NL}
\bibinfo{author}{\bibnamefont{Saikin}, \bibfnamefont{S.}},
  \bibinfo{author}{\bibfnamefont{D.}~\bibnamefont{Mozyrsky}}, and
  \bibinfo{author}{\bibfnamefont{V.}~\bibnamefont{Privman}},
  \bibinfo{year}{2002}, {``}\bibinfo{title}{Relaxation of shallow donor
  electron spin due to interaction with nuclear spin bath},{''}
  \bibinfo{journal}{Nano Lett.} \textbf{\bibinfo{volume}{2}},
  \bibinfo{pages}{651--655}.

\bibitem[{\citenamefont{Saikin} \emph{et~al.}(2003)\citenamefont{Saikin, Shen,
  Cheng, and Privman}}]{Saikin2003:JAP}
\bibinfo{author}{\bibnamefont{Saikin}, \bibfnamefont{S.}},
  \bibinfo{author}{\bibfnamefont{M.}~\bibnamefont{Shen}},
  \bibinfo{author}{\bibfnamefont{M.~C.} \bibnamefont{Cheng}}, and
  \bibinfo{author}{\bibfnamefont{V.}~\bibnamefont{Privman}},
  \bibinfo{year}{2003}, {``}\bibinfo{title}{Semiclassical {Monte}-{Carlo} model
  for in-plane transport of spin-polarized electrons in {III-V}
  heterostructures},{''} \bibinfo{journal}{J. Appl. Phys.}
  \textbf{\bibinfo{volume}{94}},  \bibinfo{pages}{1769--1775}.

\bibitem[{\citenamefont{Saito} \emph{et~al.}(2002)\citenamefont{Saito, Zaets,
  Yamagata, Suzuki, and Ando}}]{Saito2002:JAP}
\bibinfo{author}{\bibnamefont{Saito}, \bibfnamefont{H.}},
  \bibinfo{author}{\bibfnamefont{W.}~\bibnamefont{Zaets}},
  \bibinfo{author}{\bibfnamefont{S.}~\bibnamefont{Yamagata}},
  \bibinfo{author}{\bibfnamefont{Y.}~\bibnamefont{Suzuki}}, and
  \bibinfo{author}{\bibfnamefont{K.}~\bibnamefont{Ando}}, \bibinfo{year}{2002},
  {``}\bibinfo{title}{Ferromagnetism in {II-VI} diluted magnetic semiconductor
  {Zn$_{1-x}$Cr$_x$Te}},{''} \bibinfo{journal}{J. Appl. Phys.}
  \textbf{\bibinfo{volume}{91}},  \bibinfo{pages}{8085--8087}.

\bibitem[{\citenamefont{Saito} \emph{et~al.}(2003)\citenamefont{Saito, Zayets,
  Yamagata, and Ando}}]{Saito2003:PRL}
\bibinfo{author}{\bibnamefont{Saito}, \bibfnamefont{H.}},
  \bibinfo{author}{\bibfnamefont{V.}~\bibnamefont{Zayets}},
  \bibinfo{author}{\bibfnamefont{S.}~\bibnamefont{Yamagata}}, and
  \bibinfo{author}{\bibfnamefont{K.}~\bibnamefont{Ando}}, \bibinfo{year}{2003},
  {``}\bibinfo{title}{Room-temperature ferromagnetism in a {II-VI} diluted
  magnetic semiconductor {Zn$_{1-x}$Cr$_x$Te}},{''} \bibinfo{journal}{Phys.
  Rev. Lett.} \textbf{\bibinfo{volume}{90}},  \bibinfo{pages}{207202}.

\bibitem[{\citenamefont{Sakharov} \emph{et~al.}(1981)\citenamefont{Sakharov,
  Titkov, Ermakova, Komova, Mironov, and Chaikina}}]{Sakharov1981:SPSS}
\bibinfo{author}{\bibnamefont{Sakharov}, \bibfnamefont{V.~I.}},
  \bibinfo{author}{\bibfnamefont{A.~N.} \bibnamefont{Titkov}},
  \bibinfo{author}{\bibfnamefont{N.~G.} \bibnamefont{Ermakova}},
  \bibinfo{author}{\bibfnamefont{E.~M.} \bibnamefont{Komova}},
  \bibinfo{author}{\bibfnamefont{I.~F.} \bibnamefont{Mironov}}, and
  \bibinfo{author}{\bibfnamefont{E.~I.} \bibnamefont{Chaikina}},
  \bibinfo{year}{1981}, {``}\bibinfo{title}{Spin relaxation of conduction
  electrons in degenerate p-type gallium antimonide crystals. The
  {Bir}-{Aronov}-{Pikus} mechanism},{''} \bibinfo{journal}{Fiz. Tverd. Tela}
  \textbf{\bibinfo{volume}{23}},  \bibinfo{pages}{3337--3342}
  \bibinfo{note}{[Sov. Phys. Solid State {\bf 23}, 1938-1940 (1981)]}.

\bibitem[{\citenamefont{Salis}
  \emph{et~al.}(2001{\natexlab{a}})\citenamefont{Salis, Fuchs, Kikkawa,
  Awschalom, Ohno, and Ohno}}]{Salis2001:PRL}
\bibinfo{author}{\bibnamefont{Salis}, \bibfnamefont{G.}},
  \bibinfo{author}{\bibfnamefont{D.~T.} \bibnamefont{Fuchs}},
  \bibinfo{author}{\bibfnamefont{J.~M.} \bibnamefont{Kikkawa}},
  \bibinfo{author}{\bibfnamefont{D.~D.} \bibnamefont{Awschalom}},
  \bibinfo{author}{\bibfnamefont{Y.}~\bibnamefont{Ohno}}, and
  \bibinfo{author}{\bibfnamefont{H.}~\bibnamefont{Ohno}},
  \bibinfo{year}{2001}{\natexlab{a}}, {``}\bibinfo{title}{Optical manipulation
  of nuclear spin by a two-dimensional electron gas},{''}
  \bibinfo{journal}{Phys. Rev. Lett.} \textbf{\bibinfo{volume}{86}},
  \bibinfo{pages}{2677--2680}.

\bibitem[{\citenamefont{Salis}
  \emph{et~al.}(2001{\natexlab{b}})\citenamefont{Salis, Kato, Ensslin,
  Driscoll, Gossard, and Awschalom}}]{Salis2001:N}
\bibinfo{author}{\bibnamefont{Salis}, \bibfnamefont{G.}},
  \bibinfo{author}{\bibfnamefont{Y.}~\bibnamefont{Kato}},
  \bibinfo{author}{\bibfnamefont{K.}~\bibnamefont{Ensslin}},
  \bibinfo{author}{\bibfnamefont{D.~C.} \bibnamefont{Driscoll}},
  \bibinfo{author}{\bibfnamefont{A.~C.} \bibnamefont{Gossard}}, and
  \bibinfo{author}{\bibfnamefont{D.~D.} \bibnamefont{Awschalom}},
  \bibinfo{year}{2001}{\natexlab{b}}, {``}\bibinfo{title}{Electrical control of
  spin coherence in semiconductor nanostructures},{''} \bibinfo{journal}{{\sl
  Nature}} \textbf{\bibinfo{volume}{414}},  \bibinfo{pages}{619--622}.

\bibitem[{\citenamefont{Samarth} \emph{et~al.}(2003)\citenamefont{Samarth,
  Chun, Ku, Potashnik, and Schiffer}}]{Samarth2003:SSC}
\bibinfo{author}{\bibnamefont{Samarth}, \bibfnamefont{N.}},
  \bibinfo{author}{\bibfnamefont{S.~H.} \bibnamefont{Chun}},
  \bibinfo{author}{\bibfnamefont{K.~C.} \bibnamefont{Ku}},
  \bibinfo{author}{\bibfnamefont{S.~J.} \bibnamefont{Potashnik}}, and
  \bibinfo{author}{\bibfnamefont{P.}~\bibnamefont{Schiffer}},
  \bibinfo{year}{2003}, {``}\bibinfo{title}{Hybrid ferromagnetic/semiconductor
  heterostructures for spintronics},{''} \bibinfo{journal}{Solid State Commun.}
  \textbf{\bibinfo{volume}{127}},  \bibinfo{pages}{173--179}.

\bibitem[{\citenamefont{Sanada} \emph{et~al.}(2002)\citenamefont{Sanada, Arata,
  Ohno, Chen, Kayanuma, Oka, Matsukura, and Ohno}}]{Sanada2002:APL}
\bibinfo{author}{\bibnamefont{Sanada}, \bibfnamefont{H.}},
  \bibinfo{author}{\bibfnamefont{I.}~\bibnamefont{Arata}},
  \bibinfo{author}{\bibfnamefont{Y.}~\bibnamefont{Ohno}},
  \bibinfo{author}{\bibfnamefont{Z.}~\bibnamefont{Chen}},
  \bibinfo{author}{\bibfnamefont{K.}~\bibnamefont{Kayanuma}},
  \bibinfo{author}{\bibfnamefont{Y.}~\bibnamefont{Oka}},
  \bibinfo{author}{\bibfnamefont{F.}~\bibnamefont{Matsukura}}, and
  \bibinfo{author}{\bibfnamefont{H.}~\bibnamefont{Ohno}}, \bibinfo{year}{2002},
  {``}\bibinfo{title}{Relaxation of photoinjected spins during drift transport
  in {GaAs}},{''} \bibinfo{journal}{Appl. Phys. Lett.}
  \textbf{\bibinfo{volume}{81}},  \bibinfo{pages}{2788--2790}.

\bibitem[{\citenamefont{Sanderfeld}
  \emph{et~al.}(2000)\citenamefont{Sanderfeld, Jantsch, Wilamowski, and
  {Sch\"{a}ffler}}}]{Sandersfeld2000:TSF}
\bibinfo{author}{\bibnamefont{Sanderfeld}, \bibfnamefont{N.}},
  \bibinfo{author}{\bibfnamefont{W.}~\bibnamefont{Jantsch}},
  \bibinfo{author}{\bibfnamefont{Z.}~\bibnamefont{Wilamowski}}, and
  \bibinfo{author}{\bibfnamefont{F.}~\bibnamefont{{Sch\"{a}ffler}}},
  \bibinfo{year}{2000}, {``}\bibinfo{title}{{ESR} investigations of
  modulation-doped {Si/SiGe} quantum wells},{''} \bibinfo{journal}{Thin Solid
  Films} \textbf{\bibinfo{volume}{369}},  \bibinfo{pages}{312--315}.

\bibitem[{\citenamefont{Sandhu} \emph{et~al.}(2001)\citenamefont{Sandhu,
  Heberle, Baumberg, and Cleaver}}]{Sandhu2001:PRL}
\bibinfo{author}{\bibnamefont{Sandhu}, \bibfnamefont{J.~S.}},
  \bibinfo{author}{\bibfnamefont{A.~P.} \bibnamefont{Heberle}},
  \bibinfo{author}{\bibfnamefont{J.~J.} \bibnamefont{Baumberg}}, and
  \bibinfo{author}{\bibfnamefont{J.~R.~A.} \bibnamefont{Cleaver}},
  \bibinfo{year}{2001}, {``}\bibinfo{title}{Gateable suppression of spin
  relaxation in semiconductors},{''} \bibinfo{journal}{Phys. Rev. Lett.}
  \textbf{\bibinfo{volume}{86}},  \bibinfo{pages}{2150--2153}.

\bibitem[{\citenamefont{Sanvito} \emph{et~al.}(2002)\citenamefont{Sanvito,
  Theurich, and Hill}}]{Sanvito2002:JS}
\bibinfo{author}{\bibnamefont{Sanvito}, \bibfnamefont{S.}},
  \bibinfo{author}{\bibfnamefont{G.}~\bibnamefont{Theurich}}, and
  \bibinfo{author}{\bibfnamefont{N.~A.} \bibnamefont{Hill}},
  \bibinfo{year}{2002}, {``}\bibinfo{title}{Density functional calculations for
  III-V diluted ferromagnetic semiconductors: {A} review.},{''}
  \bibinfo{journal}{J. Supercond.} \textbf{\bibinfo{volume}{15}},
  \bibinfo{pages}{85--104}.

\bibitem[{\citenamefont{Sato and Mizushima}(2001)}]{Sato2001:APL}
\bibinfo{author}{\bibnamefont{Sato}, \bibfnamefont{R.}}, and
  \bibinfo{author}{\bibfnamefont{K.}~\bibnamefont{Mizushima}},
  \bibinfo{year}{2001}, {``}\bibinfo{title}{Spin-valve transistor with an
  {Fe/Au/Fe(001)} base},{''} \bibinfo{journal}{Appl. Phys. Lett.}
  \textbf{\bibinfo{volume}{79}},  \bibinfo{pages}{1157--1159}.

\bibitem[{\citenamefont{{Sch\"{a}pers}}
  \emph{et~al.}(1998)\citenamefont{{Sch\"{a}pers}, Engels, Lange, Klocke,
  Hollfelder, and L{\"{u}}th}}]{Schapers1998:JAP}
\bibinfo{author}{\bibnamefont{{Sch\"{a}pers}}, \bibfnamefont{T.}},
  \bibinfo{author}{\bibfnamefont{G.}~\bibnamefont{Engels}},
  \bibinfo{author}{\bibfnamefont{J.}~\bibnamefont{Lange}},
  \bibinfo{author}{\bibfnamefont{T.}~\bibnamefont{Klocke}},
  \bibinfo{author}{\bibfnamefont{M.}~\bibnamefont{Hollfelder}}, and
  \bibinfo{author}{\bibnamefont{L{\"{u}}th}}, \bibinfo{year}{1998},
  {``}\bibinfo{title}{Effect of the heterointerface on the spin splitting in
  modulation doped {In$_x$Ga$_{1-x}$As/InP} quantum wells for {$B\rightarrow
  0$}},{''} \bibinfo{journal}{J. Appl. Phys.} \textbf{\bibinfo{volume}{83}},
  \bibinfo{pages}{4324--4333}.

\bibitem[{\citenamefont{Schep} \emph{et~al.}(1997)\citenamefont{Schep, {van
  Hoof}, Kelly, and Bauer}}]{Schep1997:PRB}
\bibinfo{author}{\bibnamefont{Schep}, \bibfnamefont{K.~M.}},
  \bibinfo{author}{\bibfnamefont{J.~B. A.~N.} \bibnamefont{{van Hoof}}},
  \bibinfo{author}{\bibfnamefont{P.~J.} \bibnamefont{Kelly}}, and
  \bibinfo{author}{\bibfnamefont{G.~E.~W.} \bibnamefont{Bauer}},
  \bibinfo{year}{1997}, {``}\bibinfo{title}{Interface resistances of magnetic
  multilayers},{''} \bibinfo{journal}{Phys. Rev. B}
  \textbf{\bibinfo{volume}{56}},  \bibinfo{pages}{10805--10808}.

\bibitem[{\citenamefont{Schliemann}
  \emph{et~al.}(2003)\citenamefont{Schliemann, Egues, and
  Loss}}]{Schliemann2002:P}
\bibinfo{author}{\bibnamefont{Schliemann}, \bibfnamefont{J.}},
  \bibinfo{author}{\bibfnamefont{J.~C.} \bibnamefont{Egues}}, and
  \bibinfo{author}{\bibfnamefont{D.}~\bibnamefont{Loss}}, \bibinfo{year}{2003},
  {``}\bibinfo{title}{Non-ballistic spin field-effect transistor},{''}
  \bibinfo{journal}{Phys. Rev. Lett.} \textbf{\bibinfo{volume}{90}},
  \bibinfo{pages}{146801}.

\bibitem[{\citenamefont{Schmidt} \emph{et~al.}(2000)\citenamefont{Schmidt,
  Ferrand, Molenkamp, Filip, and {van Wees}}}]{Schmidt2000:PRB}
\bibinfo{author}{\bibnamefont{Schmidt}, \bibfnamefont{G.}},
  \bibinfo{author}{\bibfnamefont{D.}~\bibnamefont{Ferrand}},
  \bibinfo{author}{\bibfnamefont{L.~W.} \bibnamefont{Molenkamp}},
  \bibinfo{author}{\bibfnamefont{A.~T.} \bibnamefont{Filip}}, and
  \bibinfo{author}{\bibfnamefont{B.~J.} \bibnamefont{{van Wees}}},
  \bibinfo{year}{2000}, {``}\bibinfo{title}{Fundamental obstacle for electrical
  spin injection from a ferromagnetic metal into a diffusive
  semiconductor},{''} \bibinfo{journal}{Phys. Rev. B}
  \textbf{\bibinfo{volume}{62}},  \bibinfo{pages}{R4790--R4793}.

\bibitem[{\citenamefont{Schmidt} \emph{et~al.}(2002)\citenamefont{Schmidt,
  Gould, Grabs, Lunde, Richter, Slobodskyy, and Molenkamp}}]{Schmidt2002:P}
\bibinfo{author}{\bibnamefont{Schmidt}, \bibfnamefont{G.}},
  \bibinfo{author}{\bibfnamefont{C.}~\bibnamefont{Gould}},
  \bibinfo{author}{\bibfnamefont{P.}~\bibnamefont{Grabs}},
  \bibinfo{author}{\bibfnamefont{A.}~\bibnamefont{Lunde}},
  \bibinfo{author}{\bibfnamefont{G.}~\bibnamefont{Richter}},
  \bibinfo{author}{\bibfnamefont{A.}~\bibnamefont{Slobodskyy}}, and
  \bibinfo{author}{\bibfnamefont{L.~W.} \bibnamefont{Molenkamp}},
  \bibinfo{year}{2002}, {``}\bibinfo{title}{Spin injection in the non-linear
  regime: band bending effects},{''} \eprint{cond-mat/0206347}.

\bibitem[{\citenamefont{Schmidt and Molenkamp}(2002)}]{Schmidt2002:SST}
\bibinfo{author}{\bibnamefont{Schmidt}, \bibfnamefont{G.}}, and
  \bibinfo{author}{\bibfnamefont{L.~W.} \bibnamefont{Molenkamp}},
  \bibinfo{year}{2002}, {``}\bibinfo{title}{Spin injection into semiconductors,
  physics and experiments},{''} \bibinfo{journal}{Semicond. Sci. Technol.}
  \textbf{\bibinfo{volume}{17}},  \bibinfo{pages}{310--321}.

\bibitem[{\citenamefont{Schmidt} \emph{et~al.}(2001)\citenamefont{Schmidt,
  Richter, Grabs, Gould, Ferrand, and Molenkamp}}]{Schmidt2001:PRL}
\bibinfo{author}{\bibnamefont{Schmidt}, \bibfnamefont{G.}},
  \bibinfo{author}{\bibfnamefont{G.}~\bibnamefont{Richter}},
  \bibinfo{author}{\bibfnamefont{P.}~\bibnamefont{Grabs}},
  \bibinfo{author}{\bibfnamefont{C.}~\bibnamefont{Gould}},
  \bibinfo{author}{\bibfnamefont{D.}~\bibnamefont{Ferrand}}, and
  \bibinfo{author}{\bibfnamefont{L.~W.} \bibnamefont{Molenkamp}},
  \bibinfo{year}{2001}, {``}\bibinfo{title}{Large magnetoresistance effect due
  to spin injection into a nonmagnetic semiconductor},{''}
  \bibinfo{journal}{Phys. Rev. Lett.} \textbf{\bibinfo{volume}{87}},
  \bibinfo{pages}{227203}.

\bibitem[{\citenamefont{Schultz and Latham}(1965)}]{Schultz1965:PRL}
\bibinfo{author}{\bibnamefont{Schultz}, \bibfnamefont{S.}}, and
  \bibinfo{author}{\bibfnamefont{C.}~\bibnamefont{Latham}},
  \bibinfo{year}{1965}, {``}\bibinfo{title}{Observation of electron spin
  resonance in copper},{''} \bibinfo{journal}{Phys. Rev. Lett.}
  \textbf{\bibinfo{volume}{15}},  \bibinfo{pages}{148--151}.

\bibitem[{\citenamefont{Schultz and Shanabarger}(1966)}]{Schultz1966:PRL}
\bibinfo{author}{\bibnamefont{Schultz}, \bibfnamefont{S.}}, and
  \bibinfo{author}{\bibfnamefont{M.~R.} \bibnamefont{Shanabarger}},
  \bibinfo{year}{1966}, {``}\bibinfo{title}{Observation of electron spin
  resonance in rubidium and cesium},{''} \bibinfo{journal}{Phys. Rev. Lett.}
  \textbf{\bibinfo{volume}{16}},  \bibinfo{pages}{178--181}.

\bibitem[{\citenamefont{Schumacher}
  \emph{et~al.}(2003{\natexlab{a}})\citenamefont{Schumacher, Chappert, Crozat,
  Sousa, Freitas, Miltat, Fassbender, and Hillebrands}}]{Schumacher2003:PRLa}
\bibinfo{author}{\bibnamefont{Schumacher}, \bibfnamefont{H.~W.}},
  \bibinfo{author}{\bibfnamefont{C.}~\bibnamefont{Chappert}},
  \bibinfo{author}{\bibfnamefont{P.}~\bibnamefont{Crozat}},
  \bibinfo{author}{\bibfnamefont{R.~C.} \bibnamefont{Sousa}},
  \bibinfo{author}{\bibfnamefont{P.~P.} \bibnamefont{Freitas}},
  \bibinfo{author}{\bibfnamefont{J.}~\bibnamefont{Miltat}},
  \bibinfo{author}{\bibfnamefont{J.}~\bibnamefont{Fassbender}}, and
  \bibinfo{author}{\bibfnamefont{B.}~\bibnamefont{Hillebrands}},
  \bibinfo{year}{2003}{\natexlab{a}}, {``}\bibinfo{title}{Phase coherent
  precessional magnetization reversal in microscopic spin valve elements},{''}
  \bibinfo{journal}{Phys. Rev. Lett.} \textbf{\bibinfo{volume}{90}},
  \bibinfo{pages}{017201}.

\bibitem[{\citenamefont{Schumacher}
  \emph{et~al.}(2003{\natexlab{b}})\citenamefont{Schumacher, Chappert, Sousa,
  Freitas, and Miltat}}]{Schumacher2003:PRLb}
\bibinfo{author}{\bibnamefont{Schumacher}, \bibfnamefont{H.~W.}},
  \bibinfo{author}{\bibfnamefont{C.}~\bibnamefont{Chappert}},
  \bibinfo{author}{\bibfnamefont{R.~C.} \bibnamefont{Sousa}},
  \bibinfo{author}{\bibfnamefont{P.~P.} \bibnamefont{Freitas}}, and
  \bibinfo{author}{\bibfnamefont{J.}~\bibnamefont{Miltat}},
  \bibinfo{year}{2003}{\natexlab{b}}, {``}\bibinfo{title}{Quasiballistic
  magnetization reversal},{''} \bibinfo{journal}{Phys. Rev. Lett.}
  \textbf{\bibinfo{volume}{90}},  \bibinfo{pages}{017204}.

\bibitem[{\citenamefont{Scifres} \emph{et~al.}(1973)\citenamefont{Scifres,
  B.~A.~Huberman, and Bauer}}]{Scifres1973:SSC}
\bibinfo{author}{\bibnamefont{Scifres}, \bibfnamefont{D.~R.}},
  \bibinfo{author}{\bibfnamefont{R.~M.~W.} \bibnamefont{B.~A.~Huberman}}, and
  \bibinfo{author}{\bibfnamefont{R.~S.} \bibnamefont{Bauer}},
  \bibinfo{year}{1973}, {``}\bibinfo{title}{A new scheme for measuring
  itinerant spin polarization},{''} \bibinfo{journal}{Solid State Commun.}
  \textbf{\bibinfo{volume}{13}},  \bibinfo{pages}{1615--1617}.

\bibitem[{\citenamefont{Seck} \emph{et~al.}(1997)\citenamefont{Seck, Potemski,
  and Wyder}}]{Seck1997:PRB}
\bibinfo{author}{\bibnamefont{Seck}, \bibfnamefont{M.}},
  \bibinfo{author}{\bibfnamefont{M.}~\bibnamefont{Potemski}}, and
  \bibinfo{author}{\bibfnamefont{P.}~\bibnamefont{Wyder}},
  \bibinfo{year}{1997}, {``}\bibinfo{title}{High-field spin resonance of weakly
  bound electrons in {GaAs}},{''} \bibinfo{journal}{Phys. Rev. B}
  \textbf{\bibinfo{volume}{56}},  \bibinfo{pages}{7422--7427}.

\bibitem[{\citenamefont{Semenov and Kim}(2003)}]{Semenov2002:P}
\bibinfo{author}{\bibnamefont{Semenov}, \bibfnamefont{Y.~G.}}, and
  \bibinfo{author}{\bibfnamefont{K.~W.} \bibnamefont{Kim}},
  \bibinfo{year}{2003}, {``}\bibinfo{title}{Effect of external magnetic field
  on electron spin dephasing induced by hyperfine interaction in quantum
  dots},{''} \bibinfo{journal}{Phys. Rev. B} \textbf{\bibinfo{volume}{67}},
  \bibinfo{pages}{073301}.

\bibitem[{\citenamefont{Seufert} \emph{et~al.}(2004)\citenamefont{Seufert,
  Bacher, Sch{\"{o}}mig, Forchel, Hansen, Schmidt, and
  Molenkamp}}]{Seufert2004:PRB}
\bibinfo{author}{\bibnamefont{Seufert}, \bibfnamefont{J.}},
  \bibinfo{author}{\bibfnamefont{G.}~\bibnamefont{Bacher}},
  \bibinfo{author}{\bibfnamefont{H.}~\bibnamefont{Sch{\"{o}}mig}},
  \bibinfo{author}{\bibfnamefont{A.}~\bibnamefont{Forchel}},
  \bibinfo{author}{\bibfnamefont{L.}~\bibnamefont{Hansen}},
  \bibinfo{author}{\bibfnamefont{G.}~\bibnamefont{Schmidt}}, and
  \bibinfo{author}{\bibfnamefont{L.~W.} \bibnamefont{Molenkamp}},
  \bibinfo{year}{2004}, {``}\bibinfo{title}{Spin injection into a single
  self-assembled quantum dot},{''} \bibinfo{journal}{Phys. Rev. B}
  \textbf{\bibinfo{volume}{69}},  \bibinfo{pages}{035311}.

\bibitem[{\citenamefont{Seymour} \emph{et~al.}(1981)\citenamefont{Seymour,
  Junnarkar, and Alfano}}]{Seymour1981:PRB}
\bibinfo{author}{\bibnamefont{Seymour}, \bibfnamefont{R.~J.}},
  \bibinfo{author}{\bibfnamefont{M.~R.} \bibnamefont{Junnarkar}}, and
  \bibinfo{author}{\bibfnamefont{R.~R.} \bibnamefont{Alfano}},
  \bibinfo{year}{1981}, {``}\bibinfo{title}{Spin relaxation of photogenerated
  electron distributions in {GaAs}},{''} \bibinfo{journal}{Phys. Rev. B}
  \textbf{\bibinfo{volume}{24}},  \bibinfo{pages}{3623--3625}.

\bibitem[{\citenamefont{Sham}(1993)}]{Sham1993:JPCM}
\bibinfo{author}{\bibnamefont{Sham}, \bibfnamefont{L.~J.}},
  \bibinfo{year}{1993}, {``}\bibinfo{title}{Spin relaxation in semiconductor
  quantum wells},{''} \bibinfo{journal}{J. Phys.: Condens. Matter}
  \textbf{\bibinfo{volume}{5}},  \bibinfo{pages}{A51--A60}.

\bibitem[{\citenamefont{Sham} \emph{et~al.}(1998)\citenamefont{Sham,
  {\"{O}streich}, and {Sch\"{o}nhammer}}}]{Sham1988:PE}
\bibinfo{author}{\bibnamefont{Sham}, \bibfnamefont{L.~J.}},
  \bibinfo{author}{\bibfnamefont{T.}~\bibnamefont{{\"{O}streich}}}, and
  \bibinfo{author}{\bibfnamefont{K.}~\bibnamefont{{Sch\"{o}nhammer}}},
  \bibinfo{year}{1998}, {``}\bibinfo{title}{Spin coherence in semiconductor
  heterostructures},{''} \bibinfo{journal}{Physica E}
  \textbf{\bibinfo{volume}{2}},  \bibinfo{pages}{388--393}.

\bibitem[{\citenamefont{Shang} \emph{et~al.}(1998)\citenamefont{Shang, Nowak,
  Jansen, and Moodera}}]{Shang1998:PRB}
\bibinfo{author}{\bibnamefont{Shang}, \bibfnamefont{C.~H.}},
  \bibinfo{author}{\bibfnamefont{J.}~\bibnamefont{Nowak}},
  \bibinfo{author}{\bibfnamefont{R.}~\bibnamefont{Jansen}}, and
  \bibinfo{author}{\bibfnamefont{J.~S.} \bibnamefont{Moodera}},
  \bibinfo{year}{1998}, {``}\bibinfo{title}{Temperature dependence of
  magnetoresistance and surface magnetization in ferromagnetic tunnel
  junctions},{''} \bibinfo{journal}{Phys. Rev. B}
  \textbf{\bibinfo{volume}{58}},  \bibinfo{pages}{R2917--R2920}.

\bibitem[{\citenamefont{Sharma and Chamon}(2003)}]{Sharma2003:PRB}
\bibinfo{author}{\bibnamefont{Sharma}, \bibfnamefont{P.}}, and
  \bibinfo{author}{\bibfnamefont{C.}~\bibnamefont{Chamon}},
  \bibinfo{year}{2003}, {``}\bibinfo{title}{Adiabatic charge and spin transport
  in interacting quantum wires},{''} \bibinfo{journal}{Phys. Rev. B}
  \textbf{\bibinfo{volume}{68}},  \bibinfo{pages}{035321}.

\bibitem[{\citenamefont{Sharvin}(1965)}]{Sharvin1965:ZETF}
\bibinfo{author}{\bibnamefont{Sharvin}, \bibfnamefont{Y.~V.}},
  \bibinfo{year}{1965}, {``}\bibinfo{title}{A possible method for studying
  {Fermi} Surfaces},{''} \bibinfo{journal}{Zh. Eksp. Teor. Fiz.}
  \textbf{\bibinfo{volume}{48}},  \bibinfo{pages}{984--985}
  \bibinfo{note}{[Sov. Phys. JETP {\bf 21}, 655-656 (1965)]}.

\bibitem[{\citenamefont{Sherman}(2003{\natexlab{a}})}]{Sherman2003:PRB}
\bibinfo{author}{\bibnamefont{Sherman}, \bibfnamefont{E.~Y.}},
  \bibinfo{year}{2003}{\natexlab{a}}, {``}\bibinfo{title}{Minimum of spin-orbit
  coupling in two-dimensional structures},{''} \bibinfo{journal}{Phys. Rev. B}
  \textbf{\bibinfo{volume}{67}},  \bibinfo{pages}{035311}.

\bibitem[{\citenamefont{Sherman}(2003{\natexlab{b}})}]{Sherman2003:APL}
\bibinfo{author}{\bibnamefont{Sherman}, \bibfnamefont{E.~Y.}},
  \bibinfo{year}{2003}{\natexlab{b}}, {``}\bibinfo{title}{Random-spin-orbit
  coupling and spin relaxation in symmetric quantum wells},{''}
  \bibinfo{journal}{Appl. Phys. Lett.} \textbf{\bibinfo{volume}{82}},
  \bibinfo{pages}{209--211}.

\bibitem[{\citenamefont{Shinjo}(2002)}]{Shinjo:2002}
\bibinfo{author}{\bibnamefont{Shinjo}, \bibfnamefont{T.}},
  \bibinfo{year}{2002}, {``}\bibinfo{title}{Experiments of Giant
  Magnetoresistance},{''} in \emph{\bibinfo{booktitle}{Spin Dependent Transport
  in Magnetic Nanostructures}}, edited by
  \bibinfo{editor}{\bibfnamefont{S.}~\bibnamefont{Maekawa}} and
  \bibinfo{editor}{\bibfnamefont{T.}~\bibnamefont{Shinjo}}
  (\bibinfo{publisher}{Taylor and Francis, New York}),  \bibinfo{pages}{1--46}.

\bibitem[{\citenamefont{Shinshi} \emph{et~al.}(2003)\citenamefont{Shinshi,
  Kato, Shimokohbe, Noguchi, and Munekata}}]{Shishi2003:APL}
\bibinfo{author}{\bibnamefont{Shinshi}, \bibfnamefont{T.}},
  \bibinfo{author}{\bibfnamefont{F.}~\bibnamefont{Kato}},
  \bibinfo{author}{\bibfnamefont{A.}~\bibnamefont{Shimokohbe}},
  \bibinfo{author}{\bibfnamefont{H.}~\bibnamefont{Noguchi}}, and
  \bibinfo{author}{\bibfnamefont{H.}~\bibnamefont{Munekata}},
  \bibinfo{year}{2003}, {``}\bibinfo{title}{Light-driven microcantilever
  actuator based on photoenhanced magnetization in a {GaAs-Fe} composite
  film},{''} \bibinfo{journal}{Appl. Phys. Lett.}
  \textbf{\bibinfo{volume}{83}},  \bibinfo{pages}{3425--3427}.

\bibitem[{\citenamefont{Shockley}(1950)}]{Shockley:1950}
\bibinfo{author}{\bibnamefont{Shockley}, \bibfnamefont{W.}},
  \bibinfo{year}{1950}, \emph{\bibinfo{title}{Electrons and Holes in
  Semiconductors}} (\bibinfo{publisher}{D. {Van Nostrand}, Princeton}).

\bibitem[{\citenamefont{Si}(1997)}]{Si1997:PRL}
\bibinfo{author}{\bibnamefont{Si}, \bibfnamefont{Q.}}, \bibinfo{year}{1997},
  {``}\bibinfo{title}{Spin conductivity and spin-charge separation in the
  high-{T$_c$} cuprates},{''} \bibinfo{journal}{Phys. Rev. Lett.}
  \textbf{\bibinfo{volume}{78}},  \bibinfo{pages}{1767--1770}.

\bibitem[{\citenamefont{Si}(1998)}]{Si1998:PRL}
\bibinfo{author}{\bibnamefont{Si}, \bibfnamefont{Q.}}, \bibinfo{year}{1998},
  {``}\bibinfo{title}{Spin Injection into a {Luttinger} Liquid},{''}
  \bibinfo{journal}{Phys. Rev. Lett.} \textbf{\bibinfo{volume}{81}},
  \bibinfo{pages}{3191--3194}.

\bibitem[{\citenamefont{Sidles} \emph{et~al.}(1995)\citenamefont{Sidles,
  Garbini, Bruland, Rugar, Z{\"u}ger, Hoen, and Yannoni}}]{Sidles1995:RMP}
\bibinfo{author}{\bibnamefont{Sidles}, \bibfnamefont{J.~A.}},
  \bibinfo{author}{\bibfnamefont{J.~L.} \bibnamefont{Garbini}},
  \bibinfo{author}{\bibfnamefont{K.~J.} \bibnamefont{Bruland}},
  \bibinfo{author}{\bibfnamefont{D.}~\bibnamefont{Rugar}},
  \bibinfo{author}{\bibfnamefont{O.}~\bibnamefont{Z{\"u}ger}},
  \bibinfo{author}{\bibfnamefont{S.}~\bibnamefont{Hoen}}, and
  \bibinfo{author}{\bibfnamefont{C.~S.} \bibnamefont{Yannoni}},
  \bibinfo{year}{1995}, {``}\bibinfo{title}{Magnetic resonance force
  microscopy},{''} \bibinfo{journal}{Rev. Mod. Phys.}
  \textbf{\bibinfo{volume}{67}},  \bibinfo{pages}{249--265}.

\bibitem[{\citenamefont{Sieg} \emph{et~al.}(1998)\citenamefont{Sieg, Carlin,
  Boeckl, Ringel, Currie, Ting, Langdo, Taraschi, Fitzgerald, and
  Keyes}}]{Sieg1998:APL}
\bibinfo{author}{\bibnamefont{Sieg}, \bibfnamefont{R.~M.}},
  \bibinfo{author}{\bibfnamefont{J.~A.} \bibnamefont{Carlin}},
  \bibinfo{author}{\bibfnamefont{J.~J.} \bibnamefont{Boeckl}},
  \bibinfo{author}{\bibfnamefont{S.~A.} \bibnamefont{Ringel}},
  \bibinfo{author}{\bibfnamefont{M.~T.} \bibnamefont{Currie}},
  \bibinfo{author}{\bibfnamefont{S.~M.} \bibnamefont{Ting}},
  \bibinfo{author}{\bibfnamefont{T.~A.} \bibnamefont{Langdo}},
  \bibinfo{author}{\bibfnamefont{G.}~\bibnamefont{Taraschi}},
  \bibinfo{author}{\bibfnamefont{E.~A.} \bibnamefont{Fitzgerald}}, and
  \bibinfo{author}{\bibfnamefont{B.~M.} \bibnamefont{Keyes}},
  \bibinfo{year}{1998}, {``}\bibinfo{title}{High minority-carrier lifetimes
  grown on low-defect-density {Ge/GeSi/Si} substrates},{''}
  \bibinfo{journal}{Appl. Phys. Lett.} \textbf{\bibinfo{volume}{73}},
  \bibinfo{pages}{3111--3113}.

\bibitem[{\citenamefont{Silsbee}(1980)}]{Silsbee1980:BMR}
\bibinfo{author}{\bibnamefont{Silsbee}, \bibfnamefont{R.~H.}},
  \bibinfo{year}{1980}, {``}\bibinfo{title}{Novel method for the study of spin
  transport in conductors},{''} \bibinfo{journal}{Bull. Magn. Reson.}
  \textbf{\bibinfo{volume}{2}},  \bibinfo{pages}{284--285}.

\bibitem[{\citenamefont{Silsbee}(2001)}]{Silsbee2001:PRB}
\bibinfo{author}{\bibnamefont{Silsbee}, \bibfnamefont{R.~H.}},
  \bibinfo{year}{2001}, {``}\bibinfo{title}{Theory of the detection of
  current-induced spin polarization in a two-dimensional electron gas},{''}
  \bibinfo{journal}{Phys. Rev. B} \textbf{\bibinfo{volume}{63}},
  \bibinfo{pages}{155305{; {\bf 68}, 159902 (2003) (E).}}

\bibitem[{\citenamefont{Silsbee and Beuneu}(1983)}]{Silsbee1983:PRB}
\bibinfo{author}{\bibnamefont{Silsbee}, \bibfnamefont{R.~H.}}, and
  \bibinfo{author}{\bibfnamefont{F.}~\bibnamefont{Beuneu}},
  \bibinfo{year}{1983}, {``}\bibinfo{title}{Model calculation of the frequency
  and temperature dependence of the electron-spin-resonance linewidth of
  aluminum},{''} \bibinfo{journal}{Phys. Rev. B} \textbf{\bibinfo{volume}{27}},
   \bibinfo{pages}{2682--2692}.

\bibitem[{\citenamefont{Simon} \emph{et~al.}(2001)\citenamefont{Simon,
  {J\'{a}nossy}, {Feh\'{e}r}, {Mur\'{a}nyi}, Garaj, {Forr\'{o}}, Petrovic,
  {Bu\'{d}ko}, Lapertot, Kogan, and Canfield}}]{simon2001:PRL}
\bibinfo{author}{\bibnamefont{Simon}, \bibfnamefont{F.}},
  \bibinfo{author}{\bibfnamefont{A.}~\bibnamefont{{J\'{a}nossy}}},
  \bibinfo{author}{\bibfnamefont{T.}~\bibnamefont{{Feh\'{e}r}}},
  \bibinfo{author}{\bibfnamefont{F.}~\bibnamefont{{Mur\'{a}nyi}}},
  \bibinfo{author}{\bibfnamefont{S.}~\bibnamefont{Garaj}},
  \bibinfo{author}{\bibfnamefont{L.}~\bibnamefont{{Forr\'{o}}}},
  \bibinfo{author}{\bibfnamefont{C.}~\bibnamefont{Petrovic}},
  \bibinfo{author}{\bibfnamefont{S.~L.} \bibnamefont{{Bu\'{d}ko}}},
  \bibinfo{author}{\bibfnamefont{G.}~\bibnamefont{Lapertot}},
  \bibinfo{author}{\bibfnamefont{V.~G.} \bibnamefont{Kogan}}, and
  \bibinfo{author}{\bibfnamefont{P.~C.} \bibnamefont{Canfield}},
  \bibinfo{year}{2001}, {``}\bibinfo{title}{Anisotropy of superconducting
  {MgB$_2$} as seen in electron spin resonance and magnetization data},{''}
  \bibinfo{journal}{Phys. Rev. Lett.} \textbf{\bibinfo{volume}{87}},
  \bibinfo{pages}{047002}.

\bibitem[{\citenamefont{Skinner} \emph{et~al.}(2003)\citenamefont{Skinner,
  Davenport, and Kane}}]{Skinner2003:PRL}
\bibinfo{author}{\bibnamefont{Skinner}, \bibfnamefont{A.~J.}},
  \bibinfo{author}{\bibfnamefont{M.~E.} \bibnamefont{Davenport}}, and
  \bibinfo{author}{\bibfnamefont{B.~E.} \bibnamefont{Kane}},
  \bibinfo{year}{2003}, {``}\bibinfo{title}{Hydrogenic spin quantum computing
  in silicon: A digital approach},{''} \bibinfo{journal}{Phys. Rev. Lett.}
  \textbf{\bibinfo{volume}{90}},  \bibinfo{pages}{087901}.

\bibitem[{\citenamefont{Slichter}(1989)}]{Slichter:1989}
\bibinfo{author}{\bibnamefont{Slichter}, \bibfnamefont{C.~P.}},
  \bibinfo{year}{1989}, \emph{\bibinfo{title}{Principles of Magnetic Resonance,
  {\rm 3rd {Ed.}}}} (\bibinfo{publisher}{Springer, Berlin}).

\bibitem[{\citenamefont{Slobodskyy}
  \emph{et~al.}(2003)\citenamefont{Slobodskyy, Gould, Slobodskyy, Becker,
  Schmidt, and Molenkamp}}]{Slobodskyy2003:PRL}
\bibinfo{author}{\bibnamefont{Slobodskyy}, \bibfnamefont{A.}},
  \bibinfo{author}{\bibfnamefont{C.}~\bibnamefont{Gould}},
  \bibinfo{author}{\bibfnamefont{T.}~\bibnamefont{Slobodskyy}},
  \bibinfo{author}{\bibfnamefont{C.~R.} \bibnamefont{Becker}},
  \bibinfo{author}{\bibfnamefont{G.}~\bibnamefont{Schmidt}}, and
  \bibinfo{author}{\bibfnamefont{L.~W.} \bibnamefont{Molenkamp}},
  \bibinfo{year}{2003}, {``}\bibinfo{title}{Voltage-controlled spin selection
  in a magnetic resonant tunneling diode},{''} \bibinfo{journal}{Phys. Rev.
  Lett.} \textbf{\bibinfo{volume}{90}},  \bibinfo{pages}{246601}.

\bibitem[{\citenamefont{Slonczewski}(1976)}]{Slonczewski1976:IBM}
\bibinfo{author}{\bibnamefont{Slonczewski}, \bibfnamefont{J.~C.}},
  \bibinfo{year}{1976}, {``}\bibinfo{title}{Magnetic bubble tunnel
  detector},{''} \bibinfo{journal}{IBM Tech. Disc. Bull.}
  \textbf{\bibinfo{volume}{19}},  \bibinfo{pages}{2328--2330}.

\bibitem[{\citenamefont{Slonczewski}(1989)}]{Slonczewski1989:PRB}
\bibinfo{author}{\bibnamefont{Slonczewski}, \bibfnamefont{J.~C.}},
  \bibinfo{year}{1989}, {``}\bibinfo{title}{Conductance and exchange coupling
  of two ferromagnets separated by a tunneling barrier},{''}
  \bibinfo{journal}{Phys. Rev. B} \textbf{\bibinfo{volume}{39}},
  \bibinfo{pages}{6995--7002}.

\bibitem[{\citenamefont{Slonczewski}(1996)}]{Slonczewski1996:JMMM}
\bibinfo{author}{\bibnamefont{Slonczewski}, \bibfnamefont{J.~C.}},
  \bibinfo{year}{1996}, {``}\bibinfo{title}{Current-driven excitation of
  magnetic multilayers},{''} \bibinfo{journal}{J. Magn. Magn. Mater.}
  \textbf{\bibinfo{volume}{159}},  \bibinfo{pages}{L1--L7}.

\bibitem[{\citenamefont{Smet} \emph{et~al.}(2002)\citenamefont{Smet,
  Deutschmann, Ertl, Wegscheider, Abstreiter, and {von Klitzing}}}]{Smet2002:N}
\bibinfo{author}{\bibnamefont{Smet}, \bibfnamefont{J.~H.}},
  \bibinfo{author}{\bibfnamefont{R.~A.} \bibnamefont{Deutschmann}},
  \bibinfo{author}{\bibfnamefont{F.}~\bibnamefont{Ertl}},
  \bibinfo{author}{\bibfnamefont{W.}~\bibnamefont{Wegscheider}},
  \bibinfo{author}{\bibfnamefont{G.}~\bibnamefont{Abstreiter}}, and
  \bibinfo{author}{\bibfnamefont{K.}~\bibnamefont{{von Klitzing}}},
  \bibinfo{year}{2002}, {``}\bibinfo{title}{Gate-voltage control of spin
  interactions between electrons and nuclei in a semiconductor},{''}
  \bibinfo{journal}{{\sl Nature}} \textbf{\bibinfo{volume}{415}},
  \bibinfo{pages}{281--286}.

\bibitem[{\citenamefont{Smith} \emph{et~al.}(2000)\citenamefont{Smith,
  Kozhevnikov, Lee, and Narayanamurti}}]{Smith2000:PRB}
\bibinfo{author}{\bibnamefont{Smith}, \bibfnamefont{D.~L.}},
  \bibinfo{author}{\bibfnamefont{M.}~\bibnamefont{Kozhevnikov}},
  \bibinfo{author}{\bibfnamefont{E.~Y.} \bibnamefont{Lee}}, and
  \bibinfo{author}{\bibfnamefont{V.}~\bibnamefont{Narayanamurti}},
  \bibinfo{year}{2000}, {``}\bibinfo{title}{Scattering theory of
  ballistic-electron-emission microscopy at nonepitaxial interfaces},{''}
  \bibinfo{journal}{Phys. Rev. B} \textbf{\bibinfo{volume}{61}},
  \bibinfo{pages}{13914--13922}.

\bibitem[{\citenamefont{Smith and Silver}(2001)}]{Smith2001:PRB}
\bibinfo{author}{\bibnamefont{Smith}, \bibfnamefont{D.~L.}}, and
  \bibinfo{author}{\bibfnamefont{R.~N.} \bibnamefont{Silver}},
  \bibinfo{year}{2001}, {``}\bibinfo{title}{Electrical spin injection into
  semiconductors},{''} \bibinfo{journal}{Phys. Rev. B}
  \textbf{\bibinfo{volume}{64}},  \bibinfo{pages}{045323}.

\bibitem[{\citenamefont{Sogawa} \emph{et~al.}(2000)\citenamefont{Sogawa, Ando,
  and Ando}}]{Sogawa2000:PRB}
\bibinfo{author}{\bibnamefont{Sogawa}, \bibfnamefont{T.}},
  \bibinfo{author}{\bibfnamefont{H.}~\bibnamefont{Ando}}, and
  \bibinfo{author}{\bibfnamefont{S.}~\bibnamefont{Ando}}, \bibinfo{year}{2000},
  {``}\bibinfo{title}{Spin-transport dynamics of optically spin-polarized
  electrons in {GaAs} quantum wires},{''} \bibinfo{journal}{Phys. Rev. B}
  \textbf{\bibinfo{volume}{61}},  \bibinfo{pages}{5535--5539}.

\bibitem[{\citenamefont{Sogawa} \emph{et~al.}(2001)\citenamefont{Sogawa,
  Santos, Zhang, Eshlaghi, Wieck, and Ploog}}]{Sogawa2001:PRL}
\bibinfo{author}{\bibnamefont{Sogawa}, \bibfnamefont{T.}},
  \bibinfo{author}{\bibfnamefont{P.~V.} \bibnamefont{Santos}},
  \bibinfo{author}{\bibfnamefont{S.~K.} \bibnamefont{Zhang}},
  \bibinfo{author}{\bibfnamefont{S.}~\bibnamefont{Eshlaghi}},
  \bibinfo{author}{\bibfnamefont{A.~D.} \bibnamefont{Wieck}}, and
  \bibinfo{author}{\bibfnamefont{K.~H.} \bibnamefont{Ploog}},
  \bibinfo{year}{2001}, {``}\bibinfo{title}{Transport and lifetime enhancement
  of photoexcited spins in {GaAs} by surface acoustic waves},{''}
  \bibinfo{journal}{Phys. Rev. Lett.} \textbf{\bibinfo{volume}{87}},
  \bibinfo{pages}{276601}.

\bibitem[{\citenamefont{Solin} \emph{et~al.}(2002)\citenamefont{Solin, Hines,
  Rowe, Tsai, Pashkin, Chung, Goel, and Santos}}]{Solin2002:APL}
\bibinfo{author}{\bibnamefont{Solin}, \bibfnamefont{S.~A.}},
  \bibinfo{author}{\bibfnamefont{D.~R.} \bibnamefont{Hines}},
  \bibinfo{author}{\bibfnamefont{A.~C.~H.} \bibnamefont{Rowe}},
  \bibinfo{author}{\bibfnamefont{J.~S.} \bibnamefont{Tsai}},
  \bibinfo{author}{\bibfnamefont{Y.~A.} \bibnamefont{Pashkin}},
  \bibinfo{author}{\bibfnamefont{S.~J.} \bibnamefont{Chung}},
  \bibinfo{author}{\bibfnamefont{N.}~\bibnamefont{Goel}}, and
  \bibinfo{author}{\bibfnamefont{M.~B.} \bibnamefont{Santos}},
  \bibinfo{year}{2002}, {``}\bibinfo{title}{Nonmagnetic semiconductors as
  read-head sensors for ultra-high-density magnetic recording},{''}
  \bibinfo{journal}{Phys. Rev. Lett.} \textbf{\bibinfo{volume}{80}},
  \bibinfo{pages}{4012--4014}.

\bibitem[{\citenamefont{Solin} \emph{et~al.}(2000)\citenamefont{Solin, Thio,
  Hines, and Heremans}}]{Solin2000:S}
\bibinfo{author}{\bibnamefont{Solin}, \bibfnamefont{S.~A.}},
  \bibinfo{author}{\bibfnamefont{T.}~\bibnamefont{Thio}},
  \bibinfo{author}{\bibfnamefont{D.~R.} \bibnamefont{Hines}}, and
  \bibinfo{author}{\bibfnamefont{J.~J.} \bibnamefont{Heremans}},
  \bibinfo{year}{2000}, {``}\bibinfo{title}{Enhanced room-temperature geometric
  magnetoresistance in inhomogeneous narrow-gap semiconductors},{''}
  \bibinfo{journal}{{\sl Science}} \textbf{\bibinfo{volume}{289}},
  \bibinfo{pages}{1530--1532}.

\bibitem[{\citenamefont{Solomon}(1976)}]{Solomon1976:SSC}
\bibinfo{author}{\bibnamefont{Solomon}, \bibfnamefont{I.}},
  \bibinfo{year}{1976}, {``}\bibinfo{title}{Spin-dependent recombination in a
  silicon p-n junction},{''} \bibinfo{journal}{Solid State Commun.}
  \textbf{\bibinfo{volume}{20}},  \bibinfo{pages}{215--217}.

\bibitem[{\citenamefont{Song and Kim}(2002)}]{Song2002:PRB}
\bibinfo{author}{\bibnamefont{Song}, \bibfnamefont{P.~H.}}, and
  \bibinfo{author}{\bibfnamefont{K.~W.} \bibnamefont{Kim}},
  \bibinfo{year}{2002}, {``}\bibinfo{title}{Spin relaxation of conduction
  electrons in bulk {III-V} semiconductors},{''} \bibinfo{journal}{Phys. Rev.
  B} \textbf{\bibinfo{volume}{66}},  \bibinfo{pages}{035207}.

\bibitem[{\citenamefont{{Soulen Jr.}} \emph{et~al.}(1998)\citenamefont{{Soulen
  Jr.}, Byers, Osofsky, Nadgorny, Ambrose, Cheng, Broussard, Tanaka, Nowak,
  Moodera, Barry, and Coey}}]{Soulen1998:S}
\bibinfo{author}{\bibnamefont{{Soulen Jr.}}, \bibfnamefont{R.~J.}},
  \bibinfo{author}{\bibfnamefont{J.~M.} \bibnamefont{Byers}},
  \bibinfo{author}{\bibfnamefont{M.~S.} \bibnamefont{Osofsky}},
  \bibinfo{author}{\bibfnamefont{B.}~\bibnamefont{Nadgorny}},
  \bibinfo{author}{\bibfnamefont{T.}~\bibnamefont{Ambrose}},
  \bibinfo{author}{\bibfnamefont{S.~F.} \bibnamefont{Cheng}},
  \bibinfo{author}{\bibfnamefont{P.~R.} \bibnamefont{Broussard}},
  \bibinfo{author}{\bibfnamefont{C.~T.} \bibnamefont{Tanaka}},
  \bibinfo{author}{\bibfnamefont{J.}~\bibnamefont{Nowak}},
  \bibinfo{author}{\bibfnamefont{J.~S.} \bibnamefont{Moodera}},
  \bibinfo{author}{\bibfnamefont{A.}~\bibnamefont{Barry}}, and
  \bibinfo{author}{\bibfnamefont{J.~M.~D.} \bibnamefont{Coey}},
  \bibinfo{year}{1998}, {``}\bibinfo{title}{Measuring the spin polarization of
  a metal with a superconducting point contact},{''} \bibinfo{journal}{{\sl
  Science}} \textbf{\bibinfo{volume}{282}},  \bibinfo{pages}{85--88}.

\bibitem[{\citenamefont{Stearns}(1977)}]{Stearns1977:JMMM}
\bibinfo{author}{\bibnamefont{Stearns}, \bibfnamefont{M.~B.}},
  \bibinfo{year}{1977}, {``}\bibinfo{title}{Simple explanation of tunneling
  spin-polarization of {Fe, Co, Ni} and its alloys},{''} \bibinfo{journal}{J.
  Magn. Magn. Mater.} \textbf{\bibinfo{volume}{5}},  \bibinfo{pages}{167--171}.

\bibitem[{\citenamefont{Stein} \emph{et~al.}(1983)\citenamefont{Stein,
  v.~Klitzing, and Weimann}}]{Stein1983:PRL}
\bibinfo{author}{\bibnamefont{Stein}, \bibfnamefont{D.}},
  \bibinfo{author}{\bibfnamefont{K.}~\bibnamefont{v.~Klitzing}}, and
  \bibinfo{author}{\bibfnamefont{G.}~\bibnamefont{Weimann}},
  \bibinfo{year}{1983}, {``}\bibinfo{title}{Electron spin resonance on
  {GaAs-Al$_x$Ga$_{1-x}$As} heterostructures},{''} \bibinfo{journal}{Phys. Rev.
  Lett.} \textbf{\bibinfo{volume}{51}},  \bibinfo{pages}{130--133}.

\bibitem[{\citenamefont{Stesmans and Witters}(1981)}]{Stesmans1981:PRB}
\bibinfo{author}{\bibnamefont{Stesmans}, \bibfnamefont{A.}}, and
  \bibinfo{author}{\bibfnamefont{J.}~\bibnamefont{Witters}},
  \bibinfo{year}{1981}, {``}\bibinfo{title}{Fermi-surface structure effects
  revealed by the observation of conduction-electron-spin resonance in
  {Zn}},{''} \bibinfo{journal}{Phys. Rev. B} \textbf{\bibinfo{volume}{23}},
  \bibinfo{pages}{3159--3163}.

\bibitem[{\citenamefont{Stevens} \emph{et~al.}(2003)\citenamefont{Stevens,
  Smirl, {R. Bhat}, Najmaie, Sipe, and {van Driel}}}]{Stevens2003:PRL}
\bibinfo{author}{\bibnamefont{Stevens}, \bibfnamefont{M.~J.}},
  \bibinfo{author}{\bibfnamefont{A.~L.} \bibnamefont{Smirl}},
  \bibinfo{author}{\bibfnamefont{R.~D.} \bibnamefont{{R. Bhat}}},
  \bibinfo{author}{\bibfnamefont{A.}~\bibnamefont{Najmaie}},
  \bibinfo{author}{\bibfnamefont{J.~E.} \bibnamefont{Sipe}}, and
  \bibinfo{author}{\bibfnamefont{H.~M.} \bibnamefont{{van Driel}}},
  \bibinfo{year}{2003}, {``}\bibinfo{title}{Quantum interference control of
  ballistic pure spin currents in semiconductors},{''} \bibinfo{journal}{Phys.
  Rev. Lett.} \textbf{\bibinfo{volume}{90}},  \bibinfo{pages}{136603}.

\bibitem[{\citenamefont{Stiles}(1996)}]{Stiles1996:JAP}
\bibinfo{author}{\bibnamefont{Stiles}, \bibfnamefont{M.~D.}},
  \bibinfo{year}{1996}, {``}\bibinfo{title}{Spin-dependent interface
  transmission and reflection in magnetic multilayers},{''}
  \bibinfo{journal}{J. Appl. Phys.} \textbf{\bibinfo{volume}{79}},
  \bibinfo{pages}{5805--5810}.

\bibitem[{\citenamefont{Stiles}(2004)}]{Stiles2003:P}
\bibinfo{author}{\bibnamefont{Stiles}, \bibfnamefont{M.~D.}},
  \bibinfo{year}{2004}, {``}\bibinfo{title}{Exchange coupling},{''} in
  \emph{\bibinfo{booktitle}{Utrathin Magnetic Structures III}}, edited by
  \bibinfo{editor}{\bibfnamefont{B.}~\bibnamefont{Heinrich}} and
  \bibinfo{editor}{\bibfnamefont{J.~A.~C.} \bibnamefont{Bland}}
  (\bibinfo{publisher}{(Springer, New York, in press.)}).

\bibitem[{\citenamefont{Stiles and Penn}(2000)}]{Stiles2000:PRB}
\bibinfo{author}{\bibnamefont{Stiles}, \bibfnamefont{M.~D.}}, and
  \bibinfo{author}{\bibfnamefont{D.~R.} \bibnamefont{Penn}},
  \bibinfo{year}{2000}, {``}\bibinfo{title}{Calculation of spin-dependent
  interface resistance},{''} \bibinfo{journal}{Phys. Rev. B}
  \textbf{\bibinfo{volume}{61}},  \bibinfo{pages}{3200--3202}.

\bibitem[{\citenamefont{Stiles and Zangwill}(2002)}]{Stiles2002:PRB}
\bibinfo{author}{\bibnamefont{Stiles}, \bibfnamefont{M.~D.}}, and
  \bibinfo{author}{\bibfnamefont{A.}~\bibnamefont{Zangwill}},
  \bibinfo{year}{2002}, {``}\bibinfo{title}{Anatomy of spin-transfer
  torque},{''} \bibinfo{journal}{Phys. Rev. B} \textbf{\bibinfo{volume}{66}},
  \bibinfo{pages}{014407}.

\bibitem[{\citenamefont{Story} \emph{et~al.}(1986)\citenamefont{Story, Galazka,
  Frankel, and Wolff}}]{Story1986:PRL}
\bibinfo{author}{\bibnamefont{Story}, \bibfnamefont{T.}},
  \bibinfo{author}{\bibfnamefont{R.~R.} \bibnamefont{Galazka}},
  \bibinfo{author}{\bibfnamefont{R.~B.} \bibnamefont{Frankel}}, and
  \bibinfo{author}{\bibfnamefont{P.~A.} \bibnamefont{Wolff}},
  \bibinfo{year}{1986}, {``}\bibinfo{title}{Carrier-concentration-induced
  ferromagnetism in {PbSnMnTe}},{''} \bibinfo{journal}{Phys. Rev. Lett.}
  \textbf{\bibinfo{volume}{56}},  \bibinfo{pages}{777--779}.

\bibitem[{\citenamefont{Strand} \emph{et~al.}(2003)\citenamefont{Strand,
  Schultz, Isakovi{\'c}, Palmstr$\o$m, and Crowell}}]{Strand2003:PRL}
\bibinfo{author}{\bibnamefont{Strand}, \bibfnamefont{J.}},
  \bibinfo{author}{\bibfnamefont{B.~D.} \bibnamefont{Schultz}},
  \bibinfo{author}{\bibfnamefont{A.~F.} \bibnamefont{Isakovi{\'c}}},
  \bibinfo{author}{\bibfnamefont{C.~J.} \bibnamefont{Palmstr$\o$m}}, and
  \bibinfo{author}{\bibfnamefont{P.~A.} \bibnamefont{Crowell}},
  \bibinfo{year}{2003}, {``}\bibinfo{title}{Dynamical nuclear polarization by
  electrical spin injection in ferromagnet-semiconductor heterostructures},{''}
  \bibinfo{journal}{Phys. Rev. Lett.} \textbf{\bibinfo{volume}{91}},
  \bibinfo{pages}{036602}.

\bibitem[{\citenamefont{Stroud} \emph{et~al.}(2002)\citenamefont{Stroud,
  Hanbicki, Park, Petukhov, and Jonker}}]{Stroud2002:PRL}
\bibinfo{author}{\bibnamefont{Stroud}, \bibfnamefont{R.~M.}},
  \bibinfo{author}{\bibfnamefont{A.~T.} \bibnamefont{Hanbicki}},
  \bibinfo{author}{\bibfnamefont{Y.~D.} \bibnamefont{Park}},
  \bibinfo{author}{\bibfnamefont{A.~G.} \bibnamefont{Petukhov}}, and
  \bibinfo{author}{\bibfnamefont{B.~T.} \bibnamefont{Jonker}},
  \bibinfo{year}{2002}, {``}\bibinfo{title}{Reduction of spin injection
  efficiency by interface defect spin scattering in {ZnMnSe/AlGaAs-GaAs}
  spin-polarized light-emitting diodes},{''} \bibinfo{journal}{Phys. Rev.
  Lett.} \textbf{\bibinfo{volume}{89}},  \bibinfo{pages}{166602}.

\bibitem[{\citenamefont{Sugamo and {Kojima (Eds.)}}(2000)}]{Sugamo:2000}
\bibinfo{author}{\bibnamefont{Sugamo}, \bibfnamefont{S.}}, and
  \bibinfo{author}{\bibfnamefont{N.}~\bibnamefont{{Kojima (Eds.)}}},
  \bibinfo{year}{2000}, \emph{\bibinfo{title}{Magneto-Optics}}
  (\bibinfo{publisher}{Springer, Berlin}).

\bibitem[{\citenamefont{Suhl}(2002)}]{Suhl2002:P}
\bibinfo{author}{\bibnamefont{Suhl}, \bibfnamefont{H.}}, \bibinfo{year}{2002},
  {``}\bibinfo{title}{Proposal for {\it in situ} Enhancement of Electron Spin
  Polarization in Semiconductors},{''} \eprint{cond-mat/0206215}.

\bibitem[{\citenamefont{Sun}(2000)}]{Sun2000:PRB}
\bibinfo{author}{\bibnamefont{Sun}, \bibfnamefont{J.~Z.}},
  \bibinfo{year}{2000}, {``}\bibinfo{title}{Spin-current interaction with a
  monodomain magnetic body: A model study},{''} \bibinfo{journal}{Phys. Rev. B}
  \textbf{\bibinfo{volume}{62}},  \bibinfo{pages}{570--578}.

\bibitem[{\citenamefont{Sze}(1981)}]{Sze:1981}
\bibinfo{author}{\bibnamefont{Sze}, \bibfnamefont{S.~M.}},
  \bibinfo{year}{1981}, \emph{\bibinfo{title}{Physics of {Semiconductor}
  {Devices}}} (\bibinfo{publisher}{John Wiley, New York}).

\bibitem[{\citenamefont{Tackeuchi} \emph{et~al.}(1996)\citenamefont{Tackeuchi,
  Nishikawa, and Wada}}]{Tackeuchi1996:APL}
\bibinfo{author}{\bibnamefont{Tackeuchi}, \bibfnamefont{A.}},
  \bibinfo{author}{\bibfnamefont{Y.}~\bibnamefont{Nishikawa}}, and
  \bibinfo{author}{\bibfnamefont{O.}~\bibnamefont{Wada}}, \bibinfo{year}{1996},
  {``}\bibinfo{title}{Room-temperature electron spin dynamics in {GaAs/AlGaAs}
  quantum wells},{''} \bibinfo{journal}{Appl. Phys. Lett.}
  \textbf{\bibinfo{volume}{68}},  \bibinfo{pages}{797--799}.

\bibitem[{\citenamefont{Takahashi} \emph{et~al.}(1999)\citenamefont{Takahashi,
  Imamura, and Maekawa}}]{Takahashi1999:PRL}
\bibinfo{author}{\bibnamefont{Takahashi}, \bibfnamefont{S.}},
  \bibinfo{author}{\bibfnamefont{H.}~\bibnamefont{Imamura}}, and
  \bibinfo{author}{\bibfnamefont{S.}~\bibnamefont{Maekawa}},
  \bibinfo{year}{1999}, {``}\bibinfo{title}{Spin imbalance and
  magnetoresistance in ferromagnet/superconductor/ferromagnet double tunnel
  junctions},{''} \bibinfo{journal}{Phys. Rev. Lett.}
  \textbf{\bibinfo{volume}{82}},  \bibinfo{pages}{3911--3915}.

\bibitem[{\citenamefont{Takahashi and Maekawa}(1998)}]{Takahashi1998:PRL}
\bibinfo{author}{\bibnamefont{Takahashi}, \bibfnamefont{S.}}, and
  \bibinfo{author}{\bibfnamefont{S.}~\bibnamefont{Maekawa}},
  \bibinfo{year}{1998}, {``}\bibinfo{title}{Effect of {Coulomb} blockade on
  magnetoresistance in ferromagnetic tunnel junctions},{''}
  \bibinfo{journal}{Phys. Rev. Lett.} \textbf{\bibinfo{volume}{80}},
  \bibinfo{pages}{1758--1761}.

\bibitem[{\citenamefont{Takahashi and Maekawa}(2003)}]{Takahashi2003:PRB}
\bibinfo{author}{\bibnamefont{Takahashi}, \bibfnamefont{S.}}, and
  \bibinfo{author}{\bibfnamefont{S.}~\bibnamefont{Maekawa}},
  \bibinfo{year}{2003}, {``}\bibinfo{title}{Spin injection and detection in
  magnetic nanostructures},{''} \bibinfo{journal}{Phys. Rev. B}
  \textbf{\bibinfo{volume}{67}},  \bibinfo{pages}{052409}.

\bibitem[{\citenamefont{Takahashi} \emph{et~al.}(2001)\citenamefont{Takahashi,
  Yamashita, Koyama, and Maekawa}}]{Takahashi2001:JAP}
\bibinfo{author}{\bibnamefont{Takahashi}, \bibfnamefont{S.}},
  \bibinfo{author}{\bibfnamefont{T.}~\bibnamefont{Yamashita}},
  \bibinfo{author}{\bibfnamefont{T.}~\bibnamefont{Koyama}}, and
  \bibinfo{author}{\bibfnamefont{S.}~\bibnamefont{Maekawa}},
  \bibinfo{year}{2001}, {``}\bibinfo{title}{Joule heating generated by spin
  current through {Josephson} junctions},{''} \bibinfo{journal}{J. Appl. Phys.}
  \textbf{\bibinfo{volume}{89}},  \bibinfo{pages}{7505--7507}.

\bibitem[{\citenamefont{Tanaka}(2002)}]{Tanaka2002:SST}
\bibinfo{author}{\bibnamefont{Tanaka}, \bibfnamefont{M.}},
  \bibinfo{year}{2002}, {``}\bibinfo{title}{Ferromagnet {(MnAs)}/{III-V}
  semiconductor hybrid structures},{''} \bibinfo{journal}{Semicond. Sci.
  Technol.} \textbf{\bibinfo{volume}{17}},  \bibinfo{pages}{327--341}.

\bibitem[{\citenamefont{Tanaka and Higo}(2001)}]{Tanaka2001:PRL}
\bibinfo{author}{\bibnamefont{Tanaka}, \bibfnamefont{M.}}, and
  \bibinfo{author}{\bibfnamefont{Y.}~\bibnamefont{Higo}}, \bibinfo{year}{2001},
  {``}\bibinfo{title}{Large tunneling magnetoresistance in {GaMnAs/AlAs/GaMnAs}
  ferromagnetic semiconductor tunnel junctions},{''} \bibinfo{journal}{Phys.
  Rev. Lett.} \textbf{\bibinfo{volume}{87}},  \bibinfo{pages}{026602}.

\bibitem[{\citenamefont{Tanaka and Kashiwaya}(1995)}]{Kashiwaya1995:PRL}
\bibinfo{author}{\bibnamefont{Tanaka}, \bibfnamefont{Y.}}, and
  \bibinfo{author}{\bibfnamefont{S.}~\bibnamefont{Kashiwaya}},
  \bibinfo{year}{1995}, {``}\bibinfo{title}{Theory of tunneling spectroscopy of
  $d$-wave superconductors},{''} \bibinfo{journal}{Phys. Rev. Lett.}
  \textbf{\bibinfo{volume}{74}},  \bibinfo{pages}{3451--3454}.

\bibitem[{\citenamefont{Tang} \emph{et~al.}(2002)\citenamefont{Tang, Monzon,
  Jedema, Filip, {van Wees}, and Roukes}}]{Tang:2002}
\bibinfo{author}{\bibnamefont{Tang}, \bibfnamefont{H.~X.}},
  \bibinfo{author}{\bibfnamefont{F.~G.} \bibnamefont{Monzon}},
  \bibinfo{author}{\bibfnamefont{F.~J.} \bibnamefont{Jedema}},
  \bibinfo{author}{\bibfnamefont{A.~T.} \bibnamefont{Filip}},
  \bibinfo{author}{\bibfnamefont{B.~J.} \bibnamefont{{van Wees}}}, and
  \bibinfo{author}{\bibfnamefont{M.~L.} \bibnamefont{Roukes}},
  \bibinfo{year}{2002}, {``}\bibinfo{title}{Spin Injection and Transport in
  Micro- and Nanoscale Devices},{''} in \emph{\bibinfo{booktitle}{Semiconductor
  Spintronics and Quantum Computation}}, edited by
  \bibinfo{editor}{\bibfnamefont{D.}~\bibnamefont{Awschalom}},
  \bibinfo{editor}{\bibfnamefont{D.}~\bibnamefont{Loss}}, and
  \bibinfo{editor}{\bibfnamefont{N.}~\bibnamefont{Samarth}}
  (\bibinfo{publisher}{Springer, New York}),  \bibinfo{pages}{31--92}.

\bibitem[{\citenamefont{Tatara} \emph{et~al.}(1999)\citenamefont{Tatara,
  Garcia, Munoz, and Zhao}}]{Tatara1999:PRL}
\bibinfo{author}{\bibnamefont{Tatara}, \bibfnamefont{G.}},
  \bibinfo{author}{\bibfnamefont{N.}~\bibnamefont{Garcia}},
  \bibinfo{author}{\bibfnamefont{M.}~\bibnamefont{Munoz}}, and
  \bibinfo{author}{\bibfnamefont{Y.-W.} \bibnamefont{Zhao}},
  \bibinfo{year}{1999}, {``}\bibinfo{title}{Domain wall scattering explains
  300\% Ballistic magnetoconductance of nanocontacts},{''}
  \bibinfo{journal}{Phys. Rev. Lett.} \textbf{\bibinfo{volume}{83}},
  \bibinfo{pages}{2030--2033}.

\bibitem[{\citenamefont{Tedrow and
  Meservey}(1971{\natexlab{a}})}]{Tedrow1971:PRLb}
\bibinfo{author}{\bibnamefont{Tedrow}, \bibfnamefont{P.~M.}}, and
  \bibinfo{author}{\bibfnamefont{R.}~\bibnamefont{Meservey}},
  \bibinfo{year}{1971}{\natexlab{a}}, {``}\bibinfo{title}{Direct observation of
  spin-mixing in superconductors},{''} \bibinfo{journal}{Phys. Rev. Lett.}
  \textbf{\bibinfo{volume}{27}},  \bibinfo{pages}{919--921}.

\bibitem[{\citenamefont{Tedrow and
  Meservey}(1971{\natexlab{b}})}]{Tedrow1971:PRLa}
\bibinfo{author}{\bibnamefont{Tedrow}, \bibfnamefont{P.~M.}}, and
  \bibinfo{author}{\bibfnamefont{R.}~\bibnamefont{Meservey}},
  \bibinfo{year}{1971}{\natexlab{b}}, {``}\bibinfo{title}{Spin-dependent
  tunneling into ferromagnetic nickel},{''} \bibinfo{journal}{Phys. Rev. Lett.}
  \textbf{\bibinfo{volume}{26}},  \bibinfo{pages}{192--195}.

\bibitem[{\citenamefont{Tedrow and Meservey}(1973)}]{Tedrow1973:PRB}
\bibinfo{author}{\bibnamefont{Tedrow}, \bibfnamefont{P.~M.}}, and
  \bibinfo{author}{\bibfnamefont{R.}~\bibnamefont{Meservey}},
  \bibinfo{year}{1973}, {``}\bibinfo{title}{Spin polarization of electrons
  tunneling from films of {Fe, Co, Ni, Gd}},{''} \bibinfo{journal}{Phys. Rev.
  B} \textbf{\bibinfo{volume}{7}},  \bibinfo{pages}{318--326}.

\bibitem[{\citenamefont{Tedrow and Meservey}(1994)}]{Tedrow1994:PR}
\bibinfo{author}{\bibnamefont{Tedrow}, \bibfnamefont{P.~M.}}, and
  \bibinfo{author}{\bibfnamefont{R.}~\bibnamefont{Meservey}},
  \bibinfo{year}{1994}, {``}\bibinfo{title}{Spin-polarized electron
  tunneling},{''} \bibinfo{journal}{Phys. Rep.} \textbf{\bibinfo{volume}{238}},
   \bibinfo{pages}{173--243}.

\bibitem[{\citenamefont{Tedrow} \emph{et~al.}(1970)\citenamefont{Tedrow,
  Meservey, and Fulde}}]{Tedrow1970:PRL}
\bibinfo{author}{\bibnamefont{Tedrow}, \bibfnamefont{P.~M.}},
  \bibinfo{author}{\bibfnamefont{R.}~\bibnamefont{Meservey}}, and
  \bibinfo{author}{\bibfnamefont{P.}~\bibnamefont{Fulde}},
  \bibinfo{year}{1970}, {``}\bibinfo{title}{Magnetic field splitting of the
  quasiparticle states in superconducting aluminum films},{''}
  \bibinfo{journal}{Phys. Rev. Lett.} \textbf{\bibinfo{volume}{25}},
  \bibinfo{pages}{1270--1272}.

\bibitem[{\citenamefont{Tehrani} \emph{et~al.}(2000)\citenamefont{Tehrani,
  Engel, Slaughter, Chen, DeHerrera, Durlam, Naji, Whig, Janesky, and
  Calder}}]{Tehrani2000:IEEE}
\bibinfo{author}{\bibnamefont{Tehrani}, \bibfnamefont{S.}},
  \bibinfo{author}{\bibfnamefont{B.}~\bibnamefont{Engel}},
  \bibinfo{author}{\bibfnamefont{J.~M.} \bibnamefont{Slaughter}},
  \bibinfo{author}{\bibfnamefont{E.}~\bibnamefont{Chen}},
  \bibinfo{author}{\bibfnamefont{M.}~\bibnamefont{DeHerrera}},
  \bibinfo{author}{\bibfnamefont{M.}~\bibnamefont{Durlam}},
  \bibinfo{author}{\bibfnamefont{P.}~\bibnamefont{Naji}},
  \bibinfo{author}{\bibfnamefont{R.}~\bibnamefont{Whig}},
  \bibinfo{author}{\bibfnamefont{J.}~\bibnamefont{Janesky}}, and
  \bibinfo{author}{\bibfnamefont{J.}~\bibnamefont{Calder}},
  \bibinfo{year}{2000}, {``}\bibinfo{title}{Recent developments in Magnetic
  Tunnel Junction {MRAM}},{''} \bibinfo{journal}{IEEE Trans. Magn.}
  \textbf{\bibinfo{volume}{36}},  \bibinfo{pages}{2752--2757}.

\bibitem[{\citenamefont{Teran} \emph{et~al.}(2003)\citenamefont{Teran, Zhao,
  {Patan\`{e}}, Campion, Foxon, Eaves, and Gallagher}}]{Teran2003:APL}
\bibinfo{author}{\bibnamefont{Teran}, \bibfnamefont{F.~J.}},
  \bibinfo{author}{\bibfnamefont{L.~X.} \bibnamefont{Zhao}},
  \bibinfo{author}{\bibfnamefont{A.}~\bibnamefont{{Patan\`{e}}}},
  \bibinfo{author}{\bibfnamefont{R.~P.} \bibnamefont{Campion}},
  \bibinfo{author}{\bibfnamefont{C.~T.} \bibnamefont{Foxon}},
  \bibinfo{author}{\bibfnamefont{L.}~\bibnamefont{Eaves}}, and
  \bibinfo{author}{\bibfnamefont{B.~L.} \bibnamefont{Gallagher}},
  \bibinfo{year}{2003}, {``}\bibinfo{title}{Investigation of radiative
  recombination from {Mn}-related states in {Ga$_{1-x}$Mn$_x$As}},{''}
  \bibinfo{journal}{Appl Phys. Lett.} \textbf{\bibinfo{volume}{83}},
  \bibinfo{pages}{866--868}.

\bibitem[{\citenamefont{Te{\v{s}}anovi{\'c}}
  \emph{et~al.}(1986)\citenamefont{Te{\v{s}}anovi{\'c}, Jari{\'c}, and
  Maekawa}}]{Tesanovic1986:PRL}
\bibinfo{author}{\bibnamefont{Te{\v{s}}anovi{\'c}}, \bibfnamefont{Z.}},
  \bibinfo{author}{\bibfnamefont{M.~V.} \bibnamefont{Jari{\'c}}}, and
  \bibinfo{author}{\bibfnamefont{S.}~\bibnamefont{Maekawa}},
  \bibinfo{year}{1986}, {``}\bibinfo{title}{Quantum transport and surface
  scattering},{''} \bibinfo{journal}{Phys. Rev. Lett.}
  \textbf{\bibinfo{volume}{57}},  \bibinfo{pages}{2760--2763}.

\bibitem[{\citenamefont{Thomson}(1857)}]{Thomson1857:PRSL}
\bibinfo{author}{\bibnamefont{Thomson}, \bibfnamefont{W.}},
  \bibinfo{year}{1857}, {``}\bibinfo{title}{On the electro-dynamic qualities of
  metals:--{Effects} of magnetization on the electric conductivity of nickel
  and of iron},{''} \bibinfo{journal}{Proc. R. Soc. London}
  \textbf{\bibinfo{volume}{8}},  \bibinfo{pages}{546--550}.

\bibitem[{\citenamefont{Thruber and Smith}(2003)}]{Thruber2003:JMR}
\bibinfo{author}{\bibnamefont{Thruber}, \bibfnamefont{K.~R.}}, and
  \bibinfo{author}{\bibfnamefont{D.~D.} \bibnamefont{Smith}},
  \bibinfo{year}{2003}, {``}\bibinfo{title}{170 nm nuclear magnetic resonacnce
  imaging using magnetic resonance force spectroscopy},{''}
  \bibinfo{journal}{J. Magn. Reson.} \textbf{\bibinfo{volume}{162}},
  \bibinfo{pages}{336--340}.

\bibitem[{\citenamefont{Thurber} \emph{et~al.}(2002)\citenamefont{Thurber,
  Harrelli, Fainchtein, and Smith}}]{Thruber2002:APL}
\bibinfo{author}{\bibnamefont{Thurber}, \bibfnamefont{K.~R.}},
  \bibinfo{author}{\bibfnamefont{L.~E.} \bibnamefont{Harrelli}},
  \bibinfo{author}{\bibfnamefont{R.}~\bibnamefont{Fainchtein}}, and
  \bibinfo{author}{\bibfnamefont{D.~D.} \bibnamefont{Smith}},
  \bibinfo{year}{2002}, {``}\bibinfo{title}{Spin polarization contrast observed
  in {GaAs} by force-detected nuclear magnetic resonance},{''}
  \bibinfo{journal}{Appl. Phys. Lett.} \textbf{\bibinfo{volume}{80}},
  \bibinfo{pages}{1794--1796}.

\bibitem[{\citenamefont{Timm}(2003)}]{Timm2003:JPCM}
\bibinfo{author}{\bibnamefont{Timm}, \bibfnamefont{C.}}, \bibinfo{year}{2003},
  {``}\bibinfo{title}{Disorder effects in diluted magnetic semiconductors},{''}
  \bibinfo{journal}{J. Phys.: Condens. Matter} \textbf{\bibinfo{volume}{15}},
  \bibinfo{pages}{R1865--R1896}.

\bibitem[{\citenamefont{Ting and Cartoix\`{a}}(2003)}]{Ting2003:APL}
\bibinfo{author}{\bibnamefont{Ting}, \bibfnamefont{D.~Z.-Y.}}, and
  \bibinfo{author}{\bibfnamefont{X.}~\bibnamefont{Cartoix\`{a}}},
  \bibinfo{year}{2003}, {``}\bibinfo{title}{Bidirectional resonant tunneling
  spin pump},{''} \bibinfo{journal}{Appl. Phys. Lett.}
  \textbf{\bibinfo{volume}{83}},  \bibinfo{pages}{1391--1393}.

\bibitem[{\citenamefont{Ting and {Cartoxi\`{a}}}(2002)}]{Ting2002:APL}
\bibinfo{author}{\bibnamefont{Ting}, \bibfnamefont{D.~Z.-Y.}}, and
  \bibinfo{author}{\bibfnamefont{X.}~\bibnamefont{{Cartoxi\`{a}}}},
  \bibinfo{year}{2002}, {``}\bibinfo{title}{Resonant interband tunneling spin
  filter},{''} \bibinfo{journal}{Appl. Phys. Lett.}
  \textbf{\bibinfo{volume}{81}},  \bibinfo{pages}{4198--4200}.

\bibitem[{\citenamefont{Torrey}(1956)}]{Torrey1956:PR}
\bibinfo{author}{\bibnamefont{Torrey}, \bibfnamefont{H.~C.}},
  \bibinfo{year}{1956}, {``}\bibinfo{title}{Bloch equations with diffusion
  terms},{''} \bibinfo{journal}{Phys. Rev.} \textbf{\bibinfo{volume}{104}},
  \bibinfo{pages}{563--565}.

\bibitem[{\citenamefont{Troiani} \emph{et~al.}(2003)\citenamefont{Troiani,
  Molinari, and Hohenester}}]{Troiani2003:PRL}
\bibinfo{author}{\bibnamefont{Troiani}, \bibfnamefont{F.}},
  \bibinfo{author}{\bibfnamefont{E.}~\bibnamefont{Molinari}}, and
  \bibinfo{author}{\bibfnamefont{U.}~\bibnamefont{Hohenester}},
  \bibinfo{year}{2003}, {``}\bibinfo{title}{High-finesse optical quantum gates
  for electron spins in artificial molecules},{''} \bibinfo{journal}{Phys. Rev.
  Lett.} \textbf{\bibinfo{volume}{90}},  \bibinfo{pages}{206802}.

\bibitem[{\citenamefont{Tromp} \emph{et~al.}(2001)\citenamefont{Tromp, {van
  Gelderen}, Kelly, Brocks, and Bobbert}}]{Tromp2001:PRL}
\bibinfo{author}{\bibnamefont{Tromp}, \bibfnamefont{H.~J.}},
  \bibinfo{author}{\bibfnamefont{P.}~\bibnamefont{{van Gelderen}}},
  \bibinfo{author}{\bibfnamefont{P.~J.} \bibnamefont{Kelly}},
  \bibinfo{author}{\bibfnamefont{G.}~\bibnamefont{Brocks}}, and
  \bibinfo{author}{\bibfnamefont{P.~A.} \bibnamefont{Bobbert}},
  \bibinfo{year}{2001}, {``}\bibinfo{title}{{CaB$_6$}: A new semiconducting
  material for spin electronics},{''} \bibinfo{journal}{Phys. Rev. Lett.}
  \textbf{\bibinfo{volume}{87}},  \bibinfo{pages}{016401}.

\bibitem[{\citenamefont{Tsoi} \emph{et~al.}(1998)\citenamefont{Tsoi, Jansen,
  Bass, Chiang, Seck, Tsoi, and Wyder}}]{Tsoi1998:PRLa}
\bibinfo{author}{\bibnamefont{Tsoi}, \bibfnamefont{M.}},
  \bibinfo{author}{\bibfnamefont{A.~G.~M.} \bibnamefont{Jansen}},
  \bibinfo{author}{\bibfnamefont{J.}~\bibnamefont{Bass}},
  \bibinfo{author}{\bibfnamefont{W.-C.} \bibnamefont{Chiang}},
  \bibinfo{author}{\bibfnamefont{M.}~\bibnamefont{Seck}},
  \bibinfo{author}{\bibfnamefont{V.}~\bibnamefont{Tsoi}}, and
  \bibinfo{author}{\bibfnamefont{P.}~\bibnamefont{Wyder}},
  \bibinfo{year}{1998}, {``}\bibinfo{title}{Excitation of a magnetic multilayer
  by an electric current},{''} \bibinfo{journal}{Phys. Rev. Lett.}
  \textbf{\bibinfo{volume}{80}},  \bibinfo{pages}{4281--4284; {\bf 81}, 493
  (E)}.

\bibitem[{\citenamefont{Tsoi} \emph{et~al.}(2000)\citenamefont{Tsoi, Jansen,
  Wyder, Chiang, Tsoi, and Wyder}}]{Tsoi2000:N}
\bibinfo{author}{\bibnamefont{Tsoi}, \bibfnamefont{M.}},
  \bibinfo{author}{\bibfnamefont{A.~G.~M.} \bibnamefont{Jansen}},
  \bibinfo{author}{\bibfnamefont{P.}~\bibnamefont{Wyder}},
  \bibinfo{author}{\bibfnamefont{W.-C.} \bibnamefont{Chiang}},
  \bibinfo{author}{\bibfnamefont{V.}~\bibnamefont{Tsoi}}, and
  \bibinfo{author}{\bibfnamefont{P.}~\bibnamefont{Wyder}},
  \bibinfo{year}{2000}, {``}\bibinfo{title}{Generation and detection of
  phase-coherent current-driven magnons in magnetic multilayers},{''}
  \bibinfo{journal}{{\sl Nature}} \textbf{\bibinfo{volume}{406}},
  \bibinfo{pages}{46--49}.

\bibitem[{\citenamefont{Tsoi} \emph{et~al.}(2002)\citenamefont{Tsoi, Tsoi,
  Bass, Jansen, and Wyder}}]{Tsoi2002:PRL}
\bibinfo{author}{\bibnamefont{Tsoi}, \bibfnamefont{M.}},
  \bibinfo{author}{\bibfnamefont{V.}~\bibnamefont{Tsoi}},
  \bibinfo{author}{\bibfnamefont{J.}~\bibnamefont{Bass}},
  \bibinfo{author}{\bibfnamefont{A.~G.~M.} \bibnamefont{Jansen}}, and
  \bibinfo{author}{\bibfnamefont{P.}~\bibnamefont{Wyder}},
  \bibinfo{year}{2002}, {``}\bibinfo{title}{Current-driven resonances in
  magnetic multilayers},{''} \bibinfo{journal}{Phys. Rev. Lett.}
  \textbf{\bibinfo{volume}{89}},  \bibinfo{pages}{246803}.

\bibitem[{\citenamefont{Tsu and Esaki}(1973)}]{Tsu1973:APL}
\bibinfo{author}{\bibnamefont{Tsu}, \bibfnamefont{R.}}, and
  \bibinfo{author}{\bibfnamefont{L.}~\bibnamefont{Esaki}},
  \bibinfo{year}{1973}, {``}\bibinfo{title}{Tunneling in a finite
  superlattice},{''} \bibinfo{journal}{Appl. Phys. Lett.}
  \textbf{\bibinfo{volume}{22}},  \bibinfo{pages}{562--564}.

\bibitem[{\citenamefont{Tsubokawa}(1960)}]{Tsubokawa1960:JPSJ}
\bibinfo{author}{\bibnamefont{Tsubokawa}, \bibfnamefont{I.}},
  \bibinfo{year}{1960}, {``}\bibinfo{title}{On the magnetic properties of a
  {CrBr$_3$} single crystal},{''} \bibinfo{journal}{J. Phys. Soc. Japan}
  \textbf{\bibinfo{volume}{15}},  \bibinfo{pages}{1664--1668}.

\bibitem[{\citenamefont{Tsuei and Kirtley}(2000)}]{Tsuei2000:RMP}
\bibinfo{author}{\bibnamefont{Tsuei}, \bibfnamefont{C.~C.}}, and
  \bibinfo{author}{\bibfnamefont{J.~R.} \bibnamefont{Kirtley}},
  \bibinfo{year}{2000}, {``}\bibinfo{title}{Pairing symmetry in cuprate
  superconductors},{''} \bibinfo{journal}{Rev. Mod. Phys.}
  \textbf{\bibinfo{volume}{72}},  \bibinfo{pages}{969--1016}.

\bibitem[{\citenamefont{Tsui} \emph{et~al.}(1971)\citenamefont{Tsui, Dietz, and
  Walker}}]{Tsui1971:PRL}
\bibinfo{author}{\bibnamefont{Tsui}, \bibfnamefont{D.~C.}},
  \bibinfo{author}{\bibfnamefont{R.~E.} \bibnamefont{Dietz}}, and
  \bibinfo{author}{\bibfnamefont{L.~R.} \bibnamefont{Walker}},
  \bibinfo{year}{1971}, {``}\bibinfo{title}{Multiple magnon excitation in {NiO}
  by electron tunneling},{''} \bibinfo{journal}{Phys. Rev. Lett.}
  \textbf{\bibinfo{volume}{27}},  \bibinfo{pages}{1729--1732}.

\bibitem[{\citenamefont{Tsui} \emph{et~al.}(2003)\citenamefont{Tsui, Ma, and
  He}}]{Tsui2003:APL}
\bibinfo{author}{\bibnamefont{Tsui}, \bibfnamefont{F.}},
  \bibinfo{author}{\bibfnamefont{L.}~\bibnamefont{Ma}}, and
  \bibinfo{author}{\bibfnamefont{L.}~\bibnamefont{He}}, \bibinfo{year}{2003},
  {``}\bibinfo{title}{Magnetization-dependent rectification effect in a
  {Ge}-based magnetic heterojunction},{''} \bibinfo{journal}{Appl. Phys. Lett.}
  \textbf{\bibinfo{volume}{83}},  \bibinfo{pages}{954--956}.

\bibitem[{\citenamefont{Tsukagoshi}
  \emph{et~al.}(1999)\citenamefont{Tsukagoshi, Alphenaar, and
  Ago}}]{Tsukagoshi1999:N}
\bibinfo{author}{\bibnamefont{Tsukagoshi}, \bibfnamefont{K.}},
  \bibinfo{author}{\bibfnamefont{B.~W.} \bibnamefont{Alphenaar}}, and
  \bibinfo{author}{\bibfnamefont{H.}~\bibnamefont{Ago}}, \bibinfo{year}{1999},
  {``}\bibinfo{title}{Coherent transport of electron spin in a
  ferromagnetically contacted carbon nanotube},{''} \bibinfo{journal}{{\sl
  Nature}} \textbf{\bibinfo{volume}{401}},  \bibinfo{pages}{572--574}.

\bibitem[{\citenamefont{Tsymbal and Pettifor}(1997)}]{Tsymbal1997:JPCM}
\bibinfo{author}{\bibnamefont{Tsymbal}, \bibfnamefont{E.}}, and
  \bibinfo{author}{\bibfnamefont{D.~G.} \bibnamefont{Pettifor}},
  \bibinfo{year}{1997}, {``}\bibinfo{title}{Modeling of spin-polarized electron
  tunneling from 3d ferromagnets},{''} \bibinfo{journal}{J. Phys.: Condens.
  Matter} \textbf{\bibinfo{volume}{9}},  \bibinfo{pages}{L411--L417}.

\bibitem[{\citenamefont{Tyryshkin} \emph{et~al.}(2003)\citenamefont{Tyryshkin,
  Lyon, Astashkin, and Raitsimring}}]{Tyryshkin2003:PRB}
\bibinfo{author}{\bibnamefont{Tyryshkin}, \bibfnamefont{A.~M.}},
  \bibinfo{author}{\bibfnamefont{S.~A.} \bibnamefont{Lyon}},
  \bibinfo{author}{\bibfnamefont{A.~V.} \bibnamefont{Astashkin}}, and
  \bibinfo{author}{\bibfnamefont{A.~M.} \bibnamefont{Raitsimring}},
  \bibinfo{year}{2003}, {``}\bibinfo{title}{Electron spin relaxation times of
  phosphorus donors in silicon},{''} \bibinfo{journal}{Phys. Rev. B}
  \textbf{\bibinfo{volume}{68}},  \bibinfo{pages}{193207}.

\bibitem[{\citenamefont{Uenoyama and
  Sham}(1990{\natexlab{a}})}]{Uenoyama1990:PRB}
\bibinfo{author}{\bibnamefont{Uenoyama}, \bibfnamefont{T.}}, and
  \bibinfo{author}{\bibfnamefont{L.~J.} \bibnamefont{Sham}},
  \bibinfo{year}{1990}{\natexlab{a}}, {``}\bibinfo{title}{Carrier relaxation
  and luminescence polarization in quantum wells},{''} \bibinfo{journal}{Phys.
  Rev. B} \textbf{\bibinfo{volume}{42}},  \bibinfo{pages}{7114--7123}.

\bibitem[{\citenamefont{Uenoyama and
  Sham}(1990{\natexlab{b}})}]{Uenoyama1990:PRL}
\bibinfo{author}{\bibnamefont{Uenoyama}, \bibfnamefont{T.}}, and
  \bibinfo{author}{\bibfnamefont{L.~J.} \bibnamefont{Sham}},
  \bibinfo{year}{1990}{\natexlab{b}}, {``}\bibinfo{title}{Hole relaxation and
  luminescence polarization in doped and undoped quantum wells},{''}
  \bibinfo{journal}{Phys. Rev. Lett.} \textbf{\bibinfo{volume}{64}},
  \bibinfo{pages}{3070--3073}.

\bibitem[{\citenamefont{Upadhyay} \emph{et~al.}(1999)\citenamefont{Upadhyay,
  Louie, and Buhrman}}]{Upadhyay1999:APL}
\bibinfo{author}{\bibnamefont{Upadhyay}, \bibfnamefont{S.~K.}},
  \bibinfo{author}{\bibfnamefont{R.~N.} \bibnamefont{Louie}}, and
  \bibinfo{author}{\bibfnamefont{R.~A.} \bibnamefont{Buhrman}},
  \bibinfo{year}{1999}, {``}\bibinfo{title}{Spin filtering by ultrathin
  ferromagnetic films},{''} \bibinfo{journal}{Appl. Phys. Lett.}
  \textbf{\bibinfo{volume}{74}},  \bibinfo{pages}{3881--3883}.

\bibitem[{\citenamefont{Upadhyay} \emph{et~al.}(1998)\citenamefont{Upadhyay,
  Palanisami, Louie, and Buhrman}}]{Uphaday1998:PRL}
\bibinfo{author}{\bibnamefont{Upadhyay}, \bibfnamefont{S.~K.}},
  \bibinfo{author}{\bibfnamefont{A.}~\bibnamefont{Palanisami}},
  \bibinfo{author}{\bibfnamefont{R.~N.} \bibnamefont{Louie}}, and
  \bibinfo{author}{\bibfnamefont{R.~A.} \bibnamefont{Buhrman}},
  \bibinfo{year}{1998}, {``}\bibinfo{title}{Probing ferromagnets with {Andreev}
  reflection},{''} \bibinfo{journal}{Phys. Rev. Lett.}
  \textbf{\bibinfo{volume}{81}},  \bibinfo{pages}{3247--3250}.

\bibitem[{\citenamefont{Urazhdin} \emph{et~al.}(2003)\citenamefont{Urazhdin,
  Birge, {Pratt, Jr.}, and Bass}}]{Urazhdin2003:PRL}
\bibinfo{author}{\bibnamefont{Urazhdin}, \bibfnamefont{S.}},
  \bibinfo{author}{\bibfnamefont{N.~O.} \bibnamefont{Birge}},
  \bibinfo{author}{\bibfnamefont{W.~P.} \bibnamefont{{Pratt, Jr.}}}, and
  \bibinfo{author}{\bibfnamefont{J.}~\bibnamefont{Bass}}, \bibinfo{year}{2003},
  {``}\bibinfo{title}{Current-driven magnetic excitations in permalloy-based
  multilayer nanopillars},{''} \bibinfo{journal}{Phys. Rev. Lett.}
  \textbf{\bibinfo{volume}{91}},  \bibinfo{pages}{146803}.

\bibitem[{\citenamefont{Vagner}(2003)}]{Vagner2003:P}
\bibinfo{author}{\bibnamefont{Vagner}, \bibfnamefont{I.~D.}},
  \bibinfo{year}{2003}, {``}\bibinfo{title}{Nuclear Spintronics},{''}
  \eprint{cond-mat/0308244}.

\bibitem[{\citenamefont{Valet and Fert}(1993)}]{Valet1993:PRB}
\bibinfo{author}{\bibnamefont{Valet}, \bibfnamefont{T.}}, and
  \bibinfo{author}{\bibfnamefont{A.}~\bibnamefont{Fert}}, \bibinfo{year}{1993},
  {``}\bibinfo{title}{Theory of the perpendicular magnetoresistance in magnetic
  multilayers},{''} \bibinfo{journal}{Phys. Rev. B}
  \textbf{\bibinfo{volume}{48}},  \bibinfo{pages}{7099--7113}.

\bibitem[{\citenamefont{{van Dijken}}
  \emph{et~al.}(2002{\natexlab{a}})\citenamefont{{van Dijken}, Jiang, and
  Parkin}}]{vanDijken2002:APL}
\bibinfo{author}{\bibnamefont{{van Dijken}}, \bibfnamefont{S.}},
  \bibinfo{author}{\bibfnamefont{X.}~\bibnamefont{Jiang}}, and
  \bibinfo{author}{\bibfnamefont{S.~S.~P.} \bibnamefont{Parkin}},
  \bibinfo{year}{2002}{\natexlab{a}}, {``}\bibinfo{title}{Room temperature
  operation of a high output current magnetic tunnel transistor},{''}
  \bibinfo{journal}{Appl. Phys. Lett.} \textbf{\bibinfo{volume}{80}},
  \bibinfo{pages}{3364--3366}.

\bibitem[{\citenamefont{{van Dijken}}
  \emph{et~al.}(2002{\natexlab{b}})\citenamefont{{van Dijken}, Jiang, and
  Parkin}}]{vanDijken2002:PRB}
\bibinfo{author}{\bibnamefont{{van Dijken}}, \bibfnamefont{S.}},
  \bibinfo{author}{\bibfnamefont{X.}~\bibnamefont{Jiang}}, and
  \bibinfo{author}{\bibfnamefont{S.~S.~P.} \bibnamefont{Parkin}},
  \bibinfo{year}{2002}{\natexlab{b}}, {``}\bibinfo{title}{Spin-dependent hot
  electron transport in {Ni$_{81}$Fe$_{19}$ and Co$_{84}$Fe$_{16}$} films on
  {GaAs}(001)},{''} \bibinfo{journal}{Phys. Rev. B}
  \textbf{\bibinfo{volume}{66}},  \bibinfo{pages}{094417}.

\bibitem[{\citenamefont{{van Dijken}}
  \emph{et~al.}(2003{\natexlab{a}})\citenamefont{{van Dijken}, Jiang, and
  Parkin}}]{vanDijken2003:APLa}
\bibinfo{author}{\bibnamefont{{van Dijken}}, \bibfnamefont{S.}},
  \bibinfo{author}{\bibfnamefont{X.}~\bibnamefont{Jiang}}, and
  \bibinfo{author}{\bibfnamefont{S.~S.~P.} \bibnamefont{Parkin}},
  \bibinfo{year}{2003}{\natexlab{a}}, {``}\bibinfo{title}{Comparison of
  magnetocurrent and transfer ratio in magnetic tunnel transistors with
  spin-valve bases containing {Cu} and {Au} spacer layers},{''}
  \bibinfo{journal}{Appl. Phys. Lett.} \textbf{\bibinfo{volume}{82}},
  \bibinfo{pages}{775--777}.

\bibitem[{\citenamefont{{van Dijken}}
  \emph{et~al.}(2003{\natexlab{b}})\citenamefont{{van Dijken}, Jiang, and
  Parkin}}]{vanDijken2003:APLb}
\bibinfo{author}{\bibnamefont{{van Dijken}}, \bibfnamefont{S.}},
  \bibinfo{author}{\bibfnamefont{X.}~\bibnamefont{Jiang}}, and
  \bibinfo{author}{\bibfnamefont{S.~S.~P.} \bibnamefont{Parkin}},
  \bibinfo{year}{2003}{\natexlab{b}}, {``}\bibinfo{title}{Giant magnetocurrent
  exceeding 3400\% in magnetic tunnel transistors with spin-valve base},{''}
  \bibinfo{journal}{Appl. Phys. Lett.} \textbf{\bibinfo{volume}{83}},
  \bibinfo{pages}{951--953}.

\bibitem[{\citenamefont{{van Dijken}}
  \emph{et~al.}(2003{\natexlab{c}})\citenamefont{{van Dijken}, Jiang, and
  Parkin}}]{vanDijken2003:PRL}
\bibinfo{author}{\bibnamefont{{van Dijken}}, \bibfnamefont{S.}},
  \bibinfo{author}{\bibfnamefont{X.}~\bibnamefont{Jiang}}, and
  \bibinfo{author}{\bibfnamefont{S.~S.~P.} \bibnamefont{Parkin}},
  \bibinfo{year}{2003}{\natexlab{c}}, {``}\bibinfo{title}{Nonmonotonic bias
  voltage dependence of the magnetocurrent in {GaAs}-based magnetic tunnel
  transistors},{''} \bibinfo{journal}{Phys. Rev. Lett.}
  \textbf{\bibinfo{volume}{90}},  \bibinfo{pages}{197203}.

\bibitem[{\citenamefont{{Van Dorpe}}
  \emph{et~al.}(2003{\natexlab{a}})\citenamefont{{Van Dorpe}, Liu, Roy,
  Motsnyi, , Sawicki, Borghs, and {De Boeck}}}]{vanDorpe2003:P}
\bibinfo{author}{\bibnamefont{{Van Dorpe}}, \bibfnamefont{P.}},
  \bibinfo{author}{\bibfnamefont{Z.}~\bibnamefont{Liu}},
  \bibinfo{author}{\bibfnamefont{W.~V.} \bibnamefont{Roy}},
  \bibinfo{author}{\bibfnamefont{V.~F.} \bibnamefont{Motsnyi}}, ,
  \bibinfo{author}{\bibfnamefont{M.}~\bibnamefont{Sawicki}},
  \bibinfo{author}{\bibfnamefont{G.}~\bibnamefont{Borghs}}, and
  \bibinfo{author}{\bibfnamefont{J.}~\bibnamefont{{De Boeck}}},
  \bibinfo{year}{2003}{\natexlab{a}}, {``}\bibinfo{title}{Very high spin
  polarization in {GaAs} by injection from a {(Ga,Mn)As} {Zener} diode},{''}
  \bibinfo{note}{preprint}.

\bibitem[{\citenamefont{{Van Dorpe}}
  \emph{et~al.}(2003{\natexlab{b}})\citenamefont{{Van Dorpe}, Motsnyi, Nijboer,
  Goovaerts, Safarov, Das, {Van Roy}, Borghs, and {De Boeck}}}]{VanDorpe2002:P}
\bibinfo{author}{\bibnamefont{{Van Dorpe}}, \bibfnamefont{P.}},
  \bibinfo{author}{\bibfnamefont{V.}~\bibnamefont{Motsnyi}},
  \bibinfo{author}{\bibfnamefont{M.}~\bibnamefont{Nijboer}},
  \bibinfo{author}{\bibfnamefont{E.}~\bibnamefont{Goovaerts}},
  \bibinfo{author}{\bibfnamefont{V.~I.} \bibnamefont{Safarov}},
  \bibinfo{author}{\bibfnamefont{J.}~\bibnamefont{Das}},
  \bibinfo{author}{\bibfnamefont{W.}~\bibnamefont{{Van Roy}}},
  \bibinfo{author}{\bibfnamefont{G.}~\bibnamefont{Borghs}}, and
  \bibinfo{author}{\bibfnamefont{J.}~\bibnamefont{{De Boeck}}},
  \bibinfo{year}{2003}{\natexlab{b}}, {``}\bibinfo{title}{Highly efficient room
  temperature spin injection in a metal-insulator-semiconductor light emitting
  diode},{''} \bibinfo{journal}{Jpn. J. Appl. Phys.}
  \textbf{\bibinfo{volume}{42}},  \bibinfo{pages}{L502--L504}.

\bibitem[{\citenamefont{{Van Esch}} \emph{et~al.}(1997)\citenamefont{{Van
  Esch}, {Van Bockstal}, {De Boeck}, Verbanck, {van Steenbergen}, Wellmann,
  Grietens, Bogaerts, Herlach, and Borghs}}]{VanEsch1997:PRB}
\bibinfo{author}{\bibnamefont{{Van Esch}}, \bibfnamefont{A.}},
  \bibinfo{author}{\bibfnamefont{L.}~\bibnamefont{{Van Bockstal}}},
  \bibinfo{author}{\bibfnamefont{J.}~\bibnamefont{{De Boeck}}},
  \bibinfo{author}{\bibfnamefont{G.}~\bibnamefont{Verbanck}},
  \bibinfo{author}{\bibfnamefont{A.~S.} \bibnamefont{{van Steenbergen}}},
  \bibinfo{author}{\bibfnamefont{P.~J.} \bibnamefont{Wellmann}},
  \bibinfo{author}{\bibfnamefont{B.}~\bibnamefont{Grietens}},
  \bibinfo{author}{\bibfnamefont{R.}~\bibnamefont{Bogaerts}},
  \bibinfo{author}{\bibfnamefont{F.}~\bibnamefont{Herlach}}, and
  \bibinfo{author}{\bibfnamefont{G.}~\bibnamefont{Borghs}},
  \bibinfo{year}{1997}, {``}\bibinfo{title}{Interplay between the magnetic and
  transport properties in the {III-V} diluted magnetic semiconductors},{''}
  \bibinfo{journal}{Phys. Rev. B} \textbf{\bibinfo{volume}{56}},
  \bibinfo{pages}{13103--13112}.

\bibitem[{\citenamefont{{van Houten}} \emph{et~al.}(1989)\citenamefont{{van
  Houten}, Beenakker, Williamson, Broekaart, {van Loosdrecht}, van Wees, Mooij,
  Foxon, and Harris}}]{vanHouten1989:PRB}
\bibinfo{author}{\bibnamefont{{van Houten}}, \bibfnamefont{H.}},
  \bibinfo{author}{\bibfnamefont{C.~W.~J.} \bibnamefont{Beenakker}},
  \bibinfo{author}{\bibfnamefont{J.~G.} \bibnamefont{Williamson}},
  \bibinfo{author}{\bibfnamefont{M.~E.~I.} \bibnamefont{Broekaart}},
  \bibinfo{author}{\bibfnamefont{P.~H.~M.} \bibnamefont{{van Loosdrecht}}},
  \bibinfo{author}{\bibfnamefont{B.~J.} \bibnamefont{van Wees}},
  \bibinfo{author}{\bibfnamefont{J.~E.} \bibnamefont{Mooij}},
  \bibinfo{author}{\bibfnamefont{C.~T.} \bibnamefont{Foxon}}, and
  \bibinfo{author}{\bibfnamefont{J.~J.} \bibnamefont{Harris}},
  \bibinfo{year}{1989}, {``}\bibinfo{title}{Coherent electron focusing with
  quantum point contacts in a two-dimensional electron gas},{''}
  \bibinfo{journal}{Phys. Rev. B} \textbf{\bibinfo{volume}{39}},
  \bibinfo{pages}{8556--8575}.

\bibitem[{\citenamefont{{van Son}} \emph{et~al.}(1987)\citenamefont{{van Son},
  {van Kempen}, and Wyder}}]{vanSon1987:PRL}
\bibinfo{author}{\bibnamefont{{van Son}}, \bibfnamefont{P.~C.}},
  \bibinfo{author}{\bibfnamefont{H.}~\bibnamefont{{van Kempen}}}, and
  \bibinfo{author}{\bibfnamefont{P.}~\bibnamefont{Wyder}},
  \bibinfo{year}{1987}, {``}\bibinfo{title}{Boundary resistance of the
  ferromagnetic-nonferromagnetic metal interface},{''} \bibinfo{journal}{Phys.
  Rev. Lett.} \textbf{\bibinfo{volume}{58}},  \bibinfo{pages}{2271--2273}.

\bibitem[{\citenamefont{{van Wees}}(2000)}]{vanWees2000:PRL}
\bibinfo{author}{\bibnamefont{{van Wees}}, \bibfnamefont{B.~J.}},
  \bibinfo{year}{2000}, {``}\bibinfo{title}{Comment on `{Observation} of spin
  injection at a ferromagnet-semiconductor interface'},{''}
  \bibinfo{journal}{Phys. Rev. Lett.} \textbf{\bibinfo{volume}{84}},
  \bibinfo{pages}{5023}.

\bibitem[{\citenamefont{Vasilev} \emph{et~al.}(1993)\citenamefont{Vasilev,
  Daiminger, Straka, Forchel, Kochereshko, Sandler, and
  Uraltsev}}]{Vasilev1993:SM}
\bibinfo{author}{\bibnamefont{Vasilev}, \bibfnamefont{A.~M.}},
  \bibinfo{author}{\bibfnamefont{F.}~\bibnamefont{Daiminger}},
  \bibinfo{author}{\bibfnamefont{J.}~\bibnamefont{Straka}},
  \bibinfo{author}{\bibfnamefont{A.}~\bibnamefont{Forchel}},
  \bibinfo{author}{\bibfnamefont{V.~P.} \bibnamefont{Kochereshko}},
  \bibinfo{author}{\bibfnamefont{G.~L.} \bibnamefont{Sandler}}, and
  \bibinfo{author}{\bibfnamefont{I.~N.} \bibnamefont{Uraltsev}},
  \bibinfo{year}{1993}, {``}\bibinfo{title}{Optical orientation of holes and
  electrons in strained layer {InGaAs/GaAs} quantum-wells},{''}
  \bibinfo{journal}{Superlattices Microstruct.} \textbf{\bibinfo{volume}{13}},
  \bibinfo{pages}{97--100}.

\bibitem[{\citenamefont{Vas'ko} \emph{et~al.}(1997)\citenamefont{Vas'ko,
  Larkin, Kraus, Nikolaev, Grupp, Nordman, and Goldman}}]{Vasko1997:PRL}
\bibinfo{author}{\bibnamefont{Vas'ko}, \bibfnamefont{V.~A.}},
  \bibinfo{author}{\bibfnamefont{V.~A.} \bibnamefont{Larkin}},
  \bibinfo{author}{\bibfnamefont{P.~A.} \bibnamefont{Kraus}},
  \bibinfo{author}{\bibfnamefont{K.~R.} \bibnamefont{Nikolaev}},
  \bibinfo{author}{\bibfnamefont{D.~E.} \bibnamefont{Grupp}},
  \bibinfo{author}{\bibfnamefont{C.~A.} \bibnamefont{Nordman}}, and
  \bibinfo{author}{\bibfnamefont{A.~M.} \bibnamefont{Goldman}},
  \bibinfo{year}{1997}, {``}\bibinfo{title}{Critical current suppression in a
  superconductor by injection of spin-polarized carriers from a
  ferromagnet},{''} \bibinfo{journal}{Phys. Rev. Lett.}
  \textbf{\bibinfo{volume}{78}},  \bibinfo{pages}{1134--1137}.

\bibitem[{\citenamefont{Vas'ko} \emph{et~al.}(1998)\citenamefont{Vas'ko,
  Nikolaev, Larkin, Kraus, and Goldman}}]{Vasko1998:APL}
\bibinfo{author}{\bibnamefont{Vas'ko}, \bibfnamefont{V.~A.}},
  \bibinfo{author}{\bibfnamefont{K.}~\bibnamefont{Nikolaev}},
  \bibinfo{author}{\bibfnamefont{V.~A.} \bibnamefont{Larkin}},
  \bibinfo{author}{\bibfnamefont{P.~A.} \bibnamefont{Kraus}}, and
  \bibinfo{author}{\bibfnamefont{A.~M.} \bibnamefont{Goldman}},
  \bibinfo{year}{1998}, {``}\bibinfo{title}{Differential conductance of the
  ferromagnet/superconductor interface of
  {DyBa$_2$Cu$_3$O$_7$/La$_{2/3}$Ba$_{1/3}$MnO$_3$} heterostructures},{''}
  \bibinfo{journal}{Appl. Phys. Lett.} \textbf{\bibinfo{volume}{73}},
  \bibinfo{pages}{844--846}.

\bibitem[{\citenamefont{Vekua} \emph{et~al.}(1976)\citenamefont{Vekua, Dzhioev,
  Zakharchenya, and Fleisher}}]{Vekua1976:FTP}
\bibinfo{author}{\bibnamefont{Vekua}, \bibfnamefont{V.~L.}},
  \bibinfo{author}{\bibfnamefont{R.~I.} \bibnamefont{Dzhioev}},
  \bibinfo{author}{\bibfnamefont{B.~P.} \bibnamefont{Zakharchenya}}, and
  \bibinfo{author}{\bibfnamefont{V.~G.} \bibnamefont{Fleisher}},
  \bibinfo{year}{1976}, {``}\bibinfo{title}{Hanle effect in optical orientation
  of electrons in n-type semiconductors},{''} \bibinfo{journal}{Fiz. Tekh.
  Poluprovodn.} \textbf{\bibinfo{volume}{10}},  \bibinfo{pages}{354--357}
  \bibinfo{note}{[Sov. Phys. Semicond. {\bf 10}, 210-212 (1976)]}.

\bibitem[{\citenamefont{Versluijs} \emph{et~al.}(2001)\citenamefont{Versluijs,
  Bari, and Coey}}]{Versluijs2002:PRL}
\bibinfo{author}{\bibnamefont{Versluijs}, \bibfnamefont{J.~J.}},
  \bibinfo{author}{\bibfnamefont{M.~A.} \bibnamefont{Bari}}, and
  \bibinfo{author}{\bibfnamefont{J.~M.~D.} \bibnamefont{Coey}},
  \bibinfo{year}{2001}, {``}\bibinfo{title}{Magnetoresistance of half-metallic
  oxide nanocontacts},{''} \bibinfo{journal}{Phys. Rev. Lett.}
  \textbf{\bibinfo{volume}{87}},  \bibinfo{pages}{026601}.

\bibitem[{\citenamefont{Vescial} \emph{et~al.}(1964)\citenamefont{Vescial,
  VanderVen, and Schumacher}}]{Vescial1964:PR}
\bibinfo{author}{\bibnamefont{Vescial}, \bibfnamefont{F.}},
  \bibinfo{author}{\bibfnamefont{N.~S.} \bibnamefont{VanderVen}}, and
  \bibinfo{author}{\bibfnamefont{R.~T.} \bibnamefont{Schumacher}},
  \bibinfo{year}{1964}, {``}\bibinfo{title}{Spin-lattice relaxation time of
  conduction electrons in sodium metal},{''} \bibinfo{journal}{Phys. Rev.}
  \textbf{\bibinfo{volume}{134}},  \bibinfo{pages}{A1286--A1290}.

\bibitem[{\citenamefont{Viglin} \emph{et~al.}(1991)\citenamefont{Viglin,
  Osipov, and Samokhvalov}}]{Viglin1991:FTT}
\bibinfo{author}{\bibnamefont{Viglin}, \bibfnamefont{N.~A.}},
  \bibinfo{author}{\bibfnamefont{V.~V.} \bibnamefont{Osipov}}, and
  \bibinfo{author}{\bibfnamefont{A.~A.} \bibnamefont{Samokhvalov}},
  \bibinfo{year}{1991}, {``}\bibinfo{title}{Increase in the transmission of
  microwave power through a ferromagnetic semiconductor {HgCr$_2$Se$_4$} in a
  strong electric field},{''} \bibinfo{journal}{Fiz. Tverd. Tela}
  \textbf{\bibinfo{volume}{30}},  \bibinfo{pages}{2695--2698}
  \bibinfo{note}{[Sov. Phys. Solid State {\bf 33}, 1523-1525 (1991)]}.

\bibitem[{\citenamefont{Viglin} \emph{et~al.}(1997)\citenamefont{Viglin,
  Osipov, Samokhvalov, and Reznitskikh}}]{Viglin1997:PLDS}
\bibinfo{author}{\bibnamefont{Viglin}, \bibfnamefont{N.~A.}},
  \bibinfo{author}{\bibfnamefont{V.~V.} \bibnamefont{Osipov}},
  \bibinfo{author}{\bibfnamefont{A.~A.} \bibnamefont{Samokhvalov}}, and
  \bibinfo{author}{\bibfnamefont{O.~G.} \bibnamefont{Reznitskikh}},
  \bibinfo{year}{1997}, {``}\bibinfo{title}{Generation In the milliemter band
  for the semiconductor-ferromagnetic semiconductor heterostructure},{''}
  \bibinfo{journal}{Phys. Low.-Dimens. Semicond. Struct.}
  \textbf{\bibinfo{volume}{1/2}},  \bibinfo{pages}{89--94}.

\bibitem[{\citenamefont{Vignale and {D'Amico}}(2003)}]{Vignale2003:SST}
\bibinfo{author}{\bibnamefont{Vignale}, \bibfnamefont{G.}}, and
  \bibinfo{author}{\bibfnamefont{I.}~\bibnamefont{{D'Amico}}},
  \bibinfo{year}{2003}, {``}\bibinfo{title}{Coulomb drag, magnetoresistance,
  and spin-current injection in magnetic multilayers},{''}
  \bibinfo{journal}{Solid State Commun.} \textbf{\bibinfo{volume}{127}},
  \bibinfo{pages}{829--834}.

\bibitem[{\citenamefont{Vina} \emph{et~al.}(2001)\citenamefont{Vina, Martin,
  and Aichmayr}}]{Vina2001:PB}
\bibinfo{author}{\bibnamefont{Vina}, \bibfnamefont{L.}},
  \bibinfo{author}{\bibfnamefont{M.~D.} \bibnamefont{Martin}}, and
  \bibinfo{author}{\bibfnamefont{G.}~\bibnamefont{Aichmayr}},
  \bibinfo{year}{2001}, {``}\bibinfo{title}{Spin dynamics and spin-dependent
  interactions in semiconductor heterostructures},{''}
  \bibinfo{journal}{Physica B} \textbf{\bibinfo{volume}{298}},
  \bibinfo{pages}{376--383}.

\bibitem[{\citenamefont{Vlutters} \emph{et~al.}(2001)\citenamefont{Vlutters,
  {van 't Erve}, Jansen, Kim, and Lodder}}]{Vlutters2001:PRB}
\bibinfo{author}{\bibnamefont{Vlutters}, \bibfnamefont{R.}},
  \bibinfo{author}{\bibfnamefont{O.~M.~J.} \bibnamefont{{van 't Erve}}},
  \bibinfo{author}{\bibfnamefont{R.}~\bibnamefont{Jansen}},
  \bibinfo{author}{\bibfnamefont{S.~D.} \bibnamefont{Kim}}, and
  \bibinfo{author}{\bibfnamefont{J.~C.} \bibnamefont{Lodder}},
  \bibinfo{year}{2001}, {``}\bibinfo{title}{Modeling of spin-dependent
  hot-electron transport in the spin-valve transistor},{''}
  \bibinfo{journal}{Phys. Rev. B} \textbf{\bibinfo{volume}{65}},
  \bibinfo{pages}{024416}.

\bibitem[{\citenamefont{{von Moln\'{a}r} and
  Methfess}(1967)}]{vonMolnar1967:JAP}
\bibinfo{author}{\bibnamefont{{von Moln\'{a}r}}, \bibfnamefont{S.}}, and
  \bibinfo{author}{\bibfnamefont{S.}~\bibnamefont{Methfess}},
  \bibinfo{year}{1967}, {``}\bibinfo{title}{Giant negative magnetoresistance in
  ferromagnetic {Eu$_{1-x}$Gd$_x$Se}},{''} \bibinfo{journal}{J. Appl. Phys.}
  \textbf{\bibinfo{volume}{38}},  \bibinfo{pages}{959--964}.

\bibitem[{\citenamefont{Voskoboynikov}
  \emph{et~al.}(1998)\citenamefont{Voskoboynikov, Liu, and
  Lee}}]{Voskoboynikov1998:PRB}
\bibinfo{author}{\bibnamefont{Voskoboynikov}, \bibfnamefont{A.}},
  \bibinfo{author}{\bibfnamefont{S.~S.} \bibnamefont{Liu}}, and
  \bibinfo{author}{\bibfnamefont{C.~P.} \bibnamefont{Lee}},
  \bibinfo{year}{1998}, {``}\bibinfo{title}{Spin-dependent electronic tunneling
  at zero magnetic field},{''} \bibinfo{journal}{Phys. Rev. B}
  \textbf{\bibinfo{volume}{58}},  \bibinfo{pages}{15397--15400}.

\bibitem[{\citenamefont{Voskoboynikov}
  \emph{et~al.}(1999)\citenamefont{Voskoboynikov, Liu, and
  Lee}}]{Voskoboynikov1999:PRB}
\bibinfo{author}{\bibnamefont{Voskoboynikov}, \bibfnamefont{A.}},
  \bibinfo{author}{\bibfnamefont{S.~S.} \bibnamefont{Liu}}, and
  \bibinfo{author}{\bibfnamefont{C.~P.} \bibnamefont{Lee}},
  \bibinfo{year}{1999}, {``}\bibinfo{title}{Spin-dependent tunneling in
  double-barrier semiconductor heterostructures},{''} \bibinfo{journal}{Phys.
  Rev. B} \textbf{\bibinfo{volume}{59}},  \bibinfo{pages}{12514--12520}.

\bibitem[{\citenamefont{Vrijen} \emph{et~al.}(2000)\citenamefont{Vrijen,
  Yablonovitch, Wang, Jiang, Balandin, Roychowdhury, Mor, and
  DiVincenzo}}]{Vrijen2000:PRA}
\bibinfo{author}{\bibnamefont{Vrijen}, \bibfnamefont{R.}},
  \bibinfo{author}{\bibfnamefont{E.}~\bibnamefont{Yablonovitch}},
  \bibinfo{author}{\bibfnamefont{K.}~\bibnamefont{Wang}},
  \bibinfo{author}{\bibfnamefont{H.~W.} \bibnamefont{Jiang}},
  \bibinfo{author}{\bibfnamefont{A.}~\bibnamefont{Balandin}},
  \bibinfo{author}{\bibfnamefont{V.}~\bibnamefont{Roychowdhury}},
  \bibinfo{author}{\bibfnamefont{T.}~\bibnamefont{Mor}}, and
  \bibinfo{author}{\bibfnamefont{D.}~\bibnamefont{DiVincenzo}},
  \bibinfo{year}{2000}, {``}\bibinfo{title}{Electron-spin-resonance transistors
  for quantum computing in silicon-germanium heterostructures},{''}
  \bibinfo{journal}{Phys. Rev. A} \textbf{\bibinfo{volume}{62}},
  \bibinfo{pages}{012306}.

\bibitem[{\citenamefont{Vurgaftman and Meyer}(2003)}]{Vurgaftman2003:PRB}
\bibinfo{author}{\bibnamefont{Vurgaftman}, \bibfnamefont{I.}}, and
  \bibinfo{author}{\bibfnamefont{J.~R.} \bibnamefont{Meyer}},
  \bibinfo{year}{2003}, {``}\bibinfo{title}{Spin-polarizing properties of the
  {InAs}/{(AlSb)}/{GaMnSb}/{(AlSb)}/{InAs} ferromagnetic resonant interband
  tunneling diode},{''} \bibinfo{journal}{Phys. Rev. B}
  \textbf{\bibinfo{volume}{67}},  \bibinfo{pages}{125209}.

\bibitem[{\citenamefont{Waag} \emph{et~al.}(2001)\citenamefont{Waag, Gruber,
  Reuscher, Ossau, Schmidt, and Molenkamp}}]{Waag2001:JS}
\bibinfo{author}{\bibnamefont{Waag}, \bibfnamefont{A.}},
  \bibinfo{author}{\bibfnamefont{T.}~\bibnamefont{Gruber}},
  \bibinfo{author}{\bibfnamefont{G.}~\bibnamefont{Reuscher}},
  \bibinfo{author}{\bibfnamefont{R.~F.~W.} \bibnamefont{Ossau}},
  \bibinfo{author}{\bibfnamefont{G.}~\bibnamefont{Schmidt}}, and
  \bibinfo{author}{\bibfnamefont{L.~W.} \bibnamefont{Molenkamp}},
  \bibinfo{year}{2001}, {``}\bibinfo{title}{Spin manipulation using magnetic
  {II-VI} semiconductors},{''} \bibinfo{journal}{J. Supercond.}
  \textbf{\bibinfo{volume}{14}},  \bibinfo{pages}{291--298}.

\bibitem[{\citenamefont{Wagner} \emph{et~al.}(1993)\citenamefont{Wagner,
  Schneider, Richards, Fischer, and Ploog}}]{Wagner1993:PRB}
\bibinfo{author}{\bibnamefont{Wagner}, \bibfnamefont{J.}},
  \bibinfo{author}{\bibfnamefont{H.}~\bibnamefont{Schneider}},
  \bibinfo{author}{\bibfnamefont{D.}~\bibnamefont{Richards}},
  \bibinfo{author}{\bibfnamefont{A.}~\bibnamefont{Fischer}}, and
  \bibinfo{author}{\bibfnamefont{K.}~\bibnamefont{Ploog}},
  \bibinfo{year}{1993}, {``}\bibinfo{title}{Observation of extremely long
  electron-spin-relaxation times in p-type $\delta$-doped
  {GaAs/Al$_x$Ga$_{1-x}$As} double heterostructures},{''}
  \bibinfo{journal}{Phys. Rev. B} \textbf{\bibinfo{volume}{47}},
  \bibinfo{pages}{4786--4789}.

\bibitem[{\citenamefont{Wagoner}(1960)}]{Wagoner1960:PR}
\bibinfo{author}{\bibnamefont{Wagoner}, \bibfnamefont{G.}},
  \bibinfo{year}{1960}, {``}\bibinfo{title}{Spin resonance of charge carriers
  in graphite},{''} \bibinfo{journal}{Phys. Rev.}
  \textbf{\bibinfo{volume}{118}},  \bibinfo{pages}{647--653}.

\bibitem[{\citenamefont{Waintal} \emph{et~al.}(2000)\citenamefont{Waintal,
  Myers, Brouwer, and Ralph}}]{Wanital2000:PRB}
\bibinfo{author}{\bibnamefont{Waintal}, \bibfnamefont{X.}},
  \bibinfo{author}{\bibfnamefont{E.~B.} \bibnamefont{Myers}},
  \bibinfo{author}{\bibfnamefont{P.~W.} \bibnamefont{Brouwer}}, and
  \bibinfo{author}{\bibfnamefont{D.~C.} \bibnamefont{Ralph}},
  \bibinfo{year}{2000}, {``}\bibinfo{title}{Role of spin-dependent interface
  scattering in generating current-induced torques in magnetic
  multilayers},{''} \bibinfo{journal}{Phys. Rev. B}
  \textbf{\bibinfo{volume}{62}},  \bibinfo{pages}{12317--12327}.

\bibitem[{\citenamefont{Wald} \emph{et~al.}(1994)\citenamefont{Wald,
  Kouwenhoven, McEuen, {van der Vaart}, and Foxon}}]{Wald1994:PRL}
\bibinfo{author}{\bibnamefont{Wald}, \bibfnamefont{K.~R.}},
  \bibinfo{author}{\bibfnamefont{L.~P.} \bibnamefont{Kouwenhoven}},
  \bibinfo{author}{\bibfnamefont{P.~L.} \bibnamefont{McEuen}},
  \bibinfo{author}{\bibfnamefont{N.~C.} \bibnamefont{{van der Vaart}}}, and
  \bibinfo{author}{\bibfnamefont{C.~T.} \bibnamefont{Foxon}},
  \bibinfo{year}{1994}, {``}\bibinfo{title}{Local dynamic nuclear polarization
  using quantum point contacts},{''} \bibinfo{journal}{Phys. Rev. Lett.}
  \textbf{\bibinfo{volume}{73}},  \bibinfo{pages}{1011--1014}.

\bibitem[{\citenamefont{Walsh}
  \emph{et~al.}(1966{\natexlab{a}})\citenamefont{Walsh, Rupp, and
  Schmidt}}]{Walsh1966:PR}
\bibinfo{author}{\bibnamefont{Walsh}, \bibfnamefont{W.~M.}},
  \bibinfo{author}{\bibfnamefont{L.~W.} \bibnamefont{Rupp}}, and
  \bibinfo{author}{\bibfnamefont{P.~H.} \bibnamefont{Schmidt}},
  \bibinfo{year}{1966}{\natexlab{a}}, {``}\bibinfo{title}{g value of potassium
  conduction electrons},{''} \bibinfo{journal}{Phys. Rev.}
  \textbf{\bibinfo{volume}{142}},  \bibinfo{pages}{414--417}.

\bibitem[{\citenamefont{Walsh}
  \emph{et~al.}(1966{\natexlab{b}})\citenamefont{Walsh, Rupp, and
  Schmidt}}]{Walsh1966:PRL}
\bibinfo{author}{\bibnamefont{Walsh}, \bibfnamefont{W.~M.}},
  \bibinfo{author}{\bibfnamefont{L.~W.} \bibnamefont{Rupp}}, and
  \bibinfo{author}{\bibfnamefont{P.~H.} \bibnamefont{Schmidt}},
  \bibinfo{year}{1966}{\natexlab{b}}, {``}\bibinfo{title}{g values of rubidium
  and cesium conduction electrons},{''} \bibinfo{journal}{Phys. Rev. Lett.}
  \textbf{\bibinfo{volume}{16}},  \bibinfo{pages}{181--183}.

\bibitem[{\citenamefont{Wang}
  \emph{et~al.}(2003{\natexlab{a}})\citenamefont{Wang, Wang, and
  Guo}}]{Wang2002:P}
\bibinfo{author}{\bibnamefont{Wang}, \bibfnamefont{B.}},
  \bibinfo{author}{\bibfnamefont{J.}~\bibnamefont{Wang}}, and
  \bibinfo{author}{\bibfnamefont{H.}~\bibnamefont{Guo}},
  \bibinfo{year}{2003}{\natexlab{a}}, {``}\bibinfo{title}{Quantum spin field
  effect transistor},{''} \bibinfo{journal}{Phys. Rev. B}
  \textbf{\bibinfo{volume}{67}},  \bibinfo{pages}{092408}.

\bibitem[{\citenamefont{Wang} \emph{et~al.}(2002)\citenamefont{Wang, Tondra,
  Nordman, Qian, Daughton, Lange, Brownell, Tran, and Schuetz}}]{Wang2002:JAP}
\bibinfo{author}{\bibnamefont{Wang}, \bibfnamefont{D.}},
  \bibinfo{author}{\bibfnamefont{M.}~\bibnamefont{Tondra}},
  \bibinfo{author}{\bibfnamefont{C.}~\bibnamefont{Nordman}},
  \bibinfo{author}{\bibfnamefont{Z.}~\bibnamefont{Qian}},
  \bibinfo{author}{\bibfnamefont{J.~M.} \bibnamefont{Daughton}},
  \bibinfo{author}{\bibfnamefont{E.}~\bibnamefont{Lange}},
  \bibinfo{author}{\bibfnamefont{D.}~\bibnamefont{Brownell}},
  \bibinfo{author}{\bibfnamefont{L.}~\bibnamefont{Tran}}, and
  \bibinfo{author}{\bibfnamefont{J.}~\bibnamefont{Schuetz}},
  \bibinfo{year}{2002}, {``}\bibinfo{title}{Prototype spin-dependent tunneling
  isolators integrated with integrated circuit electronics},{''}
  \bibinfo{journal}{J. Appl. Phys.} \textbf{\bibinfo{volume}{91}},
  \bibinfo{pages}{8405--8407}.

\bibitem[{\citenamefont{Wang}
  \emph{et~al.}(2003{\natexlab{b}})\citenamefont{Wang, Khodaparast, Kono,
  Slupinski, Oiwa, and Munekata}}]{Wang2003:P}
\bibinfo{author}{\bibnamefont{Wang}, \bibfnamefont{J.}},
  \bibinfo{author}{\bibfnamefont{G.~A.} \bibnamefont{Khodaparast}},
  \bibinfo{author}{\bibfnamefont{J.}~\bibnamefont{Kono}},
  \bibinfo{author}{\bibfnamefont{T.}~\bibnamefont{Slupinski}},
  \bibinfo{author}{\bibfnamefont{A.}~\bibnamefont{Oiwa}}, and
  \bibinfo{author}{\bibfnamefont{H.}~\bibnamefont{Munekata}},
  \bibinfo{year}{2003}{\natexlab{b}}, {``}\bibinfo{title}{Ultrafast softening
  in {InMnAs}},{''} \eprint{cond-mat/0305017}.

\bibitem[{\citenamefont{Wang and Hu}(2002)}]{Wang2002b:P}
\bibinfo{author}{\bibnamefont{Wang}, \bibfnamefont{Q.}}, and
  \bibinfo{author}{\bibfnamefont{C.-R.} \bibnamefont{Hu}},
  \bibinfo{year}{2002}, {``}\bibinfo{title}{Resonant {Andreev} Reflection in a
  {FIDI'D'} system where {F} is a spin-polarized and {DI'D'} is a $d$-wave
  superconductor containing a \{100\}{$|$}\{110\} grain boundary},{''}
  \bibinfo{note}{preprint}.

\bibitem[{\citenamefont{Wangsness and Bloch}(1953)}]{Wangsness1953:PR}
\bibinfo{author}{\bibnamefont{Wangsness}, \bibfnamefont{R.~K.}}, and
  \bibinfo{author}{\bibfnamefont{F.}~\bibnamefont{Bloch}},
  \bibinfo{year}{1953}, {``}\bibinfo{title}{The dynamical theory of nuclear
  induction},{''} \bibinfo{journal}{Phys. Rev.} \textbf{\bibinfo{volume}{89}},
  \bibinfo{pages}{728--739}.

\bibitem[{\citenamefont{Watson} \emph{et~al.}(2003)\citenamefont{Watson, Potok,
  Marcus, and Umansky}}]{Watson2003:PRL}
\bibinfo{author}{\bibnamefont{Watson}, \bibfnamefont{S.~K.}},
  \bibinfo{author}{\bibfnamefont{R.~M.} \bibnamefont{Potok}},
  \bibinfo{author}{\bibfnamefont{C.~M.} \bibnamefont{Marcus}}, and
  \bibinfo{author}{\bibfnamefont{V.}~\bibnamefont{Umansky}},
  \bibinfo{year}{2003}, {``}\bibinfo{title}{Experimental realization of a
  quantum spin pump},{''} \bibinfo{journal}{Phys. Rev. Lett.}
  \textbf{\bibinfo{volume}{91}},  \bibinfo{pages}{258301}.

\bibitem[{\citenamefont{Weger}(1963)}]{Weger1963:PR}
\bibinfo{author}{\bibnamefont{Weger}, \bibfnamefont{M.}}, \bibinfo{year}{1963},
  {``}\bibinfo{title}{Nuclear polarization in homogeneous {InSb} by a direct
  current},{''} \bibinfo{journal}{Phys. Rev.} \textbf{\bibinfo{volume}{132}},
  \bibinfo{pages}{581--588}.

\bibitem[{\citenamefont{Wegrowe}(2000)}]{Wegrowe2000:PRB}
\bibinfo{author}{\bibnamefont{Wegrowe}, \bibfnamefont{J.-E.}},
  \bibinfo{year}{2000}, {``}\bibinfo{title}{Thermokinetic approach of the
  generalized {Landau-Lifshitz-Gilbert} equation with spin-polarized
  current},{''} \bibinfo{journal}{Phys. Rev. B} \textbf{\bibinfo{volume}{62}},
  \bibinfo{pages}{1067--1074}.

\bibitem[{\citenamefont{Wegrowe} \emph{et~al.}(1999)\citenamefont{Wegrowe,
  Kelly, Jaccard, Guittienne, and Ansermet}}]{Wegrowe1999:EL}
\bibinfo{author}{\bibnamefont{Wegrowe}, \bibfnamefont{J.~E.}},
  \bibinfo{author}{\bibfnamefont{D.}~\bibnamefont{Kelly}},
  \bibinfo{author}{\bibfnamefont{Y.}~\bibnamefont{Jaccard}},
  \bibinfo{author}{\bibfnamefont{P.}~\bibnamefont{Guittienne}}, and
  \bibinfo{author}{\bibfnamefont{J.~P.} \bibnamefont{Ansermet}},
  \bibinfo{year}{1999}, {``}\bibinfo{title}{Current-induced magnetization
  reversal in magnetic nanowires},{''} \bibinfo{journal}{Europhys. Lett.}
  \textbf{\bibinfo{volume}{45}},  \bibinfo{pages}{626--632}.

\bibitem[{\citenamefont{Wei}(2002)}]{Wei2002:JS}
\bibinfo{author}{\bibnamefont{Wei}, \bibfnamefont{J.~Y.~T.}},
  \bibinfo{year}{2002}, {``}\bibinfo{title}{Spin-injection quasiparticle
  nonequilibrium in cuprate/manganite heterostructures},{''}
  \bibinfo{journal}{J. Supercond.} \textbf{\bibinfo{volume}{15}},
  \bibinfo{pages}{67--70}.

\bibitem[{\citenamefont{Wei} \emph{et~al.}(1999)\citenamefont{Wei, Yeh, Fui,
  and Vasquez}}]{Wei1999:JAP}
\bibinfo{author}{\bibnamefont{Wei}, \bibfnamefont{J.~Y.~T.}},
  \bibinfo{author}{\bibfnamefont{N.-C.} \bibnamefont{Yeh}},
  \bibinfo{author}{\bibfnamefont{C.~C.} \bibnamefont{Fui}}, and
  \bibinfo{author}{\bibfnamefont{R.~P.} \bibnamefont{Vasquez}},
  \bibinfo{year}{1999}, {``}\bibinfo{title}{Tunneling spectroscopy study of
  spin-polarized quasiparticle injection effects in cuprate/manganite
  heterostructures},{''} \bibinfo{journal}{J. Appl. Phys.}
  \textbf{\bibinfo{volume}{85}},  \bibinfo{pages}{5350--5352}.

\bibitem[{\citenamefont{Wei} \emph{et~al.}(1998)\citenamefont{Wei, Yeh,
  Garrigus, and Strasik}}]{Wei1998:PRL}
\bibinfo{author}{\bibnamefont{Wei}, \bibfnamefont{J.~Y.~T.}},
  \bibinfo{author}{\bibfnamefont{N.-C.} \bibnamefont{Yeh}},
  \bibinfo{author}{\bibfnamefont{D.~F.} \bibnamefont{Garrigus}}, and
  \bibinfo{author}{\bibfnamefont{M.}~\bibnamefont{Strasik}},
  \bibinfo{year}{1998}, {``}\bibinfo{title}{Directional tunneling and {Andreev}
  reflection on {YBa$_2$Cu$_3$O$_{7-\delta}$} single crystals: Predominance of
  $d$-wave pairing symmetry verified with the generalized {Blonder}, {Tinkham},
  and {Klapwijk} theory},{''} \bibinfo{journal}{Phys. Rev. Lett.}
  \textbf{\bibinfo{volume}{81}},  \bibinfo{pages}{2542--2545}.

\bibitem[{\citenamefont{Weisbuch and Hermann}(1977)}]{Weisbuch1977:PRB}
\bibinfo{author}{\bibnamefont{Weisbuch}, \bibfnamefont{C.}}, and
  \bibinfo{author}{\bibfnamefont{C.}~\bibnamefont{Hermann}},
  \bibinfo{year}{1977}, {``}\bibinfo{title}{Optical detection of
  conduction-electron spin resonance in {GaAs}, {Ga$_{1-x}$In$_x$As}, and
  {Ga$_{1-x}$Al$_x$As}},{''} \bibinfo{journal}{Phys. Rev. B}
  \textbf{\bibinfo{volume}{15}},  \bibinfo{pages}{816--822}.

\bibitem[{\citenamefont{Weisbuch and Vinter}(1991)}]{Weisbuch:1991}
\bibinfo{author}{\bibnamefont{Weisbuch}, \bibfnamefont{C.}}, and
  \bibinfo{author}{\bibfnamefont{B.}~\bibnamefont{Vinter}},
  \bibinfo{year}{1991}, \emph{\bibinfo{title}{Quantum Semiconductor
  Structures--Fundamentals and Applications}} (\bibinfo{publisher}{Academic,
  New York}).

\bibitem[{\citenamefont{Wen} \emph{et~al.}(1968)\citenamefont{Wen, Hershenov,
  {von Philipsborn}, and Pinch}}]{Wen1968:IEEETM}
\bibinfo{author}{\bibnamefont{Wen}, \bibfnamefont{C.~P.}},
  \bibinfo{author}{\bibfnamefont{B.}~\bibnamefont{Hershenov}},
  \bibinfo{author}{\bibfnamefont{H.}~\bibnamefont{{von Philipsborn}}}, and
  \bibinfo{author}{\bibfnamefont{H.~L.} \bibnamefont{Pinch}},
  \bibinfo{year}{1968}, {``}\bibinfo{title}{Device application feasibility of
  single-crystal {CdCr$_2$Se$_4$}, a ferromagnetic semiconducting spinel.},{''}
  \bibinfo{journal}{IEEE Trans. Magn.} \textbf{\bibinfo{volume}{4}},
  \bibinfo{pages}{702--704}.

\bibitem[{\citenamefont{Werthamer}(1969)}]{Werthamer:1969}
\bibinfo{author}{\bibnamefont{Werthamer}, \bibfnamefont{N.~R.}},
  \bibinfo{year}{1969}, {``}\bibinfo{title}{The {Ginzburg-Landau} equations and
  their extensions},{''} in \emph{\bibinfo{booktitle}{Superconductivity, Vol.
  1}}, edited by \bibinfo{editor}{\bibfnamefont{R.~D.} \bibnamefont{Parks}}
  (\bibinfo{publisher}{Marcel Dekker, New York}),  \bibinfo{pages}{321--370}.

\bibitem[{\citenamefont{Wexler}(1966)}]{Wexler1966:PPSL}
\bibinfo{author}{\bibnamefont{Wexler}, \bibfnamefont{G.}},
  \bibinfo{year}{1966}, {``}\bibinfo{title}{Size effect and non-local
  {Boltzmann} transport equation in orifice and disk geometry},{''}
  \bibinfo{journal}{Proc. Phys. Soc. London} \textbf{\bibinfo{volume}{89}},
  \bibinfo{pages}{927--941}.

\bibitem[{\citenamefont{Wiesendanger}(1998)}]{Wiesendanger:1998}
\bibinfo{author}{\bibnamefont{Wiesendanger}, \bibfnamefont{R.}},
  \bibinfo{year}{1998}, {``}\bibinfo{title}{Spin-Polarized Scanning Tunneling
  Microscopy},{''} in \emph{\bibinfo{booktitle}{Scanning Probe Microscopy}},
  edited by \bibinfo{editor}{\bibfnamefont{R.}~\bibnamefont{Wiesendanger}}
  (\bibinfo{publisher}{Springer, New York}),  \bibinfo{pages}{71--95}.

\bibitem[{\citenamefont{Wiesendanger}
  \emph{et~al.}(1990)\citenamefont{Wiesendanger, G{\"{u}}ntherodt,
  G{\"{u}}ntherodt, Gambino, and Ruf}}]{Wiesendanger1990:PRL}
\bibinfo{author}{\bibnamefont{Wiesendanger}, \bibfnamefont{R.}},
  \bibinfo{author}{\bibfnamefont{H.-J.} \bibnamefont{G{\"{u}}ntherodt}},
  \bibinfo{author}{\bibfnamefont{G.}~\bibnamefont{G{\"{u}}ntherodt}},
  \bibinfo{author}{\bibfnamefont{R.~J.} \bibnamefont{Gambino}}, and
  \bibinfo{author}{\bibfnamefont{R.}~\bibnamefont{Ruf}}, \bibinfo{year}{1990},
  {``}\bibinfo{title}{Observation of vacuum tunneling of spin-polarized
  electrons with the scanning tunneling microscope},{''}
  \bibinfo{journal}{Phys. Rev. Lett.} \textbf{\bibinfo{volume}{65}},
  \bibinfo{pages}{247--250}.

\bibitem[{\citenamefont{Wilamowski and Jantsch}(2002)}]{Wilamowski2002:PE}
\bibinfo{author}{\bibnamefont{Wilamowski}, \bibfnamefont{Z.}}, and
  \bibinfo{author}{\bibfnamefont{W.}~\bibnamefont{Jantsch}},
  \bibinfo{year}{2002}, {``}\bibinfo{title}{{ESR} studies of the
  {Bychkov}-{Rashba} field in modulation doped {Si/SiGe} quantum wells},{''}
  \bibinfo{journal}{Physica E} \textbf{\bibinfo{volume}{12}},
  \bibinfo{pages}{439--442}.

\bibitem[{\citenamefont{Wilamowski and Jantsch}(2004)}]{Wilamowski2004:PRB}
\bibinfo{author}{\bibnamefont{Wilamowski}, \bibfnamefont{Z.}}, and
  \bibinfo{author}{\bibfnamefont{W.}~\bibnamefont{Jantsch}},
  \bibinfo{year}{2004}, {``}\bibinfo{title}{Suppression of spin relaxation of
  conduction electrons by cyclotron motion},{''} \bibinfo{journal}{Phys. Rev.
  B} \textbf{\bibinfo{volume}{69}},  \bibinfo{pages}{035328}.

\bibitem[{\citenamefont{Wilamowski}
  \emph{et~al.}(2002)\citenamefont{Wilamowski, Jantsch, Malissa, and
  {R\"{o}ssler}}}]{Wilamowski2002:PRB}
\bibinfo{author}{\bibnamefont{Wilamowski}, \bibfnamefont{Z.}},
  \bibinfo{author}{\bibfnamefont{W.}~\bibnamefont{Jantsch}},
  \bibinfo{author}{\bibfnamefont{H.}~\bibnamefont{Malissa}}, and
  \bibinfo{author}{\bibfnamefont{U.}~\bibnamefont{{R\"{o}ssler}}},
  \bibinfo{year}{2002}, {``}\bibinfo{title}{Evidence and evaluation of the
  {Bychkov-Rashba} effect in {SiGe/Si/SiGe} quantum wells},{''}
  \bibinfo{journal}{Phys. Rev. B} \textbf{\bibinfo{volume}{66}},
  \bibinfo{pages}{195315}.

\bibitem[{\citenamefont{Winkler}(2004)}]{Winkler2003:lanl}
\bibinfo{author}{\bibnamefont{Winkler}, \bibfnamefont{R.}},
  \bibinfo{year}{2004}, {``}\bibinfo{title}{Spin orientation and spin
  precession in inversion-asymmetric quasi two-dimensional electron
  systems},{''} \bibinfo{journal}{Phys. Rev. B} \textbf{\bibinfo{volume}{69}},
  \bibinfo{pages}{045317}.

\bibitem[{\citenamefont{Wissinger} \emph{et~al.}(1998)\citenamefont{Wissinger,
  {R\"{o}ssler}, Winkler, Jusserand, and Richards}}]{Wissinger1998:PRB}
\bibinfo{author}{\bibnamefont{Wissinger}, \bibfnamefont{L.}},
  \bibinfo{author}{\bibfnamefont{U.}~\bibnamefont{{R\"{o}ssler}}},
  \bibinfo{author}{\bibfnamefont{R.}~\bibnamefont{Winkler}},
  \bibinfo{author}{\bibfnamefont{B.}~\bibnamefont{Jusserand}}, and
  \bibinfo{author}{\bibfnamefont{D.}~\bibnamefont{Richards}},
  \bibinfo{year}{1998}, {``}\bibinfo{title}{Spin splitting in the electron
  subband of asymmetric {GaAs/Al$_x$Ga$_{1-x}$As} quantum wells: {The}
  multiband envelope function approach},{''} \bibinfo{journal}{Phys. Rev. B}
  \textbf{\bibinfo{volume}{58}},  \bibinfo{pages}{15375--15377}.

\bibitem[{\citenamefont{Wolf} \emph{et~al.}(2001)\citenamefont{Wolf, Awschalom,
  Buhrman, Daughton, {von Moln\'{a}r}, Roukes, Chtchelkanova, and
  Treger}}]{Wolf2001:S}
\bibinfo{author}{\bibnamefont{Wolf}, \bibfnamefont{S.~A.}},
  \bibinfo{author}{\bibfnamefont{D.~D.} \bibnamefont{Awschalom}},
  \bibinfo{author}{\bibfnamefont{R.~A.} \bibnamefont{Buhrman}},
  \bibinfo{author}{\bibfnamefont{J.~M.} \bibnamefont{Daughton}},
  \bibinfo{author}{\bibfnamefont{S.}~\bibnamefont{{von Moln\'{a}r}}},
  \bibinfo{author}{\bibfnamefont{M.~L.} \bibnamefont{Roukes}},
  \bibinfo{author}{\bibfnamefont{A.~Y.} \bibnamefont{Chtchelkanova}}, and
  \bibinfo{author}{\bibfnamefont{D.~M.} \bibnamefont{Treger}},
  \bibinfo{year}{2001}, {``}\bibinfo{title}{Spintronics: {A} spin-based
  electronics vision for the future},{''} \bibinfo{journal}{{\sl Science}}
  \textbf{\bibinfo{volume}{294}},  \bibinfo{pages}{1488--1495}.

\bibitem[{\citenamefont{Wolf and Treger}(2000)}]{Wolf2000:IEEE}
\bibinfo{author}{\bibnamefont{Wolf}, \bibfnamefont{S.~A.}}, and
  \bibinfo{author}{\bibfnamefont{D.}~\bibnamefont{Treger}},
  \bibinfo{year}{2000}, {``}\bibinfo{title}{Spintronics: {A} new paradigm for
  electronics for the new millenium},{''} \bibinfo{journal}{IEEE Trans. Magn.}
  \textbf{\bibinfo{volume}{36}},  \bibinfo{pages}{2748--2751}.

\bibitem[{\citenamefont{Wong} \emph{et~al.}(1999)\citenamefont{Wong, Frank,
  Solomon, Wann, and Welser}}]{Wong1999:PIEEE}
\bibinfo{author}{\bibnamefont{Wong}, \bibfnamefont{H.-S.~P.}},
  \bibinfo{author}{\bibfnamefont{D.~J.} \bibnamefont{Frank}},
  \bibinfo{author}{\bibfnamefont{P.~M.} \bibnamefont{Solomon}},
  \bibinfo{author}{\bibfnamefont{C.~H.~J.} \bibnamefont{Wann}}, and
  \bibinfo{author}{\bibfnamefont{J.~J.} \bibnamefont{Welser}},
  \bibinfo{year}{1999}, {``}\bibinfo{title}{Nanoscale {CMOS}},{''}
  \bibinfo{journal}{Proc. IEEE} \textbf{\bibinfo{volume}{87}},
  \bibinfo{pages}{537--570}.

\bibitem[{\citenamefont{Worledge and
  Geballe}(2000{\natexlab{a}})}]{Worledge2000:PRB}
\bibinfo{author}{\bibnamefont{Worledge}, \bibfnamefont{D.~C.}}, and
  \bibinfo{author}{\bibfnamefont{T.~H.} \bibnamefont{Geballe}},
  \bibinfo{year}{2000}{\natexlab{a}}, {``}\bibinfo{title}{Maki analysis of
  spin-polarized tunneling in an oxide ferromagnet},{''}
  \bibinfo{journal}{Phys. Rev. B} \textbf{\bibinfo{volume}{62}},
  \bibinfo{pages}{447--451}.

\bibitem[{\citenamefont{Worledge and
  Geballe}(2000{\natexlab{b}})}]{Worledge2000:PRL}
\bibinfo{author}{\bibnamefont{Worledge}, \bibfnamefont{D.~C.}}, and
  \bibinfo{author}{\bibfnamefont{T.~H.} \bibnamefont{Geballe}},
  \bibinfo{year}{2000}{\natexlab{b}}, {``}\bibinfo{title}{Negative
  spin-polarization of {SrRuO$_3$}},{''} \bibinfo{journal}{Phys. Rev. Lett.}
  \textbf{\bibinfo{volume}{85}},  \bibinfo{pages}{5182--5185}.

\bibitem[{\citenamefont{Wu}(2001)}]{Wu2001:JS}
\bibinfo{author}{\bibnamefont{Wu}, \bibfnamefont{M.~W.}}, \bibinfo{year}{2001},
  {``}\bibinfo{title}{Kinetic theory of spin coherence of electrons in
  semiconductors},{''} \bibinfo{journal}{J. Supercond.}
  \textbf{\bibinfo{volume}{14}},  \bibinfo{pages}{245--259}.

\bibitem[{\citenamefont{Wu and Kuwata-Gonokami}(2002)}]{Wu2002:SSC}
\bibinfo{author}{\bibnamefont{Wu}, \bibfnamefont{M.~W.}}, and
  \bibinfo{author}{\bibfnamefont{M.}~\bibnamefont{Kuwata-Gonokami}},
  \bibinfo{year}{2002}, {``}\bibinfo{title}{Can {D'yakonov-Perel'} effect cause
  spin dephasing in {GaAs}(110) quantum wells?},{''} \bibinfo{journal}{Solid
  State Commun.} \textbf{\bibinfo{volume}{121}},  \bibinfo{pages}{509--512}.

\bibitem[{\citenamefont{Wu and Metiu}(2000)}]{Wu2000:PRB}
\bibinfo{author}{\bibnamefont{Wu}, \bibfnamefont{M.~W.}}, and
  \bibinfo{author}{\bibfnamefont{H.}~\bibnamefont{Metiu}},
  \bibinfo{year}{2000}, {``}\bibinfo{title}{Kinetics of spin coherence of
  electrons in an undoped semiconductor quantum well},{''}
  \bibinfo{journal}{Phys. Rev. B} \textbf{\bibinfo{volume}{61}},
  \bibinfo{pages}{2945--2955}.

\bibitem[{\citenamefont{Wu and Ning}(2000)}]{Wu2000:PSS}
\bibinfo{author}{\bibnamefont{Wu}, \bibfnamefont{M.~W.}}, and
  \bibinfo{author}{\bibfnamefont{C.~Z.} \bibnamefont{Ning}},
  \bibinfo{year}{2000}, {``}\bibinfo{title}{D'yakonov-{Perel'} effect on spin
  dephasing in n-type {GaAs}},{''} \bibinfo{journal}{Phys. Status Solidi B}
  \textbf{\bibinfo{volume}{222}},  \bibinfo{pages}{523--534}.

\bibitem[{\citenamefont{Wunnicke} \emph{et~al.}(2002)\citenamefont{Wunnicke,
  Mavropoulos, Zeller, Dederichs, and Grundler}}]{Wunnicke2002:PRB}
\bibinfo{author}{\bibnamefont{Wunnicke}, \bibfnamefont{O.}},
  \bibinfo{author}{\bibfnamefont{P.}~\bibnamefont{Mavropoulos}},
  \bibinfo{author}{\bibfnamefont{R.}~\bibnamefont{Zeller}},
  \bibinfo{author}{\bibfnamefont{P.~H.} \bibnamefont{Dederichs}}, and
  \bibinfo{author}{\bibfnamefont{D.}~\bibnamefont{Grundler}},
  \bibinfo{year}{2002}, {``}\bibinfo{title}{Ballistic spin injection from
  {Fe(001)} into {ZnSe} and {GaAs}},{''} \bibinfo{journal}{Phys. Rev. B}
  \textbf{\bibinfo{volume}{65}},  \bibinfo{pages}{241306}.

\bibitem[{\citenamefont{Xia} \emph{et~al.}(2002)\citenamefont{Xia, Kelly,
  Bauer, and Turek}}]{Xia2002:PRL}
\bibinfo{author}{\bibnamefont{Xia}, \bibfnamefont{K.}},
  \bibinfo{author}{\bibfnamefont{P.~J.} \bibnamefont{Kelly}},
  \bibinfo{author}{\bibfnamefont{G.~E.~W.} \bibnamefont{Bauer}}, and
  \bibinfo{author}{\bibfnamefont{I.}~\bibnamefont{Turek}},
  \bibinfo{year}{2002}, {``}\bibinfo{title}{Spin-dependent transparency of
  ferromagnet/superconductor interfaces},{''} \bibinfo{journal}{Phys. Rev.
  Lett.} \textbf{\bibinfo{volume}{89}},  \bibinfo{pages}{166603}.

\bibitem[{\citenamefont{Xiang} \emph{et~al.}(2002)\citenamefont{Xiang, Zhu, Du,
  Landry, and Xiao}}]{Xiang2002:PRB}
\bibinfo{author}{\bibnamefont{Xiang}, \bibfnamefont{X.~H.}},
  \bibinfo{author}{\bibfnamefont{T.}~\bibnamefont{Zhu}},
  \bibinfo{author}{\bibfnamefont{J.}~\bibnamefont{Du}},
  \bibinfo{author}{\bibfnamefont{G.}~\bibnamefont{Landry}}, and
  \bibinfo{author}{\bibfnamefont{J.~Q.} \bibnamefont{Xiao}},
  \bibinfo{year}{2002}, {``}\bibinfo{title}{Effects of density of states on
  bias dependence in magnetic tunnel junctions},{''} \bibinfo{journal}{Phys.
  Rev. B} \textbf{\bibinfo{volume}{66}},  \bibinfo{pages}{174407}.

\bibitem[{\citenamefont{Xie} \emph{et~al.}(2003)\citenamefont{Xie, Ahn, Smith,
  Bishop, and Saxena}}]{Xie2003:PRB}
\bibinfo{author}{\bibnamefont{Xie}, \bibfnamefont{S.~J.}},
  \bibinfo{author}{\bibfnamefont{K.~H.} \bibnamefont{Ahn}},
  \bibinfo{author}{\bibfnamefont{D.~L.} \bibnamefont{Smith}},
  \bibinfo{author}{\bibfnamefont{A.~R.} \bibnamefont{Bishop}}, and
  \bibinfo{author}{\bibfnamefont{A.}~\bibnamefont{Saxena}},
  \bibinfo{year}{2003}, {``}\bibinfo{title}{Ground-state properties of
  ferromagnetic metal/conjugated polymer interfaces},{''}
  \bibinfo{journal}{Phys. Rev. B} \textbf{\bibinfo{volume}{67}},
  \bibinfo{pages}{125202}.

\bibitem[{\citenamefont{Xiong} \emph{et~al.}(2004)\citenamefont{Xiong, Wu,
  Vardeny, and Shi}}]{Xiong2004:N}
\bibinfo{author}{\bibnamefont{Xiong}, \bibfnamefont{Z.~H.}},
  \bibinfo{author}{\bibfnamefont{D.}~\bibnamefont{Wu}},
  \bibinfo{author}{\bibfnamefont{Z.~V.} \bibnamefont{Vardeny}}, and
  \bibinfo{author}{\bibfnamefont{J.}~\bibnamefont{Shi}}, \bibinfo{year}{2004},
  {``}\bibinfo{title}{Giant magnetoresistance in organic spin-valves},{''}
  \bibinfo{journal}{{\sl Nature}} \textbf{\bibinfo{volume}{427}},
  \bibinfo{pages}{821--824}.

\bibitem[{\citenamefont{Yafet}(1961)}]{Yafet1961:JPCS}
\bibinfo{author}{\bibnamefont{Yafet}, \bibfnamefont{Y.}}, \bibinfo{year}{1961},
  {``}\bibinfo{title}{Hyperfine interactions due to orbital magnetic moment of
  electrons with large g factors},{''} \bibinfo{journal}{J. Phys. Chem. Solids}
  \textbf{\bibinfo{volume}{21}},  \bibinfo{pages}{99--104}.

\bibitem[{\citenamefont{Yafet}(1963)}]{Yafet:1963}
\bibinfo{author}{\bibnamefont{Yafet}, \bibfnamefont{Y.}}, \bibinfo{year}{1963},
  {``}\bibinfo{title}{g Factors and spin-lattice relaxation of conduction
  electrons},{''} in \emph{\bibinfo{booktitle}{Solid State Physics, Vol. 14}},
  edited by \bibinfo{editor}{\bibfnamefont{F.}~\bibnamefont{Seitz}} and
  \bibinfo{editor}{\bibfnamefont{D.}~\bibnamefont{Turnbull}}
  (\bibinfo{publisher}{Academic, New York}), ~\bibinfo{pages}{2}.

\bibitem[{\citenamefont{Yamanouchi}
  \emph{et~al.}(2004)\citenamefont{Yamanouchi, Chiba, Matsukara, and
  Ohno}}]{Yamanouchi2004:N}
\bibinfo{author}{\bibnamefont{Yamanouchi}, \bibfnamefont{M.}},
  \bibinfo{author}{\bibfnamefont{D.}~\bibnamefont{Chiba}},
  \bibinfo{author}{\bibfnamefont{F.}~\bibnamefont{Matsukara}}, and
  \bibinfo{author}{\bibfnamefont{H.}~\bibnamefont{Ohno}}, \bibinfo{year}{2004},
  {``}\bibinfo{title}{Current-induced domain-wall switching in a ferromagnetic
  semiconductor structure},{''} \bibinfo{journal}{{\sl Nature}}
  \textbf{\bibinfo{volume}{428}},  \bibinfo{pages}{539--541}.

\bibitem[{\citenamefont{Yamashita}
  \emph{et~al.}(2003{\natexlab{a}})\citenamefont{Yamashita, Imamura, Takahashi,
  and Maekawa}}]{Yamashita2003:PRB}
\bibinfo{author}{\bibnamefont{Yamashita}, \bibfnamefont{T.}},
  \bibinfo{author}{\bibfnamefont{H.}~\bibnamefont{Imamura}},
  \bibinfo{author}{\bibfnamefont{S.}~\bibnamefont{Takahashi}}, and
  \bibinfo{author}{\bibfnamefont{S.}~\bibnamefont{Maekawa}},
  \bibinfo{year}{2003}{\natexlab{a}}, {``}\bibinfo{title}{Andreev reflection in
  ferromagnet/superconductor/ferromagnet double junction systems},{''}
  \bibinfo{journal}{Phys. Rev. B} \textbf{\bibinfo{volume}{67}},
  \bibinfo{pages}{094515}.

\bibitem[{\citenamefont{Yamashita} \emph{et~al.}(2002)\citenamefont{Yamashita,
  Takahashi, Imamura, and Maekawa}}]{Yamashita2002:PRB}
\bibinfo{author}{\bibnamefont{Yamashita}, \bibfnamefont{T.}},
  \bibinfo{author}{\bibfnamefont{S.}~\bibnamefont{Takahashi}},
  \bibinfo{author}{\bibfnamefont{H.}~\bibnamefont{Imamura}}, and
  \bibinfo{author}{\bibfnamefont{S.}~\bibnamefont{Maekawa}},
  \bibinfo{year}{2002}, {``}\bibinfo{title}{Spin transport and relaxation in
  superconductors},{''} \bibinfo{journal}{Phys. Rev. B}
  \textbf{\bibinfo{volume}{65}},  \bibinfo{pages}{172509}.

\bibitem[{\citenamefont{Yamashita}
  \emph{et~al.}(2003{\natexlab{b}})\citenamefont{Yamashita, Takahashi, and
  Maekawa}}]{Yamashita2003:P}
\bibinfo{author}{\bibnamefont{Yamashita}, \bibfnamefont{T.}},
  \bibinfo{author}{\bibfnamefont{S.}~\bibnamefont{Takahashi}}, and
  \bibinfo{author}{\bibfnamefont{S.}~\bibnamefont{Maekawa}},
  \bibinfo{year}{2003}{\natexlab{b}}, {``}\bibinfo{title}{Crossed {Andreev}
  reflection in structures consisting of a superconductor with ferromagnetic
  leads},{''} \bibinfo{journal}{Phys. Rev. B} \textbf{\bibinfo{volume}{68}},
  \bibinfo{pages}{174504}.

\bibitem[{\citenamefont{Yamauchi and Mizushima}(1998)}]{Yamauchi1998:PRB}
\bibinfo{author}{\bibnamefont{Yamauchi}, \bibfnamefont{T.}}, and
  \bibinfo{author}{\bibfnamefont{K.}~\bibnamefont{Mizushima}},
  \bibinfo{year}{1998}, {``}\bibinfo{title}{Theoretical approach to the
  spin-dependent hot-electron transport in a spin valve},{''}
  \bibinfo{journal}{Phys. Rev. B} \textbf{\bibinfo{volume}{58}},
  \bibinfo{pages}{1934--1939}.

\bibitem[{\citenamefont{Yan} \emph{et~al.}(2000)\citenamefont{Yan, Zhao, and
  Hu}}]{Yan2000:PRB}
\bibinfo{author}{\bibnamefont{Yan}, \bibfnamefont{X.-Z.}},
  \bibinfo{author}{\bibfnamefont{H.}~\bibnamefont{Zhao}}, and
  \bibinfo{author}{\bibfnamefont{C.-R.} \bibnamefont{Hu}},
  \bibinfo{year}{2000}, {``}\bibinfo{title}{Electron transport in
  normal-metal/superconductor junctions},{''} \bibinfo{journal}{Phys. Rev. B}
  \textbf{\bibinfo{volume}{61}},  \bibinfo{pages}{14759--14764}.

\bibitem[{\citenamefont{Yeh} \emph{et~al.}(1999)\citenamefont{Yeh, Vasquez, Fu,
  A.~V.~Samoilov, and Vakili}}]{Yeh1999:PRB}
\bibinfo{author}{\bibnamefont{Yeh}, \bibfnamefont{N.-C.}},
  \bibinfo{author}{\bibfnamefont{R.~P.} \bibnamefont{Vasquez}},
  \bibinfo{author}{\bibfnamefont{C.~C.} \bibnamefont{Fu}},
  \bibinfo{author}{\bibfnamefont{Y.~L.} \bibnamefont{A.~V.~Samoilov}}, and
  \bibinfo{author}{\bibfnamefont{K.}~\bibnamefont{Vakili}},
  \bibinfo{year}{1999}, {``}\bibinfo{title}{Nonequilibrium superconductivity
  under spin-polarized quasiparticle currents in perovskite
  ferromagnet-insulator-superconductor heterostructures},{''}
  \bibinfo{journal}{Phys. Rev. B} \textbf{\bibinfo{volume}{60}},
  \bibinfo{pages}{10522--10526}.

\bibitem[{\citenamefont{You and Bader}(2000)}]{You2000:JAP}
\bibinfo{author}{\bibnamefont{You}, \bibfnamefont{C.-Y.}}, and
  \bibinfo{author}{\bibfnamefont{S.~D.} \bibnamefont{Bader}},
  \bibinfo{year}{2000}, {``}\bibinfo{title}{Voltage controlled spintronic
  devices for logic applications},{''} \bibinfo{journal}{J. Appl. Phys.}
  \textbf{\bibinfo{volume}{87}},  \bibinfo{pages}{5215--5217}.

\bibitem[{\citenamefont{Young} \emph{et~al.}(2002)\citenamefont{Young,
  {Johnston-Halperin}, Awschalom, Ohno, and Ohno}}]{Young2002:APL}
\bibinfo{author}{\bibnamefont{Young}, \bibfnamefont{D.~K.}},
  \bibinfo{author}{\bibfnamefont{E.}~\bibnamefont{{Johnston-Halperin}}},
  \bibinfo{author}{\bibfnamefont{D.~D.} \bibnamefont{Awschalom}},
  \bibinfo{author}{\bibfnamefont{Y.}~\bibnamefont{Ohno}}, and
  \bibinfo{author}{\bibfnamefont{H.}~\bibnamefont{Ohno}}, \bibinfo{year}{2002},
  {``}\bibinfo{title}{Anisotropic electrical spin injection in ferromagnetic
  semiconductor heterostructures},{''} \bibinfo{journal}{Appl. Phys. Lett.}
  \textbf{\bibinfo{volume}{80}},  \bibinfo{pages}{1598--1600}.

\bibitem[{\citenamefont{Young} \emph{et~al.}(1999)\citenamefont{Young, Hall,
  Torelli, Fisk, Sarrao, Thompson, Ott, Oseroff, Goodrich, and
  Zysler}}]{Young1999:N}
\bibinfo{author}{\bibnamefont{Young}, \bibfnamefont{D.~P.}},
  \bibinfo{author}{\bibfnamefont{D.}~\bibnamefont{Hall}},
  \bibinfo{author}{\bibfnamefont{M.~E.} \bibnamefont{Torelli}},
  \bibinfo{author}{\bibfnamefont{Z.}~\bibnamefont{Fisk}},
  \bibinfo{author}{\bibfnamefont{J.~L.} \bibnamefont{Sarrao}},
  \bibinfo{author}{\bibfnamefont{J.~D.} \bibnamefont{Thompson}},
  \bibinfo{author}{\bibfnamefont{H.-R.} \bibnamefont{Ott}},
  \bibinfo{author}{\bibfnamefont{S.~B.} \bibnamefont{Oseroff}},
  \bibinfo{author}{\bibfnamefont{R.~G.} \bibnamefont{Goodrich}}, and
  \bibinfo{author}{\bibfnamefont{R.}~\bibnamefont{Zysler}},
  \bibinfo{year}{1999}, {``}\bibinfo{title}{High-temperature weak
  ferromagnetism in a low-density free-electron gas},{''}
  \bibinfo{journal}{{\sl Nature}} \textbf{\bibinfo{volume}{397}},
  \bibinfo{pages}{412--414}.

\bibitem[{\citenamefont{Yu and Flatt{\'e}}(2002{\natexlab{a}})}]{Yu2002:PRBa}
\bibinfo{author}{\bibnamefont{Yu}, \bibfnamefont{Z.~G.}}, and
  \bibinfo{author}{\bibfnamefont{M.~E.} \bibnamefont{Flatt{\'e}}},
  \bibinfo{year}{2002}{\natexlab{a}}, {``}\bibinfo{title}{Electric-field
  dependent spin diffusion and spin injection into semiconductors},{''}
  \bibinfo{journal}{Phys. Rev. B} \textbf{\bibinfo{volume}{66}},
  \bibinfo{pages}{201202}.

\bibitem[{\citenamefont{Yu and Flatt{\'e}}(2002{\natexlab{b}})}]{Yu2002:PRBb}
\bibinfo{author}{\bibnamefont{Yu}, \bibfnamefont{Z.~G.}}, and
  \bibinfo{author}{\bibfnamefont{M.~E.} \bibnamefont{Flatt{\'e}}},
  \bibinfo{year}{2002}{\natexlab{b}}, {``}\bibinfo{title}{Spin diffusion and
  injection in semiconductor structures: {Electric} field effects},{''}
  \bibinfo{journal}{Phys. Rev. B} \textbf{\bibinfo{volume}{66}},
  \bibinfo{pages}{235302}.

\bibitem[{\citenamefont{Yuasa} \emph{et~al.}(2002)\citenamefont{Yuasa,
  Nagahama, and Suzuki}}]{Yuasa2002:S}
\bibinfo{author}{\bibnamefont{Yuasa}, \bibfnamefont{S.}},
  \bibinfo{author}{\bibfnamefont{T.}~\bibnamefont{Nagahama}}, and
  \bibinfo{author}{\bibfnamefont{Y.}~\bibnamefont{Suzuki}},
  \bibinfo{year}{2002}, {``}\bibinfo{title}{Spin-polarized resonant tunneling
  in magnetic tunnel junctions},{''} \bibinfo{journal}{{\sl Science}}
  \textbf{\bibinfo{volume}{297}},  \bibinfo{pages}{234--237}.

\bibitem[{\citenamefont{Yusof} \emph{et~al.}(1998)\citenamefont{Yusof,
  Zasadzinski, Coffey, and Miyakawa}}]{Yusof1998:PRB}
\bibinfo{author}{\bibnamefont{Yusof}, \bibfnamefont{Z.}},
  \bibinfo{author}{\bibfnamefont{J.~F.} \bibnamefont{Zasadzinski}},
  \bibinfo{author}{\bibfnamefont{L.}~\bibnamefont{Coffey}}, and
  \bibinfo{author}{\bibfnamefont{N.}~\bibnamefont{Miyakawa}},
  \bibinfo{year}{1998}, {``}\bibinfo{title}{Modeling of tunneling spectroscopy
  in high-{T$_c$} superconductors incorporating band structure, gap symmetry,
  group velocity, and tunneling directionality},{''} \bibinfo{journal}{Phys.
  Rev. B} \textbf{\bibinfo{volume}{58}},  \bibinfo{pages}{514--521}.

\bibitem[{\citenamefont{Zakharchenya}
  \emph{et~al.}(1971)\citenamefont{Zakharchenya, Fleisher, Dzhioev, Veshchunov,
  and Rusanov}}]{Zakharchenya1971:JETPL}
\bibinfo{author}{\bibnamefont{Zakharchenya}, \bibfnamefont{B.~I.}},
  \bibinfo{author}{\bibfnamefont{V.~G.} \bibnamefont{Fleisher}},
  \bibinfo{author}{\bibfnamefont{R.~I.} \bibnamefont{Dzhioev}},
  \bibinfo{author}{\bibfnamefont{Y.~P.} \bibnamefont{Veshchunov}}, and
  \bibinfo{author}{\bibfnamefont{I.~B.} \bibnamefont{Rusanov}},
  \bibinfo{year}{1971}, {``}\bibinfo{title}{Effect of optical orientation of
  electron spins in a {GaAs} crystal},{''} \bibinfo{journal}{Zh. Eksp. Teor.
  Fiz. Pisma Red.} \textbf{\bibinfo{volume}{13}},  \bibinfo{pages}{195--197}
  \bibinfo{note}{[JETP Lett. {\bf 13}, 137-139 (1971)]}.

\bibitem[{\citenamefont{Zawadzki and Pfeffer}(2001)}]{Zawadzki2001:PRB}
\bibinfo{author}{\bibnamefont{Zawadzki}, \bibfnamefont{W.}}, and
  \bibinfo{author}{\bibfnamefont{P.}~\bibnamefont{Pfeffer}},
  \bibinfo{year}{2001}, {``}\bibinfo{title}{Average forces in bound and
  resonant quantum states},{''} \bibinfo{journal}{Phys. Rev. B}
  \textbf{\bibinfo{volume}{64}},  \bibinfo{pages}{235313}.

\bibitem[{\citenamefont{Zeng} \emph{et~al.}(2003)\citenamefont{Zeng, Li, and
  Claro}}]{Zeng2003:P}
\bibinfo{author}{\bibnamefont{Zeng}, \bibfnamefont{Z.~Y.}},
  \bibinfo{author}{\bibfnamefont{B.}~\bibnamefont{Li}}, and
  \bibinfo{author}{\bibfnamefont{F.}~\bibnamefont{Claro}},
  \bibinfo{year}{2003}, {``}\bibinfo{title}{Non-equilibrium {Green's}-function
  approach to electronic transport in hybrid mesoscopic structures},{''}
  \bibinfo{journal}{Phys. Rev. B} \textbf{\bibinfo{volume}{68}},
  \bibinfo{pages}{115319}.

\bibitem[{\citenamefont{Zerrouati} \emph{et~al.}(1988)\citenamefont{Zerrouati,
  Fabre, Bacquet, Frandon, Lampel, and Paget}}]{Zerrouati1988:PRB}
\bibinfo{author}{\bibnamefont{Zerrouati}, \bibfnamefont{K.}},
  \bibinfo{author}{\bibfnamefont{F.}~\bibnamefont{Fabre}},
  \bibinfo{author}{\bibfnamefont{G.}~\bibnamefont{Bacquet}},
  \bibinfo{author}{\bibfnamefont{J.~B.~J.} \bibnamefont{Frandon}},
  \bibinfo{author}{\bibfnamefont{G.}~\bibnamefont{Lampel}}, and
  \bibinfo{author}{\bibfnamefont{D.}~\bibnamefont{Paget}},
  \bibinfo{year}{1988}, {``}\bibinfo{title}{Spin-lattice relaxation in p-type
  gallium arsenide single crystals},{''} \bibinfo{journal}{Phys. Rev. B}
  \textbf{\bibinfo{volume}{37}},  \bibinfo{pages}{1334--1341}.

\bibitem[{\citenamefont{Zhang}(2000)}]{Zhang2000:PRL}
\bibinfo{author}{\bibnamefont{Zhang}, \bibfnamefont{S.}}, \bibinfo{year}{2000},
  {``}\bibinfo{title}{Spin {Hall} effect in the presence of spin
  diffusion},{''} \bibinfo{journal}{Phys. Rev. Lett.}
  \textbf{\bibinfo{volume}{85}},  \bibinfo{pages}{393--396}.

\bibitem[{\citenamefont{Zhang} \emph{et~al.}(1997)\citenamefont{Zhang, Levy,
  Marley, and Parkin}}]{Zhang1997:PRL}
\bibinfo{author}{\bibnamefont{Zhang}, \bibfnamefont{S.}},
  \bibinfo{author}{\bibfnamefont{P.~M.} \bibnamefont{Levy}},
  \bibinfo{author}{\bibfnamefont{A.~C.} \bibnamefont{Marley}}, and
  \bibinfo{author}{\bibfnamefont{S.~S.~P.} \bibnamefont{Parkin}},
  \bibinfo{year}{1997}, {``}\bibinfo{title}{Quenching of magnetoresistance by
  hot electrons in magnetic tunnel junction},{''} \bibinfo{journal}{Phys. Rev.
  Lett.} \textbf{\bibinfo{volume}{79}},  \bibinfo{pages}{3744--3747}.

\bibitem[{\citenamefont{Zhao} \emph{et~al.}(2002)\citenamefont{Zhao, {M\"onch},
  Vinzelberg, {M\"uhl}, and Schneider}}]{Zhao2002:APL}
\bibinfo{author}{\bibnamefont{Zhao}, \bibfnamefont{B.}},
  \bibinfo{author}{\bibfnamefont{I.}~\bibnamefont{{M\"onch}}},
  \bibinfo{author}{\bibfnamefont{H.}~\bibnamefont{Vinzelberg}},
  \bibinfo{author}{\bibfnamefont{T.}~\bibnamefont{{M\"uhl}}}, and
  \bibinfo{author}{\bibfnamefont{C.~M.} \bibnamefont{Schneider}},
  \bibinfo{year}{2002}, {``}\bibinfo{title}{Spin-coherent transport in
  ferromagnetically contacted carbon nanotubes},{''} \bibinfo{journal}{Appl.
  Phys. Lett.} \textbf{\bibinfo{volume}{80}},  \bibinfo{pages}{3144--3146}.

\bibitem[{\citenamefont{Zhao and Hershfield}(1995)}]{Zhao1995:PRB}
\bibinfo{author}{\bibnamefont{Zhao}, \bibfnamefont{H.~L.}}, and
  \bibinfo{author}{\bibfnamefont{S.}~\bibnamefont{Hershfield}},
  \bibinfo{year}{1995}, {``}\bibinfo{title}{Tunneling, relaxation of
  spin-polarized quasiparticles, and spin-charge separation in
  superconductors},{''} \bibinfo{journal}{Phys. Rev. B}
  \textbf{\bibinfo{volume}{52}},  \bibinfo{pages}{3632--3638}.

\bibitem[{\citenamefont{Zheng} \emph{et~al.}(2003)\citenamefont{Zheng, Wu,
  Wang, Wang, f.~Sun, and Guo}}]{Zheng2002:P}
\bibinfo{author}{\bibnamefont{Zheng}, \bibfnamefont{W.}},
  \bibinfo{author}{\bibfnamefont{J.}~\bibnamefont{Wu}},
  \bibinfo{author}{\bibfnamefont{B.}~\bibnamefont{Wang}},
  \bibinfo{author}{\bibfnamefont{J.}~\bibnamefont{Wang}},
  \bibinfo{author}{\bibfnamefont{Q.}~\bibnamefont{f.~Sun}}, and
  \bibinfo{author}{\bibfnamefont{H.}~\bibnamefont{Guo}}, \bibinfo{year}{2003},
  {``}\bibinfo{title}{Parametric quantum spin pump},{''}
  \bibinfo{journal}{Phys. Rev. B} \textbf{\bibinfo{volume}{68}},
  \bibinfo{pages}{113306}.

\bibitem[{\citenamefont{Zhitomirskii}
  \emph{et~al.}(1993)\citenamefont{Zhitomirskii, Kirpichev, Filin, Timofeev,
  Shepel, and {von Klitzing}}}]{Zhitomirskii1993:JETPL}
\bibinfo{author}{\bibnamefont{Zhitomirskii}, \bibfnamefont{V.~E.}},
  \bibinfo{author}{\bibfnamefont{V.~E.} \bibnamefont{Kirpichev}},
  \bibinfo{author}{\bibfnamefont{A.~I.} \bibnamefont{Filin}},
  \bibinfo{author}{\bibfnamefont{V.~B.} \bibnamefont{Timofeev}},
  \bibinfo{author}{\bibfnamefont{B.~N.} \bibnamefont{Shepel}}, and
  \bibinfo{author}{\bibfnamefont{K.}~\bibnamefont{{von Klitzing}}},
  \bibinfo{year}{1993}, {``}\bibinfo{title}{Optical detection of spin
  relaxation of {2D} electrons during photoexcitation},{''}
  \bibinfo{journal}{Zh. Eksp. Teor. Fiz. Pisma Red.}
  \textbf{\bibinfo{volume}{58}},  \bibinfo{pages}{429--434}
  \bibinfo{note}{[JETP Lett. {\bf 58}, 439-444 (1993)]}.

\bibitem[{\citenamefont{Zhu} \emph{et~al.}(2001)\citenamefont{Zhu, Ramsteiner,
  Kostial, Wassermeier, Sch{\"o}nherr, and Ploog}}]{Zhu2001:PRL}
\bibinfo{author}{\bibnamefont{Zhu}, \bibfnamefont{H.~J.}},
  \bibinfo{author}{\bibfnamefont{M.}~\bibnamefont{Ramsteiner}},
  \bibinfo{author}{\bibfnamefont{H.}~\bibnamefont{Kostial}},
  \bibinfo{author}{\bibfnamefont{M.}~\bibnamefont{Wassermeier}},
  \bibinfo{author}{\bibfnamefont{H.-P.} \bibnamefont{Sch{\"o}nherr}}, and
  \bibinfo{author}{\bibfnamefont{K.~H.} \bibnamefont{Ploog}},
  \bibinfo{year}{2001}, {``}\bibinfo{title}{Room-temperature spin injection
  from {Fe} into {GaAs}},{''} \bibinfo{journal}{Phys. Rev. Lett.}
  \textbf{\bibinfo{volume}{87}},  \bibinfo{pages}{016601}.

\bibitem[{\citenamefont{Zhu} \emph{et~al.}(1999)\citenamefont{Zhu, Friedman,
  and Ting}}]{Zhu1999:PRB}
\bibinfo{author}{\bibnamefont{Zhu}, \bibfnamefont{J.-X.}},
  \bibinfo{author}{\bibfnamefont{B.}~\bibnamefont{Friedman}}, and
  \bibinfo{author}{\bibfnamefont{C.~S.} \bibnamefont{Ting}},
  \bibinfo{year}{1999}, {``}\bibinfo{title}{Spin-polarized quasiparticle
  transport in ferromagnet-d-wave-superconductor junctions with a {110}
  interface},{''} \bibinfo{journal}{Phys. Rev. B}
  \textbf{\bibinfo{volume}{59}},  \bibinfo{pages}{9558--9563}.

\bibitem[{\citenamefont{Zhu} \emph{et~al.}(2002)\citenamefont{Zhu, f.~Sun, and
  h.~Lin}}]{Zhu2002:PRB}
\bibinfo{author}{\bibnamefont{Zhu}, \bibfnamefont{Y.}},
  \bibinfo{author}{\bibfnamefont{Q.}~\bibnamefont{f.~Sun}}, and
  \bibinfo{author}{\bibfnamefont{T.}~\bibnamefont{h.~Lin}},
  \bibinfo{year}{2002}, {``}\bibinfo{title}{Andreev reflection through a
  quantum dot coupled with two ferromagnets and a superconductor},{''}
  \bibinfo{journal}{Phys. Rev. B} \textbf{\bibinfo{volume}{65}},
  \bibinfo{pages}{024516}.

\bibitem[{\citenamefont{Ziese and {Thornton (Eds.)}}(2001)}]{Ziese:2001}
\bibinfo{author}{\bibnamefont{Ziese}, \bibfnamefont{M.}}, and
  \bibinfo{author}{\bibfnamefont{M.~J.} \bibnamefont{{Thornton (Eds.)}}},
  \bibinfo{year}{2001}, \emph{\bibinfo{title}{Spin Electronics}}
  (\bibinfo{publisher}{Springer, New York}).

\bibitem[{\citenamefont{Zudov} \emph{et~al.}(2002)\citenamefont{Zudov, Kono,
  Matsuda, Ikaida, Miura, Munekata, Sanders, Sun, and Stanton}}]{Zudov2002:PRB}
\bibinfo{author}{\bibnamefont{Zudov}, \bibfnamefont{M.~A.}},
  \bibinfo{author}{\bibfnamefont{J.}~\bibnamefont{Kono}},
  \bibinfo{author}{\bibfnamefont{Y.~H.} \bibnamefont{Matsuda}},
  \bibinfo{author}{\bibfnamefont{T.}~\bibnamefont{Ikaida}},
  \bibinfo{author}{\bibfnamefont{N.}~\bibnamefont{Miura}},
  \bibinfo{author}{\bibfnamefont{H.}~\bibnamefont{Munekata}},
  \bibinfo{author}{\bibfnamefont{G.~D.} \bibnamefont{Sanders}},
  \bibinfo{author}{\bibfnamefont{Y.}~\bibnamefont{Sun}}, and
  \bibinfo{author}{\bibfnamefont{C.~J.} \bibnamefont{Stanton}},
  \bibinfo{year}{2002}, {``}\bibinfo{title}{Ultrahigh field electron cyclotron
  resonance absorption in {In$_{1-x}$Mn$_x$As} films},{''}
  \bibinfo{journal}{Phys. Rev. B} \textbf{\bibinfo{volume}{66}},
  \bibinfo{pages}{161307}.

\bibitem[{\citenamefont{Zurek}(2003)}]{Zurek2003:RMP}
\bibinfo{author}{\bibnamefont{Zurek}, \bibfnamefont{W.~H.}},
  \bibinfo{year}{2003}, {``}\bibinfo{title}{Decoherence, einselection, and the
  quantum origins of the classical},{''} \bibinfo{journal}{Rev. Mod. Phys.}
  \textbf{\bibinfo{volume}{75}},  \bibinfo{pages}{715--775}.

\bibitem[{\citenamefont{{\v{Z}uti\'{c}}}(2002)}]{Zutic2002:JS}
\bibinfo{author}{\bibnamefont{{\v{Z}uti\'{c}}}, \bibfnamefont{I.}},
  \bibinfo{year}{2002}, {``}\bibinfo{title}{Novel aspects of spin-polarized
  transport and spin dynamics},{''} \bibinfo{journal}{J. Supercond.}
  \textbf{\bibinfo{volume}{15}},  \bibinfo{pages}{5--12}.

\bibitem[{\citenamefont{{\v{Z}uti\'c} and {Das Sarma}}(1999)}]{Zutic1999:PRBa}
\bibinfo{author}{\bibnamefont{{\v{Z}uti\'c}}, \bibfnamefont{I.}}, and
  \bibinfo{author}{\bibfnamefont{S.}~\bibnamefont{{Das Sarma}}},
  \bibinfo{year}{1999}, {``}\bibinfo{title}{Spin-polarized transport and
  {Andreev} reflection in semiconductor/superconductor hybrid structures},{''}
  \bibinfo{journal}{Phys. Rev. B} \textbf{\bibinfo{volume}{60}},
  \bibinfo{pages}{R16322--R16325}.

\bibitem[{\citenamefont{{\v{Z}uti\'{c}} and Fabian}(2003)}]{Zutic2003:P}
\bibinfo{author}{\bibnamefont{{\v{Z}uti\'{c}}}, \bibfnamefont{I.}}, and
  \bibinfo{author}{\bibfnamefont{J.}~\bibnamefont{Fabian}},
  \bibinfo{year}{2003}, {``}\bibinfo{title}{Spin-voltaic effect and its
  Implications},{''} \bibinfo{journal}{Mater. Trans., JIM}
  \textbf{\bibinfo{volume}{44}},  \bibinfo{pages}{2062--2065}.

\bibitem[{\citenamefont{{\v{Z}uti\'{c}}}
  \emph{et~al.}(2001{\natexlab{a}})\citenamefont{{\v{Z}uti\'{c}}, Fabian, and
  {Das Sarma}}}]{Zutic2001:APL}
\bibinfo{author}{\bibnamefont{{\v{Z}uti\'{c}}}, \bibfnamefont{I.}},
  \bibinfo{author}{\bibfnamefont{J.}~\bibnamefont{Fabian}}, and
  \bibinfo{author}{\bibfnamefont{S.}~\bibnamefont{{Das Sarma}}},
  \bibinfo{year}{2001}{\natexlab{a}}, {``}\bibinfo{title}{A proposal for a
  spin-polarized solar battery},{''} \bibinfo{journal}{Appl. Phys. Lett.}
  \textbf{\bibinfo{volume}{79}},  \bibinfo{pages}{1558--1560}.

\bibitem[{\citenamefont{{\v{Z}uti\'{c}}}
  \emph{et~al.}(2001{\natexlab{b}})\citenamefont{{\v{Z}uti\'{c}}, Fabian, and
  {Das Sarma}}}]{Zutic2001:PRB}
\bibinfo{author}{\bibnamefont{{\v{Z}uti\'{c}}}, \bibfnamefont{I.}},
  \bibinfo{author}{\bibfnamefont{J.}~\bibnamefont{Fabian}}, and
  \bibinfo{author}{\bibfnamefont{S.}~\bibnamefont{{Das Sarma}}},
  \bibinfo{year}{2001}{\natexlab{b}}, {``}\bibinfo{title}{Spin injection
  through the depletion layer: a theory of spin-polarized p-n junctions and
  solar cells},{''} \bibinfo{journal}{Phys. Rev. B}
  \textbf{\bibinfo{volume}{64}},  \bibinfo{pages}{121201}.

\bibitem[{\citenamefont{{\v{Z}uti\'{c}}}
  \emph{et~al.}(2002)\citenamefont{{\v{Z}uti\'{c}}, Fabian, and {Das
  Sarma}}}]{Zutic2002:PRL}
\bibinfo{author}{\bibnamefont{{\v{Z}uti\'{c}}}, \bibfnamefont{I.}},
  \bibinfo{author}{\bibfnamefont{J.}~\bibnamefont{Fabian}}, and
  \bibinfo{author}{\bibfnamefont{S.}~\bibnamefont{{Das Sarma}}},
  \bibinfo{year}{2002}, {``}\bibinfo{title}{Spin-polarized transport in
  inhomogeneous magnetic semiconductors: theory of magnetic/nonmagnetic p-n
  junctions},{''} \bibinfo{journal}{Phys. Rev. Lett.}
  \textbf{\bibinfo{volume}{88}},  \bibinfo{pages}{066603}.

\bibitem[{\citenamefont{{\v{Z}uti\'{c}}}
  \emph{et~al.}(2003)\citenamefont{{\v{Z}uti\'{c}}, Fabian, and {Das
  Sarma}}}]{Zutic2003:APL}
\bibinfo{author}{\bibnamefont{{\v{Z}uti\'{c}}}, \bibfnamefont{I.}},
  \bibinfo{author}{\bibfnamefont{J.}~\bibnamefont{Fabian}}, and
  \bibinfo{author}{\bibfnamefont{S.}~\bibnamefont{{Das Sarma}}},
  \bibinfo{year}{2003}, {``}\bibinfo{title}{Proposal for all-electrical
  measurement of {T$_1$} in semiconductors},{''} \bibinfo{journal}{Appl. Phys.
  Lett.} \textbf{\bibinfo{volume}{82}},  \bibinfo{pages}{221--223}.

\bibitem[{\citenamefont{{\v{Z}uti\'c} and Valls}(1999)}]{Zutic1999:PRBb}
\bibinfo{author}{\bibnamefont{{\v{Z}uti\'c}}, \bibfnamefont{I.}}, and
  \bibinfo{author}{\bibfnamefont{O.~T.} \bibnamefont{Valls}},
  \bibinfo{year}{1999}, {``}\bibinfo{title}{Spin polarized tunneling in
  ferromagnet/unconventional superconductor junctions},{''}
  \bibinfo{journal}{Phys. Rev. B} \textbf{\bibinfo{volume}{60}},
  \bibinfo{pages}{6320--6323}.

\bibitem[{\citenamefont{{\v{Z}uti\'c} and Valls}(2000)}]{Zutic2000:PRB}
\bibinfo{author}{\bibnamefont{{\v{Z}uti\'c}}, \bibfnamefont{I.}}, and
  \bibinfo{author}{\bibfnamefont{O.~T.} \bibnamefont{Valls}},
  \bibinfo{year}{2000}, {``}\bibinfo{title}{Tunneling spectroscopy for
  ferromagnet/superconductor junctions},{''} \bibinfo{journal}{Phys. Rev. B}
  \textbf{\bibinfo{volume}{61}},  \bibinfo{pages}{1555--1566}.

\bibitem[{\citenamefont{{\v{Z}uti\'c (Ed.)}}(2002)}]{Zutic2002:PROC}
\bibinfo{author}{\bibnamefont{{\v{Z}uti\'c (Ed.)}}, \bibfnamefont{I.}},
  \bibinfo{year}{2002}, {``}\bibinfo{title}{{in Proceedings of Spintronics
  2001} {International Conference on Novel Aspects of Spin-Polarized Transport
  and Spin Dynamics, Washington, D.C.}},{''} \bibinfo{journal}{J. Supercond.}
  \textbf{\bibinfo{volume}{15}},  \bibinfo{pages}{1--104}.

\bibitem[{\citenamefont{Zvezdin and Kotov}(1997)}]{Zvezdin:2003}
\bibinfo{author}{\bibnamefont{Zvezdin}, \bibfnamefont{A.~K.}}, and
  \bibinfo{author}{\bibfnamefont{V.~A.} \bibnamefont{Kotov}},
  \bibinfo{year}{1997}, \emph{\bibinfo{title}{Modern Magnetooptics and
  Magnetooptical Materials}} (\bibinfo{publisher}{Institute of Physics,
  Bristol/ Philadelphia}).

\bibitem[{\citenamefont{Zvezdin} \emph{et~al.}(2003)\citenamefont{Zvezdin,
  Mishchenko, and Khval'kovskii}}]{Zvezdin2003:TP}
\bibinfo{author}{\bibnamefont{Zvezdin}, \bibfnamefont{A.~K.}},
  \bibinfo{author}{\bibfnamefont{A.~S.} \bibnamefont{Mishchenko}}, and
  \bibinfo{author}{\bibfnamefont{A.~V.} \bibnamefont{Khval'kovskii}},
  \bibinfo{year}{2003}, {``}\bibinfo{title}{Current-voltage charactersitic of a
  spin half-metallic transistor},{''} \bibinfo{journal}{Zh. Tekhn. Fiz.}
  \textbf{\bibinfo{volume}{48}},  \bibinfo{pages}{53--58} \bibinfo{note}{[Tech.
  Phys. {\bf 4}, 431-436 (2003)]}.

\bibitem[{\citenamefont{Zwierzycki}
  \emph{et~al.}(2003)\citenamefont{Zwierzycki, Xia, Kelly, Bauer, and
  Turek}}]{Zwierzycki2003:PRB}
\bibinfo{author}{\bibnamefont{Zwierzycki}, \bibfnamefont{M.}},
  \bibinfo{author}{\bibfnamefont{K.}~\bibnamefont{Xia}},
  \bibinfo{author}{\bibfnamefont{P.~J.} \bibnamefont{Kelly}},
  \bibinfo{author}{\bibfnamefont{G.~E.~W.} \bibnamefont{Bauer}}, and
  \bibinfo{author}{\bibfnamefont{I.}~\bibnamefont{Turek}},
  \bibinfo{year}{2003}, {``}\bibinfo{title}{Spin injection through an {Fe/InAs}
  interface},{''} \bibinfo{journal}{Phys. Rev. B}
  \textbf{\bibinfo{volume}{67}},  \bibinfo{pages}{092401}.

\bibitem[{\citenamefont{Zwolak and {Di Ventra}}(2002)}]{Zwolak2002:APL}
\bibinfo{author}{\bibnamefont{Zwolak}, \bibfnamefont{M.}}, and
  \bibinfo{author}{\bibfnamefont{M.}~\bibnamefont{{Di Ventra}}},
  \bibinfo{year}{2002}, {``}\bibinfo{title}{{DNA} spintronics},{''}
  \bibinfo{journal}{Appl. Phys. Lett.} \textbf{\bibinfo{volume}{81}},
  \bibinfo{pages}{925--927}.

\end{thebibliography}

\end{document}